# SUPERSPACE

## *or* One thousand and one lessons in supersymmetry

**S. James Gates, Jr.**

*Massachusetts Institute of Technology, Cambridge, Massachusetts*
(Present address: *University of Maryland, College Park, Maryland)*
gatess@wam.umd.edu

**Marcus T. Grisaru**

*Brandeis University, Waltham, Massachusetts*
(Present address: *McGill University, Montreal, Quebec)*
grisaru@physics.mcgill.ca

**Martin Roček**

*State University of New York, Stony Brook, New York*
rocek@insti.physics.sunysb.edu

**Warren Siegel**

*University of California, Berkeley, California*
(Present address: *State University of New York)*
warren@wcgall.physics.sunysb.edu

*Superspace is the greatest invention since the wheel [1] .*

**Preface**

Said $\Psi$ to $\Phi$, $\Xi$, and $\Upsilon$: "Let's write a review paper." Said $\Phi$ and $\Xi$: "Great idea!" Said $\Upsilon$: "Naaa."

But a few days later $\Upsilon$ had produced a table of contents with 1001 items.

$\Xi$, $\Phi$, $\Psi$, and $\Upsilon$ wrote. Then didn't write. Then wrote again. The review grew; and grew; and grew. It became an outline for a book; it became a first draft; it became a second draft. It became a burden. It became agony. Tempers were lost; and hairs; and a few pounds (alas, quickly regained). They argued about ";" vs. ".", about "which" vs. "that", "˜" vs. "ˆ", "$\gamma$" vs. "$\Gamma$", "+" vs. "-". Made bad puns, drew pictures on the blackboard, were rude to their colleagues, neglected their duties. Bemoaned the paucity of letters in the Greek and Roman alphabets, of hours in the day, days in the week, weeks in the month. $\Xi$, $\Phi$, $\Psi$ and $\Upsilon$ wrote and wrote.

* * *

This must stop; we want to get back to research, to our families, friends and students. We want to look at the sky again, go for walks, sleep at night. Write a second volume? Never! Well, in a couple of years?

We beg our readers' indulgence. We have tried to present a subject that we like, that we think is important. We have tried to present our insights, our tools and our knowledge. Along the way, some errors and misconceptions have without doubt slipped in. There must be wrong statements, misprints, mistakes, awkward phrases, islands of incomprehensibility (but they started out as continents!). We could, probably we should, improve and improve. But we can no longer wait. Like climbers within sight of the summit we are rushing, casting aside caution, reaching towards the moment when we can shout "it's behind us".

This is not a polished work. Without doubt some topics are treated better elsewhere. Without doubt we have left out topics that should have been included. Without doubt we have treated the subject from a personal point of view, emphasizing aspects that we are familiar with, and neglecting some that would have required studying others' work. Nevertheless, we hope this book will be useful, both to those new to the subject and to those who helped develop it. We have presented many topics that are not available elsewhere, and many topics of interest also outside supersymmetry. We have

[1]. A. Oop, A supersymmetric version of the leg, Gondwanaland predraw (January 10,000,000 B.C.), to be discovered.

included topics whose treatment is incomplete, and presented conclusions that are really only conjectures. In some cases, this reflects the state of the subject. Filling in the holes and proving the conjectures may be good research projects.

Supersymmetry is the creation of many talented physicists. We would like to thank all our friends in the field, we have many, for their contributions to the subject, and beg their pardon for not presenting a list of references to their papers.

Most of the work on this book was done while the four of us were at the California Institute of Technology, during the 1982-83 academic year. We would like to thank the Institute and the Physics Department for their hospitality and the use of their computer facilities, the NSF, DOE, the Fleischmann Foundation and the Fairchild Visiting Scholars Program for their support. Some of the work was done while M.T.G. and M.R. were visiting the Institute for Theoretical Physics at Santa Barbara. Finally, we would like to thank Richard Grisaru for the many hours he devoted to typing the equations in this book, Hyun Jean Kim for drawing the diagrams, and Anders Karlhede for carefully reading large parts of the manuscript and for his useful suggestions; and all the others who helped us.

S.J.G., M.T.G., M.R., W.D.S.

Pasadena, January 1983

**Contents**

# 1. INTRODUCTION

*There is a fifth dimension beyond that which is known to man. It is a dimension as vast as space and as timeless as infinity. It is the middle ground between light and shadow, between science and superstition; and it lies between the pit of man's fears and the summit of his knowledge. This is the dimension of imagination. It is an area which we call, "the Twilight Zone."*

Rod Serling

## 1001: A superspace odyssey

Symmetry principles, both global and local, are a fundamental feature of modern particle physics. At the classical and phenomenological level, global symmetries account for many of the (approximate) regularities we observe in nature, while local (gauge) symmetries "explain" and unify the interactions of the basic constituents of matter. At the quantum level symmetries (via Ward identities) facilitate the study of the ultraviolet behavior of field theory models and their renormalization. In particular, the construction of models with local (internal) Yang-Mills symmetry that are asymptotically free has increased enormously our understanding of the quantum behavior of matter at short distances. If this understanding could be extended to the quantum behavior of gravitational interactions (quantum gravity) we would be close to a satisfactory description of micronature in terms of basic fermionic constituents forming multiplets of some unification group, and bosonic gauge particles responsible for their interactions. Even more satisfactory would be the existence in nature of a symmetry which unifies the bosons and the fermions, the constituents and the forces, into a single entity.

Supersymmetry is the supreme symmetry: It unifies spacetime symmetries with internal symmetries, fermions with bosons, and (local supersymmetry) gravity with matter. Under quite general assumptions it is the largest possible symmetry of the S-matrix. At the quantum level, renormalizable globally supersymmetric models exhibit improved ultraviolet behavior: Because of cancellations between fermionic and bosonic contributions quadratic divergences are absent; some supersymmetric models, in particular maximally extended super-Yang-Mills theory, are the only known examples of four-dimensional field theories that are finite to all orders of perturbation theory. Locally supersymmetric gravity (supergravity) may be the only way in which nature can reconcile Einstein gravity and quantum theory. Although we do not know at present if it is a finite theory, quantum supergravity does exhibit less divergent short distance behavior than ordinary quantum gravity. Outside the realm of standard quantum field theory, it is believed that the only reasonable string theories (i.e., those with fermions and without quantum inconsistencies) are supersymmetric; these include models that may be finite (the maximally supersymmetric theories).

At the present time there is no direct experimental evidence that supersymmetry is a fundamental symmetry of nature, but the current level of activity in the field indicates that many physicists share our belief that such evidence will eventually emerge. On the theoretical side, the symmetry makes it possible to build models with (super)natural hierarchies. On esthetic grounds, the idea of a superunified theory is very appealing. Even if supersymmetry and supergravity are not the ultimate theory, their study has increased our understanding of classical and quantum field theory, and they may be an important step in the understanding of some yet unknown, correct theory of nature.

We mean by (Poincaré) supersymmetry an extension of ordinary spacetime symmetries obtained by adjoining $N$ spinorial generators $Q$ whose *anticommutator* yields a translation generator: $\{ Q \, , Q \} = P$. This symmetry can be realized on ordinary fields (functions of spacetime) by transformations that mix bosons and fermions. Such realizations suffice to study supersymmetry (one can write invariant actions, etc.) but are as cumbersome and inconvenient as doing vector calculus component by component. A compact alternative to this "component field" approach is given by the *superspace--superfield* approach. Superspace is an extension of ordinary spacetime to include extra *anticommuting* coordinates in the form of $N$ two-component Weyl spinors $\theta$. Superfields $\Psi(x, \theta)$ are functions defined over this space. They can be expanded in a Taylor series with respect to the anticommuting coordinates $\theta$; because the square of an anticommuting quantity vanishes, this series has only a finite number of terms. The coefficients obtained in this way are the ordinary component fields mentioned above. In superspace, supersymmetry is manifest: The supersymmetry algebra is represented by translations and rotations involving both the spacetime and the anticommuting coordinates. The transformations of the component fields follow from the Taylor expansion of the translated and rotated superfields. In particular, the transformations mixing bosons and fermions are constant translations of the $\theta$ coordinates, and related rotations of $\theta$ into the spacetime coordinate $x$.

A further advantage of superfields is that they automatically include, in addition to the dynamical degrees of freedom, certain unphysical fields: (1) auxiliary fields (fields with nonderivative kinetic terms), needed classically for the off-shell closure of the supersymmetry algebra, and (2) compensating fields (fields that consist entirely of gauge degrees of freedom), which are used to enlarge the usual gauge transformations to an entire multiplet of transformations forming a representation of supersymmetry; together with the auxiliary fields, they allow the algebra to be field independent. The compensators are particularly important for quantization, since they permit the use of supersymmetric gauges, ghosts, Feynman graphs, and supersymmetric power-counting.

Unfortunately, our present knowledge of off-shell *extended* ($N > 1$) supersymmetry is so limited that for most extended theories these unphysical fields, and thus also the corresponding superfields, are unknown. One could hope to *find* the unphysical components directly from superspace; the essential difficulty is that, in general, a superfield is a highly reducible representation of the supersymmetry algebra, and the problem becomes one of finding *which* representations permit the construction of consistent local actions. Therefore, except when discussing the features which are common to general superspace, we restrict ourselves *in this volume* to a discussion of *simple* ($N = 1$) superfield supersymmetry. We hope to treat extended superspace and other topics that need further development in a second (and hopefully last) volume.

We introduce superfields in chapter 2 for the simpler world of three spacetime dimensions, where superfields are very similar to ordinary fields. We skip the discussion of nonsuperspace topics (background fields, gravity, etc.) which are covered in following chapters, and concentrate on a pedagogical treatment of superspace. We return to four dimensions in chapter 3, where we describe how supersymmetry is represented on superfields, and discuss all general properties of free superfields (and their relation to ordinary fields). In chapter 4 we discuss simple ($N = 1$) superfields in classical global supersymmetry. We include such topics as gauge-covariant derivatives, supersymmetric models, extended supersymmetry with unextended superfields, and superforms. In chapter 5 we extend the discussion to local supersymmetry (supergravity), relying heavily on the compensator approach. We discuss prepotentials and covariant derivatives, the construction of actions, and show how to go from superspace to component results. The quantum aspects of global theories is the topic of chapter 6, which includes a discussion of the background field formalism, supersymmetric regularization, anomalies, and many examples of supergraph calculations. In chapter 7 we make the corresponding analysis of quantum supergravity, including many of the novel features of the quantization procedure (various types of ghosts). Chapter 8 describes supersymmetry breaking, explicit and spontaneous, including the superHiggs mechanism and the use of nonlinear realizations.

We have not discussed component supersymmetry and supergravity, realistic superGUT models with or without supergravity, and some of the geometrical aspects of classical supergravity. For the first topic the reader may consult many of the excellent reviews and lecture notes. The second is one of the current areas of active research. It is our belief that superspace methods eventually will provide a framework for streamlining the phenomenology, once we have better control of our tools. The third topic is attracting increased attention, but there are still many issues to be settled; there again, superspace methods should prove useful.

We assume the reader has a knowledge of standard quantum field theory (sufficient to do Feynman graph calculations in QCD). We have tried to make this book as pedagogical and encyclopedic as possible, but have omitted some straightforward algebraic details which are left to the reader as (necessary!) exercises.

## A hitchhiker's guide

We are hoping, of course, that this book will be of interest to many people, with different interests and backgrounds. The graduate student who has completed a course in quantum field theory and wants to study superspace should:

(1) *Read* an article or two reviewing component global supersymmetry and supergravity.

(2) *Read* chapter 2 for a quick and easy (?) introduction to superspace. Sections 1, 2, and 3 are straightforward. Section 4 introduces, in a simple setting, the concept of constrained covariant derivatives, and the solution of the constraints in terms of prepotentials. Section 5 could be skipped at first reading. Section 6 does for supergravity what section 4 did for Yang-Mills; superfield supergravity in three dimensions is deceptively simple. Section 7 introduces quantization and Feynman rules in a simpler situation than in four dimensions.

(3) *Study* subsections 3.2.a-d on supersymmetry algebras, and sections 3.3.a, 3.3.b.1-b.3, 3.4.a,b, 3.5 and 3.6 on superfields, covariant derivatives, and component expansions. *Study* section 3.10 on compensators; we use them extensively in supergravity.

(4) *Study* section 4.1a on the scalar multiplet, and sections 4.2 and 4.3 on gauge theories, their prepotentials, covariant derivatives and solution of the constraints. A *reading* of sections 4.4.b, 4.4.c.1, 4.5.a and 4.5.e might be profitable.

(5) *Take a deep breath* and *slowly* study section 5.1, which is our favorite approach to gravity, and sections 5.2 to 5.5 on supergravity; this is where the action is. For an inductive approach that starts with the prepotentials and constructs the covariant derivatives section 5.2 is sufficient, and one can then go directly to section 5.5. Alternatively, one could start with section 5.3, and a deductive approach based on constrained covariant derivatives, go through section 5.4 and again end at 5.5.

(6) *Study* sections 6.1 through 6.4 on quantization and supergraphs. The topics in these sections should be fairly accessible.

(7) *Study* sections 8.1-8.4.

(8) Go back to the beginning and *skip nothing* this time.

Our particle physics colleagues who are familiar with global superspace should *skim 3.1 for notation, 3.4-6 and 4.1, read 4.2 (no, you don't know it all), and get busy on chapter 5.*

The experts should look for serious mistakes. We would appreciate hearing about them.

### A brief guide to the literature

A complete list of references is becoming increasingly difficult to compile, and we have not attempted to do so. However, the following (incomplete!) list of review articles and proceedings of various schools and conferences, and the references therein, are useful and should provide easy access to the journal literature:

For global supersymmetry, the standard review articles are:

P. Fayet and S. Ferrara, Supersymmetry, Physics Reports 32C (1977) 250.

A. Salam and J. Strathdee, Fortschritte der Physik, 26 (1978) 5.

For component supergravity, the standard review is

P. van Nieuwenhuizen, Supergravity, Physics Reports 68 (1981) 189.

The following Proceedings contain extensive and up-to-date lectures on many supersymmetry and supergravity topics:

"Recent Developments in Gravitation" (Cargesè 1978), eds. M. Levy and S. Deser, Plenum Press, N.Y.

"Supergravity" (Stony Brook 1979), eds. D. Z. Freedman and P. van Nieuwenhuizen, North-Holland, Amsterdam.

"Topics in Quantum Field Theory and Gauge Theories" (Salamanca), Phys. 77, Springer Verlag, Berlin.

"Superspace and Supergravity"(Cambridge 1980), eds. S. W. Hawking and M. Roček, Cambridge University Press, Cambridge.

"Supersymmetry and Supergravity '81" (Trieste), eds. S. Ferrara, J. G. Taylor and P. van Nieuwenhuizen, Cambridge University Press, Cambridge.

"Supersymmetry and Supergravity '82" (Trieste), eds. S. Ferrara, J. G. Taylor and P. van Nieuwenhuizen, World Scientific Publishing Co., Singapore.

**Contents of 2. A TOY SUPERSPACE**

# 2. A TOY SUPERSPACE

## 2.1. Notation and conventions

This chapter presents a self-contained treatment of supersymmetry in three spacetime dimensions. Our main motivation for considering this case is simplicity. Irreducible representations of simple ($N = 1$) global supersymmetry are easier to obtain than in four dimensions: Scalar superfields (single, real functions of the superspace coordinates) provide one such representation, and all others are obtained by appending Lorentz or internal symmetry indices. In addition, the description of local supersymmetry (supergravity) is easier.

### a. Index conventions

Our three-dimensional notation is as follows: In three-dimensional spacetime (with signature $-++$) the Lorentz group is $SL(2,R)$ (instead of $SL(2,C)$) and the corresponding fundamental representation acts on a *real* (Majorana) two-component spinor $\psi^\alpha = (\psi^+ , \psi^-)$. In general we use spinor notation for all Lorentz representations, denoting spinor indices by Greek letters $\alpha, \beta, \cdots, \mu, \nu, \cdots$. Thus a vector (the three-dimensional representation) will be described by a symmetric second-rank spinor $V^{\alpha\beta} = (V^{++} , V^{+-} , V^{--})$ or a traceless second-rank spinor $V_\alpha{}^\beta$. (For comparison, in four dimensions we have spinors $\psi^\alpha$, $\bar\psi^{\dot\alpha}$ and a vector is given by a hermitian matrix $V^{\alpha\dot\beta}$.) *All our spinors will be anticommuting (Grassmann).*

Spinor indices are raised and lowered by the second-rank antisymmetric symbol $C_{\alpha\beta}$, which is also used to define the "square" of a spinor:

$$C_{\alpha\beta} = -\,C_{\beta\alpha} = \begin{pmatrix} 0 & -i \\ i & 0 \end{pmatrix} = -\,C^{\alpha\beta} \quad , \qquad C_{\alpha\beta}C^{\gamma\delta} = \delta_{[\alpha}{}^\gamma \delta_{\beta]}{}^\delta \equiv \delta_\alpha{}^\gamma \delta_\beta{}^\delta - \delta_\beta{}^\gamma \delta_\alpha{}^\delta \quad ;$$

$$\psi_\alpha = \psi^\beta C_{\beta\alpha} \ , \quad \psi^\alpha = C^{\alpha\beta}\psi_\beta \quad , \qquad \psi^2 = \frac{1}{2}\,\psi^\alpha\psi_\alpha = i\psi^+\psi^- \quad . \tag{2.1.1}$$

We represent symmetrization and antisymmetrization of $n$ indices by ( ) and [ ], respectively (without a factor of $\frac{1}{n!}$). We often make use of the identity

$$A_{[\alpha}\, B_{\beta]} = -\,C_{\alpha\beta}\, A^\gamma\, B_\gamma \quad , \tag{2.1.2}$$

which follows from (2.1.1). We use $C_{\alpha\beta}$ (instead of the customary real $\epsilon_{\alpha\beta}$) to simplify the rules for hermitian conjugation. In particular, it makes $\psi^2$ hermitian (recall $\psi^\alpha$ and $\psi_\alpha$ anticommute) and gives the conventional hermiticity properties to derivatives (see below). Note however that whereas $\psi^\alpha$ is real, $\psi_\alpha$ is imaginary.

**b. Superspace**

Superspace for simple supersymmetry is labeled by three spacetime coordinates $x^{\mu\nu}$ and two anticommuting spinor coordinates $\theta^\mu$, denoted collectively by $z^M = (x^{\mu\nu}, \theta^\mu)$. They have the hermiticity properties $(z^M)^\dagger = z^M$. We define derivatives by

$$\partial_\mu \theta^\nu \equiv \{\partial_\mu, \theta^\nu\} \equiv \delta_\mu{}^\nu \quad ,$$

$$\partial_{\mu\nu} x^{\sigma\tau} \equiv [\partial_{\mu\nu}, x^{\sigma\tau}] \equiv \frac{1}{2}\,\delta_{(\mu}{}^\sigma \delta_{\nu)}{}^\tau \quad , \tag{2.1.3a}$$

so that the "momentum" operators have the hermiticity properties

$$(i\partial_\mu)^\dagger = -\,(i\partial_\mu)\ ,\ \ (i\partial_{\mu\nu})^\dagger = +\,(i\partial_{\mu\nu}) \quad . \tag{2.1.3b}$$

and thus $(i\partial^M)^\dagger = i\partial^M$. (Definite) integration over a single anticommuting variable $\gamma$ is defined so that the integral is translationally invariant (see sec. 3.7); hence $\int d\gamma\, 1 = 0$ , $\int d\gamma\, \gamma =$ a constant which we take to be 1. We observe that a function $f(\gamma)$ has a terminating Taylor series $f(\gamma) = f(0) + \gamma\, f'(0)$ since $\{\gamma\,, \gamma\} = 0$ implies $\gamma^2 = 0$. Thus $\int d\gamma\, f(\gamma) = f'(0)$ so that integration is equivalent to differentiation. For our spinorial coordinates $\int d\theta_\alpha = \partial_\alpha$ and hence

$$\int d\theta_\alpha\, \theta^\beta = \delta_\alpha{}^\beta \quad . \tag{2.1.4}$$

Therefore the double integral

$$\int d^2\theta\, \theta^2 = -\,1 \quad , \tag{2.1.5}$$

and we can define the $\delta$-function $\delta^2(\theta) = -\,\theta^2 = -\,\frac{1}{2}\,\theta^\alpha\,\theta_\alpha$.

* * *

We often use the notation $X|$ to indicate the quantity $X$ evaluated at $\theta = 0$.

## 2.2. Supersymmetry and superfields

### a. Representations

We define functions over superspace: $\Phi_{...}(x, \theta)$ where the dots stand for Lorentz (spinor) and/or internal symmetry indices. They transform in the usual way under the Poincaré group with generators $P_{\mu\nu}$ (translations) and $M_{\alpha\beta}$ (Lorentz rotations). We grade (or make super) the Poincaré algebra by introducing additional *spinor* supersymmetry generators $Q_\alpha$, satisfying the *supersymmetry algebra*

$$[P_{\mu\nu}\,, P_{\rho\sigma}] = 0 \quad , \tag{2.2.1a}$$

$$\{Q_\mu\,, Q_\nu\} = 2\,P_{\mu\nu} \quad , \tag{2.2.1b}$$

$$[Q_\mu\,, P_{\nu\rho}] = 0 \quad , \tag{2.2.1c}$$

as well as the usual commutation relations with $M_{\alpha\beta}$. This algebra is realized on *superfields* $\Phi_{...}(x\,, \theta)$ in terms of derivatives by:

$$P_{\mu\nu} = i\partial_{\mu\nu} \quad , \qquad Q_\mu = i(\partial_\mu - \theta^\nu i\partial_{\nu\mu}) \quad ; \tag{2.2.2a}$$

$$\psi(x^{\mu\nu}, \theta^\mu) = exp[i(\xi^{\lambda\rho}P_{\lambda\rho} + \epsilon^\lambda Q_\lambda)]\psi(x^{\mu\nu} + \xi^{\mu\nu} - \frac{i}{2}\,\epsilon^{(\mu}\theta^{\nu)}, \theta^\mu + \epsilon^\mu) \;. \tag{2.2.2b}$$

Thus $\xi^{\lambda\rho}P_{\lambda\rho} + \epsilon^\lambda Q_\lambda$ generates a supercoordinate transformation

$$x'^{\mu\nu} = x^{\mu\nu} + \xi^{\mu\nu} - \frac{i}{2}\,\epsilon^{(\mu}\theta^{\nu)} \quad , \qquad \theta'^\mu = \theta^\mu + \epsilon^\mu \quad . \tag{2.2.2c}$$

with real, constant parameters $\xi^{\lambda\rho}\,, \epsilon^\lambda$.

The reader can verify that (2.2.2) provides a representation of the algebra (2.2.1). We remark in particular that if the anticommutator (2.2.1b) vanished, $Q_\mu$ would annihilate all physical states (see sec. 3.3). We also note that because of (2.2.1a,c) and (2.2.2a), not only $\Phi$ and functions of $\Phi$, but also the space-time derivatives $\partial_{\mu\nu}\Phi$ carry a representation of supersymmetry (are superfields). However, because of (2.2.2a), this is not the case for the spinorial derivatives $\partial_\mu\Phi$. Supersymmetrically invariant derivatives can be defined by

$$D_M = (D_{\mu\nu}\,, D_\mu) = (\partial_{\mu\nu}\,, \partial_\mu + \theta^\nu\, i\,\partial_{\mu\nu}) \quad . \tag{2.2.3}$$

The set $D_M$ (anti)commutes with the generators: $[D_M\,, P_{\mu\nu}] = [D_M\,, Q_\nu\} = 0$. We use $[A\,, B\}$ to denote a graded commutator: anticommutator if both $A$ and $B$ are fermionic, commutator otherwise.

The covariant derivatives can also be defined by their graded commutation relations

$$\{D_\mu\,, D_\nu\} = 2iD_{\mu\nu} \quad , \quad [D_\mu\,, D_{\nu\sigma}] = [D_{\mu\nu}\,, D_{\sigma\tau}] = 0 \quad ; \tag{2.2.4}$$

or, more concisely:

$$[D_M\,, D_N\} = T_{MN}{}^P D_P \quad ;$$

$$T_{\mu,\nu}{}^{\sigma\tau} = i\delta_{(\mu}{}^\sigma \delta_{\nu)}{}^\tau \quad , \quad rest = 0 \quad . \tag{2.2.5}$$

Thus, in the language of differential geometry, global superspace has *torsion.* The derivatives satisfy the further identities

$$\partial^{\mu\sigma}\partial_{\nu\sigma} = \delta_\nu{}^\mu \Box \quad , \quad D_\mu D_\nu = i\partial_{\mu\nu} + C_{\nu\mu} D^2 \quad ,$$

$$D^\nu D_\mu D_\nu = 0 \quad , \quad D^2 D_\mu = -\, D_\mu D^2 = i\partial_{\mu\nu} D^\nu \quad , \quad (D^2)^2 = \Box \quad . \tag{2.2.6}$$

They also satisfy the Leibnitz rule and can be integrated by parts when inside $d^3x\ d^2\theta$ integrals (since they are a combination of $x$ and $\theta$ derivatives ). The following identity is useful

$$\int d^3x\, d^2\theta\ \Phi(x,\theta) = \int d^3x\ \partial^2\ \Phi(x,\theta) = \int d^3x\ (\,D^2\Phi(x,\theta)\,)| \tag{2.2.7}$$

(where recall that $|$ means evaluation at $\theta = 0$). The extra space-time derivatives in $D_\mu$ (as compared to $\partial_\mu$ ) drop out after $x$-integration.

## b. Components by expansion

Superfields can be expanded in a (terminating) Taylor series in $\theta$. For example,

$$\Phi_{\alpha\beta\cdots}(x,\theta) = A_{\alpha\beta\cdots}(x) + \theta^\mu \lambda_{\mu\alpha\beta\cdots}(x) - \theta^2\, F_{\alpha\beta\cdots}(x) \quad . \tag{2.2.8}$$

$A\,, B\,, F$ are the *component* fields of $\Phi$. The supersymmetry transformations of the components can be derived from those of the superfield. For simplicity of notation, we consider a scalar superfield (no Lorentz indices)

$$\Phi(x,\theta) = A(x) + \theta^\alpha \psi_\alpha(x) - \theta^2 F(x) \quad , \tag{2.2.9}$$

The supersymmetry transformation ($\xi^{\mu\nu} = 0$, $\epsilon^\mu$ infinitesimal)

$$\delta\Phi(x,\theta) = -\, \epsilon^\mu(\partial_\mu - i\theta^\nu \partial_{\mu\nu})\Phi(x,\theta)$$

$$\equiv \delta A + \theta^\alpha \delta\psi_\alpha - \theta^2 \delta F \quad , \tag{2.2.10}$$

gives, upon equating powers of $\theta$,

$$\delta A = -\, \epsilon^\alpha \psi_\alpha \quad , \tag{2.2.11a}$$

$$\delta\psi_\alpha = -\, \epsilon^\beta (C_{\alpha\beta} F + i\partial_{\alpha\beta} A) \quad , \tag{2.2.11b}$$

$$\delta F = -\, \epsilon^\alpha i \partial_\alpha{}^\beta \psi_\beta \quad . \tag{2.2.11c}$$

It is easy to verify that on the component fields the supersymmetry algebra is satisfied: The commutator of two transformations gives a translation, $[\delta_Q(\epsilon)\, , \delta_Q(\eta)] = -2i\epsilon^\alpha \eta^\beta \partial_{\alpha\beta}$, etc.

### c. Actions and components by projection

The construction of (integral) invariants is facilitated by the observation that supersymmetry transformations are coordinate transformations in superspace. Because we can ignore total $\theta$-derivatives ($\int d^3x d^2\theta\, \partial_\alpha f^\alpha = 0$, which follows from $(\partial)^3 = 0$) and total spacetime derivatives, we find that *any* superspace integral

$$S = \int d^3x\, d^2\theta\ f(\Phi, D_\alpha \Phi, \cdots) \tag{2.2.12}$$

that does not depend explicitly on the coordinates is invariant under the full algebra. If the superfield expansion in terms of components is substituted into the integral and the $\theta$-integration is carried out, the resulting component integral is invariant under the transformations of (2.2.11) (the integrand in general changes by a total derivative). This also can be seen from the fact that the $\theta$-integration picks out the $F$ component of $f$, which transforms as a spacetime derivative (see (2.2.11c)).

We now describe a technical device that can be extremely helpful. In general, to obtain component expressions by direct $\theta$-expansions can be cumbersome. A more efficient procedure is to observe that the components in (2.2.9) can be defined by *projection:*

$$A(x) = \Phi(x,\theta)| \quad ,$$
$$\psi_\alpha(x) = D_\alpha \, \Phi(x,\theta)| \quad ,$$
$$F(x) = D^2 \, \Phi(x,\theta)| \quad . \tag{2.2.13}$$

This can be used, for example, in (2.2.12) by rewriting (c.f. (2.2.7))

$$S = \int d^3x \; D^2 f(\Phi, D_\alpha\Phi, \cdots)| \quad . \tag{2.2.14}$$

After the derivatives are evaluated (using the Leibnitz rule and paying due respect to the anticommutativity of the $D$'s), the result is directly expressible in terms of the components (2.2.13). The reader should verify in a few simple examples that this is a much more efficient procedure than direct $\theta$-expansion and integration.

Finally, we can also reobtain the component transformation laws by this method. We first note the identity

$$iQ_\alpha + D_\alpha = 2\theta^\beta i\partial_{\alpha\beta} \quad . \tag{2.2.15}$$

Thus we find, for example

$$\begin{aligned} \delta A &= i\epsilon^\alpha Q_\alpha \Phi| \\ &= -\,\epsilon^\alpha\,(D_\alpha\Phi - 2\theta^\beta i\partial_{\alpha\beta}\Phi)| \\ &= -\,\epsilon^\alpha\,\psi_\alpha \quad . \end{aligned} \tag{2.2.16}$$

In general we have

$$iQ_\alpha f| = -\,D_\alpha f| \quad . \tag{2.2.17}$$

This is sufficient to obtain all of the component fields transformation laws by repeated application of (2.2.17), where $f$ is $\Phi\,, D_\alpha\Phi\,, D^2\Phi$ and we use (2.2.6) and (2.2.13).

### d. Irreducible representations

In general a theory is described by fields which in momentum space are defined for arbitrary values of $p^2$. For any fixed value of $p^2$ the fields are a representation of the Poincaré group. We call such fields, defined for *arbitrary* values of $p^2$, an *off-shell* representation of the Poincaré group. Similarly, when a set of fields is a representation of the *supersymmetry* algebra for *any* value of $p^2$, we call it an off-shell representation of supersymmetry. When the field equations are imposed, a particular value of $p^2$ (i.e., $m^2$) is picked out. Some of the components of the fields (auxiliary components) are then constrained to vanish; the remaining (physical) components form what we call an *on-shell* representation of the Poincaré (or supersymmetry) group.

A superfield $\tilde{\psi}_{\alpha\ldots}(p,\theta)$ is an irreducible representation of the Lorentz group, with regard to its external indices, if it is totally symmetric in these indices. For a representation of the (super)Poincaré group we can reduce it further. Since in three dimensions the little group is $SO(2)$, and its irreducible representations are one-component (complex), this reduction will give one-component superfields (with respect to external indices). Such superfields are irreducible representations of off-shell supersymmetry, when a reality condition is imposed in $x$-space (but the superfield is then still complex in $p$-space, where $\Phi(p)=\overline{\Phi}(-p)$ ).

In an appropriate reference frame we can assign "helicity" (i.e., the eigenvalue of the $SO(2)$ generator) $\pm\frac{1}{2}$ to the spinor indices, and the irreducible representations will be labeled by the "superhelicity" (the helicity of the superfield): half the number of $+$ external indices minus the number of $-$'s. In this frame we can also assign $\pm\frac{1}{2}$ helicity to $\theta^{\pm}$. Expanding the superfield of superhelicity $h$ into components, we see that these components have helicities $h, h\pm\frac{1}{2}, h$. For example, a *scalar multiplet,* consisting of "spins" (i.e., $SO(2,1)$ representations) $0\,,\frac{1}{2}$ (i.e., helicities $0\,,\pm\frac{1}{2}$) is described by a superfield of superhelicity 0: a scalar superfield. A *vector multiplet,* consisting of spins $\frac{1}{2}\,,1$ (helicities $0\,,\frac{1}{2}\,,\frac{1}{2}\,,1$) is described by a superfield of superhelicity $+\frac{1}{2}$: the "+" component of a spinor superfield; the "$-$" component being gauged away (in a light-cone gauge). In general, the superhelicity content of a superfield is analyzed by choosing a gauge (the supersymmetric light-cone gauge) where as many as possible Lorentz components of a superfield have been gauged to 0: the superhelicity content of any remaining component is simply $\frac{1}{2}$ the number of $+$'s minus $-$'s. Unless otherwise stated, we will automatically consider *all* three-dimensional superfields to be *real.*

## 2.3. Scalar multiplet

The simplest representation of supersymmetry is the scalar multiplet described by the real superfield $\Phi(x,\theta)$, and containing the scalars $A, F$ and the two-component spinor $\psi_\alpha$. From (2.2.1,2) we see that $\theta$ has dimension $(mass)^{-\frac{1}{2}}$. Also, the canonical dimensions of component fields in three dimensions are $\frac{1}{2}$ less than in four dimensions (because we use $\int d^3x$ instead of $\int d^4x$ in the kinetic term). Therefore, if this multiplet is to describe physical fields, we must assign dimension $(mass)^{\frac{1}{2}}$ to $\Phi$ so that $\psi_\alpha$ has canonical dimension $(mass)^1$. (Although it is not immediately obvious which scalar should have canonical dimension, there is only one spinor.) Then $A$ will have dimension $(mass)^{\frac{1}{2}}$ and will be the physical scalar partner of $\psi$, whereas $F$ has too high a dimension to describe a canonical physical mode.

Since a $\theta$ integral is the same as a $\theta$ derivative, $\int d^2\theta$ has dimension $(mass)^1$. Therefore, on dimensional grounds we expect the following expression to give the correct (massless) kinetic action for the scalar multiplet:

$$S_{kin} = -\frac{1}{2}\int d^3x\, d^2\theta\, (D_\alpha\Phi)^2 \quad , \tag{2.3.1}$$

(recall that for any spinor $\psi_\alpha$ we have $\psi^2 = \frac{1}{2}\psi^\alpha\psi_\alpha$). This expression is reminiscent of the kinetic action for an ordinary scalar field with the substitutions $\int d^3x \to \int d^3x\, d^2\theta$ and $\partial_{\alpha\beta} \to D_\alpha$. The component expression can be obtained by explicit $\theta$-expansion and integration. However, we prefer to use the alternative procedure (first integrating $D^\alpha$ by parts):

$$\begin{aligned} S_{kin} &= \frac{1}{2}\int d^3x\, d^2\theta\, \Phi D^2\Phi \\ &= \frac{1}{2}\int d^3x\, D^2[\Phi\, D^2\Phi]| \\ &= \frac{1}{2}\int d^3x\, (D^2\Phi\, D^2\Phi + D^\alpha\Phi\, D_\alpha D^2\Phi + \Phi(D^2)^2\Phi)| \\ &= \frac{1}{2}\int d^3x\, (F^2 + \psi^\alpha i\partial_\alpha{}^\beta\psi_\beta + A\Box A) \quad , \end{aligned} \tag{2.3.2}$$

where we have used the identities (2.2.6) and the definitions (2.2.13). The $A$ and $\psi$ kinetic terms are conventional, while $F$ is clearly non-propagating.

The auxiliary field $F$ can be eliminated from the action by using its equation of motion $F = 0$ (or, in a functional integral, $F$ can be trivially integrated out). The resulting action is still invariant under the bose-fermi transformations (2.2.11a,b) with $F = 0$; however, these are not supersymmetry transformations (not a representation of the supersymmetry algebra) except "on shell". The commutator of two such transformations does not close (does not give a translation) except when $\psi_\alpha$ satisfies its field equation. This "off-shell" non-closure of the algebra is typical of transformations from which auxiliary fields have been eliminated.

Mass and interaction terms can be added to (2.3.1). A term

$$S_I = \int d^3x\, d^2\theta\; f(\Phi) \quad , \tag{2.3.3}$$

leads to a component action

$$\begin{aligned} S_I &= \int d^3x\; D^2 f(\Phi)| \\ &= \int d^3x\; [f''(\Phi)\,(D_\alpha\Phi)^2 + f'(\Phi)\,D^2\Phi]| \\ &= \int d^3x\; [f''(A)\,\psi^2 + f'(A)\,F] \quad . \end{aligned} \tag{2.3.4}$$

In a renormalizable model $f(\Phi)$ can be at most quartic. In particular, $f(\Phi) = \frac{1}{2}m\Phi^2 + \frac{1}{6}\lambda\Phi^3$ gives mass terms, Yukawa and cubic interaction terms. Together with the kinetic term, we obtain

$$\begin{aligned} &\int d^3x d^2\theta [-\tfrac{1}{2}(D_\alpha\Phi)^2 + \tfrac{1}{2}m\Phi^2 + \tfrac{1}{6}\lambda\Phi^3] \\ &\quad = \int d^3x [\tfrac{1}{2}(A\Box A + \psi^\alpha i\partial_\alpha{}^\beta\psi_\beta + F^2) \\ &\qquad + m(\psi^2 + AF) + \lambda(A\psi^2 + \tfrac{1}{2}A^2F)] \quad . \end{aligned} \tag{2.3.5}$$

$F$ can again be eliminated using its (algebraic) equation of motion, leading to a

conventional mass term and quartic interactions for the scalar field $A$. More exotic kinetic actions are possible by using instead of (2.3.1)

$$S'_{kin} = \int d^3x\, d^2\theta\, \Omega\,(\zeta^\alpha\,, \Phi) \quad , \quad \zeta^\alpha = D^\alpha \Phi \quad , \tag{2.3.6}$$

where $\Omega$ is some function such that $\frac{\partial^2 \Omega}{\partial \zeta^\alpha\, \partial \zeta^\beta}\,|_{\zeta, \Phi\, =\, 0} = -\,\frac{1}{2}\, C_{\alpha\beta}$ . If we introduce more than one multiplet of scalar superfields, then, for example, we can obtain generalized supersymmetric nonlinear sigma models:

$$S = -\,\frac{1}{2} \int d^3x\, d^2\theta\; g_{ij}(\Phi)\, \frac{1}{2}\, (\, D^\alpha \Phi^i\, )\, (\, D_\alpha \Phi^j\, ) \tag{2.3.7}$$

## 2.4. Vector multiplet

### a. Abelian gauge theory

In accordance with the discussion in sec. 2.2, a real spinor gauge superfield $\Gamma_\alpha$ with superhelicity $h = \frac{1}{2}$ ($h = -\frac{1}{2}$ can be gauged away) will consist of components with helicities $0, \frac{1}{2}, \frac{1}{2}, 1$. It can be used to describe a massless gauge vector field and its fermionic partner. (In three dimensions, a gauge vector particle has one physical component of definite helicity.) The superfield can be introduced by analogy with scalar QED (the generalization to the nonabelian case is straightforward, and will be discussed below). Consider a complex scalar superfield (a doublet of real scalar superfields) transforming under a *constant* phase rotation

$$\Phi \to \Phi' = e^{iK}\,\Phi \quad ,$$

$$\bar{\Phi} \to \bar{\Phi}' = \bar{\Phi} e^{-iK} \quad . \tag{2.4.1}$$

The free Lagrangian $|D\Phi|^2$ is invariant under these transformations.

### a.1. Gauge connections

We extend this to a *local* phase invariance with $K$ a real scalar superfield depending on $x$ and $\theta$, by covariantizing the spinor derivatives $D_\alpha$:

$$D_\alpha \to \nabla_\alpha = D_\alpha \mp i\,\Gamma_\alpha \quad , \tag{2.4.2}$$

when acting on $\Phi$ or $\bar{\Phi}$, respectively. The spinor gauge potential (or connection) $\Gamma_\alpha$ transforms in the usual way

$$\delta\Gamma_\alpha = D_\alpha K \quad , \tag{2.4.3}$$

to ensure

$$\nabla'_\alpha = e^{iK}\,\nabla_\alpha\,e^{-iK} \quad . \tag{2.4.4}$$

This is required by $(\nabla\,\Phi)' = e^{iK}\,(\nabla\,\Phi)$, and guarantees that the Lagrangian $|\nabla\Phi|^2$ is locally gauge invariant. (The coupling constant can be restored by rescaling $\Gamma_\alpha \to g\Gamma_\alpha$).

It is now straightforward, by analogy with QED, to find a gauge invariant field strength and action for the multiplet described by $\Gamma_\alpha$ and to study its component couplings to the complex scalar multiplet contained in $|\nabla\Phi|^2$. However, both to understand its structure as an irreducible representation of supersymmetry, and as an introduction to more complicated gauge superfields (e.g. in supergravity), we first give a geometrical presentation.

Although the actions we have considered do not contain the spacetime derivative $\partial_{\alpha\beta}$, in other contexts we need the covariant object

$$\nabla_{\alpha\beta} = \partial_{\alpha\beta} - i\,\Gamma_{\alpha\beta} \ , \quad \delta\Gamma_{\alpha\beta} = \partial_{\alpha\beta}K \quad , \tag{2.4.5}$$

introducing a distinct (vector) gauge potential superfield. The transformation $\delta\Gamma_{\alpha\beta}$ of this connection is chosen to give:

$$\nabla'_{\alpha\beta} = e^{iK}\,\nabla_{\alpha\beta}e^{-iK} \quad . \tag{2.4.6}$$

(From a geometric viewpoint, it is natural to introduce the vector connection; then $\Gamma_\alpha$ and $\Gamma_{\alpha\beta}$ can be regarded as the components of a super 1-form $\Gamma_A = (\Gamma_\alpha, \Gamma_{\alpha\beta})$; see sec. 2.5). However, we will find that $\Gamma_{\alpha\beta}$ should not be independent, and can be expressed in terms of $\Gamma_\alpha$.

## a.2. Components

To get oriented, we examine the components of $\Gamma$ in the Taylor series $\theta$-expansion. They can be defined directly by using the spinor derivatives $D_\alpha$:

$$\chi_\alpha = \Gamma_\alpha| \quad , \qquad B = \frac{1}{2}\,D^\alpha\Gamma_\alpha| \quad ,$$

$$V_{\alpha\beta} = -\frac{i}{2}D_{(\alpha}\Gamma_{\beta)}| \quad , \qquad \lambda_\alpha = \frac{1}{2}\,D^\beta D_\alpha\Gamma_\beta| \quad , \tag{2.4.7a}$$

and

$$W_{\alpha\beta} = \Gamma_{\alpha\beta}| \quad , \qquad \rho_\beta = D^\alpha\Gamma_{\alpha\beta}| \quad ,$$

$$\psi_{\alpha\beta\gamma} = D_{(\alpha}\Gamma_{\beta\gamma)}| \quad , \qquad T_{\alpha\beta} = D^2\Gamma_{\alpha\beta}| \quad . \tag{2.4.7b}$$

We have separated the components into irreducible representations of the Lorentz group, that is, traces (or antisymmetrized pieces, see (2.1.2)) and symmetrized pieces. We also define the components of the gauge parameter $K$:

$$\omega = K| \quad , \quad \sigma_\alpha = D_\alpha K| \quad , \quad \tau = D^2 K| \tag{2.4.8}$$

The component gauge transformations for the components defined in (2.4.7) are found by repeatedly differentiating (2.4.3-5) with spinor derivatives $D_\alpha$. We find:

$$\delta\chi_\alpha = \sigma_\alpha \quad , \quad \delta B = \tau \quad ,$$

$$\delta V_{\alpha\beta} = \partial_{\alpha\beta}\omega \quad , \quad \delta\lambda_\alpha = 0 \quad , \tag{2.4.9a}$$

and

$$\delta W_{\alpha\beta} = \partial_{\alpha\beta}\omega \quad , \quad \delta\rho_\alpha = \partial_{\alpha\beta}\sigma^\beta \quad ,$$

$$\delta\psi_{\alpha\beta\gamma} = \partial_{(\beta\gamma}\sigma_{\alpha)} \quad , \quad \delta T_{\alpha\beta} = \partial_{\alpha\beta}\tau \quad . \tag{2.4.9b}$$

Note that $\chi$ and $B$ suffer arbitrary shifts as a consequence of a gauge transformation, and, in particular, can be gauged completely away; the gauge $\chi = B = 0$ is called *Wess-Zumino* gauge, and *explicitly* breaks supersymmetry. However, this gauge is useful since it reveals the physical content of the $\Gamma_\alpha$ multiplet.

Examination of the components that remain reveals several peculiar features: There are *two* component gauge potentials $V_{\alpha\beta}$ and $W_{\alpha\beta}$ for only *one* gauge symmetry, and there is a high dimension spin $\frac{3}{2}$ field $\psi_{\alpha\beta\gamma}$. These problems will be resolved below when we express $\Gamma_{\alpha\beta}$ in terms of $\Gamma_\alpha$.

We can also find supersymmetric Lorentz gauges by fixing $D^\alpha\Gamma_\alpha$; such gauges are useful for quantization (see sec. 2.7). Furthermore, in three dimensions it is possible to choose a supersymmetric light-cone gauge $\Gamma_+ = 0$. (In the abelian case the gauge transformation takes the simple form $K = D_+(i\partial_{++})^{-1}\Gamma_+$.) Eq. (2.4.14) below implies that in this gauge the superfield $\Gamma_{++}$ also vanishes. The remaining components in this gauge are $\chi_-$, $V_{+-}$, $V_{--}$, and $\lambda_-$, with $V_{++} = 0$ and $\lambda_+ \sim \partial_{++}\chi_-$.

### a.3. Constraints

To understand how the vector connection $\Gamma_{\alpha\beta}$ can be expressed in terms of the spinor connection $\Gamma_\alpha$, recall the (anti)commutation relations for the ordinary derivatives are:

$$[\, D_M \,, D_N \,\} = T_{MN}{}^P \, D_P \quad . \tag{2.4.10}$$

For the covariant derivatives $\nabla_A = (\nabla_\alpha, \nabla_{\alpha\beta})$ the graded commutation relations can be written (from (2.4.2) and (2.4.5) we see that the torsion $T_{AB}{}^C$ is unmodified):

$$[\, \nabla_A \,, \nabla_B \,\} = T_{AB}{}^C \, \nabla_C - i \, F_{AB} \quad . \tag{2.4.11}$$

The field strengths $F_{AB}$ are invariant $(F'_{AB} = F_{AB})$ due to the covariance of the derivatives $\nabla_A$. Observe that the field strengths are antihermitian matrices, $\bar{F}_{AB} = - \, F_{BA}$, so that the symmetric field strength $F_{\alpha\beta}$ is imaginary while the antisymmetric field strength $F_{\alpha\beta,\gamma\delta}$ is real. Examining a particular equation from (2.4.11), we find:

$$\{\, \nabla_\alpha \,, \nabla_\beta \,\} = 2i \, \nabla_{\alpha\beta} - i \, F_{\alpha\beta} \; = \; 2i \, \partial_{\alpha\beta} + 2\Gamma_{\alpha\beta} - i \, F_{\alpha\beta} \quad . \tag{2.4.12}$$

The superfield $\Gamma_{\alpha\beta}$ was introduced to covariantize the space-time derivative $\partial_{\alpha\beta}$. However, it is clear that an alternative choice is $\Gamma'_{\alpha\beta} = \Gamma_{\alpha\beta} - \frac{i}{2} F_{\alpha\beta}$ since $F_{\alpha\beta}$ is covariant (a field strength). The new covariant space-time derivative will then satisfy (we drop the primes)

$$\{\nabla_\alpha \,, \nabla_\beta\} = 2i\nabla_{\alpha\beta} \quad , \tag{2.4.13}$$

with the new space-time connection satisfying (after substituting in 2.4.12 the explicit forms $\nabla_A = D_A - i\Gamma_A$)

$$\Gamma_{\alpha\beta} = - \, \frac{i}{2} \, D_{(\alpha} \Gamma_{\beta)} \quad . \tag{2.4.14}$$

Thus the *conventional constraint*

$$F_{\alpha\beta} = 0 \quad , \tag{2.4.15}$$

imposed on the system (2.4.11) has allowed the vector potential to be expressed in terms of the spinor potential. This solves both the problem of two gauge fields $W_{\alpha\beta} \,, V_{\alpha\beta}$ and the problem of the higher spin and dimension components $\psi_{\alpha\beta\gamma} \,, T_{\alpha\beta}$: The gauge fields are identified with each other $(W_{\alpha\beta} = V_{\alpha\beta})$, and the extra components are expressed as derivatives of familiar lower spin and dimension fields (see 2.4.7). The independent components that remain in Wess-Zumino gauge after the constraint is imposed are $V_{\alpha\beta}$ and $\lambda_\alpha$ .

We stress the importance of the constraint (2.4.15) on the objects defined in (2.4.11). Unconstrained field strengths in general lead to reducible representations of supersymmetry (i.e., the spinor and vector potentials), and the constraints are needed to ensure irreducibility.

### a.4. Bianchi identities

In ordinary field theories, the field strengths satisfy Bianchi identities because they are expressed in terms of the potentials; they are *identities* and carry no information. For gauge theories described by covariant derivatives, the Bianchi identities are just Jacobi identities:

$$[\nabla_{[A}\,, [\nabla_B\,, \nabla_{C)}\,\}\,\} = 0 \quad , \tag{2.4.16}$$

(where $[\,)$ is the *graded* antisymmetrization symbol, identical to the usual antisymmetrization symbol but with an extra factor of $(-1)$ for each pair of interchanged fermionic indices). However, once we impose constraints such as (2.4.13,15) on some of the field strengths, the Bianchi identities imply constraints on other field strengths. For example, the identity

$$0 = [\nabla_\alpha\,, \{\nabla_\beta\,, \nabla_\gamma\}] + [\nabla_\beta\,, \{\nabla_\gamma\,, \nabla_\alpha\}] + [\nabla_\gamma\,, \{\nabla_\alpha\,, \nabla_\beta\}]$$

$$= \frac{1}{2}[\nabla_{(\alpha}\,, \{\nabla_\beta\,, \nabla_{\gamma)}\}] \tag{2.4.17}$$

gives (using the constraint (2.4.13,15))

$$0 = [\nabla_{(\alpha}\,, \nabla_{\beta\gamma)}] = -\,i\,F_{(\alpha,\beta\gamma)} \quad . \tag{2.4.18}$$

Thus the totally symmetric part of $F$ vanishes. In general, we can decompose $F$ into irreducible representations of the Lorentz group:

$$F_{\alpha,\beta\gamma} = \frac{1}{6}F_{(\alpha,\beta\gamma)} - \frac{1}{3}C_{\alpha(\beta|}F^{\delta}{}_{,\delta|\gamma)} \tag{2.4.19}$$

(where indices between $|\cdots|$ , e.g., in this case $\delta$, are not included in the symmetrization). Hence the only remaining piece is:

$$F_{\alpha,\beta\gamma} = i\,C_{\alpha(\beta}W_{\gamma)} \quad , \tag{2.4.20a}$$

where we introduce the superfield strength $W_\alpha$. We can compute $F_{\alpha,\beta\gamma}$ in terms of $\Gamma_\alpha$

and find

$$W_\alpha = \frac{1}{2} D^\beta D_\alpha \Gamma_\beta \quad . \tag{2.4.20b}$$

The superfield $W_\alpha$ is the only independent gauge invariant field strength, and is constrained by $D^\alpha W_\alpha = 0$, which follows from the Bianchi identity (2.4.16). This implies that only one Lorentz component of $W_\alpha$ is independent. The field strength describes the physical degrees of freedom: one helicity $\frac{1}{2}$ and one helicity 1 mode. Thus $W_\alpha$ is a suitable object for constructing an action. Indeed, if we start with

$$S = \frac{1}{g^2} \int d^3x \, d^2\theta \; W^2 = \frac{1}{g^2} \int d^3x \, d^2\theta \; (\tfrac{1}{2} D^\beta D_\alpha \Gamma_\beta)^2 \quad , \tag{2.4.21}$$

we can compute the component action

$$S = \frac{1}{g^2} \int d^3x \; D^2 W^2 = \frac{1}{g^2} \int d^3x \, [ W^\alpha D^2 W_\alpha - \frac{1}{2} (D^\alpha W^\beta)(D_\alpha W_\beta) ]|$$

$$= \frac{1}{g^2} \int d^3x \left[ \lambda^\alpha \, i\partial_\alpha{}^\beta \lambda_\beta - \frac{1}{2} f^{\alpha\beta} f_{\alpha\beta} \right] \quad . \tag{2.4.22}$$

Here (cf. 2.4.7) $\lambda_\alpha \equiv W_\alpha|$ while $f_{\alpha\beta} = D_\alpha W_\beta| = D_\beta W_\alpha|$ is the spinor form of the usual field strength

$$F_{\alpha\beta}{}^{\gamma\delta}| = (\partial_{\alpha\beta}\Gamma^{\gamma\delta} - \partial^{\gamma\delta}\Gamma_{\alpha\beta})| = \frac{1}{2} \delta_{(\alpha}{}^{(\gamma} f_{\beta)}{}^{\delta)}$$

$$= -\, i \, \frac{1}{2} [\partial_{\alpha\beta} D^{(\gamma}\Gamma^{\delta)} - \partial^{\gamma\delta} D_{(\alpha}\Gamma_{\beta)}]| \quad . \tag{2.4.23}$$

To derive the above component action we have used the Bianchi identity $D^\alpha W_\alpha = 0$, and its consequence $D^2 W_\alpha = i\partial_\alpha{}^\beta W_\beta$.

### a.5. Matter couplings

We now examine the component Lagrangian describing the coupling to a complex scalar multiplet. We could start with

$$S = -\, \frac{1}{2} \int d^3x d^2\theta (\nabla^\alpha \overline{\Phi})(\nabla_\alpha \Phi)$$

$$= -\frac{1}{2}\int d^3x D^2[(D^\alpha + i\Gamma^\alpha)\bar{\Phi}][(D_\alpha - i\Gamma_\alpha)\Phi] \quad , \tag{2.4.24}$$

and work out the Lagrangian in terms of components defined by projection. However, a more efficient procedure, which leads to physically equivalent results, is to define *covariant components* of $\Phi$ by *covariant* projection

$$A = \Phi(x,\theta)| \quad ,$$

$$\psi_\alpha = \nabla_\alpha \Phi(x,\theta)| \quad ,$$

$$F = \nabla^2\Phi(x,\theta)| \quad . \tag{2.4.25}$$

These components are not equal to the ordinary ones but can be obtained by a (gauge-field dependent) field redefinition and provide an equally valid description of the theory. We can also use

$$\int d^3x\, d^2\theta = \int d^3x\ D^2| = \int d^3x\ \nabla^2| \quad , \tag{2.4.26}$$

when acting on an invariant and hence

$$S = \int d^3x\ \nabla^2[\bar{\Phi}\nabla^2\Phi]|$$

$$= \int d^3x\ [\nabla^2\bar{\Phi}\nabla^2\Phi + \nabla^\alpha\bar{\Phi}\nabla_\alpha\nabla^2\Phi + \bar{\Phi}(\nabla^2)^2\Phi]|$$

$$= \int d^3x\ [\bar{F}F + \bar{\psi}^\alpha(i\partial_\alpha{}^\beta + V_\alpha{}^\beta)\psi_\beta + (i\bar{\psi}^\alpha\lambda_\alpha A + h.\,c.) + \bar{A}(\partial_{\alpha\beta} - i\,V_{\alpha\beta})^2 A]. \tag{2.4.27}$$

We have used the commutation relations of the covariant derivatives and in particular $\nabla_\alpha\nabla^2 = i\nabla_\alpha{}^\beta\nabla_\beta + iW_\alpha$, $\nabla^2\nabla_\alpha = -i\nabla_\alpha{}^\beta\nabla_\beta - 2iW_\alpha$, $(\nabla^2)^2 = \Box - iW^\alpha\nabla_\alpha$, where $\Box$ is the *covariant* d'Alembertian (covariantized with $\Gamma_{\alpha\beta}$).

### b. Nonabelian case

We now briefly consider the nonabelian case: For a multiplet of scalar superfields transforming as $\Phi' = e^{iK}\,\Phi$, where $K = K^iT_i$ and $T_i$ are generators of the Lie algebra, we introduce covariant spinor derivatives $\nabla_\alpha$ precisely as for the abelian case (2.4.2). We define $\Gamma_\alpha = \Gamma_\alpha{}^i\,T_i$ so that

$$\nabla_\alpha = D_\alpha - i\,\Gamma_\alpha = D_\alpha - i\,\Gamma_\alpha{}^i\,T_i \quad . \tag{2.4.28}$$

The spinor connection now transforms as

$$\delta\Gamma_\alpha = \nabla_\alpha K = D_\alpha K - i\,[\,\Gamma_\alpha\,,K\,] \quad , \tag{2.4.29}$$

leaving (2.4.4) unmodified. The vector connection is again constrained by requiring $F_{\alpha\beta} = 0$; in other words, we have

$$\nabla_{\alpha\beta} = -\frac{i}{2}\{\,\nabla_\alpha\,,\nabla_\beta\,\} \quad , \tag{2.4.30a}$$

$$\Gamma_{\alpha\beta} = -\,i\,\frac{1}{2}\,[D_{(\alpha}\,\Gamma_{\beta)} - i\,\{\Gamma_\alpha, \Gamma_\beta\}\,] \quad . \tag{2.4.30b}$$

The form of the action (2.4.21) is unmodified (except that we must also take a trace over group indices). The constraint (2.4.30) implies that the Bianchi identities have nontrivial consequences, and allows us to "solve" (2.4.17) for the nonabelian case as in (2.4.18,19,20a). Thus, we obtain

$$[\,\nabla_\alpha\,,\nabla_{\beta\gamma}\,] = C_{\alpha(\beta}W_{\gamma)} \tag{2.4.31a}$$

in terms of the nonabelian form of the *covariant* field strength $W$:

$$W_\alpha = \frac{1}{2}\,D^\beta D_\alpha\Gamma_\beta - \frac{i}{2}\,[\,\Gamma^\beta\,,D_\beta\Gamma_\alpha\,] - \frac{1}{6}\,[\,\Gamma^\beta\,,\{\,\Gamma_\beta\,,\Gamma_\alpha\,\}\,] \quad . \tag{2.4.31b}$$

The field strength transforms covariantly: $W'_\alpha = e^{iK}W_\alpha e^{-iK}$. The remaining Bianchi identity is

$$[\,\{\,\nabla_\alpha\,,\nabla_\beta\,\}\,,\nabla_{\gamma\delta}\,] - \{\,\nabla_{(\alpha}\,,[\nabla_{\beta)}\,,\nabla_{\gamma\delta}\,]\,\} = 0 \quad . \tag{2.4.32a}$$

Contracting indices we find $[\{\nabla^\alpha, \nabla^\beta\}, \nabla_{\alpha\beta}] = \{\nabla^{(\alpha}, [\nabla^{\beta)}, \nabla_{\alpha\beta}]\}$. However, $[\{\nabla^\alpha, \nabla^\beta\}, \nabla_{\alpha\beta}] = 2i[\nabla^{\alpha\beta}\,, \nabla_{\alpha\beta}] = 0$ and hence, using (2.4.31a),

$$0 = \{\,\nabla^{(\alpha}\,,[\,\nabla^{\beta)}\,,\nabla_{\alpha\beta}\,]\,\} = -\,6\{\,\nabla^\alpha\,,W_\alpha\,\} \quad . \tag{2.4.32b}$$

The full implication of the Bianchi identities is thus:

$$\{\,\nabla_\alpha\,,\nabla_\beta\,\} = 2i\nabla_{\alpha\beta} \tag{2.4.33a}$$

$$[\,\nabla_\alpha\,,\nabla_{\beta\gamma}\,] = C_{\alpha(\beta}W_{\gamma)} \quad , \quad \{\,\nabla^\alpha\,,W_\alpha\,\} = 0 \tag{2.4.33b}$$

$$[\,\nabla_{\alpha\beta}\,,\nabla^{\gamma\delta}\,] = -\,\frac{1}{2}\,i\delta_{(\alpha}{}^{(\gamma}\,f_{\beta)}{}^{\delta)} \quad , \quad f_{\alpha\beta} \equiv \frac{1}{2}\{\,\nabla_{(\alpha}\,,W_{\beta)}\,\} \quad . \tag{2.4.33c}$$

The components of the multiplet can be defined in analogy to (2.4.7) by *projections* of $\Gamma$:

$$\chi_\alpha = \Gamma_\alpha| \quad , \qquad B = \frac{1}{2} D^\alpha \Gamma_\alpha| \quad , \qquad V_{\alpha\beta} = \Gamma_{\alpha\beta}| \quad , \qquad \lambda_\alpha = W_\alpha| \quad . \tag{2.4.34}$$

## c. Gauge invariant masses

A curious feature which this theory has, and which makes it rather different from four dimensional Yang-Mills theory, is the existence of a gauge-invariant mass term: In the abelian case the Bianchi identity $D^\alpha W_\alpha = 0$ can be used to prove the invariance of

$$S_m = \frac{1}{g^2} \int d^3x \, d^2\theta \left[ \frac{1}{2} m \, \Gamma^\alpha W_\alpha \right] \quad . \tag{2.4.35}$$

In components this action contains the usual gauge invariant mass term for three-dimensional electrodynamics:

$$m \int d^3x \; V^{\alpha\beta} \partial_{\gamma\alpha} V_\beta{}^\gamma = m \int d^3x \; V^{\alpha\beta} f_{\alpha\beta} \quad , \tag{2.4.36}$$

which is gauge invariant as a consequence of the usual component Bianchi identity $\partial^{\alpha\beta} f_{\alpha\beta} = 0$.

The superfield equations which result from (2.4.21,35) are:

$$i\partial_\alpha{}^\beta W_\beta + m \, W_\alpha = 0 \quad , \tag{2.4.37}$$

which describes an irreducible multiplet of mass $m$. The Bianchi identity $D^\alpha W_\alpha = 0$ implies that only one Lorentz component of $W$ is independent.

For the nonabelian case, the mass term is somewhat more complicated because the field strength $W$ is covariant rather than invariant:

$$S_m = tr \frac{1}{g^2} \int d^3x \, d^2\theta \; \frac{1}{2} m \, ( \Gamma^\alpha W_\alpha + \frac{i}{6} \{ \Gamma^\alpha , \Gamma^\beta \} D_\beta \Gamma_\alpha + \frac{1}{12} \{ \Gamma^\alpha , \Gamma^\beta \} \{ \Gamma_\alpha , \Gamma_\beta \} )$$

$$= tr \frac{1}{g^2} \int d^3x \, d^2\theta \, \frac{1}{2} m \, \Gamma^\alpha \, ( \, W_\alpha - \frac{1}{6} [ \, \Gamma^\beta \, , \Gamma_{\alpha\beta} \, ] \, ) \quad . \tag{2.4.38}$$

The field equations, however, are the covariantizations of (2.4.37):

$$i \nabla_\alpha{}^\beta W_\beta + m \, W_\alpha = 0 \quad . \tag{2.4.39}$$

## 2.5. Other global gauge multiplets

### a. Superforms: general case

The gauge multiplets discussed in the last section may be described completely in terms of geometric quantities. The gauge potentials $\Gamma_A \equiv (\Gamma_\alpha, \Gamma_{\alpha\beta})$ which covariantize the derivatives $D_A$ with respect to local phase rotations of the matter superfields constitute a super 1-form. We define super $p$-forms as tensors with $p$ covariant supervector indices (i.e., supervector subscripts) that have total *graded* antisymmetry with respect to these indices (i.e., are symmetric in any pair of spinor indices, antisymmetric in a vector pair or in a mixed pair). For example, the field strength $F_{AB} \equiv (F_{\alpha,\beta}, F_{\alpha,\beta\gamma}, F_{\alpha\beta,\gamma\delta})$ constitutes a super 2-form.

In terms of supervector notation the gauge transformation for $\Gamma_A$ (from (2.4.3) and (2.4.5)) takes the form

$$\delta\Gamma_A = D_A K \quad . \tag{2.5.1}$$

The field strength defined in (2.3.6) when expressed in terms of the gauge potential can be written as

$$F_{AB} = D_{[A}\Gamma_{B)} - T_{AB}{}^C\Gamma_C \quad . \tag{2.5.2}$$

The gauge transformation law certainly takes the familiar form, but even in the abelian case, the field strength has an unfamiliar nonderivative term. One way to understand how this term arises is to make a change of basis for the components of a supervector. We can expand $D_A$ in terms of partial derivatives by introducing a matrix, $E_A{}^M$, such that

$$D_A = E_A{}^M \partial_M \ , \quad \partial_M \equiv (\ \partial_\mu \, , \partial_{\mu\nu}\ ) \quad ,$$

$$E_A{}^M = \begin{bmatrix} \delta_\alpha{}^\mu & \frac{1}{2}\, i\theta^{(\mu}\delta_\alpha{}^{\nu)} \\ & \\ 0 & \frac{1}{2}\,\delta_\alpha{}^{(\mu}\delta_\beta{}^{\nu)} \end{bmatrix} \quad . \tag{2.5.3}$$

This matrix is the *flat vielbein;* its inverse is

$$E_M{}^A = \begin{bmatrix} \delta_\mu{}^\alpha & -\frac{1}{2} i\theta^{(\alpha}\delta_\mu{}^{\beta)} \\ 0 & \frac{1}{2}\delta_\mu{}^{(\alpha}\delta_\nu{}^{\beta)} \end{bmatrix} \quad . \tag{2.5.4}$$

If we define $\Gamma_M$ by $\Gamma_A \equiv E_A{}^M \Gamma_M$, then

$$\delta\Gamma_M = \partial_M K \quad . \tag{2.5.5}$$

Similarly, if we define $F_{MN}$ by

$$F_{AB} \equiv (-)^{A(B+N)} E_B{}^N E_A{}^M F_{MN} \quad , \tag{2.5.6a}$$

then

$$F_{MN} = \partial_{[M}\Gamma_{N)} \quad . \tag{2.5.6b}$$

(In the Grassmann parity factor $(-)^{A(B+N)}$ the superscripts $A, B$, and $N$ are equal to one when these indices refer to spinorial indices and zero otherwise.) We thus see that the nonderivative term in the field strength is absent when the components of this supertensor are referred to a different coordinate basis. Furthermore, in this basis the Bianchi identities take the simple form

$$\partial_{[M}F_{NP)} = 0 \quad . \tag{2.5.7}$$

The generalization to higher-rank graded antisymmetric tensors (superforms) is now evident. There is a basis in which the gauge transformation, field strength, and Bianchi identities take the forms

$$\delta\Gamma_{M_1 \cdots M_p} = \frac{1}{(p-1)!}\partial_{[M_1}K_{M_2 \cdots M_p)} \quad ,$$

$$F_{M_1 \cdots M_{p+1}} = \frac{1}{p!}\partial_{[M_1}\Gamma_{M_2 \cdots M_{p+1})} \quad ,$$

$$0 = \partial_{[M_1}F_{M_2 \cdots M_{p+2})} \quad . \tag{2.5.8}$$

We simply multiply these by suitable powers of the flat vielbein and appropriate Grassmann parity factors to obtain

$$\delta\Gamma_{A_1 \cdots A_p} = \frac{1}{(p-1)!} D_{[A_1}K_{A_2 \cdots A_p)} - \frac{1}{2(p-2)!} T_{[A_1 A_2|}{}^B K_{B|A_3 \cdots A_p)} \quad ,$$

$$F_{A_1 \cdots A_{p+1}} = \frac{1}{p!} D_{[A_1} \Gamma_{A_2 \cdots A_{p+1})} - \frac{1}{2(p-1)!} T_{[A_1 A_2|}{}^B \Gamma_{B|A_3 \cdots A_{p+1})} \quad ,$$

$$0 = \frac{1}{(p+1)!} D_{[A_1} F_{A_2 \cdots A_{p+2})} - \frac{1}{2p!} T_{[A_1 A_2|}{}^B F_{B|A_3 \cdots A_{p+2})} \quad . \tag{2.5.9}$$

(The | 's indicate that all of the indices are graded antisymmetric except the $B$ 's.)

## b. Super 2-form

We now discuss in detail the case of a super 2-form gauge superfield $\Gamma_{AB}$ with gauge transformation

$$\delta \Gamma_{\alpha,\beta} = D_{(\alpha} K_{\beta)} - 2i\, K_{\alpha\beta} \quad ,$$

$$\delta \Gamma_{\alpha,\beta\gamma} = D_\alpha K_{\beta\gamma} - \partial_{\beta\gamma} K_\alpha \quad ,$$

$$\delta \Gamma_{\alpha\beta,\gamma\delta} = \partial_{\alpha\beta} K_{\gamma\delta} - \partial_{\gamma\delta} K_{\alpha\beta} \quad . \tag{2.5.10}$$

The field strength for $\Gamma_{AB}$ is a super 3-form:

$$F_{\alpha,\beta,\gamma} = \frac{1}{2} (D_{(\alpha} \Gamma_{\beta,\gamma)} + 2i \Gamma_{(\alpha,\beta\gamma)}) \quad ,$$

$$F_{\alpha,\beta,\gamma\delta} = D_{(\alpha} \Gamma_{\beta),\gamma\delta} + \partial_{\gamma\delta} \Gamma_{\alpha,\beta} - 2i\, \Gamma_{\alpha\beta,\gamma\delta} \quad ,$$

$$F_{\alpha,\beta\gamma,\delta\epsilon} = D_\alpha \Gamma_{\beta\gamma,\delta\epsilon} + \partial_{\delta\epsilon} \Gamma_{\alpha,\beta\gamma} - \partial_{\beta\gamma} \Gamma_{\alpha,\delta\epsilon} \quad ,$$

$$F_{\alpha\beta,\gamma\delta,\epsilon\zeta} = \partial_{\alpha\beta} \Gamma_{\gamma\delta,\epsilon\zeta} + \partial_{\epsilon\zeta} \Gamma_{\alpha\beta,\gamma\delta} + \partial_{\gamma\delta} \Gamma_{\epsilon\zeta,\alpha\beta} \quad . \tag{2.5.11}$$

All of these equations are contained in the concise supervector notation in (2.5.9).

The gauge superfield $\Gamma_A$ was subject to constraints that allowed one part ($\Gamma_{\alpha,\beta}$) to be expressed as a function of the remaining part. This is a general feature of supersymmetric gauge theories; constraints are needed to ensure irreducibility. For the tensor gauge multiplet we impose the constraints

$$F_{\alpha,\beta,\gamma} = 0 \quad , \qquad F_{\alpha,\beta,}{}^{\gamma\delta} = i\, \delta_{(\alpha}{}^{\gamma}\, \delta_{\beta)}{}^{\delta}\, G = T_{\alpha,\beta}{}^{\gamma\delta} G \quad , \tag{2.5.12}$$

which, as we show below, allow us to express all covariant quantities in terms of the single real scalar superfield $G$. These constraints can be solved as follows: we first observe that in the field strengths $\Gamma_{\alpha,\beta}$ always appears in the combination $D_{(\alpha}\, \Gamma_{\beta,\gamma)} + 2i\, \Gamma_{(\alpha,\beta\gamma)}$.

Therefore, without changing the field strengths we can redefine $\Gamma_{\alpha,\beta\gamma}$ by absorbing $D_{(\alpha}\Gamma_{\beta,\gamma)}$ into it. Thus $\Gamma_{\alpha,\beta}$ disappears from the field strengths which means it could be set to zero from the beginning (equivalently, we can make it zero by a gauge transformation). The first constraint now implies that the totally symmetric part of $\Gamma_{\alpha,\beta\gamma}$ is zero and hence we can write $\Gamma_{\alpha,\beta\gamma} = i\, C_{\alpha(\beta}\Phi_{\gamma)}$ in terms of a spinor superfield $\Phi_\gamma$. The remaining equations and constraints can be used now to express $\Gamma_{\alpha\beta,\gamma\delta}$ and the other field strengths in terms of $\Phi_\alpha$. We find a solution

$$\Gamma_{\alpha,\beta} = 0 \quad , \quad \Gamma_{\alpha,\beta\gamma} = i\, C_{\alpha(\beta}\Phi_{\gamma)} \quad ,$$

$$\Gamma_{\alpha\beta,}{}^{\gamma\delta} = \frac{1}{4}\,\delta_{(\alpha}{}^{(\gamma}\,[\, D_{\beta)}\Phi^{\delta)} + D^{\delta)}\Phi_{\beta)}\,] \quad ,$$

$$G = -\, D^\alpha \Phi_\alpha \quad . \tag{2.5.13}$$

Thus the constraints allow $\Gamma_{AB}$ to be expressed in terms of a spinor superfield $\Phi_\alpha$. (The general solution of the constraints is a gauge transform (2.5.10) of (2.5.13).)

The quantity $G$ is by definition a field strength; hence the gauge variation of $\Phi_\alpha$ must leave $G$ invariant. This implies that the gauge variation of $\Phi_\alpha$ must be (see (2.2.6))

$$\delta\Phi_\alpha = \frac{1}{2}\, D^\beta D_\alpha \Lambda_\beta \quad , \tag{2.5.14}$$

where $\Lambda_\beta$ is an arbitrary spinor gauge parameter. This gauge transformation is of course consistent with what remains of (2.5.10) after the gauge choice (2.5.13).

We expect the physical degrees of freedom to appear in the (only independent) field strength $G$. Since this is a scalar superfield, it must describe a scalar and a spinor, and $\Phi_\alpha$ (or $\Gamma_{AB}$) provides a *variant representation* of the supersymmetry algebra normally described by the scalar superfield $\Phi$. In fact $\Phi_\alpha$ contains components with helicities $0, \frac{1}{2}, \frac{1}{2}, 1$ just like the vector multiplet, but now the $\frac{1}{2}, 1$ components are auxiliary fields. ($\Phi_\alpha = \psi_\alpha + \theta_\alpha A + \theta^\beta v_{\alpha\beta} - \theta^2\chi_\alpha$). For $\Phi_\alpha$ with canonical dimension $(mass)^{\frac{1}{2}}$, on dimensional grounds the gauge invariant action must be given by

$$S = -\frac{1}{2}\int d^3x\, d^2\theta\; (D_\alpha G)^2 \quad . \tag{2.5.15}$$

Written in this form we see that in terms of the components of $G$, the action has the same form as in (2.3.2). The only differences arise because $G$ is expressed in terms of $\Phi_\alpha$. We find that only the auxiliary field $F$ is modified; it is replaced by a field $F'$. An explicit computation of this quantity yields

$$F' = -\,D^2 D^\alpha \Phi_\alpha| = i\partial^{\alpha\beta} D_\alpha \Phi_\beta| \equiv \partial^{\alpha\beta} V_{\alpha\beta}| \quad , \quad V_{\alpha\beta} \equiv \frac{1}{2}\, iD_{(\alpha}\Phi_{\beta)} \quad . \tag{2.5.16}$$

In place of $F$ the divergence of a vector appears. To see that this vector field really is a gauge field, we compute its variation under the gauge transformation (2.5.14):

$$\delta V_{\alpha\beta} = \frac{1}{4}\,\partial^\gamma{}_{(\alpha}\,[\, D_{\beta)}\,\Lambda_\gamma + D_\gamma\,\Lambda_{\beta)}] \quad . \tag{2.5.17}$$

This is not the transformation of an ordinary gauge vector (see (2.4.9)), but rather that of a second-rank antisymmetric tensor (in three dimensions a second-rank antisymmetric tensor is the same Lorentz representation as a vector). This is the component gauge field that appears at lowest order in $\theta$ in $\Gamma_{\alpha\beta,\gamma\delta}$ in eq. (2.5.13). A field of this type has no dynamics in three dimensions.

### c. Spinor gauge superfield

Superforms are not the only gauge multiplets one can study, but the pattern for other cases is similar. In general, (nonvariant) supersymmetric gauge multiplets can be described by spinor superfields carrying additional internal-symmetry group indices. (In a particular case, the additional index can be a spinor index: see below.) Such superfields contain component gauge fields and, as in the Yang-Mills case, their gauge transformations are determined by the $\theta = 0$ part of the superfield gauge parameter (cf. (2.4.9)). The gauge superfield thus takes the form of the component field with a vector index replaced by a spinor index, and the transformation law takes the form of the component transformation law with the vector derivative replaced by a spinor derivative.

For example, to describe a multiplet containing a spin $\frac{3}{2}$ component gauge field, we introduce a spinor gauge superfield with an additional spinor group index:

$$\delta\Phi_\mu{}^\alpha = D_\mu K^\alpha \quad . \tag{2.5.18}$$

The field strength has the same form as the vector multiplet field strength but with a spinor group index:

$$W_\alpha{}^\beta = \frac{1}{2} D^\gamma D_\alpha \Phi_\gamma{}^\beta \quad . \tag{2.5.19}$$

(We can, of course, introduce a supervector potential $\Gamma_M{}^\alpha$ in exact analogy with the abelian vector multiplet. The field strength here simply has an additional spinor index. The constraints are exactly the same as for the vector multiplet, i.e., $F_{\alpha\beta}{}^\gamma = 0$.)

In three dimensions massless fields of spin greater than 1 have no dynamical degrees of freedom. The kinetic term for this multiplet is analogous to the *mass term* for the vector multiplet:

$$S \sim \int d^3x d^2\theta \; W^{\alpha\beta} \Phi_{\alpha\beta} \quad . \tag{2.5.20}$$

This action describes component fields which are all auxiliary: a spin $\frac{3}{2}$ gauge field $\psi_{(\alpha\beta)\gamma}$, a vector, and a scalar, as can be verified by expanding in components. The invariance of the action in (2.5.20) is not manifest: It depends on the Bianchi identity $D^\alpha W_{\alpha\beta} = 0$. The explicit appearance of the superfield $\Phi_{\alpha\beta}$ is a general feature of supersymmetric gauge theories; it is *not* always possible to write the superspace action for a gauge theory in terms of field strengths alone.

## 2.6. Supergravity

### a. Supercoordinate transformations

Supergravity, the supersymmetric generalization of gravity, is the gauge theory of the supertranslations. The global transformations with constant parameters $\xi^{\mu\nu}, \epsilon^{\mu}$ generated by $P_{\mu\nu}$ and $Q_{\mu}$ are replaced by local ones parametrized by the supervector $K^M(x,\theta) = (K^{\mu\nu}, K^{\mu})$. For a scalar superfield $\Psi(x,\theta)$ we define the transformation

$$\Psi(z) \to \Psi'(z) = e^{iK}\, \Psi(z) = e^{iK}\, \Psi(z)\, e^{-iK} \quad , \tag{2.6.1}$$

where

$$K = K^M\, iD_M = K^{\mu\nu}\, i\partial_{\mu\nu} + K^{\mu}\, iD_{\mu} \quad . \tag{2.6.2}$$

(To exhibit the global supersymmetry, it is convenient to write $K$ in terms of $D_\mu$ rather than $Q_\mu$ (or $\partial_\mu$). This amounts to a redefinition of $K^{\mu\nu}$). The second form of the transformation of $\Psi$ can be shown to be equivalent to the first by comparing terms in a power series expansion of the two forms and noting that $iK\,\Psi = [iK, \Psi]$. It is easy to see that (2.6.1) is a general coordinate transformation in superspace: $e^{iK}\Psi(z)e^{-iK} = \Psi(e^{iK} z e^{-iK})$; defining $z' \equiv e^{-iK} z e^{iK}$, (2.6.1) becomes $\Psi'(z') = \Psi(z)$.

We may expect, by analogy to the Yang-Mills case, to introduce a gauge superfield ${H_\alpha}^M$ with (linearized) transformation laws

$$\delta {H_\alpha}^M = D_\alpha\, K^M \quad , \tag{2.6.3}$$

(we introduce ${H_{\alpha\beta}}^M$ as well, but a constraint will relate it to ${H_\alpha}^M$) and define covariant derivatives by analogy to (2.4.28):

$$E_A = D_A + {H_A}^M\, D_M = {E_A}^M\, D_M \quad . \tag{2.6.4}$$

${E_A}^M$ is the *vielbein*. The potentials ${H_\alpha}^{\mu\nu}, {H_\alpha}^{\mu}$ have a large number of components among which we identify, according to the discussion following equation (2.5.17), a second-rank tensor (the "dreibein", minus its flat-space part) describing the graviton and a spin $\frac{3}{2}$ field describing the gravitino, whose gauge parameters are the $\theta = 0$ parts of the vector and spinor gauge superparameters $K^M|$. Other components will describe gauge degrees of freedom and auxiliary fields.

**b. Lorentz transformations**

The local supertranslations introduced so far include Lorentz transformations of a scalar superfield, acting on the coordinates $z^M = (x^{\mu\nu}, \theta^\mu)$. To define their action on spinor superfields it is necessary to introduce the concept of tangent space and local frames attached at each point $z^M$ and local Lorentz transformations acting on the indices of such superfields $\Psi_{\alpha,\beta\cdots}(z^M)$. (In chapter 5 we discuss the reasons for this procedure.) The enlarged full local group is defined by

$$\Psi_{\alpha,\beta\cdots}(x,\theta) \rightarrow \Psi'_{\alpha,\beta\cdots}(x,\theta) = e^{iK}\, \Psi_{\alpha,\beta\cdots}(x,\theta)\, e^{-iK} \quad , \tag{2.6.5}$$

where now

$$K = K^M\, iD_M + K_\alpha{}^\beta\, iM_\beta{}^\alpha \quad . \tag{2.6.6}$$

Here the superfield $K_\alpha{}^\beta$ parametrizes the local Lorentz transformations and the Lorentz generators $M_\beta{}^\alpha$ act on each tangent space index as indicated by

$$[X_\beta{}^\gamma\, M_\gamma{}^\beta, \Psi_\alpha] = X_\alpha{}^\beta \Psi_\beta \quad , \tag{2.6.7}$$

for arbitrary $X_\beta{}^\gamma$. $M_{\alpha\beta}$ is symmetric, i.e., $M_\alpha{}^\beta$ is traceless (which makes it equivalent to a vector in three dimensions). Thus, $X_\alpha{}^\beta$ is an element of the Lorentz algebra $SL(2,R)$ (i.e., $SO(2,1)$). Therefore, the parameter matrix $K_\alpha{}^\beta$ is also traceless.

From now on we must distinguish tangent space and world indices; to do this, we denote the former by letters from the beginning of the alphabet, and the latter by letters from the middle of the alphabet. By definition, the former transform with $K_\alpha{}^\beta$ whereas the latter transform with $K^M$.

**c. Covariant derivatives**

Having introduced local Lorentz transformations acting on spinor indices, we now define covariant spinor derivatives by

$$\nabla_\alpha = E_\alpha{}^M\, D_M + \Phi_{\alpha\beta}{}^\gamma\, M_\gamma{}^\beta \quad , \tag{2.6.8}$$

as well as vector derivatives $\nabla_{\alpha\beta}$. However, just as in the Yang-Mills case, we impose a conventional constraint that defines

$$\nabla_{\alpha\beta} = -\, i\, \frac{1}{2} \{\nabla_\alpha, \nabla_\beta\} \quad , \tag{2.6.9}$$

The *connection coefficients* $\Phi_{A\beta}{}^{\gamma}$, which appear in

$$\nabla_A = E_A{}^M D_M + \Phi_{A\beta}{}^{\gamma} M_{\gamma}{}^{\beta} \quad , \tag{2.6.10}$$

and act as gauge fields for the Lorentz group, will be determined in terms of $H_\alpha{}^M$ by imposing further suitable constraints. The covariant derivatives transform by

$$\nabla_A \to \nabla_A{}' = e^{iK} \nabla_A e^{-iK} \quad . \tag{2.6.11a}$$

All *fields* $\Psi_{...}$ (as opposed to the operator $\nabla$) transform as

$$\Psi'_{...} = e^{iK}\Psi_{...}e^{-iK} = e^{iK}\Psi_{...} \tag{2.6.11b}$$

when all indices are flat (tangent space); we always choose to use flat indices. We can use the vielbein $E_A{}^M$ (and its inverse $E_M{}^A$) to convert between world and tangent space indices. For example, if $\Psi_M$ is a world supervector, $\Psi_A = E_A{}^M \Psi_M$ is a tangent space supervector.

The superderivative $E_A = E_A{}^M D_M$ is to be understood as a tangent space supervector. On the other hand, $D_M$ transforms under the local translations (supercoordinate transformations), and this induces transformations of $E_A{}^M$ with respect to its world index (in this case, $M$). We can exhibit this, and verify that (2.6.6) describes the familiar local Lorentz and general coordinate transformations, by considering the infinitesimal version of (2.6.11):

$$\delta\nabla_A = [iK, \nabla_A] \quad , \tag{2.6.12}$$

which implies

$$\delta E_A{}^M = E_A{}^N D_N K^M - K^N D_N E_A{}^M - E_A{}^N K^P T_{PN}{}^M - K_A{}^B E_B{}^M \quad ,$$

$$\delta\Phi_{A\gamma}{}^{\delta} = E_A K_{\gamma}{}^{\delta} - K^M D_M \Phi_{A\gamma}{}^{\delta} - K_A{}^B \Phi_{B\gamma}{}^{\delta} - K_{\gamma}{}^{\epsilon} \Phi_{A\epsilon}{}^{\delta} + K_{\epsilon}{}^{\delta} \Phi_{A\gamma}{}^{\epsilon}$$

$$= \nabla_A K_{\gamma}{}^{\delta} - K^M D_M \Phi_{A\gamma}{}^{\delta} - K_A{}^B \Phi_{B\gamma}{}^{\delta} \quad , \tag{2.6.13}$$

where $T_{MN}{}^P$ is the torsion of *flat, global* superspace (2.4.10), and $K_{\alpha\beta}{}^{\gamma\delta} \equiv \frac{1}{2} K_{(\alpha}{}^{(\gamma}\,\delta_{\beta)}{}^{\delta)}$. The first three terms in the transformation law of $E_A{}^M$ correspond to the usual form of the general coordinate transformation of a world supervector (labeled by $M$), while the last term is a local Lorentz transformation on the tangent space index $A$. The relation between $K_{\alpha\beta}{}^{\gamma\delta}$ and $K_\alpha{}^\gamma$ implies the usual reducibility of the Lorentz transformations on the tangent space, corresponding to the definition of vectors as second-rank symmetric spinors.

## d. Gauge choices

### d.1. A supersymmetric gauge

As we have mentioned above, the gauge fields (or the vielbein ${E_A}^M$) contain a large number of gauge degrees of freedom, and some of them can be gauged away using the $K$ transformations. For simplicity we discuss this only at the linearized level (where we need not distinguish world and tangent space indices); we will return later to a more complete treatment. From (2.6.13) the linearized transformation laws are

$$\delta {E_\alpha}^\mu = D_\alpha K^\mu - {K_\alpha}^\mu \quad ,$$

$$\delta {E_\alpha}^{\mu\nu} = D_\alpha K^{\mu\nu} - i {\delta_\alpha}^{(\mu} K^{\nu)} \quad . \qquad (2.6.14)$$

Thus ${K_\alpha}^\mu$ can be used to gauge away all of ${E_\alpha}^\mu$ except its trace (recall that ${K_\alpha}^\mu$ is traceless) and $K^\mu$ can gauge away part of ${E_\alpha}^{\mu\nu}$. In the corresponding gauge we can write

$${E_\alpha}^\mu = {\delta_\alpha}^\mu \Psi \quad ,$$

$${E_\alpha}^{\alpha\mu} = 0 \quad ; \qquad (2.6.15)$$

this *globally supersymmetric* gauge is maintained by further transformations restricted by

$${K_\alpha}^\beta = \frac{1}{2} D_{(\alpha} K^{\beta)} \equiv D_\alpha K^\beta - \frac{1}{2} {\delta_\alpha}^\beta D_\gamma K^\gamma \quad ,$$

$$K^\mu = -\frac{i}{3} D_\nu K^{\mu\nu} \quad . \qquad (2.6.16)$$

Under these restricted transformations we have

$$\delta\Psi = \frac{1}{6} \partial_{\mu\nu} K^{\mu\nu} \quad ,$$

$$\delta E^{(\mu,\nu\sigma)} = D^{(\mu} K^{\nu\sigma)} \quad . \qquad (2.6.17)$$

In this gauge the traceless part $h^{(\mu\nu,\rho\sigma)}$ of the ordinary dreibein (the physical graviton field) appears in $E^{(\mu,\nu\sigma)}$. The trace $h = h_{\mu\nu}{}^{\mu\nu}$ is contained in (the $\theta = 0$ part of) $\Psi$ and has an identical (linearized) transformation law. (In super "conformal" theories the vielbein also undergoes a superscale transformation whose scalar parameter can be used to gauge $\Psi$ to 1, still in a globally supersymmetric way. Thus $E^{(\mu,\nu\sigma)}$ contains the "conformal" part of the supergravity multiplet, whereas $\Psi$ contains the traces.)

### d.2. Wess-Zumino gauge

The above gauge is convenient for calculations where we wish to maintain manifest global supersymmetry. However just as in super Yang-Mills theory, we can find a non-supersymmetric Wess-Zumino gauge that exhibits the component field content of supergravity most directly. In such a gauge

$$\Psi = h + \theta^{\mu}\,\psi_{\mu} - \theta^2\,a \quad ,$$

$$E^{(\mu,\nu\rho)} = \theta_{\tau}\,h^{(\mu\nu\rho\tau)} - \theta^2\,\psi^{(\mu\nu\rho)} \quad , \tag{2.6.18}$$

where $h$ and $h^{(\mu\nu\rho\tau)}$ are the remaining parts of the dreibein, $\psi_{\mu}$ and $\psi^{(\mu\nu\rho)}$ of the gravitino, and $a$ is a scalar auxiliary field. The residual gauge invariance (which maintains the above form) is parametrized by

$$K^{\mu\nu} = \xi^{\mu\nu} + \theta^{(\mu}\,\epsilon^{\nu)} \quad , \tag{2.6.19}$$

where $\xi^{\mu\nu}(x)$ parametrizes general spacetime coordinate transformations and $\epsilon^{\nu}(x)$ parametrizes local (component) supersymmetry transformations.

## e. Field strengths

We now return to a study of the geometrical objects of the theory. The field strengths for supergravity are supertorsions $T_{AB}{}^{C}$ and supercurvatures $R_{AB\gamma}{}^{\delta}$, defined by

$$[\nabla_A\,,\nabla_B\} \equiv T_{AB}{}^{C}\nabla_C + R_{AB\gamma}{}^{\delta}M_{\delta}{}^{\gamma} \quad . \tag{2.6.20}$$

Our determination of $\nabla_{\alpha\beta}$ in terms of $\nabla_{\alpha}$ (see (2.6.9) ), is equivalent to the constraints

$$T_{\alpha\beta}{}^{\gamma\delta} = i\delta_{(\alpha}{}^{\gamma}\delta_{\beta)}{}^{\delta} \quad , \quad T_{\alpha\beta}{}^{\gamma} = R_{\alpha\beta\gamma}{}^{\delta} = 0 \quad . \tag{2.6.21}$$

We need one further constraint to relate the connection $\Phi_{\alpha\beta}{}^{\gamma}$ (the gauge field for the local Lorentz transformations) to the gauge potential $H_\alpha{}^M$ (or vielbein $E_\alpha{}^M$). It turns out that such a constraint is

$$T_{\alpha,\beta\gamma}{}^{\delta\epsilon} = 0 \ . \tag{2.6.22}$$

To solve this constraint, and actually find $\Phi$ in terms of $E_\alpha{}^M$ it is convenient to make some additional definitions:

$$\check{E}_\alpha \equiv E_\alpha \quad , \qquad \check{E}_{\alpha\beta} \equiv -\frac{i}{2}\{\check{E}_\alpha\, , \check{E}_\beta\} \quad ,$$

$$[\check{E}_A\, , \check{E}_B\} \equiv \check{C}_{AB}{}^C \check{E}_C \quad . \tag{2.6.23}$$

The constraint (2.6.22) is then solved for $\Phi_{\alpha\beta}{}^\gamma$ as follows: First, express $[\nabla_\alpha, \nabla_{\beta\gamma}]$ in terms of $\Phi_{\alpha\beta}{}^\gamma$ and the "check" objects of (2.6.23) using (2.6.9). Then, find the coefficient of $\check{E}_{\alpha\beta}$ in this expression. The corresponding coefficient of the right-hand side of (2.6.20) is $T_{\alpha,\beta\gamma}{}^{\delta\epsilon}$. This gives us the equation

$$\begin{aligned} T_{\alpha,\beta\gamma}{}^{\delta\epsilon} &= \check{C}_{\alpha,\beta\gamma}{}^{\delta\epsilon} - \frac{1}{2}\Phi_{(\beta\gamma)}{}^{(\delta}\delta_\alpha{}^{\epsilon)} + \frac{1}{2}\Phi_{\alpha(\beta}{}^{(\delta}\delta_{\gamma)}{}^{\epsilon)} \\ &= \check{C}_{\alpha,\beta\gamma}{}^{\delta\epsilon} - \frac{1}{2}C_{\alpha(\beta}\Phi^{(\delta\epsilon)}{}_{\gamma)} = 0 \quad . \end{aligned} \tag{2.6.24}$$

(From the Jacobi identity $[\check{E}_{(\alpha}\, , \{\check{E}_\beta\, , \check{E}_{\gamma)}\}] = 0$, we have, independent of (2.6.21,22), $\check{C}_{(\alpha,\beta\gamma)}{}^{\delta\epsilon} = 0$.) We then solve for $\Phi_{\alpha\beta}{}^\gamma$: We multiply (2.6.24) by $C^{\alpha\beta}$ and use the identity $\Phi_{\alpha\beta}{}^\gamma = \frac{1}{2}(\Phi_{(\alpha\beta)}{}^\gamma - C_{\alpha\beta}\Phi^{(\gamma\delta)}{}_\delta)$. We find

$$\Phi_{\alpha\beta\gamma} = \frac{1}{3}(\check{C}^{\delta,}{}_{\delta\alpha,\beta\gamma} - \check{C}^{\delta,}{}_{\delta(\beta,\gamma)\alpha}) \quad , \tag{2.6.25}$$

the $\check{C}$'s being calculable from (2.6.23) as derivatives of $E_\alpha{}^M$.

## f. Bianchi identities

The torsions and curvatures are covariant and must be expressible only in terms of the physical gauge invariant component field strengths for the graviton and gravitino and auxiliary fields. We proceed in two steps: First, we express all the $T$'s and $R$'s in (2.6.20) in terms of a small number of independent field strengths; then, we analyze the content of these superfields.

The Jacobi identities for the covariant derivatives explicitly take the form:

$$[\,[\,\nabla_{[A}\,,\nabla_B\,\}\,,\nabla_{C)}\,\}=0 \quad . \tag{2.6.26}$$

The presence of the constraints in (2.6.21,22) allows us to express all of the nontrivial torsion and curvature tensors completely in terms of two superfields $R$ and $G_{\alpha\beta\gamma}$ (where $G_{\alpha\beta\gamma}$ is totally symmetric), and their spinorial derivatives. This is accomplished by algebraically solving the constraints plus Jacobi identities (which are the Bianchi identities for the torsions and curvatures). We either repeat the calculations of the Yang-Mills case, or we make use of the results there, as follows:

We observe that the constraint (2.6.21) $\{\nabla_\alpha,\nabla_\beta\}=2i\nabla_{\alpha\beta}$ is identical to the Yang-Mills constraint (2.4.13,30a). The Jacobi identity $[\nabla_{(\alpha}\{\nabla_\beta,\nabla_{\gamma)}\}]=0$ has the same solution as in (2.4.17-20a,31a):

$$[\nabla_\alpha,\nabla_{\beta\gamma}]=C_{\alpha(\beta}W_{\gamma)} \quad , \tag{2.6.27}$$

where $W_\alpha$ is expanded over the supergravity "generators" $i\nabla$ and $iM$ (the factor $i$ is introduced to make the generators hermitian):

$$W_\alpha=W_\alpha{}^\beta i\nabla_\beta+\widetilde{W}_\alpha{}^{\beta\gamma}i\nabla_{\beta\gamma}+W_{\alpha\beta}{}^\gamma iM_\gamma{}^\beta \quad . \tag{2.6.28}$$

The solution to the Bianchi identities is thus (2.4.33), with the identification (2.6.28). The constraint (2.6.22) implies $\widetilde{W}_\alpha{}^{\beta\gamma}=0$, and we can "solve" $\{\nabla^\alpha,W_\alpha\}=0$ (see (2.4.33b)) explicitly:

$$W_{\alpha\beta}=-\,C_{\alpha\beta}R \quad , \quad W_{\alpha\beta\gamma}=G_{\alpha\beta\gamma}+\frac{1}{3}\,C_{\alpha(\beta}\nabla_{\gamma)}R \quad , \quad \nabla^\alpha G_{\alpha\beta\gamma}=-\,\frac{2}{3}\,i\nabla_{\beta\gamma}R \quad , \tag{2.6.29}$$

where we have introduced a scalar $R$ and a totally symmetric spinor $G_{\alpha\beta\gamma}$. The full solution of the Bianchi identities is thus the Yang-Mills solution (2.4.33) with the substitutions

$$-iW_\alpha=-\,R\nabla_\alpha+\frac{2}{3}\,(\nabla_\beta R)M_\alpha{}^\beta+G_{\alpha\beta}{}^\gamma M_\gamma{}^\beta$$

$$\nabla^\alpha G_{\alpha\beta\gamma}=-\,\frac{2}{3}\,i\nabla_{\beta\gamma}R$$

$$-if_{\alpha\beta}=-\,\frac{1}{3}\,(\nabla_{(\alpha}R)\nabla_{\beta)}+G_{\alpha\beta}{}^\gamma\nabla_\gamma-2R\,i\nabla_{\alpha\beta}+\frac{2}{3}\,(\nabla^2R)M_{\alpha\beta}$$

$$+\frac{1}{2}(i\nabla_{\gamma(\alpha}R)M_{\beta)}{}^{\gamma}+W_{\alpha\beta\gamma}{}^{\delta}M_{\delta}{}^{\gamma} \qquad (2.6.30)$$

where $W_{\alpha\beta\gamma\delta}\equiv\frac{1}{4!}\nabla_{(\alpha}G_{\beta\gamma\delta)}$. We have used $\nabla_\alpha\nabla_\beta=i\nabla_{\alpha\beta}-C_{\alpha\beta}\nabla^2$ to find (2.6.30). Individual torsions and curvatures can be read directly from these equations by comparing with the definition (2.6.20). Thus, for example, we have

$$R_{\alpha\beta,}{}^{\gamma\delta,}{}_{\epsilon}{}^{\zeta}=\frac{1}{2}\delta_{(\alpha}{}^{(\gamma}\, r_{\beta)}{}^{\delta)}{}_{\epsilon}{}^{\zeta} \quad ,$$

$$r_{\alpha\beta}{}^{\gamma\delta}\equiv W_{\alpha\beta}{}^{\gamma\delta}-\frac{1}{3}\delta_{(\alpha}{}^{\gamma}\delta_{\beta)}{}^{\delta}\,\nabla^2R+\frac{1}{4}\delta_{(\alpha}{}^{(\gamma}\,i\nabla_{\beta)}{}^{\delta)}\,R \quad . \qquad (2.6.31)$$

The $\theta$-independent part of $r$ is the Ricci tensor in a spacetime geometry with ($\theta$-independent) torsion.

In sec. 2.4.a.3 we discussed covariant shifts of the gauge potential. In *any* gauge theory such shifts do not change the transformation properties of the covariant derivatives and thus are perfectly acceptable; the shifted gauge fields provide an equally good description of the theory. In sec. 2.4.a.3 we used the redefinitions to eliminate a field strength. Here we redefine the connection $\Phi_{\alpha\beta,\gamma}{}^{\delta}$ to eliminate $T_{\alpha\beta,\gamma\delta}{}^{\epsilon\zeta}$ by

$$\nabla'_{\alpha\beta}=\nabla_{\alpha\beta}-iRM_{\alpha\beta} \quad . \qquad (2.6.32)$$

(This corresponds to shifting $\Phi_{\underline{abc}}$ by a term $\sim\epsilon_{\underline{abc}}R$ to cancel $T_{\underline{abc}}$; we temporarily make use of vector indices "$\underline{a}$" to represent traceless bispinors since this makes it clear that the shift (2.6.32) is possible only in three dimensions.) The shifted $r_{\alpha\beta}{}^{\gamma\delta}$, dropping primes, is

$$r_{\alpha\beta}{}^{\gamma\delta}=W_{\alpha\beta}{}^{\gamma\delta}-\frac{1}{4}\delta_{(\alpha}{}^{\gamma}\delta_{\beta)}{}^{\delta}\,r \quad , \quad r\equiv\frac{4}{3}\nabla^2R+2R^2 \quad . \qquad (2.6.33)$$

This redefinition of $\Phi_{\alpha\beta,\gamma}{}^{\delta}$ is equivalent to replacing the constraint (2.6.9) with

$$\{\nabla_\alpha\,,\nabla_\beta\}=2i\,\nabla_{\alpha\beta}-2RM_{\alpha\beta} \quad . \qquad (2.6.34)$$

We will find that the analog of the "new" term appears in the constraints for four dimensional supergravity (see chapter 5). This is because we can obtain the three dimensional theory from the four dimensional one, and there is no shift analogous to (2.6.32) possible in four dimensions.

The superfields $R$ and $G_{\alpha\beta\gamma}$ are the variations of the supergravity action (see below) with respect to the two unconstrained superfields $\Psi$ and $E^{(\mu,\nu\sigma)}$ of (2.6.15-17).

The field equations are $R = G_{\alpha\beta\gamma} = 0$; these are solved only by flat space (just as for ordinary gravity in three-dimensional spacetime), so three-dimensional supergravity has no dynamics (all fields are auxiliary).

## g. Actions

We now turn to the construction of actions and their expansion in terms of component fields. As we remarked earlier, in flat superspace the integral of *any* (scalar) superfield expression with the $d^3xd^2\theta$ measure is *globally* supersymmetric. This is no longer true for locally supersymmetric theories. (The new features that arise are not specifically limited to local supersymmetry, but are a general consequence of local coordinate invariance).

We recall that in our formalism an arbitrary "matter" superfield $\Psi$ transforms according to the rule

$$\Psi' = e^{iK}\,\Psi\,e^{-iK} = e^{-i\overleftarrow{K}}\;\Psi\;e^{i\overleftarrow{K}} \quad ,$$

$$\overleftarrow{K} = K^M\, i\overleftarrow{D}_M + K_\alpha{}^\beta i\overleftarrow{M}_\beta{}^\alpha \quad , \tag{2.6.35}$$

where $\overleftarrow{D}_M$ means that we let the differential operator act on everything to its left. (The various forms of the transformation law can be seen to be equivalent after power series expansion of the exponentials, or by multiplying by a test function and integrating by parts). Lagrangians are scalar superfields, and since any Lagrangian $I\!\!L$ is constructed from superfields and $\nabla$ operators, a Lagrangian transforms in the same way.

$$I\!\!L' = e^{iK}\, I\!\!L e^{-iK} = e^{-i\overleftarrow{K}}\; I\!\!L\, e^{i\overleftarrow{K}} \quad . \tag{2.6.36}$$

Therefore the integral $\int d^3x\, d^2\theta\; I\!\!L$ is *not* invariant with respect to our gauge group. To find invariants, we consider the vielbein as a square supermatrix in its indices and compute its superdeterminant $E$. The following result will be derived in our discussion of four-dimensions (see sec. 5.1):

$$\begin{aligned}(E^{-1})' &= e^{iK} E^{-1} e^{-iK} (1 \cdot e^{i\overleftarrow{K}}) \\ &= E^{-1}\, e^{i\overleftarrow{K}} \quad .\end{aligned} \tag{2.6.37}$$

Therefore the product $E^{-1}\, I\!\!L$ transforms in exactly the same way as $E^{-1}$:

$$(E^{-1}\, I\!\!L)' = E^{-1}\, I\!\!L\, e^{i\overleftarrow{K}} \quad . \tag{2.6.38}$$

Since every term but the first one in the power series expansion of the $e^{i\overleftarrow{K}}$ is a total derivative, we conclude that up to surface terms

$$S = \int d^3x\, d^2\theta\; E^{-1}\, I\!\!L \quad , \tag{2.6.39}$$

is invariant. We therefore have a simple prescription for turning any globally supersymmetric action into a locally supersymmetric one:

$$[\, I\!\!L(D_A\Phi\, , \Phi)]_{global} \rightarrow E^{-1}\, I\!\!L(\nabla_A\Phi\, , \Phi) \quad , \tag{2.6.40}$$

in analogy to ordinary gravity. Thus, the action for the scalar multiplet described by eq. (2.3.5) takes the covariantized form

$$S_\Phi = \int d^3x\, d^2\theta\; E^{-1}\, [\, -\frac{1}{2}\, (\nabla_\alpha\Phi)^2 + \frac{1}{2}\, m\Phi^2 + \frac{\lambda}{3!}\, \Phi^3] \quad . \tag{2.6.41}$$

For vector gauge multiplets the simple prescription of replacing flat derivatives $D_A$ by gravitationally covariant ones $\nabla_A$ is sufficient to convert global actions into local actions, if we include the Yang-Mills generators in the covariant derivatives, so that they are covariant with respect to both supergravity and super-Yang-Mills invariances. However, such a procedure is not sufficient for more general gauge multiplets, and in particular the superforms of sec. 2.5. On the other hand, it is possible to formulate *all* gauge theories within the superform framework, at least at the abelian level (which is all that is relevant for $p$-forms for $p > 1$). Additional terms due to the geometry of the space will automatically appear in the definitions of field strengths. Specifically, the curved-space formulation of superforms is obtained as follows: The definitions (2.5.8) hold in arbitrary superspaces, independent of any metric structure. Converting (2.5.8) to a tangent-space basis with the curved space $E_A{}^M$, we obtain equations that differ from (2.5.9) only by the replacement of the flat-space covariant derivatives $D_A$ with the curved-space ones $\nabla_A$.

To illustrate this, let us return to the abelian vector multiplet, now in the presence of supergravity. The field strength for the vector multiplet is a 2-form:

$$F_{\alpha\beta} = \nabla_\alpha \Gamma_\beta + \nabla_\beta \Gamma_\alpha - 2i\Gamma_{\alpha\beta} \ ,$$

$$F_{\alpha\beta,\gamma} = \nabla_{\alpha\beta}\Gamma_\gamma - \nabla_\gamma \Gamma_{\alpha\beta} - T_{\alpha\beta,\gamma}{}^\epsilon \Gamma_\epsilon \ ,$$

$$F_{\alpha\beta,\gamma\delta} = \nabla_{\alpha\beta}\Gamma_{\gamma\delta} - \nabla_{\gamma\delta}\Gamma_{\alpha\beta} - T_{\alpha\beta,\gamma\delta}{}^E \Gamma_E \ . \tag{2.6.42}$$

We again impose the constraint $F_{\alpha\beta} = 0$, which implies

$$F_{\alpha,\beta\gamma} = iC_{\alpha(\beta}W_{\gamma)} \ , \quad W_\alpha = \frac{1}{2}\nabla^\beta \nabla_\alpha \Gamma_\beta + R\Gamma_\alpha \ ; \tag{2.6.43}$$

where we have used (2.6.30) substituted into (2.4.33). Comparing this to the global field strength defined in (2.4.20), we see that a new term proportional to $R$ appears. The extra term in $W_\alpha$ is necessary for gauge invariance due to the identity $\nabla^\alpha \nabla_\beta \nabla_\alpha = i\,\frac{2}{3}\,[\nabla^\alpha, \nabla_{\alpha\beta}]$. In the global limit the commutator vanishes, but in the local case it gives a contribution that is precisely canceled by the contribution of the $R$ term. These results can also be obtained by use of derivatives that are covariant with respect to both supergravity and super-Yang-Mills.

We turn now to the action for the gauge fields of local supersymmetry. We expect to construct it out of the field strengths $G_{\alpha\beta\gamma}$ and $R$. By dimensional analysis (noting that $\kappa$ has dimensions $(mass)^{-\frac{1}{2}}$ in three dimensions), we deduce for the Poincaré supergravity action the supersymmetric generalization of the Einstein-Hilbert action:

$$S_{SG} = -\frac{2}{\kappa^2}\int d^3x\, d^2\theta\ E^{-1}\, R \ . \tag{2.6.44}$$

We can check that (2.6.44) leads to the correct component action as follows: $\int d^2\theta\, E^{-1}\, R \simeq \nabla^2 R \simeq \frac{3}{4}\, r$ (see (2.6.33)), and thus the gravitational part of the action is correct. We can also add a supersymmetric cosmological term

$$S_{cosmo} = \frac{\lambda}{\kappa^2}\int d^3x\, d^2\theta\ E^{-1} \ , \tag{2.6.45}$$

which leads to an equations of motion $R = \lambda\,,\ G_{\alpha\beta\gamma} = 0$. The only solution to this equation (in three dimensions) is empty anti-deSitter space: From (2.6.33), $r = 2\lambda^2\,, W_{\alpha\beta\gamma\delta} = 0$.

Higher-derivative actions are possible by using other functions of $G_{\alpha\beta\gamma}$ and $R$. For example, the analog of the gauge-invariant mass term for the Yang-Mills multiplet exists here and is obtained by the replacements in (2.4.38) (along with, of course, $\int d^3x\, d^2\theta \to \int d^3x\, d^2\theta\ E^{-1}$):

$$\Gamma_A{}^i T_i \to \Phi_{A\beta}{}^\gamma\, iM_\gamma{}^\beta \quad , \qquad W_\alpha{}^i T_i \to G_{\alpha\beta}{}^\gamma iM_\gamma{}^\beta + \frac{2}{3}(\nabla_\beta R) iM_\alpha{}^\beta \quad . \tag{2.6.46}$$

This gives

$$\mathbb{L}_{mass} = \int d^3x\, d^2\theta\ E^{-1}\Phi^\alpha{}_\gamma{}^\delta \left( G_{\alpha\delta}{}^\gamma + \frac{2}{3}\delta_\alpha{}^\gamma \nabla_\delta R - \frac{1}{6}\Phi^\epsilon{}_\delta{}^\eta \Phi_{(\alpha\epsilon)\eta}{}^\gamma \right) \quad . \tag{2.6.47}$$

## 2.7. Quantum superspace

### a. Scalar multiplet

In this section we discuss the derivation of the Feynman rules for three-dimensional superfield perturbation theory. Since the starting point, the superfield action, is so much like a component (ordinary field theory) action, it is possible to read off the rules for doing Feynman supergraphs almost by inspection. However, as an introduction to the four-dimensional case we use the full machinery of the functional integral. After deriving the rules we apply them to some one-loop graphs. The manipulations that we perform on the graphs are typical and illustrate the manner in which superfields handle the cancellations and other simplifications due to supersymmetry. For more details, we refer the reader to the four-dimensional discussion in chapter 6.

#### a.1. General formalism

The Feynman rules for the scalar superfield can be read directly from the Lagrangian: The propagator is defined by the quadratic terms, and the vertices by the interactions. The propagator is an operator in both $x$ and $\theta$ space, and at the vertices we integrate over both $x$ and $\theta$. By Fourier transformation we change the $x$ integration to loop-momentum integration, but we leave the $\theta$ integration alone. ($\theta$ can also be Fourier transformed, but this causes little change in the rules: see sec. 6.3.) We now derive the rules from the functional integral.

We begin by considering the generating functional for the massive scalar superfield $\Phi$ with arbitrary self-interaction :

$$
\begin{aligned}
Z(J) &= \int I\!D\Phi \; exp \int d^3x d^2\theta \; [\tfrac{1}{2}\Phi D^2\Phi + \tfrac{1}{2} m\Phi^2 + f(\Phi) + J\Phi] \\
&= \int I\!D\Phi \; exp \; [S_0(\Phi) + S_{INT}(\Phi) + \int J\Phi] \\
&= exp\; [S_{INT}(\frac{\delta}{\delta J})] \int I\!D\Phi \; exp \; [\int \tfrac{1}{2}\Phi(D^2 + m)\Phi + J\Phi] \quad . \qquad (2.7.1)
\end{aligned}
$$

In the usual fashion we complete the square, do the (functional) Gaussian integral over $\Phi$, and obtain

$$Z(J) = exp\ [S_{INT}(\frac{\delta}{\delta J})]\ exp\ [-\int d^3xd^2\theta\,\frac{1}{2}\,J\,\frac{1}{D^2+m}\,J] \quad . \tag{2.7.2}$$

Using eq.(2.2.6) we can write

$$\frac{1}{D^2+m} = \frac{D^2-m}{\square - m^2} \quad . \tag{2.7.3}$$

(Note $D^2$ behaves just as $\not\partial$ in conventional field theory.) We obtain, in momentum space, the following Feynman rules:

Propagator:

$$\frac{\delta}{\delta J(k,\theta)} \cdot \frac{\delta}{\delta J(-k,\theta')} \int \frac{d^3k}{(2\pi)^3}\, d^2\theta\, \frac{1}{2}\, J(k,\theta)\, \frac{D^2-m}{k^2+m^2}\, J(-k,\theta)$$

$$= \frac{D^2-m}{k^2+m^2}\, \delta^2(\theta-\theta') \quad . \tag{2.7.4}$$

Vertices: An interaction term, e.g. $\int d^3xd^2\theta\, \Phi D^\alpha \Phi D^\beta \Phi \cdots$ , gives a vertex with $\Phi$ lines leaving it, with the appropriate operators $D^\alpha$, $D^\beta$, etc. acting on the corresponding lines, and an integral over $d^2\theta$. The operators $D_\alpha$ which appear in the propagators, or are coming from a vertex and act on a specific propagator with momentum $k$ *leaving* that vertex, depend on that momentum:

$$D_\alpha = \frac{\partial}{\partial\theta^\alpha} + \theta^\beta k_{\alpha\beta} \quad . \tag{2.7.5}$$

In addition we have loop-momentum integrals to perform.

In general we find it convenient to calculate the effective action. It is obtained in standard fashion by a Legendre transformation on the generating functional for connected supergraphs $W(J)$ and it consists of a sum of one-particle-irreducible contributions obtained by amputating external line propagators, replacing them by external field factors $\Phi(p_i,\theta_i)$, and integrating over $p_i\,,\theta_i$. Therefore, it will have the form

$$\Gamma(\Phi) = \sum_n \frac{1}{n!} \int \frac{d^3p_1 \cdots d^3p_n}{(2\pi)^{3n}}\, d^2\theta_1 \cdots d^2\theta_n\, \Phi(p_1,\theta_1) \cdots \Phi(p_n,\theta_n)$$

$$\times\ (2\pi)^3\, \delta(\sum p_i) \prod_{loops} \int \frac{d^3k}{(2\pi)^3} \prod_{internal\ vertices} \int d^2\theta \prod propagators \prod vertices \tag{2.7.6}$$

As we have already mentioned, all of this can be read directly from the action, by analogy with the derivation of the usual Feynman rules.

The integrand in the effective action is *a priori* a nonlocal function of the $x$'s (nonpolynomial in the $p$'s) and of the $\theta_1, \cdots \theta_n$ . However, we can manipulate the $\theta$-integrations so as to exhibit it explicitly as a functional of the $\Phi$'s all evaluated at a single common $\theta$ as follows: A general multiloop integral consists of vertices labeled $i$, $i+1$, connected by propagators which contain factors $\delta(\theta_i - \theta_{i+1})$ with operators $D_\alpha$ acting on them. Consider a particular loop in the diagram and examine one line of that loop. The factors of $D$ can be combined by using the result ("transfer" rule):

$$D_\alpha(\theta_i, k)\delta(\theta_i - \theta_{i+1}) = -\, D_\alpha(\theta_{i+1}, -k)\delta(\theta_i - \theta_{i+1}) \quad , \tag{2.7.7}$$

as well as the rules of eq.(2.2.6), after which we have at most two factors of $D$ acting at one end of the line. At the vertex where this end is attached these $D$'s can be integrated by parts onto the other lines (or external fields) using the Leibnitz rule (and some care with minus signs since the $D$'s anticommute). Then the particular $\delta$-function no longer has any derivatives acting on it and can be used to do the $\theta_i$ integration, thus effectively "shrinking" the $(\theta_i, \theta_{i+1})$ line to a point in $\theta$-space. We can repeat this procedure on each line of the loop, integrating by parts one at a time and shrinking. This will generate a sum of terms, from the integration by parts. The procedure stops when in each term we are left with exactly two lines, one with $\delta(\theta_1 - \theta_m)$ which is free of any derivatives, and one with $\delta(\theta_m - \theta_1)$ which may carry zero, one, or two derivatives. We now use the rules (which follow from the definition $\delta^2(\theta) = -\,\theta^2$),

$$\delta^2(\theta_1 - \theta_m)\delta^2(\theta_m - \theta_1) = 0 \quad ,$$

$$\delta^2(\theta_1 - \theta_m)\, D^\alpha \delta^2(\theta_m - \theta_1) = 0 \quad ,$$

$$\delta^2(\theta_1 - \theta_m)\, D^2 \delta^2(\theta_m - \theta_1) = \delta^2(\theta_1 - \theta_m) \quad . \tag{2.7.8}$$

Thus, in those terms where we are left with no $D$ or one $D$ we get zero, while in the terms in which we have a $D^2$ acting on one of the $\delta$-functions, multiplied by the other $\delta$-function, we use the above result. We are left with the single $\delta$-function, which we can use to do one more $\theta$ integration, thus finally reducing the $\theta$-space loop to a point.

The procedure can be repeated loop by loop, until the whole multiloop diagram has been reduced to *one* point in $\theta$-space, giving a contribution to the effective action

$$\Gamma(\Phi) = \int \frac{d^3p_1 \cdots d^3p_n}{(2\pi)^{3n}}\, d^2\theta$$

$$\times\, G(p_1, \cdots, p_n)\, \Phi(p_1, \theta) \cdots D^\alpha \Phi(p_i, \theta) \cdots D^2\Phi(p_j, \theta) \cdots , \qquad (2.7.9)$$

where $G$ is obtained by doing ordinary loop-momentum integrals, with some momentum factors in the numerators coming from anticommutators of $D$'s arising in the previous manipulation.

### a.2. Examples

We give now two examples, in a massless model with $\Phi^3$ interactions, to show how the $\theta$ manipulation works. The first one is the calculation of a self-energy correction represented by the graph in Fig. 2.7.1

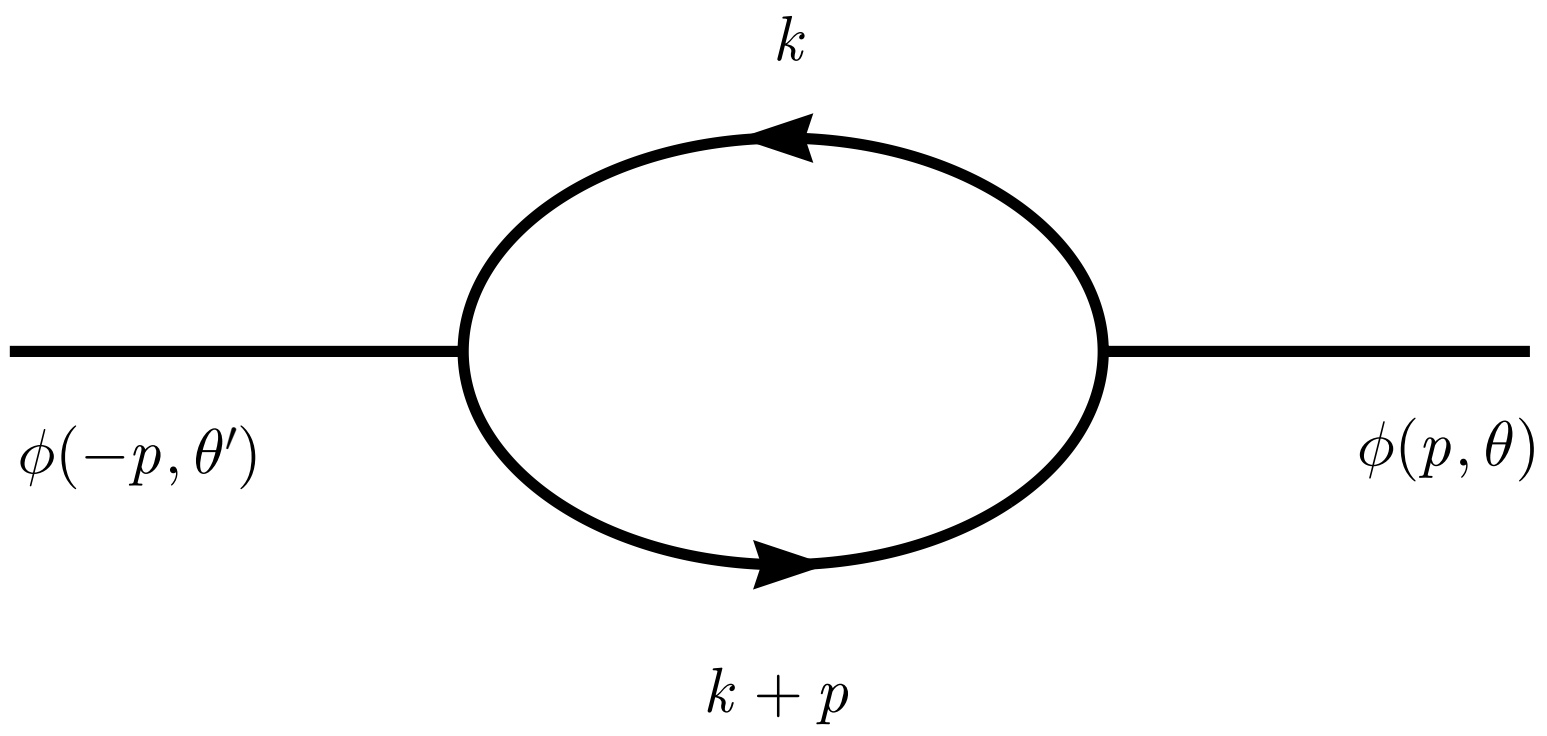

*Fig. 2.7.1*

$$\Gamma_2 = \int \frac{d^3p}{(2\pi)^3}\, d^2\theta d^2\theta' \Phi(-p, \theta')\Phi(p, \theta)\, \frac{d^3k}{(2\pi)^3}\, \frac{D^2\delta(\theta - \theta')}{k^2}\, \frac{D^2\delta(\theta' - \theta)}{(k+p)^2}\ . \qquad (2.7.10)$$

The terms involving $\theta$ can be manipulated as follows, using integration by parts:

$$D^2\delta(\theta - \theta')\, D^2\delta(\theta' - \theta)\, \Phi(p, \theta)$$

$$= -\frac{1}{2}\, D^\alpha \delta(\theta - \theta')\, [D_\alpha D^2\delta(\theta' - \theta)\Phi(p, \theta) + D^2\delta(\theta' - \theta) D_\alpha \Phi(p, \theta)]$$

$$= \delta(\theta - \theta')[(D^2)^2 \delta(\theta' - \theta)\Phi(p,\theta) + D^\alpha D^2 \delta(\theta' - \theta) D_\alpha \Phi(p,\theta)$$

$$+ D^2 \delta(\theta' - \theta) D^2 \Phi(p,\theta)] \quad . \tag{2.7.11}$$

However, using $(D^2)^2 = -k^2$ and $D^\alpha D^2 = k^{\alpha\beta} D_\beta$ we see that according to the rules in eq. (2.7.8) only the last term contributes. We find

$$\Gamma_2 = \int \frac{d^3p}{(2\pi)^3} d^2\theta \, \Phi(-p,\theta) D^2 \Phi(p,\theta) \int \frac{d^3k}{(2\pi)^3} \frac{1}{k^2(k+p)^2} \quad . \tag{2.7.12}$$

Doing the integration by parts explicitly can become rather tedious and it is preferable to perform it by indicating $D$'s and moving them directly on the graphs. We show this in Fig. 2.7.2:

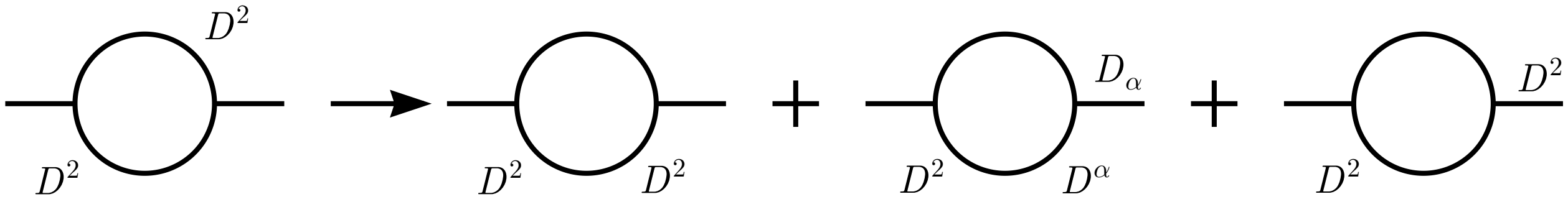

*Fig. 2.7.2*

Only the last diagram gives a contribution. One further rule is useful in this procedure: In general, after integration by parts, various $D$-factors end up in different places in the final expression and one has to worry about minus signs introduced in moving them past each other. The overall sign can be fixed at the end by realizing that we start with a particular ordering of the $D$'s and we can examine what happened to this ordering at the end of the calculation. For example, we may start with an expression such as $D^2 \cdots D^2 \cdots D^2 \cdots = \frac{1}{2} D^\alpha D_\alpha \cdots \frac{1}{2} D^\beta D_\beta \cdots \frac{1}{2} D^\gamma D_\gamma \cdots$ and end up with $D^\alpha \cdots D^\beta \cdots D^\gamma \cdots D_\alpha \cdots D_\gamma \cdots D_\beta \cdots$ where the various $D$'s act on different fields. The overall sign can obviously be determined by just counting the number of transpositions. For example, in the case above we would end up with a plus sign. Note that this rule also applies if factors such as $k_{\alpha\gamma}$ arise, provided one pays attention to the manner in which they were produced (e.g., at which end of the line were the $D$'s acting? Did it come from $D_\alpha D_\gamma$ or from $D_\gamma D_\alpha$?).

Our second example is the three-point diagram below, which we manipulate as shown in the sequence of Fig. 2.7.3:

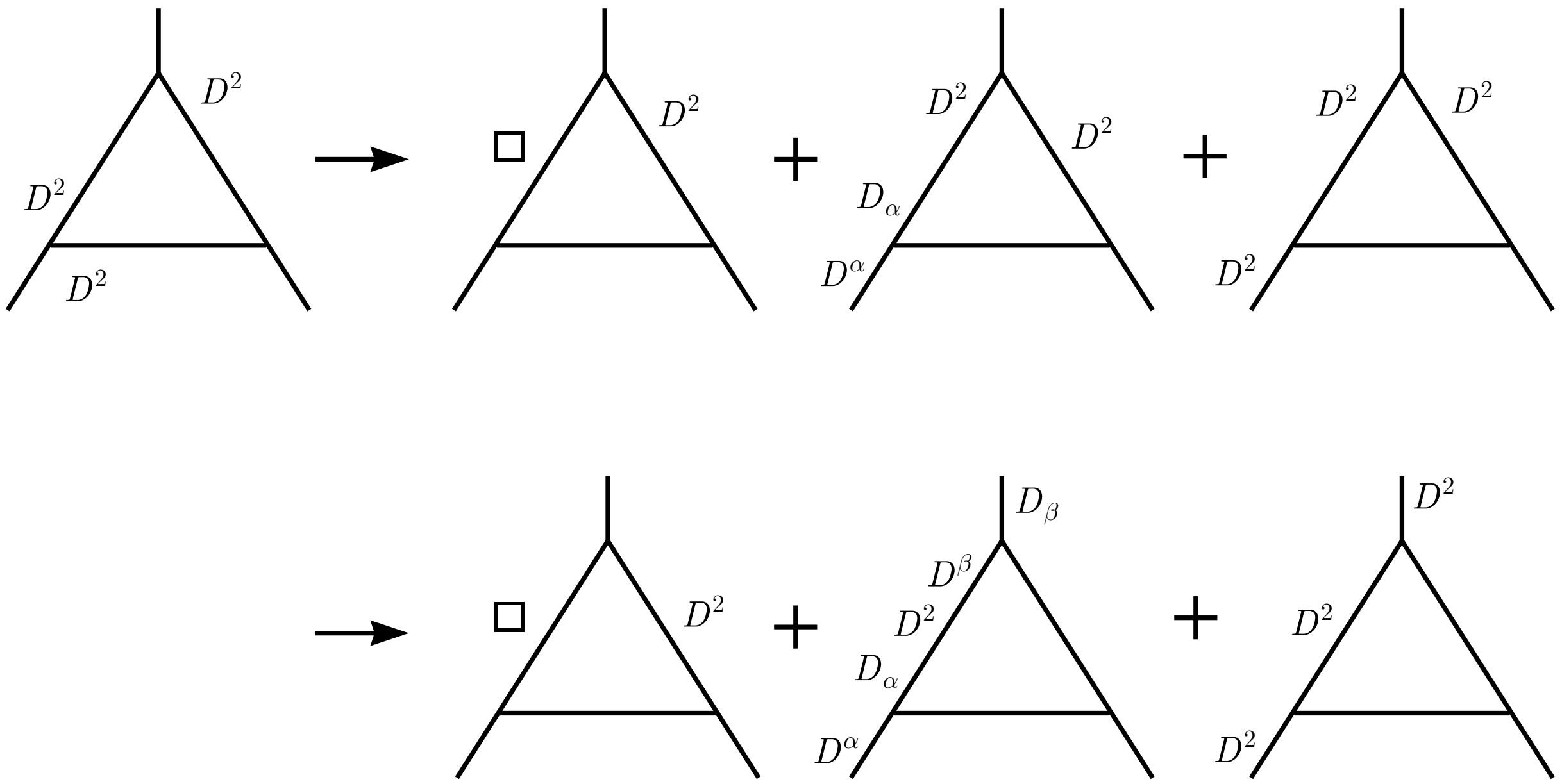

*Fig. 2.7.3*

At the first stage we have integrated by parts the $D^2$ off the bottom line and immediately replaced $(D^2)^2$ by $\Box = -\,k^2$ . At the second stage we have integrated by parts the $D^2$ off the right side, but kept only those terms that are not zero: The bottom line has already been shrunk to a point by the corresponding $\delta$-function (but we need not indicate this explicitly; any line that has no $D$'s on it can be considered as having been shrunk) and in the end we keep only terms with *exactly* two factors of $D$ in the loop. For the middle diagram this means using $D_\alpha D^2 D_\beta = D_\alpha k_{\beta\gamma} D^\gamma = -\,k_{\beta\alpha} D^2 +$ a term with no $D$'s which may be dropped. The integrand in the effective action can be written then as

$$\int \frac{d^3k}{(2\pi)^3} \frac{1}{k^2(k+p_1)^2(k-p_3)^2} \Phi(p_3,\theta)[\,-\,\Phi(p_1,\theta)\Phi(p_2,\theta)k^2$$

$$-\,D^\alpha\Phi(p_1,\theta)D^\beta\Phi(p_2,\theta)k_{\alpha\beta} + D^2\Phi(p_1,\theta)D^2\Phi(p_2,\theta)] \ , \qquad (2.7.13)$$

and only the $k$-momentum integral remains to be done.

In general, the loop-momentum integrals may have to be regularized. The procedure we use, *which is guaranteed to preserve supersymmetry,* is to do all the $D$-manipulations first, until we reduce the effective action to an integral over a single $\theta$ of an expression that is a product of superfields, and therefore manifestly supersymmetric. The remaining loop-momentum integrals may then be regularized in any manner whatsoever, e.g., by using dimensional regularization. We shall discuss the issues involved in this kind of regularization in sec. 6.6. An alternative procedure, somewhat cumbersome in its application but better understood, is supersymmetric Pauli-Villars regularization. In three dimensions this is applicable even to gauge theories, since gauge invariant mass terms exist.

**b. Vector multiplet**

Nothing new is encountered in the derivation or application of the Feynman rules. However, the derivation must be preceded by quantization, i.e., introduction of gauge-fixing terms and Faddeev-Popov ghosts, which we now discuss.

We begin with the classical action

$$S_C = \frac{1}{g^2}\, tr \int d^3x d^2\theta \, W^2 \quad . \tag{2.7.14}$$

The gauge invariance is $\delta\Gamma_\alpha = \nabla_\alpha K$ and, by direct analogy with the ordinary Yang-Mills case, we can choose the gauge-fixing function $F = \frac{1}{2}\, D^\alpha \Gamma_\alpha$ . We use an averaging procedure which leads to a gauge-fixing term without dimensional parameters, $FD^2F$, and obtain, for the quadratic action,

$$S_2 = \frac{1}{g^2}\, tr \int d^3x d^2\theta \, [\frac{1}{2}\,(\frac{1}{2}\, D^\beta D^\alpha \Gamma_\beta)\,(\frac{1}{2}\, D^\gamma D_\alpha \Gamma_\gamma)$$

$$- \frac{1}{\alpha}\,(\frac{1}{2}\, D^\beta \Gamma_\beta) D^2 (\frac{1}{2}\, D^\gamma \Gamma_\gamma)]$$

$$= \frac{1}{2}\,\frac{1}{g^2}\, tr \int d^3x d^2\theta [\frac{1}{2}\,(1 + \frac{1}{\alpha})\Gamma^\alpha \Box \Gamma_\alpha + \frac{1}{2}\,(1 - \frac{1}{\alpha})\Gamma^\alpha i\partial_\alpha{}^\beta D^2 \Gamma_\beta] \quad . \tag{2.7.15}$$

Various choices of the gauge parameter $\alpha$ are possible: The choice $\alpha = -1$ gives the kinetic term $\frac{1}{2}\,\Gamma^\alpha i\partial_\alpha{}^\beta D^2 \Gamma_\beta$ , while the choice $\alpha = 1$ gives $\frac{1}{2}\,\Gamma^\alpha \Box \Gamma_\alpha$, which results in the simplest propagator.

The Faddeev-Popov action is simply

$$S_{FP} = \frac{1}{g^2}\, tr \int d^3x d^2\theta \, c'(x,\theta)\, \frac{1}{2}\, D^\alpha \nabla_\alpha c(x,\theta) \quad , \tag{2.7.16}$$

with two scalar multiplet ghosts. (Note that in a background-field formulation of the theory, similar to the one we discuss in sec. 6.5, one would replace the operator $D^2$ in the gauge fixing term by the background-covariant operator $\boldsymbol{\nabla}^2$, and this would give rise to a *third,* Nielsen-Kallosh, ghost as well.)

The Feynman rules are now straightforward to obtain. The ghost propagator is conventional, following from the quadratic ghost kinetic term $c' D^2 c$, while the gauge field propagator is

$$\frac{\delta_\alpha{}^\beta}{k^2}\,\delta^2(\theta-\theta') \quad . \tag{2.7.17}$$

Vertices can be read off from the interaction terms. The gauge-field self-interactions (in the nonabelian case ) are

$$\begin{aligned} g^2 L_{INT} = & -\frac{i}{4}\,D^\gamma D^\alpha \Gamma_\gamma\,[\,\Gamma^\beta\,,D_\beta \Gamma_\alpha\,] - \frac{1}{12}\,D^\gamma D^\alpha \Gamma_\gamma\,[\,\Gamma^\beta\,,\{\,\Gamma_\beta\,,\Gamma_\alpha\,\}\,] \\ & - \frac{1}{8}\,[\,\Gamma^\gamma\,,D_\gamma \Gamma^\alpha\,]\,[\,\Gamma^\beta\,,D_\beta \Gamma_\alpha\,] + \frac{i}{12}\,[\,\Gamma^\gamma\,,D_\gamma \Gamma^\alpha\,]\,[\,\Gamma^\beta\,,\{\,\Gamma_\beta\,,\Gamma_\alpha\,\}\,] \\ & + \frac{1}{72}\,[\,\Gamma^\gamma\,,\{\,\Gamma_\gamma\,,\Gamma^\alpha\,\}\,]\,[\,\Gamma^\beta\,,\{\,\Gamma_\beta\,,\Gamma_\alpha\,\}\,] \quad , \end{aligned} \tag{2.7.18}$$

those of the ghosts are

$$g^2 L_{INT} = -\,i\,\frac{1}{2}\,c' D^\alpha [\Gamma_\alpha\,, c] \quad , \tag{2.7.19}$$

while those of a complex scalar field are

$$g^2 L_{INT} = \bar{\Phi}(\nabla^2 - D^2)\Phi = \bar{\Phi}[-i\Gamma^\alpha D_\alpha - i\,\frac{1}{2}\,(D^\alpha \Gamma_\alpha) - \Gamma^2]\Phi \quad . \tag{2.7.20}$$

**Contents of 3. REPRESENTATIONS OF SUPERSYMMETRY**

## 3. REPRESENTATIONS OF SUPERSYMMETRY

### 3.1. Notation

*An i for an i, and a 2 for a 2.*

We now turn to four dimensions. Our treatment will be entirely self-contained; it will not assume familiarity with our three-dimensional toy. Although supersymmetry is more complicated in four dimensions than in three, because we give a more detailed discussion, some general aspects of the theory may be easier to understand. We begin by giving the notation and conventions we use throughout the rest of the work.

#### a. Index conventions

Our index conventions are as follows: The simplest nontrivial representation of the Lorentz group, the two-component complex (Weyl) spinor representation $(\frac{1}{2},0)$ of $SL(2,C)$, is labeled by a two-valued (+ or -) lower-case Greek index (e.g., $\psi^\alpha = (\psi^+, \psi^-)$), and the complex-conjugate representation $(0,\frac{1}{2})$ is labeled by a dotted index ($\psi^{\dot\alpha} = (\psi^{\dot +}, \psi^{\dot -})$). A four-component Dirac spinor is the combination of an undotted spinor with a dotted one $(\frac{1}{2},0)\oplus(0,\frac{1}{2})$, and a Majorana spinor is a Dirac spinor where the dotted spinor is the complex conjugate of the undotted one. An arbitrary irreducible representation $(A,B)$ is then conveniently represented by a quantity with $2A$ undotted indices and $2B$ dotted indices, totally symmetric in its undotted indices and in its dotted indices. An example is the self-dual second-rank antisymmetric tensor $(1,0)$, which is represented by a second-rank symmetric spinor $f^{\alpha\beta}$. (The choice of self-dual vs. anti-self-dual follows from Wick rotation from Euclidean space, where the sign is unambiguous.)

Another example is the vector $(\frac{1}{2},\frac{1}{2})$, labeled with one undotted and one dotted index, e.g., $V^{\alpha\dot\alpha}$. A real vector satisfies the hermiticity condition $V^{\alpha\dot\beta} = \bar{V}^{\alpha\dot\beta} = \overline{V^{\beta\dot\alpha}}$. As a shorthand notation, we often use an underlined lower-case Roman index to indicate a vector index which is a composite of the corresponding undotted and dotted spinor indices: e.g., $V^{\underline{a}} \equiv V^{\alpha\dot\alpha}$. We consider such an index merely as an abbreviation: It may appear on one side of an equation while the explicit pair of spinor indices appears on the other, or it may be contracted with an explicit pair of spinor indices. When discussing Lorentz noncovariant quantities (as, e.g., in light-cone formalisms), we sometimes label the values of a vector index as follows:

$$V^{\underline{a}} = (V^{+\dot{+}}, V^{+\dot{-}}, V^{-\dot{+}}, V^{-\dot{-}}) \equiv (V^{+}, V^{T}, \overline{V}^{T}, -V^{-}) \quad , \tag{3.1.1}$$

where $\overline{V}^T$ is the complex conjugate of $V^T$, and $V^\pm$ are real in Minkowski space (but $V^+$ is the complex conjugate of $V^-$ in Wick-rotated Euclidean space). More generally, we can relate a vector label $\underset{\sim}{a}$ in an *arbitrary* basis, where $\underset{\sim}{a} \neq \alpha\dot{\alpha}$, to the $\alpha\dot{\alpha}$ basis by a set of Clebsch-Gordan coefficients, the Pauli matrices: We define

$$\textit{for fields:}\quad V^{\alpha\dot{\alpha}} = \frac{1}{\sqrt{2}}\sigma_{\underset{\sim}{b}}{}^{\alpha\dot{\alpha}}V^{\underset{\sim}{b}} \quad , \quad V^{\underset{\sim}{b}} = \frac{1}{\sqrt{2}}\sigma^{\underset{\sim}{b}}{}_{\alpha\dot{\alpha}}V^{\alpha\dot{\alpha}} \quad ;$$

$$\textit{for derivatives:}\quad \partial_{\alpha\dot{\alpha}} = \sigma^{\underset{\sim}{b}}{}_{\alpha\dot{\alpha}}\partial_{\underset{\sim}{b}} \quad , \quad \partial_{\underset{\sim}{b}} = \frac{1}{2}\sigma_{\underset{\sim}{b}}{}^{\alpha\dot{\alpha}}\partial_{\alpha\dot{\alpha}} \quad ;$$

$$\textit{for coordinates:}\quad x^{\alpha\dot{\alpha}} = \frac{1}{2}\sigma_{\underset{\sim}{b}}{}^{\alpha\dot{\alpha}}x^{\underset{\sim}{b}} \quad , \quad x^{\underset{\sim}{b}} = \sigma^{\underset{\sim}{b}}{}_{\alpha\dot{\alpha}}x^{\alpha\dot{\alpha}} \quad . \tag{3.1.2a}$$

The Pauli matrices satisfy

$$\sigma_{\underset{\sim}{b}}{}^{\alpha\dot{\alpha}}\sigma^{\underset{\sim}{c}}{}_{\alpha\dot{\alpha}} = 2\delta_{\underset{\sim}{b}}{}^{\underset{\sim}{c}} \quad , \quad \sigma^{\underset{\sim}{b}}{}_{\alpha\dot{\alpha}}\sigma_{\underset{\sim}{b}}{}^{\beta\dot{\beta}} = 2\delta_{\alpha}{}^{\beta}\delta_{\dot{\alpha}}{}^{\dot{\beta}} \quad . \tag{3.1.2b}$$

These conventions lead to an unusual normalization of the Yang-Mills gauge coupling constant $g$, since

$$\nabla_{\alpha\dot{\alpha}} \equiv \partial_{\alpha\dot{\alpha}} - igV_{\alpha\dot{\alpha}} = \sigma^{\underset{\sim}{b}}{}_{\alpha\dot{\alpha}}\partial_{\underset{\sim}{b}} - ig\,\frac{1}{\sqrt{2}}\sigma^{\underset{\sim}{b}}{}_{\alpha\dot{\alpha}}V_{\underset{\sim}{b}} \equiv \sigma^{\underset{\sim}{b}}{}_{\alpha\dot{\alpha}}(\partial_{\underset{\sim}{b}} - i\tilde{g}V_{\underset{\sim}{b}})$$

and hence our $g$ is $\sqrt{2}$ times the usual one $\tilde{g}$. (We use the summation convention: Any index of any type appearing twice in the same term, once contravariant (as a superscript) and once covariant (as a subscript), is summed over.)

Next to Lorentz indices, the type of indices we most frequently use are *isospin* indices: internal symmetry indices, usually for the group $SU(N)$ or $U(N)$. These are represented by lower-case Roman letters, without underlining. We use an *underlined* index only to indicate a composite index, an abbreviation for a pair of indices. In addition to the vector index defined above, we define a composite spinor-isospinor index by an underlined lower-case Greek index (undotted or dotted): $\psi^{\underline{\alpha}} \equiv \psi^{a\alpha}$, $\psi^{\underline{\dot{\alpha}}} \equiv \psi_{a}{}^{\dot{\alpha}}$, $\psi_{\underline{\alpha}} \equiv \psi_{a\alpha}$, $\psi_{\underline{\dot{\alpha}}} \equiv \psi^a{}_{\dot{\alpha}}$.

## b. Superspace

We define $N$-extended superspace to be a space with both the usual real commuting spacetime coordinates $x^{\alpha\dot{\alpha}} = x^{\underline{a}} = \overline{x}^{\underline{a}}$, and anticommuting coordinates $\theta^{a\alpha} = \theta^{\underline{\alpha}}$ (and their complex conjugates $\overline{\theta}^{\underline{\dot{\alpha}}} = (\theta^{\underline{\alpha}})^\dagger$) which transform as a spinor and an $N$-component isospinor. To denote these coordinates collectively we introduce *supervector* indices, using upper-case Roman letters:

$$z^A = (x^{\underline{a}}, \theta^{\underline{\alpha}}, \overline{\theta}^{\underline{\dot{\alpha}}}) \quad , \tag{3.1.3a}$$

and the corresponding partial derivatives

$$\partial_A = (\partial_{\underline{a}}, \partial_{\underline{\alpha}}, \overline{\partial}_{\underline{\dot{\alpha}}}) \quad , \quad \partial_A z^B \equiv \delta_A{}^B \quad , \tag{3.1.3b}$$

where the nonvanishing parts of $\delta_A{}^B$ are $\delta_{\underline{a}}{}^{\underline{b}}$, $\delta_{\underline{\alpha}}{}^{\underline{\beta}} \equiv \delta_a{}^b \delta_\alpha{}^\beta$, and $\delta_{\underline{\dot{\alpha}}}{}^{\underline{\dot{\beta}}} \equiv \delta_b{}^a \delta_{\dot{\alpha}}{}^{\dot{\beta}}$. The derivatives are defined to satisfy a *graded* Leibnitz rule, given by expressing differentiation as graded commutation:

$$(\partial_A XY) \equiv [\partial_A , XY\} = [\partial_A , X\}Y + (-)^{XA}\, X[\partial_A , Y\} \quad , \tag{3.1.4a}$$

where $(-)^{XA}$ is $-$ when both $X$ and $\partial_A$ are anticommuting, and $+$ otherwise, and the *graded* commutator $[A , B\} \equiv AB - (-)^{AB}\, BA$ is the anticommutator $\{A , B\}$ when $A$ and $B$ are both operators with fermi statistics, and the commutator $[A , B]$ otherwise. Eq. (3.1.4a) follows from writing each (anti)commutator as a difference (sum) of two terms. The partial derivatives also satisfy graded commutation relations:

$$[\partial_A , \partial_B\} = 0 \quad . \tag{3.1.4b}$$

## c. Symmetrization and antisymmetrization

Our notation for symmetrizing and antisymmetrizing indices is as follows: Symmetrization is indicated by parentheses ( ), while antisymmetrization is indicated by brackets [ ]. By symmetrization we mean simply the sum over all permutations of indices, without additional factors (and similarly for antisymmetrization, with the appropriate permutation signs). All indices between parentheses (brackets) are to be (anti)symmetrized except those between vertical lines | |. For example, $A_{(\alpha|\beta}B_{\gamma|\delta)} =$

$A_{\alpha\beta}B_{\gamma\delta} + A_{\delta\beta}B_{\gamma\alpha}$. In addition, just as it is convenient to define the graded commutator $[A\,, B\}$, we define graded antisymmetrization $[\ )$ to be a sum of permutations with a plus sign for any transposition of two spinor indices, and a minus sign for any other kind of pair.

## d. Conjugation

When working with operators with fermi statistics, the only type of complex conjugation that is usually defined is hermitian conjugation. It is defined so that the hermitian conjugate of a product is the product of the hermitian conjugates of the factors in reverse order. For anticommuting c-numbers hermitian conjugation again is the most natural form of complex conjugation. We denote the operation of hermitian conjugation by a dagger $^\dagger$, and indicate the hermitian conjugate of a given spinor by a bar: $(\psi^\alpha)^\dagger \equiv \bar\psi^{\dot\alpha}$, or $(\chi^{\dot\alpha})^\dagger \equiv \bar\chi^\alpha$. In particular, this applies to the coordinates $\theta$ and $\bar\theta$ introduced above. Hermitian conjugation of an object with many (upper) spinor indices is defined as for a product of spinors:

$$(\psi_1{}^{\alpha_1}\cdots\psi_j{}^{\alpha_j}\bar\chi_1{}^{\dot\beta_1}\cdots\bar\chi_k{}^{\dot\beta_k})^\dagger = \chi_k{}^{\beta_k}\cdots\chi_1{}^{\beta_1}\bar\psi_j{}^{\dot\alpha_j}\cdots\bar\psi_1{}^{\dot\alpha_1}$$

$$= (-1)^{\frac{1}{2}[j(j-1)+k(k-1)]}\chi_1{}^{\beta_1}\cdots\chi_k{}^{\beta_k}\bar\psi_1{}^{\dot\alpha_1}\cdots\bar\psi_j{}^{\dot\alpha_j} \quad , \tag{3.1.5a}$$

and hence

$$(\psi^{\alpha_1\ldots\alpha_j\dot\beta_1\ldots\dot\beta_k})^\dagger \equiv (-1)^{\frac{1}{2}[j(j-1)+k(k-1)]}\bar\psi^{\beta_1\ldots\beta_k\dot\alpha_1\ldots\dot\alpha_j} \quad . \tag{3.1.5b}$$

In addition, isospin indices for $SU(N)$ go from upper to lower, or vice versa, upon hermitian conjugation. Hermitian conjugation of partial derivatives follows from the reality of $\delta_A{}^B = (\partial_A z^B) = [\partial_A\,, z^B\}$:

$$(\partial_A)^\dagger = -\,(-)^A\partial_A \quad , \tag{3.1.6a}$$

where $(-)^A$ is $-1$ for spinor indices and $+1$ otherwise:

$$(\partial_{\underline{a}})^\dagger = -\,\partial_{\underline{a}} \quad , \quad (\partial_{\underline{\alpha}})^\dagger = +\,\bar\partial_{\underline{\dot\alpha}} \quad . \tag{3.1.6b}$$

Hermitian conjugation as applied to general operators is defined by

$$\int \bar\chi O\psi \equiv \int \overline{(O^\dagger\chi)}\psi \quad , \tag{3.1.7}$$

where the integration is over the appropriate space (as will be described in sec. 3.7) and $\overline{\chi}$ is the hermitian conjugate of the function $\chi$, as defined above.

Since integration defines not only a sesquilinear (hermitian) metric $\int \overline{\chi}\psi$ on the space of functions, as used to define a Hilbert space, but also a bilinear metric $\int \chi\psi$, we can also define the transpose of an operator:

$$\int \chi O \psi \equiv \int (\pm\, O^t \chi)\psi \quad , \tag{3.1.8}$$

where $\pm$ is $-$ for $O$ and $\chi$ anticommuting, $+$ otherwise. When the operator is expressed as a matrix, the hermitian conjugate and transpose take their familiar forms. We can also define complex conjugation of an operator:

$$\overline{O\chi} \equiv \pm\, O\,{}^{*}\, \overline{\chi} \quad , \tag{3.1.9}$$

with $\pm$ as in (3.1.8). For c-numbers we have $\psi^t = \psi$ and $\psi\,{}^* = \overline{\psi}$. For partial derivatives, integration by parts implies $(\partial_A)^t = -\,\partial_A$. In general, we also have the relation $O\,{}^{*\,t} = O^\dagger$, and the ordering relations $(O_1 O_2)^t = \pm\, O_2{}^t O_1{}^t$ and $(O_1 O_2)\,{}^* = \pm\, O_1\,{}^*\, O_2\,{}^*$, as well as the usual $(O_1 O_2)^\dagger = O_2{}^\dagger O_1{}^\dagger$.

**e. Levi-Civita tensors and index contractions**

There is only one nontrivial invariant matrix in $SL(2,C)$, the antisymmetric symbol $C_{\alpha\beta}$ (and its complex conjugate and their inverses), due to the volume-preserving nature of the group (unit determinant). Similarly, for $SU(N)$ we have the antisymmetric symbol $C_{a_1 \ldots a_N}$ (and its complex conjugate). In addition we find it useful to introduce the antisymmetric symbol of $SL(2N,C) \supset SU(N) \otimes SL(2,C)$, $C_{\underline{\alpha}_1 \cdots \underline{\alpha}_{2N}}$. Because of anticommutativity, it appears in the antisymmetric product of the $2N$ $\theta$'s of $N$-extended supersymmetry. These objects satisfy the following relations:

$$\overline{C_{\alpha\beta}} = C_{\dot{\beta}\dot{\alpha}} \quad , \quad C_{\alpha\beta} C^{\gamma\delta} = \delta_{[\alpha}{}^{\gamma} \delta_{\beta]}{}^{\delta} \quad ; \tag{3.1.10a}$$

$$\overline{C_{a_1 \ldots a_N}} = C^{a_1 \ldots a_N} \quad , \quad C_{a_1 \ldots a_N} C^{b_1 \ldots b_N} = \delta_{[a_1}{}^{b_1} \cdots \delta_{a_N]}{}^{b_N} \quad ; \tag{3.1.10b}$$

$$\overline{C_{\underline{\alpha}_1 \cdots \underline{\alpha}_{2N}}} = C_{\underline{\dot{\alpha}}_{2N} \cdots \underline{\dot{\alpha}}_1} \; , \; C_{\underline{\alpha}_1 \cdots \underline{\alpha}_{2N}} C^{\underline{\beta}_1 \cdots \underline{\beta}_{2N}} = \delta_{[\underline{\alpha}_1}{}^{\underline{\beta}_1} \cdots \delta_{\underline{\alpha}_{2N}]}{}^{\underline{\beta}_{2N}} \quad . \tag{3.1.10c}$$

The $SL(2N,C)$ symbol can be expressed in terms of the others:

$$C_{\underline{\alpha}_1 \cdots \underline{\alpha}_{2N}} = \frac{1}{N!(N+1)!} (C_{a_1 \ldots a_N} C_{a_{N+1} \ldots a_{2N}} C_{\alpha_1 \alpha_{N+1}} \cdots C_{\alpha_N \alpha_{2N}}$$

$$\pm \textit{ permutations of } \underline{\alpha}_i) \quad . \tag{3.1.11}$$

The magnitudes of the $C$'s are fixed by the conventions

$$C_{\alpha\beta} = C^{\dot\beta\dot\alpha} \quad , \quad C_{\underline{\alpha}_1 \cdots \underline{\alpha}_{2N}} = C^{\dot{\underline{\alpha}}_{2N} \cdots \dot{\underline{\alpha}}_1} \quad , \tag{3.1.12}$$

which set the absolute values of their components to 0 or 1.

We have the following relation for the product of all the $\theta$'s (because $\{\theta^\alpha, \theta^\beta\} = 0$, the square of any one component of $\theta$ vanishes):

$$\theta^{\underline{\alpha}_1} \cdots \theta^{\underline{\alpha}_{2N}} = C^{\underline{\alpha}_{2N} \cdots \underline{\alpha}_1} \left( \frac{1}{(2N)!} C_{\underline{\beta}_{2N} \cdots \underline{\beta}_1} \theta^{\underline{\beta}_1} \cdots \theta^{\underline{\beta}_{2N}} \right)$$

$$\equiv C^{\underline{\alpha}_{2N} \cdots \underline{\alpha}_1} \theta^{2N} \quad , \tag{3.1.13}$$

and a similar relation for $\bar\theta$, where, up to a phase factor, $\theta^{2N}$ is simply the product of all the $\theta$'s. Our conventions for complex conjugation of the $C$'s imply $\theta^{2N\dagger} = \bar\theta^{2N}$. Although seldom needed (except for expressing the $SL(2,C) \otimes SU(N)$ covariants in terms of covariants of a subgroup, as, e.g., when performing dimensional reduction or using a light-cone formalism), we can fix the phases (up to signs) in the definition of the $C$'s by the following conventions:

$$C_{\alpha\beta} = C_{\dot\alpha\dot\beta} \quad , \quad C_{\underline{\alpha}_1 \cdots \underline{\alpha}_{2N}} = C_{\dot{\underline{\alpha}}_1 \cdots \dot{\underline{\alpha}}_{2N}} \quad \rightarrow \quad C_{a_1 \ldots a_N} = C^{a_1 \ldots a_N} \quad . \tag{3.1.14}$$

In particular, we take

$$C_{\alpha\beta} = \begin{pmatrix} 0 & -i \\ i & 0 \end{pmatrix} \tag{3.1.15}$$

For $N = 1$ we have $\theta^2 = \frac{1}{2} C_{\beta\alpha} \theta^\alpha \theta^\beta = i\theta^+ \theta^-$. $C_{\alpha\beta}$ is thus the $SL(2,C)$ metric, and can be used for raising and lowering spinor indices:

$$\psi_\alpha = \psi^\beta C_{\beta\alpha} \quad , \quad \psi^\alpha = C^{\alpha\beta} \psi_\beta \quad , \tag{3.1.16a}$$

$$\psi \cdot \chi \equiv \psi^\alpha \chi_\alpha = \chi \cdot \psi \quad , \quad \psi^2 \equiv \frac{1}{2} C_{\beta\alpha} \psi^\alpha \psi^\beta = \frac{1}{2} \psi^\alpha \psi_\alpha = \frac{1}{2} \psi \cdot \psi = i\psi^+ \psi^- \quad ; \tag{3.1.16b}$$

$$\bar\psi_{\dot\alpha} = \bar\psi^{\dot\beta} C_{\dot\beta\dot\alpha} \quad , \quad \bar\psi^{\dot\alpha} = C^{\dot\alpha\dot\beta} \bar\psi_{\dot\beta} \quad , \tag{3.1.16c}$$

$$\bar\psi\cdot\bar\chi \equiv \bar\psi^{\dot\alpha}\bar\chi_{\dot\alpha} = \bar\chi\cdot\bar\psi \quad , \quad \bar\psi^2 \equiv \frac{1}{2} C_{\dot\beta\dot\alpha}\bar\psi^{\dot\alpha}\bar\psi^{\dot\beta} = \frac{1}{2}\bar\psi^{\dot\alpha}\bar\psi_{\dot\alpha} = \frac{1}{2}\bar\psi\cdot\bar\psi = i\bar\psi^{\dot +}\bar\psi^{\dot -} \quad ; \tag{3.1.16d}$$

$$V_{\underline a} = V^{\underline b} C_{\beta\alpha} C_{\dot\beta\dot\alpha} \equiv V^{\underline b}\eta_{\underline{ba}} \quad , \quad V^{\underline a} = C^{\alpha\beta}C^{\dot\alpha\dot\beta}V_{\underline b} \equiv \eta^{\underline{ab}}V_{\underline b} \quad , \tag{3.1.16e}$$

$$V\cdot W \equiv V^{\underline a}W_{\underline a} = W\cdot V \quad ,$$

$$V^2 \equiv \frac{1}{2}\eta_{\underline{ba}}V^{\underline a}V^{\underline b} = \frac{1}{2}V^{\underline a}V_{\underline a} = \frac{1}{2}V\cdot V = V^+V^- + \bar V^T V^T = -\, det\ V^{\alpha\dot\beta} \ . \tag{3.1.16f}$$

(As indicated by these equations, we contract indices with the contravariant index first.) Our unusual definition of the square of a vector is useful for spinor algebra, but we caution the reader not to confuse it with the standard definition. In particular, we define $\Box \equiv \frac{1}{2}\partial^{\underline a}\partial_{\underline a}$. (However, when we transform (with a *nonunimodular* transformation) to a cartesian basis, then we have the usual $\Box = \partial^{\underset{\sim}{a}}\partial_{\underset{\sim}{a}}$. For the coordinates, we have $x^2 = \frac{1}{4}x^{\underset{\sim}{a}}x_{\underset{\sim}{a}}$. Our conventions are convenient for superfield calculations, but may lead to a few unusual component normalizations.)

Defining

$$\partial_{\underline{\alpha\dot\beta}} \equiv \delta_a{}^b\partial_{\alpha\dot\beta} \quad , \quad \partial^{\underline{\alpha\dot\beta}} \equiv \delta_b{}^a\partial^{\alpha\dot\beta} \quad ; \tag{3.1.17}$$

we have the identities

$$\partial^{\alpha\dot\gamma}\partial_{\beta\dot\gamma} = \delta_\beta{}^\alpha\Box \quad , \quad \partial^{\underline{\alpha\dot\gamma}}\partial_{\underline{\beta\dot\gamma}} = \delta_{\underline\beta}{}^{\underline\alpha}\Box \quad . \tag{3.1.18}$$

From (3.1.10a) we obtain the frequently used relation

$$\psi_{[\alpha}\chi_{\beta]} = C_{\beta\alpha}(C^{\delta\gamma}\psi_\gamma\chi_\delta) = C_{\beta\alpha}(\psi^\delta\chi_\delta) \quad , \tag{3.1.19}$$

which is the Weyl-spinor form of the Fierz identities. Similar relations follow from (3.1.10b,c).

The complex conjugation properties of $C_{\alpha\beta}$ imply that the complex conjugates of *covariant* (lower index) spinors, including spinor partial derivatives (cf. (3.1.6)), have an additional minus sign:

$$(\psi_\alpha)^\dagger = -\bar\psi_{\dot\alpha} \quad . \tag{3.1.20}$$

From (3.1.11) and (3.1.18), or directly from the fact that antisymmetric symbols define determinants ($det\, V^{\underline{\alpha}\underline{\dot{\beta}}} = (det\, V^{\alpha\dot{\beta}})^N = (-V^2)^N$), we have the following identity:

$$C_{\underline{\dot{\beta}}_{2N}\cdots\underline{\dot{\beta}}_1}\partial^{\underline{\alpha}_1\underline{\dot{\beta}}_1}\cdots\partial^{\underline{\alpha}_{2N}\underline{\dot{\beta}}_{2N}} = C^{\underline{\alpha}_1\cdots\underline{\alpha}_{2N}}(-\Box)^N \quad . \tag{3.1.21}$$

Finally, we define the $SO(3,1)$ Levi-Civita tensor as

$$\epsilon_{\underline{abcd}} = i(C_{\alpha\delta}C_{\beta\gamma}C_{\dot{\alpha}\dot{\beta}}C_{\dot{\gamma}\dot{\delta}} - C_{\alpha\beta}C_{\gamma\delta}C_{\dot{\alpha}\dot{\delta}}C_{\dot{\beta}\dot{\gamma}}) \quad ,$$

$$\epsilon_{\underline{abcd}}\,\epsilon^{\underline{efgh}} = -\,\delta_{[\underline{a}}{}^{\underline{e}}\delta_{\underline{b}}{}^{\underline{f}}\delta_{\underline{c}}{}^{\underline{g}}\delta_{\underline{d}]}{}^{\underline{h}} \quad . \tag{3.1.22}$$

## 3.2. The supersymmetry groups

Lie algebras and Lie groups play an important role in field theory; groups such as the Poincaré group $ISO(3,1)$, the Lorentz group $SO(3,1)$, $SU(3)$ and $SU(2)\otimes U(1)$ are familiar. The new feature needed for supersymmetry is a generalization of Lie algebras to super-Lie algebras (also called graded Lie algebras; however, this term is sometimes used in a different way).

### a. Lie algebras

A Lie algebra consists of a set of generators $\{\Omega_{\mathrm{A}}\}$, $\mathbf{A} = 1, \ldots, M$. These objects close under an antisymmetric binary operation called a Lie bracket; we write it as a commutator:

$$[\Omega_{\mathrm{A}}, \Omega_{\mathrm{B}}] = \Omega_{\mathrm{A}}\Omega_{\mathrm{B}} - \Omega_{\mathrm{B}}\Omega_{\mathrm{A}} \quad . \tag{3.2.1}$$

The Lie algebra is defined by its structure constants $f_{\mathrm{AB}}{}^{\mathrm{C}}$:

$$[\Omega_{\mathrm{A}}, \Omega_{\mathrm{B}}] = i f_{\mathrm{AB}}{}^{\mathrm{C}}\,\Omega_{\mathrm{C}} \quad . \tag{3.2.2}$$

The structure constants are restricted by the Jacobi identities

$$f_{\mathrm{AB}}{}^{\mathrm{D}} f_{\mathrm{DC}}{}^{\mathrm{E}} + f_{\mathrm{BC}}{}^{\mathrm{D}} f_{\mathrm{DA}}{}^{\mathrm{E}} + f_{\mathrm{CA}}{}^{\mathrm{D}} f_{\mathrm{DB}}{}^{\mathrm{E}} = 0 \tag{3.2.3}$$

which follow from

$$[[\Omega_{\mathrm{A}}, \Omega_{\mathrm{B}}], \Omega_{\mathrm{C}}] + [[\Omega_{\mathrm{B}}, \Omega_{\mathrm{C}}], \Omega_{\mathrm{A}}] + [[\Omega_{\mathrm{C}}, \Omega_{\mathrm{A}}], \Omega_{\mathrm{B}}] = 0 \quad . \tag{3.2.4}$$

The generators form a basis for vectors of the form $K = \lambda^{\mathrm{A}}\Omega_{\mathrm{A}}$, where the $\lambda^{\mathrm{A}}$ are coordinates in the Lie algebra which are usually taken to commute with the generators $\Omega_{\mathrm{A}}$. In most physics applications they are taken to be real, complex, or quaternionic numbers. Because the structure constants satisfy the Jacobi identities, it is always possible to represent the generators as matrices. We can then exponentiate the Lie algebra into a Lie group with elements $g = e^{iK}$; in general, different representations of the Lie algebra will give rise to Lie groups with different topological structures. If a set of fields $\Phi(x)$ transforms linearly under the action of the Lie group, we say $\Phi(x)$ is in or carries a representation of the group. Abstractly, we write

$$\Phi'(x) = e^{iK}\Phi(x)e^{-iK} \quad ; \tag{3.2.5}$$

to give this meaning, we must specify the action of the generators on $\Phi$, i.e.,$[\Omega_A , \Phi]$. For example, if $K$ is a matrix representation and $\Phi$ is a column vector, the expression above is to be interpreted as $\Phi' = e^{iK}\Phi$.

### b. Super-Lie algebras

For supersymmetry we generalize and consider super-Lie algebras. The essential new feature is that now the Lie bracket of some generators is symmetric. Those generators whose bracket is symmetric are called fermionic; the rest are bosonic. We write the bracket as a graded commutator

$$[\Omega_A , \Omega_B \} = \Omega_A \Omega_B - (-)^{AB} \Omega_B \Omega_A \equiv \Omega_{[A} \Omega_{B)} \quad . \tag{3.2.6}$$

The structure constants of the super-Lie algebra obey super-Jacobi identities that follow from:

$$0 = \frac{1}{2} (-)^{AC} [[\Omega_{[A} , \Omega_B\} , \Omega_{C)}\}$$

$$\equiv (-)^{AC} [[\Omega_A , \Omega_B \} , \Omega_C \} + (-)^{AB} [[\Omega_B , \Omega_C \} , \Omega_A \} + (-)^{BC} [[\Omega_C , \Omega_A \} , \Omega_B \} \quad . \tag{3.2.7}$$

Again, we can define a vector space with the generators $\Omega_A$ acting as a basis; however, in this case the coordinates $\lambda^A$ associated with the fermionic generators are *anticommuting* numbers or Grassmann parameters that anticommute with each other and with the fermionic generators. Grassmann parameters commute with ordinary numbers and bosonic generators; these properties ensure that $K = \lambda^A \Omega_A$ is bosonic. Formally, we obtain super-Lie group elements by exponentiation of the algebra as we do for Lie groups.

### c. Super-Poincaré algebra

Field theories in ordinary spacetime are usually symmetric under the action of a spacetime symmetry group: the Poincaré group for massive theories in flat space, the conformal group for massless theories, and the deSitter group for theories in spaces of constant curvature. For supersymmetry, we consider extensions of these groups to supergroups. These were investigated by Haag, Łopuszański, and Sohnius, who classified the most general symmetries possible (actually, they considered symmetries of the S-matrix and generalized the Coleman-Mandula theorem on unified internal and spacetime symmetries to include super-Lie algebras). They proved that the most general super-Poincaré algebra contains, in addition to $\{J_{\alpha\beta}, \bar{J}_{\dot{\alpha}\dot{\beta}}, P_{\alpha\dot{\beta}}\}$ (the generators of the Poincaré group), $N$ fermionic spinorial generators $Q_{a\alpha}$ (and their hermitian conjugates $-\bar{Q}^a{}_{\dot{\alpha}}$), where $a = 1, \ldots, N$ is an isospin index, and at most $\frac{1}{2}N(N-1)$ complex central charges (called central because they commute with all generators in the theory) $Z_{ab} = -Z_{ba}$. The algebra is:

$$\{Q_{a\alpha}, \bar{Q}^b{}_{\dot{\beta}}\} = \delta_a{}^b P_{\alpha\dot{\beta}} \ , \tag{3.2.8a}$$

$$\{Q_{a\alpha}, Q_{b\beta}\} = C_{\alpha\beta} Z_{ab} \ , \tag{3.2.8b}$$

$$[Q_{a\alpha}, P_{\beta\dot{\beta}}] = [P_{\alpha\dot{\alpha}}, P_{\beta\dot{\beta}}] = [\bar{J}_{\dot{\alpha}\dot{\beta}}, Q_{c\gamma}] = 0 \ , \tag{3.2.8c}$$

$$[J_{\alpha\beta}, Q_{c\gamma}] = \frac{1}{2} i C_{\gamma(\alpha} Q_{c\beta)} \ , \tag{3.2.8d}$$

$$[J_{\alpha\beta}, P_{\gamma\dot{\gamma}}] = \frac{1}{2} i C_{\gamma(\alpha} P_{\beta)\dot{\gamma}} \ , \tag{3.2.8e}$$

$$[J_{\alpha\beta}, J^{\gamma\delta}] = -\frac{1}{2} i \delta_{(\alpha}{}^{(\gamma} J_{\beta)}{}^{\delta)} \ , \tag{3.2.8f}$$

$$[J_{\alpha\beta}, \bar{J}_{\dot{\alpha}\dot{\beta}}] = [Z_{ab}, Z_{cd}] = [Z_{ab}, \bar{Z}^{cd}] = 0 \ . \tag{3.2.8g}$$

The essential ingredients in the proof are the Coleman-Mandula theorem (which restricts the bosonic parts of the algebra), and the super-Jacobi identities. The $N = 1$ case is called simple supersymmetry, whereas the $N > 1$ case is called extended supersymmetry. Central charges can arise only in the case of extended ($N > 1$) supersymmetry. The supersymmetry generators $Q$ act as "square roots" of the momentum generators $P$.

## d. Positivity of the energy

A direct consequence of the algebra is the positivity of the energy in supersymmetric theories. The simplest way to understand this result is to note that the total energy $\epsilon$ can be written as

$$\epsilon = \frac{1}{2}(P^+ - P^-) = \frac{1}{2}\delta_{\alpha\dot{\beta}} P^{\alpha\dot{\beta}} = -\frac{1}{2}\delta^{\alpha\dot{\beta}} P_{\alpha\dot{\beta}} \ . \tag{3.2.9}$$

Since $P_{\alpha\dot\beta}$ can be obtained from the anticommutator of spinor charges, we have

$$\epsilon = -\frac{1}{2N}\,\delta^{\alpha\dot\beta}\,\{Q_{a\alpha}\,,\bar Q^a{}_{\dot\beta}\} = \frac{1}{2N}\,\{Q_{a\alpha}\,,(Q_{a\alpha})^\dagger\} \tag{3.2.10}$$

(we use $\bar Q^a{}_{\dot\alpha} = -\,(Q_{a\alpha})^\dagger$). The right hand side of eq. (3.2.10) is manifestly non-negative: For any operator $A$ and any state $|\psi>$,

$$<\psi|\{A\,,A^\dagger\}|\psi> = \sum_n(<\psi|A|n><n|A^\dagger|\psi> + <\psi|A^\dagger|n><n|A|\psi>)$$

$$= \sum_n(|<n|A^\dagger|\psi>|^2 + |<n|A|\psi>|^2)\ . \tag{3.2.11}$$

Hence, $\epsilon$ is also nonnegative. Further, if supersymmetry is unbroken, $Q$ must annihilate the vacuum; in this case, (3.2.10) leads to the conclusion that the vacuum energy vanishes. Although this argument is formal, it can be made more precise; indeed, it is possible to characterize supersymmetric theories by the condition that the vacuum energy vanish.

**e. Superconformal algebra**

For massless theories, Haag, Łopuszański, and Sohnius showed what form extensions of the conformal group can take: The generators of the superconformal groups consist of the generators of the conformal group $(P_{\alpha\dot\beta}\,,J_{\alpha\beta}\,,\bar J_{\dot\alpha\dot\beta}\,,K_{\alpha\dot\beta}\,,\Delta)$ (these are the generators of the Poincaré algebra, the special conformal boost generators, and the dilation generator), $2N$ spinor generators $(Q_{a\alpha}, S^{a\alpha})$ (and their hermitian conjugates $-\bar Q^a{}_{\dot\alpha}, -\bar S_a{}^{\dot\alpha}$ with a total of $8N$ components), and $N^2$ further bosonic charges $(A, T_a{}^b)$ where $T_a{}^a = 0$. The algebra has structure constants defined by the following (anti)commutators:

$$\{Q_{a\alpha}\,,\bar Q^b{}_{\dot\beta}\} = \delta_a{}^b P_{\alpha\dot\beta} \quad , \quad \{S^{a\alpha}\,,\bar S_b{}^{\dot\beta}\} = \delta_b{}^a K^{\alpha\dot\beta}\ , \tag{3.2.12a}$$

$$\{Q_{a\alpha}\,,S^{b\beta}\} = -\,i\delta_a{}^b(J_\alpha{}^\beta + \frac{1}{2}\,\delta_\alpha{}^\beta\Delta) - \frac{1}{2}\,\delta_\alpha{}^\beta\delta_a{}^b(1-\frac{4}{N})A + 2\delta_\alpha{}^\beta T_a{}^b \tag{3.2.12b}$$

$$[T_a{}^b\,,S^{c\gamma}] = \frac{1}{2}\,(\delta_a{}^c S^{b\gamma} - \frac{1}{N}\,\delta_a{}^b S^{c\gamma})\ , \tag{3.2.12c}$$

$$[A\,,S^{c\gamma}]=\frac{1}{2}\,S^{c\gamma}\quad,\quad[\Delta\,,S^{c\gamma}]=-\,i\,\frac{1}{2}\,S^{c\gamma}\ ,\tag{3.2.12d}$$

$$[J_\alpha{}^\beta\,,S^{c\gamma}]=-\,\frac{1}{2}\,i\delta_{(\alpha}{}^{|\gamma|}S^{c\beta)}\ ,\qquad[P_{\alpha\dot\alpha}\,,S^{c\gamma}]=-\,\delta_\alpha{}^\gamma\bar Q^c{}_{\dot\alpha}\ ,\tag{3.2.12e}$$

$$[T_a{}^b\,,Q_{c\gamma}]=-\,\frac{1}{2}\,(\delta_c{}^bQ_{a\gamma}-\frac{1}{N}\,\delta_a{}^bQ_{c\gamma})\ ,\tag{3.2.12f}$$

$$[A\,,Q_{c\gamma}]=-\,\frac{1}{2}\,Q_{c\gamma}\quad,\quad[\Delta\,,Q_{c\gamma}]=i\,\frac{1}{2}\,Q_{c\gamma}\ ,\tag{3.2.12g}$$

$$[J_\alpha{}^\beta\,,Q_{c\gamma}]=\frac{1}{2}\,i\delta_\gamma{}^{(\beta}Q_{c\alpha)}\quad,\quad[K^{\alpha\dot\alpha}\,,Q_{c\gamma}]=\delta_\gamma{}^\alpha\bar S_c{}^{\dot\alpha}\ ,\tag{3.2.12h}$$

$$[T_a{}^b\,,T_c{}^d]=\frac{1}{2}\,(\delta_a{}^dT_c{}^b-\delta_c{}^bT_a{}^d)\ ,\tag{3.2.12i}$$

$$[\Delta\,,K^{\alpha\dot\alpha}]=-\,iK^{\alpha\dot\alpha}\quad,\quad[\Delta\,,P_{\alpha\dot\alpha}]=iP_{\alpha\dot\alpha}\ ,\tag{3.2.12j}$$

$$[J_\alpha{}^\beta\,,K^{\gamma\dot\gamma}]=-\,\frac{1}{2}\,i\delta_{(\alpha}{}^{|\gamma|}K^{\beta)\dot\gamma}\ ,\qquad[J_\alpha{}^\beta\,,P_{\gamma\dot\gamma}]=\frac{1}{2}\,i\delta_\gamma{}^{(\beta}P_{\alpha)\dot\gamma}\ ,\tag{3.2.12k}$$

$$[J_{\alpha\beta}\,,J^{\gamma\delta}]=-\,\frac{1}{2}\,i\delta_{(\alpha}{}^{(\gamma}J_{\beta)}{}^{\delta)}\ ,\tag{3.2.12l}$$

$$[P_{\alpha\dot\alpha}\,,K^{\beta\dot\beta}]=i(\delta_{\dot\alpha}{}^{\dot\beta}J_\alpha{}^\beta+\delta_\alpha{}^\beta\bar J_{\dot\alpha}{}^{\dot\beta}+\delta_\alpha{}^\beta\delta_{\dot\alpha}{}^{\dot\beta}\Delta)=i(J_{\underline a}{}^{\underline b}+\delta_{\underline a}{}^{\underline b}\Delta)\ .\tag{3.2.12m}$$

All other (anti)commutators vanish or are found by hermitian conjugation.

The superconformal algebra contains the super-Poincaré algebra as a subalgebra; however, in the superconformal case, there are *no* central charges (this is a direct consequence of the Jacobi identities). In the same way that the supersymmetry generators $Q$ act as "square roots" of the translation generators $P$, the $S$-supersymmetry generators $S$ act as "square roots" of the special conformal generators $K$. The new bosonic charges $A$ and $T_a{}^b$ generate phase rotations of the spinors (axial or $\gamma_5$ rotations) and $SU(N)$ transformations respectively (all but the $SO(N)$ subgroup of the $SU(N)$ is axial). For $N=4$, the axial charge $A$ drops out of the $\{Q\,,S\}$ anticommutator whereas the $[Q\,,A]$ and $[S\,,A]$ commutators are $N$ independent. The normalization of $A$ is chosen such that $T_a{}^b+\frac{1}{N}\,\delta_a{}^bA$ generates $U(N)$ (e.g., $[T_a{}^b+\frac{1}{N}\,\delta_a{}^bA\,,Q_{c\gamma}]=-\,\frac{1}{2}\,\delta_c{}^bQ_{a\gamma}$).

**f. Super-deSitter algebra**

Finally, we turn to the supersymmetric extension of the deSitter algebra. The generators of this algebra are the generators of the deSitter algebra $(\tilde{P}, \tilde{J}, \tilde{\bar{J}})$, spinorial generators $(\tilde{Q}, \tilde{\bar{Q}})$, and $\frac{1}{2}N(N-1)$ bosonic $SO(N)$ charges $\tilde{T}_{ab} = -\tilde{T}_{ba}$. They can be constructed out of the superconformal algebra (just as the super-Poincaré algebra is a subalgebra of the superconformal algebra, so is the super-deSitter algebra). We can define the generators of the super-deSitter algebra as the following linear combinations of the superconformal generators:

$$\tilde{P}_{\alpha\dot{\alpha}} = P_{\alpha\dot{\alpha}} + |\lambda|^2 K_{\alpha\dot{\alpha}} \ , \quad \tilde{Q}_{a\alpha} = Q_{a\alpha} + \lambda\delta_{ab} S^b{}_\alpha \ ,$$

$$\tilde{J}_{\alpha\beta} = J_{\alpha\beta} \ , \quad \tilde{T}_{ab} = \delta_{c[b} T_{a]}{}^c \ , \tag{3.2.13}$$

where, since we break $SU(N)$ to $SO(N)$, we have lowered the isospin indices of the superconformal generators with a kronecker delta. (We could also formally maintain $SU(N)$ invariance by using instead $\lambda_{ab}$ satisfying $\lambda_{ab} = \lambda_{ba}$ and $\lambda_{ac}\bar{\lambda}^{bc} \sim \delta_a{}^b$, with $\lambda_{ab} = \lambda\delta_{ab}$ in an appropriate $SU(N)$ frame.) Thus we find the following algebra:

$$\{\tilde{Q}_{a\alpha}, \tilde{Q}_{b\beta}\} = 2\lambda(-i\delta_{ab}\tilde{J}_{\alpha\beta} + C_{\alpha\beta}\tilde{T}_{ab}) \ , \tag{3.2.14a}$$

$$\{\tilde{Q}_{a\alpha}, \tilde{\bar{Q}}^b{}_{\dot{\beta}}\} = \delta_a{}^b \tilde{P}_{\alpha\dot{\beta}} \ , \tag{3.2.14b}$$

$$[\tilde{Q}_{a\alpha}, \tilde{P}_{\beta\dot{\beta}}] = -\lambda C_{\alpha\beta}\delta_{ab}\tilde{\bar{Q}}^b{}_{\dot{\beta}} \ , \tag{3.2.14c}$$

$$[\tilde{J}_{\alpha\beta}, \tilde{Q}_{c\gamma}] = \frac{1}{2} iC_{\gamma(\alpha}\tilde{Q}_{c\beta)} \ , \tag{3.2.14d}$$

$$[\tilde{J}_{\alpha\beta}, \tilde{P}_{\gamma\dot{\gamma}}] = \frac{1}{2} iC_{\gamma(\alpha}\tilde{P}_{\beta)\dot{\gamma}} \ , \tag{3.2.14e}$$

$$[\tilde{P}_{\alpha\dot{\alpha}}, \tilde{P}_{\beta\dot{\beta}}] = -i2|\lambda|^2(C_{\dot{\alpha}\dot{\beta}}\tilde{J}_{\alpha\beta} + C_{\alpha\beta}\tilde{\bar{J}}_{\dot{\alpha}\dot{\beta}}) \ , \tag{3.2.14f}$$

$$[\tilde{J}_{\alpha\beta}, \tilde{J}^{\gamma\delta}] = -\frac{1}{2} i\delta_{(\alpha}{}^{(\gamma}\tilde{J}_{\beta)}{}^{\delta)} \ . \tag{3.2.14g}$$

$$[\tilde{T}_{ab}, \tilde{Q}_{c\gamma}] = \frac{1}{2}\delta_{c[a}\tilde{Q}_{b]\gamma} \ , \tag{3.2.14h}$$

$$[\tilde{T}_{ab}\,,\tilde{T}_{cd}] = \frac{1}{2}(\delta_{b[c}\tilde{T}_{d]a} - \delta_{a[c}\tilde{T}_{d]b}) \tag{3.2.14i}$$

This algebra, in contrast to the superconformal and super-Poincaré cases, depends on a dimensional constant $\lambda$. Physically, $|\lambda|^2$ is the curvature of the deSitter space. (Actually, the sign is such that the relevant space is the space of constant *negative* curvature, or anti-deSitter space. This is a consequence of supersymmetry: The algebra determines the relative sign in the combination $P + |\lambda|^2 K$ above.)

## 3.3. Representations of supersymmetry

### a. Particle representations

Before discussing field representations of supersymmetry, we study the particle content of Poincaré supersymmetric theories. We analyze representations of the supersymmetry group in terms of representations of its Poincaré subgroup. Because $P^2$ is a Casimir operator of supersymmetry (it commutes with all the generators), all elements of a given irreducible representation will have the same mass.

### a.1. Massless representations

We first consider massless representations. We then can choose a Lorentz frame where the only nonvanishing component of the momentum $p_{\underline{a}}$ is $p_+$. In this frame the anticommutation relations of the supersymmetry generators are

$$\{Q_{a+}\,,Q_{b+}\}=0 \quad , \quad \{Q_{a+}\,,\bar{Q}^b{}_{\dot{+}}\}=p_+\delta_a{}^b \quad ,$$

$$\{Q_{a-}\,,Q_{b-}\}=0 \quad , \quad \{Q_{a-}\,,\bar{Q}^b{}_{\dot{-}}\}=0 \quad ,$$

$$\{Q_{a+}\,,Q_{b-}\}=0 \quad , \quad \{Q_{a+}\,,\bar{Q}^b{}_{\dot{-}}\}=0 \quad . \tag{3.3.1}$$

Since the anticommutator of $Q_{a-}$ with its hermitian conjugate vanishes, $Q_{a-}$ must vanish identically on all physical states: From (3.2.11) we have the result that

$$0=<\psi|\{A\,,A^\dagger\}|\psi>=\sum_n(|<n|A^\dagger|\psi>|^2+|<n|A|\psi>|^2)$$

$$\rightarrow\ <n|A|\psi>=<n|A^\dagger|\psi>=0\ \ . \tag{3.3.2}$$

On the other hand, $Q_{a+}$ and its hermitian conjugate satisfy the standard anticommutation relations for annihilation and creation operators, up to normalization factors (with the exception of the case $p_+=0$, which in this frame means $p_{\underline{a}}=0$ and describes the physical vacuum). We can thus consider a state, the *Clifford vacuum* $|C>$, which is annihilated by all the annihilation operators $Q_{a+}$ (or construct such a state from a given state by operating on it with a sufficient number of annihilation operators) and generate all other states by action of the creation operators $\bar{Q}^a{}_{\dot{+}}$. Since, as usual, an annihilation operator acting on any state produces another with one less creation operator acting on the Clifford vacuum, this set of states is closed under the action of the supersymmetry generators, and thus forms a representation of the supersymmetry algebra. Furthermore, if the Clifford vacuum is an irreducible representation of the Poincaré group, this set of states is an irreducible representation of the supersymmetry group, since any attempt to reduce the representation by imposing a constraint on a state (or a linear combination of states) would also constrain the Clifford vacuum (after applying an appropriate number of annihilation operators; see also sec. 3.8.a). The Clifford vacuum may also carry representations of isospin and other internal symmetry groups.

The Clifford vacuum, being an irreducible representation of the Poincaré group, is also an eigenstate of helicity. In this frame, $\bar{Q}^a{}_{\dot{+}}$ has helicity $-\frac{1}{2}$, thus determining the helicities of the other states in terms of that of the Clifford vacuum. (In general frames, the helicity $-\frac{1}{2}$ component of $\bar{Q}_{\dot{\underline{\alpha}}}$ is the creation operator, and the helicity $+\frac{1}{2}$ component, which is the linearly independent Lorentz component of $P^{\alpha\dot{\alpha}}\bar{Q}^a{}_{\dot{\alpha}}$, vanishes: $\{P^{\alpha\dot{\alpha}}\bar{Q}^a{}_{\dot{\alpha}}, P^{\beta\dot{\beta}}Q_{b\beta}\} = \delta_b{}^a P^2 P_{\alpha\dot{\beta}} = 0$, since $p^2 = 0$ in the massless case.) The representations of the states under isospin are also determined from the transformation properties of the Clifford vacuum and the $Q$'s: We take the tensor product of the Clifford vacuum's representation with that of the creation operators (namely, that formed by multiplying the representations of the individual operators and antisymmetrizing).

As examples, we consider the cases of the massless scalar multiplet ($N = 1, 2$), super-Yang-Mills ($N = 1, \ldots, 4$), and supergravity ($N = 1, \ldots, 8$), defined by Clifford vacua which are isoscalars and have helicity $+\frac{1}{2}$, $+1$, and $+2$, respectively. (In the scalar and Yang-Mills cases, the states may carry a representation of a separate internal symmetry group.) The states are listed in Table 3.3.1. Each state is totally antisymmetric in the isospin indices, and thus, for a given $N$, states with more than $N$ isospin indices vanish. The scalar multiplet contains helicities $(\frac{1}{2}, \ldots, \frac{1}{2} - \frac{N}{2})$, super Yang-Mills contains helicities $(1, \ldots, 1 - \frac{N}{2})$, and supergravity contains helicities $(2, \ldots, 2 - \frac{N}{2})$. In addition, any representation of an internal symmetry group that commutes with supersymmetry (such as the gauge group of super Yang-Mills) carried by the Clifford vacuum is carried by all states (so in super Yang-Mills all states are in the adjoint representation of the gauge group). Thus the total number of states in a massless representation is $2^N k$, where $k$ is the number of states in the Clifford vacuum.

| helicity | scalar multiplet | super-Yang-Mills | supergravity |
|---|---|---|---|
| +2 | | | $\psi = \|C>$ |
| +3/2 | | | $\psi^a$ |
| +1 | | $\psi = \|C>$ | $\psi^{ab}$ |
| +1/2 | $\psi = \|C>$ | $\psi^a$ | $\psi^{abc}$ |
| 0 | $\psi^a$ | $\psi^{ab}$ | $\psi^{abcd}$ |
| -1/2 | $\psi^{ab}$ | $\psi^{abc}$ | $\psi^{abcde}$ |
| -1 | | $\psi^{abcd}$ | $\psi^{abcdef}$ |
| -3/2 | | | $\psi^{abcdefg}$ |
| -2 | | | $\psi^{abcdefgh}$ |

*Table 3.3.1. States in theories of physical interest*

The CPT conjugate of a state transforms as the complex conjugate representation. Just as for representations of the Poincaré group, one may identify a supersymmetry representation with its conjugate if it has the same quantum numbers: i.e., if it is a real representation. (In terms of classical fields, or fields in a functional integral, this self-conjugacy condition relates fields to their complex conjugates: see (3.12.4c) or (3.12.11). Thus, in a functional integral formalism, self-conjugacy is with respect to a type of charge conjugation: A charge conjugation is complex conjugation times a matrix (see sec. 3.3.b.5).) For the above examples, this self-conjugacy occurs for $N = 4$ super Yang-Mills and $N = 8$ supergravity. (This is not true for the $N = 2$ scalar multiplet, since an $SU(2)$ isospinor cannot be identified with its complex conjugate, unless an extra isospin index of the internal $SU(2)$ symmetry, independent of the supersymmetry $SU(2)$, is added. The self-conjugacy then simply cancels the doubling introduced by the extra index.)

## a.2. Massive representations and central charges

The massive case is treated similarly, except that we can no longer choose the Lorentz frame above; instead, we choose the rest frame, $p_{\alpha\dot{\alpha}} = -m\delta_{\alpha\dot{\alpha}}$:

$$\{Q_{\underline{\alpha}}, Q_{\underline{\beta}}\} = 0 \quad , \quad \{Q_{\underline{\alpha}}, \overline{Q}_{\underline{\dot{\beta}}}\} = -m\delta_{\underline{\alpha\dot{\beta}}} \quad . \tag{3.3.3}$$

Now we have twice as many creation and annihilation operators, the $Q_-$'s as well as the $Q_+$'s. Therefore the number of states in a massive representation is $2^{2N}k$. (For example, an $N = 1$ massive vector multiplet has helicity content $(1, \frac{1}{2}, \frac{1}{2}, 0)$.)

The case with central charges can be analyzed by similar methods, but it is simpler to understand if we realize that supersymmetry algebras with central charges can be obtained from supersymmetry algebras without central charges in higher-dimensional spacetimes by interpreting some of the extra components of the momentum as the central charge generators (they will commute with all the four-dimensional generators). The analysis of the state content is then the same as for the cases without central charges, since both cases are obtained from the same higher-dimensional set of states (except that we do not keep the full higher-dimensional Lorentz group). However, the two distinguishing cases are now, in terms of $P^2{}_{higher-dimensional} = P^2 + Z^2 = \frac{1}{2}(P^{\underline{a}}P_{\underline{a}} + \bar{Z}^{ab}Z_{ab})$: (1) $P^2 + Z^2 = 0$, which has the same set of states as the massless $Z = 0$ case (though the states are now massive, have a smaller internal symmetry group, and transform somewhat differently under supersymmetry), and (2) $P^2 + Z^2 < 0$, which has the same set of states as the massive $Z = 0$ case. By this same analysis, we see that $P^2 + Z^2 > 0$ is not allowed (just as for $Z = 0$ we never have $P^2 > 0$).

### a.3. Casimir operators

We can construct other Casimir operators than $P^2$. We first define the supersymmetric generalization of the Pauli-Lubanski vector

$$W_{\alpha\dot{\alpha}} = i(P^{\beta}{}_{\dot{\alpha}}J_{\alpha\beta} - P_{\alpha}{}^{\dot{\beta}}\bar{J}_{\dot{\alpha}\dot{\beta}}) - \frac{1}{2}[Q_{a\alpha}, \bar{Q}^{a}{}_{\dot{\alpha}}] \ , \tag{3.3.4}$$

where the last term is absent in the nonsupersymmetric case. This vector is not invariant under supersymmetry transformations, but satisfies

$$[W_{\underline{a}}, Q_{\underline{\beta}}] = -\frac{1}{2}P_{\underline{a}}Q_{\underline{\beta}} \ , \quad [W_{\underline{a}}, \bar{Q}_{\underline{\dot{\beta}}}] = \frac{1}{2}P_{\underline{a}}\bar{Q}_{\underline{\dot{\beta}}} \ . \tag{3.3.5}$$

As a result, $P_{[\underline{a}}W_{\underline{b}]}$ commutes with $Q_{\underline{\alpha}}$, and thus its square $P^2W^2 - \frac{1}{4}(P \cdot W)^2$ commutes with all the generators of the super-Poincaré algebra and is a Casimir operator. In the massive case this Casimir operator defines a quantum number $s$, the *superspin*. The generalization of the nonsupersymmetric relation $W^2 = m^2 s(s+1)$ is

$$P^2 W^2 - \frac{1}{4}(P \cdot W)^2 = -\, m^4 s(s+1) \ . \tag{3.3.6}$$

In the massless case, not only $P^2 = 0$, but also $P^{\alpha\dot{\alpha}} Q_{a\alpha} = P^{\alpha\dot{\alpha}} \bar{Q}^a{}_{\dot{\alpha}} = 0$, and hence $P \cdot W = P_{[\underline{a}} W_{\underline{b}]} = 0$. However, using the generator $A$ of the superconformal group (3.2.12), we can construct an object that commutes with $Q$ and $\bar{Q}$: $W_{\underline{a}} - AP_{\underline{a}}$. Thus we can define a quantum number $\lambda$, the *superhelicity,* that generalizes helicity $\lambda_0$ (defined by $W_{\underline{a}} = \lambda_0 P_{\underline{a}}$):

$$W_{\underline{a}} - AP_{\underline{a}} = \lambda P_{\underline{a}} \ . \tag{3.3.7}$$

We also can construct supersymmetry invariant generalizations of the axial generator $A$ and of the $SU(N)$ generators:

$$W_5 \equiv P^2 A + \frac{1}{4} P^{\alpha\dot{\alpha}} [Q_{a\alpha} \,, \bar{Q}^a{}_{\dot{\alpha}}] \ ,$$

$$W_a{}^b \equiv P^2 T_a{}^b + \frac{1}{4} P^{\alpha\dot{\alpha}} ([Q_{a\alpha} \,, \bar{Q}^b{}_{\dot{\alpha}}] - \frac{1}{N} \delta_a{}^b [Q_{c\alpha} \,, \bar{Q}^c{}_{\dot{\alpha}}]) \ . \tag{3.3.8}$$

In the massive case, the *superchiral charge* and the *superisospin* quantum numbers can then be defined as the usual Casimir operators of the modified group generators $-m^{-2} W_5 \,, -\, m^{-2} W_a{}^b$. In the massless case, we define the operators

$$W_{5\alpha\dot{\alpha}} \equiv P_{\alpha\dot{\alpha}} A + \frac{1}{4} [Q_{a\alpha} \,, \bar{Q}^a{}_{\dot{\alpha}}] \ ,$$

$$W_a{}^b{}_{\alpha\dot{\alpha}} \equiv P_{\alpha\dot{\alpha}} T_a{}^b + \frac{1}{4} ([Q_{a\alpha} \,, \bar{Q}^b{}_{\dot{\alpha}}] - \frac{1}{N} \delta_a{}^b [Q_{c\alpha} \,, \bar{Q}^c{}_{\dot{\alpha}}]) \ . \tag{3.3.9}$$

These commute with $Q$ and $\bar{Q}$ when the condition $P^{\alpha\dot{\alpha}} Q_{a\alpha} = 0$ holds, which is precisely the massless case. Since $P^{\beta\dot{\gamma}} W_{5\gamma\dot{\gamma}} = P^{\beta\dot{\gamma}} W_a{}^b{}_{\gamma\dot{\gamma}} = 0$, we can find matrix representations $g_5 \,, g_a{}^b$ such that

$$W_{5\underline{c}} = g_5 \, P_{\underline{c}} \ , \quad W_a{}^b{}_{\underline{c}} = g_a{}^b \, P_{\underline{c}} \ . \tag{3.3.10}$$

The superchiral charge is $g_5$, and superisospin quantum numbers can be defined from the traceless matrices $g_a{}^b$. All supersymmetrically invariant operators that we have constructed can be reexpressed in terms of covariant derivatives defined in sec. 3.4.a; see sec 3.4.d.

## b. Representations on superfields

We turn now to field (off-shell) representations of the supersymmetry algebras. These can be described in superspace, which is an extension of spacetime to include extra anticommuting coordinates. To discover the action of supersymmetry transformations on superspace, we use the method of induced representations. We discuss only simple $N = 1$ supersymmetry for the moment.

### b.1. Superspace

Ordinary spacetime can be defined as the coset space (Poincaré group)/(Lorentz group). Similarly, *global flat superspace* can be defined as the coset space (super-Poincaré group)/(Lorentz group): Its points are the orbits which the Lorentz group sweeps out in the super-Poincaré group. Relative to some origin, this coset space can be parametrized as:

$$h(x, \theta, \bar{\theta}) = e^{i( x^{\alpha\dot{\beta}} \hat{P}_{\alpha\dot{\beta}} + \theta^{\alpha} \hat{Q}_{\alpha} + \bar{\theta}^{\dot{\alpha}} \hat{\bar{Q}}_{\dot{\alpha}} )} \tag{3.3.11}$$

where $x, \theta, \bar{\theta}$ are the coordinates of superspace: $x$ is the coordinate of spacetime, and $\theta, \bar{\theta}$ are new fermionic spinor coordinates. The "hat" on $\hat{P}$ and $\hat{Q}$ indicates that they are abstract group generators, *not* to be confused with the differential operators $P$ and $Q$ used to represent them below. The statistics of $\theta, \bar{\theta}$ are determined by those of $\hat{Q}, \hat{\bar{Q}}$:

$$\{\theta, \theta\} = \{\theta, \bar{\theta}\} = \{\hat{Q}, \theta\} = [\theta, x] = [\theta, \hat{P}] = 0 \ , \tag{3.3.12}$$

etc., that is, $\theta, \bar{\theta}$ are Grassmann parameters.

### b.2. Action of generators on superspace

We define the action of the super-Poincaré group on superspace by left multiplication:

$$h(x', \theta', \bar{\theta}') = \hat{g}^{-1} \cdot h(x, \theta, \bar{\theta}) \ mod \ SO(3,1) \tag{3.3.13}$$

where $\hat{g}$ is a group element, and "$mod\ SO(3,1)$" means that any terms involving Lorentz generators are to be pushed through to the right and then dropped. To find the action of the generators $(J, P, Q)$ on superspace, we consider

$$\hat{g} = \left( e^{-i(\omega_\alpha{}^\beta \hat{J}_\alpha{}^\beta + \bar{\omega}_{\dot{\alpha}}{}^{\dot{\beta}} \hat{\bar{J}}_{\dot{\alpha}}{}^{\dot{\beta}})} \ , \ e^{-i(\xi^{\alpha\dot{\beta}} \hat{P}_{\alpha\dot{\beta}})} \ , \ e^{-i(\epsilon^\alpha \hat{Q}_\alpha + \bar{\epsilon}^{\dot{\alpha}} \hat{\bar{Q}}_{\dot{\alpha}})} \right) \ , \qquad (3.3.14)$$

respectively. Using the Baker-Hausdorff theorem ($e^A e^B = e^{A+B+\frac{1}{2}[A,B]}$ if $[A,[A,B]] = [B,[A,B]] = 0$) to rearrange the exponents, we find:

$$J \ \& \ \bar{J}\colon \quad x'^{\alpha\dot{\alpha}} = [e^\omega]_\beta{}^\alpha \, [e^{\bar{\omega}}]_{\dot{\beta}}{}^{\dot{\alpha}} \, x^{\beta\dot{\beta}} \ , \quad \theta'^\alpha = [e^\omega]_\beta{}^\alpha \, \theta^\beta \ , \quad \bar{\theta}'^{\dot{\alpha}} = [e^{\bar{\omega}}]_{\dot{\beta}}{}^{\dot{\alpha}} \, \bar{\theta}^{\dot{\beta}} \ ,$$

$$P\colon \quad x'^{\underline{a}} = x^{\underline{a}} + \xi^{\underline{a}} \ , \quad \theta'^\alpha = \theta^\alpha \ , \quad \bar{\theta}'^{\dot{\alpha}} = \bar{\theta}^{\dot{\alpha}} \ ,$$

$$Q \ \& \ \bar{Q}\colon \quad x'^{\underline{a}} = x^{\underline{a}} - i\,\frac{1}{2}\,(\epsilon^\alpha \bar{\theta}^{\dot{\alpha}} + \bar{\epsilon}^{\dot{\alpha}} \theta^\alpha) \ , \ \theta'^\alpha = \theta^\alpha + \epsilon^\alpha \ , \ \bar{\theta}'^{\dot{\alpha}} = \bar{\theta}^{\dot{\alpha}} + \bar{\epsilon}^{\dot{\alpha}} \ . \qquad (3.3.15)$$

Thus the generators are realized as coordinate transformations in superspace. The Lorentz group acts *reducibly:* Under its action the $x$'s and $\theta$'s do *not* transform into each other.

## b.3. Action of generators on superfields

To get representations of supersymmetry on physical fields, we consider *superfields* $\Psi_{\alpha\cdots}(x, \theta, \bar{\theta})$: (generalized) multispinor functions over superspace. Under the supersymmetry algebra they are defined to transform as coordinate scalars and Lorentz multispinors. They may also be in a matrix representation of an internal symmetry group. The simplest case is a scalar superfield, which transforms as: $\Phi'(x', \theta', \bar{\theta}') = \Phi(x, \theta, \bar{\theta})$ or, infinitesimally, $\delta\Phi \equiv \Phi'(z) - \Phi(z) = -\,\delta z^M \partial_M \Phi(z)$. Using (3.3.13), we write the transformation as $\delta\Phi = -\,i[(\epsilon^\alpha \hat{Q}_\alpha + \bar{\epsilon}^{\dot{\alpha}} \hat{\bar{Q}}_{\dot{\alpha}}) , \Phi] \ = \ i[(\epsilon^\alpha Q_\alpha + \bar{\epsilon}^{\dot{\alpha}} \bar{Q}_{\dot{\alpha}}) , \Phi]$, etc. Hence, just as in the ordinary Poincaré case, the generators $\hat{Q}$, etc., are represented by differential operators $Q$, etc.:

$$J_{\alpha\beta} = -\,i\,\frac{1}{2}\,(x_{(\alpha}{}^{\dot{\gamma}} \partial_{\beta)\dot{\gamma}} + \theta_{(\alpha} \partial_{\beta)}) - iM_{\alpha\beta} \ ,$$

$$P_{\alpha\dot{\beta}} = i\partial_{\alpha\dot{\beta}} \ ,$$

$$Q_\alpha = i(\partial_\alpha - \frac{1}{2}\,\bar{\theta}^{\dot{\alpha}} i\partial_{\alpha\dot{\alpha}}) \ , \quad \bar{Q}_{\dot{\alpha}} = i(\bar{\partial}_{\dot{\alpha}} - \frac{1}{2}\,\theta^\alpha i\partial_{\alpha\dot{\alpha}}) \ ; \qquad (3.3.16)$$

where $M_{\alpha\beta}$ generates the *matrix* Lorentz transformations of the superfield $\Psi$:

$[M_{\alpha\beta}, \Psi_{\gamma\cdots}] = \frac{1}{2} C_{\gamma(\alpha} \Psi_{\beta)\cdots} + \cdots$.

For future use, we write $Q$ and $\bar{Q}$ as

$$Q_\alpha = e^{\frac{1}{2}U} i\partial_\alpha e^{-\frac{1}{2}U} \quad , \quad \bar{Q}_{\dot\alpha} = e^{-\frac{1}{2}U}\, i\bar\partial_{\dot\alpha} e^{\frac{1}{2}U} \quad , \tag{3.3.17a}$$

where

$$U = \theta^\alpha \bar\theta^{\dot\beta} i\partial_{\alpha\dot\beta} \quad . \tag{3.3.17b}$$

Finally, from the relation $\{Q, \bar{Q}\} = P$, we conclude that the dimension of $\theta$ and $\bar\theta$ is $(m)^{-\frac{1}{2}}$.

## b.4. Extended supersymmetry

We now generalize to extended Poincaré supersymmetry. In principle, the results we present could be derived by methods similar to the above, or by using a systematic differential geometry procedure. In practice the simplest procedure is to start with the $N = 1$ Poincaré results and generalize them by dimensional analysis and $U(N)$ symmetry.

For general $N$, superspace has coordinates $z^A = (x^{\alpha\dot\alpha}, \theta^{a\alpha}, \bar\theta_a{}^{\dot\alpha}) \equiv (x^{\underline{a}}, \theta^{\underline{\alpha}}, \bar\theta^{\underline{\dot\alpha}})$. Superfields $\Psi_{\alpha\beta\cdots ab\cdots}(x, \theta, \bar\theta)$ transform as multispinors and isospinors, and as coordinate scalars. Including central charges, the super-Poincaré generators act on superfields as the following differential operators:

$$Q_{a\alpha} = i(\partial_{a\alpha} - \frac{1}{2}\bar\theta_a{}^{\dot\beta} i\partial_{\alpha\dot\beta} - \frac{1}{2}\theta^b{}_\alpha Z_{ba}) \quad , \tag{3.3.18a}$$

$$\bar{Q}^a{}_{\dot\alpha} = i(\bar\partial^a{}_{\dot\alpha} - \frac{1}{2}\theta^{a\beta} i\partial_{\beta\dot\alpha} - \frac{1}{2}\bar\theta_{b\dot\alpha}\bar{Z}^{ba}) \quad , \tag{3.3.18b}$$

$$J_{\alpha\beta} = -\,i\,\frac{1}{2}\,(x_{(\alpha}{}^{\dot\gamma}\partial_{\beta)\dot\gamma} + \theta^a{}_{(\alpha}\partial_{a\beta)}) - iM_{\alpha\beta} \quad , \tag{3.3.18c}$$

$$\bar{J}_{\dot\alpha\dot\beta} = -\,i\,\frac{1}{2}\,(x^\gamma{}_{(\dot\alpha}\partial_{\gamma\dot\beta)} + \bar\theta_{a(\dot\alpha}\bar\partial^a{}_{\dot\beta)}) - i\bar{M}_{\dot\alpha\dot\beta} \quad , \tag{3.3.18d}$$

$$P_{\alpha\dot\alpha} = i\partial_{\alpha\dot\alpha} \quad . \tag{3.3.18e}$$

Central charges are discussed in section 4.6.

## b.5. CPT in superspace

Poincaré supersymmetry is compatible with the discrete invariances CP (charge conjugation × parity) and T (time reversal). We begin by reviewing C, P, and T in ordinary spacetime. We describe the transformations as acting on $c$-number fields, i.e., we use the functional integral formalism, rather than acting on $q$-number fields or Hilbert space states.

Under a *reflection* with respect to an arbitrary (but not lightlike) axis $u_{\underline{a}}$, ($u = \overline{u}$, $u^2 = \pm 1$) the coordinates transform as

$$x'^{\underline{a}} = R(u)x^{\underline{a}} = -\, u^{-2}\, u^{\alpha}{}_{\dot{\beta}} u_{\beta}{}^{\dot{\alpha}} x^{\beta\dot{\beta}}$$

$$= x^{\underline{a}} - u^{-2}\, u^{\underline{a}} u \cdot x \quad , \qquad R^2 = I \tag{3.3.19}$$

($u \cdot x$ changes sign, while the components of $x$ orthogonal to $u$ are unchanged.) T then acts on the coordinates as $R(\delta_{\underline{a}}{}^0)$ while a space reflection can be represented by $R(\delta_{\underline{a}}{}^1)R(\delta_{\underline{a}}{}^2)R(\delta_{\underline{a}}{}^3)$ (in terms of a timelike vector $\delta_{\underline{a}}{}^0$, and three orthogonal spacelike vectors $\delta_{\underline{a}}{}^i$, $i = 1, 2, 3$).

We define the action of the discrete symmetries on a real scalar field by $\phi'(x') = \phi(x)$. The action on a Weyl spinor is

$$\psi'^{\alpha}(x') = i u^{\alpha}{}_{\dot{\alpha}} \overline{\psi}^{\dot{\alpha}}(x) \quad , \qquad \overline{\psi}'^{\dot{\alpha}}(x') = i u_{\alpha}{}^{\dot{\alpha}} \psi^{\alpha}(x) \quad ;$$

$$\psi''^{\alpha}(x) = u^2 \psi^{\alpha}(x) \quad . \tag{3.3.20}$$

Since this transformation involves complex conjugation, we interpret $R$ as giving CP and T. Indeed, since under complex conjugation $e^{-ipx} \to e^{+ipx}$, we have $p'^{\underline{a}} = -\,(p^{\underline{a}} - u^{-2} u^{\underline{a}} u \cdot p)$. Therefore $p^0$ changes sign for spacelike $u$, and this is consistent with our interpretation. The combined transformation CPT is simply $x \to -\,x$ and the fields transform without any factors (except for irrelevant phases). The transformation of an arbitrary Lorentz representation is obtained by treating each spinor index as in (3.3.20).

The definition of C, and thus P and CT, requires the existence of an additional, internal, discrete symmetry, e.g., a symmetry involving only sign changes: For the photon field $\mathrm{C}A_{\underline{a}} = -\,A_{\underline{a}}$; for a pair of real scalars, $\mathrm{C}\phi_1 = +\,\phi_1$, $\mathrm{C}\phi_2 = -\,\phi_2$ gives $C(\phi_1 + i\phi_2) = (\phi_1 + i\phi_2)^\dagger$. For a pair of spinors, $C\psi_1{}^\alpha = \psi_2{}^\alpha$, $C\psi_2{}^\alpha = \psi_1{}^\alpha$ gives, for the Dirac spinor $(\psi_1{}^\alpha, \bar\psi_2{}^{\dot\alpha})$, the transformation $C(\psi_1{}^\alpha, \bar\psi_2{}^{\dot\alpha}) = (\bar\psi_2{}^{\dot\alpha}, \psi_1{}^\alpha)^\dagger$, i.e., complex conjugation times a matrix. Therefore, C generally involves complex conjugation of a field, as do CP and T, whereas P and CT do not. (However, note that the definition of complex conjugation depends on the definition of the fields, e.g., combining $\phi_1$ and $\phi_2$ as $\phi_1 + i\phi_2$.)

The generalization to superspace is straightforward: In addition to the transformation $R(u)x$ given above, we have (as for any spinor)

$$\theta'^{a\alpha} = iu^\alpha{}_{\dot\alpha}\bar\theta_a{}^{\dot\alpha} \quad , \quad \bar\theta'_a{}^{\dot\alpha} = iu_\alpha{}^{\dot\alpha}\theta^{a\alpha} \quad . \tag{3.3.21}$$

A real scalar superfield and a Weyl spinor superfield thus transform as the corresponding component fields, but now with all superspace coordinates transforming under $R(u)$. To preserve the chirality of a superspace or superfield (see below), we define $R(u)$ to *always* complex conjugate the superfields. We thus have, e.g.,

$$\Phi'(z') = \overline{\Phi(z)} \quad , \quad \Psi'^\alpha(z') = iu^\alpha{}_{\dot\beta}\overline{\Psi^\beta(z)} = iu^\alpha{}_{\dot\beta}\bar\Psi^{\dot\beta}(\bar z) \quad . \tag{3.3.22}$$

As for components, C can be defined as an additional (internal) discrete symmetry which can be expressed as a matrix times hermitian conjugation.

We remark that $R(u)$ transforms the supersymmetry generators covariantly only for $u^2 = +1$. For $u^2 = -1$ there is a relative sign change between $\partial_{\underline\alpha}$ and $\bar\theta^{\underline{\dot\beta}} i\partial_{\underline{\alpha\dot\beta}}$. This is because CP changes the sign of $p^0$, which is needed to maintain the positivity of the energy (see (3.2.10)).

## b.6. Chiral representations of supersymmetry

As in the $N = 1$ case (see (3.3.17)), $Q, \bar Q$ can be written compactly for higher $N$, even in the presence of central charges:

$$Q_{\underline\alpha} = e^{\frac{1}{2}U} i(\partial_{\underline\alpha} - \frac{1}{2}\theta^b{}_\alpha Z_{ba})e^{-\frac{1}{2}U} \quad ,$$

$$\bar Q_{\underline{\dot\alpha}} = e^{-\frac{1}{2}U} i(\bar\partial_{\underline{\dot\alpha}} - \frac{1}{2}\bar\theta_{b\dot\alpha}\bar Z^{ba})e^{\frac{1}{2}U} \quad , \tag{3.3.23a}$$

$$U \equiv \theta^{\underline\gamma}\bar\theta^{\underline{\dot\delta}} i\partial_{\underline{\gamma\dot\delta}} \quad , \quad \partial_{\underline{\gamma\dot\delta}} = \delta_c{}^d\partial_{\gamma\dot\delta} \quad . \tag{3.3.23b}$$

This allows us to find other representations of the super-Poincaré algebra in which $Q$ (or $\bar{Q}$) take a very simple form. We perform nonunitary similarity transformations on *all* generators $\Omega_A$:

$$\Omega_A{}^{(\pm)} = e^{\mp\frac{1}{2}U}\Omega_A e^{\pm\frac{1}{2}U} \quad , \tag{3.3.24}$$

which leads to:

$$Q_{\underline{\alpha}}{}^{(+)} = i(\partial_{\underline{\alpha}} - \frac{1}{2}\theta^b{}_\alpha Z_{ba}) \quad ,$$

$$\bar{Q}_{\underline{\dot{\alpha}}}{}^{(+)} = e^{-U} i(\bar{\partial}_{\underline{\dot{\alpha}}} - \frac{1}{2}\bar{\theta}_{b\dot{\alpha}}\bar{Z}^{ba})e^U \quad , \tag{3.3.25}$$

or

$$Q_\alpha{}^{(-)} = e^U i(\partial_\alpha - \frac{1}{2}\theta^b{}_\alpha Z_{ba})e^{-U} \quad ,$$

$$\bar{Q}_{\underline{\dot{\alpha}}}{}^{(-)} = i(\bar{\partial}_{\underline{\dot{\alpha}}} - \frac{1}{2}\bar{\theta}_{b\dot{\alpha}}\bar{Z}^{ba}) \quad . \tag{3.3.26a}$$

The generators act on transformed superfields

$$\Psi^{(\pm)}(z) = e^{\mp\frac{1}{2}U}\Psi(z)e^{\pm\frac{1}{2}U} \tag{3.3.26b}$$

These representations are called *chiral* or *antichiral* representations, whereas the original one is called the *vector* representation. They can also be found directly by the method of induced representations by using a slightly different parametrization of the coset space manifold (superspace) (cf. (3.3.11)):

$$h^{(+)} = e^{i\theta\hat{Q}}e^{ix^{(+)}\hat{P}}e^{i\bar{\theta}\hat{\bar{Q}}} \quad , \quad h^{(-)} = e^{i\bar{\theta}\hat{\bar{Q}}}e^{ix^{(-)}\hat{P}}e^{i\theta\hat{Q}} \quad , \tag{3.3.27a}$$

where

$$x^{(\pm)} = x \pm i\,\frac{1}{2}\,\theta\bar{\theta} = e^{\pm\frac{1}{2}U}x\,e^{\mp\frac{1}{2}U} \tag{3.3.27b}$$

are complex (nonhermitian) coordinates. The corresponding superspaces are called chiral and antichiral, respectively. The similarity transformations (3.3.26b) can be regarded as complex coordinate transformations:

$$\Psi(z) = e^{\pm\frac{1}{2}U}\Psi^{(\pm)}(z)e^{\mp\frac{1}{2}U} = \Psi^{(\pm)}(z^{(\pm)}) \ ,$$

$$z^{(\pm)} = e^{\pm\frac{1}{2}U} z e^{\mp\frac{1}{2}U} = (x^{(\pm)}\,, \theta\,, \bar\theta) \quad . \tag{3.3.28}$$

Hermitian conjugation takes us from a chiral representation to an antichiral one: $\overline{(V^{(+)})} = \bar V^{(-)}$. Consequently, a hermitian quantity $V = \bar V$ in the vector representation satisfies

$$V = e^{-U}\,\bar V\,e^{U} \tag{3.3.29}$$

in the chiral representation.

## b.7. Superconformal representations

The method of induced representations can be used to find representations for the superconformal group. However, we use a different procedure. The representations of $Q$, $P$, and $J$ are as in the super-Poincaré case. The representations of the remaining generators are found as follows: In ordinary spacetime, the conformal boost generators $K$ can be constructed by first performing an inversion, then a translation ($P$ transformation), and finally performing another inversion; a similar sequence of operations can be used in superspace to construct $K$ from $P$ and $S$ from $Q$.

We define the inversion operation as the following map between chiral and antichiral superspace:

$$\begin{aligned}
x'^{(\pm)\alpha\dot\alpha} &= (x^{(\mp)})^{-2}\,x^{(\mp)\alpha\dot\alpha} &&= \overline{(x^{(\pm)})^{-2}\,x^{(\pm)\alpha\dot\alpha}} \quad , \\
\theta'^{a\alpha} &= i(x^{(-)})^{-2}\,x^{(-)\alpha}{}_{\dot\alpha}\bar\theta_a{}^{\dot\alpha} &&= -\,i\,\overline{(x^{(+)})^{-2}\,x^{(+)}{}_\alpha{}^{\dot\alpha}\theta^{a\alpha}} \quad , \\
\bar\theta'_a{}^{\dot\alpha} &= i(x^{(+)})^{-2}\,x^{(+)}{}_\alpha{}^{\dot\alpha}\theta^{a\alpha} &&= -\,i\,\overline{(x^{(-)})^{-2}\,x^{(-)\alpha}{}_{\dot\alpha}\bar\theta_a{}^{\dot\alpha}} \quad ;
\end{aligned} \tag{3.3.30}$$

we have $z'' = z$. The essential property of this mapping is that it scales a supersymetrically invariant extension of the line element. We write $ds^2 = \frac{1}{2}\,s_{\alpha\dot\beta}s^{\alpha\dot\beta}$, where

$$s^{\alpha\dot\beta} = dx^{\alpha\dot\beta} + \frac{i}{2}\,(\theta^{a\alpha}d\bar\theta_a{}^{\dot\beta} + \bar\theta_a{}^{\dot\beta}d\theta^{a\alpha}) \quad , \tag{3.3.31}$$

is a supersymmetrically invariant 1-form (invariance follows at once from (3.3.15)). Under inversions (3.3.30), we find

$$s'^{\alpha\dot\beta} = -\,(x^{(+)})^{-2}\,(x^{(-)})^{-2}\,x^{(+)\beta\dot\beta}x^{(-)\alpha\dot\alpha}\,s_{\beta\dot\alpha} \quad ,$$

$$ds'^2 = (x^{(+)})^2\,(x^{(-)})^2\,ds^2 \quad . \tag{3.3.32}$$

Superfields transform as

$$I\!I\Psi^{\alpha\cdots\dot{\beta}\cdots}(z) = (x^{(+)})^{-2d^+}(x^{(-)})^{-2d^-}\,f^\alpha{}_{\dot{\alpha}}\cdots\bar{f}^{\dot{\beta}}{}_\beta\,\overline{\Psi^{\alpha\cdots\dot{\beta}\cdots}(z')} \quad , \tag{3.3.33a}$$

$$f^\alpha{}_{\dot{\alpha}} \equiv i(x^{(+)})^{-1}x^{(+)\alpha}{}_{\dot{\alpha}} \quad , \qquad \bar{f}^{\dot{\alpha}}{}_\alpha \equiv i(x^{(-)})^{-1}x^{(-)}{}_\alpha{}^{\dot{\alpha}} \quad . \tag{3.3.33b}$$

Here $d \equiv d^+ + d^-$ is the canonical dimension (Weyl weight) of $\Psi$, and $d^- - d^+$ is proportional to the chiral $U(1)$ weight $w$. Note that *chiral* superfields (fields depending only on and $x^{(+)}$ and $\theta$, not $\bar{\theta}$; see sec. 3.5) with $d^- = 0$ and only undotted indices remain chiral after an inversion.

We can calculate $S^{\underline{\alpha}}$ as described above: We use the inversion operator $I\!I$ and compute $S^a{}_\alpha = I\!I\bar{Q}_{\underline{\dot{\alpha}}}I\!I$ and $\bar{S}_{a\dot{\alpha}} = I\!I Q_{\underline{\alpha}}I\!I$. Using the superconformal commutator algebra we then compute $K$, $A$, $T$, and $\Delta$. We find

$$A = \frac{1}{2}\,(\theta^{\underline{\alpha}}\partial_{\underline{\alpha}} - \bar{\theta}^{\underline{\dot{\alpha}}}\bar{\partial}_{\underline{\dot{\alpha}}}) - Y \quad , \tag{3.3.34a}$$

$$T_a{}^b = \frac{1}{2}\,(\theta^{b\alpha}\partial_{a\alpha} - \bar{\theta}_a{}^{\dot{\alpha}}\bar{\partial}^b{}_{\dot{\alpha}} - \frac{1}{N}\,\delta_a{}^b(\theta^{\underline{\alpha}}\partial_{\underline{\alpha}} - \bar{\theta}^{\underline{\dot{\alpha}}}\bar{\partial}_{\underline{\dot{\alpha}}})) + t_a{}^b \quad , \tag{3.3.34b}$$

$$\begin{aligned} S^{a\alpha} = {} & i(x^{\alpha\dot{\alpha}} - i\,\frac{1}{2}\,\theta^{b\alpha}\bar{\theta}_b{}^{\dot{\alpha}})\bar{Q}^a{}_{\dot{\alpha}} + \theta^{a\beta}\theta^{b\alpha}i(\partial_{b\beta} + i\,\frac{1}{2}\,\bar{\theta}_b{}^{\dot{\gamma}}\partial_{\beta\dot{\gamma}}) \\ & - 2i\theta^{b\beta}[\delta_\beta{}^\alpha(t_b{}^a + \frac{1}{4}\,\delta_b{}^a(1 - \frac{4}{N})Y) - \frac{1}{2}\,\delta_b{}^a(M_\beta{}^\alpha + \frac{1}{2}\,\delta_\beta{}^\alpha\boldsymbol{d})] \quad , \end{aligned} \tag{3.3.34c}$$

$$\begin{aligned} \bar{S}_a{}^{\dot{\alpha}} = {} & i(x^{\alpha\dot{\alpha}} + i\,\frac{1}{2}\,\theta^{b\alpha}\bar{\theta}_b{}^{\dot{\alpha}})Q_{a\alpha} + \bar{\theta}_a{}^{\dot{\beta}}\bar{\theta}_b{}^{\dot{\alpha}}i(\bar{\partial}^b{}_{\dot{\beta}} + i\,\frac{1}{2}\,\theta^{b\beta}\partial_{\beta\dot{\beta}}) \\ & -2i\bar{\theta}_b{}^{\dot{\beta}}[-\delta_{\dot{\beta}}{}^{\dot{\alpha}}(t_a{}^b + \frac{1}{4}\,\delta_a{}^b(1 - \frac{4}{N})Y) - \frac{1}{2}\,\delta_a{}^b(\bar{M}_{\dot{\beta}}{}^{\dot{\alpha}} + \frac{1}{2}\,\delta_{\dot{\beta}}{}^{\dot{\alpha}}\boldsymbol{d})] \quad , \end{aligned} \tag{3.3.34d}$$

$$\Delta = -\,i\,\frac{1}{2}\,(\{x^{\alpha\dot{\alpha}}\,,\partial_{\alpha\dot{\alpha}}\} + \frac{1}{2}\,([\theta^{\underline{\alpha}}\,,\partial_{\underline{\alpha}}] + [\bar{\theta}^{\underline{\dot{\alpha}}}\,,\bar{\partial}_{\underline{\dot{\alpha}}}])) - i\boldsymbol{d} \quad , \tag{3.3.34e}$$

$$\begin{aligned} K^{\alpha\dot{\alpha}} = {} & -\,i(x^{\alpha\dot{\beta}}x^{\beta\dot{\alpha}}\partial_{\beta\dot{\beta}} + x^{\alpha\dot{\beta}}\bar{\theta}_a{}^{\dot{\alpha}}\bar{\partial}^a{}_{\dot{\beta}} + x^{\beta\dot{\alpha}}\theta^{a\alpha}\partial_{a\beta} - \frac{1}{4}\,\theta^{a\alpha}\bar{\theta}_a{}^{\dot{\beta}}\theta^{b\beta}\bar{\theta}_b{}^{\dot{\alpha}}\partial_{\beta\dot{\beta}}) \\ & + \frac{1}{2}\,(\theta^{a\beta}\bar{\theta}_a{}^{\dot{\alpha}}\theta^{b\alpha}\partial_{b\beta} - \theta^{a\alpha}\bar{\theta}_a{}^{\dot{\beta}}\bar{\theta}_b{}^{\dot{\alpha}}\bar{\partial}^b{}_{\dot{\beta}}) \end{aligned}$$

$$- i(x^{\beta\dot{\alpha}} + i\,\frac{1}{2}\,\theta^{a\beta}\bar{\theta}_a{}^{\dot{\alpha}})M_\beta{}^\alpha - i(x^{\alpha\dot{\beta}} - i\,\frac{1}{2}\,\theta^{a\alpha}\bar{\theta}_a{}^{\dot{\beta}})\bar{M}_{\dot{\beta}}{}^{\dot{\alpha}}$$

$$- x^{\alpha\dot{\alpha}} i\boldsymbol{d} - 2\theta^{a\alpha}\bar{\theta}_b{}^{\dot{\alpha}}(t_a{}^b + \frac{1}{4}\,\delta_a{}^b(1 - \frac{4}{N})Y) \quad . \tag{3.3.34f}$$

Here $\boldsymbol{d}$ is the matrix piece of the generator $\Delta$; its eigenvalue is the canonical dimension $d$. Similarly, $Y$ , $t_a{}^b$ are the matrix pieces of the axial generator $A$ and the $SU(N)$ generators $T_a{}^b$; the eigenvalue of $Y$ is $\frac{1}{2}\,w$. The terms in $S$ , $\bar{S}$ proportional to $Y$ and $t_a{}^b$ do not follow from the inversion (3.3.33), but are determined by the commutation relations and (3.3.34a,b).

### b.8. Super-deSitter representations

To construct the generators of the super-deSitter algebra, we use the expressions for the conformal generators and take the linear combinations prescribed in (3.2.13).

To summarize, for general $N$, in each of the cases we have considered the generators act as differential operators. In addition the superfields may carry a nontrivial matrix representation of all the generators except for $P$ and $Q$ in the Poincaré and deSitter cases, and $P$, $Q$, $K$, and $S$ in the superconformal case. They may also carry a representation of some arbitrary internal symmetry group.

## 3.4. Covariant derivatives

In ordinary flat spacetime, the usual coordinate derivative $\partial_{\underline{a}}$ is translation invariant: the translation generator $P_{\underline{a}}$, which is represented by $\partial_{\alpha\dot{\alpha}}$, commutes with itself. In supersymmetric theories, the supertranslation generator $Q_\alpha$ has a nontrivial anticommutator, and hence is not invariant under supertranslations; a simple computation reveals that the fermionic coordinate derivatives $\partial_\alpha$, $\partial_{\dot{\alpha}}$ are not invariant either. There is, however, a simple way to construct derivatives that are invariant under supersymmetry transformations generated by $Q_\alpha$ , $\bar{Q}_{\dot{\alpha}}$ (and are covariant under Lorentz, chiral, and isospin rotations generated by $J_{\alpha\beta}$ , $\bar{J}_{\dot{\alpha}\dot{\beta}}$, $A$, and $T_a{}^b$).

### a. Construction

In the preceding section we used the method of induced representations to find the action of the super-Poincaré generators in superspace. The same method can be used to find covariant derivatives. We define the operators $D_\alpha$ and $\bar{D}_{\dot{\alpha}}$ by the equation

$$(e^{\epsilon D + \bar{\epsilon}\bar{D}})(e^{i(x\hat{P} + \theta\hat{Q} + \bar{\theta}\hat{\bar{Q}})}) \equiv (e^{i(x\hat{P} + \theta\hat{Q} + \bar{\theta}\hat{\bar{Q}})})(e^{i(\epsilon\hat{Q} + \bar{\epsilon}\hat{\bar{Q}})}) \quad . \tag{3.4.1}$$

The anticommutator of $\hat{Q}$ with $D$ can be examined as follows:

$$\begin{aligned}
&(e^{-i(\epsilon\hat{Q} + \bar{\epsilon}\hat{\bar{Q}})})(e^{\zeta D + \bar{\zeta}\bar{D}})(e^{i(\epsilon\hat{Q} + \bar{\epsilon}\hat{\bar{Q}})})(e^{i(x\hat{P} + \theta\hat{Q} + \bar{\theta}\hat{\bar{Q}})}) \\
&\qquad = (e^{-i(\epsilon\hat{Q} + \bar{\epsilon}\hat{\bar{Q}})})\Big((e^{i(\epsilon\hat{Q} + \bar{\epsilon}\hat{\bar{Q}})})(e^{i(x\hat{P} + \theta\hat{Q} + \bar{\theta}\hat{\bar{Q}})})(e^{i(\zeta\hat{Q} + \bar{\zeta}\hat{\bar{Q}})})\Big) \\
&\qquad = (e^{i(x\hat{P} + \theta\hat{Q} + \bar{\theta}\hat{\bar{Q}})})(e^{i(\zeta\hat{Q} + \bar{\zeta}\hat{\bar{Q}})}) \\
&\qquad = (e^{\zeta D + \bar{\zeta}\bar{D}})(e^{i(x\hat{P} + \theta\hat{Q} + \bar{\theta}\hat{\bar{Q}})}) \quad .
\end{aligned} \tag{3.4.2}$$

Thus the $D$'s are invariant under supertranslations (and also under ordinary translations):

$$\{Q, D\} = \{\bar{Q}, D\} = [P, D] = 0 \quad . \tag{3.4.3}$$

We can use the Baker-Hausdorff theorem, (3.4.1), and (3.3.11,13) to compute the explicit forms of the $D$'s from the $Q$'s. We find

$$D_\alpha = -iQ_\alpha + \bar{\theta}^{\dot{\alpha}} P_{\alpha\dot{\alpha}} \quad , \quad \bar{D}_{\dot{\alpha}} = -i\bar{Q}_{\dot{\alpha}} + \theta^\alpha P_{\alpha\dot{\alpha}} \quad . \tag{3.4.4}$$

For $N = 1$, when acting on superfields, they have the form

$$D_\alpha = \partial_\alpha + \frac{1}{2}\bar{\theta}^{\dot{\alpha}} i\partial_{\alpha\dot{\alpha}} \quad , \quad \bar{D}_{\dot{\alpha}} = \bar{\partial}_{\dot{\alpha}} + \frac{1}{2}\theta^\alpha i\partial_{\alpha\dot{\alpha}} \quad , \tag{3.4.5}$$

and are covariant generalizations of the ordinary spinor derivative $\partial_\alpha$, $\bar{\partial}_{\dot{\alpha}}$. For general $N$, with central charges, the covariant derivatives have the form:

$$D_{\underline{\alpha}} = D_{a\alpha} = \partial_{\underline{\alpha}} + \frac{1}{2}\bar{\theta}^{\underline{\dot{\alpha}}} i\partial_{\underline{\alpha}\underline{\dot{\alpha}}} + \frac{1}{2}\theta^b{}_\alpha Z_{ba} \quad ,$$

$$\bar{D}_{\underline{\dot{\alpha}}} = \bar{D}^a{}_{\dot{\alpha}} = \bar{\partial}_{\underline{\dot{\alpha}}} + \frac{1}{2}\theta^{\underline{\alpha}} i\partial_{\underline{\alpha}\underline{\dot{\alpha}}} + \frac{1}{2}\bar{\theta}_{b\dot{\alpha}}\bar{Z}^{ba} \quad . \tag{3.4.6}$$

They can be rewritten using $e^U$ as:

$$D_{\underline{\alpha}} = e^{-\frac{1}{2}U}(\partial_{\underline{\alpha}} + \frac{1}{2}\theta^b{}_\alpha Z_{ba})e^{\frac{1}{2}U} \quad ,$$

$$\bar{D}_{\underline{\dot{\alpha}}} = e^{\frac{1}{2}U}(\bar{\partial}_{\underline{\dot{\alpha}}} + \frac{1}{2}\bar{\theta}_{b\dot{\alpha}}\bar{Z}^{ba})e^{-\frac{1}{2}U} \quad . \tag{3.4.7}$$

Consequently, just as the generators $Q$ simplify in the chiral (antichiral) representation, the covariant derivatives have the simple but asymmetric form:

$$D_{\underline{\alpha}}{}^{(+)} = e^{-U}(\partial_{\underline{\alpha}} + \frac{1}{2}\theta^b{}_\alpha Z_{ab})e^U \quad , \quad \bar{D}_{\underline{\dot{\alpha}}}{}^{(+)} = \bar{\partial}_{\underline{\dot{\alpha}}} + \frac{1}{2}\bar{\theta}_{b\dot{\alpha}}\bar{Z}^{ab} \quad ,$$

$$D_{\underline{\alpha}}{}^{(-)} = \partial_{\underline{\alpha}} + \frac{1}{2}\theta^b{}_\alpha Z_{ab} \quad , \quad \bar{D}_{\underline{\dot{\alpha}}}{}^{(-)} = e^U(\bar{\partial}_{\underline{\dot{\alpha}}} + \frac{1}{2}\bar{\theta}_{b\dot{\alpha}}\bar{Z}^{ab})e^{-U} \quad . \tag{3.4.8}$$

In any representation, they have the following (anti)commutation relations:

$$\{D_{\underline{\alpha}}, D_{\underline{\beta}}\} = C_{\alpha\beta}Z_{ab} \quad , \quad \{D_{\underline{\alpha}}, \bar{D}_{\underline{\dot{\beta}}}\} = i\partial_{\underline{\alpha}\underline{\dot{\beta}}} \quad . \tag{3.4.9}$$

It is also possible to derive deSitter covariant derivatives by these methods. However, there is an easier, more useful, and more physical way to derive them within the framework of supergravity, since deSitter space is simply a curved space with constant curvature. This will be described in sec. 5.7.

## b. Algebraic relations

The covariant derivatives satisfy a number of useful algebraic relations. For $N = 1$, the only possible power of $D$ is $D^2 = \frac{1}{2}D^\alpha D_\alpha$. (Because of anticommutativity higher powers vanish: $(D)^3 = 0$.) From the anticommutation relations we also have

$$[D^\alpha, \bar{D}^2] = i\partial^{\alpha\dot{\alpha}}\bar{D}_{\dot{\alpha}} \ , \qquad D^2\bar{D}^2D^2 = \Box D^2 \ ,$$

$$D^\alpha D_\beta = \delta_\beta{}^\alpha D^2 \ , \qquad D^2\theta^2 = -1 \ . \tag{3.4.10}$$

For $N > 1$ we have similar relations; for vanishing central charges:

$$D^n{}_{\underline{\alpha}_1\cdots\underline{\alpha}_n} \equiv D_{\underline{\alpha}_1}\cdots D_{\underline{\alpha}_n} \ ,$$

$$D^{n\,\underline{\alpha}_{n+1}\cdots\underline{\alpha}_{2N}} \equiv \frac{1}{n!}C^{\underline{\alpha}_{2N}\cdots\underline{\alpha}_1}D^n{}_{\underline{\alpha}_1\cdots\underline{\alpha}_n} \ ,$$

$$D^n{}_{\underline{\alpha}_1\cdots\underline{\alpha}_n} = \frac{1}{(2N-n)!}C_{\underline{\alpha}_{2N}\cdots\underline{\alpha}_1}D^{n\,\underline{\alpha}_{n+1}\cdots\underline{\alpha}_{2N}} \ ,$$

$$D^{2N-n\,\underline{\alpha}_1\cdots\underline{\alpha}_n}D^n{}_{\underline{\beta}_1\cdots\underline{\beta}_n} = \delta_{[\underline{\beta}_1}{}^{\underline{\alpha}_1}\cdots\delta_{\underline{\beta}_n]}{}^{\underline{\alpha}_n}D^{2N} \ ,$$

$$(D^n{}_{\underline{\alpha}_1\cdots\underline{\alpha}_n})^\dagger = \bar{D}^n{}_{\underline{\dot{\alpha}}_n\cdots\underline{\dot{\alpha}}_1} \ ,$$

$$(D^{2N-n\,\underline{\alpha}_1\cdots\underline{\alpha}_n})^\dagger = (-1)^n\bar{D}^{2N-n\,\underline{\dot{\alpha}}_n\cdots\underline{\dot{\alpha}}_1} \ ,$$

$$D^{2N}\theta^{2N} = (-1)^N \ ,$$

$$\bar{D}^{2N}D^{2N}\bar{D}^{2N} = \Box^N\bar{D}^{2N} \ . \tag{3.4.11}$$

It is often necessary to reduce the product of $D$'s or $\bar{D}$'s with respect to $SU(N)$, as well as with respect to $SL(2,C)$. For each, the reduction is done by symmetrizing and antisymmetrizing the indices. Specifically, we find the irreducible representations as follows: A product $D_{\underline{\alpha}}D_{\underline{\beta}}\ldots D_{\underline{\lambda}}$ is totally antisymmetric in its combined indices since the $D$'s anticommute; however, antisymmetry in $\underline{\alpha}$, $\underline{\beta}$ implies opposite symmetries between $a$, $b$ and $\alpha$, $\beta$, (one pair symmetric, the other antisymmetric), and hence a Young tableau for the $SU(N)$ indices is paired with the same Young tableau *reflected about the diagonal* for the $SL(2,C)$ indices. The latter is actually an $SU(2)$ tableau since if we have only $D$'s then only undotted indices appear, and has at most two rows. (Actually, for $SU(2)$ a column of 2 is equivalent to a column of 0, and hence the $SL(2C)$ tableau can be reduced to a single row.) Therefore, the only $SU(N)$ tableaux that appear have two columns or less. The $SL(2,C)$ representation can be read directly from the $SU(N)$ tableau (if we keep columns of height $N$): The general $SU(N)$ tableau consists of a first column of height $p$ and a second of height $q$, where $p+q$ is the number of $D$'s; the corresponding $SL(2C)$ representation is a $(p-q)$-index totally symmetric undotted spinor. Therefore this representation of $SL(2,C)\otimes SU(N)$ has dimensionality

$$(p-q+1)\,\frac{p-q+1}{p+1}\binom{N}{p}\binom{N+1}{q} \quad . \tag{3.4.12}$$

### c. Geometry of flat superspace

The covariant derivatives *define* the geometry of "flat" superspace. We write them as a supervector:

$$D_A = (D_{\underline{\alpha}}\,, \bar{D}_{\underline{\dot{\alpha}}}\,, \partial_{\underline{a}}) \quad . \tag{3.4.13}$$

In general, in flat or curved space, a covariant derivative can be written in terms of coordinate derivatives $\partial_M \equiv \frac{\partial}{\partial z^M}$ and connections $\Gamma_A$:

$$D_A \equiv D_A{}^M \partial_M + \Gamma_A(M) + \Gamma_A(T) + \Gamma_A(Z) \quad . \tag{3.4.14}$$

The connections are the Lorentz connection

$$\Gamma_A(M) = \Gamma_{A\beta}{}^{\gamma} M_{\gamma}{}^{\beta} + \Gamma_{A\dot{\beta}}{}^{\dot{\gamma}} \bar{M}_{\dot{\gamma}}{}^{\dot{\beta}} \quad , \tag{3.4.15a}$$

isospin connection

$$\Gamma_A(T) = \Gamma_{Ab}{}^{c} T_c{}^{b} \quad , \tag{3.4.15b}$$

and central charge connection

$$\Gamma_A(Z) = \frac{1}{2}\,(\Gamma_A{}^{bc} Z_{bc} + \Gamma_{Abc} \bar{Z}^{bc}) \quad . \tag{3.4.15c}$$

The Lorentz generators $M$ act only on *tangent space* indices. (Although the distinction is unimportant in flat space, we distinguish "curved", or coordinate indices $M, N, \cdots$ from covariant or tangent space indices $A, B, \cdots$. In curved superspace we usually write the covariant derivatives as $\nabla_A = E_A{}^M D_M + \Gamma_A$, $D_M \equiv \delta_M{}^A D_A$, i.e., we use the *flat* superspace covariant derivatives instead of coordinate derivatives: see chapter 5 for details.)

In flat superspace, in the vector representation, from (3.4.6) we find the flat vielbein

$$D_A{}^M \;=\; \begin{pmatrix} \delta_{\underline{\alpha}}{}^{\underline{\mu}} & 0 & \frac{1}{2}\,i\delta_\alpha{}^\mu\delta_a{}^m\bar{\theta}_m{}^{\dot{\mu}} \\ 0 & \delta_{\underline{\dot{\alpha}}}{}^{\underline{\dot{\mu}}} & \frac{1}{2}\,i\delta_{\dot{\alpha}}{}^{\dot{\mu}}\delta_m{}^a\theta^{m\mu} \\ 0 & 0 & \delta_{\underline{a}}{}^{\underline{m}} \end{pmatrix} \quad , \qquad (3.4.16)$$

and the flat central charge connection

$$\Gamma_A{}^{bc} = -\,\frac{1}{2}\,(C_{\alpha\beta}\theta^{[b\beta}\delta_a{}^{c]}\ ,\, 0\ ,\, 0) \quad ,$$

$$\Gamma_{Abc} = -\,\frac{1}{2}\,(0\ ,\, C_{\dot{\alpha}\dot{\beta}}\bar{\theta}_{[b}{}^{\dot{\beta}}\delta_{c]}{}^a\ ,\, 0) \quad , \qquad (3.4.17)$$

all other flat connections vanishing. We can describe the geometry of superspace in terms of covariant *torsions* $T_{AB}{}^C$, *curvatures* $R_{AB}(M)$, and *field strengths* $F_{AB}(T)$ and $F_{AB}(Z)$:

$$[D_A\,, D_B\} = T_{AB}{}^C D_C + R_{AB}(M) + F_{AB}(T) + F_{AB}(Z) \qquad (3.4.18)$$

From (3.4.16-17), we find that *flat* superspace has nonvanishing torsion

$$T_{\underline{\alpha}\underline{\dot{\beta}}}{}^{\underline{c}} = i\delta_a{}^b\delta_\alpha{}^\gamma\delta_{\dot{\beta}}{}^{\dot{\gamma}} \qquad (3.4.19)$$

and nonvanishing central charge field strength

$$F_{\underline{\alpha}\underline{\beta}}{}^{cd} = C_{\alpha\beta}\delta_a{}^{[c}\delta_b{}^{d]} \quad , \quad F_{\underline{\dot{\alpha}}\underline{\dot{\beta}}cd} = C_{\dot{\alpha}\dot{\beta}}\delta_{[c}{}^a\delta_{d]}{}^b \quad , \qquad (3.4.20)$$

all other torsions, curvatures, and field strengths vanishing. Hence flat superspace has a nontrivial geometry.

## d. Casimir operators

The complete set of operators that commute with $P_{\underline{a}}$, $Q_{\underline{\alpha}}$ and $\bar{Q}_{\underline{\dot{\alpha}}}$ (and transform covariantly under $J_{\alpha\beta}$ and $\bar{J}_{\dot{\alpha}\dot{\beta}}$) is $\{D_A\,, M_\alpha{}^\beta\,, \bar{M}_{\dot{\alpha}}{}^{\dot{\beta}}\,, Y\,, t_a{}^b\,, \boldsymbol{d}\}$. (Except for $D_A$, which is only covariant with respect to the super-Poincaré algebra, all these operators are covariant with respect to the entire superconformal algebra. Note that the matrix operators $M\,, Y\,, t\,, \boldsymbol{d}$ act only on tangent space indices.) Thus the Casimir operators (group invariants) can all be expressed in terms of these operators. Following the discussion of subsec. 3.3.a.3, it is sufficient to construct:

$$P_{[\underline{a}}W_{\underline{b}]} = P_{[\underline{a}}f_{\underline{b}]} \quad , \quad f_{\underline{a}} \equiv \frac{1}{2}[D_{a\alpha}\,, \bar{D}^a{}_{\dot{\alpha}}] - i(\partial_{\beta\dot{\alpha}}M_{\alpha}{}^{\beta} - \partial_{\alpha\dot{\beta}}\bar{M}_{\dot{\alpha}}{}^{\dot{\beta}}) \quad ,$$

$$W_{\underline{a}} - AP_{\underline{a}} = f_{\underline{a}} + Yi\partial_{\underline{a}} \quad , \tag{3.4.21}$$

$$W_a{}^b = -\, m^2 t_a{}^b - \frac{i}{4}\partial^{\alpha\dot{\alpha}}([D_{a\alpha}\,, \bar{D}^b{}_{\dot{\alpha}}] - \frac{1}{N}\delta_a{}^b[D_{c\alpha}\,, \bar{D}^c{}_{\dot{\alpha}}]) \quad ,$$

$$W_5 = m^2 Y - \frac{i}{4}\partial^{\alpha\dot{\alpha}}[D_{a\alpha}\,, \bar{D}^a{}_{\dot{\alpha}}] \tag{3.4.22}$$

$$W_a{}^b{}_{\underline{a}} = t_a{}^b i\partial_{\underline{a}} - \frac{1}{4}([D_{a\alpha}\,, \bar{D}^b{}_{\dot{\alpha}}] - \frac{1}{N}\delta_a{}^b[D_{c\alpha}\,, \bar{D}^c{}_{\dot{\alpha}}]) \quad ;$$

$$W_{5\underline{a}} = -\, Yi\partial_{\underline{a}} - \frac{1}{4}[D_{a\alpha}\,, \bar{D}^a{}_{\dot{\alpha}}] \tag{3.4.23}$$

where we have used $P^2 = -\, m^2$ for $W_a{}^b$, and $P^{\alpha\dot{\alpha}}Q_{\underline{\alpha}} = \partial^{\alpha\dot{\alpha}}D_{\underline{\alpha}} = \partial^{\alpha\dot{\alpha}}\partial_{\underline{\alpha}} = 0$ for $W_a{}^b{}_{\underline{c}}$ (the massless case: see subsec. 3.3.a.3).

## 3.5. Constrained superfields

The existence of covariant derivatives allows us to consider constrained superfields; the simplest (and for many applications the most useful) is a chiral superfield defined by

$$\bar{D}_{\underline{\dot\alpha}}\Phi = 0 \quad . \tag{3.5.1}$$

We observe that the constraint (3.5.1) implies that on a chiral superfield $\bar{D}\Phi = 0$ and therefore $\{\bar{D}\,,\bar{D}\}\Phi = 0 \rightarrow \bar{Z}\Phi = 0$.

In a chiral representation, the constraint is simply the statement that $\Phi^{(+)}$ *is independent of* $\bar\theta$, that is $\Phi^{(+)}(x,\theta,\bar\theta) = \Phi^{(+)}(x,\theta)$. Therefore, in a vector representation,

$$\Phi(x,\theta,\bar\theta) = e^{\frac{1}{2}U}\Phi^{(+)}(x,\theta)e^{-\frac{1}{2}U} = \Phi^{(+)}(x^{(+)},\theta) \quad , \tag{3.5.2}$$

where $x^{(+)}$ is the chiral coordinate of (3.3.27b). Alternatively, one can write a chiral superfield in terms of a general superfield by using $\bar{D}^{2N+1} = 0$:

$$\Phi = \bar{D}^{2N}\,\Psi(x,\theta,\bar\theta) \tag{3.5.3}$$

This form of the solution to the constraint (3.5.1) is valid in any representation. It is the most general possible; see sec. 3.11.

Similarly, we can define antichiral superfields; these are annihilated by $D_{\underline{\alpha}}$. Note that $\bar\Phi$, the hermitian conjugate of a chiral superfield $\Phi$, is antichiral. These superfields may carry external indices.

* * *

The supersymmetry generators are represented much more simply when they act on chiral superfields, particularly in the chiral representation (3.3.25), than when they act on general superfields. For the super-Poincaré case we have:

$$Q_{\underline{\alpha}} = i(\partial_{\underline{\alpha}} - \frac{1}{2}\theta^b{}_\beta Z_{ba}) \ , \quad \bar{Q}_{\underline{\dot\alpha}} = \theta^{a\alpha}\partial_{\alpha\dot\alpha} \ , \quad P_{\underline{a}} = i\partial_{\underline{a}} \ ,$$

$$J_{\alpha\beta} = -\,i\,\frac{1}{2}\,(x_{(\alpha}{}^{\dot\alpha}\partial_{\beta)\dot\alpha} + \theta^a{}_{(\alpha}\partial_{a\beta)}) - iM_{\alpha\beta} \ ,$$

$$\bar{J}_{\dot\alpha\dot\beta} = -\,i\,\frac{1}{2}\,x^\alpha{}_{(\dot\alpha}\partial_{\alpha\dot\beta)} - i\bar{M}_{\dot\alpha\dot\beta} \ , \tag{3.5.4}$$

where $\bar{Z}^{ab}\Phi = 0$ (as explained above) but $Z_{ab}\Phi$ is unrestricted. If we think of $Z_{ab}$ as a partial derivative with respect to complex coordinates $\zeta^{ab}$, i.e., $Z_{ab} = i\,\frac{\partial}{\partial\zeta^{ab}}$, then a chiral superfield is a function of $x\,,\theta\,,\zeta$ and is independent of $\bar{\theta}\,,\bar{\zeta}$. In the superconformal case, $Z_{ab}$ must vanish, and, for consistency with the algebra, a chiral superfield must have *no* dotted indices (i.e., $\bar{M}_{\dot{\alpha}\dot{\beta}} = 0$). On chiral superfields, the inversion (3.3.33) takes the form

$$I\!I\Phi^{\alpha\cdots}(x\,,\theta) = x^{-2d} f^{\alpha}{}_{\dot{\alpha}}\cdots\overline{\Phi^{\alpha\cdots}(x'\,,\theta')} = x^{-2d} f^{\alpha}{}_{\dot{\alpha}}\cdots\bar{\Phi}^{\dot{\alpha}\cdots}(\bar{x}'\,,\bar{\theta}') \ ,$$

$$f^{\alpha}{}_{\dot{\alpha}} = i(x)^{-1}x^{\alpha}{}_{\dot{\alpha}} \ , \quad \overline{x'^{\underline{a}}} = x^{-2}x^{\underline{a}} \ , \quad \overline{\theta'^{\underline{\alpha}}} = ix^{-2}x_{\alpha}{}^{\dot{\alpha}}\theta^{a\alpha} \ ; \tag{3.5.5}$$

(note that $d^{-} = 0$ and hence $d = d^{+}$). The generators of the superconformal algebra are now just (3.5.4),

$$\bar{S}^{\underline{\dot{\alpha}}} = -\,x^{\alpha\dot{\alpha}}\partial_{a\alpha} \ , \quad S^{\underline{\alpha}} = -\,\theta^{a\beta}G_{\beta}{}^{\alpha} \ , \quad K^{\underline{a}} = x^{\beta\dot{\alpha}}G_{\beta}{}^{\alpha} \ ; \tag{3.5.6a}$$

with

$$G_{\alpha}{}^{\beta} = J_{\alpha}{}^{\beta} + \delta_{\alpha}{}^{\beta}[\Delta + i(\tfrac{1}{2}x^{\underline{a}}\partial_{\underline{a}} + 2 - N)] \ ,$$

$$\Delta = -\,i(x^{\underline{a}}\partial_{\underline{a}} + \tfrac{1}{2}\theta^{\underline{\alpha}}\partial_{\underline{\alpha}} + 2 - N + \boldsymbol{d}) \ ,$$

$$A = \tfrac{1}{2}\theta^{\underline{\alpha}}\partial_{\underline{\alpha}} - (4 - N)^{-1}N\boldsymbol{d} \ ,$$

$$T_{a}{}^{b} = \tfrac{1}{4}\,([\theta^{b\alpha}\,,\partial_{a\alpha}] - \tfrac{1}{N}\,\delta_{a}{}^{b}[\theta^{\underline{\alpha}}\,,\partial_{\underline{\alpha}}]) \ . \tag{3.5.6b}$$

The commutator algebra is, of course, unchanged. Note that the expression for $A$ contains a term $(1 - \frac{1}{4}N)^{-1}\boldsymbol{d}$; this implies that for $N = 4$, either $\boldsymbol{d}$ vanishes, or the axial charge must be dropped from the algebra (see sec. 3.2.e). The only known $N = 4$ theories are consistent with this fact: $N = 4$ Yang-Mills has no axial charge and $N = 4$ conformal supergravity has $\boldsymbol{d} = 0$. We further note that consistency of the algebra forbids the addition of the matrix operator $t_{a}{}^{b}$ to $T_{a}{}^{b}$ in the case of conformal chiral superfields. This means that conformal chiral superfields must be isosinglets, i.e., cannot carry external isospin indices.

* * *

For $N = 1$, a complex field satisfying the constraint $\bar{D}^2\Sigma = 0$ is called a linear superfield. A real linear superfield satisfies the constraint $D^2 G = \bar{D}^2 G = 0$. While such objects appear in some theories, they are less useful for describing interacting particle multiplets than chiral superfields. A complex linear superfield can always be written as $\Sigma = \bar{D}^{\dot{\alpha}}\Psi_{\dot{\alpha}}$, whereas a real linear superfield can be written as $G = D^{\alpha}\bar{D}^2\Psi_{\alpha} + h.c.$.

## 3.6. Component expansions

### a. $\theta$-expansions

Because the square of any anticommuting number vanishes, any function of a finite number of anticommuting variables has a terminating Taylor expansion with respect to them. This allows us to expand a superfield in terms of a finite number of ordinary spacetime dependent fields, or *components.* For general $N$, there are $4N$ independent anticommuting numbers in $\theta$, and thus $\sum_{i=0}^{4N} \binom{4N}{i} = 2^{4N}$ components in an unconstrained scalar superfield. For example, for $N = 1$, a real scalar superfield has the expansion

$$V = C + \theta^\alpha \chi_\alpha + \bar{\theta}^{\dot{\alpha}} \bar{\chi}_{\dot{\alpha}} - \theta^2 M - \bar{\theta}^2 \bar{M}$$

$$+ \theta^\alpha \bar{\theta}^{\dot{\alpha}} A_{\underline{a}} - \bar{\theta}^2 \theta^\alpha \lambda_\alpha - \theta^2 \bar{\theta}^{\dot{\alpha}} \bar{\lambda}_{\dot{\alpha}} + \theta^2 \bar{\theta}^2 \mathrm{D}' \qquad (3.6.1)$$

with 16 real components. Similarly, a chiral scalar superfield in vector representation has the expansion:

$$\Phi = e^{\frac{1}{2}U}(A + \theta^\alpha \psi_\alpha - \theta^2 F) e^{-\frac{1}{2}U}$$

$$= A + \theta^\alpha \psi_\alpha - \theta^2 F + i\,\frac{1}{2}\,\theta^\alpha \bar{\theta}^{\dot{\alpha}} \partial_{\underline{a}} A$$

$$+ i\,\frac{1}{2}\,\theta^2 \bar{\theta}^{\dot{\alpha}} \partial_{\underline{a}} \psi^\alpha + \frac{1}{4}\,\theta^2 \bar{\theta}^2 \Box A \qquad (3.6.2)$$

with 4 independent complex components.

These expansions become complicated for $N > 1$ superfields but fortunately are not needed. However, we give some examples to familiarize the reader with the component content of such superfields. For instance, for $N = 2$, in addition to carrying Lorentz spinor indices, superfields are representations of $SU(2)$. A real scalar-isoscalar superfield has the expansion

$$V(x,\theta,\bar{\theta}) = C(x) + \theta^{\underline{\alpha}} \chi_{\underline{\alpha}} + \bar{\theta}^{\underline{\dot{\alpha}}} \bar{\chi}_{\underline{\dot{\alpha}}} - \theta^{2\,\alpha\beta} M_{(\alpha\beta)} - \theta^{2\,ab} M_{(ab)}$$

$$- \bar{\theta}^{2\,\dot{\alpha}\dot{\beta}} \bar{M}_{(\dot{\alpha}\dot{\beta})} - \bar{\theta}^2{}_{ab} \bar{M}^{(ab)} + \theta^{a\alpha} \bar{\theta}_b{}^{\dot{\alpha}} (W_a{}^b{}_{\alpha\dot{\alpha}} + \delta_a{}^b V_{\alpha\dot{\alpha}}) + \cdots$$

$$+ \theta^4 N + \bar{\theta}^4 \bar{N} + \cdots + \theta^{2\alpha\beta}\bar{\theta}^{2\dot{\alpha}\dot{\beta}} h_{\alpha\beta\dot{\alpha}\dot{\beta}} + \cdots + \theta^4\bar{\theta}^4 \mathrm{D}'(x) \qquad (3.6.3)$$

where $W_a{}^a{}_{\alpha\dot{\alpha}} = 0$, while a chiral scalar isospinor superfield has the expansion (in the chiral representation)

$$\Phi^{(+)}{}_a(x,\theta) = A_a + \theta^{b\alpha}(C_{ab}\psi_\alpha + \psi_{(ab)\alpha})$$

$$- \theta^{2\,\alpha\beta} F_{a(\alpha\beta)} - \theta^{2\,bc}(F_{(abc)} + C_{ab}F_c)$$

$$- \theta^3{}_b{}^\alpha(\delta_a{}^b\lambda_\alpha + \lambda_a{}^b{}_\alpha) + \theta^4 \mathrm{D}'_a \quad , \qquad (3.6.4)$$

where $\lambda_a{}^a{}_\alpha = 0$. The spin and isospin of the component fields can be read from these expressions.

General superfields are not irreducible representations of extended supersymmetry. As we discuss in sec. 3.11, chiral superfields are irreducible under supersymmetry (except for a possible further decomposition into real and imaginary parts); we present there a systematic way of decomposing any superfield into its irreducible parts.

The supersymmetry transformations of the component fields follow straightforwardly from the transformations of the superfields. Thus, for example, for $N = 1$, from $\delta V = [i(\epsilon Q + \bar{\epsilon}\bar{Q}), V\,] = \delta C + \theta^\alpha \delta\chi_\alpha + \cdots$ (with a *constant* spinor parameter $\epsilon^\alpha$) we find:

$$\delta C(x) = -\,(\epsilon^\alpha\chi_\alpha + \bar{\epsilon}^{\dot{\alpha}}\bar{\chi}_{\dot{\alpha}}) \quad ,$$

$$\delta\chi_\alpha(x) = \epsilon_\alpha M - \bar{\epsilon}^{\dot{\alpha}}(i\,\tfrac{1}{2}\,\partial_{\underline{a}}C + A_{\underline{a}}) \quad ,$$

$$\delta\bar{\chi}_{\dot{\alpha}}(x) = \bar{\epsilon}_{\dot{\alpha}}\bar{M} - \epsilon^\alpha(i\,\tfrac{1}{2}\,\partial_{\underline{a}}C - A_{\underline{a}}) \quad ,$$

$$\delta M(x) = -\,\bar{\epsilon}^{\dot{\alpha}}(\bar{\lambda}_{\dot{\alpha}} + i\,\tfrac{1}{2}\,\partial_{\underline{a}}\chi^\alpha) \quad ,$$

$$\delta A_{\underline{a}}(x) = -\,\epsilon^\beta(C_{\beta\alpha}\bar{\lambda}_{\dot{\alpha}} + i\,\tfrac{1}{2}\,\partial_{\beta\dot{\alpha}}\chi_\alpha) + \bar{\epsilon}^{\dot{\beta}}(C_{\dot{\beta}\dot{\alpha}}\lambda_\alpha + i\,\tfrac{1}{2}\,\partial_{\alpha\dot{\beta}}\bar{\chi}_{\dot{\alpha}}) \quad ,$$

$$\delta\lambda_\alpha(x) = \epsilon^\beta(C_{\beta\alpha}\mathrm{D}' + i\,\tfrac{1}{2}\,\partial_{\beta\dot{\alpha}}A_\alpha{}^{\dot{\beta}}) - i\,\tfrac{1}{2}\,\bar{\epsilon}^{\dot{\alpha}}\partial_{\alpha\dot{\alpha}}\bar{M} \quad ,$$

$$\delta \mathrm{D}'(x) = i\,\frac{1}{2}\,\partial_{\underline{a}}(\epsilon^{\alpha}\bar{\lambda}^{\dot{\alpha}} + \bar{\epsilon}^{\dot{\alpha}}\lambda^{\alpha}) \quad , \qquad etc.\,, \tag{3.6.5}$$

Similarly, for a chiral superfield $\Phi$ we find:

$$\delta A = -\,\epsilon^{\alpha}\psi_{\alpha} \quad ,$$

$$\delta\psi_{\alpha} = -\,\bar{\epsilon}^{\dot{\alpha}} i\partial_{\alpha\dot{\alpha}} A + \epsilon_{\alpha} F \quad ,$$

$$\delta F = -\,\bar{\epsilon}^{\dot{\alpha}} i\partial^{\alpha}{}_{\dot{\alpha}}\psi_{\alpha} \quad . \tag{3.6.6}$$

## b. Projection

For many applications, the $\theta$-expansions just considered are inconvenient; an alternative is to define "components by projection" of an expression as the *$\theta$-independent* parts of its successive spinor derivatives. We introduce the notation $X|$ to indicate the $\theta$ independent part of an expression $X$. Then, for example, we can define the components of a chiral superfield by

$$A(x) = \Phi(x,\theta,\bar{\theta})| \quad ,$$

$$\psi_{\alpha}(x) = D_{\alpha}\Phi(x,\theta,\bar{\theta})| \quad ,$$

$$F(x) = D^2\Phi(x,\theta,\bar{\theta})| \quad . \tag{3.6.7}$$

The supersymmetry transformations of the component fields follow from the algebra of the covariant derivatives $D$; we use $(-iQ\Psi)| = (D\Psi)|$ and $\{D_{\alpha}\,,\bar{D}_{\dot{\beta}}\} = i\partial_{\alpha\dot{\beta}}$ to find

$$\delta A = i(\epsilon\cdot Q + \bar{\epsilon}\cdot\bar{Q})\Phi| = -\,(\epsilon\cdot D + \bar{\epsilon}\cdot\bar{D})\Phi|$$

$$= -\,\epsilon\cdot D\Phi| = -\,\epsilon^{\alpha}\psi_{\alpha} \quad ,$$

$$\delta\psi_{\alpha} = i(\epsilon\cdot Q + \bar{\epsilon}\cdot\bar{Q})D_{\alpha}\Phi| = -\,(\epsilon\cdot D + \bar{\epsilon}\cdot\bar{D})D_{\alpha}\Phi|$$

$$= (\epsilon_{\alpha}D^2 - \bar{\epsilon}^{\dot{\alpha}} i\partial_{\alpha\dot{\alpha}})\Phi| = \epsilon_{\alpha}F - \bar{\epsilon}^{\dot{\alpha}} i\partial_{\alpha\dot{\alpha}}A \quad ,$$

$$\delta F = i(\epsilon\cdot Q + \bar{\epsilon}\cdot\bar{Q})D^2\Phi| = -\,\bar{\epsilon}\cdot\bar{D}D^2\Phi| = -\,\bar{\epsilon}^{\dot{\alpha}} i\partial^{\alpha}{}_{\dot{\alpha}}\psi_{\alpha} \quad . \tag{3.6.8}$$

Explicit computation of the components shows that, in this particular case, the components in the $\theta$-expansion are identical to those defined by projection. This is not necessarily the case: For superfields that are not chiral, some components are defined with both $D$'s and $\bar{D}$'s; for these components, there is an ambiguity stemming from how the $D$'s and $\bar{D}$'s are ordered. For example, the $\theta^2\bar{\theta}$ component of a real scalar superfield $V$ could be defined as $D^2\bar{D}V|$, $D\bar{D}DV|$, or $\bar{D}D^2V|$. These definitions differ only by spacetime derivatives of components lower down in the $\theta$-expansion (defined with fewer $D$'s). In general, they will also differ from components defined by $\theta$-expansions by the same derivative terms. These differences are just field redefinitions and have no physical significance.

Usually, one particular definition of components is preferable. For example, one model that we will consider (see sec. 4.2.a) depends on a real scalar superfield $V$ which transforms as $V' = V + i\,(\bar{\Lambda} - \Lambda)$ under a gauge transformation that leaves all the physics invariant (here $\Lambda$ is a chiral field). In this case, if possible, we select components that are gauge invariant; in the example above, $D^2\bar{D}V|$ is the preferred choice.

If the superfield carries an external Lorentz index, the separation into components requires reduction with respect to the Lorentz group. Thus, for example, a chiral spinor superfield has the expansion in the chiral representation (where it only depends on $\theta$):

$$\Phi_\alpha{}^{(+)}(x,\theta) = \lambda_\alpha + \theta^\beta(C_{\beta\alpha}\mathrm{D}' + f_{\alpha\beta}) - \theta^2\chi_\alpha \quad . \tag{3.6.9}$$

Using projections, we would define the components by

$$\begin{aligned}
\lambda_\alpha &= \Phi_\alpha| \ , \\
\mathrm{D}' &= \frac{1}{2}\,D_\alpha\Phi^\alpha| \ , \\
f_{\alpha\beta} &= \frac{1}{2}\,D_{(\alpha}\Phi_{\beta)}| \ , \\
\chi_\alpha &= D^2\Phi_\alpha| \ .
\end{aligned} \tag{3.6.10}$$

For $N > 1$ a similar definition of components by projection is possible. In this case, in addition to reduction with respect to the external Lorentz indices, one can further reduce with respect to $SU(N)$ indices.

The projection method is also convenient for finding components of a product of superfields. For example, the product $\Phi = \Phi_1 \Phi_2$ is chiral, and has components

$$\Phi| = \Phi_1\Phi_2| = A_1 A_2 \quad ,$$

$$D_\alpha \Phi| = (D_\alpha \Phi_1)\Phi_2| + \Phi_1(D_\alpha \Phi_2)| = \psi_{1\alpha} A_2 + A_1 \psi_{2\alpha} \quad ,$$

$$D^2\Phi| = (D^2\Phi_1)\Phi_2| + (D^\alpha\Phi_1)(D_\alpha\Phi_2)| + \Phi_1(D^2\Phi_2)|$$

$$= F_1 A_2 + \psi_1{}^\alpha \psi_{2\alpha} + A_1 F_2 \quad . \tag{3.6.11}$$

Similarly, the components of the product $\Psi = \overline{\Phi}_1 \Phi_2$ can be worked out in a straightforward manner, using the Leibnitz rule for derivatives.

### c. The transformation superfield

The transformations of Poincaré supersymmetry (translations and $Q$-supersymmetry transformations) are parametrized by a 4-vector $\xi_{\underline{a}}$ and a spinor $\epsilon_\alpha$ respectively. It is possible to view these, along with the parameter $r$ of "R-symmetry" transformations generated by $A$ in (3.2.12, 3.3.34a), as components of an $x$-independent real superfield $\zeta$

$$\xi_{\underline{a}} \equiv \frac{1}{2}[\overline{D}_{\dot\alpha}, D_\alpha]\zeta| \quad , \quad \epsilon_\alpha \equiv i\overline{D}^2 D_\alpha \zeta| \quad , \quad r \equiv \frac{1}{2} D^\alpha \overline{D}^2 D_\alpha \zeta| \quad , \tag{3.6.12}$$

and to write the supersymmetry transformations in terms of $\zeta$ and the covariant derivatives $D_A$:

$$\delta\Psi = i(\xi^{\underline{a}} P_{\underline{a}} + \epsilon^\alpha Q_\alpha + \overline{\epsilon}^{\dot\alpha}\overline{Q}_{\dot\alpha} + 2rA)\Psi$$

$$= -\,[(i\overline{D}^2 D^\alpha\zeta)D_\alpha + (-\,iD^2\overline{D}^{\dot\alpha}\zeta)\overline{D}_{\dot\alpha} + (\frac{1}{2}[\overline{D}^{\dot\alpha}, D^\alpha]\zeta)\partial_{\alpha\dot\alpha}$$

$$+ iw(\frac{1}{2} D^\alpha \overline{D}^2 D_\alpha \zeta)]\Psi \quad , \tag{3.6.13}$$

where $\frac{1}{2}w$ is the eigenvalue of the operator $Y$ (the matrix part of the axial generator $A$). These transformations are invariant under "gauge transformations" $\delta\zeta = i(\overline{\lambda} - \lambda)$, $\lambda$ chiral and $x$-independent. Consequently, they depend only on $\xi_{\underline{a}}$, $\epsilon_\alpha$, and the component $r$. The R-transformations with parameter $r$ are axial rotations

$$\Psi'(x, \theta, \overline{\theta}) = e^{-iwr}\Psi(x, e^{ir}\theta, e^{-ir}\overline{\theta}) \quad . \tag{3.6.14}$$

## 3.7. Superintegration

### a. Berezin integral

To construct manifestly supersymmetrically invariant actions, it is useful to have a notion of (definite) integration with respect to $\theta$. The essential properties we require of the *Berezin* integral are translation invariance and linearity. Consider a 1-dimensional anticommuting space; then the most general form a function can take is $a + \theta b$. The most general form that the integral can take has the same form: $\int d\theta' (a + \theta b) = A + \theta B$ where $A, B$ are functions of $a, b$. Imposing linearity and invariance under translations $\theta' \to \theta' + \epsilon$ leads uniquely to the conclusion that $\int d\theta\, (a + \theta b) \sim b$. The normalization of the integral is arbitrary. We choose

$$\int d\theta\, \theta = 1 \tag{3.7.1}$$

and, as we found above,

$$\int d\theta\, 1 = 0 \ . \tag{3.7.2}$$

We can define a $\delta$-function: We require

$$\int d\theta\, \delta(\theta - \theta')\, (a + \theta b) = a + \theta' b \tag{3.7.3}$$

and find

$$\delta(\theta - \theta') = \theta - \theta' \tag{3.7.4}$$

These concepts generalize in an obvious way to higher dimensional anticommuting spaces; for $N$-extended supersymmetry, $\int d^{2N}\theta\, d^{2N}\bar{\theta}$ picks out the highest $\theta$ component of the integrand, and a $\delta$-function has the form

$$\delta^{4N}(\theta - \theta') = (\theta - \theta')^{2N} (\bar{\theta} - \bar{\theta}')^{2N} \quad . \tag{3.7.5}$$

We define $\delta^{4+4N}(z - z') \equiv \delta^4(x - x')\delta^{4N}(\theta - \theta')$. We thus have

$$\int d^{4+4N}z\ \delta^{4+4N}(z - z')\Psi(z)$$

$$= \int d^4x d^{4N}\theta \; \delta^4(x-x')\delta^{4N}(\theta-\theta')\Psi(x,\theta) = \Psi(z') \tag{3.7.6}$$

We note that all the properties of the Berezin integral can be characterized by saying it is identical to differentiation:

$$\int d\theta_{\underline{\beta}} \; f(\theta) = \partial_{\underline{\beta}} \; f(\theta) \; . \tag{3.7.7}$$

This has an important consequence in the context of supersymmetry: Because superspace actions are integrated over spacetime as well as over $\theta$, any spacetime total derivative added to the integrand is irrelevant (modulo boundary terms). Consequently, inside a spacetime integral, in the absence of central charges we can replace $\int d\theta_{\underline{\beta}} = \partial_{\underline{\beta}}$ by $D_{\underline{\beta}}$. This allows us to expand superspace actions directly in terms of components defined by projection (see chap. 4, where we consider specific models). Inside superspace integrals, we can integrate $D$ by parts, because $\int d^{4N}\theta\partial_{\underline{\alpha}} = \partial^{2N}\overline{\partial}^{2N}\partial_{\underline{\alpha}} = 0$ (since $\partial^{2N+1} = 0$).

Since supersymmetry variations are also total derivatives (in superspace), we have $\int d^4x d^{2N}\theta \; Q_{\underline{\alpha}}\Psi = \int d^4x d^{2N}\overline{\theta}\overline{Q}_{\underline{\dot{\alpha}}}\Psi = 0$, and thus for any general superfield $\Psi$ the following is a supersymmetry invariant:

$$S_\Psi = \int d^4x d^{4N}\theta \; \Psi \quad . \tag{3.7.8}$$

In the case of chiral superfields we can define invariants in the chiral representation by

$$S_\Phi = \int d^4x d^{2N}\theta \; \Phi \quad , \tag{3.7.9}$$

since $\Phi$ is a function of only $x^{\underline{a}}$ and $\theta^{\underline{\alpha}}$. In fact, this definition is representation independent, since the operator $U$ used to change representations is a spacetime derivative, so only the 1 part of $e^{\frac{1}{2}U}$ contributes to $S_\Phi$. Furthermore, if we express $\Phi$ in terms of a general superfield $\Psi$ by $\Phi = \overline{D}^{2N}\Psi$, we find

$$S_\Phi = \int d^4x d^{2N}\theta \; \overline{D}^{2N}\Psi = \int d^4x d^{4N}\theta \; \Psi = S_\Psi \quad , \tag{3.7.10}$$

since $\overline{D}_{\underline{\dot{\alpha}}} = \int d\overline{\theta}_{\underline{\dot{\alpha}}}$ when inside a $d^4x$ integral.

Similarly, the chiral delta function, which we define as $\delta^4(x-x')\delta^{2N}(\theta-\theta') \equiv \delta^4(x-x')(-1)^N(\theta-\theta')^{2N}$ in the chiral representation, takes the following form in arbitrary representations:

$$\bar{D}^{2N}\delta^{4+4N}(z-z') \quad , \tag{3.7.11}$$

which is equivalent in the chiral representation ($\bar{D}_{\dot{\underline{\alpha}}} = \bar{\partial}_{\dot{\underline{\alpha}}}$), and in general representations gives

$$\int d^4x d^{2N}\theta \, [\bar{D}^{2N}\delta^{4+4N}(z-z')]\Phi(z)$$

$$= \int d^4x d^{4N}\theta \, \delta^{4+4N}(z-z')\Phi(z)$$

$$= \Phi(z) \tag{3.7.12}$$

## b. Dimensions

Since the Berezin integral acts like a derivative (3.7.7), it also *scales* like a derivative; thus it has dimension $[\int d\theta] = [D]$. However, from (3.4.9), we see that the dimensions of $D_{\underline{\alpha}} \sim m^{\frac{1}{2}}$, and consequently, a general integral has dimension $\int d^4x\, d^{4N}\theta \sim m^{2N-4}$ and a chiral integral has dimension $\int d^4x\, d^{2N}\theta \sim m^{N-4}$. In particular, for $N=1$, we have $\int d^4x\, d^4\theta \ \equiv \int d^8z \ \sim m^{-2}$ and $\int d^4x\, d^2\theta \ \equiv \int d^6z \ \sim m^{-3}$.

## c. Superdeterminants

Finally, we use superspace integrals to define superdeterminants (Berezinians). Consider a $(k,n)$ by $(k,n)$ dimensional supermatrix $M$ with a $k$ by $k$ dimensional even-even part $A$, a $k$ by $n$ dimensional even-odd part $B$, an $n$ by $k$ dimensional odd-even part $C$, and an $n$ by $n$ dimensional odd-odd part $D$:

$$M = \begin{pmatrix} A & B \\ C & D \end{pmatrix} \tag{3.7.13}$$

where the entries of $A, D$ are bosonic and those of $B, C$ are fermionic. We define the superdeterminant by analogy with the usual determinant:

$$(sdet\, M)^{-1} = K \int d^k x\, d^k x'\, d^n\theta\, d^n\theta'\, e^{-z'^t M\, z} \quad , \tag{3.7.14a}$$

where

$$z'^t = (x'\; \theta') \quad , \quad z = \begin{pmatrix} x \\ \theta \end{pmatrix} \quad , \tag{3.7.14b}$$

and $K$ is a normalization factor chosen to ensure that $sdet(1) = 1$. The exponent $x'Ax + x'B\theta + \theta'Cx + \theta'D\theta$ can be written, after shifts of integration variables either in $x$ or in $\theta$, in two equivalent forms: $x'Ax + \theta'(D - C\, A^{-1}\, B)\theta$ or $x'(A - B\, D^{-1}\, C)x + \theta'D\theta$. Integration over the bosonic variables gives us an inverse determinant factor, and integration over the fermionic variables gives a determinant factor. We obtain $sdet\, M$ in terms of ordinary determinants:

$$sdet(M) = \frac{det\ A}{det(D - CA^{-1}B)} = \frac{det(A - BD^{-1}C)}{det\ D} \quad . \tag{3.7.15}$$

This formula has a number of useful properties. Just as with the ordinary determinant, the superdeterminant of the product of several supermatrices is equal to the product of the superdeterminants of the supermatrices. Furthermore,

$$ln\ (sdet\, M) = \ str(ln\, M) \ , \tag{3.7.16a}$$

where the supertrace of a supermatrix $M$ is the trace of the even-even matrix $A$ *minus* the trace of the odd-odd matrix $D$:

$$strM \equiv trA - trD \tag{3.7.16b}$$

An arbitrary infinitesimal variation of $M$ induces a variation of the superdeterminant:

$$\begin{aligned} \delta(sdet\, M) &= \delta exp[str(ln\, M)] \\ &= (sdet\, M) str(M^{-1}\delta M) \end{aligned} \tag{3.7.17}$$

## 3.8. Superfunctional differentiation and integration

### a. Differentiation

In this section we discuss functional calculus for superfields. We begin by reviewing functional differentiation for component fields: By analogy with ordinary differentiation, functional differentiation of a functional $F$ of a field $A$ can be defined as

$$\frac{\delta F[A]}{\delta A(x)} = \lim_{\epsilon \to 0} \frac{F[A + \delta_{\epsilon,x} A] - F[A]}{\epsilon} \quad , \tag{3.8.1}$$

where

$$\delta_{\epsilon,x} A(x') = \epsilon \delta^4(x - x') \quad . \tag{3.8.2}$$

This is *not* the same as dividing $\delta F$ by $\delta A$. The derivative can also be defined for arbitrary variations by a Taylor expansion:

$$F[A + \delta A] = F[A] + \left( \delta A \, , \frac{\delta F[A]}{\delta A} \right) + O((\delta A)^2) \quad , \tag{3.8.3}$$

where the product $( \, , )$ of two arbitrary functions is given by

$$(C \, , B) = \int d^4x \, C(x) B(x) \quad . \tag{3.8.4}$$

In particular, from (3.8.2) we find

$$(\delta_{\epsilon,x} A \, , B) = \epsilon B(x) \quad . \tag{3.8.5}$$

This definition allows a convenient prescription for generalized differentiation. For example, in curved space, where the invariant product is $(C, B) = \int d^4x \; g^{1/2} C B$, the normalization $(\delta A, B) = \epsilon B(x)$ corresponds to the functional variation $\delta A(x') = \epsilon g^{-1/2}(x) \delta^4(x - x')$. Generally, a choice of $\delta_{\epsilon,x}$ is equivalent to a choice of the product $( \, , )$. In particular, for (3.8.2,4) we have the functional derivative

$$\frac{\delta A(x)}{\delta A(x')} = \delta^4(x - x') \quad . \tag{3.8.6}$$

In curved space, using the invariant product, we would obtain $g^{-1/2}(x) \delta^4(x - x')$. Note that the inner product is not always symmetric: In $(C, B)$, $C$ transforms contragrediently to $B$. For example, if $A$ is a covariant vector, the quantity on the left-hand side of the inner-product is a covariant vector, while that on the right is a contravariant vector; if $A$ is an isospinor, $\frac{\delta F}{\delta A}$ is a complex-conjugate isospinor; etc.

In superspace, the definitions for general superfields are analogous. The product $(\Psi, \Psi')$ is $\int d^{4+4N}z\ \Psi(z)\Psi'(z) = \int d^4x d^{4N}\theta\ \Psi(x,\theta)\Psi'(x,\theta)$, and thus

$$\frac{\delta\Psi(z)}{\delta\Psi(z')} = \delta^{4+4N}(z - z') = \delta^4(x - x')\delta^{4N}(\theta - \theta') \quad . \tag{3.8.7}$$

(Appropriate modifications will be made in curved superspace.) However, for chiral superfields we have

$$(\Phi\,, \Phi') = \int d^{4+2N}z\ \Phi\Phi' = \int d^4x d^{2N}\theta\ \Phi\Phi' \quad , \tag{3.8.8}$$

since $\Phi$ and $\Phi'$ essentially depend on only $x^{\underline{a}}$ and $\theta^{\underline{\alpha}}$, not $\bar{\theta}^{\underline{\dot{\alpha}}}$. The variation is therefore defined in terms of the chiral delta function:

$$\delta_{\epsilon,z}\Phi(z') = \epsilon\bar{D}^{2N}\delta^{4+4N}(z - z') \tag{3.8.9}$$

so that

$$\frac{\delta\Phi(z)}{\delta\Phi(z')} = \bar{D}^{2N}\delta^{4+4N}(z - z') \quad , \tag{3.8.10}$$

and the complex conjugate relation

$$\frac{\delta\bar{\Phi}(z)}{\delta\bar{\Phi}(z')} = D^{2N}\delta^{4+4N}(z - z') \quad . \tag{3.8.11}$$

(Again, appropriate modifications will be made in curved superspace.) Furthermore, variations of chiral integrals give the expected result

$$\frac{\delta}{\delta\Phi(z')}\int d^4x d^{2N}\theta\ f(\Phi(z)) = \int d^4x d^{2N}\theta\ f'(\Phi(z))\bar{D}^{2N}\delta^{4+4N}(z - z')$$

$$= \int d^4x d^{4N}\theta\ f'(\Phi(z))\delta^{4+4N}(z - z') = f'(\Phi(z')) \quad . \tag{3.8.12}$$

When the functional differentiation is on an expression appearing in a chiral integral with $d^{2N}\theta$, the $\bar{D}^{2N}$ can always be used to convert it to a $d^{4N}\theta$ integral, after which the full $\delta$-function can be used as in (3.8.12).

This result can also be obtained by expressing $\Phi$ in terms of a general superfield, as $\Phi = \bar{D}^{2N}\Psi$: we have

$$\frac{\delta\Phi(z)}{\delta\Psi(z')} = \frac{\delta\bar{D}^{2N}\Psi(z)}{\delta\Psi(z')} = \bar{D}^{2N}\,\frac{\delta\Psi(z)}{\delta\Psi(z')} = \bar{D}^{2N}\delta^{4+4N}(z-z') \quad . \tag{3.8.13}$$

We can thus identify $\dfrac{\delta}{\delta\Phi}$ with $\dfrac{\delta}{\delta\Psi}$ for $\Phi = \bar{D}^{2N}\Psi$.

These definitions can be analyzed in terms of components and correspond to ordinary functional differentiation of the component fields. We cannot define functional differentiation for constrained superfields other than chiral or antichiral ones. For example, for a linear superfield $\Upsilon$ (which can be written as $\Upsilon = \bar{D}^{\dot\alpha}\bar{\Psi}_{\dot\alpha}$) there is no functional derivative which is both linear and a scalar.

## b. Integration

In chapters 5 and 6 we discuss quantization of superfield theories by means of functional integration. We need to define only integrals of Gaussians, as all other functional integrals in perturbation theory are defined in terms of these by introducing sources and differentiating with respect to them. The basic integrals are

$$\int I\!\!DV\; e^{\int d^4x d^{4N}\theta\, \frac{1}{2}V^2} = 1 \quad , \tag{3.8.14a}$$

$$\int I\!\!D\Phi\; e^{\int d^4x d^{2N}\theta\, \frac{1}{2}\Phi^2} = 1 \quad , \tag{3.8.14b}$$

$$\int I\!\!D\bar{\Phi}\; e^{\int d^4x d^{2N}\bar{\theta}\, \frac{1}{2}\bar{\Phi}^2} = 1 \quad , \tag{3.8.14c}$$

where, e.g., $I\!\!DV = \prod_i I\!\!DV^i$, for $V^i$ the components of $V$. Because a superfield has the same number of bose and fermi components, many factors that appear in ordinary functional integrals cancel for superfields. Thus we can make any change of variables that does not involve *both* explicit $\theta$'s and $\dfrac{\partial}{\partial\theta}$'s without generating any Jacobian factor, because unless the bosons and fermions mix nontrivially, the superdeterminant (3.7.14) is equal to one. For example, a change of variables $V \to f(V\,,X)$ where $X$ is an arbitrary external superfield generates no Jacobian factor; the same is true for the change of variables $V \to \square V$ as long as $\square$ is a purely bosonic operator. Nontrivial Jacobian determinants arise for changes of variables such as $V \to D^2 V$ or $V \to \widehat{\square} V$ where $\widehat{\square}$ is background covariant, e.g., in supergravity or super-Yang-Mills theory, and hence contains spinor derivatives.

To prove the preceding assertions, we consider the case with one $\theta$; the general case can be proven by choosing one particular $\theta$ and proceeding inductively. We expand the superfield with respect to $\theta$ as $V = A + \theta\psi$; similarly, we expand the arbitrary external superfield as $X = C + \theta\chi$. Then we can expand the new variable $f(V\,, X)$ as

$$f(V\,,X) = f(A\,,C) + \theta[\psi f_V(A\,,C)| + \chi f_X(A\,,C)|] \tag{3.8.15}$$

where $f_V| \equiv f_A \equiv \frac{\partial(f|)}{\partial A}$, etc. The Jacobian of this transformation is

$$sdet\,\frac{\partial f}{\partial V} \;=\; sdet\begin{pmatrix} f_A(A\,,C) & \psi f_{AA} + \chi f_{AC} \\ 0 & f_A(A\,,C) \end{pmatrix} \;=\; \frac{det(f_A)}{det(f_A)} \;=\; 1 \quad . \tag{3.8.16}$$

In particular, the external superfield $X$ can be a nonlocal operator such as $\square^{-1}$.

An immediate consequence of the preceding result is that superfield $\delta$-functions

$$\delta(V - V\,') \equiv \prod_i \delta(V^i - V'^i) \tag{3.8.17a}$$

are invariant under "$\theta$-nonmixing" changes of variables:

$$\delta(f(V)) = \sum_{f(c_i)=0} \delta(V - c_i) \quad . \tag{3.8.17b}$$

In general, if nontrivial operators appear in the actions, the functional integrals are no longer constant. We first introduce the following convenient notation:

$$\Xi \equiv \begin{pmatrix} V \\ \Phi \\ \overline{\Phi} \end{pmatrix} \quad ,$$

$$\int \Xi^t \equiv \int d^4x \left( \int d^{4N}\theta\; V^t \quad \int d^{2N}\theta\; \Phi^t \quad \int d^{2N}\bar{\theta}\; \overline{\Phi}^t \right) \quad , \tag{3.8.18}$$

where $V$, $\Phi$, and $\overline{\Phi}$ themselves can stand for several superfields arranged as column vectors. We next consider actions of the form

$$S = \frac{1}{2} \int \Xi^t \mathbf{O} \Xi \quad , \tag{3.8.19}$$

where the *nonsingular* operator $\mathbf{O}$ is such that the components of the column vector $\mathbf{O}\Xi$ have the same chirality as the corresponding components of $\Xi$. These actions give the field equations

$$\frac{\delta S}{\delta \Xi} = \mathbf{O}\Xi \quad , \tag{3.8.20}$$

due to the integration measures chosen for the definition of the integrals (3.8.18).

We define, for commuting $\Xi$,

$$(det\, \mathbf{O})^{-\frac{1}{2}} \equiv \int I\!\!D\Xi e^S \quad , \tag{3.8.21}$$

with $S$ given by (3.8.19). For anticommuting $\Xi$ we obtain $(det\, \mathbf{O})^{\frac{1}{2}}$. Then (3.8.14) can be written as

$$det\, \mathbf{I} = 1 \quad . \tag{3.8.22}$$

From the definition (3.8.21) we have

$$\int I\!\!D\Xi_1 I\!\!D\Xi_2 e^{\int \Xi_1{}^T \mathbf{O} \Xi_2} = (det \mathbf{O})^{-1} \quad . \tag{3.8.23}$$

We also have

$$(det\, \mathbf{O}_1)\ (det \mathbf{O}_2) = det\ (\mathbf{O}_1 \mathbf{O}_2) \quad . \tag{3.8.24}$$

This can be proven as follows: We consider the action

$$\int (\Xi_1{}^t \mathbf{O}_1 \Xi_2 + \Xi_3{}^t \mathbf{O}_2 \Xi_4) \quad . \tag{3.8.25}$$

The functional integral of the exponential of this action is equal to that of

$$\int (\Xi_1{}^t \mathbf{O}_1 \mathbf{O}_2 \Xi_2 + \Xi_3{}^t \Xi_4) \quad , \tag{3.8.26}$$

as can be seen from the field redefinitions

$$\Xi_2 \to \mathbf{O}_2 \Xi_2 \quad , \qquad \Xi_4 \to \mathbf{O}_2{}^{-1} \Xi_4 \quad , \tag{3.8.27}$$

whose Jacobians cancel.

As an important example we consider the $N = 1$ case with one $\Phi$ and one $\bar{\Phi}$ and no $V$:

$$\Xi = \begin{pmatrix} \Phi \\ \\ \bar{\Phi} \end{pmatrix} \quad , \quad \mathbf{O} = \begin{pmatrix} 0 & \bar{D}^2 \\ \\ D^2 & 0 \end{pmatrix} \quad . \tag{3.8.28}$$

This operator satisfies the identity

$$\mathbf{O}^2 \Xi = \Box \Xi \quad . \tag{3.8.29}$$

Therefore, from (3.8.24) we have

$$(det\ \mathbf{O})^2 = det\ \Box \tag{3.8.30}$$

and hence the integral of the exponential of the action

$$S = \frac{1}{2}[\int d^4x\, d^2\theta\, (\Phi_1 \bar{D}^2 \bar{\Phi}_1 + \Phi_2 \bar{D}^2 \bar{\Phi}_2) + h.\, c.]$$

$$= \int d^4x\, d^4\theta\, (\bar{\Phi}_1 \Phi_1 + \bar{\Phi}_2 \Phi_2) \tag{3.8.31}$$

is equal to that of

$$S = \frac{1}{2}[\int d^4x\, d^2\theta\, \Phi \Box \Phi + h.\, c.] \quad . \tag{3.8.32}$$

In the same manner we have the following equivalence:

$$\int d^4x\, d^4\theta\, \bar{\Phi} \Box^m \Phi \quad \longleftrightarrow \quad \sum_{i=1}^{2m+1} \int d^4x\, d^4\theta\, \bar{\Phi}_i \Phi_i \quad . \tag{3.8.33}$$

As another example we consider the case of a chiral spinor $\Phi_\alpha$:

$$\Xi = \begin{pmatrix} \Phi_\alpha \\ \\ \bar{\Phi}_{\dot{\alpha}} \end{pmatrix} \quad , \quad \mathbf{O} = \begin{pmatrix} 0 & i\partial_\alpha{}^{\dot{\beta}} \bar{D}^2 \\ \\ i\partial^\beta{}_{\dot{\alpha}} D^2 & 0 \end{pmatrix} \quad , \tag{3.8.34}$$

with

$$\mathbf{O}^2 \Xi = \Box^2 \Xi \quad . \tag{3.8.35}$$

Therefore

$$\int d^4x\, d^4\theta\; \bar{\Phi}^{\dot{\alpha}} i\partial^{\alpha}{}_{\dot{\alpha}}\Phi_{\alpha} \quad \longleftrightarrow \quad \frac{1}{2}[\int d^4x\, d^2\theta\; \Phi^{\alpha}\Box\Phi_{\alpha} + h.\,c.] \quad . \tag{3.8.36}$$

## 3.9. Physical, auxiliary, and gauge components

In section 3.6 we discussed the component field content of supersymmetric theories. However, the field content of a theory does not determine its physical states. Conversely, a given set of physical states can be described by different sets of fields.

Given a set of fields and their free Lagrangian, we can classify any *component* of a field as one of three types: (1) *physical,* with a propagating degree of freedom; (2) *auxiliary,* with an equation of motion that sets it identically equal to zero; and (3) *gauge,* not appearing in the Lagrangian. (Super)Fields can contain all three kinds of components; off-shell representations (of the Poincaré or supersymmetry group) contain only the first two; and on-shell representations contain only the first. We also classify any *field* as one of three types: (1) physical, containing physical components, but perhaps also auxiliary and/or gauge components; (2) auxiliary, containing auxiliary, but perhaps also gauge, components; and (3) *compensating,* containing only gauge components.

The simplest example of this is the conventional vector gauge field of electromagnetism. The explicit separation is necessarily non(Poincaré)covariant, and is most conveniently performed in a *light-cone* formalism. In the notation of (3.1.1) we treat $x^- = x^{-\dot{-}}$ as the "time" coordinate, and $x^+, x^T, \bar{x}^T$ as "space" coordinates. We are thus free to construct expressions that are nonlocal in $x^+$ (i.e. containing inverse powers of $\partial_+$), since the dynamics is described by evolution in $x^{-\dot{-}}$. (In fact, the formalism closely resembles *nonrelativistic* field theory, with $x^-$ acting as the time and $\partial_+$ as the *mass.)* The vector gauge field $A_{\alpha\dot{\alpha}}$ transforms as

$$\delta A_{\alpha\dot{\alpha}} = \partial_{\alpha\dot{\alpha}}\lambda \ . \tag{3.9.1}$$

By making the field redefinitions (by $A \to f(A)$ we mean $A = f(A')$ and then drop all $'$'s)

$$A_+ \to A_+ \quad ,$$

$$A_T \to A_T + (\partial_+)^{-1}\partial_T A_+ \quad ,$$

$$A_- \to A_- + (\partial_+)^{-1}(\partial_- A_+ - \partial_T \bar{A}_T - \bar{\partial}_T A_T) \quad ; \tag{3.9.2}$$

we obtain the new transformation laws

$$\delta A_T = \delta A_- = 0 \quad , \quad \delta A_+ = \partial_+ \lambda \quad . \tag{3.9.3}$$

Furthermore, the Lagrangian

$$I\!L = -\frac{1}{2} F^{\alpha\beta} F_{\alpha\beta} \ , \tag{3.9.4}$$

where

$$F_{\alpha\beta} = \frac{1}{2} \partial_{(\alpha\dot{\gamma}} A_{\beta)}{}^{\dot{\gamma}} \tag{3.9.5}$$

in terms of the *old* $A$, becomes

$$I\!L = \bar{A}_T \Box A_T - \frac{1}{4} A_- (\partial_+)^2 A_- \quad . \tag{3.9.6}$$

Thus, the complex component $A_T$ describes the two physical (propagating) polarizations, the real component $A_-$ is auxiliary (it has no dynamics; its equation of motion sets it equal to zero) , and the real component $A_+$ is gauge. In this formalism the obvious gauge choice is $A_+ = 0$ (the light-cone gauge), since $A_+$ does not appear in $I\!L$. However, gauge components are important for Lorentz covariant gauge fixing: For example, $(\partial^{\alpha\dot{\beta}} A_{\alpha\dot{\beta}})^2 \to (\partial_+ A_- + 2(\partial_+)^{-1} \Box \, A_+)^2$.

We can perform similar redefinitions to separate arbitrary fields into physical, auxiliary, and gauge components. Any original component that transforms under a gauge transformation with a $\partial_+$ or a nonderivative term corresponds to a gauge component of the redefined field. Any component that transforms with a $\partial_-$ term corresponds to an auxiliary component. Of the remaining components, some will be auxiliary and some physical (depending on the action), organized in a way that preserves the "transverse" $SO(2)$ Lorentz covariance. For the known fields appearing in interacting theories, the components with highest spin are physical and the rest (when there are any: i.e., for physical spin 2 or $\frac{3}{2}$) are auxiliary. These arguments can be applied in all dimensions.

An example that illustrates the separation between physical and auxiliary (but not gauge) components without the use of nonlocal, noncovariant redefinitions is that of a massive spinor field:

$$I\!L = \bar{\psi}^{\dot{\alpha}} i \partial^{\alpha}{}_{\dot{\alpha}} \psi_{\alpha} - \frac{1}{2} m (\psi^{\alpha} \psi_{\alpha} + \bar{\psi}^{\dot{\alpha}} \bar{\psi}_{\dot{\alpha}}) \ . \tag{3.9.7}$$

Since $\psi$ and $\bar{\psi}$ may be considered as independent fields in the functional integral (and,

in fact, must be considered independent locally after Wick rotation to Euclidean space), we can make the following nonunitary (but local and covariant) redefinition:

$$\psi_\alpha \to \psi_\alpha \ ,$$

$$\bar{\psi}_{\dot{\alpha}} \to \bar{\psi}_{\dot{\alpha}} - \frac{1}{m} i\partial_{\alpha\dot{\alpha}}\psi^\alpha \ . \tag{3.9.8}$$

The Lagrangian becomes

$$I\!\!L = \frac{1}{2m}\psi^\alpha(\Box - m^2)\psi_\alpha - \frac{1}{2}m\bar{\psi}^{\dot{\alpha}}\bar{\psi}_{\dot{\alpha}} \ . \tag{3.9.9}$$

We thus find that $\psi$ represents two physical polarizations, while $\bar{\psi}$ contains two auxiliary components.

The same analysis can be made for the simplest supersymmetric multiplet: the massive scalar multiplet, described by a chiral scalar superfield (see section 4.1). The action is

$$S = \int d^4x d^4\theta\, \bar{\Phi}\Phi - \frac{1}{2}m(\int d^4x d^2\theta\, \Phi^2 + \int d^4x d^2\bar{\theta}\, \bar{\Phi}^2) \ . \tag{3.9.10}$$

We now redefine

$$\Phi \to \Phi \ ,$$

$$\bar{\Phi} \to \bar{\Phi} + \frac{1}{m}D^2\Phi \ ; \tag{3.9.11}$$

and, using $\int d^2\bar{\theta} = \bar{D}^2$, we obtain the action

$$S = \frac{1}{2m}\int d^4x d^2\theta\, \Phi(\Box - m^2)\Phi - \frac{1}{2}m\int d^4x d^2\bar{\theta}\, \bar{\Phi}^2 \ . \tag{3.9.12}$$

(Note that the redefinition of $\bar{\Phi}$ preserves its antichirality $D_\alpha\bar{\Phi} = 0$.) Now $\Phi$ contains only physical and $\bar{\Phi}$ contains only auxiliary components; each contains two Bose components and two Fermi. As can be checked using the component expansion of $\Phi$, the original action (3.9.10) contains the spinor Lagrangian of (3.9.7), whereas (3.9.12) contains the Lagrangian (3.9.9). It also contains two scalars and two pseudoscalars, one of each being a physical field (with kinetic operator $\Box - m^2$) and the other an auxiliary field (with kinetic operator 1). For more detail of the component analysis, see sec. 4.1.

As we discuss in sec. 4.1, auxiliary fields are needed in interacting supersymmetric theories for several reasons: (1) They facilitate the construction of actions, since without them the kinetic and various interaction terms are not separately supersymmetric; (2) because of this, actions without auxiliary fields have supersymmetry transformations that are nonlinear and coupling dependent, and make difficult the application of supersymmetry Ward identities (e.g., to prove renormalizability); and (3) auxiliary fields are necessary for manifestly supersymmetric quantization. Compensating fields (see following section) are also necessary for the latter two reasons. Although they disappear from the classical action, they appear in supersymmetric gauge-fixing terms.

## 3.10. Compensators

In our subsequent discussions, we will often use "compensating" fields or compensators. These are fields that enter a theory in such a way that they can be *algebraically* gauged away. Thus, in a certain sense, they are trivial: The theory can always be written without them. However, they frequently simplify the structure of the theory; in particular, they can be used to write nonlinearly realized symmetries in a linear way. This is often important for quantization. Another application, which is particularly relevant to supergravity, arises in situations where one knows how to write invariant actions for systems transforming under a certain symmetry group $G$ (e.g., the superconformal group): If one wants to write actions for systems transforming only under a subgroup $H$ (e.g., the super-Poincaré group), one can enlarge the symmetry of such systems to the full group by introducing compensators. After writing the action for the systems with the enlarged symmetry, one simply chooses a gauge, thus breaking the symmetry of the action down to the subgroup $H$.

A simple example in ordinary field theory is "fake" scalar electrodynamics. The usual kinetic action for a complex scalar $z(x)$

$$S = \frac{1}{2}\int d^4x \; \bar{z}\partial^{\underline{a}}\partial_{\underline{a}} z \tag{3.10.1}$$

has a global $U(1)$ symmetry: $z' = e^{i\lambda} z$. This symmetry can be gauged trivially by introducing a real compensating scalar $\phi$, assumed to transform under a local $U(1)$ transformation as $\phi' = \phi - \lambda$. We can then construct a covariant derivative $\nabla_{\underline{a}} = e^{-i\phi}\partial_{\underline{a}} e^{i\phi} = \partial_{\underline{a}} + i\partial_{\underline{a}}\phi$ that can be used to define a locally $U(1)$ invariant action

$$S = \frac{1}{2}\int d^4x \; \bar{z}\nabla^{\underline{a}}\nabla_{\underline{a}} z \tag{3.10.2}$$

Fake spinor electrodynamics can be obtained by an obvious generalization.

### a. Stueckelberg formalism

In the previous example, the compensator served no useful purpose. The Stueckelberg formalism provides a familiar example of a compensator that simplifies the theory. We begin with the Lagrangian for a massive vector $A_{\underline{a}}$:

$$I\!\!L = -\,\frac{1}{8}F^{\underline{ab}}F_{\underline{ab}} - m^2(A_{\underline{a}})^2 \quad , \tag{3.10.3}$$

$$F_{\underline{ab}} = \partial_{[\underline{a}} A_{\underline{b}]} \quad . \tag{3.10.4}$$

The propagator for this theory is:

$$\mathrm{D}_{\underline{ab}} = -\, \frac{1}{\Box - m^2} \left( \eta_{\underline{ab}} - \frac{1}{2m^2}\, \partial_{\underline{a}} \partial_{\underline{b}} \right) \tag{3.10.5}$$

We can recast the theory in an improved form by introducing a $U(1)$ compensator $\phi$ that makes the action (3.10.3) gauge invariant. We define

$$A'_{\underline{a}} = A_{\underline{a}} + \frac{1}{m}\, \partial_{\underline{a}} \phi \tag{3.10.6}$$

where $A'_{\underline{a}}$ and $\phi$ transform under $U(1)$ gauge transformations:

$$\delta A'_{\underline{a}} = \partial_{\underline{a}} \lambda \quad , \quad \delta\phi = m\lambda \quad . \tag{3.10.7}$$

In terms of these fields, the *gauge invariant* Lagrangian is (dropping the prime):

$$\begin{aligned} I\!\!L = &-\, \frac{1}{8}\, F^{\underline{ab}} F_{\underline{ab}} - m^2 (A_{\underline{a}})^2 \\ &-\, m\phi \partial^{\underline{a}} A_{\underline{a}} - (\partial_{\underline{a}} \phi)^2 \quad . \end{aligned} \tag{3.10.8}$$

We now choose a gauge by adding the gauge fixing term

$$I\!\!L_{GF} = -\, \frac{1}{4} \left( \partial^{\underline{a}} A_{\underline{a}} - 2m\phi \right)^2 \tag{3.10.9}$$

and find:

$$I\!\!L + I\!\!L_{GF} = \frac{1}{2}\, A^{\underline{a}} (\Box - m^2) A_{\underline{a}} + \phi (\Box - m^2) \phi \quad . \tag{3.10.10}$$

The propagators can be trivially read off from (3.10.10): for $A_{\underline{a}}$, $\mathrm{D}_{\underline{ab}} = -\, \eta_{\underline{ab}} (\Box - m^2)^{-1}$, and for $\phi$, $\mathrm{D} = -\, \frac{1}{2} (\Box - m^2)^{-1}$. They have better high energy behavior than (3.10.5). Thus, by introducing the compensator $\phi$, we have simplified the structure of the theory. We note that the compensator decouples whenever $A_{\underline{a}}$ is coupled to a conserved source (i.e., in a gauge invariant way).

## b. CP(1) model

Another familiar example is the $CP(1)$ nonlinear $\sigma$-model, which describes the Goldstone bosons of an $SU(2)$ gauge theory spontaneously broken down to $U(1)$. It consists of a real scalar field $\rho$ and a complex field $y$ subject to the constraint

$$|y|^2 + \rho^2 = 1 \tag{3.10.11}$$

The group $SU(2)$ can be realized *nonlinearly* on these fields by

$$\delta\rho = \frac{1}{2}\,(\beta y + \overline{\beta}\overline{y})$$

$$\delta y = -\,2i\alpha y - \overline{\beta}\rho - \frac{1}{2}\,\rho^{-1}(\beta y - \overline{\beta}\overline{y})y \quad . \tag{3.10.12}$$

where $\alpha$, $\beta$, and $\overline{\beta}$ are the (constant) parameters of the global $SU(2)$ transformations. These transformations leave the Lagrangian

$$I\!L = -\,[(\partial_{\underline{a}}\rho)^2 + |\partial_{\underline{a}}y|^2 + \frac{1}{4}\,(\overline{y}\overleftrightarrow{\partial}_{\underline{a}}y)^2] \tag{3.10.13}$$

invariant, but because the transformations are nonlinear this is far from obvious.

We can give a description of the theory where the $SU(2)$ is represented linearly by introducing a *local* $U(1)$ invariance which is realized by a compensating field $\phi$. Under this local $U(1)$, $\phi$ transforms as

$$\phi'(x) = \phi(x) - \lambda(x) \quad . \tag{3.10.14}$$

We define fields $z_i$ by

$$z_1 = e^{-i\phi}\rho \ , \qquad z_2 = e^{-i\phi}y \quad . \tag{3.10.15}$$

Because of this definition they transform under the local $U(1)$ as

$$z'_i = e^{i\lambda}z_i \quad . \tag{3.10.16}$$

The constraint (3.10.11) becomes

$$|z_1|^2 + |z_2|^2 = 1 \quad . \tag{3.10.17}$$

Ignoring the constraint the $SU(2)$ acts linearly on these fields (see below):

$$\delta z_1 = i\alpha z_1 + \beta z_2$$

$$\delta z_2 = -\,i\alpha z_2 - \overline{\beta}z_1 \quad . \tag{3.10.18}$$

The complicated nonlinear transformations (3.10.12) arise in the following manner: when we fix the $U(1)$ gauge

$$z_1 = \overline{z}_1 \equiv \rho \tag{3.10.19}$$

the linear $SU(2)$ transformations (3.10.18) do not preserve the condition (3.10.19). Thus we must add a "gauge-restoring" $U(1)$ transformation with parameter

$$i\lambda(x) = -\frac{1}{2}\rho^{-1}(\delta z_1 - \delta\overline{z}_1) = -i\alpha - \frac{1}{2}\rho^{-1}(\beta z_2 - \overline{\beta}\overline{z}_2) \quad . \tag{3.10.20}$$

The combined linear $SU(2)$ transformation and gauge transformation (3.10.16) with nonlinear parameter (3.10.20) preserves the gauge condition (3.10.19) and are equivalent to (3.10.12).

To write an action invariant under both the global $SU(2)$ and the local $U(1)$ transformations we need a covariant derivative for the latter. By analogy with our first example we could write

$$\nabla_{\underline{a}} = e^{-i\phi}\partial_{\underline{a}}e^{i\phi} = \partial_{\underline{a}} + i\partial_{\underline{a}}\phi \quad . \tag{3.10.21}$$

A manifestly $SU(2)$ invariant choice in terms of the new variables is

$$\begin{aligned} \nabla_{\underline{a}} &= \partial_{\underline{a}} - \frac{1}{2}\overline{z}^i\overleftrightarrow{\partial}_{\underline{a}}z_i \\ &= \partial_{\underline{a}} + i\partial_{\underline{a}}\phi - \frac{1}{2}\overline{y}\overleftrightarrow{\partial}_{\underline{a}}y \quad . \end{aligned} \tag{3.10.22}$$

This differs from (3.10.21) by the $U(1)$ *gauge invariant* term $\overline{y}\overleftrightarrow{\partial}_{\underline{a}}y$; one is always free to change a covariant derivative by adding covariant terms to the connection. (This is similar to adding contortion to the Lorentz connection in (super)gravity; see sec. 5.3.a.3.) Then a manifestly covariant Lagrangian is

$$\begin{aligned} I\!\!L &= -|\nabla_{\underline{a}}z_i|^2 \\ &= -|\partial_{\underline{a}}z_i|^2 - \frac{1}{4}(\overline{z}^i\overleftrightarrow{\partial}_{\underline{a}}z_i)^2 \quad . \end{aligned} \tag{3.10.23}$$

In the gauge (3.10.19) this Lagrangian becomes that of (3.10.13).

We consider now another application of compensators: The constraint (3.10.17) is awkward: It makes the transformations (3.10.18) implicitly nonlinear. We can avoid this by introducing a *second* compensating field. We observe that neither the constraint nor the Lagrangian are invariant under scale transformations. However, we can introduce a scale invariance into the theory by writing

$$z_i = e^{-\zeta}Z_i \tag{3.10.24}$$

in terms of new fields $Z_i$ and the compensator $\zeta(x)$. The constraint and the action, written in terms of $Z_i$, $\zeta$, will be invariant under the scale transformations

$$Z'_i = e^{\tau} Z_i \quad , \qquad \zeta' = \zeta + \tau \quad . \tag{3.10.25}$$

The $SU(2)$ transformations of $Z_i$ are now the (truly) linear transformations (3.10.18). The $U(1)$ and the scale transformations can be combined into a single complex scale transformation with parameter

$$\sigma = \tau + i\lambda \tag{3.10.26}$$

$$Z'_i = e^{\sigma} Z_i \quad , \qquad \zeta' = \zeta + \tau \quad . \tag{3.10.27}$$

The constraint (3.10.17) becomes

$$Z\overline{Z} = e^{2\zeta} \tag{3.10.28}$$

where we write $Z\overline{Z} \equiv |Z_1|^2 + |Z_2|^2$. In terms of the new variables the Lagrangian is

$$\begin{aligned} I\!\!L &= -\, |\nabla_{\underline{a}}(e^{-\zeta} Z_i)|^2 \\ &= -\, |\partial_{\underline{a}}(e^{-\zeta} Z_i)|^2 - \frac{1}{4} e^{-4\zeta} (\overline{Z}^i \overleftrightarrow{\partial}_{\underline{a}} Z_i)^2 \quad . \end{aligned} \tag{3.10.29}$$

Substituting for $\zeta$ the solution of the constraint (3.10.28), a manifestly $SU(2)$ invariant procedure, leads to

$$\begin{aligned} I\!\!L &= -\left| \partial_{\underline{a}} \frac{Z_i}{\sqrt{Z\overline{Z}}} \right|^2 - \frac{1}{4} \frac{(\overline{Z}^i \overleftrightarrow{\partial}_{\underline{a}} Z_i)^2}{(Z\overline{Z})^2} \\ &= -\, \frac{1}{Z\overline{Z}} (\delta_i{}^k - \frac{Z_i \overline{Z}^k}{Z\overline{Z}}) \frac{1}{2} (\partial_{\underline{a}} \overline{Z}^i)(\partial^{\underline{a}} Z_k) \end{aligned} \tag{3.10.30}$$

This last form of the Lagrangian is expressed in terms of unconstrained fields $Z_i$ only. It is manifestly globally $SU(2)$ invariant and also invariant under the local complex scale transformations (3.10.27). We can use this invariance to choose a convenient gauge. For example, we can choose the gauge $Z_1 = 1$; or we can choose a gauge in which we obtain (3.10.13). Once we choose a gauge, the $SU(2)$ transformations become nonlinear again.

These two compensators allowed us to realize a global symmetry ($SU(2)$) of the system linearly. However, they play different roles: $\phi(x)$, the $U(1)$ compensator, gauges a global symmetry of the system, whereas $\zeta(x)$, the scale compensator introduces an altogether new symmetry. For the $U(1)$ invariance we introduced a connection, whereas for the scale invariance we introduced $\zeta(x)$ directly, without a connection. In the former case, the connection consisted of a pure gauge part, and a covariant part chosen to make it manifestly covariant under a symmetry ($SU(2)$) of the system; had we tried to introduce $\phi(x)$ directly, we would have found it difficult to maintain the $SU(2)$ invariance. In the case of the scale transformations no such difficulties arise, and a connection is unnecessary. As we shall see, both kinds of compensators appear in supersymmetric theories.

**c. Coset spaces**

Compensators also simplify the description of more general nonlinear $\sigma$-models. We consider a model with fields $y(x)$ that are points of a coset space $G/H$; they transform nonlinearly under the global action of a group $G$, but linearly with respect to a subgroup $H$. By introducing local transformations of the subgroup $H$ via compensators $\phi(x)$, we realize $G$ linearly, and thus easily find an invariant action.

The generators of $G$ are $T, S$, where $S$ are the generators of $H$ and $T$ are the remaining generators, with $T\,, S$ antihermitian. Since $H$ is a subgroup, the generators $S$ close under commutation:

$$[S, S] \sim S \quad . \tag{3.10.31}$$

We require in addition that the generators $T$ carry a representation of the $H$, that is

$$[T, S] \sim T \quad . \tag{3.10.32}$$

(This is always true when the structure constants are totally antisymmetric, since then the absence of $[S, S] \sim T$ terms implies the absence of $[T, S] \sim S$ terms.)

We could write $y(x) = e^{\zeta(x)T}\, mod\, H$, but instead we introduce compensating fields $\phi(x)$, and define fields $z(x)$ that are elements of the whole group $G$:

$$z = e^{\zeta(x)T} e^{\phi(x)S} \equiv e^{\Phi} \tag{3.10.33}$$

(where $\Phi = \hat{\zeta}(x)T + \hat{\phi}(x)S$ provides an equivalent parametrization of the group). The new fields $z$ transform under *global* $G$-transformations and *local* $H$-transformations:

$$z' = g\, z\, h^{-1}(x) \quad , \quad g\epsilon G \quad , \quad h\epsilon H \tag{3.10.34}$$

(where again we can use an exponential parametrization for $g$ and $h(x)$ if we wish).

The local $H$ transformations can be used to gauge away the compensators $\phi$ and reduce $z$ to the coset variables $y$. If we choose the gauge $\phi = 0$, then the global $G$-transformations will induce local gauge-restoring $H$-transformations needed to maintain $\phi = 0$: For $g \epsilon H$, due to (3.10.32), we use $h(x) = g$:

$$e^{\zeta' T} = g\, e^{\zeta T}\, g^{-1} \tag{3.10.35}$$

and thus the fields $y$ transform linearly under $H$. For $g \epsilon G/H$, the gauge restoring transformation is complicated and depends nonlinearly on $\zeta$, and thus the fields $y$ transform nonlinearly under $G/H$.

To find a globally $G$- and locally $H$-invariant Lagrangian, we consider the following quantity:

$$z^{-1}\partial_{\underline{a}} z \equiv \partial_{\underline{a}} + A_{\underline{a}} S + B_{\underline{a}} T \equiv \nabla_{\underline{a}} + B_{\underline{a}} T \quad . \tag{3.10.36}$$

Under global $G$-transformations, both $\nabla_{\underline{a}}$ and $B_{\underline{a}}$ are invariant; under local $H$-transformations we have

$$\begin{aligned} (z^{-1}\partial_{\underline{a}} z)' &= h\, z^{-1}\partial_{\underline{a}}(z\, h^{-1}) \\ &= h\partial_{\underline{a}} h^{-1} + h\, z^{-1}(\partial_{\underline{a}} z) h^{-1} \\ &= h\partial_{\underline{a}} h^{-1} + h(A_{\underline{a}} S + B_{\underline{a}} T) h^{-1} = h(\nabla_{\underline{a}} + B_{\underline{a}} T) h^{-1} \end{aligned} \tag{3.10.37}$$

Because of (3.10.31), $h\, S\, h^{-1} \sim S$ and $h\partial_{\underline{a}} h^{-1} \sim S$; because of (3.10.32), $h\, T\, h^{-1} \sim T$; hence $A_{\underline{a}}$ transforms as a connection for local $H$ transformations ($\nabla_{\underline{a}}$ transforms as a covariant derivative), and $B_{\underline{a}}$ transforms covariantly. Therefore, an invariant Lagrangian is

$$I\!L = -\frac{1}{4}\, tr(B_{\underline{a}} B^{\underline{a}}) \tag{3.10.38}$$

If we choose the gauge $\phi(x) = 0$, this becomes a complicated nonlinear Lagrangian for the fields $y(x)$. We can also couple this system to other fields transforming linearly under $H$ by replacing all derivatives with $\nabla_{\underline{a}}$.

Finally, from (3.10.33) we have

$$z^{-1}\partial_{\underline{a}} z = \partial_{\underline{a}} + e^{-\phi S}(\partial_{\underline{a}} e^{\phi S}) + e^{-\phi S}\big(e^{-\zeta T}\partial_{\underline{a}} e^{\zeta T}\big) e^{\phi S}$$

$$= \partial_{\underline{a}} + (\partial_{\underline{a}}\phi)S + (\partial_{\underline{a}}\zeta)T + \cdots \qquad (3.10.39)$$

and hence $\nabla_{\underline{a}} = \partial_{\underline{a}} + \partial_{\underline{a}}\phi S + \cdots$ and $B_{\underline{a}} = \partial_{\underline{a}}\zeta T + \cdots = \partial_{\underline{a}} y T + \cdots$. This is what we expect: The covariant derivative has the usual dependence on the compensator, and the Lagrangian (3.10.38) has a term $-\frac{1}{2} tr(\partial_{\underline{a}} y)^2$, which is appropriate for a physical field.

## 3.11. Projection operators

### a. General

The analysis of many aspects of the superspace formulation of supersymmetric theories requires an understanding of the irreducible representations of (off-shell) supersymmetry (physical and auxiliary components). We need to know how to decompose an arbitrary superfield or product of superfields into such representations. In this section we describe a procedure for constructing projection operators onto irreducible representations of supersymmetry for general $N$.

The basic idea is that a general superfield can be expanded into a sum of chiral superfields. A chiral superfield that is irreducible under the Poincaré and internal symmetry groups is also irreducible under off-shell supersymmetry (except for possible separation into real and imaginary parts, which we call *bisection).* Thus, this expansion performs the decomposition.

To show that chiral superfields are irreducible under supersymmetry up to bisection, we try to reduce a chiral superfield $\Phi$ by imposing some covariant constraint $\mathbf{\Omega}\Phi = 0$. If we do not consider reality conditions (bisection), we cannot allow constraints relating $\Phi$ to $\bar{\Phi}$. The only covariant operators available for writing constraints are the spinor derivatives $D_{a\alpha}\, , \bar{D}^a{}_{\dot{\alpha}}$ and the spacetime derivative $\partial_{\alpha\dot{\beta}}$. In momentum space, since we are off-shell, all relations must be true for arbitrary momentum, and hence we can freely divide out any spacetime derivative factors. Therefore, any constraint we write down can be reduced to a constraint that is free of spacetime derivatives. If the constraint contained any $\bar{D}$ spinor derivatives, since $\Phi$ is chiral, $\bar{D}\Phi = 0$, by moving the $\bar{D}$'s to the right we could convert them to spacetime derivatives, which we have just argued can be removed. (For example $D\bar{D}D\Phi = iD\partial\Phi$.)

We thus conclude that any possible constraint on $\Phi$ involves only products of the spinor derivatives $D_{a\alpha}$. However, by applying a sufficient number of $\bar{D}$'s to the constraint, we can convert all of the $D$'s to spacetime derivatives; hence, any constraint on $\Phi$ independent of $\bar{\Phi}$ would set $\Phi$ itself to zero (off-shell!). Therefore $\Phi$ must be irreducible. This argument is analogous to the proof in section 3.3 that irreducible representations of supersymmetry can be obtained by repeatedly applying the generators $\bar{Q}^a{}_{\dot{\alpha}}$ to the Clifford vacuum $|C>$ defined by $Q_{a\alpha}|C> = 0$: instead of $|C>$, $\bar{Q}$, and $Q$ with $Q|C>=0$, we have $\Phi$, $D$, and $\bar{D}$ with $\bar{D}\Phi=0$, respectively.

The only further reduction we can perform is to impose a reality condition on the superfield. A chiral superfield of superspin $s$ (the spin content of its external Lorentz indices) has a *single* maximum spin component of spin $s_{\max}=s+\frac{1}{2}N$ residing at the $\theta^N{}_{[a_1\cdots a_N](\alpha_1\cdots\alpha_N)}$ level of the superfield. (This is most easily seen in the chiral representation, where a chiral superfield depends only on $\theta$. The reduction of products of $\theta$'s into irreducible representations is done by the method described for the reduction of products of spinor derivatives in sec. 3.4. Since the maximum spin component has the maximum number of *symmetrized* $SL(2C)$ indices, it must have the maximum number of *antisymmetrized* $SU(N)$ indices, i.e., it must have $N$ indices of each type. Terms with fewer $\theta$'s have fewer $SL(2C)$ indices, whereas terms with more $\theta$'s cannot be antisymmetric in $N$ $SU(N)$ indices, and hence cannot be symmetric in $N$ $SL(2C)$ indices. For examples see (3.6.1-4)). Only if we can impose a reality condition on the highest spin component can we impose a reality condition on the entire superfield. This is possible when $s_{\max}$ is an integer. (A component field with an odd number of Weyl indices cannot satisfy a local reality condition.)

## a.1. Poincaré projectors

We begin with the decomposition of an arbitrary spinor into irreducible representations of the Poincaré group in ordinary spacetime, both because it is one of the steps in the superspace decomposition, and because it illustrates some of the superspace features. This reduction is most easily performed by converting dotted indices into undotted ones with the formal operator $\Delta_{\alpha\dot{\beta}}=-i\partial_{\alpha\dot{\beta}}\Box^{-\frac{1}{2}}$, reducing under $SU(2)$ (by symmetrizing and antisymmetrizing, i.e., taking traces), and converting formerly dotted indices back with $\Delta$. (This insures that no fractional powers of $\Box$ remain. We generally consider $\Box^{-\frac{1}{2}}$ to be hermitian, since we mainly are concerned with $\Box=m^2>0$.) Explicitly, we write for each index

$$\hat{\Psi}_\alpha=\Delta_\alpha{}^{\dot{\beta}}\Psi_{\dot{\beta}} \quad , \qquad \hat{\Psi}_{\dot{\alpha}}=\Delta^\beta{}_{\dot{\alpha}}\Psi_\beta \quad ,$$

$$(\widehat{\Psi_\alpha{}^\dagger})=(\widehat{\Psi_\alpha})^\dagger \quad , \qquad \hat{\hat{\Psi}}_\alpha=\Psi_\alpha \quad . \tag{3.11.1}$$

Thus, for example, a vector $\Psi_{\underline{a}}$ decomposes in the following manner:

$$\Psi_{\underline{a}} = \Delta^{\gamma}{}_{\dot{\alpha}}\hat{\Psi}_{\alpha\gamma} = \Delta^{\gamma}{}_{\dot{\alpha}}\frac{1}{2}(C_{\alpha\gamma}\hat{\Psi}_{\delta}{}^{\delta} + \hat{\Psi}_{(\alpha\gamma)})$$

$$= \Delta^{\gamma}{}_{\dot{\alpha}}\frac{1}{2}(\Delta^{\delta\dot{\epsilon}}C_{\alpha\gamma}\Psi_{\delta\dot{\epsilon}} + \Delta_{(\gamma}{}^{\dot{\delta}}\Psi_{\alpha)\dot{\delta}})$$

$$= \frac{1}{2}\Box^{-1}[\partial_{\alpha\dot{\alpha}}(\partial^{\gamma\dot{\delta}}\Psi_{\gamma\dot{\delta}}) - \partial^{\gamma}{}_{\dot{\alpha}}(\partial_{(\alpha}{}^{\dot{\delta}}\Psi_{\gamma)\dot{\delta}})]$$

$$= [(\Pi^{L} + \Pi^{T})\Psi]_{\underline{a}} \quad , \tag{3.11.2}$$

where $\Pi^L$ and $\Pi^T$ are the longitudinal and transverse projection operators for a four-vector.

The projections can be written in terms of *field strengths* $S$ and $F_{\alpha\gamma}$:

$$(\Pi^{L}\Psi)_{\underline{a}} = \Box^{-1}\partial_{\alpha\dot{\alpha}}S \quad , \qquad S = \frac{1}{2}\partial^{\gamma\dot{\delta}}\Psi_{\gamma\dot{\delta}} \quad ,$$

$$(\Pi^{T}\Psi)_{\underline{a}} = \Box^{-1}\partial^{\gamma}{}_{\dot{\alpha}}F_{\alpha\gamma} \quad , \qquad F_{\alpha\gamma} = \frac{1}{2}\partial_{(\alpha\dot{\delta}}\Psi_{\gamma)}{}^{\dot{\delta}} \quad . \tag{3.11.3}$$

The field strengths are themselves irreducible representations of the Poincaré group. The projections $\Psi^L = \Pi^L\Psi$ and $\Psi^T = \Pi^T\Psi$ are invariant under gauge transformations $\delta\Psi = \Pi^T\chi$ and $\delta\Psi = \Pi^L\chi$ respectively. The field strengths have the same gauge invariance as the projections: $\delta\Psi^{L}{}_{\underline{a}} = \Box^{-1}\partial_{\underline{a}}\delta S = 0$ implies $\delta S = 0$, and similarly $\delta\Psi^{T}{}_{\underline{a}} = \Box^{-1}\partial^{\gamma}{}_{\dot{\alpha}}\delta F_{\alpha\gamma} = 0$ implies $\delta F_{\alpha\gamma} = 0$.

### a.2. Super-Poincaré projectors

Projections of superfields can be written in terms of field strengths in superspace as well. We will find that projections of a general superfield can be expressed in terms of *chiral* field strengths with gauge invariances determined by the projection operators. Thus, for a superfield with decomposition $\Psi = (\sum_n \Pi_n)\Psi$, any single term $\Psi_n = \Pi_n\Psi$ has a gauge invariance $\delta\Psi = \sum_{i\neq n}\Pi_i\chi_i$. Each projection can be written in the form $\Psi_n = D^{2N-n}\Phi^{(n)}$ where the chiral field strengths $\Phi^{(n)} = \bar{D}^{2N}D^n\Psi$ are Poincaré and $SU(N)$ irreducible and have the same gauge invariance as $\Psi_n$: $0 = \delta\Psi_n = D^{2N-n}\delta\Phi^{(n)}$ implies $\delta\Phi^{(n)} = 0$ because $\Phi$ and hence $\delta\Phi$ are irreducible.

The same index conversion used in (3.11.1) can be used to define the operation of *rest-frame conjugation* on a component field or general superfield $\Psi_{\alpha_1 \cdots \alpha_i \dot{\beta}_{i+1} \cdots \dot{\beta}_{2s}}$ by

$$\tilde{\Psi}_{\alpha_1 \cdots \alpha_i \dot{\beta}_{i+1} \cdots \dot{\beta}_{2s}} = \Delta_{\alpha_1}{}^{\dot{\gamma}_1} \cdots \Delta_{\alpha_i}{}^{\dot{\gamma}_i} \Delta^{\delta_{i+1}}{}_{\dot{\beta}_{i+1}} \cdots \Delta^{\delta_{2s}}{}_{\dot{\beta}_{2s}} \bar{\Psi}_{\delta_{2s} \cdots \delta_{i+1} \dot{\gamma}_i \cdots \dot{\gamma}_1} \quad ,$$

$$\tilde{\tilde{\Psi}} = \Psi \quad . \tag{3.11.4}$$

For example, we have:

$$\tilde{\Psi} = \bar{\Psi} \quad , \quad \tilde{\Psi}_\alpha = \Delta_\alpha{}^{\dot{\beta}} \bar{\Psi}_{\dot{\beta}} \quad , \quad \tilde{H}_{\alpha \dot{\beta}} = \Delta_\alpha{}^{\dot{\gamma}} \Delta^\delta{}_{\dot{\beta}} \bar{H}_{\delta \dot{\gamma}} \quad . \tag{3.11.5}$$

We extend this to chiral superfields and define a rest-frame conjugation operator $\mathbf{K}$ which preserves chirality, by using an extra factor $\Box^{-\frac{1}{2}N} \bar{D}^{2N}$ to convert the antichiral (complex conjugated chiral) superfield back to a chiral one (and similarly for antichiral superfields). We define

$$\mathbf{K}\,\Phi_{\alpha_1 \cdots \alpha_i \dot{\beta}_{i+1} \cdots \dot{\beta}_{2s} a_1 \cdots a_i}{}^{b_1 \cdots b_i} = \bar{D}^{2N} \Box^{-\frac{1}{2}N} \tilde{\bar{\Phi}}_{\alpha_1 \cdots \alpha_i \dot{\beta}_{i+1} \cdots \dot{\beta}_{2s}}{}^{b_1 \cdots b_i}{}_{a_1 \cdots a_i} \quad ,$$

$$\mathbf{K}\,\bar{\Phi}_{\cdots} = D^{2N} \Box^{-\frac{1}{2}N} \tilde{\bar{\Phi}}_{\cdots} \quad , \quad \mathbf{K}\overline{(\Phi_{\cdots})} = \overline{(\mathbf{K}\Phi_{\cdots})} \quad ,$$

$$\mathbf{K}^2 = 1 \quad , \quad [\tfrac{1}{2}(1 \pm \mathbf{K})]^2 = \tfrac{1}{2}(1 \pm \mathbf{K}) \quad , \tag{3.11.6}$$

where $\bar{D}_{\underline{\dot{\alpha}}} \Phi_{\cdots} = D_{\underline{\alpha}} \bar{\Phi}_{\cdots} = 0$. For example, for an $N = 1$ chiral spinor $\Phi_\alpha$,

$$\mathbf{K}\,\Phi_\alpha = \frac{-\bar{D}^2 i\partial_\alpha{}^{\dot{\beta}}}{\Box} \bar{\Phi}_{\dot{\beta}} \quad , \tag{3.11.7}$$

We can define *self-conjugacy* or *reality* under $\mathbf{K}$ if we restrict ourselves to superfields that are real representations of $SU(N)$ with $s_{\max} = s + \frac{N}{2}$ integral (the latter is required to insure that only integral powers of $\Box$ appear). The reality condition is $\mathbf{K}\Phi_{\cdots} = \pm\,\Phi_{\cdots}$ and the splitting of a chiral superfield into real and imaginary parts is simply

$$\Phi_{\pm \cdots} = \frac{1}{2}(1 \pm \mathbf{K})\Phi_{\cdots} \quad . \tag{3.11.8}$$

In the previous example, if we impose the reality condition $\mathbf{K}\Phi_\alpha = \Phi_\alpha$, contract both sides with $D^\alpha$ and use the antichirality of $\bar{\Phi}_{\dot{\alpha}}$, we find the equivalent condition:

$$D^\alpha \Phi_\alpha = \bar{D}^{\dot{\alpha}} \bar{\Phi}_{\dot{\alpha}} \tag{3.11.9}$$

These "real" chiral superfields appear in many models of interest. For example, isoscalar "real" chiral superfields with $2-N$ undotted spinor indices describe $N \leq 2$ Yang-Mills gauge multiplets. Similar superfields with $4-N$ undotted spinor indices describe the *conformal field strength* of $N \leq 4$ supergravity.

To decompose a general superfield into irreducible representations, we first expand it in terms of chiral superfields. In the *chiral representation* ($\bar{D}_{\underline{\dot\alpha}} = \bar{\partial}_{\underline{\dot\alpha}}$) a Taylor series in $\bar\theta$ gives

$$\Psi(x,\theta,\bar\theta) = \sum_{n=0}^{2N} \frac{1}{n!}\,\bar\theta^{n\underline{\dot\alpha}_1\cdots\underline{\dot\alpha}_n}\Phi^{(n)}{}_{\underline{\dot\alpha}_n\cdots\underline{\dot\alpha}_1}(x,\theta) \quad , \tag{3.11.10a}$$

where $\Phi^n$ can be rewritten as

$$\Phi^{(n)}{}_{\underline{\dot\alpha}_1\cdots\underline{\dot\alpha}_n}(x,\theta) = \bar{D}^n{}_{\underline{\dot\alpha}_1\cdots\underline{\dot\alpha}_n}\Psi(x,\theta,\bar\theta)|_{\bar\theta\,=\,0} \quad , \tag{3.11.10b}$$

or, using $\{\bar{D}_{\underline{\dot\alpha}}, \bar\theta^{\underline{\dot\beta}}\} = \delta_{\underline{\dot\alpha}}{}^{\underline{\dot\beta}}$ and $\{\bar\theta^{\underline{\dot\alpha}}, \bar\theta^{\underline{\dot\beta}}\} = 0$ (which implies $\bar\theta^{2N+1} = 0$),

$$\Phi^{(n)}{}_{\underline{\dot\alpha}_1\cdots\underline{\dot\alpha}_n}(x,\theta) = (-1)^N \bar{D}^{2N}\bar\theta^{2N}\bar{D}^n{}_{\underline{\dot\alpha}_1\cdots\underline{\dot\alpha}_n}\Psi(x,\theta,\bar\theta) \quad . \tag{3.11.10c}$$

However, $\bar\theta$ is not covariant, and hence neither is the expansion (3.11.10). We can generalize (3.11.10): For *any operator* $\bar\zeta^{\underline{\dot\alpha}}(\bar\theta)$ which obeys

$$\{\bar{D}_{\underline{\dot\alpha}}, \bar\zeta^{\underline{\dot\beta}}\} = \delta_{\underline{\dot\alpha}}{}^{\underline{\dot\beta}} \quad , \quad \{\bar\zeta^{\underline{\dot\alpha}}, \bar\zeta^{\underline{\dot\beta}}\} = 0 \quad , \tag{3.11.11}$$

we can write

$$\hat\Psi(x,\theta,\bar\zeta) = \sum_{n=0}^{2N} \frac{1}{n!}\,\bar\zeta^{n\underline{\dot\alpha}_1\cdots\underline{\dot\alpha}_n}\hat\Phi^{(n)}{}_{\underline{\dot\alpha}_n\cdots\underline{\dot\alpha}_1}(x,\theta) \quad , \tag{3.11.12a}$$

where

$$\hat\Phi^{(n)}{}_{\underline{\dot\alpha}_n\cdots\underline{\dot\alpha}_1}(x,\theta) = (-1)^N \bar{D}^{2N}\bar\zeta^{2N}\bar{D}^n{}_{\underline{\dot\alpha}_n\cdots\underline{\dot\alpha}_1}\hat\Psi(x,\theta,\bar\zeta) \quad . \tag{3.11.12b}$$

If we choose $\Psi(x,\theta,\bar\theta) \equiv \hat\Psi(x,\theta,\bar\zeta(\bar\theta))$, we obtain, substituting (3.11.12b) into (3.11.12a):

$$\Psi(x,\theta,\bar\theta) = (-1)^N \sum_{n=0}^{2N} \frac{1}{n!}\,\bar\zeta^{n\underline{\dot\alpha}_1\cdots\underline{\dot\alpha}_n}\bar{D}^{2N}\bar\zeta^{2N}\bar{D}^n{}_{\underline{\dot\alpha}_n\cdots\underline{\dot\alpha}_1}\Psi(x,\theta,\bar\theta) \quad , \tag{3.11.13}$$

for any $\bar\zeta$ satisfying (3.11.11). A manifestly supersymmetric operator satisfying (3.11.11)

is

$$\bar{\zeta}^{\dot{\underline{\alpha}}} = -\, i\, \frac{\partial^{\underline{\beta}\dot{\underline{\alpha}}}}{\Box}\, D_{\underline{\beta}} = \bar{\theta}^{\dot{\underline{\alpha}}} - i\, \frac{\partial^{\underline{\beta}\dot{\underline{\alpha}}}}{\Box}\, \partial_{\underline{\beta}} \quad . \tag{3.11.14}$$

Substituting (3.11.14) into (3.11.13), we find

$$\Psi(x,\theta,\bar{\theta}) = \Box^{-N} \sum_{n=0}^{2N} \frac{1}{n!}\, D^{n}{}_{\underline{\alpha}_1\cdots\underline{\alpha}_n} \left( \frac{-i\partial^{\underline{\alpha}_1\dot{\underline{\beta}}_1}}{\Box} \right) \cdots \left( \frac{-i\partial^{\underline{\alpha}_n\dot{\underline{\beta}}_n}}{\Box} \right)$$

$$\times\, \bar{D}^{2N} D^{2N} \bar{D}^{n}{}_{\dot{\underline{\beta}}_n\cdots\dot{\underline{\beta}}_1} \Psi(x,\theta,\bar{\theta}) \quad , \tag{3.11.15}$$

where we have used $(\frac{-i\partial^{\underline{\beta}\dot{\underline{\alpha}}}}{\Box} D_{\underline{\beta}})^{2N} = (-\Box)^{-N} D^{2N}$. Pushing $\bar{D}^{n}$ through $D^{2N}$ to the $\bar{D}^{2N}$, we find (reordering the sum by replacing $n \to 2N - n$)

$$\Psi = \Box^{-N}\, \frac{1}{(2N)!}\, C^{\underline{\alpha}_1\cdots\underline{\alpha}_{2N}} \sum_{n=0}^{2N} (-1)^n \binom{2N}{n} D^{2N-n}{}_{\underline{\alpha}_{2N}\cdots\underline{\alpha}_{n+1}} \bar{D}^{2N} D^{n}{}_{\underline{\alpha}_n\cdots\underline{\alpha}_1} \Psi(x,\theta,\bar{\theta})$$

$$= \Box^{-N} \sum_{n=0}^{2N} \frac{1}{n!}\, (-1)^n D^{2N-n\,\underline{\alpha}_1\cdots\underline{\alpha}_n} \bar{D}^{2N} D^{n}{}_{\underline{\alpha}_1\cdots\underline{\alpha}_n} \Psi(x,\theta,\bar{\theta}) \quad . \tag{3.11.16}$$

This final expression can be compared to the noncovariant $\bar{\theta}$ expansion in (3.11.10). The chiral fields $\bar{D}^{2N} D^{n} \Psi$ are the covariant analogs of the $\Phi^{(2N-n)}$'s. We thus obtain

$$1 = \sum_{n=0}^{2N} \frac{1}{n!}\, (-1)^n \Box^{-N} D^{2N-n\,\underline{\alpha}_1\cdots\underline{\alpha}_n} \bar{D}^{2N} D^{n}{}_{\underline{\alpha}_1\cdots\underline{\alpha}_n} \quad . \tag{3.11.17}$$

For example, in $N = 1$ this is the relation

$$1 = \frac{D^2 \bar{D}^2}{\Box} - \frac{D^{\alpha} \bar{D}^2 D_{\alpha}}{\Box} + \frac{\bar{D}^2 D^2}{\Box} \quad . \tag{3.11.18}$$

Each term in the sum is a (reducible) projection operator which picks out the part of a superfield $\Psi$ appearing in the chiral field strength $\bar{D}^{2N} D^{n} \Psi$ (which is irreducible under $SL(2N, C)$ but reducible under $SU(N)\otimes$Poincaré, and possibly also under $\mathbf{K}$). We thus have the projection operators $\Pi_n$, $n = 0, 1, \ldots, 2N$:

$$\Pi_n = \frac{1}{n!}\, (-1)^n \Box^{-N} D^{2N-n\,\underline{\alpha}_1\cdots\underline{\alpha}_n} \bar{D}^{2N} D^{n}{}_{\underline{\alpha}_1\cdots\underline{\alpha}_n} \quad ,$$

$$\sum_{n=0}^{2N} \Pi_n = 1 \quad . \tag{3.11.19}$$

In particular, $\Pi_0 = \Box^{-N} D^{2N} \bar{D}^{2N}$ and $\Pi_{2N} = \Box^{-N} \bar{D}^{2N} D^{2N}$ project out the antichiral and chiral parts of $\Psi$ respectively. The projectors (3.11.19) satisfy a number of relations: Orthonormality

$$\Pi_m \Pi_n = \delta_{mn} \Pi_m \quad \text{(not summed)} \tag{3.11.20}$$

follows from $\bar{D}^{2N} D^n \bar{D}^{2N} = 0$ unless $n = 2N$ and hence $\Pi_m \Pi_n = 0$ for $m \neq n$; then $\sum \Pi_m = 1$ implies $\Pi_n = \Pi_n \sum \Pi_m = \Pi_n^2$. There are relations between the $\Pi$'s: $\Pi_n$ is equal to the transpose and to the complex conjugate of $\Pi_{2N-n}$

$$\begin{aligned} \Pi_n &= \Pi^t{}_{2N-n} \\ &= \frac{1}{(2N-n)!} \Box^{-N} D^{2N-n}{}_{\underline{\alpha}_1 \cdots \underline{\alpha}_{2N-n}} \bar{D}^{2N} D^{n\, \underline{\alpha}_1 \cdots \underline{\alpha}_{2N-n}} \quad , \end{aligned} \tag{3.11.21}$$

$$\begin{aligned} \Pi_n &= \Pi^*{}_{2N-n} \\ &= \frac{1}{(2N-n)!} (-1)^n \Box^{-N} \bar{D}^{n\, \underline{\dot{\alpha}}_1 \cdots \underline{\dot{\alpha}}_{2N-n}} D^{2N} \bar{D}^{2N-n}{}_{\underline{\dot{\alpha}}_1 \cdots \underline{\dot{\alpha}}_{2N-n}} \quad . \end{aligned} \tag{3.11.22}$$

Combining (3.11.21) and (3.11.22), we find another form of $\Pi_n$:

$$\Pi_n = \Pi^\dagger{}_n = \frac{1}{n!} \Box^{-N} \bar{D}^n{}_{\underline{\dot{\alpha}}_1 \cdots \underline{\dot{\alpha}}_n} D^{2N} \bar{D}^{2N-n\, \underline{\dot{\alpha}}_1 \cdots \underline{\dot{\alpha}}_n} \quad . \tag{3.11.23}$$

The complex conjugation relation (3.11.22) implies that half of the $\Pi$'s are redundant for real superfields: $V = \bar{V} \rightarrow \Pi_{2N-n} V = \Pi^*{}_n \bar{V} = \overline{(\Pi_n V)}$.

Reduction of the $\Pi$'s into irreducible projection operators is now easy:

(1) Algebraically reduce $\bar{D}^{2N} D^n \Psi$ under $SU(N) \otimes$Poincaré (where $\Psi$ may have further isospinor and Weyl spinor indices);

(2) When the reduced chiral field strength $\bar{D}^{2N} D^n \Psi$ is in a real representation of $SU(N)$ and has $s + \frac{1}{2} N$ integral, further reduce by bisection, i.e. multiplication by $\frac{1}{2}(1 \pm \mathbf{K})$.

To perform (1) it is convenient to first reduce $\bar{D}^{2N} D^n$ by using the total antisymmetry of the $D$'s (see sec. 3.4), and then reduce the tensor product of the irreducible representations of $\bar{D}^{2N} D^n$ with the representation of the superfield $\Psi$ as usual. If we only want to preserve $SO(N)$, further reduction is performed in step 1; for step 2, $\bar{D}^{2N} D^n \Psi$ is always in a real representation of $SO(N)$.

Although $\Pi_n$ contains the product of $2N$ $D$'s and $2N$ $\bar{D}$'s and is thus in its simplest form, $\Pi_{n\pm}$, obtained by directly introducing $\frac{1}{2}(1 \pm \mathbf{K})$ in front of the $\bar{D}^{2N}$, contains $2N$ $D$'s and $4N$ $\bar{D}$'s in the $\mathbf{K}$ term, and can be further simplified. After some algebra we find:

For $n \geq N$:

$$\mathbf{K} \bar{D}^{2N} D^n{}_{\underline{\alpha}_1 \cdots \underline{\alpha}_n} \Psi^{b_1 \cdots b_n} = (-1)^{2\hat{s}n} \Box^{\frac{1}{2}(n-N)} \bar{D}^{2N} D^{2N-n\, \underline{\beta}_1 \cdots \underline{\beta}_n} C_{\beta_1 \alpha_1} \cdots C_{\beta_n \alpha_n} \tilde{\bar{\Psi}}_{a_1 \cdots a_n} \quad , \tag{3.11.24}$$

or

$$\mathbf{K} \bar{D}^{2N} D^{n\, \underline{\alpha}_1 \cdots \underline{\alpha}_{2N-n}} \Psi_{b_1 \cdots b_{2N-n}}$$

$$= (-1)^{2\hat{s}n} \Box^{\frac{1}{2}(n-N)} \bar{D}^{2N} C^{\alpha_1 \beta_1} \cdots C^{\alpha_{2N-n} \beta_{2N-n}} D^{2N-n}{}_{\underline{\beta}_1 \cdots \underline{\beta}_{2N-n}} \tilde{\bar{\Psi}}^{a_1 \cdots a_{2N-n}} \quad , \tag{3.11.25}$$

where $2\hat{s}$ extra Weyl spinor indices, and extra isospinor indices, reduced as in step 1, are implicit on $\Psi$.

For $n \leq N$:

$$\mathbf{K} D^{2N} \bar{D}^{2N-n}{}_{\underline{\dot{\alpha}}_1 \cdots} \Psi_{b_1 \cdots} = (-1)^{2\hat{s}n} \Box^{\frac{1}{2}(N-n)} D^{2N} \bar{D}^{n\, \dot{\beta}_1 \cdots} C_{\dot{\beta}_1 \dot{\alpha}_1 \cdots} \tilde{\bar{\Psi}}^{a_1 \cdots} \quad , \tag{3.11.26}$$

or

$$\mathbf{K} D^{2N} \bar{D}^{2N-n\, \underline{\dot{\alpha}}_1 \cdots} \Psi^{b_1 \cdots} = (-1)^{2\hat{s}n} \Box^{\frac{1}{2}(N-n)} D^{2N} C^{\dot{\alpha}_1 \dot{\beta}_1 \cdots} \bar{D}^n{}_{\underline{\dot{\beta}}_1 \cdots} \tilde{\bar{\Psi}}_{a_1 \cdots} \quad . \tag{3.11.27}$$

As an example of this simplification, we consider the $N = 1$ chiral field above (3.11.7) for the special case when it is a field strength of a real superfield $V$: $\Phi_\alpha = \bar{D}^2 D_\alpha V$

$$\mathbf{K} \Phi_\alpha = \frac{-\bar{D}^2 i \partial_\alpha{}^{\dot{\beta}}}{\Box} D^2 \bar{D}_{\dot{\beta}} V = \bar{D}^2 D_\alpha V = \Phi_\alpha \quad . \tag{3.11.28}$$

We now collect our results: The superprojectors take the final form

If bisection is possible:

$$n \leq N: \quad \Pi_{n,i\pm} = \frac{1}{n!} \Box^{-N} \bar{D}^n{}_{\underline{\dot{\alpha}}_1 \cdots \underline{\dot{\alpha}}_n} \frac{1}{2} (1 \pm \mathbf{K}) I\!\!P_i D^{2N} \bar{D}^{2N-n\, \underline{\dot{\alpha}}_1 \cdots \underline{\dot{\alpha}}_n} \quad ,$$

$$n \geq N: \quad \Pi_{n,i\pm} = \frac{1}{(2N-n)!} \Box^{-N} D^{2N-n}{}_{\underline{\alpha}_1 \cdots \underline{\alpha}_{2N-n}} \frac{1}{2}(1 \pm \mathbf{K}) I\!P_i \bar{D}^{2N} D^{n\underline{\alpha}_1 \cdots \underline{\alpha}_{2N-n}} \quad , \qquad (3.11.29)$$

If bisection is not possible:

$$\Pi_{n,i} = \text{either of the above with } \frac{1}{2}(1 \pm \mathbf{K}) \text{ dropped} \quad , \qquad (3.11.30)$$

where the $I\!P_i$ are $SU(N)\otimes$Poincaré projectors acting on the explicit indices (including those of the superfield). We have chosen the particular forms of $\Pi_n$ from (3.11.19,21-23) that minimize the number of indices that the $I\!P_i$ act on. The chiral expansion, besides its simplicity, has the advantage that the chiral field strengths appear explicitly, and the superspin and superisospin of the representation onto which $\Pi$ projects are those of the chiral field strength.

## b. Examples

### b.1. N=0

We begin by giving a few Poincaré projection operators. The procedure for finding them was discussed in considerable detail in subsec. 3.11.a.1, so here we simply list results. Scalars and spinors are irreducible (no bisection is possible for a spinor: $s + \frac{N}{2} = \frac{1}{2}$ is not an integer). A (real) vector decomposes into a spin 1 and a spin 0 projection (see (3.11.2,3)). For a spinor-vector $\psi_{\alpha\dot{\beta},\gamma} = \Delta^{\beta}{}_{\dot{\beta}}\psi_{\alpha\beta,\gamma}$ we have:

$$\psi_{\alpha\beta,\gamma} = \frac{1}{3!}\psi_{(\alpha\beta,\gamma)} + \frac{1}{3!}(\psi_{(\alpha\delta)}{}^{\delta}C_{\beta\gamma} + \psi_{(\beta\delta)}{}^{\delta}C_{\alpha\gamma}) + \frac{1}{2}C_{\alpha\beta}\psi_{\delta}{}^{\delta}{}_{,\gamma}$$

and hence

$$\psi_{\alpha\dot{\beta},\gamma} = (\Pi_{\frac{3}{2}} + \Pi^{T}{}_{\frac{1}{2}} + \Pi^{L}{}_{\frac{1}{2}})\psi_{\alpha\dot{\beta},\gamma} \quad ,$$

where

$$\Pi_{\frac{3}{2}}\psi_{\alpha\dot{\beta},\gamma} = -\Box^{-1}\partial^{\beta}{}_{\dot{\beta}}w_{\alpha\beta\gamma} \quad , \qquad w_{\alpha\beta\gamma} = \frac{1}{6}\partial_{(\alpha}{}^{\dot{\delta}}\psi_{\beta\dot{\delta}\gamma)} \quad ;$$

$$\Pi^{T}{}_{\frac{1}{2}}\psi_{\alpha\dot{\beta},\gamma} = -\Box^{-1}\partial^{\beta}{}_{\dot{\beta}}[C_{\beta\gamma}r_{\alpha} + C_{\alpha\gamma}r_{\beta}] \quad , \quad r_{\alpha} = \frac{1}{6}\partial_{(\alpha}{}^{\dot{\delta}}\psi_{\beta)\dot{\delta}}{}^{\beta} \quad ;$$

$$\Pi^{L}{}_{\frac{1}{2}}\psi_{\alpha\dot{\beta},\gamma} = \Box^{-1}\partial_{\alpha\dot{\beta}}s_{\gamma} \quad , \qquad s_{\gamma} = \frac{1}{2}\partial^{\underline{a}}\psi_{\underline{a},\gamma} \quad . \qquad (3.11.31)$$

For a real two-index tensor $h_{\underline{ac}} = \Delta^{\beta}{}_{\dot{\alpha}}\Delta^{\delta}{}_{\dot{\gamma}} h_{\alpha\beta,\gamma\delta}$ we have

$$h_{\alpha\beta,\gamma\delta} = \frac{1}{4!}\, h_{(\alpha\beta,\gamma\delta)} + (C_{\gamma(\alpha} q_{\beta)\delta} + C_{\delta(\alpha} q_{\beta)\gamma}) + C_{\alpha\beta} C_{\gamma\delta} q + C_{\gamma(\alpha} C_{\beta)\delta} r$$

$$+ \frac{1}{4}\,(C_{\alpha\beta} h_{\epsilon(\gamma,}{}^{|\epsilon|}{}_{\delta)} + C_{\gamma\delta} h_{(\alpha|\epsilon|,\beta)}{}^{\epsilon})\ , \tag{3.11.32}$$

where

$$q_{\alpha\beta} = -\frac{1}{32}\,(\, h_{(\alpha}{}^{|\alpha|}{}_{,\beta)\delta} + h_{(\alpha|\delta|,\beta)}{}^{\alpha} + h_{(\alpha}{}^{|\alpha|}{}_{,\delta)\beta} + h_{(\alpha|\beta|,\delta)}{}^{\alpha}\,)\ \ ,$$

$$r = -\frac{1}{24}\, h_{(\alpha}{}^{(\alpha}{}_{,\beta)}{}^{\beta)} \quad , \quad q = \frac{1}{4}\, h_{\epsilon}{}^{\epsilon}{}_{,\zeta}{}^{\zeta}\ \ . \tag{3.11.33}$$

Therefore the complete decomposition of the the two-index tensor is given by

$$h_{\alpha\dot{\beta},\gamma\dot{\delta}} = (\Pi_{2,S} + \Pi_{1,S} + \Pi^{L}{}_{0,S} + \Pi^{T}{}_{0,S} + \Pi_{1,A}{}^{+} + \Pi_{1,A}{}^{-}) h_{\alpha\dot{\beta},\gamma\dot{\delta}}\ \ ,$$

where the projectors are labeled by the spin $(2,1,0)$, the symmetric and antisymmetric part of $h_{\underline{ab}}$ ($S$ and $A$), longitudinal and transverse parts ($L$ and $T$), and self-dual and anti-self-dual parts ($+$ and $-$). The explicit form of the projection operators is

$$\Pi_{2,S} h_{\alpha\dot{\beta},\gamma\dot{\delta}} = \Box^{-2} \partial^{\beta}{}_{\dot{\beta}} \partial^{\delta}{}_{\dot{\delta}} w_{\alpha\beta\gamma\delta}\ \ ,\quad w_{\alpha\beta\gamma\delta} = \frac{1}{4!}\,\partial_{(\alpha}{}^{\dot{\beta}} \partial_{\beta}{}^{\dot{\delta}} h_{\gamma\dot{\beta},\delta)\dot{\delta}}\ \ ;$$

$$\Pi_{1,S} h_{\alpha\dot{\beta},\gamma\dot{\delta}} = \Box^{-2} \partial^{\beta}{}_{\dot{\beta}} \partial^{\delta}{}_{\dot{\delta}} [\, C_{\gamma(\alpha} w_{\beta)\delta} + C_{\delta(\alpha} w_{\beta)\gamma}\,]\ \ ,$$

$$w_{\beta\delta} = -\frac{1}{32}\,\partial^{\alpha(\dot{\beta}}[\partial_{\delta}{}^{\dot{\delta})} h_{(\alpha\dot{\beta},\beta)\dot{\delta}} + \partial_{\beta}{}^{\dot{\delta})} h_{(\alpha\dot{\beta},\delta)\dot{\delta}}]\ \ ;$$

$$\Pi^{L}{}_{0,S} h_{\alpha\dot{\beta},\gamma\dot{\delta}} = \Box^{-2} \partial_{\alpha\dot{\beta}} \partial_{\gamma\dot{\delta}} S \quad , \quad S = \frac{1}{4}\,\partial^{\alpha\dot{\beta}} \partial^{\gamma\dot{\delta}} h_{\alpha\dot{\beta},\gamma\dot{\delta}}\ \ ;$$

$$\Pi^{T}{}_{0,S} h_{\alpha\dot{\beta},\gamma\dot{\delta}} = \Box^{-2}(C_{\gamma\alpha} C_{\dot{\beta}\dot{\delta}} \Box + \partial_{\gamma\dot{\beta}} \partial_{\alpha\dot{\delta}}) T\ \ ,\ T = -\frac{1}{12}\,\partial^{(\alpha\dot{\beta}} \partial^{\gamma)\dot{\delta}} h_{\alpha\dot{\beta},\gamma\dot{\delta}}\ \ ;$$

$$\Pi_{1,A}{}^{+} h_{\alpha\dot{\beta},\gamma\dot{\delta}} = \Box^{-1} \partial^{\beta}{}_{\dot{\beta}} \partial_{\gamma\dot{\delta}} l^{+}{}_{(\alpha\beta)}\ \ ,\ l^{+}{}_{(\alpha\beta)} = -\frac{1}{4}\, h_{(\alpha\dot{\gamma},\beta)}{}^{\dot{\gamma}}\ \ ;$$

$$\Pi_{1,A}{}^{-} h_{\alpha\dot{\beta},\gamma\dot{\delta}} = \Box^{-2} \partial_{\alpha\dot{\beta}} \partial^{\delta}{}_{\dot{\delta}} l^{-}{}_{(\gamma\delta)}\ \ ,\ l^{-}{}_{(\gamma\delta)} = -\frac{1}{4}\,\partial_{(\gamma}{}^{\dot{\beta}} \partial_{\delta)}{}^{\dot{\delta}} h_{\alpha\dot{\beta},}{}^{\alpha}{}_{\dot{\delta}}\ \ . \tag{3.11.34}$$

(The field strengths $w_{\alpha\beta\gamma\delta}$ and $T$ are proportional to the linearized Weyl tensor and scalar curvatures respectively.) From this decomposition, we see that the two-index tensor field consists of irreducible spins 2⊕1⊕1⊕1⊕0⊕0.

**b.2. N=1**

We construct the irreducible projection operators for a complex scalar superfield $\Psi$. From (3.9.26-32) we have, for the cases without bisection ($s + \frac{1}{2}N = s + \frac{1}{2}$ is half-integral, so that $s$ is integral)

$$\Pi_0 = \Box^{-1} D^2 \bar{D}^2 \quad , \quad \Pi_1 = - \Box^{-1} D^\alpha \bar{D}^2 D_\alpha \quad , \quad \Pi_2 = \Box^{-1} \bar{D}^2 D^2 \quad . \qquad (3.11.35)$$

Since $\Psi$ has no external indices we can go directly to step 2. The chiral field strengths $\bar{D}^2 \Psi$ and $\bar{D}^2 D^2 \Psi$ do not satisfy the condition that $s + \frac{1}{2}$ is integral, whereas $\bar{D}^2 D_\alpha \Psi$ does. For $N = 1$, the condition of being in a real isospin representation is trivially satisfied, and that means that $\Pi_1$ needs to be bisected:

$$\Pi_1 = \Pi_{1+} + \Pi_{1-} \quad ,$$

$$\Pi_{1\pm} = - \Box^{-1} D^\alpha \frac{1}{2} (1 \pm \mathbf{K}) \bar{D}^2 D_\alpha \quad . \qquad (3.11.36)$$

Therefore, from (3.11.26-7),

$$\Pi_{1\pm} \Psi = - \Box^{-1} D^\alpha \bar{D}^2 D_\alpha \frac{1}{2} (\Psi \pm \bar{\Psi}) \quad , \qquad (3.11.37)$$

and thus $\Pi_0$, $\Pi_{1\pm}$ and $\Pi_2$ completely reduce $\Psi$. These irreducible representations turn out to describe two scalar and two vector multiplets, respectively.

We give next the decomposition of the spinor superfield $\Psi_\alpha$. To find the irreducible parts of $\Pi_1 \Psi_\alpha$ we Poincaré reduce the chiral field strength $\bar{D}^2 D_\alpha \Psi_\beta = \frac{1}{2} [C_{\alpha\beta} \bar{D}^2 D_\gamma \Psi^\gamma + \bar{D}^2 D_{(\alpha} \Psi_{\beta)}]$. This gives the projections $\Pi_{n,s}$ for superspin $s$ of this chiral field strength:

$$\Pi_{1,0} \Psi_\alpha = \frac{1}{2} \Box^{-1} D_\alpha \bar{D}^2 D^\beta \Psi_\beta \quad , \quad \Pi_{1,1} \Psi_\alpha = - \frac{1}{2} \Box^{-1} D^\beta \bar{D}^2 D_{(\alpha} \Psi_{\beta)} \quad . \qquad (3.11.38)$$

The latter irreducible representation is a "conformal" submultiplet of the $(\frac{3}{2}, 1)$ multiplet (see section 4.5). For $\Pi_0$ and $\Pi_2$ we must bisect:

$$\Pi_{0\frac{1}{2}\pm} \Psi_\alpha = \Box^{-1} \frac{1}{2} (1 \pm \mathbf{K}) D^2 \bar{D}^2 \Psi_\alpha = \Box^{-1} D^2 \frac{1}{2} (\bar{D}^2 \Psi_\alpha \pm i \partial_{\alpha\dot{\beta}} \bar{\Psi}^{\dot{\beta}}) \quad ,$$

$$\Pi_{2\frac{1}{2}\pm}\Psi_\alpha = \Box^{-1}\,\frac{1}{2}\,(1 \pm \mathbf{K})\bar{D}^2 D^2\Psi_\alpha = \Box^{-1}\bar{D}^2\,\frac{1}{2}\,(D^2\Psi_\alpha \pm i\partial_{\alpha\dot{\beta}}\bar{\Psi}^{\dot{\beta}}) \quad . \tag{3.11.39}$$

Equivalent forms are:

$$\Pi_{0\frac{1}{2}\pm}\Psi_\alpha = -\,\Box^{-1}D_\alpha\,\frac{1}{2}\,(D^\beta\bar{D}^2\Psi_\beta \pm \bar{D}^{\dot{\beta}}D^2\bar{\Psi}_{\dot{\beta}}) \quad ,$$

$$\Pi_{2\frac{1}{2}\pm}\Psi_\alpha = \Box^{-1}\bar{D}^2 D_\alpha\,\frac{1}{2}\,(D_\beta\Psi^\beta \pm \bar{D}_{\dot{\beta}}\bar{\Psi}^{\dot{\beta}}) \quad . \tag{3.11.40}$$

Finally we decompose the real vector superfield $H_{\alpha\dot{\beta}}$. Because of its reality bisection is unnecessary. Poincaré projection is performed by writing $H_{\alpha\dot{\beta}} = \Delta^\gamma{}_{\dot{\beta}}H_{\alpha\gamma}$ and (anti)symmetrizing in the indices of the chiral field strengths. To ensure that the projection operators maintain the reality of $H_{\alpha\dot{\beta}}$, we combine the $\Pi_2$'s with the $\Pi_0$'s, since from (3.9.24) $\Pi_2 H_{\alpha\dot{\beta}} = (\Pi_0 H_{\beta\dot{\alpha}})^\dagger$. We obtain

$$\Pi^T{}_{0,1}H_{\alpha\dot{\beta}} = \frac{1}{2}\,\Box^{-1}\Delta^\gamma{}_{\dot{\beta}}\{D^2, \bar{D}^2\}H_{(\alpha\gamma)} \quad ,$$

$$\Pi^L{}_{0,0}H_{\alpha\dot{\beta}} = \frac{1}{2}\,\Box^{-1}\Delta_{\alpha\dot{\beta}}\{D^2, \bar{D}^2\}H^\gamma{}_\gamma \quad ,$$

$$\Pi^T{}_{1,\frac{3}{2}}H_{\alpha\dot{\beta}} = -\,\frac{1}{6}\,\Box^{-1}\Delta^\gamma{}_{\dot{\beta}}D^\delta\bar{D}^2 D_{(\alpha}H_{\gamma\delta)} \quad ,$$

$$\Pi^T{}_{1,\frac{1}{2}}H_{\alpha\dot{\beta}} = \frac{1}{6}\,\Box^{-1}\Delta^\gamma{}_{\dot{\beta}}(D_\alpha\bar{D}^2 D^\delta H_{(\gamma\delta)} + D_\gamma\bar{D}^2 D^\delta H_{(\alpha\delta)}) \quad ,$$

$$\Pi^L{}_{1,\frac{1}{2}}H_{\alpha\dot{\beta}} = -\,\frac{1}{2}\,\Box^{-1}\Delta_{\alpha\dot{\beta}}D^\gamma\bar{D}^2 D_\gamma H^\delta{}_\delta \quad , \tag{3.11.41}$$

where $T$ and $L$ denote transverse and longitudinal. Reexpressing $H_{\alpha\beta}$ in terms of $H_{\alpha\dot{\beta}}$, we find

$$\Pi^T{}_{0,1}H_{\alpha\dot{\beta}} = \frac{1}{2}\,\Box^{-2}\partial^\gamma{}_{\dot{\beta}}\{D^2, \bar{D}^2\}\partial_{(\alpha\dot{\delta}}H_{\gamma)}{}^{\dot{\delta}} \quad ,$$

$$\Pi^L{}_{0,0}H_{\alpha\dot{\beta}} = \frac{1}{2}\,\Box^{-2}\partial_{\alpha\dot{\beta}}\{D^2, \bar{D}^2\}\partial_{\underline{c}}H^{\underline{c}} \quad ,$$

$$\Pi^T{}_{1,\frac{3}{2}}H_{\alpha\dot{\beta}} = \frac{1}{6}\,\Box^{-2}\partial^\gamma{}_{\dot{\beta}}D^\delta\bar{D}^2 D_{(\alpha}\partial_\gamma{}^{\dot{\epsilon}}H_{\delta)\dot{\epsilon}} \quad ,$$

$$\Pi^T{}_{1,\frac{1}{2}} H_{\alpha\dot{\beta}} = \frac{1}{6}\,\Box^{-2}\partial^{\gamma}{}_{\dot{\beta}}(D_{\alpha}\bar{D}^2 D^{\delta}\partial_{(\gamma\dot{\epsilon}}H_{\delta)}{}^{\dot{\epsilon}} + D_{\gamma}\bar{D}^2 D^{\delta}\partial_{(\alpha\dot{\epsilon}}H_{\delta)}{}^{\dot{\epsilon}}) \quad ,$$

$$\Pi^L{}_{1,\frac{1}{2}} H_{\alpha\dot{\beta}} = -\,\frac{1}{2}\,\Box^{-2}\partial_{\alpha\dot{\beta}}D^{\gamma}\bar{D}^2 D_{\gamma}\partial_{\underline{d}}H^{\underline{d}} \quad . \qquad (3.11.42)$$

## b.3. N=2

We begin by giving the expressions for $SL(4,C)$ $C$'s in terms of those of $SU(2)$ and $SL(2,C)$:

$$C_{\underline{\alpha\beta\gamma\delta}} = C_{ab}C_{cd}C_{\alpha\delta}C_{\beta\gamma} - C_{ad}C_{cb}C_{\alpha\beta}C_{\delta\gamma} \quad ,$$

$$C_{\underline{\dot{\alpha}\dot{\beta}\dot{\gamma}\dot{\delta}}} = C^{ab}C^{cd}C_{\dot{\alpha}\dot{\delta}}C_{\dot{\beta}\dot{\gamma}} - C^{ad}C^{cb}C_{\dot{\alpha}\dot{\beta}}C_{\dot{\delta}\dot{\gamma}} \quad ,$$

$$C^{\underline{\alpha\beta\gamma\delta}} = C^{ab}C^{cd}C^{\alpha\delta}C^{\beta\gamma} - C^{ad}C^{cb}C^{\alpha\beta}C^{\delta\gamma} \quad ,$$

$$C^{\underline{\dot{\alpha}\dot{\beta}\dot{\gamma}\dot{\delta}}} = C_{ab}C_{cd}C^{\dot{\alpha}\dot{\delta}}C^{\dot{\beta}\dot{\gamma}} - C_{ad}C_{cb}C^{\dot{\alpha}\dot{\beta}}C^{\dot{\delta}\dot{\gamma}} \quad . \qquad (3.11.43)$$

We define the $SU(2)\otimes$Poincaré reduction of $D^2{}_{\underline{\alpha\beta}}$ as follows:

$$D^2{}_{\underline{\alpha\beta}} = C_{\beta\alpha}D^2{}_{ab} + C_{ba}D^2{}_{\alpha\beta} \quad ,$$

$$\bar{D}^2{}_{\underline{\dot{\alpha}\dot{\beta}}} = C_{\underline{\dot{\beta}\dot{\alpha}}}\bar{D}^{2\,ab} + C^{ba}\bar{D}^2{}_{\dot{\alpha}\dot{\beta}} \quad ,$$

$$D^{2\,\underline{\alpha\beta}} = C^{\beta\alpha}C^{ac}C^{db}D^2{}_{cd} + C^{ba}D^{2\,\alpha\beta} \quad ,$$

$$D^2{}_{ab} = \frac{1}{2}\,D_a{}^{\alpha}D_{b\alpha} = D^2{}_{ba} = (\bar{D}^{2\,ab})^{\dagger} \quad ,$$

$$D^2{}_{\alpha\beta} = \frac{1}{2}\,C^{ba}D_{a\alpha}D_{b\beta} = D^2{}_{\beta\alpha} = -\,(\bar{D}^2{}_{\dot{\alpha}\dot{\beta}})^{\dagger} \quad . \qquad (3.11.44)$$

The set of (possibly) reducible projection operators is:

$$\Pi_{0,0} = \Box^{-2}D^4\bar{D}^4 \quad , \quad \Pi_{4,0} = \Box^{-2}\bar{D}^4 D^4 \quad ,$$

$$\Pi_{3,\frac{1}{2}} = \Box^{-2}D_{\underline{\alpha}}\bar{D}^4 D^{3\,\underline{\alpha}} \quad , \quad \Pi_{1,\frac{1}{2}} = -\,\Box^{-2}D^{3\,\underline{\alpha}}\bar{D}^4 D_{\underline{\alpha}} \quad ,$$

$$\Pi_{2,0} = \Box^{-2}C^{ca}C^{bd}D^2{}_{ab}\bar{D}^4 D^2{}_{cd} \quad ,$$

$$\Pi_{2,1} = \Box^{-2} D^{2\,\alpha\beta} \bar{D}^4 D^2{}_{\alpha\beta} \quad . \tag{3.11.45}$$

In writing $\Pi_{2,0}$ and $\Pi_{2,1}$ we have taken $\Pi_2$ defined by (3.11.19) and used (3.11.44) to further reduce it. We can now decompose $N = 2$ superfields.

We start with a complex $N = 2$ scalar superfield $\Psi$. We need not bisect the terms obtained from $\Pi_{1,\frac{1}{2}}$ and $\Pi_{3,\frac{1}{2}}$. Bisecting the rest, we find eight more irreducible projections.

$$\Pi_{0,0\pm}\Psi = \Box^{-2}\,\frac{1}{2}\,(1 \pm \mathbf{K}) D^4 \bar{D}^4 \Psi = \Box^{-2} D^4\,\frac{1}{2}\,(\bar{D}^4\Psi \pm \Box\bar{\Psi}) \quad ,$$

$$\Pi_{4,0\pm}\Psi = \Box^{-2}\,\frac{1}{2}\,(1 \pm \mathbf{K}) \bar{D}^4 D^4 \Psi = \Box^{-2} \bar{D}^4\,\frac{1}{2}\,(D^4\Psi \pm \Box\bar{\Psi}) \quad ,$$

$$\Pi_{2,0\pm}\Psi = \Box^{-2} C^{ca} D^2{}_{ab}\,\frac{1}{2}\,(1 \pm \mathbf{K}) \bar{D}^4 C^{bd} D^2{}_{cd}\Psi = \Box^{-2} C^{ca} C^{bd} D^2{}_{ab} \bar{D}^4 D^2{}_{cd}\,\frac{1}{2}\,(\Psi \pm \bar{\Psi}) \,,$$

$$\Pi_{2,1\pm}\Psi = \Box^{-2} D^{2\alpha\beta}\,\frac{1}{2}\,(1 \pm \mathbf{K}) \bar{D}^4 D^2{}_{\alpha\beta}\Psi = \Box^{-2} D^{2\alpha\beta} \bar{D}^4 D^2{}_{\alpha\beta}\,\frac{1}{2}\,(\Psi \pm \bar{\Psi}) \quad . \tag{3.11.46}$$

We give two more results without details: For the $N = 2$ vector multiplet, described by a real scalar-isovector superfield $V_a{}^b$ we find

$$\Pi_{0,0,1\pm} V_a{}^b = \Box^{-2} D^4 (\bar{D}^4 \pm \Box) V_a{}^b \quad ,$$

$$\Pi_{1,\frac{1}{2},\frac{3}{2}} V_a{}^b = \frac{1}{3!}\,\Box^{-2} C^{db} D^{3c\gamma} \bar{D}^4 C_{e(a} D_{c\gamma} V_{d)}{}^e \quad ,$$

$$\Pi_{1,\frac{1}{2},\frac{1}{2}} V_a{}^b = \frac{1}{3}\,\Box^{-2} C^{db} D^{3c\gamma} \bar{D}^4 D_{e\gamma} C_{c(a} V_{d)}{}^e \quad ,$$

$$\Pi_{2,1,1} V_a{}^b = \Box^{-2} D^{2\,\alpha\beta} \bar{D}^4 D^2{}_{\alpha\beta} V_a{}^b \quad ,$$

$$\Pi_{2,0,2} V_a{}^b = -\,\frac{1}{4!}\,\Box^{-2} C^{bg} C^{ce} C^{fd} D^2{}_{ef} \bar{D}^4 D^2{}_{(cd} V_a{}^h C_{g)h} \quad ,$$

$$\Pi_{2,0,1} V_a{}^b = -\,\frac{1}{4}\,\Box^{-2} C^{cd} C^{be} D^2{}_{d(a|} \bar{D}^4 (D^2{}_{|e)f} V_c{}^f + D^2{}_{cf} V_{|e)}{}^f) \quad ,$$

$$\Pi_{2,0,0} V_a{}^b = \frac{1}{3}\,\Box^{-2} C^{bc} D^2{}_{ac} \bar{D}^4 C^{fe} D^2{}_{de} V_f{}^d \quad , \tag{3.11.47}$$

where the projection operators are labeled by projector number, superspin, superisospin, and $\mathbf{K}$ conjugation $\pm$. Again, to construct real projection operators, the complex conjugate must be added for the $\Pi_0$'s and $\Pi_1$'s ($\Pi_0 \to \Pi_0 + \Pi_4$, $\Pi_1 \to \Pi_1 + \Pi_3$). Finally, for the spinor-isospinor superfield $\Psi^{a\alpha}$ (the unconstrained prepotential of $N = 2$ supergravity) we find

$$\Pi_{0,\frac{1}{2},\frac{1}{2}}\Psi^a{}_\alpha = \Box^{-2} D^4 \bar{D}^4 \Psi^a{}_\alpha \quad ,$$

$$\Pi_{4,\frac{1}{2},\frac{1}{2}}\Psi^a{}_\alpha = \Box^{-2} \bar{D}^4 D^4 \Psi^a{}_\alpha \quad ,$$

$$\Pi_{2,\frac{1}{2},\frac{3}{2}}\Psi^a{}_\alpha = \frac{1}{3!}\Box^{-2} D^2{}_{de}\bar{D}^4 C^{b(d}C^{e|c}D^2{}_{bc}\Psi^{|a)}{}_\alpha \quad ,$$

$$\Pi_{2,\frac{1}{2},\frac{1}{2}}\Psi^a{}_\alpha = \frac{2}{3}\Box^{-2} C^{ad}C^{ec}D^2{}_{de}\bar{D}^4 D^2{}_{cb}\Psi^b{}_\alpha \quad ,$$

$$\Pi_{2,\frac{3}{2},\frac{1}{2}}\Psi^a{}_\alpha = \frac{1}{3!}\Box^{-2} D^{2\beta\gamma}\bar{D}^4 D^2{}_{(\alpha\beta}\Psi^a{}_{\gamma)} \quad ,$$

$$\Pi_{2,\frac{1}{2},\frac{1}{2}}\Psi^a{}_\alpha = \frac{2}{3}\Box^{-2} D^2{}_{\alpha\beta}\bar{D}^4 D^{2\beta\gamma}\Psi^a{}_\gamma \quad ,$$

$$\Pi_{1,1,1\pm}\Psi^a{}_\alpha = \frac{1}{8}\Box^{-2}\bar{D}^{b\dot{\beta}}D^4\partial_\alpha{}^{\dot{\gamma}}(\Box^{-1}\bar{D}^3{}_{(b(\dot{\beta}}\partial^\epsilon{}_{\dot{\gamma})}\Psi^{a)}{}_\epsilon \mp i\bar{D}^{(a}{}_{(\dot{\beta}}\bar{\Psi}_{b)\dot{\gamma})}) \quad ,$$

$$\Pi_{1,1,0\pm}\Psi^a{}_\alpha = \frac{1}{8}\Box^{-2}\bar{D}^{a\dot{\beta}}D^4\partial_\alpha{}^{\dot{\gamma}}(\Box^{-1}\bar{D}^3{}_{b(\dot{\beta}}\partial^\epsilon{}_{\dot{\gamma})}\Psi^b{}_\epsilon \mp i\bar{D}^b{}_{(\dot{\beta}}\bar{\Psi}_{b\dot{\gamma})}) \quad ,$$

$$\Pi_{1,0,1\pm}\Psi^a{}_\alpha = -\frac{1}{8}\Box^{-2}\bar{D}^{b\dot{\beta}}D^4\partial_{\alpha\dot{\beta}}(\Box^{-1}\partial^{\epsilon\dot{\gamma}}\bar{D}^3{}_{(b\dot{\gamma}}\Psi^{a)}{}_\epsilon \mp i\bar{D}^{(a}{}_{\dot{\gamma}}\bar{\Psi}_{b)}{}^{\dot{\gamma}}) \quad ,$$

$$\Pi_{1,0,0\pm}\Psi^a{}_\alpha = -\frac{1}{8}\Box^{-2}\bar{D}^{a\dot{\beta}}D^4\partial_{\alpha\dot{\beta}}(\Box^{-1}\partial^{\epsilon\dot{\gamma}}\bar{D}^3{}_{b\dot{\gamma}}\Psi^b{}_\epsilon \mp i\bar{D}^b{}_{\dot{\gamma}}\bar{\Psi}_b{}^{\dot{\gamma}}) \quad ,$$

$$\Pi_{3,1,1\pm}\Psi^a{}_\alpha = \frac{1}{4}\Box^{-2}D_b{}^\beta\bar{D}^4D^2{}_{\alpha\beta}C^{c(a}(D_{c\gamma}\Psi^{b)\gamma} \pm \bar{D}^{b)}{}_{\dot{\gamma}}\bar{\Psi}_c{}^{\dot{\gamma}}) \quad ,$$

$$\Pi_{3,1,0\pm}\Psi^a{}_\alpha = \frac{1}{4}\Box^{-2}C^{ab}D_b{}^\beta\bar{D}^4D^2{}_{\alpha\beta}(D_{e\gamma}\Psi^{e\gamma} \pm \bar{D}^e{}_{\dot{\gamma}}\bar{\Psi}_e{}^{\dot{\gamma}}) \quad ,$$

$$\Pi_{3,0,1\pm}\Psi^a{}_\alpha = \frac{1}{4}\Box^{-2}C^{bd}D_{b\alpha}\bar{D}^4C^{ac}D^2{}_{dc}(D_{e\gamma}\Psi^{e\gamma} \pm \bar{D}^e{}_{\dot{\gamma}}\bar{\Psi}_e{}^{\dot{\gamma}}) \quad ,$$

$$\Pi_{3,0,0\pm}\Psi^a{}_\alpha = \frac{1}{12}\Box^{-2}C^{ab}D_{b\alpha}\bar{D}^4C^{cd}D^2{}_{ce}(D_{d\gamma}\Psi^{e\gamma} \pm \bar{D}^e{}_{\dot{\gamma}}\bar{\Psi}_d{}^{\dot{\gamma}}) \quad . \qquad (3.11.48a)$$

There are 22 irreducible representations. One simplification is possible: Using (3.9.21) instead of (3.9.25) for just the first term in $1 \pm \mathbf{K}$, we find

$$\Pi_{1,1,1\pm}\Psi^a{}_\alpha = \frac{1}{8}\,\Box^{-2}(-D^{3\,b\beta}\bar{D}^4 D_{(b(\alpha}\Psi^{a)}{}_{\beta)} \mp i\partial_\alpha{}^{\dot\gamma}\bar{D}^{b\dot\beta}D^4\bar{D}^{(a}{}_{(\dot\beta}\bar\Psi_{b)\dot\gamma)}) \quad ,$$

$$\Pi_{1,1,0\pm}\Psi^a{}_\alpha = \frac{1}{8}\,\Box^{-2}(-D^{3\,a\beta}\bar{D}^4 D_{e(\alpha}\Psi^{e}{}_{\beta)} \mp i\partial_\alpha{}^{\dot\gamma}\bar{D}^{a\dot\beta}D^4\bar{D}^{e}{}_{(\dot\beta}\bar\Psi_{e\dot\gamma)}) \quad ,$$

$$\Pi_{1,0,1\pm}\Psi^a{}_\alpha = -\frac{1}{8}\,\Box^{-2}(D^{3\,b}{}_\alpha\bar{D}^4 D_{(b\gamma}\Psi^{a)\gamma} \mp i\partial_{\alpha\dot\beta}\bar{D}^{b\dot\beta}D^4\bar{D}^{(a}{}_{\dot\gamma}\Psi_{b)}{}^{\dot\gamma}) \quad ,$$

$$\Pi_{1,0,0\pm}\Psi^a{}_\alpha = -\frac{1}{8}\,\Box^{-2}(D^{3\,a}{}_\alpha\bar{D}^4 D_{e\gamma}\Psi^{e\gamma} \mp i\partial_{\alpha\dot\beta}\bar{D}^{a\dot\beta}D^4\bar{D}^{e}{}_{\dot\gamma}\Psi_{e}{}^{\dot\gamma}) \quad . \tag{3.11.48b}$$

### b.4. N=4

We begin by defining a set of irreducible $D$-operators:

$$D^2{}_{\underline{\alpha\beta}} = C_{\beta\alpha}D^2{}_{ab} + D^2{}_{[ab]\alpha\beta} \quad ,$$

$$D^3{}_{\underline{\alpha\beta\gamma}} = C_{dcba}D^{3\,d}{}_{\alpha\beta\gamma} + (C_{\alpha\beta}D^3{}_{[ac]b\gamma} - C_{\alpha\gamma}D^3{}_{[ab]c\beta})$$

$$D^4{}_{\underline{\alpha\beta\gamma\delta}} = C_{dcba}D^4{}_{\alpha\beta\gamma\delta} + \frac{1}{2}\,(C_{\alpha\delta}C_{\beta\gamma}C_{cdef}D^4{}_{[ab]}{}^{[ef]} - C_{\alpha\beta}C_{\delta\gamma}C_{bcef}D^4{}_{[ad]}{}^{[ef]})$$

$$+ \,(C_{\alpha\beta}C_{eacd}D^4{}_b{}^e{}_{\gamma\delta} - C_{\alpha\gamma}C_{eabd}D^4{}_c{}^e{}_{\beta\delta} + C_{\alpha\delta}C_{eabc}D^4{}_d{}^e{}_{\beta\gamma})$$

$$D^{5\,\underline{\alpha\beta\gamma}} = C^{dcba}D^5{}_d{}^{\alpha\beta\gamma} + (C^{\alpha\beta}D^{5[ac]b\gamma} - C^{\alpha\gamma}D^{5[ab]c\beta})$$

$$D^{6\,\underline{\alpha\beta}} = C^{\beta\alpha}D^{6\,ab} + D^{6\,[ab]\alpha\beta} \quad . \tag{3.11.49}$$

They satisfy the following algebraic relations

$$D^2{}_{ab} = D^2{}_{ba} \quad , \qquad D^{6ab} = D^{6ba} \quad ,$$

$$D^4{}_a{}^a{}_{\alpha\beta} = D^4{}_{[ab]}{}^{[cb]} = C^{abcd}D^3{}_{[ab]c\alpha} = C_{abcd}D^{5\,[ab]c}{}_\alpha = 0 \quad . \tag{3.11.50}$$

All $SL(2,C)$ indices on the $D^n$'s are totally symmetric. We also have

$$D^4{}_{\underline{\alpha}_1\cdots\underline{\alpha}_4} = \frac{1}{4!}\,C_{\underline{\alpha}_1\cdots\underline{\alpha}_8}D^{4\underline{\alpha}_5\cdots\underline{\alpha}_8} \quad ,$$

$$D^{4\,\underline{\alpha}_1\cdots\underline{\alpha}_4} = \frac{1}{4!}\,C^{\underline{\alpha}_1\cdots\underline{\alpha}_8}D^4{}_{\underline{\alpha}_5\cdots\underline{\alpha}_8} \quad , \tag{3.11.51}$$

and these imply

$$D^{4\,\underline{\alpha\beta\gamma\delta}} = C^{dcba}D^{4\,\alpha\beta\gamma\delta} + \frac{1}{2}\,(C^{\alpha\delta}C^{\beta\gamma}C^{cdef}D^4{}_{[ef]}{}^{[ab]} - C^{\alpha\beta}C^{\delta\gamma}C^{bcef}D^4{}_{[ef]}{}^{[ad]})$$

$$+ \,(C^{\alpha\beta}C^{eacd}D^4{}_e{}^{b\gamma\delta} - C^{\alpha\gamma}C^{eabd}D^4{}_e{}^{c\beta\delta} + C^{\alpha\delta}C^{eabc}D^4{}_e{}^{d\beta\gamma}) \quad , \tag{3.11.52}$$

as can be verified by substituting explicit values for the indices.

We consider now a complex scalar $N = 4$ superfield $\Psi$ and find first

$$\Pi_{0,0,1} = \Box^{-4}D^8\bar{D}^8 \quad , \qquad \Pi_{8,0,1} = \Box^{-4}\bar{D}^8D^8 \quad ,$$

$$\Pi_{1,\frac{1}{2},4} = \Box^{-4}\bar{D}_{\underline{\dot\alpha}}D^8\bar{D}^{7\,\underline{\dot\alpha}} \quad , \qquad \Pi_{7,\frac{1}{2},\bar{4}} = \Box^{-4}D_{\underline{\alpha}}\bar{D}^8D^{7\,\underline{\alpha}}$$

$$\Pi_{2,0,10} = \Box^{-4}\bar{D}^{2\,ab}D^8\bar{D}^6{}_{ab} \quad ,$$

$$\Pi_{2,1,6} = \frac{1}{2}\,\Box^{-4}\bar{D}^{2\,[ab]}{}_{\dot\alpha\dot\beta}D^8\bar{D}^6{}_{[ab]}{}^{\dot\alpha\dot\beta} \quad ,$$

$$\Pi_{3,\frac{3}{2},\bar{4}} = \Box^{-4}\bar{D}^3{}_{a\dot\alpha\dot\beta\dot\gamma}D^8\bar{D}^{5\,a\dot\alpha\dot\beta\dot\gamma} \quad ,$$

$$\Pi_{3,\frac{1}{2},20} = \frac{1}{3!}\,\Box^{-4}\bar{D}^{3\,[ab]c}{}_{\dot\alpha}D^8\bar{D}^5{}_{[ab]c}{}^{\dot\alpha} \quad ,$$

$$\Pi_{4,2,1} = \Box^{-4}\bar{D}^4{}_{\dot\alpha\dot\beta\dot\gamma\dot\delta}D^8\bar{D}^{4\dot\alpha\dot\beta\dot\gamma\dot\delta} \quad ,$$

$$\Pi_{4,1,15} = \Box^{-4}\bar{D}^4{}_a{}^b{}_{\dot\alpha\dot\beta}D^8\bar{D}^4{}_b{}^{a\dot\alpha\dot\beta} \quad ,$$

$$\Pi_{4,0,20'} = \Box^{-4}\bar{D}^4{}_{[cd]}{}^{[ab]}D^8\bar{D}^{4[cd]}{}_{[ab]} \quad ,$$

$$\Pi_{5,\frac{3}{2},4} = \Box^{-4}D^{3a}{}_{\alpha\beta\gamma}\bar{D}^8D^5{}_a{}^{\alpha\beta\gamma} \quad ,$$

$$\Pi_{5,\frac{1}{2},\overline{20}} = \frac{1}{3!}\,\Box^{-4}D^3{}_{[ab]c\alpha}\bar{D}^8D^{5\,[ab]c\alpha} \quad ,$$

$$\Pi_{6,0,\overline{10}} = \Box^{-4}D^2{}_{ab}\bar{D}^8D^{6\,ab} \quad ,$$

$$\Pi_{6,1,6} = \frac{1}{2}\Box^{-4} D^2{}_{[ab]\alpha\beta} \bar{D}^8 D^{6\,[ab]\alpha\beta} \quad , \tag{3.11.53}$$

where the superisospin quantum number here refers to the dimensionality of the $SU(4)$ representation. The only projectors that need bisection are the real representations of $SU(4)$: the 1, 6, 15, and 20'. We find:

$$\Pi_{0,0,1\pm}\Psi = \Box^{-4} D^8 \frac{1}{2}(\bar{D}^8\Psi \pm \Box^2\bar{\Psi}) \quad ,$$

$$\Pi_{8,0,1\pm}\Psi = \Box^{-4} \bar{D}^8 \frac{1}{2}(D^8\Psi \pm \Box^2\bar{\Psi}) \quad ,$$

$$\Pi_{4,2,1\pm}\Psi = \Box^{-4} \bar{D}^4{}_{\dot{\alpha}\dot{\beta}\dot{\gamma}\dot{\delta}} D^8 \bar{D}^{4\,\dot{\alpha}\dot{\beta}\dot{\gamma}\dot{\delta}} \frac{1}{2}(\Psi \pm \bar{\Psi}) \quad ,$$

$$\Pi_{2,1,6\pm}\Psi = \frac{1}{2}\Box^{-4} \bar{D}^{2\,[ab]}{}_{\dot{\alpha}\dot{\beta}} D^8 \frac{1}{2}(\bar{D}^6{}_{[ab]}{}^{\dot{\alpha}\dot{\beta}}\Psi \pm \frac{1}{2} C_{abcd}\Box\bar{D}^{2\,[cd]\dot{\alpha}\dot{\beta}}\bar{\Psi}) \quad ,$$

$$\Pi_{6,1,6\pm}\Psi = \frac{1}{2}\Box^{-4} D^2{}_{[ab]\alpha\beta} \bar{D}^8 \frac{1}{2}(D^{6\,[ab]\alpha\beta}\Psi \pm \frac{1}{2} C^{abcd}\Box D^2{}_{[cd]}{}^{\alpha\beta}\bar{\Psi}) \quad ,$$

$$\Pi_{4,1,15\pm}\Psi = \Box^{-4} \bar{D}^4{}_a{}^b{}_{\dot{\alpha}\dot{\beta}} D^8{}_b{}^{a\dot{\alpha}\dot{\beta}} \frac{1}{2}(\Psi \pm \bar{\Psi}) \quad ,$$

$$\Pi_{4,0,20'\pm}\Psi = \Box^{-4} \bar{D}^4{}_{[cd]}{}^{[ab]} D^8 \bar{D}^4{}_{[ab]}{}^{[cd]} \frac{1}{2}(\Psi \pm \bar{\Psi}) \quad , \tag{3.11.54}$$

and a total of 22 irreducible representations. The 6 is a real representation only if we use a "duality" transformation in the rest-frame conjugation (3.11.4): $\tilde{X}_{[ab]} = \frac{1}{2} C_{abcd}\bar{X}^{[cd]}$. This occurs for rank $\frac{1}{2}N$ antisymmetric tensors of $SU(N)$ when $N$ is a multiple of 4.

## 3.12. On-shell representations and superfields

In section 3.9 we discussed irreducible representations of off-shell supersymmetry in terms of superfields; here we give the corresponding analysis of on-shell representations. We first discuss the description of on-shell physical components by means of field strengths. We then describe a (non-Lorentz-covariant) subgroup of supersymmetry, which we call *on-shell* supersymmetry, under which (reducible or irreducible) off-shell representations of ordinary (or off-shell) supersymmetry decompose into multiplets that contain only one of the three types of components discussed in sec. 3.9. By considering representations of this smaller group in terms of *on-shell* superfields (defined in a superspace which is a non-Lorentz-covariant subspace of the original superspace), we can concentrate on just the physical components, and thus on the physical content of the theory.

### a. Field strengths

For simplicity we restrict ourselves to massless fields. (Massive fields may be treated similarly.) It is more convenient to describe the physical components in terms of field strengths rather than gauge fields: Every irreducible representation of the Lorentz group, when considered as a *field strength,* satisfies certain unique constraints (Bianchi identities) plus field equations, and corresponds to a unique nontrivial irreducible representation of the Poincaré group (a zero mass single helicity state). On the other hand, a given irreducible representation of the Lorentz group, when considered as a gauge field, may correspond to several representations of the Poincaré group, depending on the form of its gauge transformation.

Specifically, any field strength $\psi_{\alpha_1\ldots\alpha_{2A}\dot{\alpha}_1\ldots\dot{\alpha}_{2B}}$, *totally symmetric* in its $2A$ undotted indices and in its $2B$ dotted indices, has mass dimension $A+B+1$ and satisfies the constraints plus field equations

$$\partial^{\alpha_1\dot{\beta}}\psi_{\alpha_1\ldots\alpha_{2A}\dot{\alpha}_1\ldots\dot{\alpha}_{2B}} = \partial^{\beta\dot{\alpha}_1}\psi_{\alpha_1\ldots\alpha_{2A}\dot{\alpha}_1\ldots\dot{\alpha}_{2B}} = 0 \quad , \tag{3.12.1a}$$

$$\Box\psi_{\alpha_1\ldots\alpha_{2A}\dot{\alpha}_1\ldots\dot{\alpha}_{2B}} = 0 \quad . \tag{3.12.1b}$$

The Klein-Gordon equation (3.12.1b) projects onto the mass zero representation, while (3.12.1a) project onto the helicity $A-B$ state. The Klein-Gordon equation is a consequence of the others except when $A=B=0$. To solve these equations we go to momentum space: Then (3.12.1b) sets $p^2$ to 0 (i.e., $\psi \sim \delta(p^2)$), and we may choose the Lorentz frame $p^{+\dot{+}} = p^{+\dot{-}} = 0$, $p^{-\dot{-}} \neq 0$. In this frame (3.12.1a) states that only one component of $\psi$ is nonvanishing: $\psi_{+\cdots+\dot{+}\cdots\dot{+}}$. Since each "+" index has a helicity $\frac{1}{2}$ and each "$\dot{+}$" has helicity $-\frac{1}{2}$, the total helicity of $\psi$ is $A - B$, and of its complex conjugate $B - A$. In the cases where $A = B$ we may choose $\psi$ real (since it has an equal number of dotted and undotted indices), so that it describes a single state of helicity 0.

The most familiar examples of field strengths have $B = 0$: $A = 0$ is the usual description of a scalar, $A = \frac{1}{2}$ a Weyl spinor, $A = 1$ describes a vector (e.g., the photon), $A = \frac{3}{2}$ the gravitino, and the case $A = 2$ is the Weyl tensor of the graviton. Since we are describing only the on-shell components, we do not see field strengths that vanish on shell: e.g., in gravity the Ricci tensor vanishes by the equations of motion, leaving the Weyl tensor as the only nonvanishing part of the Riemann curvature tensor. (This happens because, although these theories are irreducible on shell, they may be reducible off shell; i.e., the field equations may eliminate Poincaré representations not eliminated by (off-shell) constraints.) The most familiar example of $A, B \neq 0$ is the field strength of the second-rank antisymmetric tensor gauge field: $(A, B) = (\frac{1}{2}, \frac{1}{2})$ (see sec. 4.4.c). Some less familiar examples are the spin-$\frac{3}{2}$ representation of spin $\frac{1}{2}$, $(A, B) = (1, \frac{1}{2})$, the spin-2 representation of spin 0, $(A, B) = (1, 1)$, and the higher-derivative representation of spin 1, $(A, B) = (\frac{3}{2}, \frac{1}{2})$. Generally, the off-shell theory contains maximum spin indicated by the indices of $\psi$: $A + B$.

Although the analogous analysis for supersymmetric multiplets is not yet completely understood, the on-shell content of superfields can be analyzed by component projection. In particular, a complete superfield analysis has been made of on-shell multiplets that contain only component field strengths of type $(A, 0)$. This is sufficient to describe all on-shell multiplets: Theories with field strengths $(A, B)$ describe the same on-shell helicity states as theories with $(A - B, 0)$, and are physically equivalent. They only differ by their auxiliary field content. Furthermore, type $(A, 0)$ theories allow the most general interactions, whereas theories with $B \neq 0$ fields are generally more restricted in the form of their self-interactions and interactions with external fields. (In some cases, they cannot even couple to gravity.)

Before discussing the general case, we consider a specific example in detail. The multiplet of $N=2$ supergravity (see sec. 3.3.a.1) with helicities $2, \frac{3}{2}, 1$, is described by *component* field strengths $\psi_{\alpha\beta\gamma\delta}(x)$, ${\psi^a}_{\alpha\beta\gamma}(x)$, ${\psi^{ab}}_{\alpha\beta}(x)$. They have dimension $3, \frac{5}{2}, 2$ respectively, and satisfy the *component* Bianchi identities and field equations (3.12.1). We introduce a *superfield* strength ${F_{(0)}}^{ab}{}_{\alpha\beta}(x,\theta)$ that contains the lowest dimension component field strength at the $\theta = 0$ level:

$$ {F_{(0)}}^{ab}{}_{\alpha\beta}(x,\theta)| = C^{ab} F_{(0)\alpha\beta}(x,\theta)| = {\psi^{ab}}_{\alpha\beta}(x) \quad . \tag{3.12.2} $$

We require that *all* the higher components of $F_{(0)}$ are component field strengths of the theory (or their spacetime derivatives; superfield strengths contain no gauge components and, on shell, no auxiliary fields). Thus, for example, we must have $D_{a(\gamma}{F_{(0)}}^{ab}{}_{\alpha\beta)}| \neq 0$, whereas $C^{c(d} D_{c\gamma} {F_{(0)}}^{ab)}{}_{\alpha\beta}| = \bar{D}_{\underline{\dot{\gamma}}} {F_{(0)}}^{ab}{}_{\alpha\beta}| = 0$. Since a superfield that vanishes at $\theta = 0$ vanishes identically (as follows from the supersymmetry transformations, e.g., (3.6.5-6)) $C^{c(d} D_{c\gamma} {F_{(0)}}^{ab)}{}_{\alpha\beta} = \bar{D}_{\underline{\dot{\gamma}}} {F_{(0)}}^{ab}{}_{\alpha\beta} = 0$. From these arguments it follows that the superfield equations and Bianchi identities are:

$$ \bar{D}_{\underline{\dot{\beta}}} {F_{(0)}}^{ab}{}_{\alpha\beta} = 0 \quad , $$

$$ D_{\underline{\gamma}} {F_{(0)}}^{ab}{}_{\alpha\beta} = \delta_c{}^{[a} \, {F_{(1)}}^{b]}{}_{\alpha\beta\gamma} \quad , $$

$$ D_{\underline{\delta}} D_{\underline{\gamma}} {F_{(0)}}^{ab}{}_{\alpha\beta} = \delta_c{}^{[a} \delta_d{}^{b]} \, F_{(2)\alpha\beta\gamma\delta} \quad , $$

$$ D_{\underline{\epsilon}} D_{\underline{\delta}} D_{\underline{\gamma}} {F_{(0)}}^{ab}{}_{\alpha\beta} = 0 \quad ; \tag{3.12.3} $$

where $F_{(1)}(x,\theta)$ and $F_{(2)}(x,\theta)$ are superfields containing the field strengths ${\psi^b}_{\alpha\beta\gamma}(x)$ and $\psi_{\alpha\beta\gamma\delta}(x)$ at the $\theta = 0$ level. By applying powers of $D_{\underline{\alpha}}$ and $\bar{D}_{\underline{\dot{\alpha}}}$ to these equations we recover the component field equations and Bianchi identities, and verify that ${F_{(0)}}^{ab}{}_{\alpha\beta}$ contains no extra components.

Generalization to the rest of the supermultiplets in Table 3.12.1 is straightforward: We introduce a set of superfields which at $\theta = 0$ are the component field strengths (as in (3.12.1)) that describe the states appearing in Table 3.3.1: These superfields satisfy a set of Bianchi identities plus field equations (as in the example (3.12.3)) that are uniquely determined by dimensional analysis and Lorentz $\otimes SU(N)$ covariance.

| helicity | scalar multiplet | super-Yang-Mills | supergravity |
|---|---|---|---|
| +2 | | | $F_{\alpha\beta\gamma\delta}$ |
| +3/2 | | | $F^{a}{}_{\alpha\beta\gamma}$ |
| +1 | | $F_{\alpha\beta}$ | $F^{ab}{}_{\alpha\beta}$ |
| +1/2 | $F_{\alpha}$ | $F^{a}{}_{\alpha}$ | $F^{abc}{}_{\alpha}$ |
| 0 | $F^{a}$ | $F^{ab}$ | $F^{abcd}$ |
| -1/2 | $F^{ab}{}_{\dot\alpha}$ | $F^{abc}{}_{\dot\alpha}$ | $F^{abcde}{}_{\dot\alpha}$ |
| -1 | | $F^{abcd}{}_{\dot\alpha\dot\beta}$ | $F^{abcdef}{}_{\dot\alpha\dot\beta}$ |
| -3/2 | | | $F^{abcdefg}{}_{\dot\alpha\dot\beta\dot\gamma}$ |
| -2 | | | $F^{abcdefgh}{}_{\dot\alpha\dot\beta\dot\gamma\dot\delta}$ |

*Table 3.12.1. Field strengths in theories of physical interest*

We now consider arbitrary supermultiplets of type $(A, 0)$. There are two cases: For an on-shell multiplet with lowest spin $s = 0$, the superfield strength has the form ${F_{(0)}}^{a_1 \cdots a_m}$, $\frac{N}{2} \le m \le N$, and is *totally antisymmetric* in its $m$ *$SU(N)$ isospin* indices. If the lowest spin $s > 0$, the superfield strength has the form $F_{(0)\alpha_1 \cdots \alpha_{2s}}$ and is *totally symmetric* in its $2s$ *Weyl spinor* indices. To treat both cases together, for $s > 0$ we write ${F_{(0)}}^{a_1 \cdots a_N}{}_{\alpha_1 \cdots \alpha_{2s}} = C^{a_1 \cdots a_N} F_{(0)\alpha_1 \cdots \alpha_{2s}}$. Then the superfield strength has the form ${F_{(0)}}^{a_1 \cdots a_m}{}_{\alpha_1 \cdots \alpha_{2s}}$ and is totally antisymmetric in its isospinor indices and totally symmetric in its spinor indices. It has (mass) dimension $s + 1$.

This superfield contains all the on-shell component field strengths; in particular, at $\theta = 0$, it contains the field strength of lowest dimension (and therefore of lowest spin). For $s = 0$, the superfield strength describes helicities $\frac{m-N}{2}, \frac{m-N+1}{2}, \cdots, \frac{m}{2}$, and its hermitian conjugate describes helicities $-\frac{m}{2}, \frac{-m+1}{2}, \cdots, \frac{N-m}{2}$. Since $m \le N$, some helicities appear in both $F_{(0)}$ and $\bar{F}_{(0)}$. For $s \ge 0$, the superfield strength describes helicities $s\,,\ s + \frac{1}{2}, \cdots, s + \frac{N}{2}$, and its hermitian conjugate describes helicities $-(s + \frac{N}{2}), \cdots, -s$. In this case, positive helicities appear only in $F_{(0)}$ and negative helicities only in $\bar{F}_{(0)}$. For both cases the superfield strength together with its conjugate describe (perhaps multiple) helicities $\pm s\,, \pm(s + \frac{1}{2}), \cdots, \pm\,(s + \frac{m}{2})$.

The higher-spin component field strengths occur at $\theta = 0$ in the superfields $F_{(n)}$ obtained by applying $n$ $D$'s (for $n > 0$) or $-n$ $\bar{D}$'s (for $n < 0$) to $F_{(0)}$. They are totally antisymmetric in their $m - n$ isospinor indices and totally symmetric in their $2s + n$ spin indices, and satisfy the following Bianchi identities and field equations:

$$n > 0 : \; D^n{}_{\underline{\beta}_n \ldots \underline{\beta}_1} F_{(0)}{}^{a_1 \ldots a_m}{}_{\alpha_1 \ldots \alpha_{2s}} = \frac{1}{(m-n)!} \delta_{b_1}{}^{[a_1} \ldots \delta_{b_n}{}^{a_n} F_{(n)}{}^{a_{n+1} \ldots a_m]}{}_{\alpha_1 \ldots \alpha_{2s} \beta_1 \ldots \beta_n} , \qquad (3.12.4a)$$

$$n < 0 : \; \bar{D}^n{}_{\underline{\dot{\beta}}_{-n} \ldots \underline{\dot{\beta}}_1} F_{(0)}{}^{a_1 \ldots a_m} = F_{(n)}{}^{a_1 \ldots a_m b_1 \ldots b_{-n}}{}_{\dot{\beta}_1 \ldots \dot{\beta}_{-n}} \quad , \qquad (3.12.4b)$$

with $m - N \leq n \leq m$; in particular, for $s > 0$, $\bar{D} F_{(0)} = 0$. These equations follow from the requirement that *all* components of the on-shell *superfield* strength (defined by projection) are on-shell *component* field strengths. The $\theta = 0$ component of the superfield $F_{(0)}$ is the *lowest dimension* component field strength; this determines the dimension and index structure of the superfield. The higher components of the superfield are either higher dimension component field strengths, or vanish; this determines the superfield equations and Bianchi identities. Note that the difference between maximum and minimum helicities in the $F_{(n)}$ is always $\frac{1}{2} N$.

In the special case $s = 0$, $m$ even, and $m = \frac{1}{2} N$ we have in addition to (3.12.4a,b) the self-conjugacy relation

$$F_{(0)}{}^{a_1 \ldots a_{\frac{1}{2}N}} = \frac{1}{(\frac{1}{2} N)!} C^{a_1 \ldots a_N} \bar{F}_{(0) a_{\frac{1}{2}N} \ldots a_N} \quad . \qquad (3.12.4c)$$

For this case only, $F_{(+n)}$ is related to $\overline{F_{(-n)}}$; this relation follows from (3.12.4c) for $n = 0$, and from spinor derivatives of (3.12.4c), using (3.12.4a,b), for $n > 0$. Eqs. (3.12.4a,b) are $U(N)$ covariant, whereas, because the antisymmetric tensor $C^{a_1 \ldots a_N}$ is not phase invariant, (3.12.4c) is only $SU(N)$ covariant; thus, *self-conjugate multiplets have a smaller symmetry.*

## b. Light-cone formalism

When studying only the on-shell properties of a free, massless theory it is simpler to represent the fields in a form where just the physical components appear. As described in sec. 3.9, we use a light-cone formalism, in which an irreducible representation of the Poincaré group is given by a single component (complex except for zero helicity). For superfields we make a light-cone decomposition of $\theta$ as well as $x$. We use the notation (see (3.1.1)):

$$(x^{+\dot{+}}, x^{+\dot{-}}, x^{-\dot{+}}, x^{-\dot{-}}) \equiv (x^{+}, x^{T}, \overline{x}^{T}, -x^{-}) \quad , \quad (\theta^{a+}, \theta^{a-}) \equiv (\theta^{a}, \zeta^{a}) \ , \tag{3.12.5a}$$

$$(\partial_{+\dot{+}}, \partial_{+\dot{-}}, \partial_{-\dot{+}}, \partial_{-\dot{-}}) \equiv (\partial_{+}, \partial_{T}, \overline{\partial}_{T}, -\partial_{-}) \quad , \quad (\partial_{a+}, \partial_{a-}) \equiv (\partial_{a}, \delta_{a}) \ . \tag{3.12.5b}$$

(The *spinor* derivative $\partial_a$ should not be confused with the *spacetime* derivative $\partial_{\underline{a}}$). Under the transverse $SO(2)$ part of the Lorentz group the coordinates transform as $x^{\pm\prime} = x^{\pm}$, $x^{T\prime} = e^{2i\eta} x^{T}$, $\theta^{a\prime} = e^{i\eta}\theta^{a}$, $\zeta^{a\prime} = e^{-i\eta}\zeta^{a}$, and the corresponding derivatives transform in the opposite way.

In sec. 3.11 we described the decomposition of general superfields in terms of chiral field strengths, which are irreducible gauge invariant representations of supersymmetry off shell. Although they contain no gauge components, they may contain auxiliary fields that only drop out on shell. To analyze the decomposition of an irreducible *off-shell* representation of supersymmetry into irreducible *on-shell* representations, we perform a nonlocal, nonlinear, nonunitary similarity transformation on the field strengths $\Phi$ and all operators $X$:

$$\Phi' = e^{iH}\Phi \quad , \quad X' = e^{iH} X e^{-iH} \quad ; \quad H = (\zeta^{a} i\partial_{a}) \frac{\overline{\partial}_{T}}{\partial_{+}} \ . \tag{3.12.6a}$$

We use this transformation because it makes some of the supersymmetry generators *independent* of $\zeta^a$ in the chiral representation. Dropping primes, we have

$$Q_{a+} = i\partial_{a} \quad , \quad \overline{Q}^{a}{}_{\dot{+}} = i(\overline{\partial}^{a} - \theta^{a} i\partial_{+}) \quad ,$$

$$Q_{a-} = i(\delta_{a} + \partial_{a}\frac{\overline{\partial}_{T}}{\partial_{+}}) \quad , \quad \overline{Q}^{a}{}_{\dot{-}} = i(\overline{\delta}^{a} - \theta^{a} i\partial_{T} + i\zeta^{a}\frac{\Box}{\partial_{+}}) \ . \tag{3.12.6b}$$

Thus $Q_{+}$ and $\overline{Q}_{\dot{+}}$ are local and depend only on $\partial_a$, $\theta^a$, and $\partial_+$, but *not* on $\delta_a$, $\zeta^a$, $\partial_-$, and $\partial_T$, whereas $Q_-$ and $\overline{Q}_{\dot{-}}$ are nonlocal and depend on all $\partial_{\underline{\alpha}}$ and $\partial_{\underline{a}}$. We expand the transformed superfield strength $\Phi$ in powers of $\zeta^a$ (the external indices of $\Phi$ are suppressed):

$$\Phi(x^{\alpha\dot{\alpha}}, \theta^{a}, \zeta^{a}) = \sum_{m=0}^{N} \frac{1}{m!} \zeta^{m\, a_1 \ldots a_m} \phi_{(m) a_m \ldots a_1}(x^{\alpha\dot{\alpha}}, \theta^{a}) \quad , \tag{3.12.7}$$

where the $m^{th}$ power $\zeta^m$, and thus $\phi_{(m)}$, is totally antisymmetric in isospinor indices.

Each $\phi_{(m)}$ is a representation of a subgroup of supersymmetry that we call "on-shell" supersymmetry, and that includes the $Q_+$ transformations, the transverse $SO(2)$ part of Lorentz transformations (and a corresponding conformal boost), $SU(N)$ (or $U(N)$), and all four translations (as well as scale transformations in the massless case). Although each $\phi_{(m)}$ is a realization of the full supersymmetry group off-shell as well as on-shell, on-shell supersymmetry is the maximal subgroup that can be realized locally (and in the interacting case, linearly).

The remaining generators of the full supersymmetry group (including the other Lorentz generators, that mix $\theta^a$ with $\zeta^a$) mix the various $\phi_{(m)}$'s. In particular, $Q_-$ and $\bar{Q}_{\dot{-}}$ allow us to distinguish physical and auxiliary on-shell superfields: Auxiliary fields vanish on-shell, and hence must have transformations proportional to field equations. We go to a Lorentz frame where $\partial_T = 0$. In this frame, $Q_{a-} = i\delta_a$ and $\bar{Q}_{\dot{-}} = i(\bar{\delta}^a + i\zeta^a \frac{\Box}{\partial_+})$. The $Q_-$ and $\bar{Q}_{\dot{-}}$ supersymmetry variation of the highest $\zeta_a$ component of $\Phi$,

$$\phi'_{(N)} - \phi_{(N)} \sim i\,\frac{\Box}{\partial_+}\,\phi_{(N-1)} \tag{3.12.8a}$$

is proportional to $\Box$, which identifies it as an auxiliary field. Setting $\phi_{(N)}$ to zero on-shell, we iterate the argument: the variation of the next component of $\Phi$,

$$\phi'_{(N-1)} - \phi_{(N-1)} \sim i\,\frac{\Box}{\partial_+}\,\phi_{(N-2)} \tag{3.12.8b}$$

is again proportional to $\Box$, etc. We find that only $\phi_{(0)}$ has a variation *not* proportional to $\Box$. This identifies it as the physical on-shell superfield.

Thus, on-shell, $\Phi$ reduces to $\phi_{(0)}$. (All other $\phi_{(m)}$ vanish.) In the Lorentz frame chosen above ($\bar{\partial}_T = 0$), $Q_-$ and $\bar{Q}_{\dot{-}}$ vanish when acting on $\phi_{(0)}$, and thus this superfield is a local representation of the *full* supersymmetry algebra on shell, namely, it describes the multiplet of physical polarizations. By expanding actions in $\zeta$, it can be shown that $\phi_{(0)}$ represents the multiplet of physical components while the other $\phi_{(m)}$'s represent multiplets of auxiliary components.

We can also define (chiral representation) spinor derivatives $D_a$ , $\bar{D}^a$ that are covariant under the on-shell supersymmetry:

$$D_a = \partial_a + \bar{\theta}_a i\partial_+ \quad , \quad \bar{D}^a = \bar{\partial}^a \quad ; \quad \{D_a , \bar{D}^b\} = \delta_a{}^b i\partial_+ \quad . \tag{3.12.9}$$

When a bisection condition is imposed on the chiral field strength $\Phi$ (i.e., $\Phi$ is real, as discussed in sec. 3.11), we can express the condition in terms of the on-shell superfields. For superspin $s = 0$, the condition

$$\bar{D}^{2N}\bar{\Phi} = \Box^{\frac{1}{2}N}\Phi \tag{3.12.10}$$

becomes

$$\bar{D}^N \bar{\phi}_{(m)}{}^{a_1 \ldots a_m} = i^N \Box^{m-\frac{1}{2}N} (i\partial_+)^{N-m} \frac{1}{(N-m)!} C^{a_N \ldots a_1} \phi_{(N-m)\, a_N \ldots a_{m+1}} \tag{3.12.11}$$

(where $D^N \equiv \frac{1}{N!} C^{a_N \ldots a_1} D^N{}_{a_1 \ldots a_N}$) and similarly for superspin $s > 0$. In general, an on-shell representation can be reduced by a reality condition of the form $\bar{D}^N \bar{\phi} \sim (i\partial_+)^{\frac{1}{2}N}\phi$ when the "middle" ($\theta^{\frac{1}{2}N}$) component of $\phi$ has helicity 0 (i.e., is invariant under transverse $SO(2)$ Lorentz transformations ). (Compare the discussion of reality of off-shell representations in sec. 3.11.)

Putting together the results of sec. 3.11 and this section, we have the following reductions: general superfields ($4N$ $\theta$'s; physical + auxiliary + gauge) $\rightarrow$ chiral field strengths ($2N$ $\theta$'s; physical + auxiliary = irreducible off-shell representations) $\rightarrow$ chiral on-shell superfields ($N$ $\theta$'s; physical = irreducible on-shell representations). All three types of superfields can satisfy reality conditions; therefore, the smallest type of each has $2^{4N}$, $2^{2N}$, and $2^N$ components, respectively (when the reality condition is allowed), and is a "real" scalar superfield. All other representations are (real or complex) superfields with (Lorentz or internal) indices, and thus have an integral multiple of this number of components. These counting arguments for off-shell and on-shell components can also be obtained by the usual operator arguments (off-shell, the counting is the same as for on-shell massive theories, since $p^2 \neq 0$), but superfields allow an explicit construction, and are thus more useful for applications.

Similar arguments apply to higher dimensions: We can use the same numbers there (but taking into account the difference in external indices), if we understand "$4N$" to mean the number of anticommuting coordinates in the higher dimensional superspace.

For simple supersymmetry in $D < 4$, because chirality cannot be defined, the counting of states is different.

## 3.13. Off-shell field strengths and prepotentials

We have shown how superfields can be reduced to irreducible off-shell representations (sec. 3.11), which can be reduced further to on-shell superfield strengths (sec. 3.12). To find a superfield description of a given multiplet of physical states, we need to reverse the procedure: Starting with an *on*-shell superfield strength $F_{(0)}$ that describes the multiplet, we need to find the *off*-shell superfield strength $W$ that reduces to $F_{(0)}$ on shell, and then find a superfield *prepotential* $\Psi$ in terms of which $W$ can be expressed. There is no unambiguous way to do this: The same $F_{(0)}$ is described by different $W$'s, and the same $W$ is described by different $\Psi$'s. However, for a class of theories that includes many of the models that are understood, we impose additional requirements to reduce the ambiguity and find a unique chiral field strength and a family of prepotentials for a given multiplet.

The multiplets we consider have on-shell superfield strengths of Lorentz representation type $(A,0)$ (superspin $s=A$) and are *isoscalars:* $F_{(0)\alpha_1\cdots\alpha_{2s}}$. From (3.12.4b), this implies that the $F_{(0)}$'s are *chiral* and therefore can be generalized to off-shell irreducible (up to bisection) field strengths $W_{\alpha_1\cdots\alpha_{2s}}$, $\bar{D}_{\underline{\dot\beta}}W_{\alpha_1\cdots\alpha_{2s}}=0$. Physically, the $W$'s correspond to field strengths of conformally invariant models. (They transform in the same way as $C^{a_1\cdots a_N}$: as $SU(N)$ scalars, but not $U(N)$ scalars). In the physical models where these superfields arise, the chirality and bisection conditions on $W$ are linearized Bianchi identities. We can use the projection operator analysis of sec. 3.11 to solve the identities by expressing the $W$'s in terms of appropriate prepotentials.

When there is no bisection ($s+\frac{1}{2}N$ not an integer), the $W$'s are general chiral superfields: $W_{\alpha_1\cdots\alpha_{2s}}=\bar{D}^{2N}\tilde{\Psi}_{\alpha_1\cdots\alpha_{2s}}$. The $\tilde{\Psi}_{\alpha_1\cdots\alpha_{2s}}$'s may be expressed in terms of more fundamental superfields. An interesting class of prepotentials are those that contain the lowest superspins: In that case, the $W$'s have the form

$$N\le 2s-1:\ W_{\alpha_1\cdots\alpha_{2s}}=\frac{1}{(2s)!}\bar{D}^{2N}D^N{}_{(\alpha_1\cdots\alpha_N}\partial_{\alpha_{N+1}}{}^{\dot\beta_1}\cdots\partial_{\alpha_{N+M}}{}^{\dot\beta_M}\Psi_{\alpha_{N+M+1}\cdots\alpha_{2s})\dot\beta_1\cdots\dot\beta_M}\quad ,$$

$$N\ge 2s-1:\ W_{\alpha_1\cdots\alpha_{2s}}=\frac{1}{(-M)!(2s)!}\bar{D}^{2N}D^N{}_{[a_1\cdots a_{-M}]}{}^{[b_1\cdots b_{-M}]}{}_{(\alpha_1\cdots\alpha_{2s-1}}\Psi_{\alpha_{2s})b_1\cdots b_{-M}}{}^{a_1\cdots a_{-M}}\qquad(3.13.1)$$

where $\Psi$ is an arbitrary (complex) superfield and $M=s-\frac{1}{2}(N+1)$.

If $W$ is bisected ($s + \frac{1}{2}N$ integer, $(1 - \mathbf{K})W = 0$), then $\tilde{\Psi}$ *must* be expressed in terms of a *real* prepotential $V$ that has maximum superspin $s$. $W$ has a form similar to (3.13.1):

$$N \leq 2s:\; W_{\alpha_1 \cdots \alpha_{2s}} = \frac{1}{(2s)!} \bar{D}^{2N} D^N{}_{(\alpha_1 \cdots \alpha_N} \partial_{\alpha_{N+1}}{}^{\dot{\beta}_1} \cdots \partial_{\alpha_{N+M}}{}^{\dot{\beta}_M} V_{\alpha_{N+M+1} \cdots \alpha_{2s})\dot{\beta}_1 \cdots \dot{\beta}_M} \quad ,$$

$$N \geq 2s:\; W_{\alpha_1 \cdots \alpha_{2s}} = \frac{1}{(-M)!} \bar{D}^{2N} D^N{}_{[a_1 \cdots a_{-M}]}{}^{[b_1 \cdots b_{-M}]}{}_{\alpha_1 \cdots \alpha_{2s}} V_{b_1 \cdots b_{-M}}{}^{a_1 \cdots a_{-M}} \qquad (3.13.2)$$

where $M = s - \frac{1}{2}N$.

Whether or not $W$ is bisected, ambiguity remains in the prepotentials $\tilde{\Psi}, V$, since they may still be expressed as derivatives of more fundamental superfields: This leads to "variant off-shell multiplets" (see sec. 4.5.c). Our expression (3.13.1) for $\tilde{\Psi}$ in terms of $\Psi$ is an example of such an ambiguity: There is no a priori reason why $\tilde{\Psi}$ must take the special form, *unless* it is obtained as a submultiplet of a bisected higher-$N$ multiplet (as, for example, in the case of the $N = 1$ spin $\frac{3}{2}, 1$ multiplet (sec. 4.5.e), which is a submultiplet of the $N = 2$ supergravity multiplet). Modulo such ambiguities, the expressions for $W$ in terms of $\tilde{\Psi}$ and $V$ are the most general local solutions to the Bianchi identities constraining $W$ (i.e., chirality, and if possible, bisection).

**Contents of 4. CLASSICAL, GLOBAL, SIMPLE (N=1) SUPERFIELDS**

# 4. CLASSICAL, GLOBAL, SIMPLE (N=1) SUPERFIELDS

In this chapter we discuss interacting field theories that can be built out of the superfields of global $N=1$ Poincaré supersymmetry. This restricts us to theories describing particles with spins no higher than 1. The simplest description of such theories is in terms of chiral scalar superfields for particles of the scalar multiplet (spins 0 and $\frac{1}{2}$), and real scalar gauge superfields for particles of the vector multiplet (spins $\frac{1}{2}$ and 1). However, other descriptions are possible; we treat some of these in a general framework provided by superforms. We describe $N=1$ theories and also extended $N \leq 4$ theories in terms of $N=1$ superfields. Our primary goal is to explain the structure of these theories in superspace. We do not discuss phenomenological models.

## 4.1. The scalar multiplet

### a. Renormalizable models

The lowest superspin representation of the $N=1$ supersymmetry algebra is carried by a chiral scalar superfield. In sec. 3.6 we described its components and transformations. In the chiral representation we have $\Phi^{(+)} = A + \theta^\alpha \psi_\alpha - \theta^2 F$, with *complex* scalar component fields $A = 2^{-\frac{1}{2}}(\mathrm{A}+\mathrm{iB})$, $F = 2^{-\frac{1}{2}}(\mathrm{F}+\mathrm{iG})$, and the transformations of (3.6.6).

### a.1. Actions

To find superspace actions for the chiral superfield we use dimensional analysis: The superfield contains two complex scalars differing by one unit of dimension (recall that $\theta$ has dimension $-\frac{1}{2}$); however, it contains only one spinor, and we require this spinor to have the usual physical dimension $\frac{3}{2}$. Therefore, we should assign the superfield dimension 1. This leads us to a unique choice for a free (quadratic) massless action with no dimensional parameters:

$$S_{kin} = \int d^4x \, d^4\theta \; \Phi\bar{\Phi} \tag{4.1.1}$$

(see sec. 3.7.a for a description of the Berezin integral). Up to an irrelevant phase there is a unique mass term and a unique interaction term with dimensionless coupling constant:

$$S_{(m,\lambda)} = \int d^4x\, d^2\theta\, (\frac{1}{2}m\Phi^2 + \frac{\lambda}{3!}\Phi^3) + \int d^4x\, d^2\bar\theta\, (\frac{1}{2}m\bar\Phi^2 + \frac{\lambda}{3!}\bar\Phi^3) \quad . \qquad (4.1.2)$$

The resulting action describes the "Wess-Zumino model".

All of the integrals are independent of the representation (vector, chiral or antichiral) in which the fields are given; the integrands in different representations differ by total $x$-derivatives (from the $e^U$ factors, see (3.3.26)), that vanish upon $x$-integration.

We can express the action in its component form either by straightforward $\theta$-expansion and integration, or by $D$-projection. In the former approach, we write for example, in the antichiral representation for $\bar\Phi = \bar\Phi^{(-)}$, and $\Phi^{(-)} = e^U\Phi^{(+)}$:

$$S_{kin} = \int d^4x\, d^4\theta\, \bar\Phi^{(-)}e^U\Phi^{(+)}$$

$$= \int d^4x\, d^4\theta\, [\bar A + \bar\theta^{\dot\alpha}\bar\psi_{\dot\alpha} - \bar\theta^2\bar F]\, e^{\theta^\alpha\bar\theta^{\dot\alpha}i\partial_{\alpha\dot\alpha}}\, [A + \theta^\alpha\psi_\alpha - \theta^2 F] \quad , \qquad (4.1.3a)$$

and after some algebra obtain

$$S_{kin} = \int d^4x\, [\bar A\Box A + \bar\psi^{\dot\alpha}i\partial^\alpha{}_{\dot\alpha}\psi_\alpha + \bar F F] \quad . \qquad (4.1.3b)$$

It is simpler to use the projection technique; we write $\int d^4xd^4\theta = \int d^4x\bar D^2D^2$ and

$$S_{kin} = \int d^4x\, d^4\theta\, \bar\Phi\Phi = \int d^4x\, \bar D^2[\bar\Phi D^2\Phi]|$$

$$= \int d^4x\, [\bar\Phi\bar D^2D^2\Phi + (\bar D^2\bar\Phi)(D^2\Phi) + (\bar D^{\dot\alpha}\bar\Phi)(\bar D_{\dot\alpha}D^2\Phi)]| \quad . \qquad (4.1.4)$$

Using the identities $\bar DD^2\Phi = Di\partial\Phi$ and $\bar D^2D^2\Phi = \Box\Phi$, which follow from the chirality of $\Phi$, and the definition of the components (3.6.7), we obtain (4.1.3b).

To evaluate chiral integrals by projection we write, for any function $f(\Phi)$

$$\int d^4x\, d^2\theta\, f(\Phi) = \int d^4x\, D^2f(\Phi)$$

$$= \int d^4x \; [f''(\Phi)(D\Phi)^2 + f'(\Phi)D^2\Phi]|$$

$$= \int d^4x \; [f''(A)\psi^2 + f'(A)F] \quad . \tag{4.1.5}$$

In particular we obtain for the mass and interaction terms

$$S_{(m,\lambda)} = \int d^4x \; \{m[\psi^2 + AF] + \lambda\,[A\psi^2 + \frac{1}{2}\,FA^2] + h.\,c.\} \quad . \tag{4.1.6}$$

(Without loss of generality, we can choose $m$ and $\lambda$ real.)

We could add a linear term $\xi\Phi$ and its hermitian conjugate to the action (4.1.2). Such a term would add to the component action a linear $\xi F + \overline{\xi}\overline{F}$ term. However, in the Wess-Zumino model such a term can always be eliminated from the action by a constant shift $\Phi \rightarrow \Phi + c$. Linear terms do however play an important role in constructing models with spontaneous supersymmetry breaking (see sec. 8.3).

### a.2. Auxiliary fields

The component field $F$ does not describe an independent degree of freedom; its equation of motion is algebraic:

$$F = -\, m\overline{A} - \frac{1}{2}\,\lambda\overline{A}^2 \quad . \tag{4.1.7}$$

If we eliminate the *auxiliary* field $F$ from the action and the transformation laws, we find

$$S = \int d^4x \; [\overline{A}(\Box - m^2)A + \overline{\psi}^{\dot\alpha} i\partial^\alpha{}_{\dot\alpha}\psi_\alpha + m(\psi^2 + \overline{\psi}^2)$$

$$- \frac{1}{2}\,m\lambda(A\overline{A}^2 + \overline{A}A^2) - \frac{1}{4}\,\lambda^2 A^2\overline{A}^2 + \lambda(A\psi^2 + \overline{A}\overline{\psi}^2)] \quad , \tag{4.1.8}$$

and

$$\delta A = -\, \epsilon^\alpha \psi_\alpha \quad ,$$

$$\delta\psi^\alpha = -\, \overline{\epsilon}^{\dot\alpha} i\partial^\alpha{}_{\dot\alpha}A - \epsilon^\alpha(m\overline{A} + \frac{1}{2}\,\lambda\overline{A}^2) \quad . \tag{4.1.9}$$

Therefore the Wess-Zumino action gives equal masses to the scalars and the spinor, cubic and quartic self-interactions for the scalars, and Yukawa couplings between the scalars and the spinor, all governed by a common coupling constant.

After eliminating $F$, the supersymmetry transformations of the spinor are nonlinear; this makes an analysis of the supersymmetry Ward identities *without* the auxiliary fields difficult. This is not the only problem caused by eliminating auxiliary component fields: The transformations are not only nonlinear, but also dependent on parameters in the Lagrangian, and it is difficult to discover further supersymmetric terms that could be added to the component Lagrangian (e.g., gauge couplings). Furthermore, equation (4.1.7) is not itself supersymmetric *unless* the equation of motion of the spinor is satisfied; only then is

$$\delta F = -\,\bar{\epsilon}^{\dot{\alpha}} i\partial^{\alpha}{}_{\dot{\alpha}}\psi_{\alpha} \tag{4.1.10a}$$

the same as

$$\delta F(A) = \delta(-m\bar{A} - \frac{1}{2}\lambda\bar{A}^2) = (m + \lambda\bar{A})\bar{\epsilon}^{\dot{\alpha}}\bar{\psi}_{\dot{\alpha}} \quad . \tag{4.1.10b}$$

For this reason, formulations of supersymmetric theories that lack the component auxiliary fields are often called *"on-shell* supersymmetric". Indeed, if we calculate the commutator of two supersymmetry transformations acting on the spinor, we find that the fields $A$, $\psi$, form a representation of the algebra (i.e., the algebra closes) only if the spinor equation of motion is satisfied.

The Wess-Zumino model can be generalized to include several chiral superfields. The most general action that leads to a conventional renormalizable theory is

$$S = \int d^4x\, d^4\theta\; \bar{\Phi}_i\Phi^i + \int d^4x\, d^2\theta\; \mathrm{P}(\Phi^i) + h.\,c. \quad , \tag{4.1.11}$$

where P is a polynomial of maximum degree 3 in the fields. The component action has the form

$$S = \int d^4x\; [\bar{A}_i\Box A^i + \bar{\psi}_i{}^{\dot{\alpha}} i\partial^{\alpha}{}_{\dot{\alpha}}\psi_{\alpha}{}^i + \bar{F}_i F^i]$$

$$+ \int d^4x\; [\mathrm{P}_i(A)F^i + \mathrm{P}_{ij}(A)\,\frac{1}{2}\,\psi^{\alpha i}\psi_{\alpha}{}^j + h.\,c.] \quad , \tag{4.1.12}$$

where

$$\mathrm{P}_i = \frac{\partial \mathrm{P}}{\partial A^i} \quad , \quad \mathrm{P}_{ij} = \frac{\partial^2 \mathrm{P}}{\partial A^i \partial A^j} \quad . \tag{4.1.13}$$

In particular, elimination of the auxiliary fields gives the scalar interaction terms (the scalar potential $U$):

$$-U(A^i) = -\sum_i |\mathrm{P}_i|^2 \quad . \tag{4.1.14}$$

As a consequence of supersymmetry (see (3.2.10)) the potential is positive semidefinite. The action (4.1.11) can also be invariant under a global internal symmetry group carried by the $\Phi$'s.

### a.3. R-invariance

An additional tool used to study these models is R-symmetry (3.6.14). This is the chiral symmetry generated by rotating $\theta$ and $\bar{\theta}$ by opposite phases (so that $\int d^4\theta$ is invariant but $\int d^2\theta$ is not) and by rotating different chiral superfields by related phases:

$$\Phi(x, \theta, \bar{\theta}) \to e^{-iwr}\Phi(x, e^{ir}\theta, e^{-ir}\bar{\theta}) \quad . \tag{4.1.15}$$

It may be, but is not always, possible to assign appropriate weights $w$ to the various superfields to make the total action R-invariant. For example, with only one chiral multiplet, R-invariance holds if either a mass or a dimensionless self-coupling is present, but not both: The appropriate transformations weights are $w = 1$ and $w = \frac{2}{3}$ respectively. With more than one chiral multiplet, it is possible to write R-symmetric Lagrangians having both mass and interaction terms: A chiral self-interaction term is R-invariant if its total R-weight $w = 2$ (i.e., the sum of the R-weights of each superfield factor is 2).

### a.4. Superfield equations

From the action for a chiral superfield, we obtain the equations of motion by functional differentiation (see (3.8.10,11)). For example, including sources, we have

$$S = \int d^4x\, d^4\theta\, \bar{\Phi}\Phi + \{ \int d^4x\, d^2\theta\, [\mathrm{P}(\Phi) + J\Phi] + h.c. \} \quad , \tag{4.1.16}$$

from which, using (3.8.9-12), we derive the equations

$$\bar{D}^2\bar{\Phi} + \mathrm{P}'(\Phi) + J = 0 \quad , \tag{4.1.17a}$$

$$D^2\Phi + \overline{\mathrm{P}}'(\overline{\Phi}) + \overline{J} = 0 \quad . \tag{4.1.17b}$$

We consider first the massive noninteracting case $\mathrm{P}(\Phi) = \frac{1}{2}m\Phi^2$. Multiplying (4.1.17a) by $D^2$, we find

$$D^2\overline{D}^2\overline{\Phi} + mD^2\Phi + D^2J = 0 \quad . \tag{4.1.18}$$

Substituting (4.1.17b) into (4.1.18) and using the chirality of $\Phi$ $(D^2\overline{D}^2\overline{\Phi} = \Box\overline{\Phi})$, we obtain

$$(\Box - m^2)\overline{\Phi} = m\overline{J} - D^2J \quad . \tag{4.1.19}$$

Similarly, we find

$$(\Box - m^2)\Phi = mJ - \overline{D}^2\overline{J} \quad . \tag{4.1.20}$$

and these equations can be readily solved.

For arbitrary $\mathrm{P}(\Phi)$, we derive the equations of motion for the component fields by projection from the superfield equations. Successively applying $D$'s to (4.1.17a) we find

$$\overline{F} + \mathrm{P}'(A) + J_A = 0$$

$$i\partial^{\alpha\dot{\alpha}}\overline{\psi}_{\dot{\alpha}} + \mathrm{P}''(A)\psi^\alpha + J_\psi{}^\alpha = 0$$

$$\Box\overline{A} + \mathrm{P}'''(A)\psi^2 + \mathrm{P}''F + J_F = 0 \tag{4.1.21}$$

as would be obtained from the component Lagrangian.

## b. Nonlinear $\sigma$-models

If renormalizability is not an issue, we can construct general supersymmetric actions by taking arbitrary functions of $\Phi$, $\overline{\Phi}$, and their derivatives, and integrating over superspace. An interesting class of supersymmetric models that can be constructed out of chiral superfields is the generalized nonlinear $\sigma$-model. In ordinary spacetime, a generalized nonlinear $\sigma$-model is described by fields $\phi^i$ that are the coordinates of an arbitrary manifold. The action of such a model is

$$S_\sigma = -\frac{1}{4}\int d^4x\ g_{ij}(\partial_{\underline{a}}\phi^i)(\partial^{\underline{a}}\phi^j) \ , \tag{4.1.22}$$

where $g_{ij}(\phi^i)$ is the metric tensor defined on the manifold. The supersymmetric generalization of these models is described by chiral superfields $\Phi^i$ and their conjugates $\bar{\Phi}_i$ which are the complex coordinates of an arbitrary Kähler manifold (see below). (We use a group theoretic convention: Upper and lower indices are related by complex conjugation, and all factors of the metric are kept explicit.) The action depends on a single real function $I\!K(\Phi, \bar{\Phi})$ defined up to arbitrary additive chiral and antichiral terms that do not contribute:

$$S_\sigma = \int d^4x\, d^4\theta\; I\!K(\Phi^i, \bar{\Phi}_j) \quad . \tag{4.1.23}$$

The component content of this action can be worked out straightforwardly using the projection technique; we find

$$S_\sigma = -\tfrac{1}{2}\int d^4x\, \frac{\partial^2 I\!K}{\partial A^i \partial \bar{A}_j}\,(\partial_{\underline{a}} A^i)(\partial^{\underline{a}} \bar{A}_j) + \cdots \quad . \tag{4.1.24}$$

This has the form (4.1.22) if we identify $\frac{\partial^2 I\!K}{\partial A^i \partial \bar{A}_j}$ as the metric $g_{ij}$. A complex manifold whose metric can be written (locally) in terms of a potential $I\!K$ is called Kähler; thus all four-dimensional supersymmetric nonlinear $\sigma$-models are defined on Kähler manifolds. Conversely, any bosonic nonlinear $\sigma$-model whose fields reside on a Kähler manifold can be extended to a supersymmetric model. The remaining terms in (4.1.24) provide couplings between the scalar fields and the spinor fields.

Kähler geometry is an interesting branch of complex manifold theory that mathematicians have investigated extensively. Here we discuss only those aspects relevant to subsequent topics (e.g., sec. 8.3.b). We define

$$I\!K_{i_1 \cdots i_m}^{j_1 \cdots j_n} = \frac{\partial}{\partial \Phi^{i_1}} \cdots \frac{\partial}{\partial \Phi^{i_m}}\, \frac{\partial}{\partial \bar{\Phi}_{j_1}} \cdots \frac{\partial}{\partial \bar{\Phi}_{j_n}}\, I\!K \quad . \tag{4.1.25a}$$

In particular, the metric is

$$I\!K_i{}^j = \frac{\partial^2 I\!K}{\partial \Phi^i \partial \bar{\Phi}_j} \quad . \tag{4.1.25b}$$

Equivalently, we can write the line element as

$$ds^2 = I\!K_i{}^j\, d\Phi^i\, d\bar{\Phi}_j \quad . \tag{4.1.25c}$$

The metric, like the action (4.1.23), is invariant under Kähler gauge transformations

$$I\!K \to I\!K + \Lambda(\Phi) + \overline{\Lambda}(\overline{\Phi}) \tag{4.1.26}$$

of the Kähler potential $I\!K$. Field redefinitions $\Phi' = f(\Phi)$ define *holomorphic* coordinate transformations on the manifold; under these, the form of the metric (4.1.25b,c) is preserved, whereas under arbitrary *nonholomorphic* coordinate transformations, in general terms of the form $g_{ij}\, d\Phi^i\, d\Phi^j$ and $\overline{g}^{ij}\, d\overline{\Phi}_i\, d\overline{\Phi}_j$ are generated in the line element. The nonhermitian metric coefficients $g_{ij}\,, \overline{g}^{ij}$ are *not* related to $I\!K_{ij}$ and $I\!K^{ij}$. When working with superfields, since $\Phi^i$ is chiral, only holomorphic coordinate transformations make obvious sense; however, we can perform arbitrary coordinate transformations on the scalar fields $A^i$.

Using the gauge transformations (4.1.26) and holomorphic coordinate transformations, it is possible to go to a *normal gauge* where, at any given point $\Phi_0$, $\overline{\Phi}_0$, evaluated at $\theta = \overline{\theta} = 0$,

$$I\!K_{i_1 \cdots i_m} = I\!K^{j_1 \cdots j_n} = 0 \qquad \textit{for all } n, m \quad , \tag{4.1.27a}$$

$$I\!K^{j}_{i_1 \cdots i_m} = I\!K_i^{j_1 \cdots j_n} = 0 \qquad \textit{for all } n, m > 1 \quad , \tag{4.1.27b}$$

$$I\!K_i{}^j = \eta_i{}^j \quad , \tag{4.1.27c}$$

with $\eta_i{}^j = (1, 1, \ldots -1, -1, \ldots)$ depending on the signature of the manifold. If the $\Phi^i$ describe physical matter multiplets, $\eta_i{}^j = \delta_i{}^j$. In a normal gauge, all the connections vanish at the point $\Phi_0$, the Riemann curvature tensor has the form:

$$R_i{}^j{}_k{}^l = I\!K_{ik}{}^{jl} \quad , \tag{4.1.28a}$$

with all other components related by the usual symmetries of the Riemann tensor or zero, and hence the Ricci tensor is simply:

$$R_k{}^j = I\!K_{ik}{}^{ji} \quad . \tag{4.1.28b}$$

In a general gauge, the connection is

$$\Gamma_{ij}{}^k = I\!K_{ij}{}^l\, (I\!K^{-1})_l{}^k \tag{4.1.29a}$$

where $(I\!K^{-1})_l{}^k$ is the inverse of the metric $I\!K_k{}^l$; all other components are related by complex conjugation or are zero. The contracted connection is, as always,

$$\Gamma_i \equiv \Gamma_{ij}{}^j = [\, ln\, det\, I\!K_k{}^l\,]_i \quad . \tag{4.1.29b}$$

The Riemann tensor in a general gauge is

$$R_i{}^j{}_k{}^l = I\!K_{ik}{}^{jl} - (I\!K^{-1})_m{}^n\, I\!K_{ik}{}^m\, I\!K_n{}^{jl} \tag{4.1.30a}$$

and the Ricci tensor has the simple form

$$R_k{}^j \equiv R_i{}^j{}_k{}^l\, I\!K_l{}^i = [\, ln\, det\, I\!K_i{}^l\,]_k{}^j \quad . \tag{4.1.30b}$$

Manifolds can have symmetries, or *isometries.* On a Kähler manifold, an isometry of the metric is, in general, an invariance of the Kähler potential $I\!K$ up to a Kähler gauge transformation (4.1.26). One can require the isometry to be an *invariance* of the potential. (Actually, this is only true if there is a point on the manifold where the isometry group is unbroken, i.e., the transformations do not shift the point.) This (partially) fixes the Kähler gauge invariance: It is no longer possible to go to a normal gauge (4.1.27). However, holomorphic coordinate transformations still make it possible to choose normal *coordinates,* where the metric $I\!K_i{}^j$ satisfies (4.1.27c), and its holomorphic derivatives $(I\!K_{i_1}{}^j)_{i_2\cdots i_m} \equiv I\!K^j_{i_1\cdots i_m}$ satisfy (4.1.27b) (likewise for the antiholomorphic derivatives) but the conditions (4.1.27a) are *not* satisfied.

In arbitrary coordinate systems, the isometries act on the coordinates as

$$\delta\Phi^i = \Lambda^{\rm A}\, k_{\rm A}{}^i \quad , \qquad \delta\bar\Phi_i = \bar\Lambda^{\rm A}\, k_{{\rm A}i} \tag{4.1.31}$$

where the $\Lambda$'s are infinitesimal parameters ($\Lambda = \bar\Lambda$ are constant unless we introduce gauge fields and gauge the isometry group; supersymmetric gauge theories are discussed in the remainder of this chapter), and the $k(\Phi, \bar\Phi)$'s are *Killing vectors.* These satisfy Killing's equations:

$$k_{\rm A}{}^{i;j} + k_{\rm A}{}^{j;i} = k_{{\rm A}i;j} + k_{{\rm A}j;i} = 0 \tag{4.1.32a}$$

$$k_{\rm A}{}^i{}_{;j} + (I\!K^{-1})^i{}_k k_{{\rm A}l}{}^{;k} I\!K^l{}_j = 0 \quad . \tag{4.1.32b}$$

where

$$k^{i;j} = k^{i,j} = \frac{\partial k^i}{\partial \bar\Phi_j} \tag{4.1.32c}$$

and

$$k^i{}_{;j} = k^i{}_{,j} + k^k\,\Gamma_{jk}{}^i = \frac{\partial k^i}{\partial \Phi^j} + k^k\,I\!K_{jk}{}^l (I\!K^{-1})_l{}^i \tag{4.1.32d}$$

For *holomorphic* Killing vectors $k^i = k^i(\Phi)\,, k_i = k_i(\overline{\Phi})$, (4.1.32a) is a triviality and (4.1.32b) follows directly from

$$I\!K_i k_{\mathrm{A}}{}^i + I\!K^i k_{\mathrm{A}\,i} = 0 \ , \tag{4.1.33}$$

which is just the statement that the Kähler potential is invariant under the isometries. (Actually, invariance up to gauge transformations (4.1.26) suffices to imply (4.1.32b).) We can also write the transformations (4.1.31) as

$$\delta\Phi^i = \Lambda^{\mathrm{A}}\ k_{\mathrm{A}}{}^j\,\frac{\partial}{\partial\Phi^j}\,\Phi^i \quad , \qquad \delta\overline{\Phi}_i = \overline{\Lambda}^{\mathrm{A}}\ k_{\mathrm{A}\,j}\,\frac{\partial}{\partial\overline{\Phi}_j}\,\overline{\Phi}_i \quad ; \tag{4.1.34a}$$

This form exponentiates to give the finite transformation:

$$\Phi'^i = exp\,(\,\Lambda^{\mathrm{A}}\ k_{\mathrm{A}}{}^j\,\frac{\partial}{\partial\Phi^j}\,)\,\Phi^i \quad , \qquad \overline{\Phi}'_i = exp\,(\,\overline{\Lambda}^{\mathrm{A}}\ k_{\mathrm{A}\,j}\,\frac{\partial}{\partial\overline{\Phi}_j}\,)\,\overline{\Phi}_i \quad . \tag{4.1.34b}$$

For the cases when there exists a fixed point on the manifold, we can choose a special coordinate system (that in general is *not* compatible with normal coordinates) where the transformations (4.1.31,34) take the familiar form

$$\delta\Phi^i = i\Lambda^{\mathrm{A}}\,(T_{\mathrm{A}})^i{}_j\,\Phi^j \quad , \qquad \delta\overline{\Phi}_i = \ -\,i\,\overline{\Phi}_j\overline{\Lambda}^{\mathrm{A}}\,(T_{\mathrm{A}})^j{}_i \tag{4.1.35a}$$

or, for finite transformations,

$$\Phi'^i = (e^{i\Lambda^{\mathrm{A}}\,T_{\mathrm{A}}})^i{}_j\,\Phi^j \quad , \qquad \overline{\Phi}'_i = \overline{\Phi}_j(e^{-i\overline{\Lambda}^{\mathrm{A}}\,T_{\mathrm{A}}})^j{}_i \ . \tag{4.1.35b}$$

In arbitrary coordinates, the notion of multiplying vectors by $i$ is represented by multiplication by a two index tensor called the *complex structure.* It has the property that its square is $-1\,\times$ a Kronecker delta. For a Kähler manifold, the complex structure is covariantly constant and preserves the metric.

It may happen that there exist nontrivial *nonholomorphic* coordinate transformations that *do* preserve the form of the metric (4.1.25b,c); then one can show that the manifold is *hyperKähler.* Such manifolds have three linearly independent complex structures and are locally quaternionic. They are even (complex) dimensional; all hyperKähler manifolds are Ricci flat, though the converse is true only in four (real) dimensions (two complex dimensions).

## 4.2. Yang-Mills gauge theories

### a. Prepotentials

In general, we can find a formulation of any supersymmetric gauge theory either by studying off-shell representations to derive the free (linear) theory in terms of *unconstrained* gauge superfields or *prepotentials,* or by postulating *covariant* derivatives and imposing *covariant* constraints on them until all quantities can be expressed in terms of a single irreducible representation of supersymmetry. In the former case, we must construct covariantly transforming derivatives out of the unconstrained fields and generalize to the nonlinear case, whereas in the latter case we must solve the covariant constraints in terms of prepotentials. We study both approaches and exhibit the relation between them.

#### a.1. Linear case

From the analysis of sec. 3.3.a.1, the $N=1$ vector multiplet consists of massless spin $\frac{1}{2}$ and spin 1 physical states. We denote the corresponding component field strengths by $\lambda_\alpha$, $f_{\alpha\beta}$. According to the discussion of sec. 3.12.a, these lie in an irreducible on-shell chiral superfield strength $\Psi_{(0)\alpha}$, which satisfies the field equations and Bianchi identities $D^\alpha \Psi_{(0)\alpha} = 0$. The corresponding irreducible off-shell field strength is a chiral superfield $W_\alpha$, $\bar{D}_{\dot{\alpha}} W_\alpha = 0$, satisfying the bisection condition ($s + \frac{1}{2}N = \frac{1}{2} + \frac{1}{2}$ is an integer) $\mathbf{K} W_\alpha = -\, W_\alpha$, which can be written (see (3.11.9))

$$D^\alpha W_\alpha = -\, \bar{D}^{\dot{\alpha}} \bar{W}_{\dot{\alpha}} \quad . \tag{4.2.1}$$

(We have a $-$ sign in the bisection condition to obtain usual parity assignments for the components.) Therefore, by (3.13.2), it can be expressed in terms of an unconstrained real scalar superfield by

$$W_\alpha = i\bar{D}^2 D_\alpha V \quad , \quad \bar{W}_{\dot{\alpha}} = -\, iD^2 \bar{D}_{\dot{\alpha}} V \quad , \quad V = \bar{V} \quad , \tag{4.2.2}$$

and this turns out to be the simplest description of the corresponding multiplet.

The definition of $W^\alpha$ is invariant under *gauge transformations* with a chiral parameter $\Lambda$

$$V' = V + i(\bar{\Lambda} - \Lambda) \quad , \quad \bar{D}_{\dot{\alpha}}\Lambda = D_\alpha \bar{\Lambda} = 0 \quad . \tag{4.2.3}$$

Later we generalize this to a nonabelian gauge invariance, but for the moment we analyze the simplest case. The prepotential $V$ can be expanded in components by projection:

$$C = V| \quad , \quad \chi_\alpha = iD_\alpha V| \quad , \quad \bar{\chi}_{\dot{\alpha}} = -\, i\bar{D}_{\dot{\alpha}}V| \quad ,$$

$$M = D^2 V| \quad , \quad \bar{M} = \bar{D}^2 V| \quad , \quad A_{\alpha\dot{\alpha}} = \frac{1}{2}\,[\bar{D}_{\dot{\alpha}}\,, D_\alpha]V| \quad ,$$

$$\lambda_\alpha = i\bar{D}^2 D_\alpha V| \quad , \quad \bar{\lambda}_{\dot{\alpha}} = -\, iD^2\bar{D}_{\dot{\alpha}}V| \quad , \quad \mathrm{D}' = \frac{1}{2}\,D^\alpha \bar{D}^2 D_\alpha V| \quad . \tag{4.2.4a}$$

(To avoid confusion with $D_\alpha$, we denote the "D" auxiliary field by D′.) As discussed in sec. 3.6.b, there is some choice in the order of the $D$'s which simply amounts to field redefinitions. The particular form we chose in (4.2.4a) is such that the physical components are invariant under the $\Lambda$ gauge transformations (except for an ordinary gauge transformation of the vector component field). By making a similar component expansion

$$\Lambda_1 = \Lambda| \quad , \quad \Lambda_\alpha = D_\alpha\Lambda| \quad , \quad \Lambda_2 = D^2\Lambda| \quad , \tag{4.2.4b}$$

we find

$$\delta C = i(\bar{\Lambda}_1 - \Lambda_1) \quad ,$$

$$\delta\chi_\alpha = \Lambda_\alpha \quad ,$$

$$\delta M = -\, i\Lambda_2 \quad ,$$

$$\delta A_{\alpha\dot{\alpha}} = \frac{1}{2}\,\partial_{\alpha\dot{\alpha}}(\Lambda_1 + \bar{\Lambda}_1) \quad ,$$

$$\delta\lambda_\alpha = 0 \quad ,$$

$$\delta\mathrm{D}' = 0 \quad . \tag{4.2.5}$$

Thus, all the components of $V$ can be gauged away by *nonderivative* gauge transformations except for $A_{\underline{a}}$, $\lambda_\alpha$ and D′. The vector and spinor are the physical component fields of the multiplet; D′ is an auxiliary field. They (and their derivatives) are the only components appearing in $W_\alpha$:

$$\lambda_\alpha = W_\alpha| \quad ,$$

$$f_{\alpha\beta} = \frac{1}{2} D_{(\alpha} W_{\beta)}| \quad , \qquad \mathrm{D}' = -\frac{1}{2} i D^\alpha W_\alpha| \quad ,$$

$$i\partial_\alpha{}^{\dot\alpha} \bar\lambda_{\dot\alpha} = D^2 W_\alpha| \quad . \tag{4.2.6}$$

The symmetric bispinor $f_{\alpha\beta}$ and its conjugate $\bar f_{\dot\alpha\dot\beta}$ are the self-dual and anti self-dual parts of the component field strength of the gauge field $A_{\underline{a}}$. The gauge in which $A, \lambda, \mathrm{D}'$ are the only nonzero components of $V$ is called the *Wess-Zumino gauge.* The remaining gauge freedom is the usual abelian gauge transformation of the vector component field.

The Wess-Zumino ("WZ") gauge *breaks* supersymmetry: The supersymmetry variations of $\chi_\alpha$ and $M$ violate the gauge condition $C = \chi_\alpha = M = 0$, e.g.,

$$\delta\chi_\alpha = i\epsilon_\alpha M + i\bar\epsilon^{\dot\alpha}(i\frac{1}{2}\partial_{\alpha\dot\alpha}C - A_{\alpha\dot\alpha}) \quad , \tag{4.2.7}$$

does *not* vanish in the WZ gauge. We can define transformations that preserve the WZ gauge by augmenting the usual supersymmetry transformations with "gauge-restoring" gauge transformations. Thus, instead of

$$\delta_\epsilon V = i(\bar\epsilon^{\dot\alpha}\bar Q_{\dot\alpha} + \epsilon^\alpha Q_\alpha)V \quad , \tag{4.2.8}$$

we take

$$\begin{aligned} \delta_\epsilon{}^{WZ} V &= i(\bar\epsilon^{\dot\alpha}\bar Q_{\dot\alpha} + \epsilon^\alpha Q_\alpha)V + i(\bar\Lambda - \Lambda)^{WZ} \\ &= i(\bar\epsilon^{\dot\alpha}\bar Q_{\dot\alpha}{}^{WZ} + \epsilon^\alpha Q_\alpha{}^{WZ})V \quad , \end{aligned} \tag{4.2.9}$$

where $\Lambda^{WZ}$ is chosen to restore the WZ gauge condition by canceling the terms in $\delta_\epsilon V$ that violate it. Specifically, $\delta_\epsilon{}^{WZ}\chi_\alpha = 0$ requires

$$\delta_\epsilon{}^{WZ}(D_\alpha V)| = 0 \quad . \tag{4.2.10}$$

Using $D^2 V| = 0$ and $\{\bar D_{\dot\alpha}, D_\alpha\}V| = \partial_{\underline{a}} V| = 0$ (in the WZ gauge), we have

$$\Lambda_\alpha{}^{WZ} = D_\alpha \Lambda^{WZ}| = i\bar\epsilon^{\dot\alpha}\bar D_{\dot\alpha} D_\alpha V| = i\bar\epsilon^{\dot\alpha} A_{\alpha\dot\alpha} \quad . \tag{4.2.11}$$

Similarly, from $\delta_\epsilon{}^{WZ} M = 0$ we find

$$\Lambda_2{}^{WZ} = D^2\Lambda^{WZ}| = -\,\bar\epsilon^{\dot\alpha}\bar\lambda_{\dot\alpha} \quad . \tag{4.2.12}$$

Finally, from $\delta_\epsilon{}^{WZ}C = 0$, we find that $\Lambda_1{}^{WZ} = \bar\Lambda_1{}^{WZ}$. The remaining real scalar in $\Lambda^{WZ}$ is the usual component gauge parameter for the vector gauge field (see (4.2.5)).

The WZ gauge preserving "supersymmetry" transformations are

$$\delta A_{\underline{a}} = -\,i(\bar\epsilon_{\dot\alpha}\lambda_\alpha + \epsilon_\alpha\bar\lambda_{\dot\alpha}) \quad ,$$

$$\delta\lambda_\alpha = -\,\epsilon^\beta f_{\beta\alpha} + i\epsilon_\alpha \mathrm{D}' \quad ,$$

$$\delta\mathrm{D}' = \frac{1}{2}\,\partial_{\alpha\dot\alpha}(\bar\epsilon^{\dot\alpha}\lambda^\alpha - \epsilon^\alpha\bar\lambda^{\dot\alpha}) \quad . \tag{4.2.13}$$

The commutator algebra of these transformations closes only up to gauge transformations of the vector field. *The need for gauge-restoring* $\Lambda$ *transformations makes supersymmetric quantization in the WZ gauge impossible.* The (vector) gauge-fixing procedure, by breaking gauge invariance, also breaks supersymmetry.

From the requirement that the physical components $A_{\underline{a}}$ and $\lambda_\alpha$ have canonical dimension, we conclude that $V$ has dimension zero. By dimensional analysis and gauge invariance under the $\Lambda$ transformations we find the action

$$S = \int d^4x\, d^2\theta\, W^2 = \frac{1}{2}\int d^4x\, d^4\theta\, V D^\alpha \bar D^2 D_\alpha V \quad . \tag{4.2.14}$$

Replacing $d^2\theta$ by $D^2$ and using (4.2.6), we obtain the component action

$$S = \int d^4x\, [-\,\frac{1}{2}\, f^{\alpha\beta} f_{\alpha\beta} + \bar\lambda^{\dot\alpha} i\partial^\alpha{}_{\dot\alpha}\lambda_\alpha + \mathrm{D}'^2] \quad . \tag{4.2.15}$$

We have not added the hermitian conjugate to $S$; $Im\, S$ is a total derivative and contributes only a surface term ($\sim \int d^4x\, \epsilon^{\underline{abcd}} f_{\underline{ab}} f_{\underline{cd}}$ + spinorial terms). The field $\mathrm{D}'$ is clearly auxiliary.

## a.2. Nonlinear case

The nonabelian generalization can be motivated by starting with a global internal symmetry and making it local. For this purpose we consider a multiplet of chiral scalar fields $\Phi$ transforming according to some representation of a global group with generators $T_{\mathrm{A}}$ and constant parameters $\lambda^{\mathrm{A}}$:

$$\Phi' = e^{i\lambda}\Phi \quad , \quad \lambda = \lambda^{A} T_{A} \quad , \quad T_{A} = T_{A}{}^{\dagger} \quad . \tag{4.2.16}$$

We extend this to a local transformation in superspace. Clearly, to maintain the chirality of $\Phi$ the local parameters should be chiral. We therefore consider transformations of the form

$$\Phi' = e^{i\Lambda}\Phi \quad , \quad \Lambda = \Lambda^{A} T_{A} \quad , \quad \bar{D}_{\dot{\alpha}}\Lambda = 0 \quad , \tag{4.2.17}$$

and correspondingly, for the antichiral $\bar{\Phi}$, transforming with the complex conjugate representation,

$$\bar{\Phi}' = \bar{\Phi} e^{-i\bar{\Lambda}} \quad , \quad \bar{\Lambda} = \bar{\Lambda}^{A} T_{A} \quad , \quad D_{\alpha}\bar{\Lambda} = 0 \quad . \tag{4.2.18}$$

The Lagrangian $\bar{\Phi}\Phi$ is invariant if the parameters $\lambda^{A}$ are real. For local transformations $\Lambda \neq \bar{\Lambda}$ and we must introduce a gauge field to covariantize the action. The simplest procedure is to introduce a multiplet of real scalar superfields $V^{A}$ transforming in the following fashion:

$$e^{V'} = e^{i\bar{\Lambda}} e^{V} e^{-i\Lambda} \quad , \quad V = V^{A} T_{A} \quad . \tag{4.2.19}$$

In the abelian case, this transformation is just (4.2.3). We covariantize the action by

$$\int d^4x\, d^4\theta\, \bar{\Phi} e^{V} \Phi \quad . \tag{4.2.20}$$

The gauge field $V$ acts as a "converter", changing a $\Lambda$ representation to a $\bar{\Lambda}$ representation of the group. Thus,

$$(e^{V}\Phi)' = e^{i\bar{\Lambda}}(e^{V}\Phi) \quad , \tag{4.2.21a}$$

and similarly

$$(\bar{\Phi} e^{V})' = (\bar{\Phi} e^{V}) e^{-i\Lambda} \quad . \tag{4.2.21b}$$

In the nonabelian case, even the infinitesimal gauge transformations of $V$ are highly nonlinear. Nonetheless, as in the abelian case they can be used to algebraically gauge away all but the physical components of $V^{A}$ and take us to the Wess-Zumino gauge: Starting with an arbitrary $V$, we perform successive gauge transformations to gauge away $C$, $\chi_{\alpha}$, and $M$. Requiring that the first transformation gauge away $C$ we find, by evaluating (4.2.19) at $\theta = 0$:

$$1 = e^{V'}| = e^{i\bar{\Lambda}^{(1)}} e^V e^{-i\Lambda^{(1)}}| = (e^{i\bar{\Lambda}^{(1)}}|) e^C (e^{-i\Lambda^{(1)}}|) \quad , \tag{4.2.22}$$

and hence we must choose ${\Lambda^{(1)}}_1 = \Lambda^{(1)}| = -\, i\,\frac{1}{2}\,C$. The gauge $C' = 0$ is preserved by all further transformations with $Im\Lambda_1 = 0$. To gauge away $\chi_\alpha$ we choose a second gauge transformation $\Lambda^{(2)}$ (with ${\Lambda^{(2)}}_1 = 0$) by requiring

$$0 = D_\alpha e^{V''}| = D_\alpha (e^{i\bar{\Lambda}^{(2)}} e^{V'} e^{-i\Lambda^{(2)}})|$$

$$= D_\alpha V'| - iD_\alpha \Lambda^{(2)}| = -\, i{\chi'}_\alpha - i{\Lambda^{(2)}}_\alpha \quad , \tag{4.2.23}$$

and hence ${\Lambda^{(2)}}_\alpha = D_\alpha \Lambda^{(2)}| = -\, \chi'_\alpha$. Finally, we can find a third transformation $\Lambda^{(3)}$ to gauge away $M$. In the WZ gauge, the only gauge freedom left corresponds to ordinary gauge transformations of the vector field $A_{\underline{a}}$, with parameter $\Lambda = \bar{\Lambda} = \omega(x)$.

As in the abelian case, the WZ gauge is not supersymmetric, and gauge-restoring transformations are required to define the WZ gauge "supersymmetry" transformations. The parameter of the transformations is still (4.2.11-12), but the transformations now become nonabelian and hence nonlinear. To find them, we compute the infinitesimal gauge transformations of $V$: We begin by defining the symbol

$$L_V X = [V, X] \quad , \tag{4.2.24}$$

so that

$$e^V X\, e^{-V} = e^{L_V} X \quad . \tag{4.2.25}$$

From $[V, e^V] = 0$ we obtain

$$(\delta V) e^V + V(\delta e^V) - e^V(\delta V) - (\delta e^V) V = 0 \quad , \tag{4.2.26a}$$

or

$$e^{-\frac{1}{2}V}(\delta V) e^{\frac{1}{2}V} - e^{\frac{1}{2}V}(\delta V) e^{-\frac{1}{2}V} + e^{-\frac{1}{2}V}[V, \delta e^V] e^{-\frac{1}{2}V} = 0 \quad , \tag{4.2.26b}$$

and hence

$$2\, sinh\big(\frac{1}{2}\, L_V\big)(\delta V) = e^{-\frac{1}{2}V} L_V(\delta e^V) e^{-\frac{1}{2}V}$$

$$= L_V[e^{-\frac{1}{2}V} i\bar{\Lambda} e^{\frac{1}{2}V} - e^{\frac{1}{2}V} i\Lambda e^{-\frac{1}{2}V}]$$

$$= iL_V[cosh(\frac{1}{2}L_V)(\bar{\Lambda} - \Lambda) - sinh(\frac{1}{2}L_V)(\bar{\Lambda} + \Lambda)] \ , \qquad (4.2.27)$$

from which it follows

$$\delta V = -\frac{1}{2} iL_V[\bar{\Lambda} + \Lambda + coth(\frac{1}{2}L_V)(\Lambda - \bar{\Lambda})]$$

$$= i\ (\bar{\Lambda} - \Lambda) - \frac{1}{2} i[V\ ,\bar{\Lambda} + \Lambda] + O(V^2)\ \ . \qquad (4.2.28)$$

From the transformations (4.2.28) and the parameter (4.2.11-12) we find the non-abelian WZ gauge-preserving "supersymmetry" transformations:

$$\delta A_{\underline{a}} = -i(\bar{\epsilon}_{\dot{\alpha}}\lambda_\alpha + \epsilon_\alpha \bar{\lambda}_{\dot{\alpha}}) \quad ,$$

$$\delta\lambda_\alpha = -\epsilon^\beta f_{\beta\alpha} + i\epsilon_\alpha \mathrm{D}' \quad ,$$

$$\delta\mathrm{D}' = \frac{1}{2}\nabla_{\alpha\dot{\alpha}}(\bar{\epsilon}^{\dot{\alpha}}\lambda^\alpha - \epsilon^\alpha \bar{\lambda}^{\dot{\alpha}}) \quad , \qquad (4.2.29)$$

where now $f_{\alpha\beta}$ is the self-dual part of the *nonabelian* field strength and $\nabla_{\alpha\dot{\alpha}} = \partial_{\alpha\dot{\alpha}} - iA_{\alpha\dot{\alpha}}$. The nonlinearity comes from the gauge-covariantization of the linear transformations (4.2.13). The components of the nonabelian vector multiplet are covariant generalizations of the abelian components; in the WZ gauge, they are the same as (4.2.4a) (see also (4.3.5)).

### a.3. Covariant derivatives

The gauge field $V$ can be used to construct derivatives, gauge covariant *with respect to* $\Lambda$ *transformations*

$$\nabla_A = D_A - i\Gamma_A = (\nabla_\alpha\, , \nabla_{\dot{\alpha}}\, , \nabla_{\alpha\dot{\alpha}}) \quad , \qquad (4.2.30)$$

defined by the requirement

$$(\nabla_A \Phi)' = e^{i\Lambda}(\nabla_A \Phi) \quad , \qquad (4.2.31)$$

i.e.,

$$\nabla'_A = e^{i\Lambda}\nabla_A e^{-i\Lambda} \quad , \qquad (4.2.32a)$$

or

$$\delta \nabla_A = i[\Lambda , \nabla_A] \quad . \tag{4.2.32b}$$

Since $\Lambda$ is chiral, $\nabla_{\dot\alpha} \equiv \bar{D}_{\dot\alpha}$ is covariant without further modification:

$$\nabla'_{\dot\alpha} = e^{i\Lambda} \bar{D}_{\dot\alpha} e^{-i\Lambda} = \nabla_{\dot\alpha} \quad . \tag{4.2.33}$$

The undotted spinor derivative $D_\alpha$ is covariant with respect to $\bar\Lambda$ transformations. We can use $e^V$ to convert it into a derivative covariant with respect to $\Lambda$ (see (4.2.21)); $\nabla_\alpha \equiv e^{-V} D_\alpha e^V$ transforms correctly:

$$\begin{aligned} \nabla'_\alpha &= (e^{i\Lambda}\, e^{-V} e^{-i\bar\Lambda}) D_\alpha (e^{i\bar\Lambda} e^V e^{-i\Lambda}) \\ &= e^{i\Lambda}\, e^{-V} D_\alpha e^V e^{-i\Lambda} \\ &= e^{i\Lambda}\, \nabla_\alpha e^{-i\Lambda} \quad . \end{aligned} \tag{4.2.34}$$

Finally, we construct $\nabla_{\underline{a}}$ by analogy with (3.4.9): $\nabla_{\underline{a}} = \nabla_{\alpha\dot\alpha} \equiv -\, i\{\nabla_\alpha , \nabla_{\dot\alpha}\}$. Its covariance follows from that of $\nabla_\alpha$ and $\nabla_{\alpha\dot\alpha}$.

We summarize:

$$\nabla_A = (e^{-V} D_\alpha e^V , \bar{D}_{\dot\alpha} , -\, i\{\nabla_\alpha , \nabla_{\dot\alpha}\}) \quad . \tag{4.2.35}$$

These derivatives are not hermitian. Their conjugates $\bar\nabla_A$ are covariant with respect to $\bar\Lambda$ transformations:

$$\begin{aligned} \bar\nabla_A &= (D_\alpha , e^V \bar{D}_{\dot\alpha} e^{-V} , -\, i\{\bar\nabla_\alpha , \bar\nabla_{\dot\alpha}\}) \quad , \\ \bar\nabla'_A &= e^{i\bar\Lambda} \bar\nabla_A e^{-i\bar\Lambda} \quad . \end{aligned} \tag{4.2.36}$$

The derivatives $\nabla_A$ ($\bar\nabla_A$) are called *gauge chiral (antichiral) representation* covariant derivatives. They are related by a nonunitary similarity transformation

$$\bar\nabla_A = e^V \nabla_A e^{-V} \quad . \tag{4.2.37}$$

This is analogous to the relation between global supersymmetry chiral and antichiral representations

$$D_A{}^{(-)} = e^U D_A{}^{(+)} e^{-U} \tag{4.2.38}$$

of (3.4.8).

The gauge covariant derivatives are usually defined in terms of vector representation $D_A$'s; if we express these in terms of ordinary derivatives, (4.2.35) becomes

$$\nabla_A = (e^{-V} e^{-\frac{1}{2}U} \partial_\alpha e^{\frac{1}{2}U} e^V , e^{\frac{1}{2}U} \bar{\partial}_{\dot{\alpha}} e^{-\frac{1}{2}U} , - i\{\nabla_\alpha , \nabla_{\dot{\alpha}}\}) \quad . \tag{4.2.39}$$

By a further similarity transformation $\nabla_A \to e^{-\frac{1}{2}U} \nabla_A e^{\frac{1}{2}U}$, we go to a new representation that is chiral with respect to both global supersymmetry and gauge transformations:

$$\nabla_A = (e^{-\frac{1}{2}U} e^{-V} e^{-\frac{1}{2}U} \partial_\alpha e^{\frac{1}{2}U} e^V e^{\frac{1}{2}U} , \bar{\partial}_{\dot{\alpha}} , - i\{\nabla_\alpha , \nabla_{\dot{\alpha}}\}) \quad . \tag{4.2.40}$$

We define $\tilde{V}$ by

$$e^{\frac{1}{2}U} e^V e^{\frac{1}{2}U} = e^{U+\tilde{V}} \quad . \tag{4.2.41}$$

In this form, it is clear that $\tilde{V}$ gauge covariantizes $U$: $i\theta^\alpha \bar{\theta}^{\dot{\alpha}} \partial_{\alpha\dot{\alpha}} \to \cdots + i\theta^\alpha \bar{\theta}^{\dot{\alpha}} (\partial_{\alpha\dot{\alpha}} - iA_{\alpha\dot{\alpha}}) + \cdots$. This combination transforms as

$$(e^{U+\tilde{V}})' = e^{i\bar{\Lambda}} (e^{U+\tilde{V}}) e^{-i\Lambda} \quad , \quad \partial_\alpha \bar{\Lambda} = \bar{\partial}_{\dot{\alpha}} \Lambda = 0 \quad . \tag{4.2.42}$$

There also exists a symmetric *gauge vector representation* that treats chiral and antichiral fields on the same footing. Such a representation uses a complex scalar gauge field $\Omega$, and requires a larger gauge group. We discuss the vector representation in subsec. 4.2.b, where the covariant derivatives are defined abstractly, and where it enters naturally.

### a.4. Field strengths

The covariant derivatives define field strengths by commutation:

$$[\nabla_A , \nabla_B\} = T_{AB}{}^C \nabla_C - iF_{AB} \quad , \tag{4.2.43}$$

with $V = V^{\mathbf{A}} T_{\mathbf{A}}$, and $T_{\mathbf{A}}$ in the adjoint representation. From the explicit form of the covariant derivatives (4.2.35) we find that the torsion $T_{AB}{}^C$ is the same one as in flat global superspace (3.4.19), and some field strengths vanish:

$$F_{\alpha\beta} = F_{\dot{\alpha}\dot{\beta}} = F_{\alpha\dot{\beta}} = 0 \quad . \tag{4.2.44}$$

The remaining field strengths are

$$F_{\dot{\alpha},\beta\dot{\beta}} = C_{\dot{\beta}\dot{\alpha}} \bar{D}^2 (e^{-V} D_\beta e^V) = iC_{\dot{\alpha}\dot{\beta}} W_\beta \quad ,$$

$$F_{\alpha\dot{\alpha},\beta\dot{\beta}} = \frac{1}{2}(C_{\dot{\alpha}\dot{\beta}}\nabla_{(\alpha}W_{\beta)} + C_{\alpha\beta}\nabla_{(\dot{\alpha}}W_{\dot{\beta})}) \ ,$$

$$W_\alpha \equiv i\bar{D}^2(e^{-V}D_\alpha e^V) \ ,$$

$$W_{\dot{\alpha}} \equiv e^{-V}\bar{W}_{\dot{\alpha}}e^V \equiv e^{-V}(-W_\alpha)^\dagger e^V \ . \tag{4.2.45}$$

(Recall that $\bar{W}^{\dot{\alpha}} \equiv (W^\alpha)^\dagger$ implies $\bar{W}_{\dot{\alpha}} = (-W_\alpha)^\dagger$ (3.1.20).) Thus all the field strengths of the theory are expressed in terms of a single spinor $W_\alpha$ that is the nonlinear version of (4.2.2). It satisfies Bianchi identities analogous to (4.2.1):

$$\nabla^\alpha W_\alpha = -\,\nabla^{\dot{\alpha}}W_{\dot{\alpha}} \ . \tag{4.2.46}$$

It is chiral, has dimension $\frac{3}{2}$, and can be used to construct a gauge invariant action

$$S = \frac{1}{g^2}\,tr\int d^4x\,d^2\theta\,W^2 = -\,\frac{1}{2g^2}\,tr\int d^4x\,d^4\theta\,(e^{-V}D^\alpha e^V)\bar{D}^2(e^{-V}D_\alpha e^V) \ ,$$

$$V = V^{\mathbf{A}}T_{\mathbf{A}} \quad , \quad trT_{\mathbf{A}}T_{\mathbf{B}} = \delta_{\mathbf{AB}} \ . \tag{4.2.47}$$

As in the abelian case, this action is hermitian up to a surface term (see discussion following (4.2.15)).

### a.5. Covariant variations

To derive the field equations from the action (4.2.47), we need to vary the action with respect to $V$. However, since $V$ is not a covariant object, this results in noncovariant field equations (although multiplication by a suitable (but complicated) invertible operator covariantizes them). In addition, variation with respect to $V$ is complicated because $V$ appears in $e^V$ factors. We therefore define a covariant variation of $V$ by

$$\Delta V \equiv e^{-V}\delta e^V = \frac{1-e^{-L_V}}{L_V}\,\delta V = \delta V + \ldots \ . \tag{4.2.48}$$

$\Delta V$ satisfies the chiral representation hermiticity condition as in (4.2.37). In practice, we always vary an action with respect to $V$ by expressing its variation in terms of $\delta e^V$, and then rewriting that in terms of $\Delta V$. We thus define a covariant functional derivative $\frac{\Delta F[V]}{\Delta V}$ by (cf. (3.8.3))

$$F[V+\delta V]-F[V]\equiv\big(\Delta V\,,\frac{\Delta F[V]}{\Delta V}\big)+O((\Delta V)^2)\quad . \tag{4.2.49}$$

We now obtain the equations of motion from:

$$g^2\delta S = i\,tr\int d^4x\,d^4\theta\;\delta(e^{-V}D^\alpha e^V)W_\alpha$$

$$= i\,tr\int d^4x\,d^4\theta\;[e^{-V}D^\alpha e^V\,,\Delta V]W_\alpha$$

$$= -\,i\,tr\int d^4xd^4\theta\,\Delta V\nabla^\alpha W_\alpha \quad , \tag{4.2.50}$$

which gives

$$g^2\,\frac{\Delta S}{\Delta V} = -\,i\nabla^\alpha W_\alpha = 0 \quad . \tag{4.2.51}$$

* * *

At the end of sec. 3.6 we expressed supersymmetry transformations in terms of the spinor derivatives $D_\alpha$. Using the covariant derivatives that we have constructed, we can write manifestly gauge covariant supersymmetry transformations by using the form (3.6.13) (for $w=0$) and adding the gauge transformation

$$\Lambda = i\bar{D}^2(\Gamma^\alpha D_\alpha\zeta) \quad , \tag{4.2.52a}$$

where $\Gamma_A$ is defined in (4.2.30). We then find

$$e^{-V}\delta_\zeta e^V = (W^\alpha\nabla_\alpha + W^{\dot\alpha}\bar\nabla_{\dot\alpha})\zeta = (W^\alpha e^{-V}D_\alpha e^V + e^{-V}\bar{W}_{\dot\alpha}e^V\bar{D}_{\dot\alpha})\zeta \tag{4.2.52b}$$

(where $\zeta$ is a real $x$-independent superfield that commutes with the group generators, e.g., $\nabla_\alpha\zeta = D_\alpha\zeta$). Since (4.2.52b) is manifestly gauge covariant, it preserves the Wess-Zumino gauge (but it is not a symmetry of the action after gauge-fixing). The corresponding supersymmetry transformations for *covariantly* chiral superfields $\Phi$, $\bar\nabla_{\dot\alpha}\Phi=0$ with arbitrary R-weight $w$ are

$$\delta\Phi = -\,i\bar\nabla^2[(\nabla^\alpha\zeta)\nabla_\alpha + w(\nabla^2\zeta)]\Phi \quad . \tag{4.2.52c}$$

## b. Covariant approach

In this subsection we discuss another approach to supersymmetric Yang-Mills theory that reverses the direction of the previous section. We postulate derivatives transforming covariantly under a gauge group, impose constraints on them, and discover that they can be expressed in terms of prepotentials. This procedure will prove especially useful in studying supergravity and extended super-Yang-Mills, so we give a detailed analysis for the simpler case of $N = 1$ super-Yang-Mills.

We start with the ordinary superspace derivatives $D_A$ satisfying $[D_A, D_B\} = T_{AB}{}^C D_C$, where $T_{AB}{}^C$ is the torsion and has only one nonzero component $T_{\alpha\dot\beta}{}^{\underline{c}}$ (see(3.4.19)). For a Lie algebra with generators $T_{\mathrm{A}}$ we covariantize the derivatives by introducing connection fields

$$\nabla_A = D_A - i\Gamma_A \quad , \tag{4.2.53}$$

where $\Gamma_A = \Gamma_A{}^{\mathrm{B}} T_{\mathrm{B}}$ is hermitian and $\overline{\nabla}_A = -\,(-)^A \nabla_A$. At the component level we have

$$\Gamma_\alpha = v_\alpha + \frac{i}{2}\,\bar\theta^{\dot\alpha} v_{\alpha\dot\alpha} + \cdots \quad , \qquad \Gamma_{\underline{a}} = w_{\underline{a}} + \cdots \quad , \tag{4.2.54a}$$

and hence

$$\nabla_\alpha = \partial_\alpha - iv_\alpha + \frac{i}{2}\,\bar\theta^{\dot\alpha}(\partial_{\alpha\dot\alpha} - iv_{\alpha\dot\alpha}) + \cdots \quad ,$$

$$\nabla_{\dot\alpha} = \bar\partial_{\dot\alpha} - i\bar v_{\dot\alpha} + \frac{i}{2}\,\theta^{\alpha}(\partial_{\alpha\dot\alpha} - i\bar v_{\alpha\dot\alpha}) + \cdots \quad ,$$

$$\nabla_a = \partial_{\underline{a}} - iw_{\underline{a}} + \cdots \quad , \tag{4.2.54b}$$

so that the component derivatives are covariantized.

Under gauge transformations the covariant derivatives are postulated to transform as

$$\nabla'_A = e^{iK}\nabla_A e^{-iK} \quad , \tag{4.2.55}$$

where the parameter $K = K^{\mathrm{A}} T_{\mathrm{A}}$ is a real superfield.

$$K = \omega(x) + \theta^\alpha K^{(1)}{}_\alpha(x) + \bar\theta^{\dot\alpha}\bar K^{(1)}{}_{\dot\alpha}(x) + \cdots \quad . \tag{4.2.56}$$

This is very different from what emerged in the previous section: Instead of chiral representation derivatives transforming with the chiral parameter $\Lambda$, we have *vector representation* hermitian derivatives, transforming with the hermitian parameter $K$. The asymmetric form of the previous section will emerge when we make a similarity transformation to go to the chiral representation.

For infinitesimal $K$, we find the component transformations:

$$\delta v_{\alpha\dot{\alpha}} = [\partial_{\alpha\dot{\alpha}} - iv_{\alpha\dot{\alpha}}\,, \omega] - i\omega_{\alpha\dot{\alpha}} \quad ,$$

$$\delta w_{\alpha\dot{\alpha}} = [\partial_{\alpha\dot{\alpha}} - iw_{\alpha\dot{\alpha}}\,, \omega] \quad , \tag{4.2.57}$$

where $\omega \equiv K|$, $\omega_{\alpha\dot{\alpha}} \equiv [\bar{D}_{\dot{\alpha}}\,, D_\alpha]K| = \bar{\omega}_{\alpha\dot{\alpha}}$. The component gauge parameter $\omega_{\alpha\dot{\alpha}}$ can be used to gauge away $Im\, v_{\alpha\dot{\alpha}}$ algebraically; however, the component fields $Re\, v_{\alpha\dot{\alpha}}$ and $w_{\alpha\dot{\alpha}}$ both remain as two a priori independent gauge fields for the *same* component gauge transformation. To avoid this we impose *constraints* on the covariant derivatives.

## b.1. Conventional constraints

Field strengths $F_{AB}$ are defined by (4.2.43). Substituting (4.2.53) we find

$$F_{AB} = D_{[A}\Gamma_{B)} - i[\Gamma_A\,, \Gamma_B\} - T_{AB}{}^C\Gamma_C \quad . \tag{4.2.58}$$

In particular,

$$F_{\alpha\dot{\alpha}} = D_\alpha\bar{\Gamma}_{\dot{\alpha}} + \bar{D}_{\dot{\alpha}}\Gamma_\alpha - i\{\Gamma_\alpha\,, \bar{\Gamma}_{\dot{\alpha}}\} - i\Gamma_{\alpha\dot{\alpha}} \quad . \tag{4.2.59}$$

If we impose the constraint

$$F_{\alpha\dot{\alpha}} = 0 \quad , \tag{4.2.60}$$

(4.2.59) *defines* the vector connection $\Gamma_{\alpha\dot{\alpha}}$ in terms of the spinor connections. (In components, this expresses $w_{\alpha\dot{\alpha}}$ in terms of $v_{\alpha\dot{\alpha}}$ and $v_\alpha$.)

In any theory one can add covariant terms to the connections (e.g., (3.10.22)) without changing the transformation of the covariant derivatives. If we did not impose the constraint (4.2.60) on the connections $\Gamma_A$, we could define equally satisfactory new connections $\Gamma'_A = (\Gamma_\alpha, \bar{\Gamma}_{\dot{\alpha}}, \Gamma_{\alpha\dot{\alpha}} - iF_{\alpha\dot{\alpha}})$ that identically satisfy the constraints. For this reason (4.2.60) is called a *conventional* constraint. It implies

$$\nabla_A = (\nabla_\alpha\,, \bar{\nabla}_{\dot{\alpha}}\,, -i\{\nabla_\alpha\,, \bar{\nabla}_{\dot{\alpha}}\}) \quad . \tag{4.2.61}$$

The theory now is expressed entirely in terms of the connection $\Gamma_\alpha$. However, it contains spin $s > 1$ gauge covariant component fields, for example

$$\psi_{(\alpha\beta)\dot{\beta}} \equiv F_{(\alpha,\beta)\dot{\beta}}| = i[\bar{D}_{\dot{\beta}}\, , D_{(\alpha}]\Gamma_{\beta)}| + \cdots \quad . \tag{4.2.62}$$

It also contains a superfield strength $F_{\alpha\beta}$ whose $\theta$-independent component

$$f_{\alpha\beta} = F_{\alpha\beta}| = D_{(\alpha}\Gamma_{\beta)}| + \cdots \quad , \tag{4.2.63}$$

is a dimension one symmetric spinor (equivalent to an antisymmetric second rank tensor). Because of its dimension, it cannot be the Yang-Mills field strength. Although in principle the theory might contain such fields (as auxiliary, not physical, components), in the covariant approach there are generally further types of constraints that eliminate (many) such components.

## b.2. Representation-preserving constraints

To couple scalar multiplets described by chiral scalar superfields to super-Yang-Mills theory, we must define *covariantly chiral* superfields $\Phi$: The covariant derivatives transform with the hermitian parameter $K$, and all fields must either be neutral or transform with the same parameter. However, $K$ is not chiral, and gauge transformations will not preserve chirality defined with $\bar{D}_{\dot{\alpha}}$. Instead we define a covariantly chiral superfield by

$$\bar{\nabla}_{\dot{\alpha}}\Phi = 0 \quad , \qquad \Phi' = e^{iK}\Phi \quad ,$$

$$\nabla_\alpha\bar{\Phi} = 0 \quad , \qquad \bar{\Phi}' = \bar{\Phi}e^{-iK} \quad . \tag{4.2.64}$$

This implies

$$0 = \{\bar{\nabla}_{\dot{\alpha}}\, , \bar{\nabla}_{\dot{\beta}}\}\Phi = -\, iF_{\dot{\alpha}\dot{\beta}}\Phi \quad . \tag{4.2.65}$$

Consistency requires that we impose the *representation-preserving* constraint

$$F_{\alpha\beta} = \bar{F}_{\dot{\alpha}\dot{\beta}} = 0 \quad . \tag{4.2.66}$$

This can be written as

$$\{\nabla_\alpha\, , \nabla_\beta\} = 0 \quad . \tag{4.2.67}$$

The most general solution is

$$\nabla_\alpha = e^{-\Omega} D_\alpha e^{\Omega} \quad , \quad \Omega = \Omega^{\mathrm{A}} T_{\mathrm{A}} \quad , \tag{4.2.68}$$

where $\Omega^{\mathrm{A}}$ is an arbitrary *complex* superfield. Eq. (4.2.67) states that $\nabla_\alpha$ satisfies the same algebra as $D_\alpha$, and the solution expresses the fact that they are equivalent up to a *complex* gauge transformation. Hermitian conjugation yields

$$\overline{\nabla}_{\dot\alpha} = e^{\overline{\Omega}} \overline{D}_{\dot\alpha} e^{-\overline{\Omega}} \quad . \tag{4.2.69}$$

Thus $\nabla_A$ is completely expressed in terms of the unconstrained *prepotential* $\Omega$ by the solutions (4.2.61,68,69) to the constraints (4.2.60,66).

The $K$ gauge transformations are realized by

$$(e^{\Omega})' = e^{\Omega} e^{-iK} \quad . \tag{4.2.70}$$

However, the solution to the constraint (4.2.67) has introduced an additional gauge invariance: The covariant derivatives (4.2.68) are *invariant* under the transformation

$$(e^{\Omega})' = e^{i\overline{\Lambda}} e^{\Omega} \quad , \quad \overline{D}_{\dot\alpha} \Lambda = 0 \quad . \tag{4.2.71}$$

Therefore, the gauge group of $\Omega$ is larger than that of $\Gamma_A$.

We define the $K$-invariant hermitian part of $\Omega$ by

$$e^{V} = e^{\Omega} e^{\overline{\Omega}} \quad . \tag{4.2.72}$$

The $K$ gauge transformations can be used to gauge away the antihermitian part of $\Omega$. In this gauge, $\Omega = \overline{\Omega} = \frac{1}{2} V$, and $\Lambda$ transformations must be accompanied by gauge-restoring $K$ transformations:

$$(e^{\Omega})' = e^{i\overline{\Lambda}} e^{\Omega} e^{-iK(\Lambda)} \quad ,$$

$$e^{-iK(\Lambda)} = e^{-\Omega} e^{-i\overline{\Lambda}} (e^{i\overline{\Lambda}} e^{2\Omega} e^{-i\Lambda})^{\frac{1}{2}} \quad . \tag{4.2.73}$$

In *any* gauge, the transformation of $V$ is

$$(e^{V})' = e^{i\overline{\Lambda}} e^{V} e^{-i\Lambda} \quad . \tag{4.2.74}$$

We have defined covariantly chiral superfields $\Phi$ by (4.2.64). We can use $\Omega$ (see (4.2.69)) to express them in terms of *ordinary* chiral superfields $\Phi_0$ (which we called $\Phi$ in sect. 4.2.a):

$$\Phi = e^{\bar{\Omega}}\Phi_0 \quad , \quad \bar{D}_{\dot{\alpha}}\Phi_0 = 0 \quad . \tag{4.2.75}$$

The factor $e^{\bar{\Omega}}$ converts $K$-transforming fields into $\Lambda$-transforming fields:

$$(\Phi_0)' = (e^{-\bar{\Omega}}\Phi)' = e^{i\Lambda}\Phi_0 \quad . \tag{4.2.76}$$

* * *

A useful identity that follows from the explicit form (4.2.68) expresses $\delta\nabla_\alpha$ in terms of an arbitrary variation $\delta\Omega$:

$$\delta\nabla_\alpha = (\delta e^{-\Omega})e^{\Omega}\nabla_\alpha + \nabla_\alpha e^{-\Omega}\delta e^{\Omega} = [\nabla_\alpha \, , e^{-\Omega}\delta e^{\Omega}] \quad . \tag{4.2.77}$$

### b.3. Gauge chiral representation

We can also use $\Omega$ to go to gauge chiral representation in which *all* quantities are $K$-inert and transform only under $\Lambda$. This is analogous to and not to be confused with the supersymmetry chiral representation (3.3.24-27), (3.4.8). We make a similarity transformation

$$\nabla_{0A} = e^{-\bar{\Omega}}\nabla_A e^{\bar{\Omega}} = (e^{-V}D_\alpha e^V \, , \bar{D}_{\dot{\alpha}} \, , -i\{\nabla_{0\alpha} \, , \nabla_{0\dot{\alpha}}\}) \quad ,$$

$$\Phi_0 = e^{-\bar{\Omega}}\Phi \quad ,$$

$$\tilde{\bar{\Phi}}_0 = \bar{\Phi}e^{\bar{\Omega}} = \overline{(\Phi_0)}e^V \quad . \tag{4.2.78}$$

The quantities $\nabla_{0A}$ and $\Phi_0$ are the chiral representation $\nabla_A$ and $\Phi$ of the previous subsection. We sometimes write the chiral representation hermitian conjugate of $\Phi_0$ as $\tilde{\bar{\Phi}}_0$ to avoid confusion with the ordinary hermitian conjugate $\bar{\Phi}_0 \equiv \overline{(\Phi_0)}$.

In the chiral representation we see no trace of $\Omega$ or $K$: Only $V$ and $\Lambda$ appear. However, we necessarily have an asymmetry between chiral and antichiral objects.

### c. Bianchi identities

In subsection 4.2.a we analyzed the physical content of the theory using component expansions and the Wess-Zumino gauge. Alternatively, we can find the field content of the theory by "solving" the Bianchi identities. These follow from the Jacobi

identities:

$$[\nabla_{[A}[\nabla_B\,,\nabla_{C)}\}\} = 0 \quad , \tag{4.2.79a}$$

which imply

$$\nabla_{[A}F_{BC)} - T_{[AB|}{}^D F_{D|C)} = 0 \quad . \tag{4.2.79b}$$

Normally these equations are trivial identities. However, once constraints have been imposed on some field strengths, they give information about the remaining ones, and in particular allow one to express all the fields strengths in terms of a basic set. We now describe the procedure.

We solve the equations (4.2.79) subject to the constraints (4.2.60,66) starting with the ones of lowest dimension. For each equation, we consider various pieces irreducible under the Lorentz group, and see what relations are implied among the field strengths. Thus, for example, the relation $[\{\nabla_{(\alpha}\,,\nabla_\beta\}\,,\nabla_{\gamma)}] = 0$ is identically satisfied when $F_{\alpha,\beta} = 0$. From $[\{\nabla_\alpha\,,\nabla_\beta\}\,,\bar\nabla_{\dot\gamma}] + [\{\bar\nabla_{\dot\gamma}\,,\nabla_{(\alpha}\}\,,\nabla_{\beta)}] = 0$, we find

$$F_{(\alpha,\beta)\dot\beta} = 0 \quad , \tag{4.2.80}$$

which implies, for some spinor superfield $W_\beta$,

$$F_{\alpha,\beta\dot\beta} = -\,iC_{\beta\alpha}\bar W_{\dot\beta} \quad . \tag{4.2.81}$$

From $[\{\nabla_\alpha\,,\nabla_\beta\}\,,\nabla_{\underline{c}}] + \{[\nabla_{\underline{c}}\,,\nabla_{(\alpha}],\nabla_{\beta)}\} = 0$ we find

$$C_{\gamma(\alpha}\nabla_{\beta)}\bar W_{\dot\gamma} = 0 \quad , \tag{4.2.82}$$

which implies

$$\nabla_\alpha \bar W_{\dot\beta} = 0 \quad . \tag{4.2.83}$$

From $[\{\nabla_\alpha\,,\bar\nabla_{\dot\beta}\}\,,\nabla_{\underline{c}}] + \{[\nabla_{\underline{c}}\,,\nabla_\alpha]\,,\bar\nabla_{\dot\beta}\} + \{[\nabla_{\underline{c}}\,,\bar\nabla_{\dot\beta}]\,,\nabla_\alpha\} = 0$ we obtain

$$F_{\alpha\dot\beta,\gamma\dot\gamma} + C_{\gamma\alpha}\bar\nabla_{\dot\beta}\bar W_{\dot\gamma} + C_{\dot\gamma\dot\beta}\nabla_\alpha W_\gamma = 0 \quad , \tag{4.2.84}$$

which separates into two equations:

$$F_{\alpha\dot\alpha,\beta\dot\beta} = \frac{1}{2}\,(C_{\alpha\beta}\bar\nabla_{(\dot\alpha}\bar W_{\dot\beta)} + C_{\dot\alpha\dot\beta}\nabla_{(\alpha}W_{\beta)}) \equiv C_{\alpha\beta}\bar f_{\dot\alpha\dot\beta} + C_{\dot\alpha\dot\beta}f_{\alpha\beta} \quad , \tag{4.2.85}$$

and

$$\nabla^{\alpha} W_{\alpha} + \bar{\nabla}^{\dot{\alpha}} \bar{W}_{\dot{\alpha}} = 0 \quad . \tag{4.2.86}$$

These can be reexpressed as

$$\nabla_{\alpha} W_{\beta} = i C_{\beta\alpha} \mathrm{D}' + f_{\alpha\beta} \quad , \qquad \mathrm{D}' = \bar{\mathrm{D}}' = -\frac{i}{2} \nabla^{\alpha} W_{\alpha} \quad . \tag{4.2.87}$$

Finally, $[[\nabla_{\alpha}, \nabla_{[\underline{b}}], \nabla_{\underline{c}]}] + [[\nabla_{\underline{b}}, \nabla_{\underline{c}}], \nabla_{\alpha}] = 0$ and $[[\nabla_{[\underline{a}}, \nabla_{\underline{b}}], \nabla_{\underline{c}]}] = 0$ are automatically satisfied as a consequence of the previous identities. From (4.2.87) we also obtain

$$\bar{\nabla}_{\dot{\alpha}} \mathrm{D}' = \frac{1}{2} \nabla^{\beta}{}_{\dot{\alpha}} W_{\beta} \quad ,$$

$$\bar{\nabla}_{\dot{\alpha}} f_{\alpha\beta} = i \frac{1}{2} \nabla_{(\alpha\dot{\alpha}} W_{\beta)} \quad , \tag{4.2.88}$$

and

$$\nabla_{\alpha} f_{\beta\gamma} = \frac{1}{2} C_{\alpha(\beta} i \nabla_{\gamma)\dot{\delta}} \bar{W}^{\dot{\delta}} \quad . \tag{4.2.89}$$

Therefore, all the field strengths are expressed in terms of the chiral field strength $W^{\alpha}$. In particular, the commutators of the covariant derivatives can be written as:

$$\{\nabla_{\alpha}, \nabla_{\beta}\} = 0 \quad ,$$

$$\{\nabla_{\alpha}, \bar{\nabla}_{\dot{\beta}}\} = i \nabla_{\alpha\dot{\beta}} \quad ,$$

$$[\bar{\nabla}_{\dot{\alpha}}, i\nabla_{\beta\dot{\beta}}] = -i C_{\dot{\beta}\dot{\alpha}} W_{\beta} \quad ,$$

$$[i\nabla_{\underline{a}}, i\nabla_{\underline{b}}] = i(C_{\dot{\alpha}\dot{\beta}} f_{\alpha\beta} + C_{\alpha\beta} \bar{f}_{\dot{\alpha}\dot{\beta}}) \quad . \tag{4.2.90}$$

Furthermore, the set

$$F = \{W_{\alpha}, \mathrm{D}', f_{\alpha\beta}\} \quad , \tag{4.2.91}$$

is closed under the operation of applying $\nabla_{\alpha}$ and $\bar{\nabla}_{\dot{\alpha}}$: Only spacetime derivatives $\nabla_{\underline{a}}$ of $F$ are generated. These superfields are the nonlinear off-shell extension of the superfield strengths $\Psi_{(n)}$ of sec. 3.12. The covariant components are the $\theta = 0$ projections of these superfields. Thus the constraints and the Bianchi identities directly determine the field content of the theory.

* * *

The existence of a "geometric" superspace formulation in terms of a (constrained) connection $\Gamma_A$ is important. For quantized super Yang-Mills theories, the geometric (or covariant) formulation can be combined with the background field method to derive improved superfield power-counting laws. We can also use $\Gamma_A$ to generalize the concept of the path-ordered phase factor to superspace:

$$I\!P\ [e^{(i\int dz^A\,\Gamma_A)}] \quad , \tag{4.2.92}$$

where the differential superspace element $dz^A$ is to be interpreted as $d\tau\,\frac{\partial dz^A}{\partial\tau}$ for $\tau$ some parametrization of the path. (In particular, $\int d\theta^\alpha$ is *not* a Berezin integral.) If we choose a closed path, this quantity defines a supersymmetric Wilson loop. Thus nonperturbative studies of ordinary Yang-Mills theories based on the properties of the Wilson loop should be extendible into superspace. (There is also a manifestly covariant form of path ordering, expressed directly in terms of covariant derivatives: see sec. 6.6.)

## 4.3. Gauge-invariant models

### a. Renormalizable models

In this subsection we consider properties of systems of interacting chiral and real gauge superfields with actions of the form

$$S = \int d^4x\, d^4\theta\, \overline{\Phi}_j (e^V)^j{}_i \Phi^i + tr \int d^4x\, d^2\theta\, W^2 + \left[ \int d^4x\, d^2\theta\, \mathrm{P}(\Phi^i) + h.\,c. \right] \quad (4.3.1)$$

(in the gauge-chiral representation), invariant under a group $G$. Here $V^i{}_j = V^{\mathrm{A}}(T_{\mathrm{A}})^i{}_j$ and $(T_{\mathrm{A}})^i{}_j$ is a (in general reducible) matrix representation of the generators $T_{\mathrm{A}}$ of $G$. In the vector representation, (4.3.1) takes the form

$$S = \int d^4x\, d^4\theta\, \overline{\Phi}_i \Phi^i + tr \int d^4x\, d^2\theta\, W^2 + \left[ \int d^4x\, d^2\theta\, \mathrm{P}(\Phi^i) + h.\,c. \right] \quad (4.3.2)$$

where we have used $tr\, e^{-\overline{\Omega}} f e^{\overline{\Omega}} = tr\, f$ in the chiral integral, and rewritten the action in terms of covariantly chiral superfields. The gauge coupling has been set to 1, but can be restored by the rescalings $W_\alpha \to g^{-1} W_\alpha$. $S$ may be R-symmetric, with the gauge superfield transforming as $V'(x, \theta, \overline{\theta}) = V(x, e^{-ir}\theta, e^{ir}\overline{\theta})$.

Another term can be added to the action: If $G$ is abelian, or has an abelian subgroup, the *Fayet-Iliopoulos* term

$$S_{FI} = tr \int d^4x\, d^4\theta\, \nu V = tr \int d^4x\, \nu \mathrm{D}' \quad , \quad (4.3.3)$$

is gauge invariant.

Component actions can be obtained by the projection techniques we have discussed before. A more efficient and, up to field redefinitions, totally equivalent procedure is to define *covariant components* by projecting with covariant derivatives. Thus, for a covariantly chiral superfield we define

$$A = \Phi| \quad , \quad \psi_\alpha = \nabla_\alpha \Phi| \quad , \quad F = \nabla^2 \Phi| \quad . \quad (4.3.4)$$

Similarly, the covariant components of the gauge multiplet can be obtained by projection from $W_\alpha$ (here $f_{\alpha\beta}$ denotes the component field strength):

$$\lambda_\alpha = W_\alpha| \quad , \quad f_{\alpha\beta} = \frac{1}{2} \{ \nabla_{(\alpha}, W_{\beta)} \}| \quad ,$$

$$i\nabla_\alpha{}^{\dot\alpha}\bar\lambda_{\dot\alpha} = \frac{1}{2}[\nabla^\beta, \{\nabla_\beta, W_\alpha\}]| \quad , \qquad \mathrm{D}' = -\frac{i}{2}\{\nabla^\alpha, W_\alpha\}| \quad . \tag{4.3.5}$$

The covariant derivative $\nabla_{\alpha\dot\beta}|$ is the covariant space-time derivative. To obtain component actions by covariant projection, we use the fact that on a gauge *invariant* quantity $\bar D^2 D^2 = \bar\nabla^2\nabla^2$.

The component action that results from (4.3.1) plus (4.3.3) takes the form

$$S = \int d^4x \, [A^i \Box \bar A_i + \Psi^{\alpha i}\, i\nabla_\alpha{}^{\dot\alpha}\bar\Psi_{\dot\alpha i} + i\bar A_i(\lambda^\alpha)^i{}_j\Psi_\alpha{}^j - i\bar\Psi^{\dot\alpha}{}_i(\bar\lambda_{\dot\alpha})^i{}_j A^j$$

$$+ \bar A_i(\mathrm{D}')^i{}_j A^j + F^i\bar F_i + tr\,(\,\lambda^\alpha[i\nabla_\alpha{}^{\dot\alpha}, \bar\lambda_{\dot\alpha}] - \frac{1}{2}f^{\alpha\beta}f_{\alpha\beta} + \mathrm{D}'^2\,)$$

$$+ tr\nu\mathrm{D}' + (\mathrm{P}_i F^i + \frac{1}{2}\mathrm{P}_{ij}\Psi^{\alpha i}\Psi_\alpha{}^j + h.\,c.\,)] \tag{4.3.6}$$

where $\Box \equiv \frac{1}{2}\nabla^{\underline{a}}\nabla_{\underline{a}}$, $\mathrm{P}_i\,, \mathrm{P}_{ij}$ are defined in (4.1.13), $(\lambda)^i{}_j = \lambda^{\mathrm{A}}T_{\mathrm{A}}$, etc. The auxiliary field $\mathrm{D}'$ can be eliminated algebraically using its field equations. This leads to interaction terms for the spin-zero fields of the chiral multiplets:

$$-U_{\mathrm{D}'} = -\frac{1}{4}[\bar A_i(T_{\mathrm{A}})^i{}_j A^j + \nu tr T_{\mathrm{A}}]^2 \tag{4.3.7}$$

in addition to those obtained by eliminating $F$ (see (4.1.14)).

## b. CP(n) models

In sec. 4.1.b we discussed supersymmetric nonlinear $\sigma$-models written in terms of chiral and antichiral superfields that are the complex coordinates of a Kähler manifold. Some nonlinear $\sigma$-models can be written linearly if we introduce a (classically) non-propagating gauge field. We consider here supersymmetric extensions of the bosonic $CP(n)$ models. The bosonic models are straightforward generalizations of the $CP(1)$ model of sec. 3.10. They are written in terms of $(n+1)$ complex scalar fields $z^i$ constrained by $z^i\bar z_i = c$; the action is written by introducing an abelian gauge field with no kinetic term:

$$S = \int d^4x \, [|(\partial_{\alpha\dot\beta} - i\,A_{\alpha\dot\beta})z^i|^2 + \mathrm{D}'(|z^i|^2 - c)] \ , \tag{4.3.8}$$

where $\mathrm{D}'$ is a Lagrange multiplier field. Eliminating $A_{\alpha\dot\beta}$ by its classical field equation,

we find the action given in (3.10.23). This action is still invariant under the local U(1) gauge transformation $z \to e^{i\alpha} z$, $\bar{z} \to e^{-i\alpha}$ where $\alpha(x)$ is a real parameter. It can be rewritten in terms of $n+1$ unconstrained fields $Z^i$ in the form (3.10.30).

In the supersymmetric case, the model is most conveniently described in terms of $(n+1)$ chiral fields $\Phi$ (and their complex conjugates $\bar{\Phi}$), and a single abelian gauge field $V$. The action, which is globally supersymmetric, $SU(n+1)$ invariant, and locally gauge invariant, is:

$$S = \int d^4x \, d^4\theta \, (\Phi^i \bar{\Phi}_i \, e^V - c\, V) \quad . \tag{4.3.9}$$

Note the presence of the Fayet-Iliopoulos term. Upon eliminating the gauge field $V$ by its field equation we find

$$S = \int d^4x \, d^4\theta \; c \, ln(\Phi^i \bar{\Phi}_i) \quad . \tag{4.3.10}$$

This action is still invariant under the (local) abelian gauge transformation $\Phi \to e^{i\Lambda}\Phi$. We can use this invariance to choose a gauge, e.g., $\Phi^i = (c \,, u_a)$. In components, (4.3.10) gives the action generalizing (3.10.30) for the $CP(n)$ nonlinear $\sigma$-model coupled to a spinor field.

The action (4.3.9) has a straightforward generalization:

$$S = \int d^4x \, d^4\theta \, (\bar{\Phi}_i (e^V)^i{}_j \Phi^j - c\, tr\, V) \quad , \tag{4.3.11}$$

where, as in (4.3.1), $V = V^{\mathrm{A}} T_{\mathrm{A}}$ and $(T_{\mathrm{A}})^i{}_j$ is a (in general reducible) matrix representation of the generators $T_{\mathrm{A}}$ of some group. However, in contrast to (4.3.9), when we vary (4.3.11) with respect to $V$, we get an equation that in general does not have an explicit solution:

$$\bar{\Phi} \, e^V T_{\mathrm{A}} \, \Phi - c\, tr\, T_{\mathrm{A}} = 0 \quad . \tag{4.3.12}$$

(To derive (4.3.12), we use the covariant variation (4.2.48) $\Delta V = \Delta V^{\mathrm{A}} T_{\mathrm{A}} \equiv e^{-V} \delta e^V$, and $tr \Delta V = tr \delta V$.)

## 4.4. Superforms

### a. General

In ordinary spacetime, there is a family of gauge theories that can be constructed systematically; these theories are expressed in terms of $p$-forms $\Gamma_p = \frac{1}{p!} dx^{\underline{m}_1} \wedge dx^{\underline{m}_2} \wedge \cdots \wedge dx^{\underline{m}_p} \Gamma_{\underline{m}_1 \underline{m}_2 \cdots \underline{m}_p}$ where the differentials satisfy $dx^{\underline{m}} \wedge dx^{\underline{n}} = -dx^{\underline{n}} \wedge dx^{\underline{m}}$. The "tower" of theories based on forms is: $\Gamma_0$ = scalar, $\Gamma_1$ = vector gauge field, $\Gamma_2$ = tensor gauge field, $\Gamma_3$ = auxiliary field, and $\Gamma_4$ = "nothing" field. Their gauge transformations, field strengths, and Bianchi identities are given by

$$\text{gauge transformation:} \qquad \delta\Gamma_p = dK_{p-1} \quad ,$$

$$\text{field strength:} \qquad F_{p+1} = d\Gamma_p \quad ,$$

$$\text{Bianchi identity:} \qquad dF_{p+1} = 0 \quad . \tag{4.4.1}$$

Here $K_p$, $\Gamma_p$, $F_p$ are $p$-form gauge parameters, gauge fields, and field strengths respectively, and $d = dx^{\underline{m}} \partial_{\underline{m}}$. By definition, $-1$-forms vanish, and 5-forms (or (D+1)-forms in D dimensions) vanish by antisymmetry. The Bianchi identities and the gauge invariance of the field strengths are automatic consequences of the Poincaré lemma $dd = 0$.

In superspace the same construction is possible, using super $p$-forms:

$$\Gamma_p = (-1)^{\frac{1}{2}p(p-1)} \frac{1}{p!} dz^{M_1} \wedge \cdots \wedge dz^{M_p} \Gamma_{M_p \cdots M_1} \tag{4.4.2a}$$

(note the ordering of the indices), where now

$$dz^M \wedge dz^N = -(-)^{MN} dz^N \wedge dz^M \quad , \tag{4.4.2b}$$

the coefficients of the form are superfields, and $d = dz^M \partial_M$. The same tower of gauge parameters, gauge fields, field strengths, and Bianchi identities can be built up (now using the *super*Poincaré lemma $dd = 0$). An advantage of this description of flat superspace theories is that it generalizes immediately to curved superspace and determines the coupling of these global multiplets to supergravity.

However, superforms do not describe irreducible representations of supersymmetry unless we impose constraints. To maintain gauge invariance, these constraints should be imposed on the coefficients of the field strength form; when the constraints are solved, the coefficients of the (gauge) potential form are expressed in terms of prepotentials. In table 4.4.1 the prepotentials $\hat{A}_p$ correspond to the *constrained* super $p$-form $A_p$ and the expressions $\widehat{dA}_p$ correspond to $dA_p$.

| $p$ | $\hat{A}_p$ | $\widehat{dA}_p$ |
|---|---|---|
| 0 | $\Phi$ | $i(\bar{\Phi} - \Phi)$ |
| 1 | $V$ | $i\bar{D}^2 D^\alpha V$ |
| 2 | $\Phi^\alpha$ | $\frac{1}{2}(D_\alpha \Phi^\alpha + \bar{D}_{\dot{\alpha}} \bar{\Phi}^{\dot{\alpha}})$ |
| 3 | $V$ | $\bar{D}^2 V$ |
| 4 | $\Phi$ | 0 |

*Table 4.4.1. Simple superfields (prepotentials) corresponding to superforms*

In this Table $\Phi$ and $\Phi^\alpha$ are chiral and $V$ is real. The relation $\hat{A}_p = \hat{A}_{4-p}$ corresponds to Hodge duality of the component forms.

The constrained super $p$-forms correspond to particular prepotentials $\hat{A}_p$ whether $A_p$ is a gauge parameter $K_p$, a potential $\Gamma_p$, a field strength $F_p$, or a Bianchi identity $(dF)_p$. The explicit expressions for $A_p$ in terms of $\hat{A}_p$ take the same form whether $A$ is $K$,$\Gamma$,$F$,or $dF$. Thus the prepotentials give rise to a tower of theories that mimics (4.4.1): The gauge field strength and Bianchi identities at one level are the gauge parameter and field strength at the next level. If $A_{p-1}$, $A_p$, and $A_{p+1}$ are the gauge parameter $K_{p-1}$, the gauge field $\Gamma_p$, and the field strength $F_{p+1}$ superforms, respectively, then the gauge transformation, field strength, and Bianchi identities of the *prepotentials* are

$$\text{gauge transformation:} \qquad \delta\hat{\Gamma}_p = \widehat{dK}_{p-1} \quad ,$$

$$\text{field strength:} \qquad \hat{F}_{p+1} = \widehat{d\Gamma}_p \quad ,$$

$$\text{Bianchi identity:} \qquad \widehat{dF}_{p+1} = 0 \quad . \tag{4.4.3}$$

The Lagrangians for all $p$-form theories are quadratic in the field strengths, without extra derivatives. We discuss details in the subsections that follow.

Under a supersymmetry transformation the superforms are defined to transform as

$$\Gamma'(z', dz') = \Gamma(z, dz) \quad , \tag{4.4.4}$$

where (cf. (3.3.15))

$$dz' = (d\theta'^{\mu}, d\bar{\theta}'^{\dot{\mu}}, dx'^{\mu\dot{\mu}}) = (d\theta^{\mu}, d\bar{\theta}^{\dot{\mu}}, dx^{\mu\dot{\mu}} - \frac{i}{2}(\bar{\epsilon}^{\dot{\mu}} d\theta^{\mu} + \epsilon^{\mu} d\bar{\theta}^{\dot{\mu}})) \,. \tag{4.4.5}$$

Consequently, the coefficients $\Gamma_{MN\cdots}$ mix under supersymmetry transformations and this makes it difficult to impose supersymmetric constraints on them. To maintain manifest supersymmetry, we therefore go to a "tangent space" basis, parametrized by the duals of the covariant derivatives $D_A$ rather than the duals of $\partial_M$. We use the flat superspace vielbeins $D_A{}^M$ (3.4.16):

$$D_A = D_A{}^M \partial_M = (D_\alpha, \bar{D}_{\dot{\alpha}}, \partial_{\alpha\dot{\alpha}}) \quad , \tag{4.4.6}$$

and the dual forms

$$\omega^A \equiv dz^M (D^{-1})_M{}^A \quad . \tag{4.4.7}$$

From (3.4.18), the $D$'s satisfy

$$D_{[A} D_{B)}{}^M = T_{AB}{}^C D_C{}^M \quad , \tag{4.4.8}$$

and hence

$$d\omega^A = \frac{1}{2} \omega^C \wedge \omega^B T_{BC}{}^A \quad . \tag{4.4.9}$$

In this $\omega$-basis we write a superform as

$$\Gamma_p = (-1)^{\frac{1}{2}p(p-1)} \frac{1}{p!} \omega^{A_1} \wedge \cdots \wedge \omega^{A_p} \Gamma_{A_p \cdots A_1} \quad . \tag{4.4.10}$$

We also have $d \equiv dz^M \partial_M = \omega^A D_A$. The tangent space coefficients $\Gamma_{A_p \cdots A_1}$ of the $p$-form do not mix under supersymmetry transformations because $\omega^A$ is invariant. We can now impose supersymmetric constraints on individual coefficients of a form.

In this basis, the coefficients of the field strength form (on which we impose the constraints) $F_{p+1} = d\Gamma_p$ have the following expression in terms of the gauge fields:

$$F_{A_1 \cdots A_{p+1}} = \frac{1}{p!} D_{[A_1} \Gamma_{A_2 \cdots A_{p+1})} - \frac{1}{2(p-1)!} T_{[A_1 A_2|}{}^B \Gamma_{B|A_3 \cdots A_{p+1})} \,, \tag{4.4.11}$$

where the torsion terms come from (4.4.9). The Bianchi identity on $F$ takes a similar appearance. Equation (4.4.11) is the essential result we need for the discussion of subsecs. 4.4.b-e.

We now summarize some of the results of subsecs. 4.4.b-e. In particular, we give the explicit expressions for the coefficients of the superforms $A_p$ in terms of the prepotentials $\hat{A}_p$ (of table 4.4.1) for all $p$. In the case of $\Gamma_p$, these expressions are found by solving the constraints on certain coefficients of $F_{p+1}$ and choosing a suitable $K$-gauge ($\delta\Gamma = dK$). The expressions for $K$ follow from the new invariance found when solving these constraints. The expressions for $F$ follow from solving those Bianchi identities $dF$ that *explicitly* express one part of $F$ in terms of another in the presence of the constraints. Finally, for $dF$, the explicit expressions correspond to the remaining part of the Bianchi identities that are not algebraically soluble. (For clarification, see subsecs. 4.4.b-e, where the expressions are worked out in detail.) We find:

$$p=0: \quad A = \frac{1}{2}(\hat{A} + \bar{\hat{A}}) \quad ;$$

$$p=1: \quad A_\alpha = i\frac{1}{2}D_\alpha \hat{A} \quad , \quad A_{\underline{a}} = \frac{1}{2}[\bar{D}_{\dot{\alpha}}, D_\alpha]\hat{A} \quad ;$$

$$p=2: \quad A_{\alpha\beta} = A_{\alpha\dot{\beta}} = 0 \quad , \quad A_{\alpha\underline{b}} = iC_{\alpha\beta}\bar{\hat{A}}_{\dot{\beta}} \quad ,$$

$$A_{\underline{ab}} = \frac{1}{2}(C_{\dot{\alpha}\dot{\beta}}D_{(\alpha}\hat{A}_{\beta)} + C_{\alpha\beta}\bar{D}_{(\dot{\alpha}}\bar{\hat{A}}_{\dot{\beta})}) \quad ;$$

$$p=3: \quad A_{\alpha\beta\gamma} = A_{\alpha\beta\dot{\gamma}} = A_{\alpha\beta\underline{c}} = 0 \quad , \quad A_{\alpha\dot{\beta}\underline{c}} = T_{\alpha\dot{\beta}\underline{c}}\hat{A} \quad ,$$

$$A_{\alpha\underline{bc}} = -\,C_{\dot{\beta}\dot{\gamma}}C_{\alpha(\beta}D_{\gamma)}\hat{A} \quad , \quad A_{\underline{abc}} = \epsilon_{\underline{dabc}}[\bar{D}^{\dot{\delta}}, D^{\delta}]\hat{A} \quad ;$$

$$p=4: \quad A_{\alpha\beta\gamma\delta} = A_{\alpha\beta\gamma\dot{\delta}} = A_{\alpha\beta\dot{\gamma}\dot{\delta}} = A_{\alpha\beta\gamma\underline{d}} = A_{\alpha\beta\dot{\gamma}\underline{d}} = A_{\alpha\dot{\beta}\underline{cd}} = 0 \quad ,$$

$$A_{\alpha\beta\underline{cd}} = 2\,C_{\dot{\gamma}\dot{\delta}}C_{\alpha(\gamma}C_{\delta)\beta}\,\bar{\hat{A}} \quad ,$$

$$A_{\alpha\underline{bcd}} = 2\epsilon_{\underline{abcd}}\bar{D}^{\dot{\alpha}}\,\bar{\hat{A}} \quad , \quad A_{\underline{abcd}} = 2i\epsilon_{\underline{abcd}}(\bar{D}^2\bar{\hat{A}} - D^2\hat{A}) \,, \qquad (4.4.12)$$

where for even $p$, $\bar{D}_{\dot{\alpha}}\hat{A} = 0$, and for odd $p$, $\hat{A} = \bar{\hat{A}}$.

For example, in the case of the vector multiplet of sec. 4.2, we found the vector representation potentials $\Gamma_A$ given by the case $p=1$ above, with $\hat{A}=V$ (in the vector representation, and in the gauge where $\Omega=\overline{\Omega}=\frac{1}{2}V$; then $K=\frac{1}{2}(\Lambda+\overline{\Lambda})$, as given above by $p=0$); the field strengths $F_{AB}$ by $p=2$ above, with $\hat{A}_\alpha=W_\alpha=i\overline{D}^2 D_\alpha V$ (by table 4.4.1); and the remaining Bianchi identity $dF$ on $W_\alpha$ by $p=3$ above, with $\hat{A}=\frac{1}{2}(D_\alpha W^\alpha+\overline{D}_{\dot{\alpha}}\overline{W}^{\dot{\alpha}})$ (again by table 4.4.1; $dF=0$ thus reduces to $\hat{A}(W^\alpha)=0$). Further examples will be derived in the remainder of this section. (Note that an action written in terms of a super 0-form does *not* describe the *most general* chiral multiplet theory: The field strength $F_A=D_A\Gamma$ always has the invariance $\delta\Gamma=k$, where $k$ is a real constant. Here $\delta\hat{F}=\delta[i(\overline{\Phi}-\Phi)]=0$ for $\delta\Phi=k$. This invariance excludes mass terms, and has consequences even for the free massless multiplet when it is coupled to supergravity.)

## b. Vector multiplet

As an introduction, we describe the abelian vector multiplet in the language of superforms. We begin with a real super 1 −form

$$\Gamma_1=\omega^\alpha\Gamma_\alpha+\overline{\omega}^{\dot{\alpha}}\overline{\Gamma}_{\dot{\alpha}}+\omega^{\underline{a}}\Gamma_{\underline{a}} \quad , \tag{4.4.13}$$

with gauge transformation $\delta\Gamma_1=dK_0$, where $K_0$ is a 0 −form (scalar). The field strength is a super 2 −form $F_2=d\Gamma_1$, with superfield coefficients that follow from (4.4.11):

$$F_{\alpha,\beta}=D_{(\alpha}\Gamma_{\beta)} \quad ,$$

$$F_{\alpha,\dot{\beta}}=D_\alpha\overline{\Gamma}_{\dot{\beta}}+\overline{D}_{\dot{\beta}}\Gamma_\alpha-i\Gamma_{\alpha\dot{\beta}} \quad ,$$

$$F_{\alpha,\underline{b}}=D_\alpha\Gamma_{\underline{b}}-\partial_{\underline{b}}\Gamma_\alpha \quad ,$$

$$F_{\underline{a},\underline{b}}=-\frac{1}{2}C_{\dot{\alpha}\dot{\beta}}\partial_{(\alpha}{}^{\dot{\gamma}}\Gamma_{\beta)\dot{\gamma}}+h.c. \quad . \tag{4.4.14}$$

We impose a conventional constraint $F_{\alpha,\dot{\beta}}=0$ which algebraically determines $\Gamma_{\alpha\dot{\alpha}}$. We further restrict the form by imposing the constraint (4.2.44) $F_{\alpha,\beta}=0$. The solution to the constraints is

$$\Gamma_\alpha = iD_\alpha\Omega \quad , \quad \bar{\Gamma}_{\dot\alpha} = -\, i\bar{D}_{\dot\alpha}\bar{\Omega} \quad ,$$

$$\Gamma_{\underline{a}} = -\, i(D_\alpha\bar{\Gamma}_{\dot\alpha} + \bar{D}_{\dot\alpha}\Gamma_\alpha) \quad , \tag{4.4.15}$$

and the prepotential $\Omega$ transforms as

$$\delta\Omega = -\, iK_0 + \bar{\Lambda} \quad . \tag{4.4.16}$$

The $\Lambda$ transformations are an invariance of $\Gamma_1$ introduced by solving the constraints.

It is always obvious, by examining equations such as (4.4.14), what conventional constraints can be imposed. Finding additional constraints is more difficult. In general, if we wish to describe a multiplet that contains a *component* $p$-form, we require that it be the $\theta = 0$ component of a super $p$-form coefficient with only vector indices (e.g., in (4.4.14) $F_{\underline{a},\underline{b}}|$ is the Yang-Mills field strength), and therefore we will not constrain this coefficient. For the same reason we assign dimension 2 to this coefficient, and this determines the dimension of the superform. As a consequence, coefficients with more than two spinor indices have too low dimension to contain component field strengths (or auxiliary fields), and must be constrained to zero. We also constrain to zero coefficients that contain at the $\theta = 0$ level component forms that are not present in the multiplet.

## c. Tensor multiplet

### c.1. Geometric formulation

The antisymmetric-tensor gauge multiplet contains among its component fields a second-rank antisymmetric tensor (2-form). To describe it in superspace we consider a super 2-form $\Gamma_2$:

$$-\Gamma_2 = \frac{1}{2}\,\omega^\beta\wedge\omega^\alpha\Gamma_{\alpha,\beta} + \bar{\omega}^{\dot\beta}\wedge\omega^\alpha\Gamma_{\alpha,\dot\beta} + \omega^{\underline{b}}\wedge\omega^\alpha\Gamma_{\alpha,\underline{b}} + \frac{1}{2}\,\omega^{\underline{b}}\wedge\omega^{\underline{a}}C_{\dot\alpha\dot\beta}\Gamma_{(\alpha\beta)} + h.\,c. \quad , \tag{4.4.17}$$

where we have used the symmetries of $\Gamma$ to write $\Gamma_{\underline{ab}} = C_{\dot\alpha\dot\beta}\Gamma_{(\alpha\beta)} + h.\,c.$. The gauge variations $\delta\Gamma_2 = dK_1$ are

$$\delta\Gamma_{\alpha,\beta} = D_{(\alpha}K_{\beta)} \quad ,$$

$$\delta\Gamma_{\alpha,\dot\beta} = D_\alpha\bar{K}_{\dot\beta} + \bar{D}_{\dot\beta}K_\alpha - iK_{\alpha\dot\beta} \quad ,$$

$$\delta\Gamma_{\alpha,\underline{b}} = D_\alpha K_{\underline{b}} - \partial_{\underline{b}} K_\alpha \quad ,$$

$$\delta\Gamma_{(\alpha\beta)} = -\frac{1}{2}\partial_{(\alpha}{}^{\dot\gamma} K_{\beta)\dot\gamma} \quad . \qquad (4.4.18)$$

The field strengths follow from the definition (4.4.11):

$$F_{\alpha,\beta,\gamma} = \frac{1}{2} D_{(\alpha}\Gamma_{\beta,\gamma)} \quad ,$$

$$F_{\alpha,\beta,\dot\gamma} = D_{(\alpha}\Gamma_{\beta)\dot\gamma} + \bar{D}_{\dot\gamma}\Gamma_{\alpha,\beta} + i\Gamma_{(\alpha,\beta)\dot\gamma} \quad ,$$

$$F_{\alpha,\beta,\underline{c}} = D_{(\alpha}\Gamma_{\beta)\underline{c}} + \partial_{\underline{c}}\Gamma_{\alpha,\beta} \quad ,$$

$$F_{\alpha,\dot\beta,\underline{c}} = D_\alpha\bar\Gamma_{\dot\beta,\underline{c}} + \bar{D}_{\dot\beta}\Gamma_{\alpha,\underline{c}} + \partial_{\underline{c}}\Gamma_{\alpha,\dot\beta} - iC_{\dot\beta\dot\gamma}\Gamma_{(\alpha\gamma)} - iC_{\alpha\gamma}\Gamma_{(\dot\beta\dot\gamma)} \quad ,$$

$$F_{\alpha,\underline{b},\underline{c}} = C_{\dot\beta\dot\gamma}(D_\alpha\Gamma_{(\beta\gamma)} - \frac{1}{2}\partial_{(\beta}{}^{\dot\delta}\Gamma_{\gamma)\dot\delta,\alpha}) + C_{\beta\gamma}(D_\alpha\Gamma_{(\dot\beta\dot\gamma)} + \frac{1}{2}\partial_{\delta(\dot\beta}\Gamma^{\delta}{}_{\dot\gamma),\alpha}) \quad ,$$

$$F_{\underline{a},\underline{b},\underline{c}} \equiv -\epsilon_{\underline{abcd}}F^{\underline{d}} = -i(C_{\dot\alpha\dot\gamma}C_{\beta\gamma}F_{\alpha\dot\beta} - C_{\alpha\gamma}C_{\dot\beta\dot\gamma}F_{\beta\dot\alpha}) \quad ,$$

$$F_{\alpha\dot\beta} = -i(\partial_\alpha{}^{\dot\gamma}\bar\Gamma_{(\dot\beta\dot\gamma)} - \partial^\gamma{}_{\dot\beta}\Gamma_{(\alpha\gamma)}) \quad . \qquad (4.4.19)$$

where we have used (3.1.22).

We can impose two conventional constraints. The first,

$$F_{\alpha,\beta,\dot\gamma} = 0 \quad , \qquad (4.4.20)$$

gives

$$\Gamma_{(\alpha,\beta)\dot\beta} = i[D_{(\alpha}\Gamma_{\beta)\dot\beta} + \bar{D}_{\dot\beta}\Gamma_{\alpha,\beta}] \quad , \qquad (4.4.21)$$

which implies

$$\Gamma_{\alpha,\beta\dot\beta} = i\,C_{\alpha\beta}\bar\Phi_{\dot\beta} + i[D_\alpha\Gamma_{\beta,\dot\beta} + \frac{1}{2}\bar{D}_{\dot\beta}\Gamma_{\alpha,\beta}] \quad , \qquad (4.4.22)$$

for an arbitrary spinor $\bar\Phi_{\dot\gamma}$. The second conventional constraint,

$$F_{(\alpha,\dot\beta,\beta)}{}^{\dot\beta} = 0 \quad , \qquad (4.4.23)$$

gives

$$\Gamma_{(\alpha\beta)} = -\, i\,\frac{1}{4}\,[D_{(\alpha}\bar{\Gamma}_{\dot{\beta},\beta)}{}^{\dot{\beta}} + \bar{D}_{\dot{\beta}}\Gamma_{(\alpha,\beta)}{}^{\dot{\beta}} - \partial_{(\alpha\dot{\beta}}\Gamma_{\beta)}{}^{\dot{\beta}}] \quad , \tag{4.4.24}$$

which implies

$$\Gamma_{(\alpha\beta)} = \frac{1}{2}\,D_{(\alpha}\Phi_{\beta)} - \frac{1}{2}\,\bar{D}^2\Gamma_{\alpha,\beta} - \frac{1}{2}\,i\partial_{(\alpha}{}^{\dot{\beta}}\Gamma_{\beta),\dot{\beta}} \quad . \tag{4.4.25}$$

The potential $\Gamma_{\alpha,\dot{\beta}}$ is pure gauge: It can be gauged to zero using (4.4.18). To eliminate the remaining unwanted physical states we choose two additional constraints

$$F_{\alpha,\beta,\gamma} = F_{\alpha,\beta,\underline{c}} = 0 \quad . \tag{4.4.26}$$

The first implies $\Gamma_{\alpha,\beta}$ is pure gauge, and the second imposes

$$D_{\alpha}\bar{\Phi}_{\dot{\beta}} = 0 \quad , \qquad \bar{D}_{\dot{\alpha}}\Phi_{\beta} = 0 \quad . \tag{4.4.27}$$

In the gauge $\Gamma_{\alpha,\beta} = \Gamma_{\alpha,\dot{\beta}} = 0$, all of $\Gamma_{AB}$ is expressed in terms of $\Phi_{\alpha}$; thus the superfield $\Phi_{\alpha}$ is the chiral spinor prepotential that describes the tensor gauge multiplet.

The constraints also imply that all the nonvanishing field strengths can be expressed in terms of a single independent field strength

$$G = -\,\frac{1}{2}\,(D^{\alpha}\Phi_{\alpha} + \bar{D}^{\dot{\alpha}}\bar{\Phi}_{\dot{\alpha}}) \quad . \tag{4.4.28}$$

For example,

$$F_{\alpha,\dot{\beta}}{}^{\underline{c}} = i\,\delta_{\alpha}{}^{\gamma}\delta_{\dot{\beta}}{}^{\dot{\gamma}}G = T_{\alpha\dot{\beta}}{}^{\underline{c}}G \quad . \tag{4.4.29}$$

$G$ is a linear superfield: $D^2G = 0$. It is invariant under gauge transformations of the prepotential

$$\delta\Phi_{\alpha} = i\bar{D}^2D_{\alpha}L \quad , \quad L = \bar{L} \quad . \tag{4.4.30}$$

Projecting the components of $\Phi_{\alpha}$ we have:

$$\begin{aligned} &\chi_{\alpha} = \Phi_{\alpha}| \quad , && \mathrm{A} + i\mathrm{B} = -\,D^{\alpha}\Phi_{\alpha}| \quad , \\ &t_{\alpha\beta} = \frac{1}{2}\,D_{(\alpha}\Phi_{\beta)}| = \Gamma_{(\alpha\beta)}| \quad , && \tilde{\psi}_{\alpha} = D^2\Phi_{\alpha}| \quad . \end{aligned} \tag{4.4.31}$$

The components of the gauge parameter that enter $\delta\Phi_{\alpha}$ are:

$$L_{\alpha} = i\bar{D}^2D_{\alpha}L| \quad ,$$

$$L^{(1)} = D^\alpha \bar{D}^2 D_\alpha L| = \bar{L}^{(1)} \quad ,$$

$$L_{(\alpha,\beta)} = i\,\frac{1}{2}\,D_{(\alpha}\bar{D}^2 D_{\beta)}L| = \frac{1}{2}\,\partial_{(\beta}{}^{\dot\alpha}[D_{\alpha)}\,,\bar{D}_{\dot\alpha}]L|$$

$$\equiv -\,\frac{1}{2}\,\partial_{(\beta}{}^{\dot\alpha}L_{\alpha)\dot\alpha} \quad , \quad L_{\alpha\dot\alpha} = \bar{L}_{\alpha\dot\alpha} \quad . \tag{4.4.32}$$

The components $\chi_\alpha$ and B can be algebraically gauged away by $L_\alpha$ and $L^{(1)}$ respectively, whereas $L_{\alpha\dot\alpha}$ is the parameter of the usual gauge transformation for the tensor gauge field $t_{\alpha\beta}$. The spinor $\tilde{\psi}_\alpha$ is the physical spinor of the theory (up to terms that vanish in the WZ gauge). The gauge invariant components are found by projecting from the field strength $G$:

$$\mathrm{A} = G| \quad ,$$

$$\psi_\alpha = D_\alpha G| = \frac{1}{2}\,(\tilde{\psi}_\alpha - i\partial_\alpha{}^{\dot\alpha}\bar{\chi}_{\dot\alpha}) \quad ,$$

$$f_{\underline{a}} = F_{\underline{a}}| = [\bar{D}_{\dot\alpha}\,, D_\alpha]G| = i(\partial^\beta{}_{\dot\alpha}t_{\alpha\beta} - \partial_\alpha{}^{\dot\beta}\bar{t}_{\dot\alpha\dot\beta}) \quad ,$$

$$D^2 G = \bar{D}^2 G = 0 \quad . \tag{4.4.33}$$

Since there is only one physical spinor in the multiplet, $G$ has dimension one. This determines the kinetic action uniquely:

$$S_k = -\,\frac{1}{2}\int d^4x\, d^4\theta\; G^2 \quad . \tag{4.4.34}$$

The corresponding component action is

$$S_k = \int d^4x\,[\frac{1}{4}\,\mathrm{A}\Box\mathrm{A} + \frac{1}{4}\,(f^{\underline{a}})^2 + \bar{\psi}^{\dot\alpha}i\partial^\alpha{}_{\dot\alpha}\psi_\alpha] \quad . \tag{4.4.35}$$

Note that none of the fields is auxiliary. The physical degrees of freedom are those of the scalar multiplet. On shell, the only difference is the replacement of the physical pseudoscalar by the field strength of the antisymmetric tensor.

### c.2. Duality transformation to chiral multiplet

We can write two first order actions that are equivalent to $S_k$. Introducing an auxiliary superfield $X$, we define

$$S'_k = \int d^4x\, d^4\theta\, [\tfrac{1}{2} X^2 - GX] \quad . \tag{4.4.36a}$$

Varying $X$ and substituting the result back into $S'_k$, we reobtain $S_k$. We also see that the tensor multiplet is classically equivalent to a chiral scalar multiplet: Varying $\Phi^\alpha$, we obtain $\bar{D}^2 D_\alpha X = 0$, which is solved by $X = \chi + \bar{\chi}$, $\bar{D}_{\dot{\alpha}}\chi = 0$. Substitution back into $S'_k$ yields the usual kinetic action for a chiral scalar $\chi$ (because $\chi$ is chiral and $G$ is linear, $\int d^4x\, d^4\theta\, \chi G = 0$). Because the same first order action can be used to describe the tensor multiplet and the chiral scalar multiplet, we say that they are *dual* to each other.

Alternatively, we can write

$$S''_k = \int d^4x\, d^4\theta\, [-\tfrac{1}{2} X^2 + (\chi + \bar{\chi})X] \quad . \tag{4.4.36b}$$

Varying $X$ and substituting the result back into $S''_k$, we obtain the usual kinetic action for the chiral scalar $\chi$; varying $\chi, \bar{\chi}$, we find $D^2 X = \bar{D}^2 X = 0$, which is solved by $X = G$. Substitution back into $S''$ yields $S_k$ (4.4.34).

The tensor multiplet admits arbitrary (nonrenormalizable) self-interactions with a dimensional coupling constant $\mu$:

$$S = \mu^2 \int d^4x\, d^4\theta\, f(\mu^{-1}G) \quad . \tag{4.4.37}$$

The component action contains quartic fermion self-interactions and "Yukawa" terms $\psi^\alpha \bar{\psi}^{\dot{\alpha}} F_{\alpha\dot{\alpha}}$, multiplied by derivatives of $f(\mu^{-1}A)$. Remarkably, we can perform the duality transformation to a chiral scalar multiplet even in the interacting theory. The first order action equivalent to $S$ is:

$$S' = \mu^2 \int d^4x\, d^4\theta\, [f(X) - \mu^{-1}(\chi + \bar{\chi})X] \quad . \tag{4.4.38}$$

Varying $\chi, \bar{\chi}$, we find $X = \mu^{-1}G$ (the normalization can be chosen arbitrarily), and reobtain the interacting action (4.4.37). Varying $X$, we find the dual action in terms of $\chi, \bar{\chi}$:

$$\tilde{S} = \mu^2 \int d^4x\, d^4\theta\; I\!K(\mu^{-1}(\chi + \bar{\chi})) \tag{4.4.39}$$

where $I\!K$ is the *Legendre transform* of $f$:

$$I\!K(\mu^{-1}(\chi + \bar{\chi})) = f(X(\mu^{-1}(\chi + \bar{\chi}))) - \mu^{-1}(\chi + \bar{\chi})X(\mu^{-1}(\chi + \bar{\chi})) \quad ,$$

$$\frac{\partial f(X)}{\partial X} \equiv \mu^{-1}(\chi + \bar{\chi}) \quad . \tag{4.4.40}$$

The dual action (4.4.39) is recognizable as the action for a nonlinear $\sigma$-model (see sec. 4.1.b, e.g. (4.1.23)).

We can also perform the reverse duality transformation, that is, start with a theory described by a chiral scalar superfield and find an equivalent theory described by a tensor multiplet. Although we can find the model dual to an arbitrary tensor multiplet model, the reverse is not true: For a chiral scalar model, possibly with interactions to other chiral and/or gauge multiplets, we can find the dual tensor model only if the original action depends only on $\chi + \bar{\chi}$, or equivalently, defining $\eta \equiv \mu e^{\mu^{-1}\chi}$, on $\eta\bar{\eta}$. Thus, starting with an action

$$S_\chi = \mu^2 \int d^4x\, d^4\theta\; I\!K(\mu^{-1}(\chi + \bar{\chi})) \tag{4.4.41}$$

we can write the first order action

$$S'' = \mu^2 \int d^4x\, d^4\theta\; [I\!K(X) + \mu^{-1}GX\,] \tag{4.4.42}$$

Varying $G$ yields $X = \mu^{-1}(\chi + \bar{\chi})$ and (4.4.41), whereas varying $X$ leads to (4.4.37), where now $f$ is the (inverse) Legendre transform of $I\!K$:

$$f(\mu^{-1}G) = I\!K(X(\mu^{-1}G)) + \mu^{-1}GX(\mu^{-1}G) \quad ,$$

$$\frac{\partial I\!K(X)}{\partial X} = -\,\mu^{-1}G \quad . \tag{4.4.43}$$

We can now find a second tensor multiplet model dual to the *free* chiral scalar multiplet. We begin with

$$S_\eta = \mu^2 \int d^4x\, d^4\theta\; \bar{\eta}\eta = \int d^4x\, d^4\theta\; e^{\mu^{-1}(\chi + \bar{\chi})} \tag{4.4.44}$$

We write this in first order form as

$$S'_{imp} = \mu^2 \int d^4x\, d^4\theta\, [e^X - GX] \ , \tag{4.4.45}$$

and find the dual action

$$S_{imp} = -\mu^2 \int d^4x\, d^4\theta\, G\, ln\, G \ , \tag{4.4.46}$$

where now $G$ has a nonvanishing classical vacuum expectation value. This duality holds even in the presence of supergravity, where the equivalence is to the superconformal form of the scalar multiplet $(\bar{\eta}\eta)$, as opposed to the $(\chi + \bar{\chi})^2$ form obtained from (4.4.36); in general curved superspace, these two Lagrangians are different. The model described by the action $S_{imp}$ (4.4.46) is called the improved tensor multiplet, because, unlike the unimproved action (4.4.34), $S_{imp}$ is conformally invariant. (Both are globally scale invariant, but the action for an antisymmetric tensor by itself is not invariant under conformal boosts.)

It is interesting to study what happens to the interactions of a chiral multiplet after a duality transformation. Here we consider interactions with a gauge vector multiplet (for other examples, see secs. 4.5e, 4.6, 5.5). For an action of the form

$$S_{gauge} = \int d^4x\, d^4\theta\, I\!K(\chi + \bar{\chi} + V) + \int d^4x\, d^2\theta\, W^2 \tag{4.4.47}$$

where $V$ is an abelian gauge superfield, $W$ is its field strength, and $I\!K(\chi + \bar{\chi} + V) \equiv I\!K(ln(\bar{\eta}e^V\eta))$, we can write the first order action

$$S'_{gauge} = \int d^4x\, d^4\theta\, [I\!K(X + V) + GX] + \int d^4x\, d^2\theta\, W^2 \ . \tag{4.4.48}$$

Varying $G$ gives (4.4.47); varying $X$ gives

$$S'_G = \int d^4x\, d^4\theta\, [f(G) - GV] + \int d^4x\, d^2\theta\, W^2 \ . \tag{4.4.49}$$

Thus the gauge interactions of the original theory are described by the single term $GV$ in the dual theory (this coupling is gauge invariant because $G$ is linear). Observe that for the usual kinetic term $I\!K = \bar{\eta}e^V\eta$, the dual theory has the improved Lagrangian $-G\, ln\, G - GV$ (4.4.46) rather than $-\frac{1}{2}G^2 - GV$ (4.4.34). It is straightforward to verify that the latter theory describes a massive vector multiplet rather than a scalar coupled to a vector. Another way to describe a massive vector multiplet, but without vector fields, is in terms of the chiral spinor $\Phi_\alpha$ alone by adding a mass term (which breaks the gauge invariance (4.4.30)) to $S_k$ (4.4.34):

$$S_m = -\frac{1}{2}m^2 \int d^4x\, d^2\theta\, (\Phi_\alpha)^2 + h.\,c.\ \ . \tag{4.4.50}$$

$S_k + S_m$ describes a massive vector multiplet. The component antisymmetric tensor describes a massive spin 1 field, $\chi_\alpha$ and $\psi_\alpha$ describe a massive Dirac spinor, A is a massive scalar, and B is auxiliary.

## d. Gauge 3-form multiplets

### d.1. Real 3-form

We begin by considering a *real* 3-form. It has the following independent coefficient superfields

$$\Gamma_{\alpha,\beta,\gamma}\ \ ,\quad \Gamma_{\alpha,\beta,\dot\gamma}\ \ ,\quad \Gamma_{\alpha,\beta,\underline{c}}\ \ ,\quad \Gamma_{\alpha,\dot\beta,\underline{c}}\ \ ,$$

$$\Gamma_{\alpha,(\beta\gamma)}\ \ ,\quad \Gamma_{\alpha,(\dot\beta\dot\gamma)}\ \ ,\quad \Gamma_{\underline{a}}\ \ , \tag{4.4.51}$$

where we have used the symmetries of $\Gamma$ to write it in terms of Lorentz irreducible coefficients.

$$\Gamma_{\alpha,\underline{b},\underline{c}} = C_{\dot\beta\dot\gamma}\Gamma_{\alpha,(\beta\gamma)} + C_{\beta\gamma}\Gamma_{\alpha,(\dot\beta\dot\gamma)}\ \ ,$$

$$\Gamma_{\underline{a},\underline{b},\underline{c}} = -\,\epsilon_{\underline{abcd}}\Gamma^{\underline{d}} = -\,i(C_{\dot\alpha\dot\gamma}C_{\beta\gamma}\Gamma_{\alpha\dot\beta} - C_{\alpha\gamma}C_{\dot\beta\dot\gamma}\Gamma_{\beta\dot\alpha})\ \ , \tag{4.4.52}$$

The independent field strengths are

$$F_{\alpha,\beta,\gamma,\delta}\ \ ,\quad F_{\alpha,\beta,\gamma,\dot\delta}\ \ ,\quad F_{\alpha,\beta,\dot\gamma,\dot\delta}\ \ ,$$

$$F_{\alpha,\beta,\gamma,\underline{d}}\ \ ,\quad F_{\alpha,\beta,\dot\gamma,\underline{d}}\ \ ,$$

$$F_{\alpha,\beta,(\gamma\delta)}\ \ ,\quad F_{\alpha,\beta,(\dot\gamma\dot\delta)}\ \ ,\quad F_{\alpha,\dot\beta,(\gamma\delta)}\ \ ,$$

$$F_{\alpha,\underline{b}} \quad , \quad F \quad . \tag{4.4.53}$$

The last five field strengths are Lorentz irreducible coefficients, e.g. (see 3.1.22)),

$$F_{\underline{a},\underline{b},\underline{c},\underline{d}} = F\,\epsilon_{\underline{a},\underline{b},\underline{c},\underline{d}} = i(C_{\alpha\delta}C_{\beta\gamma}C_{\dot{\alpha}\dot{\gamma}}C_{\dot{\beta}\dot{\delta}} - C_{\alpha\gamma}C_{\beta\delta}C_{\dot{\alpha}\dot{\delta}}C_{\dot{\beta}\dot{\gamma}})F \quad . \tag{4.4.54}$$

We impose the following constraints on the field strengths:

$$F_{\alpha,\beta,\gamma,\delta} = F_{\alpha,\beta,\gamma,\dot{\delta}} = F_{\alpha,\beta,\dot{\gamma},\dot{\delta}} = 0 \quad ,$$

$$F_{\alpha,\beta,\gamma,\underline{d}} = F_{\alpha,\beta,\dot{\gamma},\underline{d}} = F_{\alpha,\beta,(\dot{\gamma}\dot{\delta})} = F_{\alpha,\dot{\beta},(\gamma\delta)} = 0 \quad ,$$

$$F_{\alpha,\beta,(\gamma\delta)} = 2C_{\alpha(\gamma}C_{\delta)\beta}\overline{\Pi} \quad , \tag{4.4.55}$$

where $\overline{\Pi}$ is an undetermined gauge invariant superfield. Solving the constraints gives

$$\Gamma_{\alpha,\beta,\gamma} = \Gamma_{\alpha,\beta,\dot{\gamma}} = \Gamma_{\alpha,\beta,\underline{c}} = \Gamma_{\alpha,(\dot{\beta}\dot{\gamma})} = 0 \quad ,$$

$$\Gamma_{\alpha,\dot{\beta},\underline{c}} = iC_{\alpha\gamma}C_{\dot{\beta}\dot{\gamma}}V \quad ,$$

$$\Gamma_{\alpha,(\beta\gamma)} = -\,C_{\alpha(\beta}D_{\gamma)}V \quad ,$$

$$\Gamma_{\alpha\dot{\alpha}} = [\overline{D}_{\dot{\alpha}}\,,D_{\alpha}]V \quad , \quad V = \overline{V} \quad , \tag{4.4.56}$$

up to a pure gauge transformation of $\Gamma_{ABC}$. Given the solution, we find

$$\overline{\Pi} = D^2V \quad . \tag{4.4.57}$$

The prepotential $V$ has gauge transformations

$$\delta V = -\,\frac{1}{2}(D^{\alpha}\omega_{\alpha} + \overline{D}^{\dot{\alpha}}\overline{\omega}_{\dot{\alpha}}) \quad , \quad D_{\alpha}\overline{\omega}_{\dot{\alpha}} = 0 \quad . \tag{4.4.58}$$

The physical component fields of this multiplet are

$$\overline{\phi} = \overline{\Pi}| = D^2V| \quad , \quad \psi_{\alpha} = D_{\alpha}\Pi| = D_{\alpha}\overline{D}^2V \quad ,$$

$$h = (D^2\Pi + \overline{D}^2\overline{\Pi})| = \{D^2\,,\overline{D}^2\}V| \quad ,$$

$$f = -\,i(D^2\Pi - \overline{D}^2\overline{\Pi})| = \frac{1}{2}\,\partial_{\underline{a}}\Gamma^{\underline{a}}| = \frac{1}{2}\,\partial^{\alpha\dot{\alpha}}[\overline{D}_{\dot{\alpha}}\,,D_{\alpha}]V| \quad . \tag{4.4.59}$$

The quantity $f$ is the field strength of the component gauge three-form $l_{\alpha\dot{\alpha}} = \Gamma_{\alpha\dot{\alpha}}|$. The component three-form transforms as (cf. (4.4.33) for $f_{\underline{a}}$)

$$\delta l_{\alpha\dot{\alpha}} = i\,\frac{1}{2}\,(\partial^{\beta}{}_{\dot{\alpha}} D_{(\beta}\omega_{\alpha)} - \partial_{\alpha}{}^{\dot{\beta}} \bar{D}_{(\dot{\beta}}\bar{\omega}_{\dot{\alpha})})| \quad , \tag{4.4.60}$$

so that its field strength $f$ is invariant.

The field strength $\Pi$ is a chiral field of dimension one (determined by $\psi_{\alpha}$), and hence the kinetic action is

$$S = \int d^4x\, d^4\theta\, \bar{\Pi}\Pi \quad . \tag{4.4.61}$$

It gives conventional kinetic terms for the components $\phi$ and $\psi_{\alpha}$; the scalar field $h$ is an auxiliary field and the gauge field $l_{\alpha\dot{\alpha}}$ enters the action through the square of its field strength $f$. Such a field does not propagate physical states in four dimensions.

The only difference between this multiplet, described by $\Pi$, and the usual chiral scalar multiplet $\Phi$ is the replacement of the imaginary part (the pseudoscalar field) of the $F$ auxiliary field by the field strength of the component gauge three-form. Mass and interaction terms for $\Phi$ can also be used for $\Pi$. However, at the component level, after elimination of the auxiliary fields the theories differ: We no longer obtain algebraic equations, since $f$ is the derivative of another field $l_{\alpha\dot{\alpha}}$. Another difference is that the super three-form gauge multiplet *cannot* be coupled to Yang-Mills multiplets.

### d.2. Complex 3-form

A complex super three-form multiplet can be treated in the same way. It has more independent coefficient superfields:

$$\Gamma_{\alpha,\beta,\gamma} \quad , \quad \Gamma_{\alpha,\beta,\dot{\gamma}} \quad , \quad \Gamma_{\alpha,\dot{\beta},\dot{\gamma}} \quad , \quad \Gamma_{\dot{\alpha},\dot{\beta},\dot{\gamma}} \quad ,$$

$$\Gamma_{\alpha,\beta,\underline{c}} \quad , \quad \Gamma_{\alpha,\dot{\beta},\underline{c}} \quad , \quad \Gamma_{\dot{\alpha},\dot{\beta},\underline{c}} \quad ,$$

$$\Gamma_{\alpha,(\beta\gamma)} \quad , \quad \Gamma_{\alpha,(\dot{\beta}\dot{\gamma})} \quad , \quad \Gamma_{\dot{\alpha},(\beta\gamma)} \quad , \quad \Gamma_{\dot{\alpha},(\dot{\beta}\dot{\gamma})} \quad ,$$

$$\Gamma_{\underline{a}} \quad . \tag{4.4.62}$$

(For example, $\Gamma_{\alpha(\dot{\beta}\dot{\gamma})} \neq \overline{\Gamma_{\dot{\alpha}(\beta\gamma)}}$.) Correspondingly, there are more independent field

strengths. These are

$$F_{\alpha,\beta,\gamma,\delta} \quad , \quad F_{\alpha,\beta,\gamma,\dot{\delta}} \quad , \quad F_{\alpha,\beta,\dot{\gamma},\dot{\delta}} \quad , \quad F_{\alpha,\dot{\beta},\dot{\gamma},\dot{\delta}} \quad , \quad F_{\dot{\alpha},\dot{\beta},\dot{\gamma},\dot{\delta}} \quad ,$$

$$F_{\alpha,\beta,\gamma,\underline{d}} \quad , \quad F_{\alpha,\beta,\dot{\gamma},\underline{d}} \quad , \quad F_{\alpha,\dot{\beta},\dot{\gamma},\underline{d}} \quad , \quad F_{\dot{\alpha},\dot{\beta},\dot{\gamma},\underline{d}} \quad ,$$

$$F_{\alpha,\beta,(\gamma\delta)} \quad , \quad F_{\alpha,\beta,(\dot{\gamma}\dot{\delta})} \quad , \quad F_{\alpha,\dot{\beta},(\gamma\delta)} \quad , \quad F_{\alpha,\dot{\beta},(\dot{\gamma}\dot{\delta})} \quad , \quad F_{\dot{\alpha},\dot{\beta},(\gamma\delta)} \quad , \quad F_{\dot{\alpha},\dot{\beta},(\dot{\gamma}\dot{\delta})} \quad ,$$

$$F_{\alpha,\underline{b}} \quad , \quad F_{\dot{\alpha},\underline{b}} \quad , \quad F \quad , \tag{4.4.63}$$

The constraints however, set more field strengths to zero. The nonzero ones are

$$F_{\alpha,\beta,(\gamma\delta)} \quad , \quad F_{\alpha,\underline{b}} \quad , \quad F \quad , \tag{4.4.64}$$

and we still impose the constraint:

$$F_{\alpha,\beta,(\gamma\delta)} = 2C_{\alpha(\gamma}C_{\delta)\beta}\overline{\Pi} \quad . \tag{4.4.65}$$

The only form coefficients that are not pure gauge are given by

$$\Gamma_{\alpha,(\beta\gamma)} = -\,C_{\alpha(\beta}D_{\gamma)}\overline{D}^{\dot{\epsilon}}\overline{\Psi}_{\dot{\epsilon}} \quad ,$$

$$\Gamma_{\dot{\alpha},(\dot{\beta}\dot{\gamma})} = -\,C_{\dot{\alpha}(\dot{\beta}}\overline{D}_{\dot{\gamma})}\overline{D}^{\dot{\epsilon}}\overline{\Psi}_{\dot{\epsilon}} = C_{\dot{\alpha}(\dot{\beta}}\overline{D}^2\overline{\Psi}_{\dot{\gamma})} \quad ,$$

$$\Gamma_{\alpha,\dot{\beta},\underline{c}} = iC_{\alpha\gamma}C_{\dot{\beta}\dot{\gamma}}\overline{D}^{\dot{\epsilon}}\overline{\Psi}_{\dot{\epsilon}} \quad ,$$

$$\Gamma_{\underline{a}} = [\overline{D}_{\dot{\alpha}}\, , D_{\alpha}]\overline{D}^{\dot{\epsilon}}\overline{\Psi}_{\dot{\epsilon}} \quad , \tag{4.4.66}$$

(up to arbitrary gauge transformation terms). These expressions allow us to compute $\Pi$; we find that it is expressed in terms of the prepotential $\Psi_\alpha$ as follows:

$$\Pi = \overline{D}^2 D^\alpha \Psi_\alpha \quad , \quad \overline{D}_{\dot{\alpha}}\Pi = 0 \quad . \tag{4.4.67}$$

The gauge transformations of the prepotential $\Psi_\alpha$ are

$$\delta\Psi_\alpha = \Lambda_\alpha + D^\beta L_{(\alpha\beta)} \quad , \quad \overline{D}_{\dot{\alpha}}\Lambda_\beta = 0 \quad . \tag{4.4.68}$$

The components contained in the field strength are

$$A = \Pi| = \overline{D}^2 D^\alpha \Psi_\alpha| \quad ,$$

$$\zeta_\alpha = D_\alpha \Pi| = D_\alpha \bar{D}^2 D^\beta \Psi_\beta| \quad ,$$

$$f = D^2 \Pi| = \frac{i}{2} \partial^{\alpha\dot{\alpha}} [\bar{D}_{\dot{\alpha}} , D_\alpha] D^\beta \Psi_\beta| \quad , \tag{4.4.69}$$

where $f$ is the field strength of a *complex* 3-form.

This multiplet can be described in terms of two real super 3-form multiplets: $\Gamma = \Gamma_1 + i\Gamma_2$. The constraints imposed above are the ones given in sec. 4.4.d.1 for $\Gamma_1$ and $\Gamma_2$, plus the additional constraint $F_{\dot{\alpha},\dot{\beta},(\dot{\gamma}\dot{\delta})} = 0$. This is simply the constraint $\Pi_1 + i\Pi_2 = \bar{D}^2(V_1 + iV_2) = 0$, which implies $V_1 + iV_2 = \bar{D}^{\dot{\epsilon}}\bar{\Psi}_{\dot{\epsilon}}$.

The field strength $\Pi$ is chiral and of dimension one. Therefore all of the action formulae for the usual chiral scalar can be used for $\Pi$. As for the real gauge three-form multiplet, the equations of motion for the auxiliary fields are no longer purely algebraic. Again, this multiplet cannot be coupled to Yang-Mills multiplets.

## e. 4-form multiplet

The final superform we consider has *no* physical degrees of freedom. It is described by a real super 4-form $\Gamma_{ABCD}$. The field strength supertensor is a super 5-form $F_{ABCDE}$. Therefore the field strength with all five vector indices vanishes by antisymmetry.

As constraints we "impose" the equations $F_{ABCDE} = 0$. This implies that all of $\Gamma_{ABCD}$ is pure gauge. Since all field strengths vanish, no gauge invariant action is possible at the classical level. However, this multiplet (and the corresponding component form) has some unusual properties at the quantum level, because its gauge fixing term is not zero.

## 4.5. Other gauge multiplets

### a. Gauge Wess-Zumino model

In (3.5.3) we noted that a chiral superfield can be expressed in terms of an unconstrained superfield

$$\Phi = \bar{D}^2 \Psi \quad . \tag{4.5.1}$$

The field $\Psi$ provides an alternate description of the scalar multiplet. The actions we considered in secs. 4.1-2 can be expressed in terms of $\Psi$. For example, the Wess-Zumino action (4.1.1-2) becomes

$$S = \int d^4x\, d^4\theta\, [(D^2\bar{\Psi})(\bar{D}^2\Psi) + \frac{1}{2}m(\Psi\bar{D}^2\Psi + \bar{\Psi}D^2\bar{\Psi})$$

$$+ \frac{\lambda}{3!}(\Psi(\bar{D}^2\Psi)^2 + \bar{\Psi}(D^2\bar{\Psi})^2)] \quad , \tag{4.5.2a}$$

where we have used

$$\int d^4x\, d^2\theta\, (\bar{D}^2\Psi)^2 = \int d^4x\, d^4\theta\, \Psi\bar{D}^2\Psi \quad , \tag{4.5.2b}$$

etc.

The solution (4.5.1) of the chirality constraint introduces the abelian gauge invariance

$$\delta\Psi = \bar{D}^{\dot{\alpha}}\bar{\omega}_{\dot{\alpha}} \tag{4.5.3}$$

where $\omega_\alpha$ is an unconstrained superfield. The gauge invariant superfield $\Phi$ is the chiral field strength of the gauge superfield $\Psi$, and the action is obviously invariant. The gauge transformation can be used to go to a WZ gauge, by algebraically removing all the components of $\Psi$ except those that appear in $\Phi$. In this formulation the coupling to super Yang-Mills can be achieved by covariantizing the derivatives: If $\Phi$ is covariantly chiral, then $\Phi = \bar{\nabla}^2\Psi$, $\delta\Psi = \bar{\nabla}^{\dot{\alpha}}\omega_{\dot{\alpha}}$. Under Yang-Mills gauge transformations $\Psi$ transforms in the same way as $\Phi$.

**b. The nonminimal scalar multiplet**

This multiplet has a number of interesting features: (a) It is a multiplet where the spin of auxiliary fields *exceeds* that of the physical fields; (b) none of the component fields (in a Wess-Zumino gauge) of this multiplet are gauge fields, even though the multiplet is described by a gauge superfield; (c) this multiplet, unlike other scalar multiplets, forms a *reducible* representation of supersymmetry.

We introduce a general spinor superfield $\Psi^\alpha$ with the gauge transformation $\delta\Psi^\alpha = D_\beta L^{(\alpha\beta)}$, $L_{(\alpha\beta)}$ arbitrary. An action that is invariant under this gauge transformation is

$$S = -\int d^4x\, d^4\theta\, \bar{\Sigma}\Sigma \quad , \qquad \Sigma = \bar{D}^{\dot{\alpha}}\bar{\Psi}_{\dot{\alpha}} \quad , \tag{4.5.4}$$

The field strength $\Sigma$ satisfies $\bar{D}^2\Sigma = 0$, so that it is a *complex* linear superfield; in contrast, the field strength of the tensor gauge multiplet is a real linear superfield.

The component fields of the multiplet are

$$A = \Sigma| \quad , \qquad \bar{\zeta}_{\dot{\alpha}} = \bar{D}_{\dot{\alpha}}\Sigma| \quad ,$$

$$\lambda_\alpha = D_\alpha\Sigma| \quad , \qquad P_{\alpha\dot{\beta}} = \bar{D}_{\dot{\beta}}D_\alpha\Sigma| \quad ,$$

$$F = D^2\Sigma| \quad , \qquad \bar{\chi}_{\dot{\alpha}} = \frac{1}{2}D^\alpha\bar{D}_{\dot{\alpha}}D_\alpha\Sigma| \quad . \tag{4.5.5}$$

The component action is

$$S = \int d^4x\, [\bar{A}\Box A + \bar{\zeta}^{\dot{\beta}} i\partial^\alpha{}_{\dot{\beta}}\zeta_\alpha - |F|^2$$

$$+ 2|P_{\alpha\dot{\alpha}}|^2 + \chi^\alpha\lambda_\alpha + \bar{\chi}^{\dot{\alpha}}\bar{\lambda}_{\dot{\alpha}}] \quad , \tag{4.5.6}$$

with propagating complex $A$ and $\zeta^\alpha$. All the other fields are auxiliary.

In terms of superfields we can see that the action (4.5.4) describes a scalar multiplet. The constraint and field equations for $\bar{\Sigma}$ are:

$$D^2\bar{\Sigma} = 0 \quad , \qquad \bar{D}_{\dot{\alpha}}\bar{\Sigma} = 0 \quad . \tag{4.5.7}$$

These are the same as those for the on-shell chiral scalar multiplet, but with constraint and field equation interchanged.

To see the reducibility of this multiplet we use the superprojectors of sec. 3.11. The action can be written

$$S = \int d^4x\, d^4\theta\, \bar{\Psi}^{\dot{\beta}} i\partial^{\alpha}{}_{\dot{\beta}}[2\Pi_{1,0} + (\Pi_{2,\frac{1}{2}+} + \Pi_{2,\frac{1}{2}-})]\Psi_{\alpha} \quad ,$$

$$\Pi_{1,0}\Psi_{\alpha} = -\frac{1}{2}\Box^{-1}D_{\alpha}\bar{D}^2 D_{\gamma}\Psi^{\gamma} \quad ,$$

$$\Pi_{2,\frac{1}{2}\pm}\Psi_{\alpha} = \Box^{-1}\bar{D}^2 D_{\alpha}\frac{1}{2}(D_{\gamma}\Psi^{\gamma} \pm \bar{D}_{\dot{\gamma}}\bar{\Psi}^{\dot{\gamma}}) \quad . \qquad (4.5.8)$$

Thus the multiplet consists of three irreducible submultiplets: one of superspin 0, and two of superspin $\frac{1}{2}$.

In contrast to the chiral scalar multiplet, it is not possible to introduce arbitrary mass and nonderivative self-interaction terms. However, we can write down the action

$$S = \int d^4x\, d^4\theta\, f(\Sigma, \bar{\Sigma}) \quad , \qquad (4.5.9)$$

where $f(z,\bar{z}) = \overline{f(z,\bar{z})}$. Thus, for example, it is possible to formulate supersymmetric nonlinear $\sigma$-models in terms of the nonminimal scalar multiplet. Furthermore, the nonminimal multiplet can be coupled to Yang-Mills multiplets by covariantizing the derivatives: $\Sigma = \nabla^{\alpha}\Psi_{\alpha}$.

* * *

Just as for the tensor multiplet (sec. 4.4.c), we can exhibit the duality of the nonminimal scalar and chiral multiplets by writing a first order action. Most of the discussion of sec. 4.4.c.2 has an analog for the nonminimal scalar multiplet, except, since the multiplet is described by a linear superfield, the Legendre transform is two dimensional and hence there is no restriction on the form of the nonlinear $\sigma$-model that can be described. The two first order actions equivalent to (4.5.9) are (see (4.4.38,42)):

$$S' = \int d^8z\, [f(X, \bar{X}) - \Phi X - \bar{\Phi}\bar{X}] \quad ,$$

$$\bar{D}_{\dot{\alpha}}\Phi = 0 \quad , \qquad (4.5.10a)$$

and

$$S'' = \int d^8z \; [I\!K(X\,,\bar{X}\,) + \Sigma X + \bar{\Sigma}\bar{X}\,] \quad ,$$

$$\bar{D}^2\Sigma = 0 \quad , \tag{4.5.10b}$$

where $X$ is a complex unconstrained superfield and $I\!K$ is the Legendre transform of $f$. Just as for the tensor multiplet, this duality transformation can be performed even in the presence of interactions with other multiplets (e.g., supergravity).

**c. More variant multiplets**

As we have seen, several inequivalent superfield formulations can describe the same set of physical states. The $(0, \frac{1}{2})$ multiplet can be described by a chiral scalar, a gauge two-form, real (or complex) gauge three-forms, or a gauge spinor. The chiral scalar provides the simplest representation. All but one of the other representations are obtained by replacing either the physical or auxiliary field by component 2-forms or 3-forms respectively. We call these "variant" representations of the scalar multiplet. In general, variant representations are very restricted in either their self-interactions or couplings to other multiplets. In this subsection we discuss variant vector and tensor multiplets.

**c.1. Vector multiplet**

We have described super Yang-Mills theories in terms of a hermitian gauge prepotential $V$. It contains a component vector as its highest spin component: $A_{\underline{a}} = \frac{1}{2}[\bar{D}_{\dot{\alpha}}\,, D_\alpha]V|$. There is, however, a smaller superfield that contains a component vector: A chiral *dotted* spinor $\Phi_{\dot{\alpha}}$ ($\bar{D}_{\dot{\alpha}}\Phi_{\dot{\beta}} = 0$), has as its highest spin component

$$A_{\underline{a}} \equiv -\, i(\bar{D}_{\dot{\alpha}}\bar{\Phi}_\alpha + D_\alpha\Phi_{\dot{\alpha}})| \quad . \tag{4.5.11}$$

The superfield $\Phi_{\dot{\alpha}}$ is reducible; it can be bisected (see (3.11.7)

$$\frac{1}{2}(1 \pm \mathbf{K})\Phi_{\dot{\alpha}} = \frac{1}{2}(\Phi_{\dot{\alpha}} \mp \Box^{-1}\bar{D}^2 i\partial^\alpha{}_{\dot{\alpha}}\bar{\Phi}_\alpha) \quad . \tag{4.5.12}$$

Since we want to describe a gauge theory, we gauge away one of the representations instead of constraining it. The transformation

$$\delta\Phi_{\dot\alpha} = \bar D_{\dot\alpha}(D_\beta\Lambda^\beta + \bar D_{\dot\beta}\bar\Lambda^{\dot\beta}) = \bar D^2\bar\Lambda_{\dot\alpha} - i\partial^\alpha{}_{\dot\alpha}\Lambda_\alpha = (1+\mathbf{K})\bar D^2\bar\Lambda_{\dot\alpha} \quad ,$$

$$D_\alpha\bar\Lambda_{\dot\beta} = 0 \quad , \tag{4.5.13}$$

can be used to gauge away $(1+\mathbf{K})\Phi_{\dot\alpha}$, and leaves $(1-\mathbf{K})\Phi_{\dot\alpha}$ inert. The gauge parameter $D_\beta\Lambda^\beta + \bar D_{\dot\beta}\Lambda^{\dot\beta}$ describes the tensor multiplet of sec. 4.4.

The field strength for $\bar\Phi_\alpha$ is the lowest dimension local gauge invariant superfield:

$$W_\alpha = \bar D^2(1-\mathbf{K})\bar\Phi_\alpha = \bar D^2\bar\Phi_\alpha + i\partial_\alpha{}^{\dot\alpha}\Phi_{\dot\alpha} \quad . \tag{4.5.14}$$

The field strength $W_\alpha$ is the familiar chiral field strength of the gauge multiplet described by $V$, but now with $\Gamma_\alpha = \bar\Phi_\alpha$, $\Gamma_{\dot\alpha} = \Phi_{\dot\alpha}$, $\Gamma_{\underline{a}} = -\,i(\bar D_{\dot\alpha}\bar\Phi_\alpha + D_\alpha\Phi_{\dot\alpha})$, and $A_{\underline{a}} = \Gamma_{\underline{a}}|$. Its components are the same, except for the auxiliary field D′:

$$\lambda_\alpha \equiv W_\alpha| \quad ,$$

$$f_{\alpha\beta} \equiv \frac{1}{2}D_{(\alpha}W_{\beta)}| = \frac{1}{2}\partial_{(\alpha\dot\gamma}A_{\beta)}{}^{\dot\gamma} \quad ,$$

$$D \equiv \frac{1}{2}iD_\alpha W^\alpha = \frac{1}{2}\partial^{\alpha\dot\alpha}(\bar D_{\dot\alpha}\bar\Phi_\alpha - D_\alpha\Phi_{\dot\alpha})| = \frac{1}{2}\partial^{\alpha\dot\alpha}B_{\alpha\dot\alpha} \quad . \tag{4.5.15}$$

We thus see the auxiliary pseudoscalar has been replaced by the field strength of a gauge three-form. The action is still (4.2.14), and in components differs from the usual vector multiplet only by the replacement $\mathrm{D}' \to \frac{1}{2}\partial^{\alpha\dot\alpha}B_{\alpha\dot\alpha}$.

This variant form of the vector multiplet can also be obtained from the covariant approach of sec. 4.2: In the abelian case, we can solve the constraint $F_{\alpha\beta} = D_{(\alpha}\Gamma_{\beta)} = 0$ by $\Gamma_\alpha = \bar\Phi_\alpha$. Just as the usual solution $\Gamma_\alpha = -\,i\frac{1}{2}D_\alpha V$ directly in terms of the real scalar prepotential fixed some of the $K$ invariance (corresponding to a gauge condition $D^\alpha\bar D^2\Gamma_\alpha = -\,\bar D^{\dot\alpha}D^2\bar\Gamma_{\dot\alpha}$, which implies $K = \frac{1}{2}(\Lambda + \bar\Lambda)$), the variant solution fixes some of the $K$ invariance with the gauge condition $D^\alpha\Gamma_\alpha = 0$ (which, together with the constraint, implies that $\Gamma_\alpha$ is antichiral), reducing it to $K = D_\alpha\Lambda^\alpha + \bar D_{\dot\alpha}\bar\Lambda^{\dot\alpha}$.

The covariant derivatives can be used to couple this abelian multiplet to matter. However, $\Gamma_\alpha = \bar\Phi_\alpha$ is not a solution to the nonabelian constraints, nor to the abelian ones in general curved superspace. Thus, like other variant multiplets, it is limited in the types of interactions it can have.

**c.2. Tensor multiplet**

The variant representation for the tensor multiplet is described by the same chiral spinor superfield $\Phi_\alpha$ as the usual one (4.4.27), but the gauge transformation is changed. In place of the real scalar parameter $L$ (4.4.30), we use the chiral dotted spinor $\Lambda_{\dot\alpha}$. Explicitly, the modified gauge variation is (cf. (4.5.14))

$$\delta\Phi_\alpha = \bar{D}^2\bar{\Lambda}_\alpha + i\partial_\alpha{}^{\dot\alpha}\Lambda_{\dot\alpha} \quad . \tag{4.5.16}$$

This leads to the usual transformations for $t_{(\alpha\beta)}$ and leaves $A$ and $\psi_\alpha$ invariant (see (4.4.31)). But the variation of the component field $B = i\,\frac{1}{2}\,(\bar{D}_{\dot\alpha}\bar{\Phi}^{\dot\alpha} - D_\alpha\Phi^\alpha)|$ is

$$\delta B = -\,\frac{1}{2}\,\partial^{\alpha\dot\alpha}v_{\alpha\dot\alpha} \quad . \tag{4.5.17}$$

Therefore this component field is a gauge four-form.

The action for $\Psi_\alpha$ is the usual one proportional to $G^2$, and the four-form does not appear in $G^2$ and in the action. However, at the quantum level, the four-form would reappear in gauge fixing terms, and chiral dotted spinors would appear as ghosts.

**d. Superfield Lagrange multipliers**

We have given a number of examples of supersymmetric theories that describe the scalar multiplet on shell (same physical states) but are inequivalent off shell. They differ primarily in the types of interactions they can have. So far, we have found that the simplest formulation of the scalar multiplet, a chiral scalar superfield (or, equivalently even off shell, $\bar{D}^2$ on a general scalar), has the most general interactions. However, in extended supersymmetry none of the known $N = 2$ theories equivalent on-shell to the $N = 2$ scalar multiplet can have all the interactions known from on-shell formulations. We now introduce a form of the $N = 1$ scalar multiplet that is a submultiplet of an off-shell formulation of the $N = 2$ scalar multiplet. Its most distinctive feature is a superfield that appears only as a Lagrange multiplier. This formulation has some drawbacks in common with the tensor multiplet (another theory equivalent to the scalar multiplet on shell), to which it is closely related: (1) It does not have renormalizable self-interactions (i.e, those corresponding to terms $\int d^4x d^2\theta\, \mathrm{P}(\Phi)$), (2) it is restricted in its couplings to supergravity, and (3) it is not an off-shell representation of the (chiral) $U(1)$ symmetry which the scalar multiplet has on shell (corresponding to $\Phi' = e^{i\lambda}\Phi$). On the other hand, unlike the tensor multiplet (and most variant forms of the scalar multiplet), it does couple to Yang-Mills. However, because of (3), it can only be a *real* representation of any internal symmetry group, and couples to Yang-Mills accordingly (e.g., it can couple to a $U(1)$ vector multiplet only as a doublet of opposite charges).

The formulation is described by a general spinor gauge superfield with a term in the action like that of the chiral spinor gauge superfield of the tensor multiplet, and a real scalar superfield Lagrange multiplier with a term in the action that constrains to zero the submultiplets in the former term that don't occur in the tensor multiplet. Explicitly, the action is

$$S = - \int d^4x \, d^4\theta \, (\tfrac{1}{2} F^2 + Y \, G) \quad ,$$

$$F = \frac{1}{2} (D^\alpha \Psi_\alpha + \bar{D}^{\dot{\alpha}} \bar{\Psi}_{\dot{\alpha}}) \quad , \quad G = i \, \frac{1}{2} (D^\alpha \Psi_\alpha - \bar{D}^{\dot{\alpha}} \bar{\Psi}_{\dot{\alpha}}) \quad ; \tag{4.5.18}$$

with gauge invariance

$$\delta \Psi^\alpha = D_\beta L^{(\alpha\beta)} \tag{4.5.19}$$

in terms of a general superfield gauge parameter. The Bianchi identities and field equations are:

$$\textit{Bianchi identities}: D^2(F - iG) = 0 \quad , \tag{4.5.20a}$$

$$\textit{field equations}: \bar{D}_{\dot{\alpha}}(F + iY) = G = 0 \quad . \tag{4.5.20b}$$

If we make a "duality" transformation by switching the Bianchi identities with the field equations, we obtain the usual formulation of the scalar multiplet, with the identifications

$$F = \frac{1}{2} (\Phi + \bar{\Phi}) \quad , \quad Y = \frac{1}{2} i(\bar{\Phi} - \Phi) \quad , \quad G = 0 \quad . \tag{4.5.21}$$

In terms of irreducible representations of supersymmetry, this theory contains superspins $\frac{1}{2} \oplus \frac{1}{2} \oplus 0$ in $\Psi^\alpha$ and $\frac{1}{2} \oplus 0$ in $Y$. The representations in $Y$ set the corresponding ones in $\Psi$ to zero on shell, leaving the remaining one as a tensor multiplet. However, unlike the tensor multiplet, the physical spin zero states are all represented by scalars: The vector obtained by projection from $[\bar{D}_{\dot{\alpha}}, D_\alpha]F$ is an unconstrained auxiliary field, appearing at $\theta^2\bar\theta$ order in $\Psi$, whereas the corresponding vector in the tensor multiplet is the transverse field strength of the tensor appearing at $\theta$ order in the chiral spinor $\Phi^\alpha$. This theory has the same component-field content as the nonminimal scalar multiplet plus an auxiliary real scalar superfield.

Coupling to Yang-Mills is straightforward; however, since both $\Psi$ and $\bar\Psi$ appear in $F$ and in $G$, $\Psi$ must transform under a *real* representation of the Yang-Mills group. We covariantize by replacing the spinor derivatives in the definitions of $F$ and $G$ by Yang-Mills covariant spinor derivatives. Invariance of the action under the Yang-Mills covariantization of (4.5.19) then requires

$$0 = \delta\nabla_\alpha\Psi^\alpha = \nabla_\alpha\nabla_\beta L^{(\alpha\beta)} = -\,i\,\frac{1}{2}\,F_{\alpha\beta}L^{(\alpha\beta)} \quad , \tag{4.5.22}$$

implying the same representation-preserving constraint $F_{\alpha\beta}=0$ as for the chiral scalar formulation. The total set of Bianchi identities and field equations is the same as for the chiral scalar.

Just as there is an improved form of the tensor multiplet, with superconformal invariance, there is an improved form of this scalar multiplet. In analogy to the tensor multiplet, it is obtained by replacing $\frac{1}{2}F^2$ in (4.5.18) with $F\,ln\,F$ (cf. (4.4.46)). Furthermore, the first-order formulation of this multiplet turns out to be equivalent to the first-order formulation of the *nonminimal* scalar multiplet. We start with (cf. (4.4.45))

$$S = \int d^4x\, d^4\theta\, [e^X - XF - Y\,G\,] \quad . \tag{4.5.23}$$

The first order form of the nonminimal scalar multiplet is usually written as (4.5.10b):

$$S = \int d^4x\, d^4\theta\, [X'\bar X' - (\bar X'D^\alpha\Psi'_\alpha + h.\,c.)] \quad . \tag{4.5.24}$$

Upon elimination of the complex scalar $X'$, this gives the second-order form of the nonminimal scalar multiplet (4.5.4). Upon elimination of the spinor $\Psi'^\alpha$, we obtain the constraint $\bar D_{\dot\alpha}X'=0$, whose solution $X'=\Phi$ in terms of a chiral scalar $\Phi$ gives the minimal scalar multiplet (proving their duality). The equivalence of the actions (4.5.23) and (4.5.24) follows from the change of variables

$$\Psi'_\alpha = \bar X'^{-1}\Psi_\alpha \quad , \quad \bar X' = e^{\frac{1}{2}(X+iY)} \quad , \tag{4.5.25}$$

(some integration by parts is necessary to show the equivalence).

## e. The gravitino matter multiplet

Thus far we have considered supermultiplets with physical fields of spin one or less. We conclude our discussion of global $N = 1$ multiplets by considering one with a spin 1 and a spin $\frac{3}{2}$ (gravitino) component field. It is possible to discuss it without introducing supergravity only if the multiplet describes a free theory. The gravitino multiplet is of interest because many of the features encountered in the superfield formulation of supergravity, such as irreducible submultiplets, compensators, and inequivalent off-shell formulations, are already present.

### e.1. Off-shell field strength and prepotential

Following the discussion of sec. 3.12.a we describe this multiplet on-shell with component field strengths $\psi_{\alpha\beta}$ (vector field strength) and $\psi_{\alpha\beta\gamma}$ (the Rarita-Schwinger field strength), totally symmetric in their indices. We denote the off-shell superfield strength corresponding to $\psi_{\alpha\beta}$ by $W_{\alpha\beta}$. It is a chiral field strength of superspin 1, no bisection is possible ($s + \frac{1}{2} N = \frac{3}{2}$ is not an integer), and therefore we write (see (3.13.1))

$$W_{\alpha\beta} = \frac{1}{2} \bar{D}^2 D_{(\alpha} \Psi_{\beta)} \quad , \tag{4.5.26}$$

in terms of a general spinor superfield. From dimensional analysis (the gravitino field has canonical dimension $\frac{3}{2}$), the dimension of $W$ is 2.

The gauge transformations that leave $W$ invariant are

$$\delta\Psi_\alpha = \Lambda_\alpha + D_\alpha \Omega \quad , \quad \bar{D}_{\dot\beta} \Lambda_\alpha = 0 \quad , \quad \Omega \neq \bar{\Omega} \quad , \tag{4.5.27}$$

To analyze the transformations of the components, we define

$$t_{\alpha\beta} = W_{\alpha\beta}| = \frac{1}{2} \bar{D}^2 D_{(\alpha} \Psi_{\beta)}| \quad ,$$

$$\psi_{\alpha\beta\gamma} = D_{(\alpha} W_{\beta\gamma)}| = \frac{1}{2} D_{(\alpha} \bar{D}^2 D_\beta \Psi_{\gamma)}| = i \, \frac{1}{2} \partial_{(\alpha\dot\delta} \, \frac{1}{2} [\bar{D}^{\dot\delta}, D_\beta] \Psi_{\gamma)}|$$

$$\chi_\alpha = D^\beta W_{\alpha\beta}| = \frac{1}{2} D^\beta \bar{D}^2 D_{(\alpha} \Psi_{\beta)}| \quad ,$$

$$f_{\alpha\beta} = D^2 W_{\alpha\beta}| = \frac{1}{2} D^2 \bar{D}^2 D_{(\alpha}\Psi_{\beta)}| = i\,\frac{1}{2}\,\partial_{(\alpha\dot{\gamma}} D^2 \bar{D}^{\dot{\gamma}}\Psi_{\beta)}| \quad . \tag{4.5.28}$$

We identify the gravitino and (complex) vector field strengths, $\psi_{\alpha\beta\gamma}$ and $f_{\alpha\beta}$. The corresponding component gauge fields appear in $\Psi_\alpha$ and are given by

$$\psi_{\underline{a}\gamma} = i\,\frac{1}{2}\,[\bar{D}_{\dot{\alpha}}\,, D_\alpha]\Psi_\gamma| \quad , \qquad A_{\underline{a}} = iD^2\bar{D}_{\dot{\alpha}}\Psi_\alpha| \quad , \tag{4.5.29}$$

Their gauge transformations are

$$\delta\psi_{\underline{a}\gamma} = [\frac{1}{2}\,\partial_{\underline{a}}(D_\gamma\Omega - \Lambda_\gamma) + iC_{\gamma\alpha}\bar{D}_{\dot{\alpha}}D^2\Omega]| \quad ,$$

$$\delta A_{\underline{a}} = -\,\partial_{\underline{a}} D^2\Omega| \quad . \tag{4.5.30}$$

The gravitino field, in addition to undergoing a Rarita-Schwinger gauge transformation described by the first term, also is translated by $iC_{\gamma\alpha}\bar{\eta}_{\dot{\beta}}$, $\bar{\eta}_{\dot{\beta}} = \bar{D}_{\dot{\beta}}D^2\Omega|$. (When coupled to $N = 1$ supergravity to give $N = 2$ supergravity, the transformation (4.5.30) is part of the $N = 2$ superconformal group: The first term becomes the second local $Q$-supersymmetry transformation, the second term, the second $S$-supersymmetry transformation.) We refer to this multiplet as the conformal gravitino multiplet.

The multiplet of component fields of $W_{\alpha\beta}$ is irreducible and gauge invariant and should appear in the gauge invariant action. However, in the absence of dimensional constants we cannot write a free action of the correct dimension in terms of $W$. We can write a nonlocal gauge invariant action in terms of $\Psi$, e.g.,

$$S = \frac{1}{2}\int d^4x\, d^4\theta\; \bar{\Psi}^{\dot{\beta}} i\partial^{\alpha}{}_{\dot{\beta}}\,\Pi_{1,1}\Psi_\alpha + h.\,c. \quad , \tag{4.5.31}$$

( $\Pi_{1,1}\Psi^\alpha$ is the correct projector onto the physical gauge invariant representation) but it leads to a nonlocal component action.

To find a local action, we add more representations. Clearly removing $\Pi$ from the action restores locality but introduces *all* the representations in $\Psi_\alpha$ and destroys the gauge invariance. We can, however, restrict the representations which appear. We begin with the general expression

$$S = \frac{1}{2}\int d^4x\, d^4\theta\; \bar{\Psi}^{\dot{\beta}} i\partial^{\alpha}{}_{\dot{\beta}}(\sum_i c_i\Pi_i)\Psi_\alpha + h.\,c. \quad , \tag{4.5.32}$$

with the sum running over all projectors (3.11.38,39), and choose $c_i$ to obtain a local action. We require that the superprojector $\Pi_{1,1}$ be present, and find two solutions containing the least number of additional superprojectors. One uses the superprojectors $-\Pi_{1,1} - 2\Pi_{0,\frac{1}{2},-} - \Pi_{2,\frac{1}{2},+}$ the other uses $-\Pi_{1,1} - 2\Pi_{0,\frac{1}{2},-} + \Pi_{1,0}$. (The overall sign is chosen to give the physical vector the correct kinetic term.) The resulting actions are

$$(1) \qquad S_{(1)} = \int d^4x\, d^4\theta\, [-(\bar{D}^{\dot{\alpha}}\Psi^\alpha)(D_\alpha \bar{\Psi}_{\dot{\alpha}}) - \frac{1}{2}(\Psi^\alpha \bar{D}^2 \Psi_\alpha + h.\,c.) + \frac{1}{4}(D^\alpha \Psi_\alpha + \bar{D}^{\dot{\alpha}}\bar{\Psi}_{\dot{\alpha}})^2]$$

$$(2) \qquad S_{(2)} = \int d^4x\, d^4\theta\, [-(\bar{D}^{\dot{\alpha}}\Psi^\alpha)(D_\alpha \bar{\Psi}_{\dot{\alpha}}) - \frac{1}{2}(\Psi^\alpha \bar{D}^2 \Psi_\alpha + h.\,c.)] \quad . \qquad (4.5.33)$$

However, the gauge group is no longer described by (4.5.27); the invariance groups associated with the actions above are smaller than the invariance group of $W_{\alpha\beta}$; they are

$$(1) \qquad \delta\Psi_\alpha = i\bar{D}^2 D_\alpha K_1 + D_\alpha K_2 \quad , \qquad K_i = \bar{K}_i \quad ,$$

$$(2) \qquad \delta\Psi_\alpha = \Lambda_{1\alpha} + D_\alpha(D_\beta \Lambda_2{}^\beta + \bar{D}_{\dot{\beta}}\bar{\Lambda}_2{}^{\dot{\beta}}) \quad , \qquad \bar{D}_{\dot{\beta}}\Lambda_{i\alpha} = 0 \quad , \qquad (4.5.34)$$

for the two actions.

As compared to (4.5.27), in the first case the invariance group has been reduced because $\Lambda_\alpha$ is restricted to the special form $i\bar{D}^2 D_\alpha K_1$, $K_1 = \bar{K}_1$, and $\Omega$ is restricted to be real. In the second case $\Lambda_\alpha$ remains unrestricted but $\Omega$ is restricted to the form $D_\beta \Lambda_2{}^\beta + \bar{D}_{\dot{\beta}}\bar{\Lambda}_2{}^{\dot{\beta}}$. In both cases the final gauge group has fewer parameters than the original one. However, for many purposes (e.g., quantization), we need to use the original gauge group; to do this, we introduce compensating multiplets (see sec. 3.10).

## e.2. Compensators

For the gravitino multiplet, two inequivalent sets of compensators can be introduced. We do this by *nonlocal* field redefinitions of the basic gauge superfield. Thus, for the two local actions we make the redefinitions

$$(1) \qquad \Psi_\alpha \to \Psi_\alpha + \Box^{-1}(\frac{1}{2}D^2 W_\alpha + \bar{D}^2 D_\alpha G) \quad ,$$

$$W_\alpha = i\bar{D}^2 D_\alpha V \quad , \qquad V = \bar{V} \quad ,$$

$$G = \frac{1}{2}(D_\alpha \Phi^\alpha + \bar{D}_{\dot{\alpha}}\bar{\Phi}^{\dot{\alpha}}) \quad , \quad \bar{D}_{\dot{\beta}}\Phi_\alpha = 0 \quad ,$$

$$(2) \qquad \Psi_\alpha \to \Psi_\alpha + \Box^{-1}(\frac{1}{2} D^2 W_\alpha + D_\alpha \bar{D}^2 \bar{\Phi}) \quad ,$$

$$\bar{D}_{\dot{\beta}} \Phi = 0 \quad . \tag{4.5.35}$$

They induce the following changes in the actions

$$(1) \qquad S_{(1)} \to S_{(1)} - \int d^6 z \, W^2 + \int d^8 z \, [ - (W^\alpha \Psi_\alpha + \bar{W}^{\dot{\alpha}} \bar{\Psi}_{\dot{\alpha}}) + G^2$$

$$- G(D^\alpha \Psi_\alpha + \bar{D}^{\dot{\alpha}} \bar{\Psi}_{\dot{\alpha}})] \quad ,$$

$$(2) \qquad S_{(2)} \to S_{(2)} - \int d^6 z \, W^2 + \int d^8 z \, [ - (W^\alpha \Psi_\alpha + \bar{W}^{\dot{\alpha}} \bar{\Psi}_{\dot{\alpha}}) - 2\bar{\Phi}\Phi$$

$$- (\Phi D^\alpha \Psi_\alpha + \bar{\Phi} \bar{D}^{\dot{\alpha}} \bar{\Psi}_{\dot{\alpha}})] \quad . \tag{4.5.36}$$

Although the redefinitions are nonlocal, the actions remain local. (Actually, in case (1) we can also use simply $\Psi_\alpha \to \Psi_\alpha + i \frac{1}{2} D_\alpha V + \Phi_\alpha$.)

In the above field redefinitions we introduced a vector multiplet $V$ and either a tensor multiplet $\Phi_\alpha$ or a chiral scalar multiplet $\Phi$. These choices are a reflection of the representations that were introduced by the additional projections: A vector multiplet $\Pi_{0,\frac{1}{2}-}\Psi_\alpha$, a tensor multiplet $\Pi_{2,\frac{1}{2}+}\Psi_\alpha$, and a chiral scalar multiplet $\Pi_{1,0}\Psi_\alpha$. In the presence of the compensating multiplets the gauge variation of $\Psi_\alpha$ is given by (4.5.27). The compensating multiplets transform as follows:

$$(1) \qquad \delta V = i(\Omega - \bar{\Omega}) \quad , \qquad \delta \Phi_\alpha = - \Lambda_\alpha + i \bar{D}^2 D_\alpha K_3 \quad ,$$

$$(2) \qquad \delta V = i(\Omega - \bar{\Omega}) \quad , \qquad \delta \Phi = - \bar{D}^2 \bar{\Omega} \quad , \tag{4.5.37}$$

Since they are compensators, they can be algebraically gauged to zero. In the resulting gauge, the transformations (4.5.27) of $\Psi_\alpha$ are restricted back to (4.5.34).

The two inequivalent formulations of the gravitino multiplet, one using a tensor multiplet compensator and the other using a chiral scalar compensator, lead to different auxiliary field structures at the component level. In the Wess-Zumino gauge for case (1) the components of the gravitino multiplet are

$$\lambda_\alpha = \bar{D}^{\dot{\beta}} D_\alpha \bar{\Psi}_{\dot{\beta}} + \bar{D}^2 \Psi_\alpha + \frac{1}{2} D_\alpha ( D^\beta \Psi_\beta + \bar{D}^{\dot{\beta}} \bar{\Psi}_{\dot{\beta}} ) + W_\alpha - D_\alpha G \ ,$$

$$P = - i ( D^\alpha \bar{D}^2 \Psi_\alpha - \bar{D}^{\dot{\alpha}} D^2 \bar{\Psi}_{\dot{\alpha}} + D^\alpha W_\alpha ) \ ,$$

$$V_{\underline{a}} = i ( D^2 \bar{D}_{\dot{\alpha}} \Psi_\alpha + \bar{D}^2 D_\alpha \bar{\Psi}_{\dot{\alpha}} ) - \partial_{\underline{a}} [ G - \frac{1}{2} ( D^\gamma \Psi_\gamma + \bar{D}^{\dot{\gamma}} \bar{\Psi}_{\dot{\gamma}} ) ] \ ,$$

$$A'_{\underline{a}} = ( D^2 \bar{D}_{\dot{\alpha}} \Psi_\alpha - \bar{D}^2 D_\alpha \bar{\Psi}_{\dot{\alpha}} ) \ ,$$

$$t'_{\alpha\beta} = D_{(\alpha} [ \Phi_{\beta)} + \Psi_{\beta)} ] \ ,$$

$$t_{\alpha\beta} = 2 W_{\alpha\beta} + \partial_{(\alpha}{}^{\dot{\gamma}} B_{\beta)\dot{\gamma}} \ ,$$

$$\chi_\alpha = \bar{D}^2 D_\alpha [ G - \frac{1}{2} ( D^\beta \Psi_\beta + \bar{D}^{\dot{\beta}} \bar{\Psi}_{\dot{\beta}} ) ] \ ,$$

$$B_{\underline{a}} = \frac{1}{2} [ D_\alpha , \bar{D}_{\dot{\alpha}} ] V + i ( D_\alpha \bar{\Psi}_{\dot{\alpha}} + \bar{D}_{\dot{\alpha}} \Psi_\alpha ) \ ,$$

$$\psi_{\underline{a}}{}^\beta = i \frac{1}{2} [ D_\alpha , \bar{D}_{\dot{\alpha}} ] \Psi^\beta + i \delta_\alpha{}^\beta [ D^\gamma \bar{D}_{\dot{\alpha}} \Psi_\gamma + D^2 \bar{\Psi}_{\dot{\alpha}} + \bar{W}_{\dot{\alpha}} ] \ . \tag{4.5.38}$$

For case (2) we have instead

$$\lambda_\alpha = \bar{D}^{\dot{\beta}} D_\alpha \bar{\Psi}_{\dot{\beta}} + \bar{D}^2 \Psi_\alpha + W_\alpha - D_\alpha \Phi \ ,$$

$$P = - i ( D^\alpha \bar{D}^2 \Psi_\alpha - \bar{D}^{\dot{\alpha}} D^2 \bar{\Psi}_{\dot{\alpha}} + D^\alpha W_\alpha ) \ ,$$

$$J = \bar{D}^2 ( D^\alpha \Psi_\alpha + 2 \bar{\Phi} ) \ ,$$

$$A_{\underline{a}} = - i \bar{D}^2 D_\alpha \bar{\Psi}_{\dot{\alpha}} + \partial_{\underline{a}} \Phi \ ,$$

$$t_{\alpha\beta} = 2 W_{\alpha\beta} + \partial_{(\alpha}{}^{\dot{\gamma}} B_{\beta)\dot{\gamma}} \ ,$$

$$\chi_\alpha = D^2 \bar{D}^2 \Psi_\alpha + i \partial_{\alpha\dot{\alpha}} D^\beta \bar{D}^{\dot{\alpha}} \Psi_\beta - i \partial_{\alpha\dot{\alpha}} \bar{D}^{\dot{\alpha}} \bar{\Phi} \ ,$$

$$B_{\underline{a}} = \frac{1}{2}[D_\alpha\, , \bar{D}_{\dot{\alpha}}]V + i(\, D_\alpha \bar{\Psi}_{\dot{\alpha}} + \bar{D}_{\dot{\alpha}}\Psi_\alpha\,)\ \ ,$$

$$\psi_{\underline{a}}{}^\beta = i\,\frac{1}{2}[D_\alpha\, , \bar{D}_{\dot{\alpha}}]\Psi^\beta + i\delta_\alpha{}^\beta[\, D^\gamma \bar{D}_{\dot{\alpha}}\Psi_\gamma + D^2\bar{\Psi}_{\dot{\alpha}} + \bar{W}_{\dot{\alpha}}\,]\ \ . \tag{4.5.39}$$

In case (1), the gauge field $t'_{\alpha\beta}$ of the tensor multiplet replaces the complex scalar component field $J$, which corresponds to the auxiliary field of the chiral scalar multiplet of case (2). In the component action $t'_{\alpha\beta}$ only appears as $t'_{\alpha\beta}\,\partial^{\alpha\dot{\gamma}}A'^{\beta}{}_{\dot{\gamma}} - \bar{t}'_{\dot{\alpha}\dot{\beta}}\,\partial^{\gamma\dot{\alpha}}A'_\gamma{}^{\dot{\beta}}$. This term is invariant under separate gauge transformations of $t'_{\alpha\beta}$ and $A'_{\alpha\dot{\beta}}$. Also it should be noted that the field $A'_{\underline{a}}$ is real. In case (2) $A_{\alpha\dot{\beta}} \neq (\bar{A}_{\alpha\dot{\beta}})$ has no gauge transformations because the physical scalars of the chiral scalar multiplet have become the longitudinal parts of $A_{\alpha\dot{\beta}}$ and cancel the transformation in (4.5.30). In both cases, the physical vector of the compensating vector multiplet has become the physical vector of the gravitino multiplet, while the physical spinor of the vector multiplet becomes the spin $\frac{1}{2}$ part of the gravitino and cancels the spinor translation in (4.5.30).

### e.3. Duality

Since the two formulations above differ in that (1) has a tensor compensator where (2) has a chiral compensator, using the approach of sec. 4.4.c.2, we can write first order actions that demonstrate the duality between the two formulations. For example, we can start with (1) and write

$$S'_{(1)} = S_{(1)}[\Psi\, , V] + \int d^8z\ [X^2 - X(D^\alpha\Psi_\alpha + \bar{D}^{\dot{\alpha}}\bar{\Psi}_{\dot{\alpha}}) - 2X(\Phi + \bar{\Phi})]\ \ . \tag{4.5.40a}$$

Varying the chiral field $\Phi$ leads to $X = G$ and formulation (1), whereas eliminating $X$ results in formulation (2). Similarly, we can start with (2) and write

$$S'_{(2)} = S_{(2)}[\Psi\, , V] + \int d^8z\ [-X^2 - X(D^\alpha\Psi_\alpha + \bar{D}^{\dot{\alpha}}\bar{\Psi}_{\dot{\alpha}}) + 2XG\,]\ \ . \tag{4.5.40b}$$

Varying the linear superfield $G$ we find $X = \Phi + \bar{\Phi}$ and formulation (2), whereas eliminating $X$ leads directly to (1).

There are other inequivalent formulations where we replace $V$ by the variant vector multiplet and/or replace $\Phi$ by either the real or complex three-form multiplets. This simply replaces some of the scalar auxiliary fields with gauge three-form field strengths.

### e.4. Geometric formulations

Finally, we give geometrical formulations of the theories. To describe a multiplet that gauges a symmetry with a spinorial gauge parameter, we introduce a super 1-form $\Gamma_A{}^\beta$ with an additional spinor "group" index. The analysis is simplified if the irreducible multiplet is considered first. The irreducible theory was described by $W_{\alpha\beta}$ in (4.5.26). To describe this multiplet geometrically, we introduce more gauge fields (in particular a complex super 1-form $\Gamma_A$ ($\neq \bar{\Gamma}_A$)) and enlarge the gauge group. When we get to supergravity we will find that this process can also be carried out. There the irreducible multiplet is the Weyl multiplet and the enlarged group is the conformal group. The final form of the $(\frac{3}{2},1)$ multiplet with more irreducible multiplets and compensators is analogous to Poincaré supergravity. We will use the words "Poincaré" and "Weyl" for the $(\frac{3}{2},1)$ multiplet to emphasize this analogy.

The complete set of gauge fields and gauge transformations that describe the Weyl $(\frac{3}{2},1)$ multiplet is:

$$\delta\Gamma_A{}^\beta = D_A K^\beta - \delta_A{}^\beta L \quad , \quad \delta\bar{\Gamma}_A{}^{\dot\beta} = D_A \bar{K}^{\dot\beta} - \delta_A{}^{\dot\beta}\bar{L} \ ,$$

$$\delta\Gamma_A = D_A L \ , \quad \delta\bar{\Gamma}_A = D_A\bar{L} \ . \qquad (4.5.41)$$

The $K$-terms are the usual gauge transformations associated with a superform and the $L$-terms are the "conformal" transformation. Recall that we found that the gauge transformations of the irreducible multiplet contain an $S$-supersymmetry term. $L$ is the superfield parameter that contains these component parameters. The vector component of the complex super 1-form $\Gamma_A$ is the component gauge field whose field strength appears in (4.5.28). The field strengths for $\Gamma_A$ are those for an ordinary (complex) vector multiplet, but those for $\Gamma_A{}^\beta$ and its conjugate $\bar{\Gamma}_A{}^{\dot\beta}$ must be $L$-covariantized:

$$F_{AB}{}^\gamma = D_{[A}\Gamma_{B)}{}^\gamma - T_{AB}{}^D\Gamma_D{}^\gamma + \Gamma_{[A}\delta_{B)}{}^\gamma \ ,$$

$$F_{AB}{}^{\dot\gamma} = D_{[A}\bar{\Gamma}_{B)}{}^{\dot\gamma} - T_{AB}{}^D\bar{\Gamma}_D{}^{\dot\gamma} + \bar{\Gamma}_{[A}\delta_{B)}{}^{\dot\gamma} \ . \qquad (4.5.42)$$

We can now impose the constraints :

$$F_{\alpha,\beta}{}^{\gamma} = F_{\alpha,\dot{\beta}}{}^{\gamma} = F_{\dot{\alpha},\dot{\beta}}{}^{\gamma} = F_{\underline{a},\beta}{}^{\gamma} = 0 \ ,$$

$$F_{\underline{a}}{}^{\dot{\alpha},\alpha} + \bar{F}_{\underline{a}}{}^{\alpha,\dot{\alpha}} = 0 \ ,$$

$$F_{\alpha,\beta} = F_{\alpha,\dot{\beta}} = 0 \quad , \quad (F_{\dot{\alpha},\dot{\beta}} \neq 0) \ . \tag{4.5.43}$$

Even with the $L$ invariance the geometrical description here does not quite reduce to the irreducible multiplet $W_{\alpha\beta}$. However, these constraints reduce the super 1-forms to the irreducible multiplet plus the compensating vector multiplet, which are the two irreducible multiplets common to both forms of the Poincaré $(\frac{3}{2}, 1)$ multiplet, and thus are sufficient for their general analysis.

The explicit solution of these constraints is in terms of prepotentials $\hat{\Psi}_\alpha$, $\tilde{\Psi}_\alpha$, $\Psi$(complex), and $V$(real):

$$\Gamma_\alpha{}^\beta = D_\alpha \hat{\Psi}^\beta - \delta_\alpha{}^\beta \Psi \quad , \quad \Gamma_{\dot{\alpha}}{}^\beta = \bar{D}_{\dot{\alpha}} \tilde{\Psi}^\beta \ ,$$

$$\Gamma_{\underline{a}}{}^\beta = -\, i[\, \bar{D}_{\dot{\alpha}} D_\alpha \hat{\Psi}^\beta + D_\alpha \bar{D}_{\dot{\alpha}} \tilde{\Psi}^\beta + \delta_\alpha{}^\beta (\, \Gamma_{\dot{\alpha}} - \bar{D}_{\dot{\alpha}} \Psi \,)] \ ;$$

$$\Gamma_\alpha = D_\alpha \Psi \ , \ \Gamma_{\dot{\alpha}} = -\, D^\alpha \bar{D}_{\dot{\alpha}} (\hat{\Psi}_\alpha - \tilde{\Psi}_\alpha) - D^2 (\bar{\hat{\Psi}}_{\dot{\alpha}} - \bar{\tilde{\Psi}}_{\dot{\alpha}}) + \bar{D}_{\dot{\alpha}} \Psi - \bar{W}_{\dot{\alpha}} \ ,$$

$$\Gamma_{\underline{a}} = -\, i\, D^2 \bar{D}_{\dot{\alpha}} (\, \hat{\Psi}_\alpha - \tilde{\Psi}_\alpha \,) + \partial_{\underline{a}} \Psi \ ; \tag{4.5.44}$$

where $W_\alpha = i\bar{D}^2 D_\alpha V$. The prepotentials transform under $K_\alpha$ and $L$, as well as under new parameters $\Omega$(complex) and $\Lambda_\alpha$(complex) under which the $\Gamma$'s are invariant (this is analogous to the $\Lambda$-group parameters in super-Yang-Mills):

$$\delta \hat{\Psi}_\alpha = K_\alpha + D_\alpha \Omega \ , \ \delta \tilde{\Psi}_\alpha = K_\alpha - \Lambda_\alpha \ , \quad \bar{D}_{\dot{\alpha}} \Lambda_\alpha = 0 \ ;$$

$$\delta \Psi = L - D^2 \Omega \ , \ \delta V = i(\, \Omega - \bar{\Omega} \,) \ . \tag{4.5.45}$$

As with the vector multiplet, we can go to a chiral representation where $\hat{\Psi}_\alpha$ and $\tilde{\Psi}_\alpha$ only appear as the combination $\Psi_\alpha = \hat{\Psi}_\alpha - \tilde{\Psi}_\alpha$, with

$$\delta \Psi_\alpha = \delta(\, \hat{\Psi}_\alpha - \tilde{\Psi}_\alpha \,) = \Lambda_\alpha + D_\alpha \Omega \ . \tag{4.5.46}$$

At this point, by comparison with (4.5.27), we can identify $\Psi_\alpha$ and $V$ with the corresponding quantities there. To recover the full Poincaré theory, we must break the $L$ invariance. To break the $L$ invariance, we introduce a "tensor" compensator $\hat{G}$ *or* $\hat{\Phi}$, to obtain the tensor-multiplet or scalar-multiplet, respectively. These "tensor" (scalars) are *not* prepotentials, and transform covariantly under *all* of the gauge transformations defined thus far. By covariant, we mean that these transform *without* derivatives $D_A$.

$$\delta\hat{G} = \, - \, ( \, L + \bar{L} \, ) \ , \tag{4.5.47}$$

$$\delta\hat{\Phi} = \, - \, \bar{L} \ . \tag{4.5.48}$$

We now impose the $L$-covariantized form of the usual constraints ($\bar{D}^2 G = 0$ and $\bar{D}_{\dot\alpha}\Phi = 0$) which describe tensor and chiral scalar multiplets;

$$\frac{1}{2}\,\bar{D}^{\dot\alpha}[\,\bar{D}_{\dot\alpha}\hat{G} + (\,\Gamma_{\dot\alpha} + \bar{\Gamma}_{\dot\alpha}\,)] + h.\,c. = 0 \ , \tag{4.5.49}$$

$$\bar{D}_{\dot\alpha}\hat{\Phi} + \bar{\Gamma}_{\dot\alpha} = 0 \ . \tag{4.5.50}$$

The invariance of these constraints follows directly (4.5.27,37,47,48). (The hermitian conjugate term above is necessary to avoid constraining $\Psi_\alpha$ itself.) These constraints can be solved in terms of prepotentials:

$$\hat{G} = G - (\,\Psi + \bar{\Psi}\,) - \frac{1}{2}\,(\,D^\alpha\Psi_\alpha + \bar{D}^{\dot\alpha}\bar{\Psi}_{\dot\alpha}\,) \ , \tag{4.5.51}$$

$$\hat{\Phi} = \Phi - \bar{\Psi} \ ; \tag{4.5.52}$$

where $G$ and $\Phi$ are given in (4.5.35) and transform as in (4.5.37). To obtain case (1) as described above, we introduce the tensor compensator $\hat{G}$, choose the $L$-gauge $\hat{G} = 0$, and solve for $\Psi + \bar{\Psi}$ in terms of $G$. The quantity $\Psi - \bar{\Psi}$ is undetermined, but can be gauged away by using the remaining invariance parametrized $L - \bar{L}$. (Recall gauging $\hat{G}$ to zero only uses the freedom in $L + \bar{L}$.) To obtain case (2), we introduce the chiral scalar compensator $\hat{\Phi}$ and gauge it to zero which gives $\Psi = \bar{\Phi}$. Thus, gauging either tensor compensator $\hat{\Phi}$ or $\hat{G}$ to zero forces the $\Gamma$'s to contain the correct and complete Poincaré multiplets. Alternatively, we could gauge $\Psi$ to zero, so that the tensor(scalar) submultiplet is contained only in $\hat{G}(\hat{\Phi})$. We should also mention that other choices could be made for tensor compensators. Any of the variant scalar or nonminimal scalar multiplets can be used by generalizing the discussion above. These will lead to a number of inequivalent off-shell formulations of the Poincaré $(\frac{3}{2}, 1)$ theory.

The field equations, obtained from the action (4.5.36), take the covariant form

$$\bar{D}_{\dot{\alpha}} X + (\Gamma_{\dot{\alpha}} + \bar{\Gamma}_{\dot{\alpha}}) = 0 \ , \quad X = \hat{G} \ \ or \ \ \hat{\Phi} + \hat{\bar{\Phi}} \quad . \tag{4.5.53}$$

(In case (2), using (4.5.50), these simplify to $\bar{D}_{\dot{\alpha}} \hat{\bar{\Phi}} + \Gamma_{\dot{\alpha}} = 0$.)

## 4.6. N-extended multiplets

So far in this chapter we have described the multiplets of $N = 1$ global supersymmetry. For interacting theories there are two such multiplets, with spins $(\frac{1}{2}, 0)$, and $(1, \frac{1}{2})$, although their superfield description may take many forms. For $N$-extended supersymmetry, global multiplets exist for $N \leq 4$. They are naturally described in terms of extended superfields. It is possible, however, to discuss these multiplets, and their interactions, in terms of $N = 1$ superfields describing their $N = 1$ submultiplets. In many cases of interest this is the most complete description that we have at the present time.

### a. N=2 multiplets

As discussed in sec. 3.3, there exist two global $N = 2$ multiplets: a vector multiplet with spins $(1, \frac{1}{2}, \frac{1}{2}, 0, 0)$, and a scalar multiplet with spins $(\frac{1}{2}, \frac{1}{2}, 0, 0, 0, 0)$. There exists only one global $N = 4$ multiplet: the $N = 4$ vector multiplet, with $SU(4)$ representation $\otimes$ spins $(1 \otimes 1, 4 \otimes \frac{1}{2}, 6 \otimes 0)$. (The only $N = 3$ multiplet is the same as that of $N = 4$.) We begin by discussing the $N = 2$ situation.

#### a.1. Vector multiplet

The $N = 2$ vector multiplet consists of an $N = 1$ Yang-Mills multiplet coupled to a scalar multiplet in the same (adjoint) representation of the internal symmetry group. The action is

$$S = \frac{1}{g^2} tr \left( \int d^4x d^4\theta \, \bar{\Phi}\Phi + \int d^4x d^2\theta \, W^2 \right) \tag{4.6.1}$$

in the vector representation. In addition to the usual gauge invariance, it is invariant under the following global transformations with parameters $\chi$, $\zeta$:

$$\begin{aligned} \delta\Phi &= - W^\alpha \nabla_\alpha \chi - i[\bar{\nabla}^2 (\nabla^\alpha \zeta) \nabla_\alpha + (\nabla^\alpha \zeta) i W_\alpha] \Phi \\ &= - W^\alpha \nabla_\alpha \chi - [(\tfrac{1}{2} [\bar{\nabla}^{\dot{\alpha}}, \nabla^\alpha] \zeta) \nabla_{\alpha\dot{\alpha}} + (i \bar{\nabla}^2 \nabla^\alpha \zeta) \nabla_\alpha] \Phi \quad , \\ e^{-\Omega} \delta e^{\Omega} &= - i \chi \bar{\Phi} + \bar{W}^{\dot{\alpha}} \bar{\nabla}_{\dot{\alpha}} \zeta \quad . \end{aligned} \tag{4.6.2}$$

(For the $\zeta$ transformation we use (4.2.52), and also a gauge transformation with $K = -i(\bar{\Gamma}^{\dot{\alpha}} D^2 \bar{D}_{\dot{\alpha}}\zeta - \Gamma^{\alpha}\bar{D}^2 D_{\alpha}\zeta) + \Gamma^{\alpha\dot{\alpha}}\frac{1}{2}[\bar{D}_{\dot{\alpha}}, D_{\alpha}]\zeta$.) Due to the identity $\delta\nabla_{\alpha} = [\nabla_{\alpha}, (e^{-\Omega}\delta e^{\Omega})]$ (4.2.77), the second transformation can be written as

$$\delta\nabla_{\alpha} = -\nabla_{\alpha}(i\chi\bar{\Phi} - \bar{W}^{\dot{\beta}}\bar{\nabla}_{\dot{\beta}}\zeta) = -i\bar{\Phi}\nabla_{\alpha}\chi + \bar{W}^{\dot{\alpha}}\frac{1}{2}[\bar{\nabla}_{\dot{\alpha}}, \nabla_{\alpha}]\zeta \quad . \tag{4.6.3}$$

Both parameters are $x$-independent superfields and commute with the group generators (e.g., $\nabla_{\alpha}\chi = D_{\alpha}\chi$). The parameter $\chi$ is chiral and mixes the two $N = 1$ multiplets, whereas $\zeta$ is the real parameter of the $N = 1$ supersymmetry transformations (3.6.13). Since $\zeta$ has the ($x$-independent) gauge invariance $\delta\zeta = i(\bar{\lambda} - \lambda)$, the global superparameters themselves form an abelian $N = 2$ vector multiplet. Referring to the components of this parameter multiplet $(\chi, \zeta)$ by the names of the corresponding components in the field multiplet $(\Phi, V)$, we find the following: The "physical bosonic fields" give translations (from the vector $\zeta_{\underline{a}} \equiv \frac{1}{2}[\bar{D}_{\dot{\alpha}}, D_{\alpha}]\zeta|$) and central charges (from the scalars $z \equiv \chi|$); the "physical fermionic fields" give supersymmetry transformations ($\epsilon^1{}_{\alpha} \equiv i\bar{D}^2 D_{\alpha}\zeta|$, $\epsilon^2{}_{\alpha} \equiv D_{\alpha}\chi|$); and the "auxiliary fields" give internal symmetry $U(2)/SO(2)$ transformations ($r \equiv \frac{1}{2}D^{\alpha}\bar{D}^2 D_{\alpha}\zeta|$, $q \equiv D^2\chi$). (The full $U(2)$ symmetry has, in addition to $(r, q, \bar{q})$ transformations, phase rotations $\delta\Phi = iu\Phi$, $\delta V = 0$).

The algebra of the $N = 2$ global transformations closes off shell; e.g., the commutator of two $\chi$ transformations gives a $\zeta$ transformation:

$$[\delta_{\chi_1}, \delta_{\chi_2}] = \delta_{\zeta_{12}} \quad , \quad \zeta_{12} = i\bar{\chi}_{[1}\chi_{2]} = i(\bar{\chi}_1\chi_2 - \bar{\chi}_2\chi_1) \quad . \tag{4.6.4}$$

The transformations take a somewhat different form in the chiral representation:

$$\delta\Phi = -W^{\alpha}\nabla_{\alpha}\chi - i\bar{\nabla}^2(\nabla^{\alpha}\zeta)\nabla_{\alpha}\Phi \quad ,$$

$$e^{-V}\delta e^{V} = i(\bar{\chi}\Phi - \chi\tilde{\bar{\Phi}}) + (W^{\beta}\nabla_{\beta} + \bar{W}^{\dot{\beta}}\bar{\nabla}_{\dot{\beta}})\zeta \quad , \tag{4.6.5a}$$

and hence

$$\delta\nabla_{\alpha} = \nabla_{\alpha}[i(\bar{\chi}\Phi - \chi\tilde{\bar{\Phi}}) + (W^{\beta}\nabla_{\beta} + \bar{W}^{\dot{\beta}}\bar{\nabla}_{\dot{\beta}})\zeta] \quad . \tag{4.6.5b}$$

Now the $i(\bar{\lambda} - \lambda)$ part of the $\zeta$ transformation does contribute, but only as a field-dependent gauge transformation $\Lambda = W^{\alpha}\nabla_{\alpha}\lambda$.

We can add an $N = 2$ Fayet-Iliopoulos term (parametrized by constants $\nu = \nu_1 + i\nu_2\,, \nu_3 = \overline{\nu}_3$)

$$\int d^4x d^4\theta\, \nu_3 V + (-i \int d^4x d^2\theta\, \nu\Phi + h.\,c.) \tag{4.6.6}$$

to the above action in the abelian (free) case. This is invariant under (4.6.5) if we restrict the global parameters by

$$-\overline{\nu}\overline{D}^2\overline{\chi} = \nu D^2\chi \quad , \quad \nu_3\overline{D}^2\overline{\chi} = i\nu(2\overline{D}^2D^2\zeta + u) \quad , \tag{4.6.7}$$

where $u$ is the real constant parameter of the phase $SO(2)$ part of $U(2)$. The constraint on the parameters implies that the $U(2)$ is broken down to $SO(2)\otimes U(1)$.

This model has some interesting quantum properties. It has gauge invariant divergences at one-loop, but explicit calculations show their absence at the two- and three-loop level. In sec. 7.7 we present an argument to establish their absence at all higher loops.

## a.2. Hypermultiplet

### a.2.i. Free theory

The $N = 2$ scalar multiplet can be described by a chiral scalar isospinor superfield $\Phi^a$ (the "$\Phi^a$ hypermultiplet") with the free action

$$S = \int d^4x d^4\theta\, \overline{\Phi}_a\Phi^a + \frac{1}{2}(\int d^4x d^2\theta\, \Phi^a m_{ab}\Phi^b + h.\,c.) \quad , \tag{4.6.8}$$

where the symmetric matrix $m$ satisfies the condition

$$m_{ac}C^{cb} = C_{ac}\overline{m}^{cb} \quad . \tag{4.6.9}$$

(The explicit form is $m_{ab} = iMC_{ab}\tau^b{}_c$, $M = \overline{M}$, with $\tau^a{}_a = 0$ and $\overline{\tau^b{}_a} = \tau^a{}_b$. Without loss of generality, $m_{ab}$ can be chosen proportional to $\delta_{ab}$.) The free action is invariant under the global symmetries

$$\delta\Phi^a = -\,(\overline{D}^2\overline{\chi}C^{ab}\overline{\Phi}_b - \chi Z\Phi^a) - i\overline{D}^2[(D^\alpha\zeta)D_\alpha\Phi^a + (D^2\zeta)\Phi^a] \quad , \tag{4.6.10}$$

where $Z$ is a central charge:

$$Z\Phi^a = C^{ab}m_{bc}\Phi^c \quad , \quad Z\overline{\Phi}_a = C_{ab}\overline{m}^{bc}\overline{\Phi}_c \quad . \tag{4.6.11}$$

On shell, we also have

$$Z\Phi^a = -\, C^{ab}\bar{D}^2\bar{\Phi}_b \quad . \tag{4.6.12}$$

We can use either of the forms (4.6.11,12) in the transformation (4.6.10), because of the local invariance

$$\delta\Phi^a = \eta C^{ab} S_b \quad , \quad S_b \equiv \frac{\delta S}{\delta\Phi^b} \quad , \tag{4.6.13}$$

for arbitrary $x$-dependent chiral $\eta$. (This is an invariance because the variation of the action is proportional to $\delta S \sim S_a C^{ab} S_b = 0$.) If we use the form (4.6.12), the variations do not depend on the parameters $m_{ab}$. An interesting feature of the algebra (4.6.10) is that it does not close off-shell if we use realization (4.6.11) for $Z$. On the other hand, if we use realization (4.6.12) instead, the symmetries (4.6.10) contain part of the field equations, and hence become nonlinear and coupling-dependent when interactions are introduced. These effects are a signal that in the decomposition of the $N = 2$ superfield that describes the theory into $N = 1$ superfields, some auxiliary $N = 1$ superfields have been discarded. We discuss further aspects of this problem below.

Without the mass term, the internal symmetries of the free scalar multiplet are the explicit $SU(2)$ that acts on the isospinor index of $\Phi^a$ and the $U(2)$ made up of the $r$ and $q$ transformations in $\zeta$ and $\chi$, and of the uniform phase rotations $\delta\Phi^a = iu\Phi^a$. The mass term breaks the explicit $SU(2)$ to the $U(1)$ subgroup that commutes with $m_{ac}$.

### a.2.ii. Interactions

The $N = 2$ scalar multiplet can interact with an $N = 2$ vector multiplet, and it can have self-interactions describing a nonlinear $\sigma$ model. A class of supersymmetric $\sigma$-models can be found by coupling an abelian $N = 2$ vector multiplet (with no kinetic term but with a Fayet-Iliopoulos term) to $n$ $N = 2$ scalar multiplets described by the $n$-vector $\vec{\Phi}$. The supersymmetry transformations of the vector multiplet are the same as those given above in (4.6.2) or (4.6.5) for the abelian case. (They are independent of the fields in the scalar multiplets.) However, the transformations of the scalar multiplets (each of which is described by a pair of chiral superfields $\Phi^a$) are gauge covariantized:

$$\delta\Phi^a = -\, \bar{D}^2\big[\bar{\chi}\bar{\Phi}_c\big(e^{\tau V}\big)^c{}_b\, C^{ab}\big] - i\bar{D}^2[(D^\alpha\zeta)\nabla_\alpha\Phi^a + (D^2\zeta)\Phi^a] - i\,\frac{1}{2}\,u\Phi^a \quad .$$

(4.6.14)

The matrix $\tau$ is an $SU(2)$ generator that breaks the explicit $SU(2)$ of the scalar multiplet down to $U(1)$. Because $SU(2)$ preserves the alternating tensor $C_{ab}$, $\left(e^{\tau V}\right)^a{}_b \left(e^{\tau V}\right)^c{}_d C_{ac} = C_{bd}$. The action that is left invariant by these transformations is:

$$S = \int d^4x\, d^4\theta\, [\vec{\overline{\Phi}}_a \left(e^{\tau V}\right)^a{}_b \cdot \vec{\Phi}^b + \nu_3 V]$$

$$+ \int d^4x\, d^2\theta\; i\Phi\big[\frac{1}{2}\vec{\Phi}^a\, C_{ab}\tau^b{}_c \cdot \vec{\Phi}^c - \nu\big] \;+\; h.c. \qquad (4.6.15)$$

provided (4.6.7) are satisfied. The theory is also invariant under local abelian gauge transformations:

$$\delta\vec{\Phi}^a = i\,\Lambda\,\tau^a{}_b\vec{\Phi}^b \quad , \quad \delta V = i(\overline{\Lambda} - \Lambda) \quad , \quad \delta\Phi = 0 \;\; ; \qquad (4.6.16)$$

as well as global $SU(n)$ rotations of $\vec{\Phi}^a$. For explicit computation, it is useful to choose a specific $\tau$: We choose $\tau = \tau_3$. We write $\vec{\Phi}^a \equiv (\vec{\Phi}_+ , \vec{\Phi}_-) \equiv (\Phi_+{}^i , \Phi_{-i})$ where $i = 1 \cdots n$ is the $SU(n)$ index, $+,-$ are the $SU(2)$ isospin indices, and $\vec{\Phi}_+$ transforms under the $SU(n)$ representation conjugate to $\vec{\Phi}_-$. The transformations (4.6.14) and the action (4.6.15) become (using (4.6.7))

$$\delta\Phi_\pm = \pm\, \bar{D}^2(\bar{\chi}\overline{\Phi}_\mp e^{\mp V}) - \frac{1}{2}\frac{\nu_3}{\nu}\,(\bar{D}^2\bar{\chi})\Phi_\pm - i\bar{D}^2(D^\alpha\zeta)\nabla_\alpha\Phi_\pm \qquad (4.6.17a)$$

$$S = \int d^4x\, d^4\theta\, [\overline{\Phi}_{+i}\, e^V \Phi_+{}^i + \Phi_{-i}\, e^{-V}\overline{\Phi}_-{}^i + \nu_3 V] \quad ,$$

$$+ \int d^4x\, d^2\theta\; i\Phi\big[\,\Phi_{-i}\Phi_+{}^i - \nu\big] \;+\; h.c. \qquad (4.6.17b)$$

We now proceed as we did in the case of the $CP(n)$ models (see (4.3.9)): We eliminate the vector multiplet by its (algebraic) equations of motion. In this case, $\Phi$ acts as a Lagrange multiplier to impose the constraint:

$$\Phi_{-i}\Phi_+{}^i = \nu \quad . \qquad (4.6.18)$$

Choosing a gauge (e.g., $\Phi_+{}^1 = \Phi_{-1}$), we can easily solve this constraint; for example, we can parametrize the solution as:

$$\Phi_+{}^i = (1 + u_+ \cdot u_-)^{-\frac{1}{2}} \nu^{\frac{1}{2}} \big( 1 \, , \, \vec{u}_+ \big) \ \ ,$$

$$\Phi_{-i} = (1 + u_+ \cdot u_-)^{-\frac{1}{2}} \nu^{\frac{1}{2}} \big( 1 \, , \, \vec{u}_- \big) \ \ . \tag{4.6.19}$$

The $V$ equation of motion gives:

$$\overline{\Phi}_{+i} \, e^V \Phi_+{}^i - \Phi_{-i} \, e^{-V} \overline{\Phi}_-{}^i + \nu_3 = 0 \ \ , \tag{4.6.20a}$$

or

$$M_\pm e^{\pm V} \ = \frac{1}{2} \left[ (\nu_3{}^2 + 4 M_+ M_-)^{\frac{1}{2}} \mp \nu_3 \right] \ \ ; \tag{4.6.20b}$$

where

$$M_\pm = \overline{\Phi}_\pm \cdot \Phi_\pm = |\Phi_\pm|^2 = |\nu| \, |1 + u_+ \cdot u_-|^{-1} (1 + |u_\pm|^2) \ \ . \tag{4.6.20c}$$

Substituting, we find the action

$$S = \int d^4x \, d^4\theta \, \big\{ (\nu_3{}^2 + 4M_+ M_-)^{\frac{1}{2}} + |\nu_3| \, ln \big[ (\nu_3{}^2 + 4M_+ M_-)^{\frac{1}{2}} - |\nu_3| \big] \big\} \ \ . \tag{4.6.21}$$

In terms of the unconstrained chiral superfields $u_\pm$, the transformations (4.6.17a) become

$$\delta u_\pm = \pm \, \bar{D}^2 \big[ \bar{\chi} e^{\mp V} (\frac{\bar{\nu}}{\nu})^{\frac{1}{2}} (1 + u_+ \cdot u_-)^{\frac{1}{2}} (1 + \bar{u}_+ \cdot \bar{u}_-)^{-\frac{1}{2}} (\bar{u}_\mp - u_\pm) \big]$$

$$- \, i \bar{D}^2 \big[ (D^\alpha \zeta) D_\alpha u_\pm \big] \ \ , \tag{4.6.22a}$$

where the auxiliary gauge field $V$ is expressed in terms of $u_\pm$ by (4.6.20). The supersymmetry transformations (4.6.22a) include a compensating gauge transformation with parameter

$$i\Lambda = - \, \bar{D}^2 [ \bar{\chi} (cosh \ V) (\frac{\bar{\nu}}{\nu})^{\frac{1}{2}} (1 + u_+ \cdot u_-)^{\frac{1}{2}} (1 + \bar{u}_+ \cdot \bar{u}_-)^{-\frac{1}{2}} ] \tag{4.6.22b}$$

that must be added to (4.6.17a) to maintain the gauge choice we made in (4.6.19).

As for the free $N = 2$ scalar multiplet, we can add an invariant mass term (which introduces a nonvanishing central charge). The mass term necessarily breaks $SU(n)$ and has the form

$$I_m = i \, \frac{1}{2} \int d^4x \, d^2\theta \, \Phi^a \, C_{ab} \tau^b{}_c \, M \, \Phi^c \ + \ h.\,c. \ \ , \tag{4.6.23}$$

where $M$ is any traceless $n \times n$ matrix ($M$'s differing by $SU(n)$ transformations are equivalent). The supersymmetry transformations that leave this term invariant are the same as before, including the $Z$ term of (4.6.10). The realization of $Z$ given in (4.6.11) is preferable, since it is linear, whereas the realization (4.6.12) must be gauge covariantized.

These nonlinear $\sigma$-models live on Kähler manifolds with three independent complex coordinate systems related by *nonholomorphic* coordinate transformations (they have three independent complex structures (see the end of sec. 4.1); the constants $\nu_3\,,\nu\,,\bar{\nu}$ parametrize the linear combination of complex structures chosen by the particular coordinate system). Thus these manifolds are hyperKähler. Just as we found that for every Kähler manifold there is an $N=1$ nonlinear $\sigma$-model (and conversely), one can show that for every hyperKähler manifold there is an $N=2$ nonlinear $\sigma$-model, and conversely, $N=2$ nonlinear $\sigma$-models are defined only on hyperKähler manifolds. An immediate consequence of this relation is a strong restriction on possible off-shell formulations of the $N=2$ scalar multiplet:

> No formulation can exist that contains as physical submultiplets two $N=1$ scalar multiplets (e.g., such as we have considered), that can be used to describe $N=2$ nonlinear $\sigma$-models, and that has supersymmetry transformations *independent* of the form of the action.

If such a formulation existed, then the sum of two $N=2$ invariant actions would necessarily be invariant; however, the sum of the Kähler potentials of two *hyper*Kähler manifolds is *not* in general the Kähler potential of a hyperKähler manifold. We will see below that we *can* give an off-shell formulation of the $N=2$ scalar multiplet that avoids this problem.

We can generalize the action (4.6.15) in the same way that we generalized the $CP(n)$ models (see (4.3.11)):

$$S = \int d^4x\, d^4\theta\, \left[\bar{\Phi}_a \left(e^{\tau V}\right)^a{}_b\, \Phi^b + \nu_3 trV\right]$$

$$+ \int d^4x\, d^2\theta\; i\big[\frac{1}{2}\,\Phi^a\,\Phi\, C_{ab}\tau^b{}_c\,\Phi^c - \nu tr\Phi\big] \;+\; h.\,c.\;\;, \qquad (4.6.24\text{a})$$

where $V = V^{\mathrm{A}} T_{\mathrm{A}}$, $\Phi = \Phi^{\mathrm{A}} T_{\mathrm{A}}$, and $T_{\mathrm{A}}$ are the generators of some group. The $N=2$ transformations that leave (4.6.24a) invariant are the obvious nonabelian generalizations of (4.6.14). The equations that result from varying (4.6.23a) with respect to $\Phi^{\mathbf{A}}$ , $V^{\mathbf{A}}$ are, choosing $\tau = \tau_3$ as above,

$$\Phi_- T_{\mathbf{A}} \Phi_+ - \nu tr T_{\mathbf{A}} = 0 \quad ,$$

$$\overline{\Phi}_+ e^V T_{\mathbf{A}} \Phi_+ - \overline{\Phi}_- T_{\mathbf{A}} e^{-V} \Phi_- + \nu_3 tr T_{\mathbf{A}} = 0 \quad . \tag{4.6.24b}$$

As in the $N = 1$ case, these do not, in general, have an explicit solution.

### a.3. Tensor multiplet

Just as the $N = 1$ scalar multiplet can be described by different superfields, we can describe the $N = 2$ scalar multiplet by superfields other than the chiral isodoublet $\Phi^a$. We now discuss the $N = 2$ tensor formulation of the scalar multiplet. This is dual to the previous description in the same way that the $N = 1$ tensor and scalar multiplets are dual (see sec. 4.4.c). We write the tensor form of the scalar multiplet in terms of one chiral scalar field $\eta$ and a chiral spinor gauge field $\phi_\alpha$ with linear field strength $G = \frac{1}{2}(D_\alpha \phi^\alpha + \overline{D}_{\dot\alpha} \phi^{\dot\alpha})$, $D^2 G = \overline{D}^2 G = 0$. The $N = 2$ supersymmetry transformations of this theory are

$$\delta\phi_\alpha = -\, 2\eta D_\alpha \chi - i\overline{D}^2[(D^\beta\zeta)D_\beta\phi_\alpha + (D^2\zeta)\phi_\alpha] \quad ,$$

$$\delta\eta = -\, \overline{D}^2(\overline{\chi}G) - i\overline{D}^2[(D^\beta\zeta)D_\beta\eta + 2(D^2\zeta)\eta] \quad . \tag{4.6.25}$$

In contrast to the $\Phi^a$ hypermultiplet realization of the $N = 2$ scalar multiplet, these transformations close *off-shell;* they have the same algebra as the transformations of the $N = 2$ vector multiplet (4.6.4) (up to a gauge transformation of $\phi_\alpha$). However, although the superfields describe a scalar multiplet, the central charge transformations $z = \chi|$ leave the fields $\phi\, , \eta$ inert; this gives one guide to understanding the duality to the hypermultiplet.

The simplest action invariant under the transformations (4.6.25) is the sum of the usual free chiral and tensor actions ((4.1.1) and (4.4.34)):

$$S_{kin} = \int d^4x\, d^4\theta\, [-\tfrac{1}{2}G^2 + \eta\overline{\eta}] \quad . \tag{4.6.26}$$

To find other actions, we consider a general ansatz, and require invariance under the transformations (4.6.25). Actually, we can consider a slightly more general case that still has full off-shell $N = 2$ invariance by restricting the chiral parameter $\chi$ by $D^2\chi = 0$, and the real parameter $\zeta$ by $\bar{D}^2 D^2\zeta = 0$. This means that we do not impose $SU(2)$ invariance. An action

$$S = \int d^4x\, d^4\theta\ f(G\,, \eta\,, \bar{\eta}) \tag{4.6.27}$$

is invariant under (4.6.25) (with $D^2\chi = 0$) if $f$ satisfies

$$\frac{\partial^2 f}{\partial G^2} + \frac{\partial^2 f}{\partial\eta\partial\bar{\eta}} \equiv f_{GG} + f_{\eta\bar{\eta}} = 0 \quad . \tag{4.6.28}$$

($f$ contains no derivatives of $G$, $\eta$, $\bar{\eta}$.) It describes a general $N = 2$ tensor multiplet interacting model. We also can consider more than one multiplet $G^i\,, \eta^i\,, \bar{\eta}^i$, each transforming as (4.6.25); then the most general invariant action is (4.6.27) where the Lagrangian $f$ satisfies

$$f_{G^i G^j} + f_{\eta^i \bar{\eta}^j} = 0 \quad . \tag{4.6.29}$$

(Actually, we can generalize (4.6.27) slightly by adding a term $\int d^4x\, d^2\theta\ h_i \eta^i + h.\,c.$ where the $h_i$'s are arbitrary constants.)

### a.4. Duality

To gain insight into the physics of these models we find the dual theories described by the $\Phi^a$ hypermultiplet. We consider the following first order action (cf. (4.4.38)):

$$S' = \int d^4x\, d^4\theta\ [f(V^i, \eta^i, \bar{\eta}^i) - V^i(\Phi_i + \bar{\Phi}_i)] \quad . \tag{4.6.30}$$

Eliminating $\Phi, \bar{\Phi}$ gives (4.6.27), while eliminating $V$ results in the dual theory. We find the $N = 2$ transformations of the resulting ${\Phi^a}_i$ hypermultiplets from the transformations that leave the first order action (4.6.43) invariant. Since $\int d^4x\, d^4\theta\ f(V, \eta, \bar{\eta})$ is invariant under (4.6.25) with $G \to V$ *except* for terms $\sim D^2V$ or $\sim \bar{D}^2V$ ($V$ differs from $G$ only because it does not satisfy the Bianchi identities $D^2G = \bar{D}^2G = 0$), we can cancel these terms by choosing the variation of $\Phi$ appropriately. The first order action (4.6.43) is invariant under

$$\delta V^i = D^\alpha \eta^i D_\alpha \chi + \bar{D}^{\dot{\alpha}} \bar{\eta}^i \bar{D}_{\dot{\alpha}} \bar{\chi} + \delta_\zeta V^i \quad , \tag{4.6.31a}$$

$$\delta \eta^i = - \bar{D}^2 (\bar{\chi} V^i) + \delta_\zeta \eta^i \quad , \tag{4.6.31b}$$

$$\delta \Phi_i = - \bar{D}^2 [\bar{\chi} (f_{\eta^i} + V^j (f_{\bar{\eta}^j V^i} - f_{\bar{\eta}^i V^j}))] + \delta_\zeta \Phi_i \quad ; \tag{4.6.31c}$$

where $\delta_\zeta$ is the usual $N = 1$ supersymmetry (3.6.13) (with $w_V = 0$, $w_\eta = -2$, $w_\Phi = 0$). (To prove the invariance of (4.6.30) under (4.6.31), we need (4.6.29) and its consequences, in particular, $f_{V^i V^{[j} \eta^{k]}} = 0$ and $f_{\bar{\eta}^i \eta^{[j} V^{k]}} = 0$ because of the antisymmetrization, and therefore, using the chain rule we find $\bar{D}_{\dot{\alpha}} f_{\bar{\eta}^{[j} V^{i]}} = 0$.) Performing the duality transformations, we can rewrite the transformations (4.6.31) and the condition (4.6.29) in terms of the dual variables $\Phi, \eta$ and the Legendre transformed Lagrangian $I\!K(\Phi + \bar{\Phi}, \eta, \bar{\eta})$. We find (dropping the uninteresting $\delta_\zeta$ terms)

$$\delta \eta^i = \bar{D}^2 (\bar{\chi} I\!K_{\Phi_i}) \quad , \tag{4.6.32a}$$

$$\delta \Phi_i = - \bar{D}^2 [\bar{\chi} (I\!K_{\eta^i} + I\!K_{\Phi_j} ((I\!K_{\Phi_k \bar{\Phi}_i})^{-1} I\!K_{\eta^j \Phi_k} - (I\!K_{\Phi_k \bar{\Phi}_j})^{-1} I\!K_{\eta^i \bar{\Phi}_k}))] \quad , \tag{4.6.32b}$$

for the transformations, and

$$I\!K_{\eta^i \bar{\eta}^j} = (I\!K_{\Phi_i \bar{\Phi}_j})^{-1} + I\!K_{\eta^i \bar{\Phi}_m} (I\!K_{\Phi_m \bar{\Phi}_n})^{-1} I\!K_{\Phi_n \bar{\eta}^j} \tag{4.6.33}$$

for the condition that the Lagrangian must satisfy to guarantee invariance. Note that in contrast with the *off-shell* transformations (4.6.25), the *on-shell* transformations (4.6.31,32) depend explicitly on the form of the action. Furthermore, the condition (4.6.29) needed for invariance of the off-shell version of the model is *linear,* and hence the sum of two invariant actions is automatically invariant, whereas the condition (4.6.33) is *nonlinear.* The Legendre transformation allows this to occur, and allows us to comply with the restriction on off-shell formulations that we discussed above.

Although it is always possible to go from the off-shell formulation (in terms of the tensor multiplet) to the on-shell formulation (in terms of the hypermultiplet), the reverse transformation is generally not so straightforward. The improved form (see (4.4.45-5)) of the free multiplet can be found by exploiting an analogy with the nonlinear $\sigma$-models discussed above (Actually, the tensor multiplet form of the interacting models can be found in this way). Alternatively, some simple models can be found by using the central charge invariance of the tensor multiplet (see below).

By analogy with (4.6.17), we can write down the following first order $N=2$ invariant action by introducing an auxiliary $N=2$ vector multiplet $V, \Phi$:

$$S' = \int d^4x\, d^4\theta\, \left[\bar{\Phi}_+ e^V \Phi_+ + \Phi_- e^{-V} \bar{\Phi}_- - GV\right]$$

$$+ \int d^4x\, d^2\theta\; i\Phi\left[\Phi_-\Phi_+ - \eta\right] \; + \; h.c. \tag{4.6.34}$$

This action is invariant under the transformations (4.6.5,14,25). Varying $G$ and $\eta$, we find the free hypermultiplet (see discussion in sec. 4.6.a.2); varying $V$ and $\Phi$, we find that $\Phi_\pm$ drop out of the action entirely, and the improved ($N=2$) tensor multiplet results:

$$S_{imp} = \int d^4x\, d^4\theta\, \left[(G^2 + 4\bar{\eta}\eta)^{\frac{1}{2}} - G\, ln(G + (G^2 + 4\bar{\eta}\eta)^{\frac{1}{2}})\right] \; . \tag{4.6.35}$$

This complicated nonlinear action corresponds to a free hypermultiplet! It is, however, an off-shell formulation, invariant under the transformations (4.6.25). It generalizes directly to give an off-shell formulation of the nonlinear $\sigma$-models we discussed above.

An alternative derivation of the improved tensor multiplet does not require an $N=2$ vector multiplet, but uses the central charge invariance of the tensor multiplet. We begin with the free hypermultiplet action (4.6.8) (without loss of generality, we take $m_{ab} = imC_{ab}(\tau_3)^b{}_c$). We wish to Legendre transform one of the chiral fields $\Phi^a = (\Phi_+, \Phi_-)$, and keep the other field as the chiral field $\eta$ of the tensor multiplet. However, though $\eta$ is inert under central charge transformations, $\Phi_\pm$ are not; we therefore define the invariant combination $\eta \equiv i\Phi_+\Phi_-$, and in terms of it write the first order action

$$S' = \int d^4x\, d^4\theta\, [\bar{\eta}\eta e^{-V} + e^V - GV] + \frac{1}{2} m[\int d^4x\, d^2\theta\; \eta + h.c.] \; . \tag{4.6.36}$$

Varying $G$, we recover the hypermultiplet action (4.6.8) with $V = ln(\bar{\Phi}_+\Phi_+)$; varying $V$, we recover the improved tensor multiplet action (4.6.35) with a *linear* $\eta$ term that acts as a mass term. The algebra of transformations that act on the massive scalar multiplet *has* a central charge; however, the description of the multiplet given by the $N=2$ tensor multiplet only involves fields that are *inert* under the central charge.

Finally, we note that the gauge interactions of the $N = 2$ tensor multiplet are analogous to the $N = 1$ case (see sec. 4.4.c).

### a.5. N=2 superfield Lagrange multiplier

Another formulation of the $N = 2$ scalar multiplet with off-shell $N = 2$ supersymmetry is the $N = 2$ Lagrange multiplier multiplet. It is the $N = 2$ generalization of the multiplet discussed in sec. 4.5.d, and contains that $N = 1$ multiplet as a submultiplet. Unlike the off-shell $N = 2$ supersymmetric scalar multiplet discussed above (the $N = 2$ tensor multiplet of secs. 4.6.a.3,4), this multiplet can be coupled to the ($N = 2$) non-abelian vector multiplet, though only in real representations. By using the adjoint representation, this allows construction of $N = 4$ Yang-Mills with off-shell $N = 2$ supersymmetry, as discussed below in sec. 4.6.b.2.

The $N = 2$ Lagrange multiplier multiplet is described by the following $N = 1$ superfields: (1) ${\Psi_1}^\alpha$ and $Y$, describing an $N = 1$ Lagrange multiplier multiplet as in (4.5.18), with the gauge invariance of (4.5.19), and field strength $\Sigma_1 = \bar{D}_{\dot\alpha}{\bar\Psi_1}{}^{\dot\alpha}$ (for which $F$ and $G$ of (4.5.18) are the real and imaginary parts); (2) a second spinor ${\Psi_2}^\alpha$, with the same dimension and gauge invariance, but which is auxiliary; (3) a complex Lagrange multiplier $\Xi$, which constrains all of $\Sigma_2$ to vanish (instead of just the imaginary part, as does $Y$ for $\Sigma_1$), and has a field strength $\bar{D}_{\dot\alpha}\Xi$ with gauge invariance $\delta\Xi = \Lambda$ (for $\Lambda$ chiral); (4) a minimal scalar multiplet, described by a complex gauge field $\Psi_1$ with chiral field strength $\Phi_1$ (see sec. 4.5.a); and (5) two more minimal scalar multiplets $\Psi_2$ and $\Psi_3$, but auxiliary. We thus have an $N = 1$ Lagrange multiplier multiplet, a minimal scalar multiplet, and assorted auxiliary superfields.

The action is

$$S = -\int d^4x\, d^4\theta\, [\, \tfrac{1}{8}(\, \Sigma_1 + \bar\Sigma_1\,)^2 + \tfrac{i}{2} Y(\, \Sigma_1 - \bar\Sigma_1\,)\,]$$

$$+ \int d^4x\, d^4\theta\, [\bar\Phi_1\Phi_1 + (\, \Xi\Sigma_2 + \bar\Xi\bar\Sigma_2\,) + (\, \Psi_2\Phi_3 + \bar\Psi_2\bar\Phi_3\,)\,] \ . \qquad (4.6.37)$$

The most interesting properties of this theory appear when it is coupled to $N = 2$ super-Yang-Mills. We do this by $N = 2$ gauge covariantizing the $N = 2$ Lagrange multiplier multiplet field strengths. (In the *absence* of Yang-Mills coupling, the $\Phi$'s can be considered as ordinary scalar multiplets, rather than field strengths.) This coupling is discussed in sec. 4.6.b.2.

## b. N=4 Yang-Mills

In many respects, the $N = 2$ nonlinear $\sigma$-models, when studied in two dimensions, are analogs of $N = 4$ Yang-Mills theory in four dimensions. Despite power-counting arguments, they are completely finite on shell, and they are the maximally supersymmetric models containing only scalar multiplets (the vector is auxiliary and can be eliminated). The $N = 4$ Yang-Mills theory is the first and best-studied 4-dimensional theory that is ultraviolet finite to all orders of perturbation theory, and thus scale invariant at the quantum as well as the classical level. (Its $\beta$-function has been calculated to vanish through three loops; arguments for total finiteness are given in sec. 7.7. Independent arguments using light-cone superfields have been given elsewhere.) It is self-conjugate and is the maximally extended globally supersymmetric theory. Two superfield formulations of the theory have been given: One uses an $N = 2$ vector multiplet coupled to a $\Phi^a$ hypermultiplet and has only $N = 1$ supersymmetry off shell, and the other uses an $N = 2$ vector multiplet coupled to an $N = 2$ Lagrange multiplier multiplet and has $N = 2$ supersymmetry off shell (however, it has a large number of auxiliary superfields).

### b.1. Minimal formulation

At the component level the theory contains a gauge vector particle, four spin $\frac{1}{2}$ Weyl spinors, and six spin 0 particles, all in the adjoint representation of the internal symmetry group. It can be described by one real scalar gauge superfield $V$ and three chiral scalar superfields $\Phi^i$, and is the same as an $N = 2$ vector multiplet coupled to an $N = 2$ scalar multiplet. If we use a matrix representation for the $\Phi^i$, the (chiral representation) conjugate can be written as $\tilde{\overline{\Phi}}_i = e^{-V}\overline{\Phi}_i e^V$. The $N = 1$ supersymmetric action (in the chiral representation) is given by

$$S = \frac{1}{g^2} tr(\int d^4x d^4\theta\, e^{-V}\overline{\Phi}_i e^V \Phi^i + \int d^4x d^2\theta\, W^2$$

$$+ \frac{1}{3!}\int d^4x d^2\theta\, iC_{ijk}\Phi^i[\Phi^j\, ,\Phi^k] + \frac{1}{3!}\int d^4x d^2\overline{\theta}\; iC^{ijk}\overline{\Phi}_i[\overline{\Phi}_j\, ,\overline{\Phi}_k])\ . \qquad (4.6.38)$$

In addition to the manifest SU(3) symmetry on the $i, j, k$ indices of $\Phi$ and $\overline{\Phi}$, it has the following global symmetries:

$$\delta\Phi^i = -\,(W^\alpha\nabla_\alpha\chi^i + C^{ijk}\overline{\nabla}^2\overline{\chi}_j\tilde{\overline{\Phi}}_k) - i\overline{\nabla}^2[(\nabla^\alpha\zeta)\nabla_\alpha\Phi^i + \frac{2}{3}(\nabla^2\zeta)\Phi^i] \quad ,$$

$$\delta\nabla_\alpha = \nabla_\alpha[i(\overline{\chi}_i\Phi^i - \chi^i\tilde{\overline{\Phi}}_i) + (W^\beta\nabla_\beta + W^{\dot\beta}\overline{\nabla}_{\dot\beta})\zeta] \quad ; \qquad (4.6.39)$$

in the chiral representation, and in the vector representation

$$\delta\Phi^i = -\,(W^\alpha\nabla_\alpha\chi^i + C^{ijk}\overline{\nabla}^2\overline{\chi}_j\overline{\Phi}_k - i[\overline{\chi}_j\Phi^j\,,\Phi^i])$$

$$- i[\overline{\nabla}^2(\nabla^\alpha\zeta)\nabla_\alpha\Phi^i + (\nabla^\alpha\zeta)iW_\alpha\Phi^i + \frac{2}{3}\overline{\nabla}^2(\nabla^2\zeta)\Phi^i] \quad ;$$

$$\delta\nabla_\alpha = -\,\nabla_\alpha(i\chi^i\overline{\Phi}_i + \overline{W}^{\dot\beta}\overline{\nabla}_{\dot\beta}\zeta) \quad . \qquad (4.6.40)$$

The $\chi^i$ are the generalization of those given for the $N = 2$ multiplets above, but now they form an $SU(3)$ isospinor, as does $\Phi^i$. The identification of the components of $\chi$ and $\zeta$ is the same: The "physical bosonic fields" are the translations and the central charge parameters (3 complex = 6 real, as follows from dimensional reduction from D=10: see sec. 10.6), the spinors are the supersymmetry parameters, and the "auxiliary fields" are internal symmetry parameters of $SU(4)/SU(3)$. The algebra does not close off-shell. Upon reduction to its $N = 2$ submultiplets, (4.6.39) (or (4.6.40)) reduces to (4.6.5) (or (4.6.2)) and (4.6.14) (but with different R-weights).

The corresponding component action has a conventional appearance, with gauge, Yukawa, and quartic scalar couplings all governed by the same coupling constant. In sec. 6.4 we discuss some of the quantum properties of this theory.

## b.2. Lagrange multiplier formulation

We now briefly describe another $N = 1$ superfield formulation of $N = 4$ super-Yang-Mills; it employs the (unimproved) type of $N = 1$ scalar multiplet of sec. 4.5.d. Although even less of the $SU(4)$ symmetry is manifest, this formulation is off-shell $N = 2$ supersymmetric: It follows from the $N = 2$ superfield formulation of the theory, as described by the coupling of $N = 2$ super Yang-Mills to an $N = 2$ (Lagrange multiplier) scalar multiplet. This formulation has a number of other novel features: (1)

renormalizable couplings between nonminimal and other scalar multiplets, (2) the necessary appearance (in interaction terms) of the minimal scalar multiplet in the form of a gauge multiplet (sec. 4.5.a), and (3) loss of (super)conformal invariance off shell (this occurs because the model includes an *unimproved* Lagrange multiplier multiplet).

The action can be written as (in the super-Yang-Mills vector representation) the sum of (4.6.1) and (4.6.37). However, the definitions of the field strengths $\Sigma_i$ and $\Phi_i$ are now modified:

$$\Sigma_i = \overline{\nabla}_{\dot\alpha}\overline{\Psi}_i{}^{\dot\alpha} - i[\Phi_0\,,\Psi_i]$$

$$\Phi_i = \overline{\nabla}^2\Psi_i \quad (i = 0,1,2) \quad , \quad \Phi_3 = \overline{\nabla}^2\Psi_3 - i[\Phi_0\,,\Xi] \ ; \tag{4.6.41}$$

where $\Phi_0$ (with prepotential $\Psi_0$) is the chiral superfield of the $N = 2$ Yang-Mills multiplet. The $\nabla_A$ in these definitions is the Yang-Mills covariant derivative. In addition to the usual (adjoint, vector representation) Yang-Mills gauge transformations, we have many new local symmetries of the action:

$$\delta\Psi_i = \overline{\nabla}_{\dot\alpha}\overline{K}_i{}^{\dot\alpha} \quad (i = 0,1,2) \quad , \quad \delta\Psi_3 = \overline{\nabla}_{\dot\alpha}\overline{K}_3{}^{\dot\alpha} + i[\Psi_0\,,\Lambda] \ ; \tag{4.6.42a}$$

$$\delta\Psi_i{}^{\alpha} = \nabla_\beta K_i{}^{(\alpha\beta)} + i[\overline{\Phi}_0\,,K_i{}^{\alpha}] \quad (i = 1,2) \ ; \ \delta\Xi = \Lambda \ ; \ \delta Y = \delta\Omega = 0 \ ; \tag{4.6.42b}$$

where $\Lambda$ is covariantly chiral ($\overline{\nabla}_{\dot\alpha}\Lambda = 0$), and $\Omega$ is the Yang-Mills vector-representation prepotential. Under these transformations the field strengths $\Phi_i$ $(i = 0,\ldots,3)$, $\Sigma_i$ $(i = 1,2)$, $\overline{\nabla}_{\dot\alpha}\Xi$, $Y$, and $W_\alpha$ are invariant. In the abelian (or linearized) case, the sum of (4.6.1) and (4.6.37) as modified by (4.6.42) describes an $N = 2$ vector multiplet ($W_\alpha$ and $\Phi_0$) plus an $N = 2$ scalar multiplet consisting of the $N = 1$ Lagrange multiplier multiplet of (4.5.18) ($\Psi_1{}^\alpha$ and $Y$), a minimal $N = 1$ scalar multiplet ($\Phi_1$), and some auxiliary superfields ($\Psi_2{}^\alpha$, $\Xi$, $\Psi_2$, and $\Psi_3$). However, in the interacting case the formulation is somewhat unusual in that $\Phi_3$ is not just $N = 1$ covariantly chiral ($\overline{\nabla}_{\dot\alpha}\Phi_3 \neq 0$) nor are $\Sigma_i$ $N = 1$ covariantly linear ($\overline{\nabla}^2\Sigma_i \neq 0$), but they satisfy the $N = 2$ covariant Bianchi identities

$$\overline{\nabla}_{\dot\alpha}\Phi_3 = -\,i[\Phi_0\,,\overline{\nabla}_{\dot\alpha}\Xi] \ ,$$

$$\overline{\nabla}^2\Sigma_i = -\,i[\Phi_0\,,\Phi_i] \ . \tag{4.6.43}$$

The interaction terms of the auxiliary superfields (introduced through the nonlinearities of the field strengths $\Sigma_2$ and $\Phi_3$) cancel among themselves: Their terms in the action can be rewritten as, in the *chiral* representation,

$$\int d^4x\, d^4\theta\, (\, \Xi \bar{D}_{\dot{\alpha}} \bar{\Psi}_2{}^{\dot{\alpha}} + h.\, c.\,) + [\int d^4x\, d^2\theta\;\; \Phi_2(\, \bar{D}^2 \Psi_3\,) + h.\, c.] \;\; . \tag{4.6.44}$$

By combining the Bianchi identities (4.6.43), the usual constraint $\overline{\nabla}_{\dot{\alpha}} \Phi_i = 0$ (for $i = 0, 1, 2$), and $\overline{\nabla}_{\dot{\alpha}} W_\alpha = 0$, $\nabla^\alpha W_\alpha + \overline{\nabla}^{\dot{\alpha}} W_{\dot{\alpha}} = 0$ with the field equations which follow from the action, we obtain the on-shell equations for all of the superfields

$$\bar{D}_{\dot{\alpha}} \Xi = \Sigma_1 - \bar{\Sigma}_1 = \Sigma_2 = \Phi_2 = \Phi_3 = 0 \;\; ,$$

$$i \nabla^\alpha W_\alpha = -\, i \overline{\nabla}^{\dot{\alpha}} \bar{W}_{\dot{\alpha}} = [\Phi_0\, , \bar{\Phi}_0] + [\Phi_1\, , \bar{\Phi}_1] + \frac{1}{4} [(\, \Sigma_1 + iY\,)\, , \overline{(\, \Sigma_1 + iY\,)}]\, ,$$

$$\overline{\nabla}_{\dot{\alpha}} \Phi_0 = \overline{\nabla}^2 \bar{\Phi}_0 + i[\Phi_1\, , \frac{1}{2} (\, \Sigma_1 + iY\,)] = 0\, ,$$

$$\overline{\nabla}_{\dot{\alpha}} \Phi_1 = \overline{\nabla}^2 \bar{\Phi}_1 + i[\frac{1}{2} (\, \Sigma_1 + iY\,)\, , \Phi_0] = 0\, ,$$

$$\overline{\nabla}_{\dot{\alpha}} (\, \Sigma_1 + iY\,) = \overline{\nabla}^2 \overline{(\, \Sigma_1 + iY\,)} + 2i[\Phi_0\, , \Phi_1] = 0\, . \tag{4.6.45}$$

We can thus identify this formulation on shell with that given above in subsec. 4.6.b.1. by the correspondences

$$W_\alpha \longleftrightarrow W_\alpha \quad , \; (\, \Phi_0\, , \Phi_1\, , \frac{1}{2} (\, \Sigma_1 + iY\,)\,) \longleftrightarrow \Phi^i \;\; . \tag{4.6.46}$$

**Contents of 5. CLASSICAL N=1 SUPERGRAVITY**

# 5. CLASSICAL N=1 SUPERGRAVITY

## 5.1. Review of gravity

### a. Potentials

Our review is intended to describe the approach to gravity that is most useful in understanding supergravity. We treat gravity as the theory of a massless spin-2 particle described by a gauge field with an additional *vector* index as a group index (so that it contains spin 2). By analogy with the theory of a massless spin 1 particle its linearized transformation law is

$$\delta h_{\underline{a}}{}^{\underline{m}} = \partial_{\underline{a}} \lambda^{\underline{m}} \quad . \tag{5.1.1}$$

Since the only global symmetry of the S-matrix with a vector generator is translations, we choose partial spacetime derivatives (momentum) as the generators appearing contracted with the gauge field's group index in the covariant derivative

$$e_{\underline{a}} \equiv \partial_{\underline{a}} - i h_{\underline{a}}{}^{\underline{m}} (i \partial_{\underline{m}})$$

$$= (\delta_{\underline{a}}{}^{\underline{m}} + h_{\underline{a}}{}^{\underline{m}}) \partial_{\underline{m}} \equiv e_{\underline{a}}{}^{\underline{m}} \partial_{\underline{m}} \quad . \tag{5.1.2a}$$

Thus, in contrast with Yang-Mills theory, we are able to combine the derivative and "group" terms into a single term. The gauge field $e_{\underline{a}}{}^{\underline{m}}$ is the *vierbein,* which reduces to a Kronecker delta in flat space. It is invertible: Its inverse $e_{\underline{m}}{}^{\underline{a}}$ is defined by

$$e_{\underline{m}}{}^{\underline{a}} \, e_{\underline{a}}{}^{\underline{n}} = \delta_{\underline{m}}{}^{\underline{n}} \quad , \qquad e_{\underline{a}}{}^{\underline{m}} \, e_{\underline{m}}{}^{\underline{b}} = \delta_{\underline{a}}{}^{\underline{b}} \quad . \tag{5.1.2b}$$

Finite gauge transformations are also defined by analogy with Yang-Mills theory:

$$e'_{\underline{a}} = e^{i\lambda} \, e_{\underline{a}} \, e^{-i\lambda} \; , \qquad \lambda \equiv \lambda^{\underline{m}} i \partial_{\underline{m}} \quad . \tag{5.1.3}$$

The linearized transformation takes the form of (5.1.1), whereas the full infinitesimal form takes the form of a *Lie derivative:*

$$(\delta e_{\underline{a}}{}^{\underline{m}}) \partial_{\underline{m}} = i[\lambda, e_{\underline{a}}] = - \, [\lambda^{\underline{n}} \partial_{\underline{n}} \, , \, e_{\underline{a}}{}^{\underline{m}} \partial_{\underline{m}}], \tag{5.1.4a}$$

or, in more conventional notation,

$$\delta e_{\underline{a}}{}^{\underline{m}} = e_{\underline{a}}{}^{\underline{n}} \partial_{\underline{n}} \lambda^{\underline{m}} - \lambda^{\underline{n}} \partial_{\underline{n}} e_{\underline{a}}{}^{\underline{m}} \quad . \tag{5.1.4b}$$

The gauge transformation of a scalar matter field is, again by analogy with Yang-Mills theory,

$$\psi' = e^{i\lambda}\psi \ \equiv \ e^{i\lambda}\psi e^{-i\lambda} \quad , \tag{5.1.5}$$

and in infinitesimal form

$$\delta\psi = i[\lambda, \psi] = -\,\lambda^{\underline{m}}\partial_{\underline{m}}\psi \quad . \tag{5.1.6}$$

Equation (5.1.5) can also be written as the more common general coordinate transformation

$$\psi'(x') \ \equiv \ \psi(x) \ , \qquad x' = e^{-i\lambda}xe^{i\lambda} \quad . \tag{5.1.7}$$

(This can be verified by a Taylor expansion.) For the case of constant $\lambda$ it takes the familiar form of global translations. Orbital (global) Lorentz transformations are obtained by choosing $\lambda^{\underline{m}} = \Omega^{\mu}{}_{\nu}x^{\nu\dot{\mu}} + \Omega^{\dot{\mu}}{}_{\dot{\nu}}x^{\mu\dot{\nu}}$ (which just equals $\delta x^{\underline{m}}$ in the infinitesimal case); $\Omega$ is traceless. Scale transformations are obtained by choosing $\lambda^{\underline{m}} = \sigma x^{\underline{m}}$.

We could at this point define field strengths in terms of the covariant derivatives (5.1.2), but the invariance group we have defined is too small for two reasons: (1) The vierbein is a reducible representation of the (global) Lorentz group, so more of it should be gauged away; and (2) there are difficulties in realizing (global) Lorentz transformations on general representations, as we now discuss.

Since under global Lorentz transformations $\psi(x)$ transforms as a scalar field, its gradient $\partial_{\underline{m}}\psi$ will transform as a covariant vector. In general, we define a covariant vector to be any object that transforms like $\partial_{\underline{m}}\psi$. We can define a contravariant vector to belong to the "adjoint" representation of our gauge group. Indeed, if we define $V \equiv V^{\underline{m}}i\partial_{\underline{m}}$ and require that $[V, \psi] = V^{\underline{m}}i\partial_{\underline{m}}\psi$ transform as a scalar, i.e.,

$$V' \equiv V'^{\underline{m}}i\partial_{\underline{m}} = e^{i\lambda}Ve^{-i\lambda} \quad , \tag{5.1.8}$$

then $V^{\underline{m}}$ transforms contravariantly under global Lorentz transformations. However, this procedure does not allow us to define objects which transform as spinors under global Lorentz transformations, and in fact it is impossible to define a field, transforming *linearly* under the $\lambda$ group, which also transforms as a spinor when the $\lambda$'s are restricted to represent global Lorentz transformations. It is possible to get around this difficulty by realizing the $\lambda$ transformations nonlinearly, but this is not a convenient solution.

Comparing (5.1.3) to (5.1.8), we see $e_{\underline{a}}{}^{\underline{m}}$ transforms as four independent contravariant vectors under the global Lorentz group: The $\lambda$ transformations do not act on the $\underline{a}$ indices.

To solve these problems we enlarge the gauge group by adjoining to the $\lambda$ transformations a group of *local* Lorentz transformations, and define spinors with respect to this group. This is a procedure familiar in treatments of nonlinear $\sigma$ models. Nonlinear realizations of a group are replaced by linear representations of an enlarged (gauge) group. The nonlinearities reappear only when a definite gauge choice is made. Similarly here, by enlarging the gauge group, we obtain linear spinor representations. The nonlinear spinor representations of the general coordinate group reappear only if we fix a gauge for the local Lorentz transformations. It will thus turn out that our final gauge group for gravity can be interpreted physically as the direct product of the translation (general coordinate) group with the spin (internal) angular momentum group.

We define the action of the local Lorentz group on the vierbein to be

$$\delta e_{\underline{a}}{}^{\underline{m}} = -\,\lambda_{\alpha}{}^{\beta}\, e_{\beta\dot{\alpha}}{}^{\underline{m}} + h.\,c. \quad , \quad \lambda_{\alpha}{}^{\alpha} = 0 \; . \tag{5.1.9}$$

These transformations act only on the free indices in the operator $e_{\underline{a}}$ (but not the hidden indices contracted with $\partial$, since we want the operator to transform *covariantly).* From now on we will indicate indices on which the local Lorentz transformation acts *(flat* or *tangent space* indices) by using letters from the beginning of the Greek and Roman alphabets $(\alpha, \beta, \ldots a, b, \ldots)$, and indices on which local translations (general coordinate transformations) act *(curved* or *world* indices) by letters from the middle $(\mu, \nu, \ldots m, n, \ldots)$. Transformations represented by a matrix multiplying the free index of $e_{\underline{a}}$ are called tangent space transformations.

The linearized form of the local Lorentz transformations is

$$\delta h_{\underline{a}}{}^{\underline{m}} = -\,\delta_{\dot{\alpha}}{}^{\dot{\mu}} \lambda_{\alpha}{}^{\mu} + h.\,c. \quad . \tag{5.1.10}$$

It is thus possible to gauge away the antisymmetric-tensor part of the vierbein (although not the scalar part) with a *nonderivative* transformation. To stay in this gauge a local coordinate transformation must be accompanied by a related local Lorentz transformation; the Lorentz parameter is determined in terms of the translation parameter. At the linearized level we find, using the combined transformations (5.1.1) and (5.1.10),

$$\delta_{\dot{\mu}}{}^{\dot{\alpha}} h_{(\alpha\dot{\alpha}}{}^{\mu)\dot{\mu}} = 0 \quad \rightarrow \quad \lambda_\alpha{}^\beta = -\,\frac{1}{2}\,\partial_{(\alpha\dot{\alpha}}\lambda^{\beta)\dot{\alpha}} \quad . \tag{5.1.11}$$

In this gauge the $\Omega$ (orbital Lorentz coordinate) transformation defined above induces the same global Lorentz transformation acting on the flat indices. We can thus define a Lorentz *spinor* by choosing its spinor index to be a flat index; flat, or tangent space, indices transform under local Lorentz transformations but not local translations *except* when a gauge is chosen, e.g., as in (5.1.11). Furthermore, we can define *all* covariant objects *except the vierbein* to have only flat indices. The curved-index vectors defined above can be related to flat-index ones by multiplying with the vierbein or its inverse.

## b. Covariant derivatives

We now define our new local group of Poincaré transformations, derivatives covariant under it, and its representation on all fields. The parameter of our enlarged local group is defined by

$$\lambda = \lambda^{\underline{m}} i\partial_{\underline{m}} + (\lambda_\alpha{}^\beta iM_\beta{}^\alpha + \bar{\lambda}_{\dot{\alpha}}{}^{\dot{\beta}} i\bar{M}_{\dot{\beta}}{}^{\dot{\alpha}}) \quad . \tag{5.1.12}$$

The generator $M_\beta{}^\alpha$ (and the parameter $\lambda_\alpha{}^\beta$) is traceless and acts only on free flat indices. Its action on such indices is defined by

$$[\lambda_\beta{}^\gamma M_\gamma{}^\beta , \psi_\alpha] = \lambda_\alpha{}^\beta \psi_\beta \ , \qquad [\lambda_\beta{}^\gamma M_\gamma{}^\beta , \psi_{\dot{\alpha}}] = 0 \quad ,$$

$$[\bar{\lambda}_{\dot{\beta}}{}^{\dot{\gamma}} \bar{M}_{\dot{\gamma}}{}^{\dot{\beta}} , \psi_\alpha] = 0 \ , \qquad [\bar{\lambda}_{\dot{\beta}}{}^{\dot{\gamma}} \bar{M}_{\dot{\gamma}}{}^{\dot{\beta}} , \psi_{\dot{\alpha}}] = \bar{\lambda}_{\dot{\alpha}}{}^{\dot{\beta}} \psi_{\dot{\beta}} \quad . \tag{5.1.13}$$

Any covariant field with only flat indices transforms under this gauge group as:

$$\psi'_{\dots} = e^{i\lambda} \psi_{\dots} e^{-i\lambda} \quad . \tag{5.1.14}$$

The covariant derivative is defined by introducing a gauge field for each group generator, so we must now add to $e_{\underline{a}}$ of (5.1.2) a new gauge field for the Lorentz generators:

$$\mathbf{D}_{\underline{a}} = e_{\underline{a}} + (\phi_{\underline{a},\beta}{}^\gamma M_\gamma{}^\beta + \bar{\phi}_{\underline{a},\dot{\beta}}{}^{\dot{\gamma}} \bar{M}_{\dot{\gamma}}{}^{\dot{\beta}}) \quad . \tag{5.1.15}$$

Its transformation law takes the covariant form

$$\mathbf{D}'_{\underline{a}} = e^{i\lambda} \mathbf{D}_{\underline{a}} e^{-i\lambda} \quad . \tag{5.1.16}$$

In defining this covariant derivative we have introduced a gauge field $\phi_{\underline{a}}$ which transforms with a derivative of the Lorentz gauge parameter ( $\delta\phi_{\underline{a}\alpha}{}^{\beta} = \partial_{\underline{a}}\lambda_{\alpha}{}^{\beta} + \cdots$). However, due to the vierbein's Lorentz transformation law (5.1.10), we can define this Lorentz gauge field to be a derivative of our fundamental field (the vierbein), just as for the nonlinear $\sigma$-model (see sec. 3.10), rather than having it as an independent field. There are two ways to find this expression for $\phi_{\underline{a}}$: (1) Compare the full transformation laws of the vierbein and the Lorentz gauge field, and construct directly from the vierbein a Lorentz connection that has the correct transformation properties; or (2) constrain some of the field strengths in such a way that the Lorentz gauge field is determined in terms of the vierbein. Because the field strengths are covariant this will automatically lead to correctly transforming gauge fields $\phi_{\underline{a}}$.

The field strengths $t_{\underline{a}\underline{b}}{}^{\underline{c}}$ and $r_{\underline{a}\underline{b}}(M)$ are defined by:

$$[\mathbf{D}_{\underline{a}}\, , \mathbf{D}_{\underline{b}}] = t_{\underline{a},\underline{b}}{}^{\underline{c}}\mathbf{D}_{\underline{c}} + (r_{\underline{a},\underline{b},\gamma}{}^{\delta} M_{\delta}{}^{\gamma} + \overline{r}_{\underline{a},\underline{b},\dot{\gamma}}{}^{\dot{\delta}} \overline{M}_{\dot{\delta}}{}^{\dot{\gamma}}) \quad . \tag{5.1.17}$$

We have expanded the right-hand side over $\mathbf{D}$ and $M$ instead of $\partial$ and $M$ because then the torsion $t$ and curvature $r$ are covariant. By examining the resultant expressions for the field strengths in terms of the gauge fields, we find

$$t_{\underline{a},\underline{b}}{}^{\underline{c}} = c_{\underline{a},\underline{b}}{}^{\underline{c}} + [(\phi_{\underline{a},\beta}{}^{\gamma}\delta_{\dot{\beta}}{}^{\dot{\gamma}} + \overline{\phi}_{\underline{a},\dot{\beta}}{}^{\dot{\gamma}}\delta_{\beta}{}^{\gamma}) - \underline{a} \longleftrightarrow \underline{b}] \quad ,$$

$$r_{\underline{a},\underline{b},\gamma}{}^{\delta} = (e_{\underline{a}}\phi_{\underline{b},\gamma}{}^{\delta} - \underline{a} \longleftrightarrow \underline{b}) - c_{\underline{a},\underline{b}}{}^{\underline{e}}\phi_{\underline{e},\gamma}{}^{\delta} + \phi_{\underline{a},(\gamma|}{}^{\epsilon}\phi_{\underline{b},\epsilon}{}^{|\delta)} \quad , \tag{5.1.18a}$$

where the *anholonomy coefficient* $c$ is defined by

$$[e_{\underline{a}}\, , e_{\underline{b}}] = c_{\underline{a},\underline{b}}{}^{\underline{c}} e_{\underline{c}} \quad . \tag{5.1.18b}$$

We see that constraining the torsion to vanish gives a suitable Lorentz gauge field:

$$\phi_{\underline{a},\beta}{}^{\gamma} = -\frac{1}{4}\,( c_{\underline{a},(\beta\dot{\beta}}{}^{\gamma)\dot{\beta}} + c^{(\gamma}{}_{\dot{\alpha},\beta)\dot{\beta},\alpha}{}^{\dot{\beta}}) \quad . \tag{5.1.19}$$

(If instead of the torsion we constrained the curvature to vanish, the connection $\phi_{\underline{a}}$ would be pure Lorentz gauge, and unrelated to the vierbein. However, the antisymmetric part of the vierbein would remain as a compensator for a second *hidden* local Lorentz group of the theory, under which $\phi_{\underline{a}}$ would transform homogeneously and not as a connection. Hence $\mathbf{D}$ defined by (5.1.15) would be *noncovariant* under the new transformations, and instead

$$\hat{\mathbf{D}}_{\underline{a}} = e_{\underline{a}} - [\frac{1}{2}(c_{\underline{a},\beta\dot{\beta}}{}^{\gamma\dot{\beta}} + c^{\gamma}{}_{\dot{\alpha},\beta\dot{\beta},\alpha}{}^{\dot{\beta}})M_{\gamma}{}^{\beta} + h.c.]$$

$$= e_{\underline{a}} - [\frac{1}{2}(t_{\underline{a},\beta\dot{\beta}}{}^{\gamma\dot{\beta}} + t^{\gamma}{}_{\dot{\alpha},\beta\dot{\beta},\alpha}{}^{\dot{\beta}})M_{\gamma}{}^{\beta} + h.c.] \quad (5.1.20)$$

would be covariant. Since Einstein theory is now described in terms of a curvature constructed out of $\hat{\mathbf{D}}$, the original $\phi_{\underline{a}}$ and its associated Lorentz invariance would be irrelevant to the theory. Constraining the curvature to vanish is gauge equivalent to not introducing any connection at all. Such a formulation of gravity is often referred to as a "teleparallelism" theory. Of course, if we were to constrain both $t$ and $r$ to vanish, $\mathbf{D}$ would be gauge equivalent to $\partial$, and we would have no gravity.)

In the absence of any constraint, we could always express the covariant derivative as the constrained covariant derivative ($t=0$) *plus* Lorentz covariant terms that contain only the torsion. The torsion could thus be considered as an independent tensor with no relation to gravity. Our torsion constraint is thus a "conventional" constraint, just like the conventional constraint (4.2.60) of super-Yang-Mills theories.

All remaining tensors (i.e., covariant objects that are not operators) can be expressed in terms of the curvature and its covariant derivatives. The curvature itself is algebraically reducible (under the Lorentz group) into three tensors:

$$r_{\underline{a},\underline{b},}{}^{\gamma\delta} = C_{\dot{\alpha}\dot{\beta}}\,(w_{\alpha\beta}{}^{\gamma\delta} - \frac{1}{2}\delta_{(\alpha}{}^{\gamma}\,\delta_{\beta)}{}^{\delta}\,r) + C_{\alpha\beta}\,r^{\gamma\delta}{}_{\dot{\alpha}\dot{\beta}} \quad , \quad (5.1.21)$$

where the tensors are totally symmetric in undotted indices and in dotted indices (which is equivalent to being algebraically Lorentz-irreducible). The tensors $r$ and $r_{\alpha\beta\dot{\alpha}\dot{\beta}}$ are the trace and traceless parts of the *Ricci tensor,* and $w_{\alpha\beta\gamma\delta}$ is the *Weyl tensor.* (Note that our normalization of the Ricci scalar differs from the standard: We use the more convenient normalization, in general spacetime dimension D, $r_{\underline{ab}}{}^{\underline{cd}} = -\,\delta_{[\underline{a}}{}^{\underline{c}}\delta_{\underline{b}]}{}^{\underline{d}}r + \cdots$, rather than $r_{\underline{ab}}{}^{\underline{ab}} = r$. The sign is chosen so that $r$ is nonnegative on shell in unbroken supersymmetric theories.) These tensors are, of course, related differentially through the Bianchi identities (the Jacobi identities of the covariant derivatives). Explicitly from

$$[[\mathbf{D}_{\underline{a}}\,,\mathbf{D}_{\underline{b}}]\,,\mathbf{D}_{\underline{c}}] + [[\mathbf{D}_{\underline{b}}\,,\mathbf{D}_{\underline{c}}]\,,\mathbf{D}_{\underline{a}}] + [[\mathbf{D}_{\underline{c}}\,,\mathbf{D}_{\underline{a}}]\,,\mathbf{D}_{\underline{b}}] = 0, \quad (5.1.22)$$

we find

$$\mathbf{D}_{[\underline{a}}t_{\underline{b},\underline{c}]}{}^{\underline{d}} - t_{[\underline{a},\underline{b}|}{}^{\underline{e}}t_{\underline{e},|\underline{c}]}{}^{\underline{d}} - r_{[\underline{a},\underline{b},\underline{c}]}{}^{\underline{d}} = 0, \quad (5.1.23a)$$

$$\mathbf{D}_{[\underline{a}}r_{\underline{b},\underline{c}]\gamma}{}^{\delta} - t_{[\underline{a},\underline{b}|}{}^{\underline{e}}r_{\underline{e},|\underline{c}]\gamma}{}^{\delta} = 0. \tag{5.1.23b}$$

These last two equations (which follow from the linear independence of $\mathbf{D}_{\underline{a}}$ and $M_{\gamma}{}^{\delta}$) are the first and second Bianchi identities, respectively.

## c. Actions

In contrast to Yang-Mills theory, in gravity one cannot trace over the group without integrating over the spacetime coordinates, since the translation group acts on the coordinates themselves. Thus, only integrated quantities can form invariants. Furthermore, gravity differs even from the group manifold approach to Yang-Mills, where the group generators are treated as translations in the group space, in that the local translation group is not unitary: Although $\lambda^{\underline{m}}$ is hermitian, an infinitesimal translation is not:

$$(\lambda^{\underline{m}} i\partial_{\underline{m}})^{\dagger} = i\partial_{\underline{m}}\lambda^{\underline{m}} = \lambda^{\underline{m}} i\partial_{\underline{m}} + (i\partial_{\underline{m}}\lambda^{\underline{m}}) \quad . \tag{5.1.24}$$

From the reordering of the two factors, we get an additional term proportional to the divergence of $\lambda$. This term arises because some coordinate transformations are not *volume-preserving:* e.g., the transformation given by $\lambda^{\underline{m}} \sim x^{\underline{m}}$ is a scale transformation. Consequently the volume element $d^4x \sim dx^{+\dot{+}} \wedge dx^{+\dot{-}} \wedge dx^{-\dot{+}} \wedge dx^{-\dot{-}}$ is not covariant. To covariantize, we simply replace $dx^{\underline{m}}$ with an object that is a scalar under coordinate transformations (a world scalar): $\omega^{\underline{a}} = dx^{\underline{m}} e_{\underline{m}}{}^{\underline{a}}$. The resulting volume element is $\omega^4 = d^4x\mathrm{e}^{-1}$, where e is the determinant of $e_{\underline{a}}{}^{\underline{m}}$.

The invariance of a scalar integrated with the covariant volume element can also be seen from the transformation law of e, which we write in the compact and convenient form

$$\mathrm{e}'^{-1} = \mathrm{e}^{-1} e^{i\overleftarrow{\lambda}} \quad . \tag{5.1.25}$$

Here $\overleftarrow{\lambda} = \lambda^{\underline{m}} i\overleftarrow{\partial}_{\underline{m}}$ means that the derivative acts on all objects to its left. (For the present discussion we may ignore Lorentz transformations.) Before deriving this transformation law, we show how it allows $\mathrm{e}^{-1}$ to form invariant integrals: For any scalar $L$,

$$\int d^4x \, (\mathrm{e}^{-1}L)' = \int d^4x \, (\mathrm{e}^{-1}e^{i\overleftarrow{\lambda}})(e^{i\lambda}Le^{-i\lambda})$$

$$= \int d^4x \, \mathrm{e}^{-1}e^{i\overleftarrow{\lambda}}(e^{-i\overleftarrow{\lambda}}Le^{i\overleftarrow{\lambda}})$$

$$= \int d^4x \, (\mathrm{e}^{-1}L) e^{i\overleftarrow{\lambda}} \quad , \tag{5.1.26a}$$

where we have used the identity (for any $X$)

$$[\lambda \, , X] = [X \, , \overleftarrow{\lambda}] \qquad \to \qquad e^{i\lambda} X e^{-i\lambda} = e^{-i\overleftarrow{\lambda}} X e^{i\overleftarrow{\lambda}} \quad . \tag{5.1.26b}$$

Finally, using $X e^{i\overleftarrow{\lambda}} = X + \text{total derivative}$, we find

$$\int d^4x \, (\mathrm{e}^{-1}L)' = \int d^4x \; \mathrm{e}^{-1}L \quad . \tag{5.1.27}$$

To derive the transformation law (5.1.25), we need the identity

$$\delta \, det \, X = det \, X \, tr(X^{-1}\delta X) \ , \tag{5.1.28}$$

which follows from $det\, X = e^{tr \, ln \, X}$. Thus we find, from (5.1.4b),

$$\delta \mathrm{e}^{-1} = - \, \mathrm{e}^{-1}(e_{\underline{m}}{}^{\underline{a}}(e_{\underline{a}}{}^{\underline{n}}\partial_{\underline{n}}\lambda^{\underline{m}} - \lambda^{\underline{n}}\partial_{\underline{n}}e_{\underline{a}}{}^{\underline{m}}))$$

$$= - \, \mathrm{e}^{-1}(\partial_{\underline{m}}\lambda^{\underline{m}} - e_{\underline{m}}{}^{\underline{a}}\lambda^{\underline{n}}\partial_{\underline{n}}e_{\underline{a}}{}^{\underline{m}})$$

$$= - \, \mathrm{e}^{-1}\partial_{\underline{m}}\lambda^{\underline{m}} - \lambda^{\underline{m}}\partial_{\underline{m}}\mathrm{e}^{-1}$$

$$= - \, \partial_{\underline{m}}(\lambda^{\underline{m}}\mathrm{e}^{-1}) = \mathrm{e}^{-1}i\overleftarrow{\lambda} \quad . \tag{5.1.29}$$

To find the finite transformation, we iterate the infinitesimal transformation (5.1.29) and use $e^x = \lim_{n\to\infty}(1+\frac{x}{n})^n$; we thus arrive at the desired result (5.1.25). An equivalent statement of our result is that $1 \cdot e^{i\overleftarrow{\lambda}}$ is the Jacobian determinant of the coordinate transformation $e^{i\lambda}$. ($1 \cdot e^{i\overleftarrow{\lambda}}$ means that derivatives act to the left until annihilating the 1.)

We can now construct invariant actions for gravity and its couplings to matter. The only possible action that gives $h_{\underline{ab}}$ a second-order kinetic operator is

$$S = -\frac{3}{\kappa^2}\int d^4x \, \mathrm{e}^{-1}r \quad , \tag{5.1.30}$$

where $r$ is the curvature scalar defined by (5.1.18) and (5.1.21). The resultant field equations are $r = r_{\alpha\beta\dot{\alpha}\dot{\beta}} = 0$. Coupling to matter is achieved by covariantization of the derivatives, as in Yang-Mills theory, but now the volume element is also covariantized (with $\mathrm{e}^{-1}$). As in Yang-Mills, we are also free to add nonminimal couplings depending on the curvature. (For the teleparallelism theory we can still use this action if $r$ is defined by the commutator of $\widehat{\mathbf{D}}_{\underline{a}}$. This action leads, by use of the first Bianchi identity, to an expression that is purely quadratic in $t_{\underline{a},\underline{b}}{}^{\underline{c}}$.)

## d. Conformal compensator

In flat space (i.e., without gravity) certain theories (e.g., massless $\phi^4$, or massless QCD) are invariant under (global) conformal transformations at the classical level. On the other hand, when gravity is present *all* theories are conformally invariant since conformal transformations are a special case of general coordinate transformations. However, this type of conformal invariance has no physical significance, and is present simply because the vierbein automatically compensates the conformal transformations of other fields. This is analogous to global orbital Lorentz transformations: *Any* nonLorentz covariant flat-space theory can be made covariant under these orbital transformations in curved space, because the antisymmetric part of the vierbein acts as a compensator (e.g., $\int d^4x(\partial_{\underline{0}}\phi)^2 \to \int d^4x\mathrm{e}^{-1}(e_{\underline{0}}{}^{\underline{m}}\partial_{\underline{m}}\phi)^2$ ). As we saw above, it is necessary to introduce additional, local, tangent-space Lorentz transformations to give a meaningful definition of Lorentz invariance in curved space. Theories that are invariant under these tangent space Lorentz transformations will automatically be invariant under the usual Lorentz transformations in flat space, or when a gauge for local Lorentz and general coordinate transformations is chosen.

Similarly, in the presence of gravity it is possible to give a meaning to global conformal invariance by observing that in curved space it corresponds to an additional invariance under *local* scale transformations

$$e'_{\underline{a}} = e^{\zeta} e_{\underline{a}} \quad , \quad \psi'_{\dots} = e^{d\zeta}\psi_{\dots} \quad . \tag{5.1.31}$$

Here $\zeta(x)$ is a local parameter and $d$ is the canonical dimension of the field $\psi_{\dots}$ (usually 1 for bosons, $\frac{3}{2}$ for fermions) when written with *flat* tangent-space indices. (Note that $e_{\underline{a}}$, which has no free curved indices and describes a boson, has canonical dimension 1 since it contains a derivative.) Any theory in curved space that has local scale invariance gives a flat space theory which is conformally invariant. The transformation in (5.1.31) is another example of a tangent space transformation.

Thus, both local Lorentz and local scale invariance reflect flat space invariance properties of matter systems. However, there is an important distinction between Lorentz and conformal transformations: Conformal invariance is not a general property of physical systems, and consequently we do not introduce local scale generators and corresponding gauge fields into our covariant derivatives.

As described above, the antisymmetric part of the vierbein can be gauged away by local Lorentz transformations. In the resulting gauge, general coordinate transformations must be accompanied by related local Lorentz transformations that restore the gauge. The local Lorentz parameter becomes a nonlinear function of the general coordinate parameter, making construction of Lorentz covariant actions more difficult. Similarly, local scale transformations can be used to gauge away the trace of the vierbein. In fact, *in locally scale invariant theories,* the determinant of the vierbein can be gauged to 1 by local scale transformations. In the resulting gauge, general coordinate transformations must be accompanied by local scale transformations with parameter $\zeta$ determined by

$$1 = (\mathrm{e}^{-1})' = \mathrm{e}^{-1} e^{i\overleftarrow{\lambda}} e^{-4\zeta} = (1 \cdot e^{i\overleftarrow{\lambda}}) e^{-4\zeta} \quad . \tag{5.1.32}$$

The local scale parameter becomes a nonlinear function of the general coordinate parameter. In particular, dimension $d$ fields $\psi_{\cdots}$ now transform as *densities* under general coordinate transformations, i.e., with an additional factor $(1 \cdot e^{i\overleftarrow{\lambda}})^{d/4}$. The formalism becomes rather cumbersome. (We note that even in theories that are not invariant under local scale transformations, e can still be gauged to 1, at least in small regions of spacetime, by some of the general coordinate transformations: $\delta\mathrm{e}{\sim}\partial_{\underline{m}}\lambda^{\underline{m}}$. However, this results in the constraint $\partial_{\underline{m}}\lambda^{\underline{m}} = 0$ on further coordinate transformations, and differentially constrained gauge parameters are undesirable when a theory is quantized (see sec. 7.3); such gauge choices are possible upon quantization, but e should not be set to 1 before quantization.)

We have already indicated that e acts as a compensator for the local scale transformations of fields $\psi_{\cdots}$ as given in (5.1.31). In fact, by making the field redefinition $\psi_{\cdots} \to \mathrm{e}^{-d/4}\psi_{\cdots}$ we can make all fields except e inert under scale transformations. In terms of the new fields local scale invariance of an action is equivalent to independence of e. However, to maintain manifest coordinate invariance, it is preferable to keep explicit the dependence on e. On the other hand, it is frequently useful to describe to what extent a theory breaks local scale invariance, both because locally scale invariant theories are interesting in their own right, and because a decomposition into locally scale invariant plus scale-breaking parts can be helpful. This can be done by introducing an *additional* compensating field into the theory, but one which unlike e is a *scalar* under general coordinate and local Lorentz transformations. To distinguish this type of compensator from the e type, we will henceforth refer to them as *tensor compensators* and *density compensators,* respectively. Density compensators (e.g., $e_{[\underline{a}}{}^{\underline{m}]}$ for local Lorentz, or e for local scale) generally occur as parts of physical fields and are not tensors under the local symmetry group (e.g., general coordinate transformations) of the action without compensators. Tensor compensators are covariant, and their presence allows the introduction of a local symmetry even in the absence of a corresponding global, flat space symmetry.

For local scale transformations we introduce a scalar compensator transforming as

$$\phi' = e^{\zeta}\phi \quad . \tag{5.1.33}$$

Starting with fields invariant under $\zeta$ transformations,we now make the replacements

$$e_{\underline{a}} \to \phi^{-1} e_{\underline{a}} \ , \qquad \psi_{...} \to \phi^{-d}\psi_{...} \quad . \tag{5.1.34}$$

The new fields still transform according to (5.1.31). The replacement (5.1.34) is just a $\phi$-dependent scale transformation. Hence local scale invariance of a given quantity is *equivalent* to independence from $\phi$. For example, after the redefinition (5.1.34), the usual gravity action (5.1.30) becomes

$$S = -\frac{3}{\kappa^2}\int d^4x \ \mathrm{e}^{-1}\,\phi(\Box + r)\phi \quad . \tag{5.1.35}$$

This action is scale invariant because $\phi$ *compensates* the transformation of e and $\Box + r$. Since it is not $\phi$-independent the original Einstein action was not scale invariant. Alternatively, (5.1.35) can be interpreted as a scale invariant action for the field $\phi$. The scale invariance allows $\phi$ to be gauged to one. In that gauge one recovers the usual Einstein action.

In contrast, the Weyl tensor, which is the only part of the curvature which is homogeneous in $\phi$ after (5.1.34), can form a $\phi$ independent, locally scale invariant (but higher-derivative) action:

$$S_{Weyl} \sim \int d^4x \, e^{-1} (w_{\alpha\beta\gamma\delta})^2 \quad . \tag{5.1.36}$$

* * *

We introduced the linearized vierbein $h_{\underline{a}}{}^{\underline{m}}$ as a gauge field for translations; alternatively, we can use the analysis of chapter 3 to *find* $h_{\underline{a}}{}^{\underline{m}}$. Linearized gravity is the theory of a massless spin 2 field. As discussed in sec. 3.12, it is described by an irreducible *on-shell* field strength $\psi_{\alpha\beta\gamma\delta}$ satisfying (see (3.12.1))

$$\partial^{\alpha\dot{\alpha}} \psi_{\alpha\beta\gamma\delta} = 0 \quad . \tag{5.1.37}$$

Using the results of sec. 3.13, the corresponding irreducible *off-shell* field strength is the linearized Weyl tensor $w_{\alpha\beta\gamma\delta}$ satisfying the bisection condition ($s + \frac{N}{2} = 2$ is an integer)

$$w_{\alpha\beta\gamma\delta} = \mathbf{K} w_{\alpha\beta\gamma\delta} = \Delta_\alpha{}^{\dot{\alpha}} \Delta_\beta{}^{\dot{\beta}} \Delta_\gamma{}^{\dot{\gamma}} \Delta_\delta{}^{\dot{\delta}} \bar{w}_{\dot{\alpha}\dot{\beta}\dot{\gamma}\dot{\delta}} \tag{5.1.38}$$

which is equivalent to

$$\partial^\alpha{}_{\dot{\alpha}} \partial^\beta{}_{\dot{\beta}} w_{\alpha\beta\gamma\delta} = \partial_\gamma{}^{\dot{\gamma}} \partial_\delta{}^{\dot{\delta}} \bar{w}_{\dot{\alpha}\dot{\beta}\dot{\gamma}\dot{\delta}} \quad . \tag{5.1.39}$$

By (3.13.2) applied to $N = 0$, the solution to this equation is

$$w_{\alpha\beta\gamma\delta} = \partial_{(\alpha}{}^{\dot{\alpha}} \partial_\beta{}^{\dot{\beta}} V_{\gamma\delta)\dot{\alpha}\dot{\beta}} \quad , \tag{5.1.40}$$

where $h_{\underline{a}\underline{b}} \equiv h_{\alpha\dot{\alpha}\beta\dot{\beta}} \equiv V_{(\alpha\beta)(\dot{\alpha}\dot{\beta})}$ is a traceless symmetric tensor. The maximal gauge invariance of (5.1.40) is:

$$\delta h_{\underline{a}\underline{b}} = \partial_{(\underline{a}} \lambda_{\underline{b})} - \frac{1}{2} \eta_{\underline{a}\underline{b}} \partial_{\underline{c}} \lambda^{\underline{c}} \quad . \tag{5.1.41}$$

These are linearized coordinate transformations identical to (5.1.1), *except* that scale transformations and Lorentz transformations are not included. They can be added by introducing compensators: the trace and antisymmetric parts of $h_{\underline{a}\underline{b}}$.

## 5.2. Prepotentials

### a. Conformal

As for superfield Yang-Mills and for gravity, one can develop a formulation for superfield supergravity in either of two ways: (1) Study off-shell representations to determine the linearized formulation in terms of unconstrained superfields *(prepotentials),* and then construct covariant derivatives, which provide the generalization to the nonlinear case; or (2) start by postulating covariant derivatives, determine what constraints they must satisfy, and solve them in terms of prepotentials. In this section we will describe the former approach, and in the following section the latter.

### a.1. Linearized theory

From the analysis in sec. 3.3.a.1, we know that the $N = 1$ supergravity multiplet consists of massless spin 2 and spin $\frac{3}{2}$ physical states. The corresponding on-shell component field strengths are $\psi_{\alpha\beta\gamma\delta}$ and $\psi_{\alpha\beta\gamma}$ (on-shell Weyl tensor and Rarita-Schwinger field strength), totally symmetric in their indices, as discussed in sec. 3.12.a. These lie in an irreducible on-shell multiplet described by a chiral superfield $\Psi_{(0)\alpha\beta\gamma}$ that satisfies the constraint

$$D_\delta \Psi_{(0)\alpha\beta\gamma} = \Psi_{(1)\alpha\beta\gamma\delta} \quad , \tag{5.2.1}$$

where $\Psi_{(1)}$ is totally symmetric and is the superfield containing the on-shell Weyl tensor $\psi_{\alpha\beta\gamma\delta}$ $(= w_{\alpha\beta\gamma\delta})$ at the $\theta = 0$ level. The constraint implies

$$D^\alpha \Psi_{(0)\alpha\beta\gamma} = 0 \quad . \tag{5.2.2}$$

By the analysis of sec. 3.13, the corresponding *irreducible off-shell* superfield strength is a chiral superfield $W_{\alpha\beta\gamma}$ satisfying the bisection condition ($s + \frac{N}{2} = \frac{3}{2} + \frac{1}{2}$ is an integer)

$$W_{\alpha\beta\gamma} = -\mathbf{K} W_{\alpha\beta\gamma} = -\Box^{-\frac{1}{2}} \bar{D}^2 \Delta_\alpha{}^{\dot\alpha} \Delta_\beta{}^{\dot\beta} \Delta_\gamma{}^{\dot\gamma} \bar{W}_{\dot\alpha\dot\beta\dot\gamma} \quad , \tag{5.2.3}$$

which can be rewritten as

$$\partial^\beta{}_{\dot\beta} D^\alpha W_{\alpha\beta\gamma} = -\partial_\gamma{}^{\dot\gamma} \bar{D}^{\dot\alpha} \bar{W}_{\dot\alpha\dot\beta\dot\gamma} \quad . \tag{5.2.4}$$

By (3.13.2) the solution to this equation is

$$W_{\alpha\beta\gamma} = -\, i\, \frac{1}{3!}\, \bar{D}^2 D_{(\alpha} \partial_{\beta}{}^{\dot\beta} H_{\gamma)\dot\beta} \quad , \tag{5.2.5}$$

where $H_{\gamma\dot\beta}$ is real. ($H$ might be expressed as a derivative of a more fundamental field; this possibility is eliminated when we examine the $N = 1$ theory at the nonlinear level.)

We remark in passing that

$$S_{conf} = \int d^4x d^2\theta W^2 = \int d^4x [W D^2 W + (DW)^2]| \tag{5.2.6}$$

contains the action for linearized conformal gravity $\int d^4x (w_{\alpha\beta\gamma\delta})^2$. Thus (5.2.6) is the extension of conformal gravity to conformal supergravity at the linearized level.

A careful examination of (5.2.4) reveals that the largest gauge invariance of $W_{\alpha\beta\gamma}$, written in a form containing the fewest derivatives (and thus the component transformations contain the fewest possible spacetime derivatives), is

$$\delta H_{\underline{a}} = D_\alpha \bar{L}_{\dot\alpha} - \bar{D}_{\dot\alpha} L_\alpha \quad . \tag{5.2.7}$$

To get insight into the physical content of $H$ and its transformation, we consider their components using $D$ projection.

The components of $H_{\underline{a}}$ are

$$h_{\underline{a}} = H_{\underline{a}}| \quad , \quad h_{\alpha\beta\dot\beta} = D_\beta H_{\alpha\dot\beta}| \quad , \quad h^{(2)}{}_{\underline{a}} = D^2 H_{\underline{a}}| \quad ,$$

$$h_{\underline{ab}} = -\, \frac{1}{2}\, [D_\alpha, \bar{D}_{\dot\alpha}] H_{\beta\dot\beta}| \quad , \quad \bar\psi_{\underline{a},\dot\beta} = -\, i\, D^2 \bar{D}_{\dot\alpha} H_{\alpha\dot\beta}| \quad ,$$

$$A_{\underline{a}} = -\, \frac{2}{3}\, D^\beta \bar{D}^2 D_\beta H_{\underline{a}}| - \frac{1}{6}\, \epsilon_{\underline{abcd}}\, \partial^{\underline{b}} [D^\gamma, \bar{D}^{\dot\gamma}] H^{\underline{d}}| \quad . \tag{5.2.8}$$

where $\epsilon_{\underline{abcd}} = i\,( C_{\alpha\delta} C_{\beta\gamma} C_{\dot\alpha\dot\beta} C_{\dot\gamma\dot\delta} - C_{\alpha\beta} C_{\gamma\delta} C_{\dot\alpha\dot\delta} C_{\dot\beta\dot\gamma} )$ (3.1.22). Although it is convenient to define the component fields $h_{\underline{ab}}$ , $\bar\psi_{\underline{a},\dot\beta}$ , $A_{\underline{a}}$ as above, these are only the *linearized, conformal* definitions of these component fields. In the final Poincaré theory additional $H_{\underline{a}}$ and compensator superfield dependent terms, as well as nonlinearities, are present.

The components of $\bar{D}_{\dot\alpha} L_\alpha$ (the rest of $L$ never enters) are:

$$\xi_{\underline{a}} = \bar{D}_{\dot\alpha} L_\alpha| \quad , \quad L^1{}_{\alpha\beta\dot\beta} = D_\alpha \bar{D}_{\dot\beta} L_\beta| \quad , \quad \epsilon_\alpha = \bar{D}^2 L_\alpha| \quad ,$$

$$L^2{}_{\underline{a}} = D^2 \bar{D}_{\dot{\alpha}} L_\alpha| \quad , \quad \sigma = D^\alpha \bar{D}^2 L_\alpha| \quad ,$$

$$\omega_{\alpha\beta} = \frac{1}{2} D_{(\alpha} \bar{D}^2 L_{\beta)}| \quad , \quad \eta_\alpha = D^2 \bar{D}^2 L_\alpha| \quad , \tag{5.2.9}$$

and similarly for the complex conjugate (but note $\bar{\xi}_{\underline{a}} = -D_\alpha \bar{L}_{\dot{\alpha}}$). The transformations of the independent components of $H$ are:

$$\delta h_{\underline{a}} = 2\, Re\, \xi_{\alpha\dot{\alpha}} \quad ,$$

$$\delta h_{\alpha\beta\dot{\beta}} = \frac{1}{2} C_{\alpha\beta} \bar{\epsilon}_{\dot{\beta}} - L^1{}_{\beta\alpha\dot{\beta}} \quad ,$$

$$\delta h^{(2)}{}_{\underline{a}} = -\, L^2{}_{\underline{a}} \;\; ,$$

$$\delta h_{\underline{ab}} = -\, (C_{\dot{\alpha}\dot{\beta}} \omega_{\alpha\beta} + C_{\alpha\beta} \omega_{\dot{\alpha}\dot{\beta}}) + C_{\alpha\beta} C_{\dot{\alpha}\dot{\beta}}\, Re\, \sigma - \partial_{\underline{a}}\, Im\, \xi_{\underline{b}} \quad ,$$

$$\delta \bar{\psi}_{\underline{a},\dot{\beta}} = \partial_{\underline{a}} \bar{\epsilon}_{\dot{\beta}} - i\, C_{\dot{\beta}\dot{\alpha}} \eta_\alpha \quad ,$$

$$\delta A_{\underline{a}} = \frac{2}{3} \partial_{\underline{a}} Im\, \sigma \;\; . \tag{5.2.10}$$

We can therefore go to a Wess-Zumino gauge by using $Re\,\xi, L^1, L^2$ to algebraically gauge away all of $H_{\underline{a}}$ except $h_{\underline{ab}}$ , $\psi_{\underline{a}\beta}$ , $A_{\underline{a}}$. These can be identified as the linearized vierbein, the spin $\frac{3}{2}$ Rarita-Schwinger field, and an axial vector auxiliary field, respectively.

We study the remaining transformations: Examining $\delta h_{\underline{ab}}$ we note that $\omega$ can be used to eliminate the antisymmetric part of the vierbein, which identifies it as an infinitesimal local Lorentz transformation. $Re\,\sigma$ removes the trace, and is therefore a local scale transformation. Finally, $Im\,\xi$ generates a coordinate transformation. Examining $\bar{\psi}_{\underline{a},\dot{\beta}}$ we identify the $\bar{\epsilon}_{\dot{\beta}}$ term as a Rarita-Schwinger gauge transformation (a linearized local supersymmetry transformation). The $\eta_\alpha$ term is a local $S$-supersymmetry transformation: it gauges away the trace $\psi_{\alpha\dot{\beta},}{}^{\dot{\beta}}$. From $\delta A_{\underline{a}}$ we identify $Im\,\sigma$ as an axial gauge transformation. (Note that the local $S$-transformation of the spin $\frac{3}{2}$ field contains no spacetime derivatives. Avoiding derivatives is important for quantization (see sec. 7.3)).

Thus, in the Wess-Zumino gauge, the $L$-gauge group that remains consists of local superconformal transformations: the super-Poincaré subgroup (coordinate, Lorentz, and local supersymmetry), and axial, $S$-supersymmetry, and scale transformations (see sec. 3.2.e). Local conformal invariance plays a more important role in supergravity than in gravity: Whereas the nonconformal part of the vierbein (its trace) can be projected out algebraically, the analogous statement does not hold for $H$ (a vector is not algebraically reducible in a Lorentz covariant way). The same distinction between the reducibility of the vierbein and $H$ applies with regard to local Lorentz invariance (which must be maintained in supergravity simply because supersymmetric theories contain spinors).

## a.2. Nonlinear theory

To generalize to the nonlinear case we examine (as in super-Yang-Mills) the appropriate transformations of the simplest multiplet, the chiral scalar superfield. Since gravity gauges translations, supergravity will gauge supertranslations. We therefore look for the most general transformation of the form

$$\eta' = e^{i\Lambda}\eta e^{-i\Lambda} \quad , \quad \Lambda = \Lambda^M iD_M \quad ; \tag{5.2.11}$$

(We choose to parametrize with $D_M$ rather than $\partial_M$ in order to keep manifest global supersymmetry. This simply amounts to a redefinition of the parameters.) We maintain the chirality of $\eta$ ($\bar{D}_{\dot\mu}\eta = 0$), by requiring $\Lambda$ to satisfy

$$[\bar{D}_{\dot\mu}, \Lambda]\eta = 0 \quad , \tag{5.2.12}$$

which implies

$$\bar{D}_{\dot\mu}\Lambda^\nu = 0 \quad , \quad \bar{D}_{\dot\mu}\Lambda^{\underline{n}} = i\,\Lambda^\nu \delta_{\dot\mu}{}^{\dot\nu} , \tag{5.2.13}$$

and has the solution

$$\Lambda^{\underline{m}} = -\,i\bar{D}^{\dot\mu}L^\mu \quad , \quad \Lambda^\mu = \bar{D}^2 L^\mu \quad , \quad \Lambda^{\dot\mu} \ \textit{arbitrary} \quad ; \tag{5.2.14a}$$

i.e.,

$$\Lambda^{\underline{m}}\partial_{\underline{m}} + \Lambda^\mu D_\mu = \frac{1}{2}\{\bar{D}^{\dot\mu}, [\bar{D}_{\dot\mu}, L^\mu D_\mu]\} \quad . \tag{5.2.14b}$$

Note that the parameter superfield $\Lambda^M$ must be complex. In particular this means that $\Lambda^{\underline{m}} \neq (\Lambda^{\underline{m}})^\dagger$ , $\Lambda^\mu \neq (\Lambda^{\dot\mu})^\dagger$ , and $\Lambda^{\dot\mu} \neq (\Lambda^\mu)^\dagger$ . Care must be taken to distinguish $\Lambda^\mu$ and $\Lambda^{\dot{\mu}}$ from the hermitian conjugated quantities $\overline{\Lambda}^{\mu}$ $(= (\Lambda^{\dot{\mu}})^{\dagger})$ and $\overline{\Lambda}^{\dot{\mu}}$ $(= (\Lambda^{\mu})^{\dagger})$. We define the transformation of an antichiral scalar in a similar fashion:

$$\overline{\eta}' = e^{i\overline{\Lambda}}\overline{\eta}e^{-i\overline{\Lambda}} \quad , \quad \overline{\Lambda} = \overline{\Lambda}^{M} iD_{M} \quad ,$$

$$\overline{\Lambda}^{\underline{m}} = -\, iD^{\mu}\overline{L}^{\dot{\mu}} \quad , \quad \overline{\Lambda}^{\dot{\mu}} = D^2\overline{L}^{\dot{\mu}} \quad , \quad \overline{\Lambda}^{\mu} \ arbitrary \quad . \tag{5.2.15}$$

The quantity $\overline{\Lambda}^{M}$ is the complex conjugate of $\Lambda^{M}$.

At this point it is clear, by analogy with super-Yang-Mills, that $H^{\underline{m}}$ is the correct field to covariantize the $\Lambda^{\underline{m}}$ part of the transformation of the scalar multiplet kinetic term, since its linearized transformation is (from (5.2.7)) $\delta H^{\underline{m}} = i\overline{\Lambda}^{\underline{m}} - i\Lambda^{\underline{m}}$. We therefore complete $H^{\underline{m}}$ to a supervector $H^{M} = (H^{\mu}, H^{\dot{\mu}}, H^{\underline{m}})$ and introduce an exponential $e^{H}$, $H = H^{M} iD_{M}$. As for Yang-Mills, the nonlinear transformation law is

$$e^{H'} = e^{i\overline{\Lambda}}\, e^{H} e^{-i\Lambda} \quad . \tag{5.2.16}$$

We note that $H^{\dot{\mu}}$ can be trivially gauged away because the parameter $\Lambda^{\dot{\mu}}$ is arbitrary; consequently $H^{\mu}$ is also gauged away by $\overline{\Lambda}^{\mu}$. To preserve this gauge choice, the $L$-gauge transformations (5.2.14,15) must be accompanied now by compensating $\Lambda^{\dot{\mu}}$ and $\overline{\Lambda}^{\mu}$ transformations. For infinitesimal $\Lambda$ we have

$$\delta(e^{H^{\underline{m}} i\partial_{\underline{m}}}) = -\,(\overline{\Lambda}^{\underline{m}}\partial_{\underline{m}} + \overline{\Lambda}^{\dot{\mu}}\overline{D}_{\dot{\mu}} + \overline{\Lambda}^{\mu}D_{\mu})(e^{H^{\underline{m}} i\partial_{\underline{m}}})$$

$$+\,(e^{H^{\underline{m}} i\partial_{\underline{m}}})(\Lambda^{\underline{m}}\partial_{\underline{m}} + \Lambda^{\mu}D_{\mu} + \Lambda^{\dot{\mu}}\overline{D}_{\dot{\mu}}) \quad . \tag{5.2.17}$$

(This equation is to be interpreted as an operator equation acting on an arbitrary superfunction to the right.) This implies that we must cancel $D_{\mu}$ and $\overline{D}_{\dot{\mu}}$ terms on the right-hand side and hence

$$\overline{\Lambda}^{\mu} = e^{H}\Lambda^{\mu}e^{-H} = e^{H}\overline{D}^2 L^{\mu}e^{-H} \quad , \quad \Lambda^{\dot{\mu}} = e^{-H}\overline{\Lambda}^{\dot{\mu}}e^{H} = e^{-H}D^2\overline{L}^{\dot{\mu}}e^{H} \quad . \tag{5.2.18}$$

However, we will not restrict ourselves to this gauge in the subsequent discussion.

### a.3. Covariant derivatives

Our next task is to construct covariant derivatives $\nabla_A = (\nabla_\alpha, \overline{\nabla}_{\dot\alpha}, \nabla_{\underline{a}})$. By analogy with Yang-Mills theory we would require $(\nabla_A \Phi)' = e^{i\Lambda}(\nabla_A \Phi)$, i.e., $\nabla_A{}' = e^{i\Lambda}\nabla_A e^{-i\Lambda}$ or $\delta\nabla_A = i[\Lambda, \nabla_A]$. However, since we expect local Lorentz transformations to be present, we can generalize to

$$(\nabla_A \Phi)' = L_A{}^B e^{i\Lambda} \nabla_B \Phi \quad ,$$

$$L_A{}^B = (L_\alpha{}^\beta \, , L_{\dot\alpha}{}^{\dot\beta} \, , L_\alpha{}^\beta \, L_{\dot\alpha}{}^{\dot\beta}) \quad , \tag{5.2.19}$$

We define, by analogy with Einstein's theory (5.1.15), covariant derivatives that take the form:

$$\nabla_A = E_A + \Phi_{A\gamma}{}^\beta M_\beta{}^\gamma + \Phi_{A\dot\gamma}{}^{\dot\beta} \bar M_{\dot\beta}{}^{\dot\gamma} \quad , \tag{5.2.20}$$

where the $M$'s are Lorentz rotation operators. Their action is defined in (5.1.13). Again in analogy with ordinary gravity, we adhere to a late-early index convention to distinguish between quantities with curved indices (that transform only under the $\Lambda$-gauge group) and quantities with flat indices (that transform only under the action of the Lorentz generators $M_\alpha{}^\beta$ and $\bar M_{\dot\alpha}{}^{\dot\beta}$). The form (5.2.19) assumes that the Lorentz transformations $L_A{}^B$ act in the usual manner: Spinors and vectors do not mix, and both rotate with the same parameter. For an infinitesimal Lorentz transformation, $L_A{}^B = \delta_A{}^B + \omega_A{}^B$, the covariant derivatives transform as

$$\delta\nabla_A = [i\Lambda \, , \nabla_A] + \omega_A{}^B \nabla_B \quad , \tag{5.2.21}$$

where $\omega_A{}^B = (\tilde\omega_\alpha{}^\beta \, , \omega_{\dot\alpha}{}^{\dot\beta} \, , \omega_{\underline{a}}{}^{\underline{b}})$ and (from (5.2.19)) $\omega_{\underline{a}}{}^{\underline{b}} = \tilde\omega_\alpha{}^\beta \delta_{\dot\alpha}{}^{\dot\beta} + \delta_\alpha{}^\beta \omega_{\dot\alpha}{}^{\dot\beta}$ ($\tilde\omega_\alpha{}^\beta$ is the chiral representation conjugate of $\omega_{\dot\alpha}{}^{\dot\beta}$: see below). This implies the following transformation laws for the connections:

$$\delta\Phi_{A\dot\beta}{}^{\dot\gamma} = [i\Lambda \, , \Phi_{A\dot\beta}{}^{\dot\gamma}] - E_A \omega_{\dot\beta}{}^{\dot\gamma} + \omega_A{}^D \Phi_{D\dot\beta}{}^{\dot\gamma} + \omega_{\dot\beta}{}^{\dot\delta} \Phi_{A\dot\delta}{}^{\dot\gamma} - \Phi_{A\dot\beta}{}^{\dot\delta} \omega_{\dot\delta}{}^{\dot\gamma} \, . \tag{5.2.22}$$

There is a certain amount of arbitrariness in defining connections that transform properly: One can always add to $\Phi_{A\beta}{}^\gamma$ any tensor $K_{A\beta}{}^\gamma$ that transforms covariantly. As will be discussed in the next section, this arbitrariness is physically irrelevant. A standard way to find connections is to compute the *anholonomy coefficients* $C_{AB}{}^C$ defined by $[E_A, E_B\} = C_{AB}{}^C E_C$. Suitable connections can be defined as linear combinations of the $C$'s.

Pushing the analogy with Yang-Mills further, we can try to construct the following *chiral-representation* "covariant" derivatives:

$$\hat{E}_{\dot{\mu}} \equiv \bar{D}_{\dot{\mu}} \quad , \quad \hat{E}_\mu \equiv e^{-H} D_\mu e^H \quad , \quad \hat{E}_{\underline{m}} \equiv -i \{ \hat{E}_\mu , \hat{E}_{\dot{\mu}} \} \quad . \tag{5.2.23}$$

However, at this point the analogy with Yang-Mills theory breaks down. These derivatives are not covariant for two reasons: (1) Acting on nontrivial representations of the Lorentz group, they are noncovariant because they have no connections (this is easily cured); and (2) more seriously, even acting on scalars they are noncovariant because $\Lambda^{\dot{\mu}}$ is not chiral. Thus,

$$\begin{aligned} \delta \hat{E}_{\dot{\mu}} &= [\, i\Lambda , \hat{E}_{\dot{\mu}} \,] - (\, \hat{E}_{\dot{\mu}} \Lambda^{\dot{\nu}} \,)\, \hat{E}_{\dot{\nu}} \\ &= [\, i\Lambda , \hat{E}_{\dot{\mu}} \,] + \omega_{\dot{\mu}}{}^{\dot{\nu}} \hat{E}_{\dot{\nu}} + \Sigma \hat{E}_{\dot{\mu}} \quad , \end{aligned} \tag{5.2.24}$$

where

$$\omega_{\dot{\mu}}{}^{\dot{\nu}} = -\frac{1}{2} \hat{E}_{(\dot{\mu}} \Lambda^{\dot{\nu})} = -\frac{1}{2} \bar{D}_{(\dot{\mu}} \Lambda^{\dot{\nu})} \quad , \quad \Sigma = -\frac{1}{2} \hat{E}_{\dot{\mu}} \Lambda^{\dot{\mu}} = -\frac{1}{2} \bar{D}_{\dot{\mu}} \Lambda^{\dot{\mu}} \quad . \tag{5.2.25}$$

The term involving $\omega$ is harmless: it is just a Lorentz rotation, and will be perfectly covariant after we introduce Lorentz connections. (There is a slight problem, however. The indices in (5.2.19,20) are flat spinor indices whereas those in (5.2.24,25) are curved indices in analogy with our discussion of ordinary gravity. Therefore it is not quite correct to identify the $\omega$ in (5.2.25) with the one in (5.2.20). We will find a solution for this shortly.)

By contrast, the term proportional to $\Sigma$ is a *superspace* scale transformation which is not part of our original gauge group as defined by (5.2.19) and (5.2.20). For the time being we introduce into the theory a (density) *compensator* $\Psi$ that transforms as:

$$\delta \Psi = [\, i\Lambda , \Psi \,] - \Sigma \Psi \quad . \tag{5.2.26}$$

Later on, $\Psi$ will be determined in terms of $H$. With this object, we can construct a covariant spinor derivative:

$$\check{E}_{\dot{\mu}} \equiv \Psi \hat{E}_{\dot{\mu}} = \Psi \bar{D}_{\dot{\mu}} \quad , \qquad \delta \check{E}_{\dot{\mu}} = [\, i\,\Lambda\,, \check{E}_{\dot{\mu}}\,] + \omega_{\dot{\mu}}{}^{\dot{\nu}} \check{E}_{\dot{\nu}} \quad . \tag{5.2.27}$$

The complex conjugate of $\check{E}_{\dot{\mu}}$ is covariant not with respect to $\Lambda$ but rather with respect to $\overline{\Lambda}$; however, just as in the Yang-Mills case, we can use $e^H$ to convert any object covariant with respect to $\overline{\Lambda}$ into an object covariant with respect to $\Lambda$. We obtain

$$\begin{aligned} \check{E}_{\mu} &\equiv e^{-H}\, \overline{\Psi} D_{\mu}\, e^{H} = (e^{-H}\, \overline{\Psi}\, e^{H})\, \hat{E}_{\mu} \\ &\equiv \tilde{\overline{\Psi}} \hat{E}_{\mu} \quad , \end{aligned} \tag{5.2.28}$$

where $\tilde{\overline{\Psi}}$ is the chiral-representation Hermitian conjugate of $\Psi$ (as in super-Yang-Mills: see (4.2.37) and (4.2.78)). The covariant transformation of $\check{E}_{\mu}$ is

$$\delta \check{E}_{\mu} = [i\Lambda\,, \check{E}_{\mu}] + \tilde{\omega}_{\mu}{}^{\nu} \check{E}_{\nu} \quad , \tag{5.2.29}$$

where $\tilde{\omega}_{\mu}{}^{\nu} = e^{-H} \bar{\omega}_{\mu}{}^{\nu} e^{H}$.

The spinor vielbeins that we have constructed transform as in (5.2.21) but with the important restriction that the parameter of Lorentz rotations, $\omega_A{}^B$, must be determined (by the definition in (5.2.25) and those following (5.2.29) and (5.2.21)) in terms of the parameter of supercoordinate transformations $\Lambda^M$. In the discussion of ordinary gravity (see (5.1.10,11)), we saw that an analogous situation occurred only if the antisymmetric part of the vierbein was gauged away. We also have the related problem that the free index on the vielbein is curved whereas the index in (5.2.19) is flat (and consequently the problem of identifying $\omega_{\dot{\alpha}}{}^{\dot{\beta}}$ with $\omega_{\dot{\mu}}{}^{\dot{\nu}}$). This situation arises because the vielbein $E_{\dot{\alpha}}{}^{\dot{\mu}}$ as defined in (5.2.27) is given by $\Psi \delta_{\dot{\alpha}}{}^{\dot{\mu}}$ and thus has no symmetric part (i.e., $E_{(\dot{\alpha}}{}^{\dot{\mu})} = 0$). The solution is to restore the "missing" part by introducing a new superfield $N_{\dot{\alpha}}{}^{\dot{\mu}}$. In a general Lorentz frame the spinor vielbein (5.2.27) is modified to

$$E_{\dot{\alpha}} = N_{\dot{\alpha}}{}^{\dot{\mu}} \Psi \bar{D}_{\dot{\mu}} \quad , \tag{5.2.30}$$

where $N_{\dot{\alpha}}{}^{\dot{\mu}}$ is an arbitrary $SL(2C)$ matrix superfield ($det\ N = 1$). It acts as a compensating field for tangent space Lorentz transformations. (This is analogous to generalizing from a frame where the usual vierbein is symmetric.) The $N$-dependence of the other equations can easily be found by simply performing the general Lorentz transformation which takes $E_{\dot{\alpha}}{}^{\dot{\mu}}$ from $\Psi \delta_{\dot{\alpha}}{}^{\dot{\mu}}$ to $\Psi N_{\dot{\alpha}}{}^{\dot{\mu}}$. The quantity $N_{\alpha}{}^{\mu}$ maps between curved and flat spinor indices. This permits us to solve the problem of identifying $\omega_{\dot\alpha}{}^{\dot\beta}$ with $\omega_{\dot\mu}{}^{\dot\nu}$. Since we began in the gauge $N_{\dot\alpha}{}^{\dot\mu} = \delta_{\dot\alpha}{}^{\dot\mu}$, the two quantities are equal. Furthermore, as long as we remain in this gauge, we need not be careful to distinguish curved and flat spinor indices. *A distinction must still be made between flat and curved vector indices.*

We now attempt to construct the vector covariant derivative by analogy with Yang-Mills theory:

$$\check{E}_{\underline{m}} = -\, i\, \{ \check{E}_{\mu} , \check{E}_{\dot\mu} \} \quad . \tag{5.2.31}$$

The transformation law follows from (5.2.27,29):

$$\delta\check{E}_{\underline{m}} = [i\Lambda , \check{E}_{\underline{m}}] + \tilde{\omega}_{\mu}{}^{\nu}\check{E}_{\nu\dot\mu} + \omega_{\dot\mu}{}^{\dot\nu}\check{E}_{\mu\dot\nu} - i(\check{E}_{\mu}\omega_{\dot\mu}{}^{\dot\nu})\check{E}_{\dot\nu} - i(\check{E}_{\dot\mu}\tilde{\omega}_{\mu}{}^{\nu})\check{E}_{\nu} \quad . \tag{5.2.32}$$

Defining $\check{E}_M \equiv (\check{E}_{\mu} , \check{E}_{\dot\mu} , \check{E}_{\underline{m}})$ we can write (5.2.27,29,32) as

$$\delta\check{E}_M = [i\Lambda , \check{E}_M] + \omega_M{}^N \check{E}_N \ . \tag{5.2.33}$$

However, because of terms like $\omega_{\underline{m}}{}^{\dot\nu} = -\, i\check{E}_{\mu}\omega_{\dot\mu}{}^{\dot\nu}$, which are not present in (5.2.19), $\check{E}_M$ is not quite covariant.

The terms we want to eliminate are (spinor) derivatives of the Lorentz transformation parameter $\omega_{\dot\mu}{}^{\dot\nu}$; therefore, the remedy is to introduce (spinor) Lorentz connections into (5.2.31). These connection terms will redefine $\check{E}_{\underline{m}}$ so that it transforms covariantly. To find the connections, we define a (noncovariant) set of anholonomy coefficients $\check{C}_{MN}{}^P$ by

$$[\check{E}_M , \check{E}_N\} = \check{C}_{MN}{}^P \check{E}_P \quad . \tag{5.2.34}$$

From the transformations in (5.2.27,29,32) we obtain

$$\delta\check{C}_{MN}{}^P = [\, i\Lambda, \check{C}_{MN}{}^P] + \check{E}_{[M}\omega_{N)}{}^P + \omega_{[M|}{}^R \check{C}_{R|N)}{}^P - \check{C}_{MN}{}^R \omega_R{}^P \ . \tag{5.2.35}$$

In particular we find

$$\delta\check{C}_{\dot\mu\dot\nu}{}^{\dot\pi} = [\, i\Lambda, \check{C}_{\dot\mu\dot\nu}{}^{\dot\pi}] + \check{E}_{(\dot\mu}\,\omega_{\dot\nu)}{}^{\dot\pi} + \omega_{(\dot\mu|}{}^{\dot\rho}\check{C}_{\dot\rho,|\dot\nu)}{}^{\dot\pi} - \check{C}_{\dot\mu,\dot\nu}{}^{\dot\rho}\omega_{\dot\rho}{}^{\dot\pi} \quad ,$$

$$\delta\check{C}_{\underline{m},\dot\nu}{}^{\underline{r}} = [\, i\Lambda, \check{C}_{\underline{m},\dot\nu}{}^{\underline{r}}] - \check{E}_{\dot\nu}\,\omega_{\underline{m}}{}^{\underline{r}} + \omega_{\underline{m}}{}^{\underline{n}}\check{C}_{\underline{n},\dot\nu}{}^{\underline{r}} - \check{C}_{\underline{m},\dot\nu}{}^{\underline{n}}\omega_{\underline{n}}{}^{\underline{r}}$$

$$+\, \omega_{\dot\nu}{}^{\dot\sigma}\check{C}_{\dot\sigma,\underline{m}}{}^{\underline{r}} + i\omega_{\underline{m}}{}^{\rho}\delta_{\dot\nu}{}^{\dot\rho} \ , \tag{5.2.36}$$

and corresponding equations for the conjugates $\check{C}_{\mu\nu}{}^{\pi}$ and $\check{C}_{\underline{m},\nu}{}^{\underline{r}}$. From (5.2.36) we see that $-\frac{1}{2}\check{C}_{(\mu\dot{\nu},\dot{\mu}}{}^{\nu)\dot{\nu}}$ transforms as the needed spinor connection:

$$\delta[-i\frac{1}{2}\check{C}_{(\mu\dot{\nu},\dot{\mu}}{}^{\nu)\dot{\nu}}\check{E}_{\nu}] = i(\check{E}_{\dot{\mu}}\tilde{\omega}_{\mu}{}^{\nu})\check{E}_{\nu} - i\frac{1}{2}\tilde{\omega}_{\mu}{}^{\rho}\check{C}_{(\rho\dot{\nu},\dot{\mu}}{}^{\nu)\dot{\nu}}\check{E}_{\nu}$$

$$- i\frac{1}{2}\omega_{\dot{\mu}}{}^{\dot{\rho}}\check{C}_{(\mu\dot{\nu},\dot{\rho}}{}^{\nu)\dot{\nu}}\check{E}_{\nu} \quad . \tag{5.2.37}$$

Therefore we define

$$E_{\underline{a}} \equiv \delta_{\underline{a}}{}^{\underline{m}}[\check{E}_{\underline{m}} - i\frac{1}{2}\check{C}_{(\mu\dot{\nu},\dot{\mu}}{}^{\nu)\dot{\nu}}\check{E}_{\nu} - i\frac{1}{2}\check{C}_{\nu(\dot{\mu},\mu}{}^{\nu\dot{\nu})}\check{E}_{\dot{\nu}}] \ , \tag{5.2.38}$$

where $\delta_{\underline{a}}{}^{\underline{m}} \equiv \tilde{N}_{\alpha}{}^{\mu}N_{\dot{\alpha}}{}^{\dot{\mu}}$ in the gauge $N_{\dot{\alpha}}{}^{\dot{\mu}}=\delta_{\dot{\alpha}}{}^{\dot{\mu}}$. The vector vielbein $E_{\underline{a}}$ transforms covariantly.

We have already constructed one of the Lorentz connection superfields $\Phi_{A\beta}{}^{\gamma}$ (as noted above, in the gauge $N_{\alpha}{}^{\mu} = \delta_{\alpha}{}^{\mu}$ we need not distinguish curved and flat spinor indices). We can construct the remaining connections in the standard way (see subsec. 5.3.b.1) from the anholonomy coefficients $C_{AB}{}^{C}$ defined by $E_A \equiv (E_{\dot{\alpha}}, E_{\alpha}, E_{\underline{a}})$, where $E_{\dot{\alpha}} = \delta_{\dot{\alpha}}{}^{\dot{\mu}}\check{E}_{\dot{\mu}}, E_{\alpha} = \delta_{\alpha}{}^{\mu}\check{E}_{\mu}$ in our particular Lorentz gauge, and $E_{\underline{a}}$ is given in (5.2.38).

Alternatively, we can use $\check{C}$ directly. As we saw above, an appropriately transforming spin connection $\Phi_{\dot{\beta}\alpha}{}^{\gamma}$ is given by $\frac{1}{2}\check{C}_{(\alpha\dot{\alpha},\dot{\beta}}{}^{\gamma)\dot{\alpha}}$. For $\Phi_{\dot{\alpha}\dot{\beta}}{}^{\dot{\gamma}}$ we have a choice: Both $-\frac{1}{4}\check{C}_{\dot{\alpha},\alpha(\dot{\beta}}{}^{\alpha\dot{\gamma})}$ and $-\frac{1}{2}[\check{C}_{\dot{\alpha},\dot{\beta}}{}^{\dot{\gamma}} + \check{C}_{\dot{\alpha},}{}^{\dot{\gamma}}{}_{,\dot{\beta}} - \check{C}_{\dot{\beta},}{}^{\dot{\gamma}}{}_{,\dot{\alpha}}]$ transform appropriately. In general, any linear combination of these can be used as a spin connection. Furthermore $\delta_{\dot{\alpha}}{}^{(\dot{\gamma}}\check{C}_{\dot{\beta}),\underline{d}}{}^{\underline{d}}$ is Lorentz covariant, and can be added to $\Phi_{\dot{\alpha}\dot{\beta}}{}^{\dot{\gamma}}$; see sec. 5.3.a.3. We choose

$$\Phi_{\dot{\alpha}\dot{\beta}}{}^{\dot{\gamma}} = \frac{1}{4}[-\check{C}_{\dot{\alpha},\alpha(\dot{\beta}}{}^{\alpha\dot{\gamma})} + \delta_{\dot{\alpha}}{}^{(\dot{\gamma}}\check{C}_{\dot{\beta}),\underline{d}}{}^{\underline{d}}] \quad . \tag{5.2.39a}$$

We already had

$$\Phi_{\alpha\dot{\beta}}{}^{\dot{\gamma}} = -\frac{1}{2}\check{C}_{\alpha,\beta(\dot{\beta}}{}^{\beta\dot{\gamma})} \quad . \tag{5.2.39b}$$

We also have corresponding expressions for the complex conjugates $\Phi_{\alpha\beta}{}^{\gamma}$, $\Phi_{\dot{\alpha}\beta}{}^{\gamma}$.

The vector connection is defined by:

$$\Phi_{\underline{a},\beta}{}^{\gamma} = -i[E_{\alpha}\Phi_{\dot{\alpha}\beta}{}^{\gamma} + \Phi_{\alpha,\dot{\alpha}}{}^{\dot{\delta}}\Phi_{\dot{\delta}\beta}{}^{\gamma} + E_{\dot{\alpha}}\Phi_{\alpha\beta}{}^{\gamma} + \Phi_{\dot{\alpha},\alpha}{}^{\delta}\Phi_{\delta\beta}{}^{\gamma} + \Phi_{\alpha(\beta}{}^{|\delta}\Phi_{\dot{\alpha}\delta|}{}^{\gamma)}] \ . \tag{5.2.40}$$

as follows from $\nabla_{\alpha\dot{\alpha}} = -i\{\nabla_\alpha , \bar{\nabla}_{\dot{\alpha}}\}$.

### a.4. Covariant actions

In supergravity Lagrangians cannot be invariant because all our quantities, including scalars, transform under $\Lambda$ transformations. At best a Lagrangian can transform as a total derivative:

$$\delta I\!L = -(-)^M D_M(\Lambda^M I\!L) \quad . \tag{5.2.41}$$

This can be rewritten as

$$\delta I\!L = i I\!L \overleftarrow{\Lambda} = i[\Lambda , I\!L] + i(1 \cdot \overleftarrow{\Lambda})I\!L \quad , \tag{5.2.42}$$

where

$$\overleftarrow{\Lambda} = i\Lambda^M \overleftarrow{D}_M = i(-)^M[\overleftarrow{D}_M \Lambda^M + (D_M \Lambda^M)] \quad . \tag{5.2.43}$$

Equation (5.2.42) is the transformation law for a *density.* It is easy to check that a scalar times a density is also a density.

By analogy with gravity, we take the vielbein superdeterminant $E = sdet\, E_A{}^M$ as a candidate for a density. Indeed, from (3.7.17)

$$\delta E^{-1} = -(-)^M E^{-1}[(E^{-1})_M{}^A \delta E_A{}^M] \quad , \tag{5.2.44}$$

where, from (5.2.21)

$$\delta E_A{}^M = i(\Lambda E_A{}^M) + (E_A \Lambda^M) + \omega_A{}^B E_B{}^M \quad . \tag{5.2.45}$$

(However the Lorentz rotation terms trivially drop out of (5.2.44).) Consequently,

$$\begin{aligned} \delta E^{-1} &= -i(-)^M E^{-1}[(E^{-1})_M{}^A \Lambda E_A{}^M] - (-)^M E^{-1}[D_M \Lambda^M] \\ &= i[\Lambda , E^{-1}] - (-)^M (D_M \Lambda^M) E^{-1} \\ &= iE^{-1}\overleftarrow{\Lambda} \quad . \end{aligned} \tag{5.2.46}$$

Therefore, invariant actions can be constructed as integrals of products of $E^{-1}$ and scalar quantities of the appropriate dimension ($E$ itself is dimensionless): $\delta(E^{-1}I\!L) = i(E^{-1}I\!L)\overleftarrow{\Lambda}$.

For supergravity we want to have the usual Einstein term in the component action; hence we require

$$S_{SG} = \frac{1}{\kappa^2} \int d^4x \, d^4\theta \; E^{-1} I\!L_{SG} = \frac{1}{\kappa^2} \int d^4x \; \mathrm{e}^{-1} I\!L_{component} \quad . \tag{5.2.47}$$

This implies that $I\!L_{SG}$ is a dimensionless scalar. In general, the only dimensionless scalars in the theory are constants, so we must take (however, see below)

$$S_{SG} \sim \frac{1}{\kappa^2} \int d^4x \, d^4\theta \; E^{-1} \tag{5.2.48}$$

as the locally supersymmetric invariant action for Poincaré supergravity.

The vielbein superdeterminant can be worked out in a straightforward manner using (3.7.15). We find

$$E = sdet[E_A{}^M] = sdet[\check{E}_A{}^M] = \Psi^2 \widetilde{\bar{\Psi}}^2 sdet[\hat{E}_A{}^M(H)] \quad . \tag{5.2.49}$$

However, variation of the action with the field $\Psi$ considered as an independent variable leads to a singular field equation: $(\Psi E)^{-1} = 0$. This is not surprising: $\Psi$ was introduced as a device to simplify the construction of the covariant derivatives, and it should be related to the fundamental prepotential $H^M$.

## b. Poincaré

We now consider specific forms for the compensator $\Psi$ in terms of $H$. As we discussed earlier, the $\Lambda$-gauge group includes superconformal transformations: Thus, the covariant derivatives we have constructed are appropriate for describing a superconformally invariant theory. The superconformal action is the nonlinear version of (5.2.6). It is a functional of $H^M$ only; $\Psi$ drops out completely. To describe Poincaré supergravity, we will have to break the extra invariance, i.e., the component superscale invariance.

To find an appropriate expression for $\Psi$ we recall that it transforms as a (noncovariant) density (5.2.25,26)

$$\delta\Psi = [i\Lambda \, , \Psi] - \Sigma \, \Psi \quad . \tag{5.2.50}$$

The only other dimensionless object that transforms as a density with respect to $\Lambda$ and *not* $\overline{\Lambda}$ is $E(\Psi, H)$ (see (5.2.46)). We therefore express $\Psi$ in terms of $H^M$ (implicitly) by writing

$$\Psi^{4n} = E^{n+1} \quad . \tag{5.2.51}$$

(This particular parametrization will prove convenient when writing the explicit action (5.2.65).) However, this relation is not preserved by the full $\Lambda$-group: If we transform both sides of (5.2.51), using (5.2.46,50) we find the restriction

$$-4n\Sigma = -(n+1)(1 \cdot i\overleftarrow{\Lambda}) \quad , \tag{5.2.52}$$

or, more explicitly,

$$(3n+1)\bar{D}_{\dot{\mu}}\Lambda^{\dot{\mu}} = (n+1)(\partial_{\underline{m}}\Lambda^{\underline{m}} - D_\mu\Lambda^\mu) \quad . \tag{5.2.53}$$

This is an acceptable restriction on the gauge group: We can show that it corresponds to reducing the *component* local superconformal group to the super-Poincaré group. We note that when $n = -\frac{1}{3}$ (5.2.53) sets the *chiral* quantity $\partial_{\underline{m}}\Lambda^{\underline{m}} - D_\mu\Lambda^\mu$ to zero: i.e., using (5.2.14) the restriction can be written as

$$n = -\frac{1}{3}: \qquad D_\mu\Lambda^\mu - \partial_{\underline{m}}\Lambda^{\underline{m}} = \bar{D}^2 D_\mu L^\mu = 0 \quad . \tag{5.2.54}$$

On the other hand, for $n \neq -\frac{1}{3}$ the condition restricts $\Lambda^{\dot{\mu}}$: (5.2.53,54) imply

$$n \neq -\frac{1}{3}: \qquad \bar{D}^2\Lambda^{\dot{\mu}} = 0 \quad . \tag{5.2.55}$$

We analyze the case $n = -\frac{1}{3}$ first. Since $\Lambda^{\dot{\mu}}$ is unrestricted, we can still use it to gauge away $H^{\dot{\mu}}$; we then need only reconsider our discussion of the Wess-Zumino gauge for $H^{\underline{m}}$ subject to the restriction (5.2.54). This restriction implies the following relations among the components of $L_\mu$ in (5.2.9):

$$\sigma = i\partial_{\underline{a}}\xi^{\underline{a}} \quad , \qquad \eta_\alpha = -i\partial^{\beta\dot{\beta}}L^1{}_{\alpha\beta\dot{\beta}} \quad ,$$

$$\partial^{\underline{a}}L^2{}_{\underline{a}} = 0 \quad . \tag{5.2.56}$$

Thus the local superconformal transformations are reduced to those of local super-Poincaré: $\sigma$ and $\eta_\alpha$ have been removed as independent parameters. Further, a *differential* constraint has been imposed on one of the parameters, the $L^2$ that we used to gauge away extra components of $H^{\underline{m}}$: $B = \partial_{\underline{m}}h^{(2)\underline{m}}$ can no longer be eliminated. Consequently (cf. the discussion following (5.2.10)), we find that the minimal set of component fields describing $N = 1$ Poincaré supergravity are the vierbein (the nonlinear extension of $h_{\underline{ab}}$), the gravitino $\psi_{\underline{a}\beta}$, an axial vector field $A_{\underline{a}}$, and in addition the complex scalar field $B$ (which could be gauged away in the superconformal case).

For $n \neq -\frac{1}{3}$, we cannot use $\Lambda^{\dot{\mu}}$ to gauge away all of $H^{\dot{\mu}}$. We can gauge away parts of it by using up all the components of $\bar{D}_{\dot{\mu}}\Lambda^{\dot{\mu}}$ and hence, because of the constraint (5.2.52), those of $\bar{D}^2 D_\mu L^\mu$. Again the local superconformal group has been reduced to the super-Poincaré group: $\sigma$ and $\eta_\alpha$ no longer enter as independent parameters. We will discuss the component content of $n \neq -\frac{1}{3}$ supergravity later.

For $n = 0$, (5.2.51) implies that $E = 1$ and hence that the action (5.2.48) vanishes. It is clear from (5.2.53) that for $n = 0$ the parameter $\Lambda^M$ satisfies $(-)^M D_M \Lambda^M = 0$ . This is precisely the condition that the supercoordinate transformation parametrized by $\Lambda$ are "supervolume preserving". However, the $n = 0$ case is unique because it contains a (constrained) dimensionless scalar $V$ (an abelian gauge prepotential) which can be used to construct an action. Furthermore, the constraint (5.2.51) is invariant under an arbitrary local phase rotation of $\Psi$: $\Psi' = e^{iK_5}\Psi$, $K_5 = \tilde{\bar{K}}_5 = e^{-H}\bar{K}_5 e^H$. This invariance can be used to choose a gauge where $\Psi = \tilde{\bar{\Psi}}$. Another consequence of (5.2.51), $E = \bar{E} = 1$, is that we have imposed a partial gauge condition on $H$: The hermiticity condition that $E^{-1}$ satisfies (see (5.2.60) below) implies that $(1 \cdot e^{-\overleftarrow{H}}) = 1$, i.e., $1 \cdot \overleftarrow{H} = 0$. This is achieved by choosing the gauge where $H^\alpha = -i\bar{D}_{\dot{\alpha}} H^{\alpha\dot{\alpha}}$ instead of zero, so that $H = D_\alpha H^{\alpha\dot{\alpha}}\bar{D}_{\dot{\alpha}} + \bar{D}_{\dot{\alpha}} H^{\alpha\dot{\alpha}} D_\alpha$ where the $D$ and $\bar{D}$ preceding $H^{\alpha\dot{\alpha}}$ act on all objects to the right. The case of $n = 0$ will be discussed in more detail in the following sections.

To find the explicit expression for $\Psi$ in terms of $H$, we use (5.2.49) and (5.2.51) and write

$$\tilde{\bar{\Psi}}^{4n} = e^{-H}\bar{\Psi}^{4n}e^H = e^{-H}\bar{E}^{n+1}e^H \quad . \tag{5.2.57}$$

(For simplicity we have assumed that $n$ is real; the generalization to complex $n$ is straightforward but not interesting). Therefore, we must compute $\bar{E}$. Although this could be done by brute force, a more elegant procedure is possible:

In (5.2.28) we defined

$$E_\alpha = e^{-H} \bar{E}_\alpha e^{H} \quad , \tag{5.2.58}$$

where $\bar{E}_\alpha$ is the hermitian conjugate of $E_{\dot\alpha}$. The right-hand side can be interpreted as a coordinate transformation with imaginary parameter $iH^M$ instead of $i\Lambda^M$. By hermitian conjugation $E_{\dot\alpha}$ is the coordinate transform of $\bar{E}_{\dot\alpha}$ (the hermitian conjugate of $E_\alpha$). Therefore, any covariant constructed from $E_\alpha$ and $E_{\dot\alpha}$ can be obtained by a complex coordinate transformation from the corresponding object constructed out of $\bar{E}_{\dot\alpha}$ and $\bar{E}_\alpha$. This is the case for $E_{\underline{a}}$, and also for the vielbein superdeterminant.

The full nonlinear transformation of the superdeterminant follows from (5.2.46):

$$(E^{-1})' = E^{-1} e^{i\overleftarrow{\Lambda}}$$

$$= (e^{i\Lambda} E^{-1} e^{-i\Lambda})(1 \cdot e^{i\overleftarrow{\Lambda}}) \quad . \tag{5.2.59}$$

By the same method used to derive this result from $E'_A = e^{i\Lambda} E_A e^{-i\Lambda}$, from

$$\bar{E}_A = e^{H} E_A e^{-H} \tag{5.2.60a}$$

we have

$$\bar{E}^{-1} = E^{-1} e^{\overleftarrow{H}} \tag{5.2.60b}$$

and hence

$$E^{-1} = e^{-H} \bar{E}^{-1} e^{H} (1 \cdot e^{-\overleftarrow{H}}) \quad . \tag{5.2.60c}$$

Substituting (5.2.60) into (5.2.57) we find

$$\tilde{\Psi}^{4n} = [E(1 \cdot e^{-\overleftarrow{H}})]^{n+1} \quad , \tag{5.2.61}$$

or, using the original constraint (5.2.51),

$$\tilde{\Psi}^{4n} = \Psi^{4n} (1 \cdot e^{-\overleftarrow{H}})^{n+1} \quad . \tag{5.2.62}$$

Finally, substituting (5.2.62) and (5.2.51) into (5.2.49) we find

$$\Psi^{4n} = \Psi^{4(n+1)} (1 \cdot e^{-\overleftarrow{H}})^{\frac{(n+1)^2}{2n}} \hat{E}^{n+1} \quad , \tag{5.2.63}$$

or

$$\Psi = [(1 \cdot e^{-\overleftarrow{H}})^{\frac{(n+1)^2}{8n}} \hat{E}^{\frac{n+1}{4}}]^{-1} \quad . \tag{5.2.64}$$

Thus we have solved for $\Psi$ in terms of $H$. Using these results, we can rewrite the $n \neq 0$ supergravity action (5.2.48) in terms of the unconstrained superfield $H^M$:

$$S_{SG} = \frac{1}{n\kappa^2} \int d^4x\, d^4\theta\, [\hat{E}^n (1 \cdot e^{-\overleftarrow{H}})^{\frac{n+1}{2}}] \quad . \tag{5.2.65}$$

(The factor $\frac{1}{n}$ gives the appropriate normalization for the physical component actions and for the supersymmetric-gauge propagators: $\hat{E}^n = (1+\Delta)^n \simeq 1 + n\Delta$; see, e.g., (7.2.26).) This action is invariant under the group of $\Lambda$ transformations restricted by (5.2.53).

## c. Density compensators

For many purposes, e.g., quantization, it is awkward to work with the constrained gauge group. As described in sec 3.10 we can enlarge the invariance group of a theory by introducing compensating fields. In this case, the constraint (5.2.53) was introduced by the relation (5.2.51), which is not covariant under the full gauge group. We choose our compensators to restore the covariance of (5.2.51) under the full $\Lambda$ group.

For the $n = -\frac{1}{3}$ case a suitable compensating field is a *chiral density* $\phi$ that transforms as

$$\delta\phi = [i\Lambda \, , \phi] + \frac{1}{3}(D_\mu \Lambda^\mu - \partial_{\underline{m}} \Lambda^{\underline{m}})\phi \quad , \quad \bar{D}_{\dot{\mu}}\phi = 0 \quad . \tag{5.2.66}$$

(The factor $\frac{1}{3}$ gives $\phi$ the same weight as a density matter multiplet: see below.) From (5.2.46,50,66), it follows that the covariant version of (5.2.51) is:

$$\Psi^{-\frac{4}{3}} = \phi^2 E^{\frac{2}{3}} \quad . \tag{5.2.67}$$

Eq. (5.2.66) can be rewritten as

$$(\phi^3)' = (\phi^3) e^{i\overleftarrow{\Lambda}_{ch}} \quad , \quad \Lambda_{ch} = \Lambda^{\underline{m}} i\partial_{\underline{m}} + \Lambda^\mu i D_\mu \quad . \tag{5.2.68}$$

The chiral parameter $(1 \cdot i\overleftarrow{\Lambda}_{ch}) = -\,\partial_{\underline{m}}\Lambda^{\underline{m}} + D_\mu \Lambda^\mu$ can be used to scale $\phi$ arbitrarily: In particular, if we *choose the gauge* $\phi = 1$, from (5.2.67,68) we recover the constraints (5.2.51,54), respectively.

In the $n \neq -\frac{1}{3}$ case, the constrained object is the linear superfield expression (cf.

(5.2.53,54)) $\bar{D}_{\dot{\mu}}\Lambda^{\dot{\mu}} + (\frac{n+1}{3n+1})\bar{D}^2 D_\mu L^\mu$. Hence, a suitable compensator is a complex linear superfield $\Upsilon$, $\bar{D}^2\Upsilon = 0$. We determine its transformation properties by requiring that its variation is linear and that $\Upsilon$ can be scaled to 1. (It should be recalled from the discussion in sec. 4 that a complex linear superfield can always be expressed in terms of an unconstrained spinor superfield: $\Upsilon = \bar{D}_{\dot{\mu}}\bar{\Phi}^{\dot{\mu}}$ (4.5.4).) Since the product of a chiral and a linear superfield is linear, we can always have a term $(D_\mu\Lambda^\mu - \partial_{\underline{m}}\Lambda^{\underline{m}})\Upsilon$ in $\delta\Upsilon$. To scale $\Upsilon$ to 1, we need a term $(\bar{D}_{\dot{\mu}}\Lambda^{\dot{\mu}})\Upsilon$, but since the product of two linear superfields is *not* linear, such a term must come from the combination $(\bar{D}_{\dot{\mu}}\Lambda^{\dot{\mu}})\Upsilon - \Lambda^{\dot{\mu}}\bar{D}_{\dot{\mu}}\Upsilon = \bar{D}_{\dot{\mu}}(\Lambda^{\dot{\mu}}\Upsilon)$. This leads uniquely to

$$\delta\Upsilon = [i\Lambda\,,\Upsilon] + (\bar{D}_{\dot{\mu}}\Lambda^{\dot{\mu}})\Upsilon + (\frac{n+1}{3n+1})(D_\mu\Lambda^\mu - \partial_{\underline{m}}\Lambda^{\underline{m}})\Upsilon \qquad (5.2.69)$$

and

$$\Psi^{4n} = \Upsilon^{3n+1}E^{n+1} \quad . \qquad (5.2.70)$$

As in the minimal $n = -\frac{1}{3}$ case, choosing the gauge $\Upsilon = 1$ leads back to the constraints (5.2.53,51). We observe that for $n = 0$, (5.2.70) implies $\Upsilon = \tilde{\Upsilon}(1 \cdot e^{-\overleftarrow{H}})$. (We can define a linear compensator $\Upsilon' \equiv \phi^{-\frac{3n+3}{3n+1}}\Upsilon$ in terms of both $\Upsilon$ and $\phi$; this enlarges the gauge group by a chiral scale transformation, and results in $n$-independent transformation laws for both $\phi$ and $\Upsilon'$).

We can repeat the computations of eqs. (5.2.57-64) including the compensators; we find

$$n \neq -\frac{1}{3} \quad , \quad \Psi = [\,\Upsilon^{n-1}\,\tilde{\Upsilon}^{n+1}\,]^{-(\frac{3n+1}{8n})}\,[\,(1\cdot e^{-\overleftarrow{H}})^{n+1}\,\hat{E}^{2n}\,]^{-(\frac{n+1}{8n})} \quad ,$$

$$n = -\frac{1}{3} \quad , \quad \Psi = \phi^{-1}\,\tilde{\phi}^{\frac{1}{2}}\,(1\cdot e^{-\overleftarrow{H}})^{\frac{1}{6}}\,\hat{E}^{-\frac{1}{6}} \quad . \qquad (5.2.71)$$

The $n \neq 0$ supergravity action (5.2.48,65) takes the form

$$n \neq -\frac{1}{3} \;,\; S_{SG}(H\,,\Upsilon) = \frac{1}{n\kappa^2}\int d^4x\,d^4\theta\;\hat{E}^n(1\cdot e^{-\overleftarrow{H}})^{\frac{n+1}{2}}\,[\tilde{\Upsilon}\Upsilon]^{\frac{3n+1}{2}} \quad ,$$

$$n = -\frac{1}{3} \;,\; S_{SG}(H\,,\phi) = -\frac{3}{\kappa^2}\int d^4x\,d^4\theta\;\hat{E}^{-\frac{1}{3}}(1\cdot e^{-\overleftarrow{H}})^{\frac{1}{3}}\tilde{\phi}\phi \quad . \qquad (5.2.72)$$

These are invariant under the full $\Lambda$ gauge group; after $\theta$-integration, they become the usual Poincaré supergravity component action for the graviton, gravitino, and auxiliary fields. The $n = 0$ case will be discussed later.

## d. Gauge choices

The component field content of the actions (5.2.72) is manifest in a Wess-Zumino gauge where the $\Lambda$ transformations have been used to remove algebraically the gauge components of the superfields. Since the $\Lambda$ group is now unconstrained, we can choose the gauge $H^{\mu} = 0$ and $H^{\underline{m}}$ with only $h\,,\psi\,,A$ as described following (5.2.10). There we found that the remaining gauge freedom is parametrized by the $\omega$, $Im\xi$, $\epsilon_{\alpha}$, $Re\sigma + i\,Im\sigma$, and $\eta_{\alpha}$ components of $L_{\alpha}$; these correspond to Lorentz, coordinate, local supersymmetry, scale $+\,i$chiral, and $S$-supersymmetry transformations respectively.

For $n = -\frac{1}{3}$, we define the (linearized) components of $\phi$ by

$$u = \phi| \quad , \quad u^{\alpha} = D^{\alpha}\phi| \quad , \quad S = \mathrm{S} + i\mathrm{P} = D^2\phi| \quad . \tag{5.2.73}$$

Under the gauge transformations that remain in the Wess-Zumino gauge (we need the compensating transformations (5.2.18) and in addition $L^1{}_{\beta\alpha\dot{\beta}} = \frac{1}{2}C_{\alpha\beta}\bar{\epsilon}_{\dot{\beta}}$ in (5.2.10)), these components transform as

$$\delta u = -\frac{1}{3}(\sigma + i\partial^{\underline{a}}\xi_{\underline{a}}) \quad ,$$

$$\delta u^{\alpha} = \frac{1}{3}(\eta^{\alpha} + i\partial_{\beta\dot{\beta}}L^{1\,\alpha\beta\dot{\beta}}) \quad ,$$

$$\delta S = i\,\frac{1}{3}\,\partial^{\underline{a}}L^2{}_{\underline{a}} \quad . \tag{5.2.74}$$

In writing these transformation laws we have linearized the full infinitesimal transformation of (5.2.66) and kept only those terms independent of $H^{\underline{a}}$ and $\phi$. (This is analogous to the approximation given in (5.2.7) as compared to the full infinitesimal transformation given in (5.2.17).) The scale and axial transformations parametrized by $\sigma$ can be used to scale $u$ to 1, and the $S$-supersymmetry transformation can be used to gauge $u^{\alpha}$ away. Thus we see that $\phi$ acts as a compensator for the constrained part of the $\Lambda$ group, i.e., $u$ and $u^{\alpha}$ are compensators for the component superscale transformations (scale, chiral $U(1)$, and $S$-supersymmetry). Setting $u - 1 = u^{\alpha} = 0$ restores the constraint and leaves only the component super-Poincaré transformations. The remaining fields $S\,,\bar{S}$ along with $h$, $\psi$, and $A$ from $H^{\underline{m}}$ are the component fields of minimal supergravity.

Within the framework of the $n=-\frac{1}{3}$ theory, the choice of compensator is not unique. Any superfield that contains only superspin 0 (components $u$ and $u_\alpha$) can be used as a compensator. In global supersymmetry we found a number of such multiplets: the variant representations of sec. 4.5.d. Thus we can replace $\phi^3-1$ in (5.2.66-72) by the chiral field strengths $\Pi=\bar{D}^2V$ of (4.5.56) or $\Pi=\bar{D}^2D^\alpha\Psi_\alpha$ of (4.5.66). The new components $u\,,u_\alpha$ of the new multiplets are completely equivalent to the corresponding components of $\phi$. On the other hand the $S$ component is changed. We thus have two cases in addition to $S=D^2\phi|=\mathrm{S}+i\mathrm{P}$

$$(2)\qquad S=D^2\Pi|=D^2\bar{D}^2V|=\frac{1}{2}\{D^2\,,\bar{D}^2\}V|+\frac{1}{2}[D^2\,,\bar{D}^2]V|$$

$$=\frac{1}{2}\{D^2\,,\bar{D}^2\}V|+i\partial^{\alpha\dot{\alpha}}[D_\alpha\,,\bar{D}_{\dot{\alpha}}]V|=\mathrm{S}+i\partial^{\underline{a}}\mathrm{P}_{\underline{a}}\quad, \tag{5.2.75a}$$

or

$$(3)\qquad S=D^2\bar{D}^2D^\alpha\Psi_\alpha|=-\,i\partial^{\alpha\dot{\alpha}}D^2\bar{D}_{\dot{\alpha}}\Psi_\alpha|=\partial^{\underline{a}}\mathrm{S}_{\underline{a}}+i\partial^{\underline{a}}\mathrm{P}_{\underline{a}}\quad. \tag{5.2.75b}$$

In case (2) the auxiliary fields are the real scalar S and the divergence of the axial vector $\mathrm{P}_{\underline{a}}$ (instead of the pseudoscalar P). Also in case (2) we can make the replacement $V=iV'$. The effect of this at the component level is that $S$ takes the form $S=\partial^{\underline{a}}S_{\underline{a}}+i\mathrm{P}$. The auxiliary fields are the divergence of a vector $S_{\underline{a}}$ and the pseudoscalar P. In case (3), both auxiliary fields are divergences.

For $n=-\frac{1}{3}$, before we introduced compensators, the restricted gauge group (5.2.54) could be used to eliminate all but the $h\,,\psi\,,A$ and $S$ components of $H^{\underline{m}}$, where $S\equiv\partial_{\underline{m}}D^2H^{\underline{m}}|$ (cf. the discussion after (5.2.56)). In the presence of the compensators, the full gauge group was used to gauge away $S$ from $H^{\underline{m}}$; $S$ appeared in the compensator instead. However, only in case (3) above is $S$ (in the compensator) a divergence; therefore this is the only case in which the theory *with* the compensator is completely equivalent to the theory *without* the compensator.

For $n\neq-\frac{1}{3}\,,0$, we can also go to the Wess-Zumino gauge and find the components of the nonminimal theory. We define the (linearized) components of $\Upsilon$:

$$u = \Upsilon| \quad , \quad u^\alpha = D^\alpha \overline{\Upsilon}| \quad , \quad S = D^2 \Upsilon| \quad ,$$

$$\lambda^\alpha = D^\alpha \Upsilon| - (\frac{4n+1}{3n}) D^\alpha \overline{\Upsilon}| \ , \ V^{\underline{a}} = D^\alpha \overline{D}^{\dot{\alpha}} \Upsilon| \ , \ \overline{\chi}^{\dot{\alpha}} = D^2 \overline{D}^{\dot{\alpha}} \Upsilon| \ . \tag{5.2.76}$$

The transformations analogous to (5.2.74) are

$$\delta u = -\sigma - (\frac{n+1}{3n+1})(\overline{\sigma} - i\partial_{\underline{m}}\xi^{\underline{m}}) \quad ,$$

$$\delta u^\alpha = \eta^\alpha \quad ,$$

$$\delta S = i(\frac{n+1}{3n+1})\partial^{\underline{m}} L^2{}_{\underline{m}} \quad ,$$

$$\delta \lambda^\alpha = -i\partial^{\alpha\dot{\alpha}}\overline{\epsilon}_{\dot{\alpha}} + i(\frac{n+1}{3n+1})\partial_{\beta\dot{\beta}} L^{1\,\alpha\beta\dot{\beta}} \quad ,$$

$$\delta V^{\underline{a}} = -i\,\frac{1}{2}\,\partial^{\underline{a}}\sigma - i\partial^\alpha{}_{\dot{\beta}}\overline{\omega}^{\dot{\alpha}\dot{\beta}} \quad ,$$

$$\delta \overline{\chi}^{\dot{\alpha}} = \Box \overline{\epsilon}^{\dot{\alpha}} \quad . \tag{5.2.77}$$

As before we can scale $u$ to 1 and gauge $u_\alpha$ to zero. The remaining components $\chi^\alpha$ , $V^{\underline{a}}$ , $\lambda^\alpha$ are the additional auxiliary fields of nonminimal supergravity.

Once again we note that the case $n = 0$ is different; the transformation $\delta u$ is independent of $Im\,\sigma$ (the axial rotations) so that this gauge invariance *survives* in the Wess-Zumino gauge for $\Lambda$. Since $\Upsilon = \tilde{\Upsilon}(1 \cdot e^{-\overleftarrow{H}})$, we have $S = \overline{\chi}^{\dot{\alpha}} = \overline{\lambda}^{\dot{\alpha}} = 0$ and $V^{\underline{a}} = \partial_{\underline{b}} T^{[\underline{ab}]}$, where $T^{[\underline{ab}]}$ is a real antisymmetric tensor gauge field.

### e. Summary

We summarize here some of the quantities that we have constructed so far:

$$\hat{E}_{\dot{\mu}} = \overline{D}_{\dot{\mu}} \quad , \quad \hat{E}_\mu = e^{-H} D_\mu e^H \quad , \quad H = H^M i D_M \quad ,$$

$$\check{E}_{\dot{\mu}} = \Psi \hat{E}_{\dot{\mu}} \quad , \quad \check{E}_\mu = \tilde{\overline{\Psi}} \hat{E}_\mu \quad , \quad \tilde{\overline{\Psi}} = e^{-H}\overline{\Psi} e^H \quad ,$$

$$\check{E}_{\mu\dot{\mu}} = -i\{\check{E}_\mu , \check{E}_{\dot{\mu}}\} \quad , \quad [\check{E}_M , \check{E}_N\} = \check{C}_{MN}{}^R \check{E}_R \quad . \tag{5.2.78a}$$

$$E_{\dot{\alpha}} = N_{\dot{\alpha}}{}^{\dot{\mu}} \check{E}_{\dot{\mu}} \quad , \qquad E_{\alpha} = \tilde{N}_{\alpha}{}^{\mu} \check{E}_{\mu} \ ,$$

$$E_{\alpha\dot{\alpha}} = N_{\alpha\dot{\alpha}}{}^{\mu\dot{\mu}} ( \check{E}_{\mu\dot{\mu}} + i \frac{1}{2} \check{C}_{\dot{\mu},(\mu\dot{\rho}}{}^{\nu)\dot{\rho}} \check{E}_{\nu} + i \frac{1}{2} \check{C}_{\mu,\rho(\dot{\mu}}{}^{\rho\dot{\nu})} \check{E}_{\dot{\nu}} ) \ ,$$

$$N_{\alpha\dot{\alpha}}{}^{\mu\dot{\mu}} = \tilde{N}_{\alpha}{}^{\mu} N_{\dot{\alpha}}{}^{\dot{\mu}} \quad , \qquad det(N_{\dot{\alpha}}{}^{\dot{\mu}}) = 1 \ . \tag{5.2.78b}$$

The factor $N_{\alpha}{}^{\mu}$ is equal to $\delta_{\alpha}{}^{\mu}$ in a suitable Lorentz frame. The superscale compensator $\Psi$ takes the form

$$n = -\frac{1}{3}: \qquad \Psi = \phi^{-1} \tilde{\phi}^{\frac{1}{2}} (1 \cdot e^{-\overleftarrow{H}})^{\frac{1}{6}} \hat{E}^{-\frac{1}{6}} \ ,$$

$$n \neq -\frac{1}{3}\,, 0: \qquad \Psi = [\, \Upsilon^{n-1} \tilde{\Upsilon}^{n+1} \,]^{-(\frac{3n+1}{8n})} [(1 \cdot e^{-\overleftarrow{H}})^{n+1} \hat{E}^{2n} \,]^{-(\frac{n+1}{8n})} \ . \tag{5.2.78c}$$

The tilde quantities are chiral representation hermitian conjugates defined by analogy with the way $\tilde{\bar{\Psi}}$ is defined from $\Psi$. The Lorentz connection superfields $\Phi_{A\gamma}{}^{\delta}$ and $\Phi_{A\dot{\gamma}}{}^{\dot{\delta}}$ take simple forms when expressed as functions of the anholonomy coefficients defined by $[\, E_A \,, E_B \} = C_{AB}{}^{C} E_C$. They are given by

$$\Phi_{\alpha\beta}{}^{\gamma} = -\frac{1}{4} [\, C_{\alpha,(\beta\dot{\beta}}{}^{\gamma)\dot{\beta}} - \delta_{\alpha}{}^{(\gamma} C_{\beta),\underline{d}}{}^{\underline{d}} \,] \ ,$$

$$\Phi_{\alpha\dot{\beta}}{}^{\dot{\gamma}} = -\frac{1}{4} C_{\alpha,\beta(\dot{\beta}}{}^{\beta\dot{\gamma})} \ ,$$

$$\Phi_{\underline{a}\beta}{}^{\gamma} = -\, i [\, E_{\alpha} \Phi_{\dot{\alpha}\beta}{}^{\gamma} + E_{\dot{\alpha}} \Phi_{\alpha\beta}{}^{\gamma} - C_{\alpha,\dot{\alpha}}{}^{\delta} \Phi_{\delta\beta}{}^{\gamma} - C_{\alpha,\dot{\alpha}}{}^{\dot{\delta}} \Phi_{\dot{\delta}\beta}{}^{\gamma}$$

$$+ \Phi_{\alpha\beta}{}^{\delta} \Phi_{\dot{\alpha}\delta}{}^{\gamma} + \Phi_{\dot{\alpha}\beta}{}^{\delta} \Phi_{\alpha\delta}{}^{\gamma} \,] \ , \tag{5.2.78d}$$

and $\Phi_{\dot{\alpha}\dot{\beta}}{}^{\dot{\gamma}}$ , $\Phi_{\dot{\alpha}\beta}{}^{\gamma}$ , $\Phi_{\underline{a}\dot{\beta}}{}^{\dot{\gamma}}$ are obtained by chiral-representation complex conjugation (i.e., the tilde operation).

### f. Torsions and curvatures

From the covariant derivatives $\nabla_A = E_A + \Phi_A(M)$ we define torsions and curvatures

$$[\nabla_A \,, \nabla_B \} = T_{AB}{}^{C} \nabla_C + R_{AB}(M) \ . \tag{5.2.79}$$

Using the explicit form, we find that the torsions and curvatures identically satisfy the following constraints:

$$T_{\alpha,\beta}{}^{\dot\gamma} = T_{\alpha,\dot\beta}{}^{\underline{c}} - i\delta_\alpha{}^\gamma \delta_{\dot\beta}{}^{\dot\gamma} = T_{\alpha,\beta}{}^{\underline{c}} = 0 \quad ,$$

$$R_{\alpha,\dot\beta\,\gamma\delta} = R_{\alpha,\beta\,\dot\gamma\dot\delta} = T_{\alpha,(\underline{b}}{}^{\underline{c})} - \frac{1}{2}\delta_{\underline{b}}{}^{\underline{c}} T_{\alpha,\underline{d}}{}^{\underline{d}} = 0 \quad . \tag{5.2.80a}$$

We also find further constraints whose form depends on the value of $n$. These are

$$n = -\frac{1}{3}: \qquad T_{\alpha,\underline{b}}{}^{\underline{c}} = 0 \quad ,$$

$$n \neq -\frac{1}{3}, 0: \qquad T_{\alpha,\underline{b}}{}^{\underline{c}} = \frac{1}{2}\delta_{\dot\beta}{}^{\dot\gamma}\delta_\alpha{}^\gamma T_{\beta,\underline{d}}{}^{\underline{d}} \quad ,$$

$$R = -\frac{n}{3n+1}(\overline{\nabla}^{\dot\alpha} + \frac{n-1}{2(3n+1)}\overline{T}^{\dot\alpha})\overline{T}_{\dot\alpha} \quad , \tag{5.2.80b}$$

where $R = i\frac{1}{4}T_{\alpha\dot\alpha,}{}^{\dot\alpha,\alpha}$ and $T_\alpha = T_{\alpha,\underline{b}}{}^{\underline{b}}$. We refer to these as *conformal breaking* constraints. In a treatment that does not use compensators, these constraints have to be imposed directly, to break conformal supergravity down to Poincaré supergravity. In the compensator approach they arise naturally when a gauge choice is made to break the conformal invariance; see sec. 5.3.b.6,7.

Another way to express the constraints on the torsions and curvatures is to write the graded commutator of the covariant derivatives. For the minimal theory ($n = -\frac{1}{3}$) we find

$$\{\nabla_\alpha , \nabla_\beta\} = -2\,\overline{R}M_{\alpha\beta} \ ,$$

$$\{\nabla_\alpha , \overline{\nabla}_{\dot\beta}\} = i\,\nabla_{\alpha\dot\beta} \ ,$$

$$[\nabla_\alpha , \nabla_{\underline{b}}] = -\,iC_{\alpha\beta}[\,\overline{R}\overline{\nabla}_{\dot\beta} - G^\gamma{}_{\dot\beta}\nabla_\gamma\,]$$

$$+\,iC_{\alpha\beta}[\,\overline{W}_{\dot\beta\dot\gamma}{}^{\dot\delta}\overline{M}_{\dot\delta}{}^{\dot\gamma} - (\nabla^\gamma G_{\delta\dot\beta})M_\gamma{}^\delta\,] - i(\overline{\nabla}_{\dot\beta}\overline{R})M_{\alpha\beta} \ ,$$

$$[\nabla_{\underline{a}} , \nabla_{\underline{b}}] = \big\{[\,C_{\dot\alpha\dot\beta}W_{\alpha\beta}{}^\gamma + C_{\alpha\beta}(\overline{\nabla}_{\dot\alpha}G^\gamma{}_{\dot\beta}) - C_{\dot\alpha\dot\beta}(\nabla_\alpha R)\delta_\beta{}^\gamma\,]\nabla_\gamma + iC_{\alpha\beta}G^\gamma{}_{\dot\beta}\nabla_{\gamma\dot\alpha}$$

$$+ [\, C_{\dot{\alpha}\dot{\beta}}(\nabla_{\alpha}W_{\beta\gamma}{}^{\delta} + (\overline{\nabla}^{2}\overline{R} + 2R\overline{R})C_{\gamma\beta}\delta_{\alpha}{}^{\delta}\,)$$

$$- C_{\alpha\beta}(\overline{\nabla}_{\dot{\alpha}}\nabla^{\delta}G_{\gamma\dot{\beta}})\,]M_{\delta}{}^{\gamma}\} + h.\,c. \;\; , \qquad (5.2.81)$$

where $W_{\alpha\beta\gamma}$, $G_{\underline{a}}$, and $R$ are independent Lorentz irreducible field strengths *defined* by these equations.

For nonminimal theories ($n \neq -\frac{1}{3}\,, 0$) the graded commutators take the forms

$$\{\nabla_{\alpha}\,, \nabla_{\beta}\} = \frac{1}{2}\,T_{(\alpha}\nabla_{\beta)} - 2\overline{R}M_{\alpha\beta} \;\; ,$$

$$\{\nabla_{\alpha}\,, \overline{\nabla}_{\dot{\beta}}\} = i\nabla_{\alpha\dot{\beta}} \;\; ,$$

$$[\nabla_{\alpha}\,, \nabla_{\underline{b}}] = \frac{1}{2}\,T_{\beta}\nabla_{\alpha\dot{\beta}} - iC_{\alpha\beta}[\,\overline{R} + \frac{1}{4}\,\nabla^{\gamma}T_{\gamma}\,]\overline{\nabla}_{\dot{\beta}}$$

$$+ i[\,C_{\alpha\beta}G^{\gamma}{}_{\dot{\beta}} - \frac{1}{2}\,C_{\alpha\beta}((\nabla^{\gamma} + \frac{1}{2}\,T^{\gamma})\overline{T}_{\dot{\beta}}) + \frac{1}{2}\,(\overline{\nabla}_{\dot{\beta}}T_{\beta})\delta_{\alpha}{}^{\gamma}\,]\nabla_{\gamma}$$

$$- i[\,C_{\alpha\beta}(\nabla^{\gamma}G_{\delta\dot{\beta}})M_{\gamma}{}^{\delta} + ((\overline{\nabla}_{\dot{\beta}} - \overline{T}_{\dot{\beta}})\overline{R})M_{\alpha\beta}\,]$$

$$+ iC_{\alpha\beta}[\,\overline{W}_{\dot{\beta}\dot{\gamma}}{}^{\dot{\delta}}\overline{M}_{\dot{\delta}}{}^{\dot{\gamma}} + i\,\frac{1}{3}\,\overline{W}_{\dot{\gamma}}\overline{M}_{\dot{\beta}}{}^{\dot{\gamma}}\,] \;\; , \qquad (5.2.82)$$

where $W_{\alpha\beta\gamma}$, $G_{\underline{a}}$, and $T_{\alpha}$ are the independent tensors. In these equations, $R$ and $W_{\alpha}$ are defined in terms of $T_{\alpha}$ . The quantity $R$ was defined in (5.2.80b) and $W_{\alpha}$ is given by

$$W_{\alpha} = i\,[\,\frac{1}{2}\,\overline{\nabla}^{\dot{\beta}}(\overline{\nabla}_{\dot{\beta}} + \frac{1}{2}\,\overline{T}_{\dot{\beta}}) + R\,]T_{\alpha} \qquad (5.2.83)$$

The expression for the commutator of two vectorial covariant derivatives in the nonminimal theory can be calculated from the Bianchi identity;

$$[\nabla_{\underline{a}}\,, \{\nabla_{\beta}\,, \overline{\nabla}_{\dot{\beta}}\}\,] = \{\overline{\nabla}_{\dot{\beta}}\,, [\nabla_{\underline{a}}\,, \nabla_{\beta}]\,\} + \{\nabla_{\beta}\,, [\nabla_{\underline{a}}\,, \overline{\nabla}_{\dot{\beta}}]\,\}\;. \qquad (5.2.84)$$

## 5.3. Covariant approach to supergravity

### a. Choice of constraints

In the previous section we showed how to start with unconstrained prepotentials and construct out of them a covariant superfield formulation of supergravity. Here we do the reverse: Starting with a manifestly covariant but highly reducible representation of supersymmetry, we impose constraints on the geometry and solve them in terms of the prepotentials. Prepotentials are essential for superfield quantization, whereas manifestly covariant formulations make coupling to matter straightforward and allow us to develop an efficient and powerful background field method for the quantum theory.

#### a.1. Compensators

As discussed in sec. 3.10, it is often useful to introduce (additional) local symmetries realized through compensators. As discussed in sec. 5.1, in gravitational theories there are two types of compensators: (1) density compensators, which transform noncovariantly under the full local symmetry group, and thus appear in covariant quantities only in combination with other fields; (2) tensor compensators, which transform covariantly, and thus allow the realization of symmetries that may not be invariances of the entire theory. Density compensators allow the linear realization of symmetries that would otherwise be realized nonlinearly (as, for example, in nonlinear $\sigma$ models, or Lorentz invariance in gravity with spinors). When such symmetries are global symmetries of some theory (at least on-shell), the (super) spacetime derivative of the density compensator can appear as a gauge connection. (If some field transforms as $\delta\sigma = \lambda$ then it is easy to construct a connection as $\partial\sigma$ since $\delta\partial\sigma = \partial\lambda$.)

On the other hand, if the symmetries one wants to realize linearly and locally are *not* even global symmetries of the entire theory, it is necessary to introduce tensor compensators to "cancel" arbitrary gauge transformations of this type in terms that are not invariant. This phenomenon was illustrated in sec. 3.10.b in the discussion of the $CP(1)$ model, and in (5.1.35), where the compensator $\phi$ permits a generalization of the Einstein-Hilbert action to an action with an additional local scale invariance. (This does *not* imply any new physics. The gauge invariance with respect to scale transformations must be fixed just as any gauge invariance and the most convenient choice is $\phi = 1$.)

In gravitational theories, some symmetries need *both* types of compensators. This is because the symmetries are realized twice. For example, in ordinary gravity, the density compensator $e^{-1}$ transforms as $\delta e^{-1} = -\partial_{\underline{m}}\lambda^{\underline{m}} - 4\zeta$, where $\partial_{\underline{m}}\lambda^{\underline{m}}$ represents the local scale transformation part of the general coordinate transformation parametrized by $\lambda^{\underline{m}}$ , and $\zeta$ is an independent local tangent space parameter. The quantity $e^{-1}$ is thus a *gauge* field for scale transformations in $\lambda^{\underline{m}}$, but also a *compensator* for the transformations parametrized by $\zeta$. It allows the full $\lambda^{\underline{m}}$ invariance to be realized linearly; it also allows local scale transformations to be realized linearly via $\zeta$. However, most gravitational theories are not locally (tangent space) scale invariant: Thus, if we still want to represent local scale transformations, we must introduce a tensor compensator $\phi$ , $\delta\phi = \zeta$ , to cancel the $\zeta$ transformation of noninvariant terms in an action. To summarize: (1) $e^{-1}$ is a *density* compensator for local scale transformations, allowing them to be realized *linearly* in locally scale-invariant theories; while (2) $\phi$ is a *tensor* compensator for local scale transformations, allowing them to be realized linearly (when $e^{-1}$ is also present) in *non-invariant* theories. Note that $e^{-1}$ transforms under both $\partial_{\underline{m}}\lambda^{\underline{m}}$ and $\zeta$ , whereas $\phi$ transforms only under $\zeta$ (in the linearized transformation). There is also the combination $e^{-1}\phi^4$ , transforming only under $\partial_{\underline{m}}\lambda^{\underline{m}}$ as $\delta e^{-1}\phi^4 = -\partial_{\underline{m}}\lambda^{\underline{m}}$, which is useful in constructing invariant actions. It should be noted that $e^{-1}\phi^4$ is, in a sense, a "field strength" for the $\zeta$ gauge transformations. Another such "field strength" is the integrand of the expression in (5.1.35).

We find a similar situation in supergravity. There we introduce not only local (real) scale invariance, but also local (chiral) $U(1)$ invariance (as a generalization of the global R-invariance of pure supergravity) to simplify the analysis of constraints and Bianchi identities as much as possible, and we include its generator in the covariant derivatives. (In extended supergravity, the corresponding extra invariance is $U(N)$: i.e., the largest internal symmetry of the on-shell theory. See secs. 3.2 and 3.12.) In the present application, the result of such an approach is that the torsions and curvatures contain fewer tensors than they would without the enlarged tangent space. (The missing tensors reappear as field strengths of a *tensor* compensator.) These latter tensors are generally the tensors of lowest dimension, so their elimination from the torsions and curvatures allows great simplification in the analysis of the Bianchi identities, as discussed below (sec. 5.4), and simplifies our analysis of constraints.

Since neither $U(1)$ (R-invariance) nor scale invariance is a symmetry of general theories of supergravity + matter, we introduce a complex scalar *tensor* compensator to compensate for both (except for $n=0$, where R-invariance is maintained, so the compensator is real). In analogy to gravity, there is also a complex scalar *density* compensator for these transformations: It is the density $\Psi$ of the previous section, now *unconstrained*, which will appear when solving the constraints. The quantity $\Psi$ is the direct analog of $\mathrm{e}^{-1}$ of gravity, and the tensor compensator is the analog of gravity's component field compensator $\phi$. We will furthermore find a special significance for the analog of gravity's combination $\mathrm{e}^{-1}\phi^4$: It is the superspace density compensator $\phi$ or $\Upsilon$, which satisfies a simple (noncovariant) constraint. The tensor compensator satisfies the direct covariantization of this constraint. The analog of gravity's $\lambda^{\underline{m}}$ is $\Lambda^M$, and that of $\zeta$ is the scale parameter $L$ and $U(1)$ parameter $K_5$. A feature of supergravity not appearing in gravity is that of a global symmetry, namely $U(1)$, with both density and tensor compensators, whose density compensator ($\Psi$) is used to construct a $U(1)$-gauge connection that trivially gauges the symmetry in superspace (as in nonlinear $\sigma$ models). (However, as in gravity, it is not useful to introduce gauge connections for scale transformations, since global scale transformations, unlike R-symmetry, are not an unbroken invariance of the classical theory (even without matter)).

After completing our analysis of the $U(1)$-covariant derivatives and tensor compensators, we will obtain the $(n \neq 0)$ $U(1)$-noncovariant derivatives of the previous section, which are more convenient for some applications. This is achieved by first gauging the tensor compensator to 1, which expresses $\Psi$ in terms of $H$ and $\phi$ or $\Upsilon$, and then by dropping the $U(1)$ connection, should it not disappear automatically.

We begin with the covariant derivatives (cf. (5.2.20))

$$\nabla_A = E_A + \Phi_A(M) - i\Gamma_A Y \quad ,$$

$$[\nabla_A , \nabla_B\} = T_{AB}{}^C \nabla_C + R_{AB}(M) - iF_{AB} Y \quad , \tag{5.3.1}$$

where $Y = Y^\dagger$ is the $U(1)$ generator, whose tangent space action can be summarized by $[Y, \nabla_A] = \frac{1}{2} w(A) \nabla_A$, or explicitly:

$$[Y , \nabla_\alpha] = -\frac{1}{2} \nabla_\alpha \quad , \quad [Y , \nabla_{\dot\alpha}] = \frac{1}{2} \nabla_{\dot\alpha} \quad , \quad [Y , \nabla_{\underline{a}}] = 0 \ . \tag{5.3.2}$$

The covariant derivatives transform as

$$\nabla'_A = e^{iK} \nabla_A e^{-iK} \quad , \quad K = K^M iD_M + (K_\alpha{}^\beta iM_\beta{}^\alpha + \bar{K}_{\dot\alpha}{}^{\dot\beta} i\bar{M}_{\dot\beta}{}^{\dot\alpha}) + K_5 Y \ . \qquad (5.3.3)$$

## a.2. Conformal supergravity constraints

The covariant derivatives define a realization of local supersymmetry. However, it is highly reducible and contains much more than the supergravity multiplet. Therefore, by analogy with Yang-Mills, we impose covariant constraints on these derivatives to eliminate unwanted representations. The supergravity constraints can be expressed in the simple form

$$\nabla_{\alpha\dot\alpha} = -i\{\nabla_\alpha , \overline{\nabla}_{\dot\alpha}\} \ , \qquad (5.3.4a)$$

$$T_{\alpha\beta}{}^\gamma = T_{\alpha\underline{b}}{}^{\underline{b}} = T_{\alpha,\beta(\dot\beta}{}^{\beta\dot\gamma)} = 0 \ , \qquad (5.3.4b)$$

$$\{\nabla_\alpha , \nabla_\beta\}\overline{\chi} = 0 \ \ when \ \ \nabla_\alpha \overline{\chi} = 0 \ ; \qquad (5.3.4c)$$

(and their hermitian conjugates) or, in terms of the field strengths,

$$T_{\alpha\dot\beta}{}^{\underline{c}} = i\delta_\alpha{}^\gamma \delta_{\dot\beta}{}^{\dot\gamma} \ , \ T_{\alpha\dot\beta}{}^\gamma = R_{\alpha\dot\beta\gamma}{}^\delta = F_{\alpha\dot\beta} = 0 \ , \qquad (5.3.5a)$$

$$T_{\alpha\beta}{}^\gamma = T_{\alpha\underline{b}}{}^{\underline{b}} = T_{\alpha,\beta(\dot\beta}{}^{\beta\dot\gamma)} = 0 \ , \qquad (5.3.5b)$$

$$T_{\alpha\beta}{}^{\underline{c}} = T_{\alpha\beta}{}^{\dot\gamma} = 0 \ . \qquad (5.3.5c)$$

We have divided the constraints into three categories: (a) conventional constraints that determine the vector Lorentz component of the covariant derivative, $\nabla_{\underline{a}}$, in terms of the spinor components $\nabla_\alpha$, $\nabla_{\dot\alpha}$; (b) conventional constraints that determine the spinor connections $\Phi_\alpha$ and $\Gamma_\alpha$ (and their hermitian conjugates) in terms of the spinor vielbein $E_\alpha$; and (c) representation-preserving constraints that are needed for consistency with the definition of chiral superfields in curved superspace. As for super-Yang-Mills and ordinary gravity, conventional constraints can be interpreted as either setting certain field strengths to zero, or as eliminating them from the theory by field redefinitions. The first set of conventional constraints is of the same form as for super-Yang-Mills theory, while the second is analogous to the constraints of ordinary gravity. The representation-preserving constraints are also of the same form as for super-Yang-Mills theory. Although we have only required the existence of chiral scalars (i.e., scalar multiplets), this set of constraints is also sufficient to allow the existence of chiral undotted spinors (for example, the field strengths of super-Yang-Mills), with arbitrary $U(1)$ charge. (The second type of constraint already determines the spinorial Lorentz and $U(1)$ connections.)

The constraints actually have a larger invariance group than that implied by (5.3.3): in addition to being invariant under the transformations generated by (5.3.3), they are invariant under local *superscale* transformations. In the compensator approach, we use constraints that determine only the *conformal* part of the Poincaré supergravity multiplet. The rest of the multiplet (the superscale part) is contained in the compensator itself, and therefore the particular form of the Poincaré supergravity multiplet depends on the choice of compensator multiplet.

To discover the explicit form of the additional invariance, we first note that the infinitesimal variation of the spinorial vielbein under scale transformations must be of the form

$$\delta_L E_\alpha = \frac{1}{2} L E_\alpha \tag{5.3.6}$$

where $L$ is a real unconstrained superfield which parametrizes the scale transformation (see (5.3.4c)). Next, to find the superscale variation of $\nabla_A$, we use (5.3.6), vary $\Phi(M)$, $\Gamma$, and $E_{\underline{a}}$ arbitrarily, and demand that (5.3.5) is satisfied. This determines the remaining variations. The results can be summarized as

$$\delta_L \nabla_\alpha = \frac{1}{2} L \nabla_\alpha + 2(\nabla_\beta L) M_\alpha{}^\beta + 3(\nabla_\alpha L) Y \ , \tag{5.3.7a}$$

$$\delta_L \nabla_{\alpha\dot\alpha} = L \nabla_{\alpha\dot\alpha} - 2i(\overline\nabla_{\dot\alpha} L)\, \nabla_\alpha - 2i(\nabla_\alpha L)\, \overline\nabla_{\dot\alpha}$$

$$- 2i(\overline\nabla_{\dot\alpha} \nabla_\beta L) M_\alpha{}^\beta - i2(\nabla_\alpha \overline\nabla_{\dot\beta} L) \bar M_{\dot\alpha}{}^{\dot\beta} + 3i([\nabla_\alpha , \overline\nabla_{\dot\alpha}] L) Y \ , \tag{5.3.7b}$$

and consequently

$$\delta_L E^{-1} = -\, 2 L\, E^{-1} \ . \tag{5.3.8}$$

The gauge symmetries of the theory, (5.3.3,7), and the constraints (5.3.4) are sufficient to reduce the covariant derivatives so that they describe an *irreducible* multiplet: conformal supergravity. To see this, we study the scaling properties of the remaining field strengths. These are not all independent; Using the Bianchi identities, as we show in sec. 5.4, all nontrivial field strengths can be expressed in terms of three tensors $R$, $G_{\underline{a}}$, and $W_{\alpha\beta\gamma}$. For convenience, we also introduce the (dependent) $U(1)$ field strength $W_\alpha$. These objects can be defined by

$$R = \frac{1}{6} R_{\dot{\alpha}\dot{\beta}}{}^{\dot{\alpha}\dot{\beta}} = \frac{1}{4} iT_{\alpha\dot{\alpha}}{}^{\dot{\alpha},\alpha} \quad ,$$

$$G_{\underline{a}} = iT_{\underline{a}\beta}{}^{\beta} \quad ,$$

$$W_{\alpha\beta\gamma} = \frac{1}{12} iR^{\dot{\alpha}}{}_{(\alpha\dot{\alpha},\beta\gamma)} = -\frac{1}{12} T_{(\alpha}{}^{\dot{\alpha},}{}_{\beta\dot{\alpha},\gamma)} \quad ,$$

$$W_\alpha = \frac{1}{2} iF^{\dot{\alpha},}{}_{\alpha\dot{\alpha}} \quad . \tag{5.3.9}$$

From (5.3.2,7) we have the $U(1)$ and superscale transformations of $\nabla_\alpha$ (and hence $\overline{\nabla}_{\dot{\alpha}}$ by hermitian conjugation), and $\nabla_{\underline{a}}$ . We can then determine the transformations of these field strengths by evaluating commutators. The result is:

$$[Y\,, R] = R \quad , \quad \delta_L R = LR - 2\overline{\nabla}^2 L \quad ;$$

$$[Y\,, G_{\underline{a}}] = 0 \quad , \quad \delta_L G_{\underline{a}} = LG_{\underline{a}} - 2[\overline{\nabla}_{\dot{\alpha}}\,, \nabla_\alpha]L \quad ;$$

$$[Y\,, W_{\alpha\beta\gamma}] = \frac{1}{2} W_{\alpha\beta\gamma} \quad , \quad \delta_L W_{\alpha\beta\gamma} = \frac{3}{2} LW_{\alpha\beta\gamma} \quad ;$$

$$[Y\,, W_\alpha] = \frac{1}{2} W_\alpha \quad , \quad \delta_L W_\alpha = \frac{3}{2} LW_\alpha + 6i(\overline{\nabla}^2 + R)\nabla_\alpha L \quad . \tag{5.3.10}$$

Thus the superscale and $U(1)$ transformations can be used to gauge away parts of these tensors, leaving only the field $W_{\alpha\beta\gamma}$ of conformal supergravity. At the linearized level, this contains the pure superspin $\frac{3}{2}$ projection of $H^{\underline{m}}$ discussed in sec. 5.2.a.1.

### a.3. Contortion

There is nothing unique about the set of conventional constraints we use. Any set that allows us to express the vector derivative and spinor connections in terms of the spinor vielbeins $E_\alpha , \bar{E}_{\dot{\alpha}}$ is equally suitable. For example, we could use $T_{\underline{ab}}{}^{\underline{c}} = 0$ instead of $R_{\alpha\dot{\beta}\underline{c}}{}^{\underline{d}} = 0$ to determine $\Phi_{\underline{ab}}{}^{\underline{c}}$. This would give a $\Phi'_{\underline{ab}}{}^{\underline{c}}$ whose corresponding $\nabla'_{\underline{a}}$ is an equally good covariant derivative. The difference $\Phi - \Phi'$ is a tensor (the *contortion* tensor). Adding contortions to connections does not change the physics and simply amounts to a redefinition of minimal coupling. Indeed, for most familiar models the connections do not enter at all: For the scalar multiplet the Lagrangian $\bar{\eta}\eta$ and the chirality constraint $\bar{\nabla}_{\dot{\alpha}}\eta = 0$ are independent of the connection. The field strengths of super-Yang-Mills theory are

$$F_{AB} = \nabla_{[A}\Gamma_{B)} + \Gamma_{[A}\Gamma_{B)} - T_{AB}{}^C\Gamma_C$$

$$= E_{[A}\Gamma_{B)} + \Gamma_{[A}\Gamma_{B)} - C_{AB}{}^C\Gamma_C \qquad (5.3.11)$$

and are also independent of the supergravity connections. Finally, the supergravity Lagrangian (for $n \neq 0$) is $E^{-1}$, also independent of the connections.

Furthermore, any other set of constraints that determines $E_{\underline{a}}$ is correct: We can always redefine $E_{\underline{a}}{}^M$ by writing

$$E'_{\underline{a}}{}^M = E_{\underline{a}}{}^M + g_{\underline{a}}{}^\gamma E_\gamma{}^M + g_{\underline{a}}{}^{\dot{\gamma}} E_{\dot{\gamma}}{}^M \quad , \qquad (5.3.12)$$

where $g_{\underline{a}}{}^\gamma$ is a covariant object constructed out of the field strengths of $\nabla_A$. Therefore, in superspace, in addition to the contortion tensor for the Lorentz connection, we have a contortion that changes $E_{\underline{a}}{}^M$. However, this does not affect the physics as, once again, it amounts simply to a redefinition of minimal coupling.

There is another ambiguity in the choice of constraints, which, however, leads to no modification of the theory at all: Since the Bianchi identities relate various field strengths, there are many ways to express any particular constraint. For example, since all curvatures and $U(1)$ field strengths can be expressed in terms of torsions (see sec. 5.4), any constraint on a curvature or $U(1)$ field strength can be expressed in terms of torsions. Which form is chosen is purely a matter of convenience.

### a.4. Poincaré supergravity constraints

The superspin $\frac{3}{2}$ superconformal multiplet with field strength $W_{\alpha\beta\gamma}$ is not sufficient to describe off-shell Poincaré supergravity. We must include lower superspin superfields to obtain a consistent action. We can do this in two ways: By introducing extra conformal representations as compensators, or by directly restricting the gauge group so that some lower superspin conformal representations contained in $H^M$ cannot be gauged away. Such restrictions on the gauge group are introduced by imposing constraints that are not invariant under the full group. These constraints appear naturally when we use the full superconformal transformations to gauge the compensators away and require that the remaining transformations preserve the resulting superconformal gauge.

There are three types of tensor compensators that can be coupled to conformal supergravity and can be used to reduce it to Poincaré supergravity. The possible compensators are restricted by the requirement that they must have dimensionless scalar field strengths to compensate for $L$ of (5.3.6,7,8,10). (Thus, the compensator field strength $X$ has the usual linearized compensator transformation $\delta X = L$. $X|$ is then a scalar with action (5.1.35). The remaining type of conformal matter multiplet, the vector multiplet, cannot be used as a compensator because its only scalar field strength $\nabla^\alpha W_\alpha$ has the wrong dimension and its prepotential is inert under superscale transformations.) They are parametrized by the complex number $n$: (1) the scalar multiplet $\Phi$ ($n = -\frac{1}{3}$), (2) the nonminimal scalar multiplet $\Sigma$ (any $n$ except 0 or $-\frac{1}{3}$), and (3) the tensor multiplet $G$ ($n = 0$). These multiplets can be defined by constraints and can be expressed explicitly in terms of unconstrained superfields (prepotentials):

$$\overline{\nabla}_{\dot\alpha}\Phi = 0 \ , \qquad \Phi = (\overline{\nabla}^2 + R)\Xi \ ; \tag{5.3.13a}$$

$$(\overline{\nabla}^2 + R)\Sigma = 0 \ , \qquad \Sigma = \overline{\nabla}^{\dot\alpha}\Xi_{\dot\alpha} \ ; \tag{5.3.13b}$$

$$(\overline{\nabla}^2 + R)G = 0 \ , \qquad G = \overline{G} = \frac{1}{2}\nabla_\alpha(\overline{\nabla}^2 + R)\Xi^\alpha + h.c. \ ; \tag{5.3.13c}$$

where $R$ is a field strength (see (5.3.9) and sec. 5.4) and $\overline{\nabla}^2 + R$ gives a chiral superfield when acting on a superfield without dotted spinor indices (see below). The $U(1)$ and superscale transformations for which they compensate are

$$[Y\,,\Phi] = \frac{1}{3}\Phi \ \ , \ \ \delta_L\Phi = L\Phi \ \ ; \tag{5.3.14a}$$

$$[Y\,,\Sigma] = -\frac{2n}{3n+1}\Sigma \ \ , \ \ \delta_L\Sigma = \frac{2}{3n+1}L\Sigma \ \ ; \tag{5.3.14b}$$

$$[Y\,,G] = 0 \ \ , \ \ \delta_L G = 2LG \ \ . \tag{5.3.14c}$$

In sec. 5.3.b.7 we will break the superconformal symmetry by fixing the compensators. In the resulting super-Poincaré theory (5.3.13) become additional, conformal breaking constraints on the covariant derivatives (i.e., on the torsions and curvatures).

The scale weight of $\Phi$ is arbitrary (since we could replace $\Phi$ by $\Phi^m$ and still satisfy (5.3.13a)). However, the ratio of the $U(1)$ charge to the scale weight for a chiral superfield is fixed. This can be seen by a simple argument. Consider an arbitrary chiral superfield $\chi$, $\overline{\nabla}_{\dot\alpha}\chi = 0$. We write its scale transformation in terms of the dilatational generator $\boldsymbol{d}$ (see (3.3.34)): $\delta_L\chi = L\,[\boldsymbol{d}\,,\chi]$. If we perform a scale variation of the defining condition for a chiral field, and use (5.3.7a), we find:

$$\begin{aligned} 0 &= (\delta_L\overline{\nabla}_{\dot\alpha})\chi + \overline{\nabla}_{\dot\alpha}(\delta_L\chi) \\ &= -3(\overline{\nabla}_{\dot\alpha}L)[\,Y\,,\chi\,] + \overline{\nabla}_{\dot\alpha}(L[\,\boldsymbol{d}\,,\chi\,]) \\ &= (\overline{\nabla}_{\dot\alpha}L)[\,-3Y + \boldsymbol{d}\,,\chi\,]\ , \end{aligned} \tag{5.3.15a}$$

and hence

$$0 = [\,-3Y + \boldsymbol{d}\,,\chi\,]\ . \tag{5.3.15b}$$

Thus the $U(1)$ charge and the dilatational charge *always* satisfy the rule $\boldsymbol{d} - 3Y = 0$ for chiral superfields. This is seen for $W_{\alpha\beta\gamma}$, $W_\alpha$, and $R$ in (5.3.10) and for $\Phi$ in (5.3.14a). (Actually for $R$ this is only clear if the transformation law is written in the form $\delta_L R = 3LR - 2(\overline{\nabla}^2 + R)L$.) The relation of the chiral charge to the dilatation charge for chiral superfields in $N = 1$ supersymmetry is a special case of the general relation noted in sec. 3.5.

In precisely the same manner, starting from the defining condition (5.3.13b,c) for a linear superfield, we can show that the condition $\boldsymbol{d} - 3Y = 2$ must be satisfied for *all* linear superfields. This is seen for $\Sigma$ and $G$ in (5.3.14b,c). In the case of $\Sigma$ in (5.3.14b) we have chosen a convenient parametrization for its scale weight. The tensor multiplet is neutral because its field strength is real ($Y = 0$), and thus the $n = 0$ theory (with the $G$ compensator) retains its local $U(1)$ invariance after superscale invariance is broken.

The relation of the dilatational charge and the chiral $U(1)$ charge for chiral and linear superfields implies that combined $L$ and $K_5$ transformations on arbitrary chiral and linear superfields, $\chi$ and $\Xi$ respectively, take the forms

$$\delta\chi = L[\boldsymbol{d}\,, \chi] + iK_5[Y\,, \chi]\ ,$$

$$= d\,(L + i\,\frac{1}{3}\,K_5)\chi\ \ , \tag{5.3.16a}$$

$$\delta\Xi = L[\boldsymbol{d}\,, \Xi] + iK_5[Y\,, \Xi]\ ,$$

$$= [d'L + i\,\frac{1}{3}\,(d' - 2)K_5]\Xi\ . \tag{5.3.16b}$$

The quantities $d$ and $d'$ are the scale weights of the superfields.

## b. Solution to constraints

### b.1. Conventional constraints

The first constraint in the form (5.3.4a) is already explicitly solved (as was the case for super-Yang-Mills).

We begin our analysis of the second constraint by extracting from (5.3.1) the explicit form of the torsion. In sections 5.1,2 we defined the coefficients of anholonomy $C_{AB}{}^C$ by

$$[E_A, E_B\} = C_{AB}{}^C E_C \tag{5.3.17}$$

They can be expressed explicitly in terms of the $E_A{}^M$ and their derivatives. We then have

$$T_{AB}{}^C = C_{AB}{}^C + \Phi_{[AB)}{}^C - i\,\frac{1}{2}\,w(C)\Gamma_{[A}\delta_{B)}{}^C\ \ ; \tag{5.3.18}$$

i.e.,

$$T_{\alpha\beta}{}^{\dot\gamma} = C_{\alpha\beta}{}^{\dot\gamma}\ \ , \qquad\qquad T_{\alpha\beta}{}^{\underline{c}} = C_{\alpha\beta}{}^{\underline{c}}\ \ ,$$

$$T_{\alpha\dot{\beta}}{}^{\underline{c}} = C_{\alpha\dot{\beta}}{}^{\underline{c}} \ , \qquad T_{\alpha\underline{b}}{}^{\dot{\gamma}} = C_{\alpha\underline{b}}{}^{\dot{\gamma}} \ ,$$

$$T_{\underline{ab}}{}^{\gamma} = C_{\underline{ab}}{}^{\dot{\gamma}} \ ,$$

$$T_{\alpha\beta}{}^{\gamma} = C_{\alpha\beta}{}^{\gamma} + \Phi_{(\alpha\beta)}{}^{\gamma} + \frac{1}{2} i\delta_{(\alpha}{}^{\gamma}\Gamma_{\beta)} \ ,$$

$$T_{\alpha\dot{\beta}}{}^{\dot{\gamma}} = C_{\alpha\dot{\beta}}{}^{\dot{\gamma}} + \Phi_{\alpha\dot{\beta}}{}^{\dot{\gamma}} - \frac{1}{2} i\delta_{\dot{\beta}}{}^{\dot{\gamma}}\Gamma_{\alpha} \ ,$$

$$T_{\underline{a}\beta}{}^{\gamma} = C_{\underline{a}\beta}{}^{\gamma} + \Phi_{\underline{a}\beta}{}^{\gamma} + \frac{1}{2} i\delta_{\beta}{}^{\gamma}\Gamma_{\underline{a}} \ ,$$

$$T_{\alpha\underline{b}}{}^{\underline{c}} = C_{\alpha\underline{b}}{}^{\underline{c}} + \Phi_{\alpha\beta}{}^{\gamma}\delta_{\dot{\beta}}{}^{\dot{\gamma}} + \Phi_{\alpha\dot{\beta}}{}^{\dot{\gamma}}\delta_{\beta}{}^{\gamma} \ ,$$

$$T_{\underline{ab}}{}^{\underline{c}} = C_{\underline{ab}}{}^{\underline{c}} + (\Phi_{\underline{a}\beta}{}^{\gamma}\delta_{\dot{\beta}}{}^{\dot{\gamma}} + h.c. - \underline{b} \longleftrightarrow \underline{a}) \ ; \tag{5.3.19}$$

as well as the complex conjugates.

By using these equations the first constraint of (5.3.4b) can be solved directly (noting that $\Phi_{A\beta}{}^{\gamma}$ is traceless in its last two indices):

$$T_{\alpha\beta}{}^{\gamma} = 0 \ \rightarrow \ \Phi_{\alpha\beta\gamma} = \frac{1}{2}(C_{\beta\gamma\alpha} - C_{\alpha(\beta\gamma)}) - \frac{1}{2} iC_{\alpha(\beta}\Gamma_{\gamma)} \ . \tag{5.3.20}$$

However, solving the last two equations of (5.3.4b) for $\Gamma_{\alpha}$ and $\Phi_{\alpha\dot{\beta}}{}^{\dot{\gamma}}$, respectively, is less straightforward, since $C_{\alpha\underline{b}}{}^{\underline{c}}$ itself depends on them through $\nabla_{\underline{a}}$. (On the other hand, (5.3.4a) introduces no dependence of $\nabla_{\underline{a}}$ on $\Phi_{\alpha\beta}{}^{\gamma}$.) To solve these constraints we introduce, as in sec. 5.2.a.3,

$$\check{E}_A = (\check{E}_{\alpha} , \check{E}_{\dot{\alpha}} , \check{E}_{\underline{a}}) \equiv (E_{\alpha} , E_{\dot{\alpha}} , -i\{E_{\alpha} , E_{\dot{\alpha}}\}) \ . \tag{5.3.21}$$

We define $\check{C}_{AB}{}^{C}$ by $[\check{E}_A , \check{E}_B\} = \check{C}_{AB}{}^{C}\check{E}_C$. We emphasize that, since $E_{\underline{a}}$ is still dependent on $\Phi_{\alpha\dot{\beta}}{}^{\dot{\gamma}}$ and $\Gamma_{\alpha}$, $C_{AB}{}^{C}$ is also. In contrast, $\check{C}_{AB}{}^{C}$ is completely determined in terms of $E_{\alpha}$ and $E_{\dot{\alpha}}$. We begin by expressing $E_{\underline{a}}$ in terms of $\check{E}_{\underline{a}}$ and (the as yet undetermined) $C_{\alpha\dot{\beta}}{}^{M}$:

$$E_{\underline{a}} = -i[\{E_{\alpha}, E_{\dot{\alpha}}\} + (\Phi_{\alpha\dot{\alpha}}{}^{\dot{\beta}} - \frac{1}{2} i\delta_{\dot{\alpha}}{}^{\dot{\beta}}\Gamma_{\alpha})E_{\dot{\beta}} + (\Phi_{\dot{\alpha}\alpha}{}^{\beta} + \frac{1}{2} i\Gamma_{\dot{\alpha}}\delta_{\alpha}{}^{\beta})E_{\beta}]$$

$$= \check{E}_{\alpha\dot{\alpha}} - i(\Phi_{\alpha\dot{\alpha}}{}^{\dot{\beta}} - \frac{1}{2}i\delta_{\dot{\alpha}}{}^{\dot{\beta}}\Gamma_\alpha)E_{\dot{\beta}} - i(\Phi_{\dot{\alpha}\alpha}{}^{\beta} + \frac{1}{2}i\Gamma_{\dot{\alpha}}\delta_\delta{}^\beta)E_\beta \ , \tag{5.3.22}$$

where we use (5.3.4a). Then computing the commutator $[E_\alpha, E_{\underline{b}}] = C_{\alpha\underline{b}}{}^D E_D$ we find

$$C_{\alpha\underline{b}}{}^{\underline{c}} = \check{C}_{\alpha\underline{b}}{}^{\underline{c}} - \delta_\alpha{}^\gamma(\Phi_{\beta\dot{\beta}}{}^{\dot{\gamma}} - \frac{1}{2}i\delta_{\dot{\beta}}{}^{\dot{\gamma}}\Gamma_\beta) + i(\Phi_{\dot{\beta}\beta}{}^{\delta} + \frac{1}{2}i\delta_\beta{}^\delta\Gamma_{\dot{\beta}})\check{C}_{\alpha\delta}{}^{\underline{c}} \ . \tag{5.3.23}$$

As we will see shortly, the next constraint we impose (eq. (5.3.4c)) will set $\check{C}_{\alpha\delta}{}^{\underline{c}} = 0$, and therefore the nonlinear term drops out. Consequently, the last two constraints of (5.3.4b) (in combined form)

$$0 = T_{\alpha,\beta\dot{\beta}}{}^{\beta\dot{\gamma}} = C_{\alpha,\beta\dot{\beta}}{}^{\beta\dot{\gamma}} + 2\Phi_{\alpha\dot{\beta}}{}^{\dot{\gamma}}$$

$$= (\check{C}_{\alpha,\beta\dot{\beta}}{}^{\beta\dot{\gamma}} - \Phi_{\alpha\dot{\beta}}{}^{\dot{\gamma}} + \frac{1}{2}i\delta_{\dot{\beta}}{}^{\dot{\gamma}}\Gamma_\alpha) + 2\Phi_{\alpha\dot{\beta}}{}^{\dot{\gamma}} \tag{5.3.24}$$

give

$$\Phi_{\alpha\dot{\beta}}{}^{\dot{\gamma}} = -\frac{1}{2}\check{C}_{\alpha,\beta(\dot{\beta}}{}^{\beta\dot{\gamma})} \ , \ \Gamma_\alpha = i\check{C}_{\alpha\underline{b}}{}^{\underline{b}} \ . \tag{5.3.25}$$

This completes the solution of the conventional constraints (5.3.4a,b). We have now determined $E_{\underline{a}}$, $\Phi_A$, and $\Gamma_A$ in terms of $E_\alpha$ and $E_{\dot{\alpha}}$. Furthermore, from the form of (5.3.22) we immediately obtain

$$E = sdet\, E_A{}^M = sdet\, \check{E}_A{}^M \ . \tag{5.3.26}$$

(The last terms in (5.3.22) give no contribution to the superdeterminant.)

**b.2. Representation preserving constraints**

Having determined all quantities in terms of $E_\alpha$, we have a realization of local supersymmetry with 512 ordinary component fields. In the Yang-Mills case, further reduction was achieved by imposing *representation-preserving* constraints: To ensure the existence of (anti)chiral scalar superfields (defined by $\nabla_\alpha\bar{\chi} = 0$), we required

$$\{\nabla_\alpha, \nabla_\beta\}\bar{\chi} = 0 \ . \tag{5.3.27}$$

In supergravity (assuming $[Y\,,\chi] = 0$ for simplicity), we find (5.3.27) implies

$$T_{\alpha\beta}{}^D\nabla_D\bar{\chi} = T_{\alpha\beta}{}^\gamma\nabla_\gamma\bar{\chi} + T_{\alpha\beta}{}^{\dot{\gamma}}\nabla_{\dot{\gamma}}\bar{\chi} + T_{\alpha\beta}{}^{\underline{c}}\nabla_{\underline{c}}\bar{\chi} = 0 \ . \tag{5.3.28}$$

Therefore, to allow the existence of chiral scalars in supergravity we must enforce the

constraints (5.3.5c):

$$T_{\alpha\beta}{}^{\dot\gamma} = T_{\alpha\beta}{}^{\underline{c}} = 0 \quad . \tag{5.3.29}$$

From (5.3.19), this implies:

$$\{ E_\alpha \, , E_\beta \} = C_{\alpha\beta}{}^{\gamma} E_\gamma \quad . \tag{5.3.30}$$

(Equivalently since $\nabla_\alpha \bar\chi = E_\alpha \bar\chi = 0$ it follows that $\{E_\alpha, E_\beta\}\bar\chi = 0$ which immediately leads to (5.3.30).) Thus $E_\alpha = E_\alpha{}^M D_M$ is a basis for tangent vectors that lie in a complex two-dimensional subspace of the full superspace: All operators $\lambda^\alpha E_\alpha$ generate *complex* translations with an algebra that closes. We can also parametrize these translations by a basis of derivatives with respect to coordinates $(\tau^1, \tau^2)$: $E_\alpha = A_\alpha{}^\mu \dfrac{\partial}{\partial \tau^\mu}$, where $A_\alpha{}^\mu$ is an arbitrary matrix or *zweibein.* We can always express the coordinates $\tau^\mu$ as *complex* supercoordinate transforms of the usual $\theta$-coordinates: $\dfrac{\partial}{\partial\tau^\mu} = e^{-\Omega} D_\mu e^{\Omega}$, where $\Omega = \Omega^M iD_M \neq \bar\Omega$ is arbitrary. Our full solution of the constraints (5.3.4c) is thus

$$E_\alpha = A_\alpha{}^\mu e^{-\Omega} D_\mu e^{\Omega} \equiv e^{-\Omega} \underset{\sim}{A}_\alpha{}^\mu D_\mu e^{\Omega} \quad ;$$

$$\Omega = \Omega^M iD_M \quad , \quad A_\alpha{}^\mu = \bar\Psi N_\alpha{}^\mu \quad ; \tag{5.3.31}$$

where we have split $A_\alpha{}^\mu$ into a complex scale factor $\bar\Psi$ and a Lorentz rotation $N_\alpha{}^\mu$ ($det\, N = 1$). This solution is closely analogous to the Yang-Mills solution $\nabla_\alpha = e^{-\Omega} D_\alpha e^{\Omega}$ to $\{\nabla_\alpha \, , \nabla_\beta\} = 0$, except for the introduction of $A_\alpha{}^\mu$. In fact, the $\Omega$ of Yang-Mills can be interpreted as a complex translation in the group manifold. We now have a description of supergravity in terms of $\Omega$, $\Psi$, and $N_\alpha{}^\mu$. However, $N_\alpha{}^\mu$ can be gauged away by a Lorentz transformation (with parameter $K_\alpha{}^\beta$ ).

## b.3. The $\Lambda$ gauge group

At this point, we can make contact with the previous section: The solution of the constraints imposed so far has introduced a new gauge group as an invariance of $E_\alpha = e^{-\Omega} \underset{\sim}{A}_\alpha{}^\mu D_\mu e^{\Omega}$. The vielbein $E_\alpha$ remains unchanged under the transformations

$$(e^{\Omega})' = e^{i\bar\Lambda} e^{\Omega} \quad ,$$

$$(\underset{\sim}{A}_\alpha{}^\mu D_\mu)' = e^{i\overline{\Lambda}}(\underset{\sim}{A}_\alpha{}^\mu D_\mu)e^{-i\overline{\Lambda}} \ . \tag{5.3.32}$$

with

$$\overline{\Lambda} = \overline{\Lambda}^M iD_M \quad , \qquad [\overline{\Lambda}, D_\mu] = -\, i(D_\mu \overline{\Lambda}^\nu)D_\nu \quad ; \tag{5.3.33}$$

provided that

$$D_\nu \overline{\Lambda}^{\dot\mu} = D_\nu \overline{\Lambda}^{\mu\dot\mu} - i\delta_\nu{}^\mu \overline{\Lambda}^{\dot\mu} = 0 \quad , \quad \overline{\Lambda}^\mu \ \textit{arbitrary} \quad ; \tag{5.3.34}$$

or

$$\overline{\Lambda}^{\dot\mu} = D^2 \overline{L}^{\dot\mu} \quad , \qquad \overline{\Lambda}^{\mu\dot\mu} = -\, iD^\mu \overline{L}^{\dot\mu} \quad . \tag{5.3.35}$$

The fact that $\overline{\Lambda}^\nu$ is completely arbitrary implies that the part of the $\overline{\Lambda}$-gauge group parametrized by $\overline{\Lambda}^\nu$ can always be "compensated away" by a redefinition of $N_\alpha{}^\mu$. Thus as in the Yang-Mills theory solving a constraint ($F_{\alpha\beta} = 0$) gives rise to a new gauge group. The transformation on $\underset{\sim}{A}_\alpha{}^\mu$ can be rewritten as

$$\underset{\sim}{\overline{\Psi}}' = e^{i\overline{\Lambda}}(1 \cdot e^{-i\overleftarrow{\Sigma}})^{-\frac{1}{2}} \underset{\sim}{\overline{\Psi}}\, e^{-i\overline{\Lambda}} \quad ,$$

$$(\,\underset{\sim}{N}_\alpha{}^\mu D_\mu\,)' = e^{i\overline{\Lambda}}(1 \cdot e^{-i\overleftarrow{\Sigma}})^{\frac{1}{2}} \underset{\sim}{N}_\alpha{}^\mu D_\mu e^{-i\overline{\Lambda}} \quad ; \tag{5.3.36}$$

where the factor $(1 \cdot e^{-i\overleftarrow{\Sigma}})$, $\overleftarrow{\Sigma} = \overline{\Lambda}^\mu i\overleftarrow{D}_\mu$, is the super-Jacobian of the transformation (c.f. (5.2.59)).

We still have the *real* $K = K^M iD_M$ coordinate transformations of the theory (see (5.3.3)), as well as the tangent space Lorentz and $U(1)$ rotations: $E'_\alpha = e^{iK}E_\alpha e^{-iK}$ is realized by

$$(e^\Omega)' = e^\Omega e^{-iK} \quad , \qquad (\,\underset{\sim}{A}_\alpha{}^\mu\,)' = \underset{\sim}{A}_\alpha{}^\mu \quad ; \tag{5.3.37}$$

while $E'_\alpha = e^{-\frac{1}{2}iK_5}K_\alpha{}^\beta E_\beta$ is realized by

$$(e^\Omega)' = e^\Omega \quad , \qquad (\,\underset{\sim}{A}_\alpha{}^\mu\,)' = (e^\Omega e^{-\frac{1}{2}iK_5}K_\alpha{}^\beta e^{-\Omega})\underset{\sim}{A}_\beta{}^\mu \quad . \tag{5.3.38}$$

The $K$ transformations can be rewritten as

$$\Omega' = \Omega - iK + O(\Omega, K) \ . \tag{5.3.39}$$

Since $K = \bar{K}$, this implies that $\Omega - \bar{\Omega}$ can be gauged away. In the resulting gauge,

$$\Omega = \frac{1}{2} H \quad , \qquad H = \bar{H} = H^M i D_M \quad . \tag{5.3.40}$$

Similarly, the Lorentz transformation can be used to gauge $N_\alpha{}^\mu$ to $\delta_\alpha{}^\mu$. In the resulting gauge we have

$$E_\alpha = e^{-\frac{1}{2}H} \underset{\sim}{\bar{\Psi}} D_\alpha e^{\frac{1}{2}H} \quad ,$$

$$E_{\dot{\alpha}} = e^{\frac{1}{2}H} \underset{\sim}{\Psi} \bar{D}_{\dot{\alpha}} e^{-\frac{1}{2}H} \quad . \tag{5.3.41}$$

However, it is more convenient to eliminate the real part of $\Omega$ by going to a chiral representation (as for super-Yang-Mills), as discused in sec. 5.3.b.5 below.

### b.4. Evaluation of $\Gamma_\alpha$ and $R$

We can now find simple forms for $\Gamma_\alpha$ (5.3.25) and $R$ (5.3.9). The results are contained in (5.3.52,53,56). The details of the derivation are not essential for further reading, but present some useful general techniques. To solve for $\Gamma_\alpha$, we use the identity

$$E^{-1} \overleftarrow{\nabla}_A = - E^{-1} (-)^B T_{AB}{}^B \quad , \tag{5.3.42}$$

which holds independently of any constraints, for any superspace, for any tangent space. In cases where $(-)^B T_{AB}{}^B$ vanishes (as here), it allows covariant integration by parts, since

$$\int dz\, E^{-1} \nabla_A X = - \int dz\, E^{-1} \overleftarrow{\nabla}_A X = 0 \quad . \tag{5.3.43}$$

To derive this identity we will save ourselves a lot of trouble by noting that at the end of a calculation the signs resulting from graded statistics can easily be determined if the indices of each contracted pair are adjacent, with the contravariant index first. The net sign change is then just that resulting from the graded reordering of the indices of the initial expression. Using this fact to ignore the signs from grading at intermediate steps of the calculation, we have (in the basis $E_A = E_A{}^M \partial_M$)

$$(-)^B T_{AB}{}^B = E_M{}^B [\nabla_A , \nabla_B \} z^M = E_M{}^B \nabla_{[A} E_{B)}{}^M$$

$$= E_M{}^B \nabla_A E_B{}^M - E_M{}^B \nabla_B E_A{}^M \quad . \tag{5.3.44}$$

We evaluate the second term by use of the identity (again ignoring grading signs) for arbitrary superfunctions $X$ and $Y$

$$XY\overleftarrow{\nabla}_A = X[Y\,,\overleftarrow{\nabla}_A] + X\overleftarrow{\nabla}_A Y = X\nabla_A Y + X\overleftarrow{\nabla}_A Y \quad , \tag{5.3.45}$$

where we have used (5.1.26b) to evaluate the commutator. (5.3.44) now becomes

$$E_M{}^B \nabla_A E_B{}^M - E_M{}^B E_A{}^M \overleftarrow{\nabla}_B + E_M{}^B \overleftarrow{\nabla}_B E_A{}^M$$

$$= \nabla_A ln\, E - 1 \cdot \overleftarrow{\nabla}_A + 0 \quad , \tag{5.3.46}$$

which leads to (5.3.42). The evaluation of the first term used the usual expression for the derivative of the logarithm of a determinant (see (5.1.28); for tangent space groups which include scale transformations, $(-)^B \Phi_{AB}{}^B \neq 0$, so that the scale generator acts non-trivially on $E$). The last term vanishes because $E_M{}^B \overleftarrow{\nabla}_B = E_M{}^B E_B{}^N (\overleftarrow{\partial}_N + \Phi_N(\overleftarrow{M}) - i\Gamma_N \overleftarrow{Y})$. Therefore $\delta_M{}^N (\overleftarrow{\partial}_N + \Phi_N(\overleftarrow{M}) - i\Gamma_N \overleftarrow{Y}) = \delta_M{}^N \overleftarrow{\nabla}_N = 0$.

Actually, it is simpler for our purposes to use the form of (5.3.42) in terms of $\check{E}_A$ instead of $\nabla_A$. Using $\check{E} = E$, we have:

$$E^{-1} \overleftarrow{\check{E}}_A = -\, E^{-1}(-)^B \check{C}_{AB}{}^B \quad . \tag{5.3.47}$$

From the expression (5.3.25) for $\Gamma_\alpha$, $\check{E}_\alpha = E_\alpha$, and $\check{C}_{\alpha\dot{\beta}}{}^{\gamma} = \check{C}_{\alpha\dot{\beta}}{}^{\dot{\gamma}} = 0$ we obtain

$$-\, i\Gamma_\alpha = (-)^B \check{C}_{\alpha B}{}^B + \check{C}_{\alpha\beta}{}^\beta = -\, E^{-1} \overleftarrow{E}_\alpha E + \check{C}_{\alpha\beta}{}^\beta \quad . \tag{5.3.48}$$

Using the expression (5.3.31) for $E_\alpha$ in the Lorentz gauge $N_\alpha{}^\mu = \delta_\alpha{}^\mu$ (the general Lorentz gauge will be easily restored at the end), we find $\check{C}_{\alpha\beta}{}^\gamma = \delta_{(\alpha}{}^\gamma E_{\beta)} ln\, \overline{\Psi}$, so this expression becomes

$$-\, i\Gamma_\alpha = -\, E^{-1} \overleftarrow{E}_\alpha E + 3E_\alpha ln\, \overline{\Psi} = -\, 1 \cdot e^{\overleftarrow{\Omega}} \overleftarrow{D}_\alpha e^{-\overleftarrow{\Omega}} \overline{\Psi} + E_\alpha ln\, E\overline{\Psi}^2 \quad , \tag{5.3.49}$$

where we have used

$$\hat{E}_\alpha = e^{-\Omega} D_\alpha e^{\Omega} \tag{5.3.50a}$$

which implies

$$\overleftarrow{\hat{E}}_{\alpha} = e^{\overleftarrow{\Omega}} \overleftarrow{D}_{\alpha} e^{-\overleftarrow{\Omega}} \quad . \tag{5.3.50b}$$

We use the identity, valid for any function $f$ and linear operator $X$,

$$\begin{aligned} f e^{\overleftarrow{X}} &= (1 \cdot e^{\overleftarrow{X}} e^{-\overleftarrow{X}}) f e^{\overleftarrow{X}} \\ &= (1 \cdot e^{\overleftarrow{X}})(e^{-\overleftarrow{X}} f e^{\overleftarrow{X}}) \\ &= (1 \cdot e^{\overleftarrow{X}})(e^{X} f e^{-X}) \\ &= (1 \cdot e^{\overleftarrow{X}})(e^{X} f) \quad , \end{aligned} \tag{5.3.51a}$$

to derive the relation

$$1 = (1 \cdot e^{-\overleftarrow{X}}) e^{\overleftarrow{X}} = (1 \cdot e^{\overleftarrow{X}})[e^{X}(1 \cdot e^{-\overleftarrow{X}})] \quad . \tag{5.3.51b}$$

These two results make it possible to rewrite (5.3.49) as

$$\begin{aligned} -i\Gamma_{\alpha} &= -\overline{\Psi}(1 \cdot e^{-\overleftarrow{\Omega}}) e^{-\Omega} D_{\alpha}(1 \cdot e^{\overleftarrow{\Omega}}) + E_{\alpha} ln\, E\overline{\Psi}^2 \\ &= -\overline{\Psi}(1 \cdot e^{-\overleftarrow{\Omega}}) e^{-\Omega} D_{\alpha} e^{\Omega}(1 \cdot e^{-\overleftarrow{\Omega}})^{-1} + E_{\alpha} ln\, E\overline{\Psi}^2 \\ &= \nabla_{\alpha} T \equiv T_{\alpha} \quad , \end{aligned} \tag{5.3.52}$$

where we have introduced a (noncovariant) scalar *density* $T$:

$$T \equiv ln\,[E\overline{\Psi}^2(1 \cdot e^{-\overleftarrow{\Omega}})] \quad . \tag{5.3.53}$$

An immediate consequence of (5.3.52) is $F_{\alpha\beta} = 0$ (see (5.3.1)).

We now solve for $R$, where

$$\{\nabla_{\alpha}, \nabla_{\beta}\} = -2\bar{R} M_{\alpha\beta} \quad , \tag{5.3.54}$$

as follows from the Bianchi identities (sec. 5.4). Using the same form for $E_{\alpha}$ as in the previous calculation, and using the result for $\Gamma_{\alpha}$, we find from (5.3.20)

$$\begin{aligned} \Phi_{\alpha\beta}{}^{\gamma} &= -\frac{1}{2}\delta_{\alpha}{}^{(\gamma}(E_{\beta)} ln\, \overline{\Psi}^2 + i\Gamma_{\beta)}) = -\frac{1}{2}\delta_{\alpha}{}^{(\gamma} E_{\beta)} ln\,(e^{-T}\overline{\Psi}^2) \\ &= -\frac{1}{2}\delta_{\alpha}{}^{(\gamma} E_{\beta)} ln\,[(1 \cdot e^{-\overleftarrow{\Omega}})^{-1} E^{-1}] \quad . \end{aligned} \tag{5.3.55}$$

We then find ($\Phi_{\alpha\beta\dot{\gamma}}{}^{\dot{\gamma}}$ does not contribute)

$$R = e^{\overline{T}}(\hat{\overline{E}}_{\dot{\alpha}})^2(e^{-\overline{T}}\Psi^2) = e^{\overline{T}}(\hat{\overline{E}}_{\dot{\alpha}})^2(1 \cdot e^{\overleftarrow{\overline{\Omega}}})^{-1}E^{-1}$$

$$= E^{-1} \cdot (\overleftarrow{\hat{\overline{E}}}_{\dot{\alpha}})^2 e^{\overline{T}}(1 \cdot e^{\overleftarrow{\overline{\Omega}}})^{-1} \quad . \tag{5.3.56}$$

where $(\overleftarrow{\hat{\overline{E}}}_{\dot{\alpha}})^2 = \frac{1}{2}(\overleftarrow{\hat{\overline{E}}}{}^{\dot{\alpha}})(\overleftarrow{\hat{\overline{E}}}_{\dot{\alpha}})$.

* * *

It is useful to derive the explicit form of the operator that gives a chiral scalar from a general scalar $f$. For the case $[Y\,, f] = 0$, a simple calculation using (5.3.55) for the connection gives the result that $(\overline{\nabla}^2 + R)f = fE^{-1}(\overleftarrow{\hat{\overline{E}}}_{\dot{\alpha}})^2 e^{\overline{T}}(1 \cdot e^{\overleftarrow{\overline{\Omega}}})^{-1}$ is covariantly chiral. This result can then most easily be extended to arbitrary $U(1)$ charge $[Y\,, f] = \frac{1}{2}wf$ by using the expression (5.3.53) for $\Gamma_\alpha$ to write

$$(\overline{\nabla}^2 + R)f = [\,e^{\frac{1}{2}w\overline{T}}(\widehat{\overline{\nabla}}{}^2 + R)e^{-\frac{1}{2}w\overline{T}}f\,](1 \cdot e^{\overleftarrow{\overline{\Omega}}})^{-1} \quad , \tag{5.3.57}$$

where $\widehat{\overline{\nabla}}{}^2$ is the form of $\overline{\nabla}^2$ on a neutral scalar (as implied by (5.3.57) for $\frac{1}{2}w = 0$). We thus obtain

$$(\overline{\nabla}^2 + R)f = fe^{-\frac{1}{2}w\overline{T}}E^{-1}(\overleftarrow{\hat{\overline{E}}}_{\dot{\alpha}})^2 e^{(1+\frac{1}{2}w)\overline{T}}(1 \cdot e^{\overleftarrow{\overline{\Omega}}})^{-1} \quad . \tag{5.3.58}$$

This quantity is covariantly chiral with $U(1)$ charge $1 + \frac{1}{2}w$.

### b.5. Chiral representation

Due to the form of $E_\alpha$ in (5.3.31), it is possible to define local representations that are chiral with respect to the supergravity fields. (These are analogous to chiral representations in super Yang-Mills (4.2.78) as well as in global supersymmetry (3.4.8).) On all quantities $F$ we perform a (nonunitary) similarity transformation

$$F^{(+)} = e^{-\overline{\Omega}}Fe^{\overline{\Omega}} \quad . \tag{5.3.59}$$

(Antichiral representations can also be defined, with $\overline{\Omega} \to -\Omega$.) In this representation, as for super-Yang-Mills, all quantities are invariant under $K^M$ transformations, and the covariant derivatives transform explicitly under $\Lambda$ transformations. Furthermore, we choose the Lorentz gauge $N_\alpha{}^\mu = \delta_\alpha{}^\mu$, which forces $K_{\dot\alpha}{}^{\dot\beta}$ to equal $\omega_{\dot\alpha}{}^{\dot\beta}$ of (5.2.27) to maintain the gauge. In the chiral representation, the vielbein becomes:

$$E^{(+)}{}_{\dot\alpha} = \Psi \bar{D}_{\dot\alpha} \ , \ E^{(+)}{}_\alpha = e^{-H}\bar{\Psi} D_\alpha e^H = \tilde{\Psi} e^{-H} D_\alpha e^H \ ;$$

$$e^H = e^\Omega e^{\bar\Omega} \ . \qquad (5.3.60)$$

This is precisely what we had constructed in the previous section ((5.2.27) and (5.2.28)). The transformation of $H$ can be obtained from that of $\Omega$:

$$(e^H)' = e^{i\bar\Lambda} e^H e^{-i\Lambda} \ . \qquad (5.3.61)$$

(Note that, as in super-Yang-Mills, (5.3.60) can be used to define $H$ in any $K$-gauge; it is $K$ invariant. Alternately the $\bar\Lambda^\mu$ and $\Lambda^{\dot\mu}$ transformations can be used to gauge away $H^\mu$ and $\bar{H}^{\dot\mu}$.)

It is possible to go to a representation that is also chiral with respect to $U(1)$. From (5.3.53) we have

$$\Gamma_\alpha = iE_\alpha T \ , \qquad \Gamma_{\dot\alpha} = -\, i\bar{E}_{\dot\alpha}\tilde{T} \ ; \qquad (5.3.62a)$$

where $\tilde{T} = e^{-H}\bar{T}e^H$ is the chiral-representation hermitian conjugate of $T$ (cf. (5.2.28)). Using (5.3.2), we can write

$$E_\alpha - i\Gamma_\alpha Y = e^{-TY} e^{-\frac{1}{2}T} E_\alpha e^{TY} \ ,$$

$$E_{\dot\alpha} - i\Gamma_{\dot\alpha} Y = e^{\tilde{T}Y} e^{-\frac{1}{2}\tilde{T}} E_{\dot\alpha} e^{-\tilde{T}Y} \ ; \qquad (5.3.62b)$$

In addition to the transformation $\delta T = -\, iK_5$, these expressions are invariant under $\delta T = i\tilde{\Lambda}_5$, where $\Lambda_5$ is chiral. We can use this gauge freedom to replace $T$ by

$$\Omega_5 = T + 3ln\tilde{\phi} \ , \qquad (5.3.62c)$$

thus introducing for subsequent use the chiral density $\phi$, $\bar{D}_{\dot\alpha}\phi = 0$. We now go to a chiral representation not only with respect to $\Omega^M D_M$ and $N_\alpha{}^\mu$, but also with respect to $\Omega_5$, by making the appropriate nonunitary $U(1)$ transformation, and obtain:

$$E_{\dot\alpha} - i\Gamma_{\dot\alpha} Y = E^{-\frac{1}{2}} \bar{D}_{\dot\alpha} \ ,$$

$$E_\alpha - i\Gamma_\alpha Y = e^{-V_5 Y} E^{-\frac{1}{2}}(1 \cdot e^{-\overleftarrow{H}})^{-\frac{1}{2}} \hat{E}_\alpha e^{V_5 Y} = E\hat{E}^{-\frac{1}{2}}\hat{E}_\alpha + (E\hat{E}^{-\frac{1}{2}}\hat{E}_\alpha V_5)Y \ ,$$

$$e^{V_5} \equiv e^{\Omega_5} e^{\tilde{\Omega}_5} = E^3 \hat{E}^{-1}(1 \cdot e^{-\overleftarrow{H}})\tilde{\phi}^3\phi^3 = (\ddot{E}^{-1}E)^3 \ ,$$

$$\ddot{E}^{-1} \equiv \hat{E}^{-\frac{1}{3}}(1 \cdot e^{-\overleftarrow{H}})^{\frac{1}{3}}\tilde{\phi}\phi \ ; \tag{5.3.63}$$

where we have used $E = \Psi^2\overline{\Psi}^2\hat{E}$ (5.2.49) to replace $\Psi$ with $\hat{E}$. This effectively replaces $\Psi$ with $E$ as superscale density compensator: In the $U(1)$-chiral representation, the $U(1)$ density compensator $\Omega_5$ no longer appears. For this reason this chiral representation is useful for $n = 0$ supergravity, where a true local $U(1)$ invariance remains, but not very useful for other $n$. However, it does bear a close relationship to the $n = -\frac{1}{3}$ results of the previous section: $n = -\frac{1}{3}$ can be obtained by constraining $\Gamma_A$ to vanish identically. In this representation, the result is simply that $V_5$ vanishes, and hence $E^{-1} = \ddot{E}^{-1}$, in agreement with (5.2.72). In section 5.5, this result will be used to write a first-order formalism for $n = 0$ combined with $n = -\frac{1}{3}$.

**b.6. Density compensators**

After gauging away $H^\mu$ and $H^{\dot{\mu}}$, we have now determined all the geometrical superfields in $\nabla_A$ in terms of $H^{\underline{m}}$ and $\overline{\Psi}$, which contain 64 and 32 component fields, respectively. The axial vector prepotential $H^{\underline{m}}$ contains the component gauge fields. The superfield $\overline{\Psi}$ is the superconformal (density type) compensator: By scaling $\overline{\Psi}$ *arbitrarily* (without transforming $H^{\underline{m}}$), we generate complex scale transformations (real scale $\otimes\, U(1)$) of the vielbein. Thus, the complex scale transformation properties of any quantity expresses its $\Psi$ dependence. These transformations must be restricted and the representation reduced further, since Einstein theory is included in Poincaré supergravity and is *not* scale invariant. We now consider the (scalar) tensor-type compensators $\Phi\,, \Sigma\,, G$ (5.3.13), which also transform under these combined transformations. Fixing the gauges of these transformations by fixing the compensator is a convenient way of determining $\Psi$, since it separates the lower superspin multiplets from the conformal supergravity multiplet in a covariant way.

It is convenient to solve the constraints (5.3.13) on the tensor compensators in terms of corresponding *density* compensators that satisfy the corresponding *flat-space* constraints; this allows us to find an explicit solution for $\Psi$.

### b.6.i. Minimal ($n = -\frac{1}{3}$) supergravity

We first consider the (covariantly) chiral scalar compensator $\Phi$. The constraint is, using (5.3.62a),

$$\overline{\nabla}_{\dot\alpha}\Phi = (\overline{E}_{\dot\alpha} - \frac{1}{3}\, i\overline{\Gamma}_{\dot\alpha})\Phi = e^{\frac{1}{3}\overline{T}}\,\overline{E}_{\dot\alpha}e^{-\frac{1}{3}\overline{T}}\Phi = 0 \qquad (5.3.64a)$$

and is solved by

$$\Phi = e^{\frac{1}{3}\overline{T}}\phi = [\Psi^4\overline{\Psi}^2\hat{E}(1\cdot e^{\overleftarrow{\overline{\Omega}}})]^{\frac{1}{3}}\phi \ \ , \ \ \overline{E}_{\dot\alpha}\phi = 0 \ \ . \qquad (5.3.64b)$$

It transforms under scale transformations as in (5.3.14). Here $\phi$ is a flat space chiral superfield, in the chiral representation, as follows from the definition of $E_\alpha$. If we choose the gauge $\Phi = 1$, then $\overline{T} = -\,3ln\phi$ and we obtain

$$\Psi = \phi^{-1}\overline{\phi}^{\frac{1}{2}}\hat{E}^{-\frac{1}{6}}(1\cdot e^{-\overleftarrow{\Omega}})^{\frac{1}{6}}(1\cdot e^{\overleftarrow{\overline{\Omega}}})^{-\frac{1}{3}} \quad . \qquad (5.3.65)$$

(We have again used $E = \Psi^2\overline{\Psi}^2\hat{E}$.) In this gauge $\overline{\nabla}_{\dot\alpha}\Phi = 0$ implies $\overline{\Gamma}_{\dot\alpha} = 0$; by complex conjugation and (5.3.4a) $\Gamma_A = 0$, and thus the field strength of the axial $U(1)$ transformations (see (5.3.9)) vanishes: $W_\alpha = (\overline{\nabla}^2 + R)\Gamma_\alpha = 0$. The relation (5.3.58) becomes

$$\phi^3(\overline{\nabla}^2 + R)f = fE^{-1}(\overleftarrow{\hat{\overline{E}}}_{\dot\alpha})^2(1\cdot e^{\overleftarrow{\overline{\Omega}}})^{-1} \quad . \qquad (5.3.66a)$$

In the chiral representation, this simplifies to

$$\phi^3(\overline{\nabla}^2 + R)f = \overline{D}^2(E^{-1}f) \quad . \qquad (5.3.66b)$$

The theory is now described by $H^{\underline{m}}$ and $\phi$ and the only superspace gauge freedom left is that of super-Poincaré transformations. As in (5.2.75), the density compensator $\phi$ can be replaced by one of its variants.

### b.6.ii. Nonminimal ($n \neq -\frac{1}{3}$) supergravity

For the nonminimal scalar multiplet, we again find a solution in terms of a density compensator $\Upsilon$. The constraint is, using (5.3.58),

$$(\overline{\nabla}^2 + R)\Sigma = \Sigma E^{-1}e^{\frac{2n}{3n+1}\overline{T}}(\overleftarrow{\hat{\overline{E}}}_{\dot\alpha})^2e^{-\frac{2n}{3n+1}\overline{T}}(1\cdot e^{\overleftarrow{\overline{\Omega}}})^{-1} = 0 \qquad (5.3.67a)$$

and is solved by

$$\Sigma = e^{-\frac{2n}{3n+1}\overline{T}} E(1 \cdot e^{\overleftarrow{\Omega}})\Upsilon = [\Psi^4 \overline{\Psi}^2 \hat{E}(1 \cdot e^{\overleftarrow{\Omega}})]^{\frac{n+1}{3n+1}} \Psi^{-2}\Upsilon \ , \ (\hat{\overline{E}}_{\dot{\alpha}})^2 \Upsilon = 0 \ . \qquad (5.3.67b)$$

Here $\Upsilon$ is a flat-space-linear superfield (as opposed to the covariantly linear superfield $\Sigma$): In the chiral representation $\overline{D}^2\Upsilon = 0$. Again, in the gauge $\Sigma = 1$, we obtain the solution for $\Psi$:

$$\Psi = [\Upsilon^{n-1}\overline{\Upsilon}^{n+1}]^{-\frac{3n+1}{8n}} [\hat{E}^{2n}(1 \cdot e^{-\overleftarrow{\Omega}})^{n+1}(1 \cdot e^{\overleftarrow{\Omega}})^{n-1}]^{-\frac{n+1}{8n}} \ . \qquad (5.3.68)$$

In this gauge we have $T_\alpha \equiv i\Gamma_\alpha$ as a new tensor (appearing in arbitrary gauges as $\sim\nabla_\alpha\Sigma$), in terms of which $R$ and $W_\alpha$ are determined. The solution (5.3.68) does not apply to the following cases: (1) $n = -\frac{1}{3}$, for which the minimal scalar multiplet is used instead; (2) $n = 0$, for which the solution of (5.3.67a) is more subtle and will be discussed next; and (3) $n = \infty$, which does not lead to a sensible theory. The parameter $n$ can also be generalized to complex values, but the constraints then violate parity (off shell), and we do not discuss them here.

### b.6.iii. Axial ($n = 0$) supergravity

The constraint (5.3.13c) for $n = 0$ is most easily solved by expressing the compensator $G$ in terms of a covariantly chiral spinor $\phi^\alpha = (\overline{\nabla}^2 + R)\Xi^\alpha$. Using the relation $\nabla^\alpha\phi_\alpha = E(\phi^\alpha E^{-1}\overleftarrow{E}_\alpha)$ (as follows from integration by parts on $\int d^4x d^4\theta \, E^{-1}\phi^\alpha \nabla_\alpha f = \int d^4x d^4\theta \, E^{-1}\phi^\alpha E_\alpha f$ for any $f$), we have in the chiral representation (using (5.3.63)):

$$G = \frac{1}{2}(\nabla_\alpha\phi^\alpha + \overline{\nabla}_{\dot{\alpha}}\overline{\phi}^{\dot{\alpha}}) = \frac{1}{2}E(\phi_\alpha \hat{E}^{-\frac{1}{2}}\overleftarrow{\hat{E}}^{\alpha} + h.c.) \equiv E\hat{G} \ , \qquad (5.3.69)$$

so that $\hat{G}$ is a function of only $H$ and $\phi^\alpha$. In the chiral representation, but in the Lorentz gauge $\Phi_{\alpha\dot{\beta}}{}^{\dot{\gamma}} = 0$, $\phi_\alpha$ is flat-space chiral: $\overline{D}_{\dot{\alpha}}\phi_\alpha = 0$. (This gauge exists because $R_{\alpha\beta\dot{\gamma}}{}^{\dot{\delta}} = 0$, as follows from the Bianchi identities, see sec. 5.4, which implies that $\Phi_{\alpha\dot{\beta}}{}^{\dot{\gamma}}$ and its conjugate are pure gauge.) In such a gauge $N_\alpha{}^\mu = X_\alpha{}^\mu \neq \delta_\alpha{}^\mu$, but depends only on $H$. On the other hand, in the Lorentz gauge $N_\alpha{}^\mu = \delta_\alpha{}^\mu$, where $\Phi_{\alpha\dot{\beta}}{}^{\dot{\gamma}}$ depends only on $H$ up to a factor of $\Psi$ (see (5.3.25)), $\phi^\alpha$ is $(X^{-1})_\mu{}^\alpha$ times a flat-space chiral spinor.

In the Weyl gauge $G = 1$ we obtain

$$E^{-1} = \hat{G}(H, \phi_\alpha) = \frac{1}{2}\,(\phi_\alpha \hat{E}^{-\frac{1}{2}} \overleftarrow{\hat{E}}^{\alpha} + h.\,c.)\ \ . \tag{5.3.70}$$

In this gauge, (5.3.13c) implies the constraint $R = 0$. Furthermore, in the chiral representation (5.3.70) may be combined with (5.3.63) to replace $\Psi$ with $E$ as the compensator for the $n = 0$ covariant derivatives:

$$\Psi = E^{-\frac{1}{2}} = \hat{G}^{\frac{1}{2}}\ \ ,\ \ \tilde{\overline{\Psi}} = E\hat{E}^{-\frac{1}{2}} = \hat{G}^{-1}\hat{E}^{-\frac{1}{2}}\ \ . \tag{5.3.71}$$

## b.7. Degauging

The theory and the covariant derivatives we have constructed so far contain explicit $U(1)$ generators and connections. For $n = 0$, the $U(1)$ symmetry is a genuine *local* symmetry of the theory at the classical level (there are anomalies at the quantum level, see sec. 7.10) and therefore $n = 0$ supergravity only couples to R-invariant matter systems (3.6.14, 4.1.15). For $n \neq 0$, the superscale compensator can be used to *remove* the $U(1)$ charge of any multiplet: By multiplying the superfield by an appropriate power of the compensator (see (5.3.12a,b)) we can always construct a $U(1)$ neutral object. If we do this to all quantities (vielbein, connections, matter), the $U(1)$ generators do not act and can be dropped from the theory. The resulting formalism is applicable to matter multiplets *without* definite $U(1)$ charge and hence to systems without global R-invariance (see sec. 5.5). The procedure we are following is similar to what one does in ordinary spontaneously broken gauge theories. One goes to a *U-gauge* either by using the Goldstone field to define gauge invariant quantities as we just did, or by gauging the Goldstone field away, as we do now: Instead of rescaling fields by the compensator, we can gauge it away, and fix the $U(1)$ (and superscale) gauge as discussed in sec. b.6. above. For $n = -\frac{1}{3}$, this sets $\Gamma_A = 0$ and the $Y$ generator drops from the covariant derivatives. For $n \neq -\frac{1}{3}, 0$, the $U(1)$ connection becomes a covariant tensor with respect to the remaining (super-Poincaré) group. Therefore we can eliminate $Y$ by adding a "contortion" term $\nabla_A \to \nabla_A - i\Gamma_A Y$ and thus, by (5.3.18), $T_{AB}{}^C \to T_{AB}{}^C - i\frac{1}{2}w(C)\Gamma_{[A}\delta_{B)}{}^C$ (but with no change in the curvatures). The only modified torsions are (where $\Gamma_\alpha \equiv iT_\alpha$, see (5.3.52)):

$$T_{\alpha\beta}{}^{\gamma} \rightarrow T_{\alpha\beta}{}^{\gamma} - \frac{1}{2}\delta_{(\alpha}{}^{\gamma}T_{\beta)} \ ,$$

$$T_{\alpha\dot{\beta}}{}^{\dot{\gamma}} \rightarrow T_{\alpha\dot{\beta}}{}^{\dot{\gamma}} + \frac{1}{2}\delta_{\dot{\beta}}{}^{\dot{\gamma}}T_{\alpha} \ ,$$

$$T_{\underline{a}\beta}{}^{\gamma} \rightarrow T_{\underline{a}\beta}{}^{\gamma} - \frac{1}{2}i\delta_{\beta}{}^{\gamma}(\nabla_{\alpha}\overline{T}_{\dot{\alpha}} - \overline{\nabla}_{\dot{\alpha}}T_{\alpha} + T_{\alpha}\overline{T}_{\dot{\alpha}}) \ . \tag{5.3.72}$$

Furthermore, $R$ and $W_{\alpha}$ have the following explicit expressions in terms of $T_{\alpha}$ and the $Y$ independent or *degauged* $\nabla$:

$$R = -\Sigma^{-1}\overline{\nabla}^2\Sigma \rightarrow -\frac{n}{3n+1}(\overline{\nabla}^{\dot{\alpha}} + \frac{n-1}{2(3n+1)}\overline{T}^{\dot{\alpha}})\overline{T}_{\dot{\alpha}} \ , \tag{5.3.73}$$

$$W_{\alpha} = (\overline{\nabla}^2 + R)\Gamma_{\alpha} \rightarrow i[\frac{1}{2}\overline{\nabla}^{\dot{\alpha}}(\overline{\nabla}_{\dot{\alpha}} + \frac{1}{2}\overline{T}_{\dot{\alpha}}) + R]T_{\alpha} \ . \tag{5.3.74}$$

We emphasize that for $n = -\frac{1}{3}$ we can simply drop all reference to $U(1)$ without any other modifications.

Although we have emphasized the compensator approach to the breaking of the superconformal invariance, we should point out that fixing the conformal gauge by setting the compensator to 1 is completely equivalent to imposing additional, conformal-breaking constraints on the covariant derivatives. After $U(1)$ degauging, the constraint equations (5.3.13) or (5.3.64a,67a) become, when the compensators are fixed, conditions on the covariant derivatives. These conditions are the constraints on torsions and curvatures given in (5.2.80b).

At this point we have a description of Poincaré supergravity in terms of $H$ and one of the density compensators. They compensate for *component* conformal transformations. For example, the $\theta$-independent of $\phi$ can be identified with the component $\phi$ of (5.1.33). Similarly, the linear $\theta$ component, a spinor, compensates for $S$-supersymmetry.

After degauging, the superconformal invariance of the supergravity constraints is destroyed for arbitrary superfields $L$ and $K_5$. Nevertheless a remnant of superconformal invariance remains. This is because the unconstrained superfields that describe Poincaré supergavity are $H^{\underline{m}}$ and some scalar superfield compensator. Thus superfield supergravity, unlike ordinary gravity, *always* contains a component $\phi$ compensator of (5.1.33). (This is the reason why at the component level superconformal symmetry is so useful.)

Obviously a redefinition of the density compensating multiplet (once its type has been specified) cannot affect the Poincaré supergravity constraints. This is realized by an invariance group of the constraints in addition to that parametrized by $K^M$ and $K_\alpha{}^\beta$. The transformations of this invariance group are exactly the same as those of the conformal group, but with the important restriction that $L$ and $K_5$ are no longer arbitrary. The simplest way to obtain the form of these restricted conformal transformations is to use the tensor type compensators $\Phi$, $\Sigma$, and $G$.

Before gauging the compensators to 1 we can simply make an arbitrary redefinition of the tensor type compensators. We have (i): $\delta\Phi = -\Delta\,\Phi$, (ii) $\delta\Sigma = -\Delta\,\Sigma$, and (iii) $\delta G = -\Delta G$, where $\Delta$ is a covariant superfield. In each case the redefinition must be such that the product of the compensator times $\Delta$ satisfies the same differential equation as the original compensator (5.3.13) (i.e., (i) $\overline{\nabla}_{\dot\alpha}\Delta = 0$, (ii) $(\overline{\nabla}^2 + R)(\Delta\Sigma) = 0$, and (iii) $(\overline{\nabla}^2 + R)(\Delta G) = 0$, $\Delta = \overline{\Delta}$). The $\Delta$ transformations affect *only* the compensators, not the covariant derivatives nor matter superfields, whereas $L$ and $K_5$ transformations affect all fields. The combined transformation of the tensor compensators under $L$,$K_5$, and $\Delta$ is thus

$$n = -\frac{1}{3}: \qquad \delta\Phi = (L + i\,\frac{1}{3}K_5)\Phi - \Delta\Phi \quad , \tag{5.3.75a}$$

$$n \neq -\frac{1}{3}\,,\,0: \qquad \delta\Sigma = [\,(\frac{2}{3n+1})L - i\,(\frac{2n}{3n+1})K_5\,]\Sigma - \Delta\Sigma \quad , \tag{5.3.75b}$$

$$n = 0: \qquad \delta G = 2L\,G - \Delta G \quad . \tag{5.3.75c}$$

We now degauge by setting the compensator to one. In order to maintain this gauge condition we must set the total variation of the compensator to zero and we find that $L$ and $K_5$ satisfy the constraints

$$L = \frac{1}{2}(\Delta + \overline{\Delta}) \ \ , \ \ K_5 = \frac{3}{2}i(\overline{\Delta} - \Delta) \ \ ; \tag{5.3.76a}$$

$$L = \frac{3n+1}{4}(\Delta + \overline{\Delta}) \ \ , \ \ K_5 = \frac{3n+1}{4n}i(\Delta - \overline{\Delta}) \ \ ; \tag{5.3.76b}$$

$$L = \frac{1}{2}\Delta \ \ , \ \ \Delta = \overline{\Delta} \ \ ; \tag{5.3.76c}$$

respectively.

## 5.4. Solution to Bianchi identities

In any gauge theory the field strengths satisfy Bianchi identities that are a consequence of the Jacobi identities for the covariant derivatives, or more generally, for superforms, a consequence of the Poincaré theorem. As explained in sec. 4.2, the Bianchi identities contain no useful information *unless* some of the field strengths have been constrained. In that case they make it possible to express all the field strengths in terms of an irreducible set (that may still satisfy *differential* constraints that are also called Bianchi identities). In sec. 4.2 we gave a detailed example of this procedure for super Yang-Mills theories; here we consider supergravity.

We begin in a general context, with covariant derivatives for arbitrary $N$ and arbitrary internal symmetry generators $\Omega_i$ (cf. (5.2.20, 5.3.1)):

$$\nabla_A = E_A{}^M D_M + \Phi_{A\delta}{}^\gamma M_\gamma{}^\delta + \Phi_{A\dot{\delta}}{}^{\dot{\gamma}} \bar{M}_{\dot{\gamma}}{}^{\dot{\delta}} + \Gamma_A{}^i \Omega_i \tag{5.4.1}$$

where $E_A{}^M$, $\Phi_A$, and $\Gamma_A$ are the vielbein, Lorentz connection, and gauge potential, respectively. We define field strengths: torsions $T_{AB}{}^C$, curvatures $R_{AB\delta}{}^\gamma$ and $R_{AB\dot{\delta}}{}^{\dot{\gamma}}$, and gauge field strengths $F_{AB}{}^i$ in terms of the graded commutator

$$[\nabla_A , \nabla_B\} = T_{AB}{}^C \nabla_C + R_{AB\delta}{}^\gamma M_\gamma{}^\delta + R_{AB\dot{\delta}}{}^{\dot{\gamma}} \bar{M}_{\dot{\gamma}}{}^{\dot{\delta}} + F_{AB}{}^i \Omega_i \quad . \tag{5.4.2}$$

The geometry of superspace implicit in (5.4.1) gives nontrivial relations among the field strengths $T, R, F$. We have chosen the action of the Lorentz group to be reducible in tangent space: It does not mix the $(V_\alpha , V_{\dot{\alpha}} , V_{\underline{a}})$ parts of a supervector $V_A$, and it rotates the vector and the spinor parts by the same transformation; i.e., $\delta_\omega V_A \equiv \omega_A{}^B V_B$, where $\omega_A{}^B = (\omega_\alpha{}^\beta , \bar{\omega}_{\dot{\alpha}}{}^{\dot{\beta}} , \omega_\alpha{}^\beta \delta_{\dot{\alpha}}{}^{\dot{\beta}} + \bar{\omega}_{\dot{\alpha}}{}^{\dot{\beta}} \delta_\alpha{}^\beta)$ (cf. (5.2.19, 5.3.3)). Consequently, there are no connections such as $\Phi_{A\beta}{}^{\underline{c}}$ or $\Phi_{A\beta}{}^{\dot{\gamma}}$, and $\Phi_{A\underline{b}}{}^{\underline{c}} = \Phi_{A\beta}{}^\gamma \delta_{\dot{\beta}}{}^{\dot{\gamma}} + h.\,c.$, etc. We can view this restriction as a constraint: It has the consequence that the Bianchi identities now give algebraic relations among the field strengths. Because we have chosen the action of the Lorentz group to be reducible, we have imposed the constraints

$$R_{AB\,\underline{c}}{}^{\delta} = R_{AB\,\underline{\gamma}}{}^{\gamma} = R_{AB\dot{\underline{\gamma}}}{}^{\delta} = R_{AB\gamma}{}^{\underline{d}} = 0 \quad ,$$

$$R_{AB\underline{\gamma}}{}^{\underline{\delta}} = R_{AB\gamma}{}^{\delta} \delta_c{}^d \quad , \quad R_{AB\underline{c}}{}^{\underline{d}} = R_{AB\gamma}{}^{\delta} \delta_{\dot{\gamma}}{}^{\dot{\delta}} + R_{AB\dot{\gamma}}{}^{\dot{\delta}} \delta_\gamma{}^\delta \quad , \tag{5.4.3}$$

and their hermitian conjugates. These constraints are sufficient to express all of the curvatures $R$ and gauge field strengths $F$ in terms of the torsions $T$ if we assume that ${E_A}^M$ transforms under the action of $\Omega_i$; otherwise $F^i$ remains as an independent object. In particular, in the presence of central charges the corresponding field strengths can remain as independent quantities. We write the transformation in terms of matrices ${(\Omega_i)_A}^B$:

$$[\Omega_i , \nabla_A] = {(\Omega_i)_A}^B \nabla_B \tag{5.4.4}$$

where the only nonvanishing ${(\Omega_i)_A}^B$ are

$$ {(\Omega_i)_a}^b {\delta_\alpha}^\beta \quad , \quad {(\Omega_i)^a}_b {\delta_{\dot\alpha}}^{\dot\beta} \quad , \quad {(\Omega_i)_{\underline{a}}}^{\underline{b}} \quad , \quad {(\Omega_i)_a}^b = \left({(\Omega_i)^a}_b\right)^\dagger \tag{5.4.5}$$

The Bianchi identities follow from the Jacobi identities $0 = [[\nabla_{[A} , \nabla_B\}, \nabla_{C)}\} = {B_{ABC}}^E \nabla_E + B_{ABC}$ and are:

$${B_{ABC}}^E \equiv - {\Delta_{ABC}}^E + {R_{[AB,C)}}^E + {F_{[AB,C)}}^E = 0 \quad , \tag{5.4.6a}$$

$$B_{ABC}(M) \equiv - \nabla_{[A} R_{BC)}(M) + {T_{[AB|}}^D R_{D|C)}(M) = 0 \quad , \tag{5.4.6b}$$

$${B_{ABC}}^i \equiv - \nabla_{[A} {F_{BC)}}^i + {T_{[AB|}}^D {F_{D|C)}}^i = 0 \tag{5.4.6c}$$

where

$${\Delta_{ABC}}^E \equiv \nabla_{[A} {T_{BC)}}^E - {T_{[AB|}}^D {T_{D|C)}}^E \quad , \quad {F_{ABC}}^E \equiv {F_{AB}}^i {(\Omega_i)_C}^E \quad . \tag{5.4.7}$$

The Bianchi identities are satisfied identically simply because the field strengths are constructed out of the potentials ${E_A}^M$, $\Phi_A$, and $\Gamma_A$. In (5.4.6a) by decomposing ${B_{ABC}}^E$ into irreducible pieces under the Lorentz and internal symmetry groups, we express $F$ and $R$ in terms of $T$; this solution automatically satisfies $B_{ABC}(M) = {B_{ABC}}^i = 0$, so that (5.4.6b-c) contain no useful information.

To organize the analysis of the Bianchi identities, we classify the identities by (mass) dimension. The lowest dimension identities have dimension $\frac{1}{2}$: ${B_{\underline{\alpha\beta\gamma}}}^{d} = {B_{\underline{\alpha\beta\dot\gamma}}}^{d} = 0$ and hermitian conjugates. The highest dimension identities have dimension 3: ${B_{\underline{abc}}}^i = B_{\underline{abc}}(M) = 0$. Ordinarily, we start with the lowest dimension identities and work our way up; however, the dimension $\frac{1}{2}$ $B$'s are relations among the torsions only (they are independent of the curvatures and field strengths). Therefore, to determine $R$ and $F$ we start with the dimension 1 $B$'s. For example, ${B_{\underline{\alpha\beta\dot\gamma}}}^{\dot\epsilon} = 0$ implies

$$R_{\underline{\alpha\beta},\dot{\gamma}}{}^{\dot{\epsilon}}\delta^c{}_e + F_{\underline{\alpha\beta},}{}^c{}_e\delta_{\dot{\gamma}}{}^{\dot{\epsilon}} = \Delta_{\underline{\alpha\beta}\underline{\dot{\gamma}}}{}^{\underline{\dot{\epsilon}}} \tag{5.4.8}$$

(The graded symmetrization drops out because of the constraints (5.4.3) and (5.4.5)). To extract $F_{\underline{\alpha\beta},}{}^c{}_e$ and $R_{\underline{\alpha\beta},\dot{\gamma}}{}^{\dot{\epsilon}}$ from (5.4.8), we decompose it into Lorentz irreducible pieces. This gives:

$$F_{\underline{\alpha\beta},}{}^c{}_e = \frac{1}{2}\,\Delta_{\underline{\alpha\beta},\dot{\epsilon}}{}^{c\dot{\epsilon}}{}_e \tag{5.4.9a}$$

$$R_{\underline{\alpha\beta},\dot{\gamma}}{}^{\dot{\epsilon}} = \frac{1}{2N}\,\Delta_{\underline{\alpha\beta},(\dot{\gamma}}{}^{c\dot{\epsilon})}{}_c \tag{5.4.9b}$$

Proceeding in a similar manner, we determine *all* the remaining curvatures and gauge field strengths in terms of the torsions:

$$R_{\underline{\alpha}\,,\,\underline{\beta}\,\gamma\delta} = \frac{1}{4}\,\Delta_{\underline{\alpha}\,,\,\underline{\beta}\,,\,(\gamma\dot{\delta}\,,\,\delta)}{}^{\dot{\delta}} \quad , \tag{5.4.9c}$$

$$F_{\underline{\alpha}\,,\,\underline{\beta}\,c}{}^d = \frac{1}{2}\,\Delta_{\underline{\alpha}\,,\,\underline{\beta}\,,}{}^d{}_{\dot{\gamma}\,,\,c}{}^{\dot{\gamma}} \quad . \tag{5.4.9d}$$

Furthermore we find

$$R_{AB\,\gamma\delta} = \frac{1}{2N}\,\Delta_{AB\,d(\gamma,}{}^d{}_{\delta)} \quad , \tag{5.4.9e}$$

$$F_{AB\,c}{}^d = \frac{1}{2}\,\Delta_{AB\,c\gamma,}{}^{d\gamma} \quad , \tag{5.4.9f}$$

for $(A, B) = (\underline{\dot{\alpha}}\,,\underline{\dot{\beta}})$, $(\underline{\dot{\alpha}}\,,\underline{b})$, and $(\underline{a}\,,\underline{b})$, and finally

$$\begin{aligned} R_{\underline{\alpha}\,,\,B\,\gamma\delta} = \frac{1}{4}\,[\,(\frac{1}{N+1})(\,\Delta_{B\,,\,(a|\alpha\,,\,|c)(\gamma\,,}{}^c{}_{\delta)} + \frac{1}{3}\,\Delta^+{}_{B\,,\,a(\gamma}C_{\delta)\alpha}\,) \\ + (\frac{1}{N-1})(\,\Delta_{B\,,\,[a|\alpha\,,\,|c](\gamma\,,}{}^c{}_{\delta)} + \Delta^-{}_{B\,,\,a(\gamma}C_{\delta)}\,)\,]\,, \end{aligned} \tag{5.4.9g}$$

$$\begin{aligned} F_{\underline{\alpha}\,,\,B\,c}{}^d = \frac{1}{2}\,[\,\frac{1}{3}\,\Delta_{B\,,\,(a|\alpha\,,\,|c)\gamma\,,}{}^{d\gamma} + \Delta_{B\,,\,[a|\alpha\,,\,|c]\gamma\,,}{}^{d\gamma}\,] \\ - \frac{1}{2}\,[\,\frac{1}{3}\,R_{(a|\epsilon\,,\,B\,\alpha}{}^{\epsilon}\delta_{|c)}{}^d - R_{[a|\epsilon\,,\,B\,\alpha}{}^{\epsilon}\delta_{|c]}{}^d\,]\,, \end{aligned} \tag{5.4.9h}$$

$$\Delta^+{}_{B\,,\,\underline{\alpha}} = \Delta_{B\,,\,(a|\alpha\,,\,|c)\gamma}{}^{c\gamma} \quad , \quad \Delta^-{}_{B\,,\,\underline{\alpha}} = \Delta_{B\,,\,[a|\alpha\,,\,|c]\gamma}{}^{c\gamma} \,,$$

for $(B) = (\underline{\dot{\beta}})$ and $(\underline{b})$. This solution automatically satisfies (5.4.6b-c), as can be verified by direct computation. From now on we need only consider the Bianchi identities in (5.4.6a) that are independent of $R$ and $F$.

We now specialize to $N=1$ supergravity: Our tangent space transformations contain only the Lorentz group and $U(1)$; i.e., $\Omega_i = Y$. We impose:

$$T_{\alpha\beta}{}^{\gamma} = T_{\alpha\beta}{}^{\dot\gamma} = T_{\alpha\beta}{}^{\underline{c}} = T_{\alpha\dot\beta}{}^{\dot\gamma} = T_{\alpha\dot\beta}{}^{\underline{c}} - i\delta_\alpha{}^\gamma\delta_{\dot\beta}{}^{\dot\gamma} = 0 \quad , \tag{5.4.10a}$$

$$T_{\alpha,(\beta\dot\beta}{}^{\gamma)\dot\beta} = T_{\alpha,\beta(\dot\beta}{}^{\beta\dot\gamma)} = T_{\alpha\underline{b}}{}^{\underline{b}} = F_{\alpha\dot\beta} = 0 \quad . \tag{5.4.10b}$$

We proceed as above, starting with the lowest dimension identities. For example,

$$B_{\alpha,\beta,\dot\gamma}{}^{\underline{d}} = \nabla_\alpha T_{\beta,\dot\gamma}{}^{\underline{d}} + \nabla_\beta T_{\dot\gamma,\alpha}{}^{\underline{d}} + \overline{\nabla}_{\dot\gamma} T_{\alpha,\beta}{}^{\underline{d}}$$

$$+ T_{\alpha,\beta}{}^{E} T_{E,\dot\gamma}{}^{\underline{d}} + T_{\beta,\dot\gamma}{}^{E} T_{E,\alpha}{}^{\underline{d}} + T_{\dot\gamma,\alpha}{}^{E} T_{E,\beta}{}^{\underline{d}} \tag{5.4.11}$$

but $T_{\alpha,\beta}{}^{E} = 0$ and $T_{\alpha,\dot\beta}{}^{E} = i\delta_\alpha{}^\epsilon\delta_{\dot\beta}{}^{\dot\epsilon}$ which implies

$$0 = B_{\alpha,\beta,\dot\gamma}{}^{\underline{d}} = -\,i\delta_{\dot\gamma}{}^{\dot\epsilon} T_{\alpha,\beta\dot\gamma}{}^{\delta\dot\delta} - i\delta_{\dot\gamma}{}^{\dot\epsilon} T_{\beta,\alpha\dot\gamma}{}^{\delta\dot\delta} \ . \tag{5.4.12}$$

Next we decompose $T_{\alpha,\underline{b}}{}^{\underline{c}}$ into irreducible representations of the Lorentz group,

$$T_{\alpha,\underline{b},\underline{c}} = C_{\dot\beta\dot\gamma}[\,\tilde f^1{}_{\alpha\beta\gamma} + C_{\alpha(\beta}\tilde f^2{}_{\gamma)} + C_{\beta\gamma}\tilde f^3{}_\alpha\,]$$

$$+ [\,\tilde f^4{}_{(\alpha\beta\gamma)(\dot\beta\dot\gamma)} + C_{\alpha(\beta}\tilde f^5{}_{\gamma)(\dot\beta\dot\gamma)} + C_{\beta\gamma}\tilde f^6{}_{\alpha(\dot\beta\dot\gamma)}\,] \tag{5.4.13}$$

and note $T_{\alpha,\underline{b}}{}^{\underline{b}} = 0$ implies $\tilde f^3 = 0$, $T_{\alpha,(\beta\dot\beta}{}^{\gamma)\dot\beta} = 0$ implies $\tilde f^2 = \tilde f^1 = 0$,and finally $T_{\alpha,\beta(\dot\beta}{}^{\beta\dot\gamma)} = 0$ implies $\tilde f^6 = 0$. Now substituting (5.4.13) into (5.4.12) leads to

$$0 = -\,i2\tilde f^4{}_{(\alpha\beta\gamma)(\dot\gamma\dot\delta)} - i2C_{\alpha(\beta}\tilde f^5{}_{\gamma)(\dot\gamma\dot\delta)} \ , \tag{5.4.14}$$

which yields $\tilde f^4{}_{(\alpha\beta\gamma)(\dot\gamma\dot\delta)} = \tilde f^5{}_{\beta(\dot\gamma\dot\delta)} = 0$. In other words $T_{\alpha,\underline{b}}{}^{\underline{c}}$ vanishes identically.

This example shows how we decompose the torsions into irreducible representations of the Lorentz group, and then solve the constraints by looking at what they imply about the various irreducible parts. For two component spinors this decomposition simply consists of symmetrizing and antisymmetrizing in all possible ways. Other examples are

$$T_{\dot{\alpha},\beta\dot{\beta},\gamma} = C_{\dot{\alpha}\dot{\beta}}C_{\beta\gamma}X^1 + C_{\beta\gamma}X^2{}_{(\dot{\alpha}\dot{\beta})} + C_{\dot{\alpha}\dot{\beta}}X^3{}_{(\beta\gamma)} + X^4{}_{(\dot{\alpha}\dot{\beta})(\beta\gamma)} ,$$

$$T_{\dot{\alpha},\beta\dot{\beta},\dot{\gamma}} = \tilde{X}^1{}_{\beta(\dot{\alpha}\dot{\beta}\dot{\gamma})} + \tilde{X}^2{}_{\beta(\dot{\alpha}}C_{\dot{\beta})\dot{\gamma}} + \tilde{X}^3{}_{\beta\dot{\gamma}}C_{\dot{\alpha}\dot{\beta}} . \qquad (5.4.15)$$

These expressions are now substituted into the Bianchi identities which separate into several equations. These are then solved, with the result that some of the irreducible parts are zero, while others are expressible in terms of a minimal set.

The complete analysis is straightforward but tedious. We find that all the Bianchi identities and constraints are satisfied by

$$\{\nabla_\alpha , \overline{\nabla}_{\dot{\beta}}\} = i\nabla_{\alpha\dot{\beta}} \ ,$$

$$\{\overline{\nabla}_{\dot{\alpha}} , \overline{\nabla}_{\dot{\beta}}\} = -\,2R\overline{M}_{\dot{\alpha}\dot{\beta}} \ ,$$

$$[\overline{\nabla}_{\dot{\alpha}} , i\nabla_{\beta\dot{\beta}}] = C_{\dot{\beta}\dot{\alpha}}[\,-\,R\nabla_\beta - G_\beta{}^{\dot{\gamma}}\overline{\nabla}_{\dot{\gamma}} + (\overline{\nabla}^{\dot{\gamma}}G_{\beta\dot{\delta}})\overline{M}_{\dot{\gamma}}{}^{\dot{\delta}} - iW_\beta Y$$

$$-\,\frac{1}{3}\,iW_\gamma M_\beta{}^\gamma + W_{\beta\gamma}{}^\delta M_\delta{}^\gamma] + (\nabla_\beta R)\overline{M}_{\dot{\alpha}\dot{\beta}} \ ,$$

$$[i\nabla_{\alpha\dot{\alpha}} , i\nabla_{\beta\dot{\beta}}] = C_{\dot{\beta}\dot{\alpha}}f_{\alpha\beta} - h.\,c. \qquad (5.4.16)$$

where the operator $f_{\alpha\beta}$ is defined by

$$f_{\alpha\beta} = -\,i\,\frac{1}{2}\,G_{(\alpha}{}^{\dot{\gamma}}\nabla_{\beta)\dot{\gamma}} - \frac{1}{2}\,(\,\nabla_{(\alpha}R - i\,\frac{1}{3}\,W_{(\alpha})\nabla_{\beta)} + W_{\alpha\beta}{}^\gamma\nabla_\gamma - \frac{1}{2}\,(\,\nabla_{(\alpha}G_{\beta)}{}^{\dot{\gamma}})\overline{\nabla}_{\dot{\gamma}}$$

$$-\,i\,\frac{1}{2}\,(\nabla_{(\alpha}W_{\beta)})Y - W_{\alpha\beta\gamma\delta}M^{\gamma\delta} - i\,\frac{1}{8}\,[(\nabla_{(\gamma}{}^{\dot{\alpha}}G_{\alpha)\dot{\alpha}})M_\beta{}^\gamma + \alpha \longleftrightarrow \beta]$$

$$+\,(\overline{\nabla}^2\overline{R} + 2R\overline{R} + i\,\frac{1}{6}\,\nabla^\gamma W_\gamma)M_{\alpha\beta} + \frac{1}{2}\,(\nabla_{(\alpha}\overline{\nabla}^{\dot{\gamma}}G_{\beta)\dot{\delta}})M_{\dot{\gamma}}{}^{\dot{\delta}} \ , \qquad (5.4.17)$$

and

$$W_{\alpha\beta\gamma\delta} \equiv \frac{1}{4!}\,\nabla_{(\alpha}W_{\beta\gamma\delta)} \ .$$

The independent tensors $R$, $G_{\underline{a}}$, and $W_{\alpha\beta\gamma}$, and the dependent one $W_\alpha$ satisfy the relations

$$G_{\underline{a}} = \overline{G}_{\underline{a}} \ , \ \overline{\nabla}_{\dot{\alpha}}R = \overline{\nabla}_{\dot{\alpha}}W_{\alpha\beta\gamma} = \overline{\nabla}_{\dot{\alpha}}W_\alpha = 0 \ ,$$

$$\overline{\nabla}^{\dot{\alpha}} G_{\alpha\dot{\alpha}} = \nabla_{\alpha} R + i W_{\alpha} \ ,$$

$$\nabla^{\alpha} W_{\alpha\beta\gamma} + \frac{1}{3} i \nabla_{(\beta} W_{\gamma)} = \frac{1}{2} i \nabla_{(\beta}{}^{\dot{\alpha}} G_{\gamma)\dot{\alpha}} \ ,$$

$$\nabla^{\alpha} W_{\alpha} + \overline{\nabla}^{\dot{\alpha}} \overline{W}_{\dot{\alpha}} = 0 \ . \tag{5.4.18}$$

Therefore, all the torsions, curvatures, and field strengths of (5.4.2) are expressible in terms of the three covariant superfields $W_{\alpha\beta\gamma}$, $G_{\alpha\dot{\alpha}}$, and $R$, and their derivatives.

By considering the coefficient of $M_{\gamma\delta}$ on both sides of the equation for the commutator of two vectorial derivatives, we conclude that the superspace analog of the decomposition in (5.1.21) takes the form

$$R_{\underline{ab}}{}^{\gamma\delta} = C_{\dot{\alpha}\dot{\beta}}[\, W_{\alpha\beta}{}^{\gamma\delta} - \frac{1}{2} \delta_{(\alpha}{}^{\gamma} \delta_{\beta)}{}^{\delta} (\overline{\nabla}^2 \overline{R} + 2R\overline{R} + i \frac{1}{6} \nabla^{\gamma} W_{\gamma}) + \frac{1}{2} \delta_{(\alpha}{}^{(\gamma} X_{\beta)}{}^{\delta)}]$$

$$+ \, C_{\beta\alpha} \frac{1}{4} (\overline{\nabla}_{(\dot{\alpha}} \nabla^{(\gamma} G^{\delta)}{}_{\dot{\beta})}) \ , \tag{5.4.19}$$

where $X_{\alpha}{}^{\beta}$ is defined by

$$X_{\alpha\beta} = -\, i \frac{1}{8} \nabla_{(\alpha}{}^{\dot{\alpha}} G_{\beta)\dot{\alpha}} \ . \tag{5.4.20}$$

By comparing this to (5.1.21), we see there is a representation $X_{\alpha\beta}$ present in the supercovariant curvature $R_{\underline{ab}}{}^{\gamma\delta}$ that was absent in the component curvature $r_{\underline{ab}}{}^{\gamma\delta}$. This occurs because the constraints that we have chosen imply there is nontrivial $x$-space torsion $T_{\underline{abc}} \sim \epsilon_{\underline{abcd}} G^{\underline{d}}$ present.

We now choose the scale $\oplus\, U(1)$ gauge where the compensator equals 1. For $n = 0$ the only resulting modification of (5.4.16) is that we set $R = 0$ (thus, for $n = 0$, the *spinor* derivatives (but not the vector derivatives!) obey the global supersymmetry algebra). For other $n$ it is necessary to drop the $Y$ part of the covariant derivatives. However, for $n = -\frac{1}{3}$, $\Gamma_A$ vanishes identically in this gauge, so the only modification of (5.4.16) is to set $W_{\alpha} = 0$ (see discussion before (5.3.72)). In this case, (5.4.16) reduces to (5.2.81). For $n \neq 0, -\frac{1}{3}$ the modifications are slightly more complicated: (1) The spinor $U(1)$ connection $\Gamma_{\alpha}$ is now covariant; to avoid confusion, we define $T_{\alpha} \equiv -\, i\Gamma_{\alpha}(\textit{degauged})$. (2) No tensors are set to zero, but now $R$ is determined by the compensator constraint (see (5.3.73)), and $W_{\alpha}$ by its explicit form (see (5.3.74)). (3)

Separating out the $Y$ parts causes shifts in a few of the torsions (see (5.3.72)) by the covariant quantity $T_\alpha$. The resulting form of (5.4.16) is

$$\{\nabla_\alpha , \overline{\nabla}_{\dot\alpha}\} = i\nabla_{\alpha\dot\alpha} - \frac{1}{2}(T_\alpha \overline{\nabla}_{\dot\alpha} + \overline{T}_{\dot\alpha}\nabla_\alpha)$$

$$\{\overline{\nabla}_{\dot\alpha} , \overline{\nabla}_{\dot\beta}\} = \frac{1}{2}\overline{T}_{(\dot\alpha}\overline{\nabla}_{\dot\beta)} - 2R\overline{M}_{\dot\alpha\dot\beta}$$

$$[\overline{\nabla}_{\dot\alpha} , i\nabla_{\beta\dot\beta}] = C_{\dot\alpha\dot\beta}[R\nabla_\beta + G_\beta{}^{\dot\gamma}\overline{\nabla}_{\dot\gamma}] - \frac{1}{2}(\nabla_\beta \overline{T}_{\dot\beta} - \overline{\nabla}_{\dot\beta}T_\beta + T_\beta \overline{T}_{\dot\beta})\overline{\nabla}_{\dot\alpha}$$

$$+ C_{\dot\alpha\dot\beta}[-(\overline{\nabla}^{\dot\gamma}G_{\beta\dot\delta})\overline{M}_{\dot\gamma}{}^{\dot\delta} - W_{\beta\gamma}{}^{\delta}M_\delta{}^\gamma + i\frac{1}{3}W_\gamma M_\beta{}^\gamma] + ((\nabla_\beta + T_\beta)R)\overline{M}_{\dot\alpha\dot\beta} \quad , \quad (5.4.21)$$

with (5.4.18) modified by $\nabla_\alpha \to \nabla_\alpha + T_\alpha Y$ and $\overline{\nabla}_{\dot\alpha} \to \overline{\nabla}_{\dot\alpha} - \overline{T}_{\dot\alpha}Y$. This is just the $U(1)$ degauging described in sec. 5.3.b.7.

This result corresponds to one of the many forms of nonminimal $n \neq -\frac{1}{3}$ $N = 1$ supergravity. As explained in sec. 5.3, covariant derivatives can be redefined by shifting with contortions. As an example, we note that the two anticommutators in (5.4.21) can be simplified by the shift

$$\nabla_{\alpha\dot\alpha} \to \nabla_{\alpha\dot\alpha} - i\frac{1}{2}(T_\alpha \overline{\nabla}_{\dot\alpha} + \overline{T}_{\dot\alpha}\nabla_\alpha) \quad , \quad (5.4.22)$$

which gives (5.2.82). For the remainder of the book, unless otherwise stated, our covariant derivatives for $N = 1$ supergravity will be in the gauge with the tensor compensator set to one, and for $n \neq 0$, with the $Y$ parts dropped.

## 5.5. Actions

In this section we construct and discuss superspace actions for matter systems coupled to supergravity, and for supergravity itself.

### a. Review of vector and chiral representations

In the vector representation (where, e.g., $\overline{\nabla}_{\dot{\alpha}} = (\nabla_\alpha)^\dagger$), a covariant superfield $X_{\alpha\dot{\beta}\cdots}$ transforms under the gauge transformations of local supersymmetry (hermitian supercoordinate and tangent space transformations) as

$$X' = e^{iK} X e^{-iK} \tag{5.5.1a}$$

with $K = \bar{K} = K^M iD_M + K_\alpha{}^\beta iM_\beta{}^\alpha + \bar{K}_{\dot{\alpha}}{}^{\dot{\beta}} i\bar{M}_{\dot{\beta}}{}^{\dot{\alpha}}$ (cf. (5.3.3)). In chiral ($\overline{\nabla}_{\dot{\alpha}} \sim \bar{D}_{\dot{\alpha}}$) or antichiral ($\nabla_\alpha \sim D_\alpha$) representations, the transformation laws are

$$X^{(+)}{}' = e^{i\Lambda} X^{(+)} e^{-i\Lambda} \quad , \tag{5.5.1b}$$

$$X^{(-)}{}' = e^{i\bar{\Lambda}} X^{(-)} e^{-i\bar{\Lambda}} \quad , \tag{5.5.1c}$$

respectively, with $\Lambda$, $\bar{\Lambda}$ given by, e.g., (5.3.33-35), and

$$X^{(+)} = e^{-H} X^{(-)} e^{H} = e^{-\bar{\Omega}} X e^{\bar{\Omega}} \quad . \tag{5.5.2}$$

(Recall that the hermitian conjugate of an object in the chiral representation is in the antichiral representation and transforms with $\bar{\Lambda}$ (5.5.1c). Therefore, just as in (5.2.28) and in Yang-Mills theory (4.2.21), we must convert the conjugate to an object transforming with $\Lambda$ (5.5.1b), and define the chiral representation conjugate $\tilde{X}^{(+)} = e^{-H}(X^{(+)})^\dagger e^{H}$.) The transformation properties of $E^{-1}$ in the vector, chiral, and antichiral representations are

$$E^{-1}{}' = E^{-1} e^{i\overleftarrow{K}} \tag{5.5.3a}$$

$$E^{(+)-1}{}' = E^{(+)-1} e^{i\overleftarrow{\Lambda}} \tag{5.5.3b}$$

$$E^{(-)-1}{}' = E^{(-)-1} e^{i\overleftarrow{\bar{\Lambda}}} \tag{5.5.3c}$$

respectively.

**b. The general measure**

Using the results of the previous sections, it is straightforward to construct locally supersymmetric actions: We covariantize all derivatives (with possibly some ambiguity in whether we use minimal coupling or add contortion terms), including the derivatives used to define constrained matter fields (such as chiral fields), and we covariantize the measure. For integrals over all superspace, by analogy with ordinary space (see 5.1.23-5), we use $E^{-1}$ as a density to define a covariant measure, and write actions of the form:

$$S = \int d^4x d^4\theta \, E^{-1} \, I\!\!L_{gen} \quad , \tag{5.5.4}$$

where $I\!\!L_{gen}$ is a general real scalar superfield constructed out of covariant matter fields, derivatives, etc. Since by construction $I\!\!L_{gen}$ must transform as in (5.5.1), and since $E^{-1}$ transforms as in (5.5.3), the expression in (5.5.4) is invariant. (Recall that coordinate invariance is defined only up to surface terms (5.1.27).) This type of expression is the integrated version of what is referred to as a "D - type" density formula.

**c. Tensor compensators**

Just as in gravity (sec. 5.1.d), we can generalize the coordinate invariant measure (5.5.4) to a scale (and $U(1)$) invariant measure by introducing tensor compensators. For an $I\!\!L_{gen}$ that has scale weight $d$ (the reality of the action implies that it must be $U(1)$ invariant), we have (see (5.3.8,14))

$$n = -\frac{1}{3}: \qquad S = \int d^4x \, d^4\theta \; E^{-1}(\Phi\bar{\Phi})^{1-\frac{d}{2}} \, I\!\!L_{gen} \quad , \tag{5.5.5a}$$

$$n \neq -\frac{1}{3}, 0: \qquad S = \int d^4x \, d^4\theta \; E^{-1}(\Sigma\bar{\Sigma})^{\frac{3n+1}{2}(1-\frac{d}{2})} I\!\!L_{gen} \quad , \tag{5.5.5b}$$

$$n = 0: \qquad S = \int d^4x \, d^4\theta \; E^{-1}G^{1-\frac{d}{2}} I\!\!L_{gen} \quad . \tag{5.5.5c}$$

These actions reduce to (5.5.4) in the gauge where the compensator is 1. (The analogous expression in ordinary gravity is $\int d^4x \, d^4\theta \; \mathrm{e}^{-1}\phi^{4-d}L$ where $\phi$ is the tensor-type component scale compensator introduced in (5.1.33) and $L$ is a scale weight $d$ Lagrangian.)

**d. The chiral measure**

In the chiral representation, a covariantly chiral scalar superfield $\Phi^{(+)}\, , \overline{\nabla}_{\dot{\alpha}}\Phi^{(+)} = 0$ is chiral in the flat superspace sense $\bar{D}_{\dot{\alpha}}\Phi^{(+)} = 0$. Therefore, its transformation (5.5.2b) can be written as

$$\Phi^{(+)\,\prime} = e^{i\Lambda}\Phi^{(+)}e^{-i\Lambda} = e^{i\Lambda_{ch}}\Phi^{(+)}e^{-i\Lambda_{ch}} \qquad (5.5.6a)$$

where

$$\Lambda_{ch} = \Lambda \,(\text{for}\, \Lambda^{\dot{\mu}} = 0) = \Lambda^{\underline{m}} i\partial_{\underline{m}} + \Lambda^{\mu} iD_{\mu} \quad . \qquad (5.5.6b)$$

Since for $n = -\frac{1}{3}$ the transformation of $\phi^3$ is (5.2.68)

$$\phi^{3\,\prime} = \phi^3 e^{i\overleftarrow{\Lambda}_{ch}} \quad , \quad \overleftarrow{\Lambda}_{ch} = \overleftarrow{\Lambda}\ (\text{for}\, \Lambda^{\dot{\mu}} = 0) = (\Lambda^{\underline{m}} i\overleftarrow{\partial}_{\underline{m}} + \Lambda^{\mu} i\overleftarrow{D}_{\mu}) \ , \qquad (5.5.7)$$

the quantity $\phi^3$ is a suitable chiral density to covariantize the flat space chiral measure (see sec. 5.2.c).

$$S_{n=-\frac{1}{3}} = \int d^4x\, d^2\theta\ \phi^3\ I\!L_{chiral} \quad . \qquad (5.5.8)$$

This is the integrated version of an "F - type" density multiplet. For other values of $n$ no dimensionless chiral density exists. We describe how this situation is handled below.

**e. Representation independent form of the chiral measure**

For $n = -\frac{1}{3}$, we can write the chiral measure in terms of the real measure. From (5.3.66b) we have, in *chiral representation,*

$$\bar{D}^2 E^{-1} I\!L = \phi^3 (\overline{\nabla}^2 + R) I\!L \quad . \qquad (5.5.9)$$

Thus

$$\int d^4x\, d^4\theta\ E^{-1} I\!L_{gen} = \int d^4x\, d^2\theta\ \phi^3 (\overline{\nabla}^2 + R) I\!L_{gen} \quad . \qquad (5.5.10)$$

From (5.3.66a) we can find the vector representation of (5.5.10):

$$\int d^4x\, d^4\theta\ E^{-1} I\!L_{gen} = \int d^4x\, d^2\theta\ e^{-\overleftarrow{\Omega}} \phi^3 (\overline{\nabla}^2 + R) I\!L_{gen} \quad . \qquad (5.5.11)$$

If we choose $I\!L_{gen} = R^{-1} I\!L_{chiral}$, since $\overline{\nabla} I\!L_{chiral} = \overline{\nabla} R = 0$,

$$\int d^4xd^2\theta\ \phi^3 I\!\!L_{chiral} = \int d^4xd^4\theta\ E^{-1}R^{-1}I\!\!L_{chiral} \quad . \tag{5.5.12}$$

The form of the chiral measure $d^4xd^4\theta\, E^{-1}R^{-1}$ is valid in *all* representations since it does not depend on the existence of a chiral density. It is manifestly covariant and, in principle, could be used for all $n \neq 0$ ($R = 0$ for $n = 0$). Thus "F - type" density multiplets also exist for *nonminimal* supergravity. However, unless $n = -\frac{1}{3}$, (5.5.12) leads to component actions containing inverse powers of the auxiliary fields (with the exception of R-invariant systems: see below).

The $U(1)$-covariant form of the chiral measure ($n = -\frac{1}{3}$) is somewhat more subtle: $U(1)$ invariance alone gives the analog of (5.5.12) as

$$S = \int d^4x\, d^4\theta\ E^{-1}R^{-1}\Phi^{3(1-\frac{1}{2}w)} I\!\!L_{chiral} \tag{5.5.13}$$

when $[Y\, , I\!\!L_{chiral}] = \frac{1}{2}\, w I\!\!L_{chiral}$. In particular, superconformal actions always have $w = 2$. However, scale invariance is not so straightforward: Using (5.3.10), we see that $R$ has an inhomogeneous term in its transformation law proportional to $\overline{\nabla}^2 L$, but because of the chirality of $R$, $\Phi$, and $I\!\!L_{chiral}$ this term vanishes upon integration by parts. Thus the chirality of the compensator is essential for constructing general chiral actions. Since in the ($U(1)$-)chiral representation $\Phi = \phi$, using (5.3.64a), we reobtain (5.5.8). The expression (5.5.13) can be used for $n \neq -\frac{1}{3}\, , 0$ if $w = 2$, since then the action is $\Phi$ independent and consequently superconformal.

### f. Scalar multiplet

To discuss specific couplings to matter, we first consider the $U(1)$-covariant form. According to our general prescription, the direct covariantization of the action (4.1.1) for the free scalar multiplet is

$$S = \int d^4x\, d^4\theta\ E^{-1}\ \eta\bar{\eta} \quad , \tag{5.5.14a}$$

with covariantly chiral $\eta$

$$\overline{\nabla}_{\dot{\alpha}}\eta = 0 \quad , \tag{5.5.14b}$$

This form of the action is valid for any $n$. If we assign scale weight $d = 1$ to $\eta$ (see

(5.5.5)), the action is superconformal since it is independent of the tensor type compensators. In particular, it is invariant under the "restricted" superconformal transformations that survive in Poincaré supergravity (5.3.76). At the component level this action leads to conformally coupled scalar fields with actions as in (5.1.35) but with the opposite sign. (Compensators generally have actions with an overall minus sign relative to physical systems). Since the action is superconformal even without the compensators, it is clear that the scalars of the multiplet are conformally coupled to gravity without the need for a component calculation.

After degauging, the action (5.5.14a) and defining condition remain unchanged for $n = -\frac{1}{3}$. For the nonminimal theories, we use the same action but if we want the component scalars to be conformally coupled to gravity we must change the defining condition ((5.5.18) with $w = \frac{2}{3}$; see below). Alternatively if the defining condition is not modified and the $\overline{\nabla}$ operator in (5.5.14b) is for a degauged nonminimal theory, then the action of (5.5.14a) does not have conformally coupled scalars.

### f.1. Superconformal interactions

For superconformally invariant actions, the density compensator $\phi$ can be gauged away (i.e., removed by a field redefinition which is a superscale transformation). In this case, the action (5.5.8) written for $n = -\frac{1}{3}$ makes sense for any $n$. For example, since $E^{-1} = \phi\tilde{\phi}\hat{E}^{-\frac{1}{3}}(1 \cdot e^{-\overleftarrow{H}})^{\frac{1}{3}}$ (5.2.72), the conformally invariant action for a covariantly chiral scalar superfield is

$$S = \int d^8z\ E^{-1}\tilde{\eta}\eta + (\lambda\frac{1}{3!}\int d^6z\ \phi^3\eta^3 + h.c.)\ . \tag{5.5.15}$$

At the component level, the terms proportional to $\lambda$ describe quartic self-interactions and Yukawa couplings for the component fields of the matter chiral multiplet just as in the global case. The rescaling $\hat{\eta} \equiv \phi\eta$ removes $\phi$ from the action entirely, and the result is valid for any $n$ ($\hat{\eta}$ is a chiral density of weight $w = \frac{2}{3}$). This can be generalized slightly: To remove $\phi$ from the chiral integrands, full superconformal invariance is not required; R-invariance (3.6.14) is sufficient. Then $\phi$ appears only in the full superspace integrand, and only in the combination $\phi\tilde{\phi}$; in that case, the $n = -\frac{1}{3}$ compensator $\phi$ can be replaced by $n = 0$ or nonminimal compensators. For example, in the case of a single chiral multiplet, we can generalize to a rescaled chiral density $\hat{\eta}$ with arbitrary weight $w$; see (5.5.20) and the paragraph after (5.5.32) below.

### f.2. Conformally noninvariant actions

Nonconformal couplings of the scalar multiplet are also possible. We can always add the supersymmetric term

$$S_{nonconf} = \int d^4x\, d^4\theta\ E^{-1}(\eta^2 + \bar{\eta}^2) \tag{5.5.16}$$

This actually vanishes for $n = 0$. (It is also possible to write a CP non-conserving term by taking $i$ time the difference instead of the sum in (5.5.16).) At the component level, (5.5.16) generates "dis-improvement" terms $r(\mathrm{A}^2 - \mathrm{B}^2) + \cdots$ with opposite contributions for the scalar and pseudoscalar fields. Therefore, (5.5.16) cannot be used to eliminate the improvement terms of both fields. For $n = -\frac{1}{3}$, we can rewrite (5.5.16) as the chiral integral

$$S_{nonconf} = \int d^4x\, d^2\theta\ \phi^3\, R\, \eta^2 + h.\,c. \tag{5.5.17}$$

For nonminimal $(n \neq -\frac{1}{3}, 0)$ supergravity (5.5.16) also introduces "dis-improvement" terms but there exists another way of introducing such nonconformal terms for the scalar multiplet. Before degauging, if the scale weight of $\eta$ is not 1, then the only way to write a superconformal kinetic action for the chiral multiplet is to introduce one of the density compensators of (5.5.5). Thus the action without the compensators is not superconformal. After degauging the $U(1)$ invariance (see sec. 5.3.b.8), we can use the "new" tensor $T_\alpha$ to define a modified chiral condition. We can replace (5.5.14b) by

$$(\overline{\nabla}_{\dot{\alpha}} + \frac{1}{2}\, w\, \overline{T}_{\dot{\alpha}})\eta = 0 \ . \tag{5.5.18}$$

The kinetic action is still given by (5.5.14a), and if $w \neq \frac{2}{3}$, it is not superconformal without one of the compensators of (5.5.5). Even though the $U(1)$ group is no longer gauged, it still exists as a global R-invariance of the action (5.5.14), and the constraint (5.5.18) is covariant even under local transformations

$$[Y\,, \eta] = \frac{1}{2}\, w\eta \quad . \tag{5.5.19}$$

The modified chirality condition of (5.5.18) in terms of unconstrained superfields leads to a more complicated action for the scalar multiplet. We can express $\eta$ in terms of a flat *chiral density* $\hat{\eta} \equiv e^{-\frac{1}{2}w\overline{T}}\eta\,, \overline{E}_{\dot{\alpha}}\hat{\eta} = 0$ (in the chiral representation $\overline{D}_{\dot{\alpha}}\hat{\eta} = 0$). In terms of unconstrained superfields, the action (5.5.14a) becomes (in the gauge with compensators set equal to one)

$$S = \int d^4x\, d^4\theta\ \hat{E}^{\tilde{n}}[(1 \cdot e^{-\overleftarrow{\Omega}})(1 \cdot e^{\overleftarrow{\overline{\Omega}}})]^{\frac{\tilde{n}+1}{2}}\,\hat{\eta}\hat{\overline{\eta}} \ \ , \tag{5.5.20}$$

where $\tilde{n}$ is defined in terms of $w$ by $w = 2\,\frac{n - \tilde{n}}{3n + 1}$. Only $\tilde{n} = -\frac{1}{3}$ is superconformal and has the conventional conformal improvement terms; then $w = \frac{2}{3}$. Thus in the nonminimal theories, conformal coupling for the scalars is achieved by replacing (5.5.14b) by (5.5.18) with $w = \frac{2}{3}$.

### f.3. Chiral self-interactions

The covariantization of any global chiral polynomial self-interaction terms $\mathrm{P}(\eta)$ is straightforward. From our general prescription (5.5.12) we have (for $n \neq 0$)

$$S_{int} = \int d^4x\, d^4\theta\ E^{-1}R^{-1}\mathrm{P}(\eta) + h.\,c. \ \ . \tag{5.5.21}$$

The expression (5.5.21) remains locally supersymmetric for fields satisfying the modified chirality condition (5.5.18) or in the $U(1)$-covariant formalism with the usual $\overline{\nabla}_{\dot{\alpha}}\eta = 0$ for arbitrary chiral weight. For $n = -\frac{1}{3}$, $S_{int}$ is polynomial in the component fields after the elimination of the supergravity auxiliary fields whenever $\mathrm{P}(\eta)$ is polynomial. For $n \neq -\frac{1}{3}$, it is in general nonpolynomial, except for $\mathrm{P}(\eta) = \eta^{\frac{2}{w}}$ (for a single chiral multiplet; for more multiplets, the condition is given below). Thus, as mentioned earlier, although "F - type" densities exist for nonminimal theories, in general these will lead to nonpolynomiality after the elimination of auxiliary fields.

In the chiral representation, for $n = -\frac{1}{3}$, (5.5.21) can be rewritten in the form (5.5.8), and for $n \neq -\frac{1}{3}$, in the form

$$S = \int d^4x\, d^2\theta\; e^{-\bar{T}}\, \mathrm{P}(\eta) + h.\,c. \tag{5.5.22}$$

when the chiral charge of $\mathrm{P}(\eta)$ is $\frac{1}{2}w = 1$. This follows from (5.5.13), since for $n \neq -\frac{1}{3}$, $S$ must be $\Phi$ independent. (For the special interaction given above, this can be written as $\int d^4x\, d^2\theta\; \hat{\eta}^{\frac{2}{w}}$, with no dependence on the supergravity fields.)

## g. Vector multiplet

The vector multiplet can be coupled to supergravity by simply defining derivatives that are covariant with respect to the local invariances of both supergravity and super-Yang-Mills:

$$\underset{\sim}{\nabla}_A = E_A + (\Phi_{A\beta}{}^{\gamma} M_{\gamma}{}^{\beta} + \Phi_{A\dot{\beta}}{}^{\dot{\gamma}} \bar{M}_{\dot{\gamma}}{}^{\dot{\beta}}) - i\, \Gamma_A{}^{\mathbf{A}} T_{\mathbf{A}} \quad ;$$

$$\underset{\sim}{\nabla}_A{}' = e^{iK} \underset{\sim}{\nabla}_A e^{-iK} \quad , \quad K = K^M iD_M + (K_{\alpha}{}^{\beta} iM_{\beta}{}^{\alpha} + h.\,c.\,) + K^{\mathbf{A}} T_{\mathbf{A}} \quad ; \tag{5.5.23}$$

where $\Gamma_A{}^{\mathbf{A}}$ is the Yang-Mills potential and $K^{\mathbf{A}}$ its gauge parameter. Field strengths for both supergravity and Yang-Mills are defined by the graded commutators of the covariant derivatives as usual, and the same supergravity and Yang-Mills constraints are imposed (see (4.2.66) and (5.3.4,5,13)). The solution to the constraints can be given by expressing $\underset{\sim}{\nabla}_A$ in terms of the pure supergravity covariant derivatives $\nabla_A$ and the usual Yang-Mills superpotential $\underset{\sim}{\Omega} = \Omega^{\mathbf{A}} T_{\mathbf{A}}$:

$$\underset{\sim}{\nabla}_\alpha = e^{-\underset{\sim}{\Omega}} \nabla_\alpha e^{\underset{\sim}{\Omega}} \quad , \quad \underset{\sim}{\nabla}_{\alpha\dot{\alpha}} = -\, i\{ \underset{\sim}{\nabla}_\alpha , \bar{\underset{\sim}{\nabla}}_{\dot{\alpha}} \} \quad . \tag{5.5.24}$$

Alternatively, the solution can be written with $\underset{\sim}{\nabla}$ of the same form as $\nabla$ but now with $\Omega = \Omega^M iD_M + \underset{\sim}{\Omega}$ (in analogy to $U \to U + V$ in the global case). The Yang-Mills field strength is

$$W_\alpha = i[\bar{\underset{\sim}{\nabla}}{}^{\dot{\alpha}} , \{ \bar{\underset{\sim}{\nabla}}_{\dot{\alpha}} , \underset{\sim}{\nabla}_\alpha \}] \tag{5.5.25}$$

and the action is

$$S = g^{-2}\, tr \int d^4x\, d^4\theta\; E^{-1} R^{-1} W^2 \quad . \tag{5.5.26}$$

For $n = -\frac{1}{3}$ in the Yang-Mills chiral representation, using the Bianchi identities (5.4.16,18) we can rewrite (5.5.25) as

$$W_\alpha = i(\overline{\nabla}^2 + R)e^{-V}\nabla_\alpha e^V \quad , \tag{5.5.27}$$

and the action is

$$S = g^{-2}\, tr \int d^4x\, d^2\theta\ \phi^3\ W^2 \quad . \tag{5.5.28}$$

As for the conformal coupling of the scalar multiplet in (5.5.8), the actions in (5.5.26,28) are invariant with respect to the conformal transformations parametrized by arbitrary $L$ and $K_5$ superfields. This ensures that the $\phi$ dependence of $W$ is such that it cancels in the action of (5.5.28). More generally, also as a consequence of conformal invariance, (5.5.26) is independent of $\phi$ or $\overline{T}$.

For $n = 0$ the form (5.5.26) cannot be used since $R = 0$ and (5.5.28) cannot be used since $\phi$ only occurs in the $n = \frac{1}{3}$ theory. The correct action is

$$S = g^{-2}\, tr \int d^4x\, d^4\theta\ E^{-1}\Gamma^\alpha(W_\alpha - \frac{1}{6}[\overline{\Gamma}^{\dot\alpha}, \Gamma_{\alpha\dot\alpha}]) + h.\,c. \tag{5.5.29}$$

where $W_\alpha$ is given by (5.5.25) and $\Gamma_\alpha, \Gamma_{\alpha\dot\alpha}$ are obtained from (5.5.24) using $-i\Gamma_A = \underset{\sim}{\nabla}_A - \nabla_A$. The gauge invariance of the action (5.5.29) follows from the Bianchi identity $\underset{\sim}{\nabla}^\alpha W_\alpha + \overline{\underset{\sim}{\nabla}}{}^{\dot\alpha}\overline{W}_{\dot\alpha} = 0$. We note that this form (valid for all $n$ and in all representations) is similar to the three-dimensional gauge invariant mass term (2.4.38). Alternatively, it is possible to use (5.5.28) for all $n$, if we are only interested in the explicit dependence on the supergravity prepotential $H^{\underline{m}}$. The $H^{\underline{m}}$ dependence and density type compensator independence of any *truly* conformal action is independent of $n$.

## h. General matter models

We now consider a general class of matter multiplets (chiral and gauge) coupled to $n = -\frac{1}{3}$ supergravity. A *globally* supersymmetric gauge invariant action, restricted only by the requirement that no bosonic terms with more than two derivatives or fermionic terms with more than one derivative appear in the component Lagrangian is

$$S = \int d^4x\, d^4\theta\, [I\!K(\Phi^i, \tilde{\Phi}_j) + \nu tr V]$$

$$+ \int d^4x\, d^2\theta\, [\mathrm{P}(\Phi^i) + \frac{1}{4} Q_{\mathrm{AB}}(\Phi^i) W^{\alpha \mathrm{A}} W_\alpha{}^{\mathrm{B}}] + h.c. \qquad (5.5.30)$$

where

$$\tilde{\Phi}_j = \overline{\Phi}_k (e^V)_j{}^k \quad , \qquad W_\alpha{}^{\mathrm{A}} = i\bar{D}^2 (e^{-V} D_\alpha e^V)^{\mathrm{A}} \qquad (5.5.31)$$

and $\mathrm{P}(\Phi^i)$ and $Q_{\mathrm{AB}}(\Phi^i) = \delta_{\mathrm{AB}} + O(\Phi)$ are chiral. The term $\nu tr V$ is the (global) Fayet-Iliopoulos term (4.3.3). As explained in sec. (4.1.b), $I\!K$ can be interpreted as the Kähler potential of an internal space manifold.

The corresponding *locally* supersymmetric action, *including* $(n = -\frac{1}{3})$ supergravity, is

$$S = -\frac{3}{\kappa^2} \int d^8z\; E^{-1} e^{-\frac{\kappa^2}{3}[I\!K(\Phi^i, \tilde{\Phi}_j) + \nu tr V]}$$

$$+ \int d^6z\; \phi^3 [\mathrm{P}(\Phi^i) + \frac{1}{4} Q_{\mathrm{AB}}(\Phi^i) W^{\alpha \mathrm{A}} W_\alpha{}^{\mathrm{B}}] + h.c. \qquad (5.5.32)$$

In the limit $\kappa \to 0$, $E$ and $\phi \to 1$, this reduces to the global action (5.5.30). The covariant Fayet-Iliopoulos term is (Yang-Mills) gauge invariant only if the chiral action is globally R-invariant. Under a gauge transformation $\delta(tr V) = i\, tr(\bar{\Lambda} - \Lambda)$, $E^{-1} exp(-\frac{1}{3}\kappa^2 \nu tr V)$ is invariant if we simultaneously perform the (restricted) complex superscale transformation discussed at the end of sec. 5.3 with chiral parameter $L + i\frac{1}{3} K_5 = -i\frac{\kappa^2}{6} tr\Lambda$. The invariance of the chiral integral in (5.5.32) follows from R-invariance of (the chiral piece of) the global action.

In general the couplings of the scalar multiplet in superspace involve conformal coupling of the spin zero component fields to gravity. There is one special choice of the Kähler potential, however, where all such conformal coupling can be eliminated for the component scalar fields. This special choice is given by $I\!K(\Phi^i, \tilde{\Phi}_j) = \Phi^i \tilde{\Phi}_i$.

For R-invariant theories superscale transformations can be used to rescale the matter fields and remove $\phi$ from the *chiral* integral; as mentioned above, the resulting action depends only on the combination $\phi\bar{\phi}$, and we can rewrite it for any $n$, e.g., using duality transformations of the compensator as will be described in sec. 5.5.i below. In particular, if we perform a duality transformation to the $n = 0$ theory, the action (5.5.32) becomes

$$S = \int d^8 z\ E^{-1} [ - \frac{1}{\kappa^2} V_5 + I\!K(\Phi^i , \tilde{\bar{\Phi}}_j) + \nu tr V]$$

$$+ \int d^4 x\, d^2\theta\, [\mathrm{P}(\Phi^i) + \frac{1}{4} Q_{\mathrm{AB}}(\Phi^i) W^{\alpha \mathrm{A}} W_\alpha{}^{\mathrm{B}}] + h.\,c. \qquad (5.5.33)$$

where $\Phi^i$ and $W_\alpha$ are suitably defined densities (the $\phi$-independent quantities we defined to make the duality transformation possible).

Although we have concentrated here on the $n = -\frac{1}{3}$ theory coupled to vector and chiral scalar multiplets, more general systems also can be considered. As we stated above, coupling of other versions of supergravity can be obtained by performing duality transformations. As described in chapter 4, there are a large number of "scalar" multiplets and many other matter multiplets. These may be coupled to supergravity by use the prescription of (5.5.4,12).

## i. Supergravity actions

### i.1. Poincaré

For $n \neq 0$, the Poincaré supergravity action is obtained from (5.5.4) (or (5.5.5), for $d = 0$) by choosing $L_{gen} = (n\kappa^2)^{-1}$. For $n = -\frac{1}{3}$, this can be rewritten as

$$S = -\, 3\kappa^{-2} \int d^4 x\, d^2\theta\, \phi^3 R \ . \qquad (5.5.34)$$

For $n = 0$, the obvious choice $S = \int d^8 z\ E^{-1}$, (or its scale invariant form with the tensor compensator (see (5.5.5c)) and $L_{gen} = \kappa^{-2}$) vanishes: With the compensator $G = 1$, the chiral curvature $R = 0$ (see sec. 5.3.b.7.iii), and (e.g., in the chiral representation) (5.3.56) implies $\bar{D}^2 E^{-1} = 0$. If the action vanishes in one gauge, it must do so in all gauges, including ones where the compensator has not been gauged away. However, the $n = 0$ theory has a dimensionless $U(1)$ prepotential $V_5$ that allows us to write an action: Since $\bar{D}^2 E^{-1} = 0$, in the gauge $G = 1$ the following action is invariant under

$U(1)$ gauge transformations $\delta E = 0\,, \delta V_5 = i(\overline{\Lambda}_5 - \Lambda_5)$:

$$S_{n=0} = -\frac{1}{\kappa^2}\int d^4x\, d^4\theta\ E^{-1}V_5 \quad . \tag{5.5.35}$$

In the chiral representation, this can be rewritten, using (5.3.63), as

$$S_{n=0} = \frac{3}{\kappa^2}\int d^4x\, d^4\theta\ E^{-1}ln[E^{-1}\hat{E}^{\frac{1}{3}}(1\cdot e^{-H})^{-\frac{1}{3}}] \quad . \tag{5.5.36}$$

Since in the gauge $G = 1$ we have $E^{-1} = \hat{G}$ (5.3.70), this is the covariantization of the flat space action (4.4.46) for the improved tensor multiplet. We saw that (4.4.46) could be written in a first-order form that made manifest the duality between the scalar and tensor multiplet. This construction carries over to the local case, and we find that $n = 0$ supergravity (with a tensor compensator) is dual to $n = -\frac{1}{3}$ supergravity (with a chiral scalar compensator).

We write a first-order action as

$$S = -\frac{3}{\kappa^2}\int d^4x\, d^4\theta\ \ddot{E}^{-1}(e^{\ddot{X}} - \ddot{G}\ddot{X}) \quad ,$$

$$\ddot{G} = \frac{1}{2}(\ddot{\nabla}_\alpha\phi^\alpha + \bar{\ddot{\nabla}}_{\dot{\alpha}}\bar{\phi}^{\dot{\alpha}}) \quad , \quad \bar{\ddot{\nabla}}_{\dot{\alpha}}\phi_\alpha = 0 \quad ; \tag{5.5.37}$$

where $\ddot{X}$ is an independent, unconstrained, real superfield, and *all* objects ($\ddot{E}^{-1}$, $\ddot{\nabla}_\alpha$) are those of $n = -\frac{1}{3}$. This is just $n = -\frac{1}{3}$ supergravity coupled to the first-order form of the improved tensor multiplet (4.4.45). If we vary with respect to $\ddot{X}$, and substitute the result back into (5.5.37), we find the $n = 0$ action; on the other hand, if we vary with respect to $\phi_\alpha$, we find the $n = -\frac{1}{3}$ action. In detail, we have, from the variation with respect to $\ddot{X}$

$$\ddot{X} = ln\ddot{G} \tag{5.5.38}$$

and hence (5.5.37) becomes

$$S_{n=0} = \frac{3}{\kappa^2}\int d^4x\, d^4\theta\ \ddot{E}^{-1}(\ddot{G}ln\ddot{G} - \ddot{G}) \tag{5.5.39}$$

Because $\ddot{G}$ is linear, the second term can be dropped. Since $\ddot{E}^{-1}\ddot{G} = \hat{G} = E^{-1}G$, (see

(5.3.63)), using (5.3.63) we obtain the action (5.5.35) with the compensator $G$ in a general gauge:

$$S_{n=0} = \frac{3}{\kappa^2} \int d^4x\, d^4\theta\ E^{-1} G(\, lnG - \frac{1}{3} V_5\,) \ . \tag{5.5.40}$$

This action is scale and $U(1)$ invariant.

Alternatively, variation with respect to $\phi^\alpha$ gives $(\ddot{\overline{\nabla}}^2 + \ddot{R})\ddot{\nabla}_\alpha \ddot{X} = 0$ and hence $\ddot{X} = ln\,\Phi + ln\,\overline{\Phi}$ , $\ddot{\overline{\nabla}}_{\dot\alpha}\Phi = 0$, so that again using the linearity of $\ddot{G}$ to eliminate the terms $\ddot{G}\, ln\,\Phi + h.\,c.$, we obtain

$$S = -\frac{3}{\kappa^2} \int d^4x\, d^4\theta\ \ddot{E}^{-1}\Phi\overline{\Phi} \ \ , \tag{5.5.41}$$

i.e., the $n = -\frac{1}{3}$ action (5.5.5a).

The duality transformation from the $n = -\frac{1}{3}$ supergravity theory to the $n = 0$ theory, as described above, can be reversed through a straightforward covariantization of the reverse dual transform (4.4.38) (compare to (4.4.42)). Both forms of the duality transform can be performed even in systems where the supergravity multiplet is coupled to matter multiplets (just as in sec. 4.4.c.2); however, though any $n = 0$ system can be "converted" to an $n = -\frac{1}{3}$ system, the reverse transformation is possible only if the $n = -\frac{1}{3}$ system is R-invariant, and hence the action can be written so that it depends on the $n = -\frac{1}{3}$ compensator (tensor or density type) in the combination $\Phi\overline{\Phi}$ or $\phi\tilde{\phi}$.

Analogous duality transformations that are the covariantization of those described at the end of sec. 4.5.b. can be used to relate $n = -\frac{1}{3}$ and nonminimal supergravity systems.

The form of the superspace action for $n = 0$ reveals a characteristic common to most extended supersymmetric theories. Naively, we might expect actions to take a geometrical form $\int dz\ E^{-1}\mathbb{L}(\textit{field strengths})$. However, we can easily see that for $N \geq 3$, even if $dz$ is a chiral measure, there are no quantities of proper dimensions to form such an action for global or local supersymmetry. Our experience with the $n = 0$ theory shows that it is possible, after solving constraints, to find quantities like $V_5$ that we may call "semiprepotentials" or "precurvatures", without which the action cannot be written. Thus, the unconstrained superfield approach becomes increasingly important, since such precurvatures are actually found as intermediate steps in solving constraints.

### i.2. Cosmological term

To the Poincaré supergravity action we can add a supersymmetric cosmological term (for a discussion of global deSitter supersymmetry, see sec. 5.7). For $n = -\frac{1}{3}$, we have

$$S_{cosmo} = \lambda\kappa^{-2} \int d^4x\, d^2\theta\, \phi^3 + h.\,c. \tag{5.5.42}$$

For $n \neq -\frac{1}{3}, 0$ we could write a type of cosmological term using the form (5.5.12), but that term contains inverse powers of the scalar auxiliary field; for $n = 0$, $R = 0$ and hence it is impossible to write a cosmological term (these difficulties arise because the cosmological term is *not* R-invariant). One other interesting feature of the cosmological term for nonminimal supergravity is that the sum of (5.5.12) (with $I\!\!L_{chiral} = 1$) and (5.5.5b) (with $I\!\!L_{gen} = 1$) leads to a spontaneous breaking of supersymmetry. The resulting field equations are such that it is not possible to construct an anti-deSitter background which is supersymmetric (see sec. 5.7).

### i.3. Conformal supergravity

Next we consider the action for conformal supergravity. It is just the covariantization of the linearized expression (5.2.6):

$$S_{conf} = \int d^4x d^2\theta\, \phi^3\, (W_{\alpha\beta\gamma})^2 \quad . \tag{5.5.43}$$

The conformal field strength $W_{\alpha\beta\gamma}$ depends on $\phi$ only through a proportionality factor $\phi^{-\frac{3}{2}}$, so all $\phi$ dependence cancels. The form of (5.5.43) is valid only for the minimal theory. It can be extended to the nonminimal theory by the use of (5.5.12). For $n = 0$ even this insufficient, again because $R = 0$. This does not imply that conformal supergravity does not exist; it is $n$-independent. Instead the action for conformal supergravity takes a form similar to the three-dimensional supergravity topological mass term with $W_{\alpha\beta\gamma}$ taking the place of the three-dimensional $G_{\alpha\beta\gamma}$, and $G_{\alpha\dot{\alpha}}$ and $R$ the place of the three-dimensional $R$ (see (2.6.47)).

## j. Field equations

To obtain covariant field equations from the action by functional differentiation with respect to the supergravity superfields, which are not covariant themselves, we define a modified functional variation, as we did for super-Yang-Mills (see (4.2.48)):

$$\Delta H \equiv e^{-H}\delta e^{H} \quad or \quad \Delta\Omega \equiv e^{-\Omega}\delta e^{\Omega} \ , \ \Delta\overline{\Omega} \equiv (\delta e^{\overline{\Omega}})e^{-\overline{\Omega}} \ ; \tag{5.5.44a}$$

$$\Delta\phi \equiv \delta(\phi^3) \ . \tag{5.5.44b}$$

The equations of motion for supergravity with action given by (5.5.4) and the cosmological term (5.5.42) can then be shown to be

$$\frac{\Delta S}{\Delta H^{\underline{a}}} = -\,\kappa^{-2}G_{\underline{a}} = 0 \ , \quad \frac{\Delta S}{\Delta\phi} = -\,\kappa^{-2}(R-\lambda) = 0 \ . \tag{5.5.45}$$

The covariantized field equation for $H^{\alpha\dot{\alpha}}$ is the same as that obtained by the background field method (the variation $\Delta$ is the same as the background-quantum splitting linearized in the quantum field). The derivation of this field equation will be described in more detail when we describe this splitting in sec. 7.2. The $\phi$ equation is easily obtained using (5.2.71,5.3.56,5.5.34,42).

To obtain covariant field equations for a covariantly chiral superfield, it is necessary to define a suitable functional derivative. This can be done in any of three ways: (1) by first using the flat-space definition for differentiation by $\eta$, and using the relation (5.5.2); (2) by covariantizing the flat-space form in a way that satisfies the correct covariant chirality condition; or (3) by expressing the chiral superfield as the field strength of a general superfield. The result is:

$$\frac{\delta\eta(z)}{\delta\eta(z')} = (\overline{\nabla}^2 + R)\delta^8(z-z') \ , \tag{5.5.46}$$

where, for $n \neq -\frac{1}{3}$, $\nabla$ is $U(1)$ covariant. The resulting field equations for a scalar multiplet are thus the same as in the global case except that $\bar{D}^2$ is replaced with $\overline{\nabla}^2 + R$. The field equations for supergravity coupled to a scalar multiplet are (for $n = -\frac{1}{3}$, and using the action (5.5.14))

$$\kappa^{-2} G_{\alpha\dot{\alpha}} = \frac{1}{3} [\bar{\eta} i \overset{\leftrightarrow}{\nabla}_{\alpha\dot{\alpha}} \eta - (\bar{\nabla}_{\dot{\alpha}} \bar{\eta})(\nabla_\alpha \eta) + \bar{\eta}\eta G_{\alpha\dot{\alpha}}] \ ,$$

$$\kappa^{-2} R = \frac{1}{3} (\bar{\nabla}^2 + R)\bar{\eta} = 0 \ . \tag{5.5.47}$$

When a self-interaction term is added, the $G$ equation is unchanged, but $R$ becomes nonzero (except for the superconformal coupling $\eta^3$). In the last equation we have used the equation of motion of the scalar multiplet. Alternatively, terms in field equations proportional to other field equations can be removed in general even off shell by field redefinitions in the action. (To remove terms proportional to the field equations of $\psi_2$ from the field equations of $\psi_1$, a field redefinition of the form $\psi_1 = \psi_1{}'$, $\psi_2 = \psi_2{}' + f$ modifies the field equations to $\frac{\delta S}{\delta \psi_1{}'} = \frac{\delta S}{\delta \psi_1} + \frac{\delta S}{\delta \psi_2} \frac{\delta f}{\delta \psi_2{}'}$.) In this case, the appropriate field redefinition is $\eta = \phi^{-1}\eta' = \eta' - \chi\eta' + \cdots$, which removes all $\phi$ dependence from the scalar-multiplet action (see sec. 7.10.c).

For the coupled supergravity-Yang-Mills system (sec. 5.5.h), the field equations for Yang-Mills are still $\{ \underset{\sim}{\nabla}^\alpha , W_\alpha \} = 0$, while the supergravity equations are

$$\kappa^{-2} G_{\alpha\dot{\alpha}} = g^{-2}\, tr\, \bar{W}_{\dot{\alpha}} W_\alpha \quad , \quad \kappa^{-2} R = 0 \quad . \tag{5.5.48}$$

We have dropped terms in the $G$ equation proportional to the Yang-Mills field equation. These terms, which in this case are not Yang-Mills gauge covariant, can again be eliminated by a field redefinition (again see sec. 7.10.c).

Although we have only considered $n = -\frac{1}{3}$ for simplicity, covariant variation with respect to the compensators for the other versions of supergravity can also be defined analogously. In both $n = 0$ and nonminimal theories the important point to note in defining the covariant variations is that the unconstrained compensators for both theories are spinors ($\Upsilon = \bar{D}_{\dot{\mu}} \Phi^{\dot{\mu}}$ for the nonminimal theory and $\phi_\alpha$ for $n = 0$). Thus functional differentiation in these cases lead to spinorial equations of motion.

## 5.6. From superspace to components

### a. General considerations

So far in our discussion of supergravity we have concentrated exclusively on superspace and superfields. On the other hand, somewhere in this formalism a supergravity theory in ordinary spacetime is being described. The question arises how to extract from a superspace formulation information about component fields. We know how to do this in the global supersymmetry case, and here we will describe the corresponding procedure in local supersymmetry, and derive the *tensor calculus* of component supergravity. We cannot use $D$ and $\bar{D}$ to define the components of superfields by projection as in global superspace, since this would not be covariant with respect to local supersymmetry.

To discuss component supergravity, we must first choose a Wess-Zumino gauge in which the $K$-transformations have been used to set to zero all supergravity components that can be gauged away algebraically. A Wess-Zumino gauge is necessary so that results for noncovariant quantities (i.e. gauge fields) can be derived along with those for covariant quantities. We can then derive transformation laws for the remaining supergravity components as well as components of other superfields and exhibit supercovariantization and the commutator algebra of local supersymmetry at the component level. We derive multiplication rules for local (covariantly chiral) scalar multiplets, and write the component form of the integration measures (density formulae), from which component actions can be obtained. All the results reflect the underlying superspace geometry and can be obtained for any $N$, imposing as few constraints as possible (preferably none). This implies that superspace geometry is more general than a component tensor calculus which follows from a choice of constraints on superspace torsions and/or curvatures. The final form of the tensor calculus is determined by which solution of the Bianchi identities is utilized.

We begin with a general superspace for $N$-extended supergravity. (We will specialize to $N = 1$ whenever needed.) In such a superspace we have a vielbein $E_A{}^M$ which describes supergravity. We also introduce a number of connection superfields $\Phi_A{}^\iota$ for tangent space symmetries such as Lorentz rotations, scale transformations, $SU(N)$-rotations, central charges, etc. These superfields are combined with operators $D_M$ and $M_\iota$ to form a supercovariant derivative

$$\nabla_A = E_A + \Phi_A{}^{\boldsymbol{\iota}} M_{\boldsymbol{\iota}} \quad , \quad E_A = E_A{}^M D_M \quad , \tag{5.6.1}$$

where $M_{\boldsymbol{\iota}}$ are the generators of the tangent space symmetries. The $\boldsymbol{\iota}$-subscript is a label that runs over all the generators of the tangent space symmetries. For instance, in $N = 1$, $n \neq 0$ supergravity $M_{\boldsymbol{\iota}} = (M_{\alpha\beta}, \overline{M}_{\dot{\alpha}\dot{\beta}})$. The realization of these generators is specified by giving their action on an arbitrary tangent vector $X_A$. Thus, for some set of matrices $(M_{\boldsymbol{\iota}})_A{}^B$ we have

$$[M_{\boldsymbol{\iota}}, X_A] = (M_{\boldsymbol{\iota}})_A{}^B X_B \quad . \tag{5.6.2}$$

We write $D_M = D_M{}^N \frac{\partial}{\partial z^N} + \Gamma_M{}^{\boldsymbol{\iota}} M_{\boldsymbol{\iota}}$ for *fixed* matrices $D_M{}^N$ and $\Gamma_M{}^{\boldsymbol{\iota}}$ where $D_M{}^N - \delta_M{}^N$ and $\Gamma_M{}^{\boldsymbol{\iota}}$ vanish at $\theta = 0$ (see sec. 3.4.c). We assume the vielbein is invertible; specifically, we assume that we can always find a coordinate system (or gauge) in which we can write

$$\nabla_A = \partial_A + \Delta_A \quad . \tag{5.6.3}$$

The gauge transformations of $\nabla_A$ are given as usual by

$$\nabla'_A = e^{iK} \nabla_A e^{-iK} \quad . \tag{5.6.4}$$

The parameter $K$ is a superfield which is also expanded over $iD_M$ and $iM_{\boldsymbol{\iota}}$

$$K = K^M iD_M + \hat{K}^{\boldsymbol{\iota}} iM_{\boldsymbol{\iota}} \quad , \tag{5.6.5a}$$

and is subject to a reality condition $K = \overline{(K)}$. We can equally well expand the parameter $K$ over the covariant derivatives $\nabla_A$ and $M_{\boldsymbol{\iota}}$:

$$K = K^A i\nabla_A + (\hat{K}^{\boldsymbol{\iota}} - K^A \Phi_A{}^{\boldsymbol{\iota}}) iM_{\boldsymbol{\iota}} \quad ,$$

$$= K^A i\nabla_A + K^{\boldsymbol{\iota}} iM_{\boldsymbol{\iota}} \quad . \tag{5.6.5b}$$

A gauge transformation of an *arbitrary* covariant superfield quantity is always generated by acting with $iK$ as in (5.6.4). For infinitesimal transformations of supercovariant quantities this implies that we simply act on the quantity with the operator $iK$.

## b. Wess-Zumino gauge for supergravity

We define components of *covariant quantities* (matter fields, torsions and curvatures) using the local generalization of the covariant projection method introduced in a global context: these components are the $\theta$, $\bar{\theta}$ independent projections of the superfields and their covariant derivatives. We define components of *gauge fields* $E_A{}^M$, $\Phi_A{}^{\boldsymbol{\iota}}$ by choosing a special gauge and then projecting as on covariant quantities. This Wess-Zumino gauge choice reduces the superspace gauge transformations to component gauge transformations: It uses all but the $\theta$, $\bar{\theta}$ independent part of $K$ to algebraically gauge away the noncovariant pieces of the higher components of the gauge fields (the lowest components remain as the spacetime component gauge fields). We use the notation $X|$ to mean the $\theta$, $\bar{\theta}$ independent part of any superfield quantity $X$; if $X$ is an operator $X^M iD_M + \hat{X}^{\boldsymbol{\iota}} iM_{\boldsymbol{\iota}}$, then $X|$ is the operator $X^M|i\partial_M + \hat{X}^{\boldsymbol{\iota}}|iM_{\boldsymbol{\iota}}$; we use $D_M| = \partial_M$ and do *not* set $\partial_{\underline{\mu}}$, $\bar{\partial}_{\underline{\dot{\mu}}}$ to zero. In particular, we define the components of the covariant derivatives by $\nabla_C|$, $\nabla_\alpha\nabla_C|$, $\nabla_\alpha\nabla_\beta\nabla_C|$, $\nabla_\alpha\bar{\nabla}_{\dot{\beta}}\nabla_C|$, etc.

We define the usual component gauge fields by

$$\nabla_{\underline{a}}| \equiv e_{\underline{a}}{}^{\underline{m}}\partial_{\underline{m}} + \psi_{\underline{a}}{}^{\underline{\mu}}\partial_{\underline{\mu}} + \bar{\psi}_{\underline{a}}{}^{\underline{\dot{\mu}}}\bar{\partial}_{\underline{\dot{\mu}}} + \phi_{\underline{a}}{}^{\boldsymbol{\iota}}M_{\boldsymbol{\iota}}$$

$$\equiv \mathbf{D}_{\underline{a}} + \psi_{\underline{a}}{}^{\underline{\mu}}\partial_{\underline{\mu}} + \bar{\psi}_{\underline{a}}{}^{\underline{\dot{\mu}}}\bar{\partial}_{\underline{\dot{\mu}}} \ , \tag{5.6.6}$$

where $e_{\underline{a}}{}^{\underline{m}}$ is the component inverse vierbein, $\psi_{\underline{a}}{}^{\underline{\mu}}, \bar{\psi}_{\underline{a}}{}^{\underline{\dot{\mu}}}$ are the component gravitino fields, and $\phi_{\underline{a}}{}^{\boldsymbol{\iota}}$ are the component gauge fields of the component tangent space symmetries. (For $M_{\boldsymbol{\iota}} = (M_\gamma{}^\beta, \bar{M}_{\dot{\gamma}}{}^{\dot{\beta}})$ these gauge fields are the Lorentz spin connections $\phi_{\underline{a}\beta}{}^\gamma$ and $\phi_{\underline{a}\dot{\beta}}{}^{\dot{\gamma}}$.) From the infinitesimal transformation law $\delta\nabla_{\underline{a}} = [iK\,, \nabla_{\underline{a}}] = -\,i\partial_{\underline{a}}K + \cdots$, we see that these components transform as spacetime gradients of the gauge parameters, which justifies the definition. We have also introduced the ordinary spacetime covariant derivative $\mathbf{D}_{\underline{a}} = e_{\underline{a}} + \phi_{\underline{a}}{}^{\boldsymbol{\iota}}M_{\boldsymbol{\iota}}$ (cf. 5.1.15). Covariantly transforming components of the supergravity multiplet (e.g. auxiliary fields) appear as components of the torsions and curvatures.

We now derive the first few components of the covariant derivatives. We begin by exploiting the existence of a Wess-Zumino gauge. From the infinitesimal transformation law $\delta\nabla_{\underline{\alpha}} = [iK\,, \nabla_{\underline{\alpha}}]$, using (5.6.3) we find

$$\delta\Delta_{\underline{\alpha}}| = -\, i\partial_{\underline{\alpha}}K + \cdots$$

$$= [iK\,, \nabla_{\underline{\alpha}}]| \equiv K^{(1)}{}_{\underline{\alpha}} \tag{5.6.7}$$

and hence, by using the $\nabla_{\underline{\alpha}}$ component of $K$, i.e., $K^{(1)}{}_{\underline{\alpha}}$, we can choose a gauge $\Delta_{\underline{\alpha}}| = 0$ or

$$\nabla_{\underline{\alpha}}| = \partial_{\underline{\alpha}} \quad . \tag{5.6.8}$$

We have thus determined $\nabla_A|$. We can proceed to find the higher-order terms in a straightforward manner. Thus to find $\nabla_{\underline{\alpha}}\nabla_{\underline{\beta}}|$ we start with

$$\delta(\nabla_{\underline{\alpha}}\nabla_{\underline{\beta}}) = [iK\,, \nabla_{\underline{\alpha}}\nabla_{\underline{\beta}}] \;. \tag{5.6.9}$$

Then

$$\delta(\nabla_{\underline{\alpha}}\nabla_{\underline{\beta}})| = -\, i\partial_{\underline{\alpha}}\partial_{\underline{\beta}}K + \cdots \;. \tag{5.6.10}$$

Since $\{\partial_{\underline{\alpha}}\,, \partial_{\underline{\beta}}\} = 0$ we can gauge away $[\nabla_{\underline{\alpha}}\,, \nabla_{\underline{\beta}}]$ but not $\{\nabla_{\underline{\alpha}}\,, \nabla_{\underline{\beta}}\}$. However, the latter is covariant: It can be expressed in terms of torsions and curvatures. Hence in this gauge we find the $\nabla_{\underline{\alpha}}$ component of $\nabla_{\underline{\beta}}$:

$$\nabla_{\underline{\alpha}}\nabla_{\underline{\beta}}| = \frac{1}{2}\{\nabla_{\underline{\alpha}}\,, \nabla_{\underline{\beta}}\}| = \frac{1}{2}T_{\underline{\alpha}\underline{\beta}}{}^{C}\nabla_{C}| + \frac{1}{2}R_{\underline{\alpha}\underline{\beta}}{}^{\iota}|M_{\iota} \quad . \tag{5.6.11}$$

In the same way, we find the $\overline{\nabla}_{\underline{\dot{\alpha}}}$ component of $\nabla_{\underline{\beta}}$:

$$\overline{\nabla}_{\underline{\dot{\alpha}}}\nabla_{\underline{\beta}}| = \frac{1}{2}\{\overline{\nabla}_{\underline{\dot{\alpha}}}\,, \nabla_{\underline{\beta}}\}| = \frac{1}{2}T_{\underline{\dot{\alpha}}\underline{\beta}}{}^{C}\nabla_{C}| + \frac{1}{2}R_{\underline{\dot{\alpha}}\underline{\beta}}{}^{\iota}|M_{\iota} \quad . \tag{5.6.12}$$

Similarly, we find the next component of $\nabla_{\underline{b}}$; we first observe that because $\nabla_{\underline{\alpha}}| = \partial_{\underline{\alpha}}$, we have

$$\nabla_{\underline{b}}| = \mathbf{D}_{\underline{b}} + \psi_{\underline{b}}{}^{\underline{\gamma}}\nabla_{\underline{\gamma}}| + \bar{\psi}_{\underline{b}}{}^{\underline{\dot{\gamma}}}\overline{\nabla}_{\underline{\dot{\gamma}}}| \;\;. \tag{5.6.13}$$

Then we compute

$$\nabla_{\underline{\alpha}}\nabla_{\underline{b}}| = [\nabla_{\underline{\alpha}}\,, \nabla_{\underline{b}}]| + \nabla_{\underline{b}}\nabla_{\underline{\alpha}}|$$

$$= [\nabla_{\underline{\alpha}}\,, \nabla_{\underline{b}}]| + \mathbf{D}_{\underline{b}}\nabla_{\underline{\alpha}}| + \psi_{\underline{b}}{}^{\underline{\gamma}}\nabla_{\underline{\gamma}}\nabla_{\underline{\alpha}}| + \bar{\psi}_{\underline{b}}{}^{\underline{\dot{\gamma}}}\overline{\nabla}_{\underline{\dot{\gamma}}}\nabla_{\underline{\alpha}}| \;. \tag{5.6.14}$$

Using (5.6.8,11,12) we obtain

$$\nabla_{\underline{\alpha}}\nabla_{\underline{b}}| = T_{\underline{\alpha b}}{}^{C}\nabla_{C}| + R_{\underline{\alpha b}}{}^{\boldsymbol{\iota}}|M_{\boldsymbol{\iota}} + \phi_{\underline{b}}{}^{\boldsymbol{\iota}}[M_{\boldsymbol{\iota}}\,,\nabla_{\underline{\alpha}}|\,]$$

$$+\,\frac{1}{2}\,\psi_{\underline{b}}{}^{\underline{\gamma}}[T_{\underline{\gamma\alpha}}{}^{C}\nabla_{C}| + R_{\underline{\gamma\alpha}}{}^{\boldsymbol{\iota}}|M_{\boldsymbol{\iota}}] + \frac{1}{2}\,\overline{\psi}_{\underline{b}}{}^{\underline{\dot\gamma}}[T_{\underline{\dot\gamma}\underline{\alpha}}{}^{C}\nabla_{C}| + R_{\underline{\dot\gamma}\underline{\alpha}}{}^{\boldsymbol{\iota}}|M_{\boldsymbol{\iota}}] \qquad (5.6.15)$$

Thus we have found $\nabla_{\underline{\alpha}}\nabla_{B}|$. We can find higher components, but what we have is sufficient for the applications we give below. We have obtained these formulae without imposing *any* constraints.

The procedure we have described uses all the higher components (projections with more $\nabla$'s) of $K$ to eliminate the noncovariant pieces of $\nabla_{\underline{\alpha}}$ and $\overline{\nabla}_{\underline{\dot\alpha}}$ and defines the Wess-Zumino gauge. The remaining gauge transformations, determined by the $\theta$ independent term $K|$, are just the usual component transformations. Coordinate transformations are determined by

$$iK_{GC}| = -\,\lambda^{\underline{m}}(x)\,\partial_{\underline{m}} \qquad (5.6.16)$$

(or equivalently covariant translations $iK_{CT}| = -\,\lambda^{\underline{a}}(x)\,\mathbf{D}_{\underline{a}} = -\,\lambda^{\underline{m}}\partial_{\underline{m}} - \lambda^{\underline{a}}(\phi_{\underline{a}}{}^{\boldsymbol{\iota}}M_{\boldsymbol{\iota}})$). Tangent space gauge transformations are determined by

$$iK_{TS}| = -\,\lambda^{\boldsymbol{\iota}}(x)\,M_{\boldsymbol{\iota}}\;, \qquad (5.6.17)$$

and supersymmetry transformations are determined by

$$iK_{Q}| = -\,\epsilon^{\underline{\alpha}}(x)\partial_{\underline{\alpha}} - \epsilon^{\underline{\dot\alpha}}(x)\overline{\partial}_{\underline{\dot\alpha}}$$

$$= -\,\epsilon^{\underline{\alpha}}(x)\nabla_{\underline{\alpha}}| - \epsilon^{\underline{\dot\alpha}}(x)\overline{\nabla}_{\underline{\dot\alpha}}|\;\;. \qquad (5.6.18)$$

However, to stay in the Wess-Zumino gauge, the $K$ transformations must be restricted: the higher components are expressed in terms of $K|$. For example, $\nabla_{\underline{\alpha}}| = \partial_{\underline{\alpha}}$ implies:

$$\delta\partial_{\underline{\alpha}} = 0 = [iK\,,\nabla_{\underline{\alpha}}]|\;, \qquad (5.6.19a)$$

so that

$$0 = -\,[K^{B}\nabla_{B}\,,\nabla_{\underline{\alpha}}]| - [K^{\boldsymbol{\iota}}M_{\boldsymbol{\iota}}\,,\nabla_{\underline{\alpha}}]|$$

$$= -\,K^{B}[\nabla_{B}\,,\nabla_{\underline{\alpha}}\}| + [\nabla_{\underline{\alpha}}\,,K^{B}\}\nabla_{B}| - K^{\boldsymbol{\iota}}[M_{\boldsymbol{\iota}}\,,\nabla_{\underline{\alpha}}]| + [\nabla_{\underline{\alpha}}\,,K^{\boldsymbol{\iota}}]|M_{\boldsymbol{\iota}}\;, \qquad (5.6.19b)$$

and hence

$$\nabla_{\underline{\alpha}} K^B| = K^C T_{C\underline{\alpha}}{}^B| + K^{\iota}|(M_{\iota})_{\underline{\alpha}}{}^B$$

$$\nabla_{\underline{\alpha}} K^{\iota}| = K^C R_{C\underline{\alpha}}{}^{\iota}| \ . \qquad (5.6.20)$$

Similarly, we can find the higher components of $K$ from the higher components of $\nabla_{\underline{\alpha}}$ and the requirement that the Wess-Zumino gauge is maintained. It turns out that in the Wess-Zumino gauge (5.6.16) holds to all orders in $\theta$ i.e., $K_{GC}$ has no higher components, whereas both $K_{TS}$ and $K_Q$ have higher components depending on the component fields, the gauge parameters, and in general, the *gradients* of the parameters. Thus, in the local case, we cannot write $-iK_Q = \epsilon^{\underline{\alpha}} Q_{\underline{\alpha}} + \bar{\epsilon}^{\dot{\underline{\alpha}}} \overline{Q}_{\dot{\underline{\alpha}}}$ for some operator $Q_{\underline{\alpha}}$. The higher order terms in $iK_{TS}$ are always proportional to the matrices $(M)_{\underline{\alpha}}{}^{\underline{\beta}}$, $(M)_{\dot{\underline{\alpha}}}{}^{\dot{\underline{\beta}}}$; hence for *internal symmetries* as compared to tangent space symmetries, $iK_{TS}$ has no higher components and (5.6.17) is exact.

## c. Commutator algebra

As another application of the use of the Wess-Zumino gauge supersymmetry generator, we derive the commutator algebra of local component supersymmetry. In this gauge we use the differential operator $iK_Q$ as the local supersymmetry generator for the component formulation of supergravity. Since the supersymmetry generator is field dependent, we can indicate this by writing $iK_Q(\epsilon;\psi)$ where $\psi$ denotes all of the $x$-space fields contained in $iK_Q$. This means that care must be taken in defining the commutator of two such transformations. Let us imagine performing sequentially on $\eta$ two supersymmetry transformations with parameters $\epsilon_2$ and $\epsilon_1$. The first transformation is obtained from $iK_Q(\epsilon_2;\psi)\eta$, where we have dropped the commutator notation, keeping in mind that $iK_Q$ is an operator. The second transformation is implemented by $iK_Q(\epsilon_1;\psi+\delta_2\psi) iK_Q(\epsilon_2;\psi)\eta$. Therefore, the correct way to compute the commutator algebra is from the definition

$$[iK_{Q_1}, iK_{Q_2}] \equiv iK_Q(\epsilon_1;\psi+\delta_2\psi) iK_Q(\epsilon_2;\psi) - (\,1 \longleftrightarrow 2\,) \equiv iK_{12} \ . \qquad (5.6.21)$$

However, by looking at the form of the supersymmetry generator in (5.6.18) we note that $iK_Q|$ has no field dependent terms. This implies $[iK_{Q_1}, iK_{Q_2}]|$ corresponds to the usual commutator $[iK_Q(\epsilon_1,\psi)\,, iK_Q(\epsilon_2,\psi)]$, and this is all we need to find the component commutator algebra. Taking the expression for $iK_Q$ from (5.6.18) and using the Wess-

Zumino gauge-preserving condition $[iK\,,\nabla_{\underline{\alpha}}]|=0$ (see (5.6.19)), we obtain

$$[iK_{Q_1}\,,iK_{Q_2}]=\epsilon_1{}^{\underline{\alpha}}\epsilon_2{}^{\underline{\beta}}\{\nabla_{\underline{\alpha}}\,,\nabla_{\underline{\beta}}\}+\bar{\epsilon}_1{}^{\underline{\dot\alpha}}\bar{\epsilon}_2{}^{\underline{\dot\beta}}\{\bar\nabla_{\underline{\dot\alpha}}\,,\bar\nabla_{\underline{\dot\beta}}\}+(\epsilon_1{}^{\underline{\alpha}}\bar{\epsilon}_2{}^{\underline{\dot\beta}}+\bar{\epsilon}_1{}^{\underline{\dot\beta}}\epsilon_2{}^{\underline{\alpha}})\{\nabla_{\underline{\alpha}}\,,\bar\nabla_{\underline{\dot\beta}}\}|\;. \tag{5.6.22}$$

Comparing the right hand side of the above equation to $iK_{GC}|$, $iK_{TS}|$ and $iK_Q|$ we find

$$iK_{12}\equiv iK_{GC}(\lambda^{\underline{m}})+iK_{TS}(\lambda^{\iota})+iK_Q(\epsilon)\;,$$

$$\lambda^{\underline{m}}=-\,[(\epsilon_1{}^{\underline{\alpha}}\bar{\epsilon}_2{}^{\underline{\dot\beta}}+\bar{\epsilon}_1{}^{\underline{\dot\beta}}\epsilon_2{}^{\underline{\alpha}})T_{\underline{\alpha}\underline{\dot\beta}}{}^{\underline{c}}+\epsilon_1{}^{\underline{\alpha}}\epsilon_2{}^{\underline{\beta}}T_{\underline{\alpha}\underline{\beta}}{}^{\underline{c}}+\bar{\epsilon}_1{}^{\underline{\dot\alpha}}\bar{\epsilon}_2{}^{\underline{\dot\beta}}T_{\underline{\dot\alpha}\underline{\dot\beta}}{}^{\underline{c}}]e_{\underline{c}}{}^{\underline{m}}\;,$$

$$\lambda^{\iota}=-\,[(\epsilon_1{}^{\underline{\alpha}}\bar{\epsilon}_2{}^{\underline{\dot\beta}}+\bar{\epsilon}_1{}^{\underline{\dot\beta}}\epsilon_2{}^{\underline{\alpha}})(R_{\underline{\alpha}\underline{\dot\beta}}{}^{\iota}+T_{\underline{\alpha}\underline{\dot\beta}}{}^{\underline{c}}\Phi_{\underline{c}}{}^{\iota})$$

$$+\,\epsilon_1{}^{\underline{\alpha}}\epsilon_2{}^{\underline{\beta}}(R_{\underline{\alpha}\underline{\beta}}{}^{\iota}+T_{\underline{\alpha}\underline{\beta}}{}^{\underline{c}}\Phi_{\underline{c}}{}^{\iota})+\bar{\epsilon}_1{}^{\underline{\dot\alpha}}\bar{\epsilon}_2{}^{\underline{\dot\beta}}(R_{\underline{\dot\alpha}\underline{\dot\beta}}{}^{\iota}+T_{\underline{\dot\alpha}\underline{\dot\beta}}{}^{\underline{c}}\Phi_{\underline{c}}{}^{\iota})]\;,$$

$$\epsilon^{\underline{\delta}}=-\,[(\epsilon_1{}^{\underline{\alpha}}\bar{\epsilon}_2{}^{\underline{\dot\beta}}+\bar{\epsilon}_1{}^{\underline{\dot\beta}}\epsilon_2{}^{\underline{\alpha}})(T_{\underline{\alpha}\underline{\dot\beta}}{}^{\underline{\delta}}+T_{\underline{\alpha}\underline{\dot\beta}}{}^{\underline{c}}\psi_{\underline{c}}{}^{\underline{\delta}})$$

$$+\,\epsilon_1{}^{\underline{\alpha}}\epsilon_2{}^{\underline{\beta}}(T_{\underline{\alpha}\underline{\beta}}{}^{\underline{\delta}}+T_{\underline{\alpha}\underline{\beta}}{}^{\underline{c}}\psi_{\underline{c}}{}^{\underline{\delta}})+\bar{\epsilon}_1{}^{\underline{\dot\alpha}}\bar{\epsilon}_2{}^{\underline{\dot\beta}}(T_{\underline{\dot\alpha}\underline{\dot\beta}}{}^{\underline{\delta}}+T_{\underline{\dot\alpha}\underline{\dot\beta}}{}^{\underline{c}}\psi_{\underline{c}}{}^{\underline{\delta}})]\;. \tag{5.6.23}$$

These results show how the commutator algebra of local supersymmetry is completely determined by superspace geometry. In particular the field dependence of the local algebra is a consequence of only considering component fields which are present in the WZ gauge. The full result for (5.6.21), to all orders in $\theta$ is given by

$$[iK_{Q_1}\,,iK_{Q_2}]=iK_{GC}(\lambda^{\underline{m}})+iK_{TS}(\lambda^{\iota};\psi')+iK_Q(\epsilon;\psi')\;,$$

$$\psi'\equiv\psi+\delta_2\psi-\delta_1\psi\;. \tag{5.6.24}$$

## d. Local supersymmetry and component gauge fields

We now derive the supersymmetry variation of the component gauge fields. We obtain these by evaluating a superfield equation at $\theta=0$

$$\delta\nabla_{\underline{a}}|=[iK_Q\,,\nabla_{\underline{a}}]|\;. \tag{5.6.25}$$

From (5.6.13) we have

$$[iK\,,\nabla_{\underline{a}}]| = iK\nabla_{\underline{a}}| - \nabla_{\underline{a}}iK|$$

$$= -\,(\epsilon^{\underline{\beta}}\nabla_{\underline{\beta}} + \bar\epsilon^{\underline{\dot\beta}}\nabla_{\underline{\dot\beta}})\nabla_{\underline{a}}| - \mathbf{D}_{\underline{a}}iK| - \psi_{\underline{a}}{}^{\underline{\beta}}\nabla_{\underline{\beta}}iK| - \bar\psi_{\underline{a}}{}^{\underline{\dot\beta}}\overline{\nabla}_{\underline{\dot\beta}}iK|\ . \quad (5.6.26)$$

Using the Wess-Zumino gauge condition (5.6.19a,17), we rewrite this as

$$[iK\,,\nabla_{\underline{a}}]| = -\,(\epsilon^{\underline{\beta}}\nabla_{\underline{\beta}} + \bar\epsilon^{\underline{\dot\beta}}\overline{\nabla}_{\underline{\dot\beta}})\nabla_{\underline{a}}| - \mathbf{D}_{\underline{a}}iK| - \psi_{\underline{a}}{}^{\underline{\beta}}iK\nabla_{\underline{\beta}}| - \psi_{\underline{a}}{}^{\underline{\dot\beta}}iK\overline{\nabla}_{\underline{\dot\beta}}|$$

$$= -\,(\epsilon^{\underline{\beta}}\nabla_{\underline{\beta}}\nabla_{\underline{a}}| + \bar\epsilon^{\underline{\dot\beta}}\overline{\nabla}_{\underline{\dot\beta}}\nabla_{\underline{a}}|) + \mathbf{D}_{\underline{a}}(\epsilon^{\underline{\beta}}\nabla_{\underline{\beta}} + \bar\epsilon^{\underline{\dot\beta}}\overline{\nabla}_{\underline{\dot\beta}})|$$

$$+\ \psi_{\underline{a}}{}^{\underline{\beta}}(\epsilon^{\underline{\gamma}}\nabla_{\underline{\gamma}} + \bar\epsilon^{\underline{\dot\gamma}}\overline{\nabla}_{\underline{\dot\gamma}})\nabla_{\underline{\beta}}| + \bar\psi_{\underline{a}}{}^{\underline{\dot\beta}}(\epsilon^{\underline{\gamma}}\nabla_{\underline{\gamma}} + \bar\epsilon^{\underline{\dot\gamma}}\overline{\nabla}_{\underline{\dot\gamma}})\overline{\nabla}_{\underline{\dot\beta}}|\ . \quad (5.6.27)$$

Expanding over $\partial_{\underline{m}}$, $\nabla_{\underline{\alpha}}|$ and $M_{\boldsymbol{\iota}}$, and using (5.6.10,11,15) we find:

$$\delta_Q e_{\underline{a}}{}^{\underline{m}} = -\,[\epsilon^{\underline{\beta}}T_{\underline{\beta}\underline{a}}{}^{\underline{d}} + \bar\epsilon^{\underline{\dot\beta}}T_{\underline{\dot\beta}\underline{a}}{}^{\underline{d}} + (\bar\epsilon^{\underline{\dot\beta}}\psi_{\underline{a}}{}^{\underline{\gamma}} + \epsilon^{\underline{\gamma}}\bar\psi_{\underline{a}}{}^{\underline{\dot\beta}})T_{\underline{\dot\beta}\underline{\gamma}}{}^{\underline{d}}$$

$$+\ \epsilon^{\underline{\beta}}\psi_{\underline{a}}{}^{\underline{\gamma}}T_{\underline{\gamma}\underline{\beta}}{}^{\underline{d}} + \bar\epsilon^{\underline{\dot\beta}}\bar\psi_{\underline{a}}{}^{\underline{\dot\gamma}}T_{\underline{\dot\beta}\underline{\dot\gamma}}{}^{\underline{d}}]e_{\underline{d}}{}^{\underline{m}}\ ,$$

$$\delta_Q\psi_{\underline{a}}{}^{\underline{\delta}} = \mathbf{D}_{\underline{a}}\epsilon^{\underline{\delta}} - \epsilon^{\underline{\beta}}(T_{\underline{\beta}\underline{a}}{}^{\underline{\delta}} + T_{\underline{\beta}\underline{a}}{}^{\underline{e}}\psi_{\underline{e}}{}^{\underline{\delta}}) - \bar\epsilon^{\underline{\dot\beta}}(T_{\underline{\dot\beta}\underline{a}}{}^{\underline{\delta}} + T_{\underline{\dot\beta}\underline{a}}{}^{\underline{e}}\psi_{\underline{e}}{}^{\underline{\delta}})$$

$$-\ (\bar\epsilon^{\underline{\dot\beta}}\psi_{\underline{a}}{}^{\underline{\gamma}} + \epsilon^{\underline{\gamma}}\bar\psi_{\underline{a}}{}^{\underline{\dot\beta}})(T_{\underline{\gamma}\underline{\dot\beta}}{}^{\underline{\delta}} + T_{\underline{\gamma}\underline{\dot\beta}}{}^{\underline{e}}\psi_{\underline{e}}{}^{\underline{\delta}})$$

$$-\ \epsilon^{\underline{\beta}}\psi_{\underline{a}}{}^{\underline{\gamma}}(T_{\underline{\beta}\underline{\gamma}}{}^{\underline{\delta}} + T_{\underline{\beta}\underline{\gamma}}{}^{\underline{e}}\psi_{\underline{e}}{}^{\underline{\delta}}) - \bar\epsilon^{\underline{\dot\beta}}\bar\psi_{\underline{a}}{}^{\underline{\dot\gamma}}(T_{\underline{\dot\beta}\underline{\dot\gamma}}{}^{\underline{\delta}} + T_{\underline{\dot\beta}\underline{\dot\gamma}}{}^{\underline{e}}\psi_{\underline{e}}{}^{\underline{\delta}})\ ,$$

$$\delta_Q\phi_{\underline{a}}{}^{\boldsymbol{\iota}} = -\,\epsilon^{\underline{\beta}}(R_{\underline{\beta}\underline{a}}{}^{\boldsymbol{\iota}} + T_{\underline{\beta}\underline{a}}{}^{\underline{e}}\phi_{\underline{e}}{}^{\boldsymbol{\iota}}) - \bar\epsilon^{\underline{\dot\beta}}(R_{\underline{\dot\beta}\underline{a}}{}^{\boldsymbol{\iota}} + T_{\underline{\dot\beta}\underline{a}}{}^{\underline{e}}\phi_{\underline{e}}{}^{\boldsymbol{\iota}})$$

$$-\ (\bar\epsilon^{\underline{\dot\beta}}\psi_{\underline{a}}{}^{\underline{\gamma}} + \epsilon^{\underline{\gamma}}\bar\psi_{\underline{a}}{}^{\underline{\dot\beta}})(R_{\underline{\gamma}\underline{\dot\beta}}{}^{\boldsymbol{\iota}} + T_{\underline{\gamma}\underline{\dot\beta}}{}^{\underline{e}}\phi_{\underline{e}}{}^{\boldsymbol{\iota}})$$

$$-\ \epsilon^{\underline{\beta}}\psi_{\underline{a}}{}^{\underline{\gamma}}(R_{\underline{\beta}\underline{\gamma}}{}^{\boldsymbol{\iota}} + T_{\underline{\beta}\underline{\gamma}}{}^{\underline{e}}\phi_{\underline{e}}{}^{\boldsymbol{\iota}}) - \bar\epsilon^{\underline{\dot\beta}}\bar\psi_{\underline{a}}{}^{\underline{\dot\gamma}}(R_{\underline{\dot\beta}\underline{\dot\gamma}}{}^{\boldsymbol{\iota}} + T_{\underline{\dot\beta}\underline{\dot\gamma}}{}^{\underline{e}}\phi_{\underline{e}}{}^{\boldsymbol{\iota}})\ . \quad (5.6.28)$$

These results can be specialized to $N = 1$, $n = -\frac{1}{3}$ superspace, and using the solution to constraints and Bianchi identities we can deduce the transformation law for $e_{\underline{a}}{}^{\underline{m}}$ and $\psi_{\underline{a}}{}^{\underline{\gamma}}$:

$$\delta_{\rm Q} e_{\underline{a}}{}^{\underline{m}} = -\, i(\epsilon^{\beta}\bar{\psi}_{\underline{a}}{}^{\dot{\beta}} + \bar{\epsilon}^{\dot{\beta}}\psi_{\underline{a}}{}^{\beta}) e_{\beta\dot{\beta}}{}^{\underline{m}} \quad , \tag{5.6.29a}$$

$$\delta_{\rm Q}\psi_{\underline{a}}{}^{\beta} = \mathbf{D}_{\underline{a}}\epsilon^{\beta} + i\bar{\epsilon}_{\dot{\alpha}} S\delta_{\alpha}{}^{\beta} - i\epsilon_{\alpha} A^{\beta}{}_{\dot{\alpha}}$$
$$-\, i(\bar{\epsilon}^{\dot{\gamma}}\psi_{\underline{a}}{}^{\gamma} + \epsilon^{\gamma}\bar{\psi}_{\underline{a}}{}^{\dot{\gamma}})\psi_{\gamma\dot{\gamma}}{}^{\beta} \quad . \tag{5.6.29b}$$

The transformation law of the gravitino can be simplified somewhat by considering the supersymmetry variation of $\psi_{\underline{m}}{}^{\beta}$. (The last term in (5.6.29b) is absent as a consequence of (5.6.29a).) The auxiliary fields $S$ and $A_{\underline{a}}$ are defined as $R|$ and $G_{\underline{a}}|$ respectively; consequently, their transformations can be found directly because $R$ and $G_{\underline{a}}$ are covariant (see below). These covariant definitions of the minimal auxiliary fields are the generalizations of the linearized expressions of (5.2.8) and (5.2.73).

We should point out that the results for $\delta_Q\phi_{\underline{a}}{}^{\boldsymbol{\iota}}$ are valid for gauged internal symmetries (such as $U(1)$, $SU(2)$, etc.) also. In this case $(M_{\boldsymbol{\iota}})_{\underline{\alpha}}{}^{\underline{\beta}} = 0$, $(M_{\boldsymbol{\iota}})_{\underline{\dot{\alpha}}}{}^{\underline{\dot{\beta}}} = 0$ and the quantities $R_{AB}{}^{\boldsymbol{\iota}}$ are the field strengths for the internal symmetry gauge superfield. Therefore the formulae in (5.6.28) contains part of the tensor calculus for a matter vector multiplet. The covariant components of such a multiplet are treated just like those of any covariant multiplet, e.g., a chiral scalar multiplet.

## e. Superspace field strengths

To simplify calculations with component gauge fields it is convenient to define supercovariant field strengths (quantities which transform without derivatives of the local supersymmetry parameter). We begin by computing

$$[\nabla_{\underline{a}}| \,, \nabla_{\underline{b}}|] = [\mathbf{D}_{\underline{a}}\,, \mathbf{D}_{\underline{b}}] + (\mathbf{D}_{[\underline{a}}\psi_{\underline{b}]}{}^{\underline{\gamma}})\partial_{\underline{\gamma}} + (\mathbf{D}_{[\underline{a}}\bar{\psi}_{\underline{b}]}{}^{\underline{\dot{\gamma}}})\bar{\partial}_{\underline{\dot{\gamma}}}$$
$$+\, (\phi_{[\underline{a}|\underline{\gamma}}{}^{\underline{\delta}}\psi_{|\underline{b}]}{}^{\underline{\gamma}})\partial_{\underline{\delta}} + (\bar{\phi}_{[\underline{a}|\underline{\dot{\gamma}}}{}^{\underline{\dot{\delta}}}\bar{\psi}_{|\underline{b}]}{}^{\underline{\dot{\gamma}}})\bar{\partial}_{\underline{\dot{\delta}}} \quad . \tag{5.6.30}$$

The ordinary spacetime torsions and curvatures are defined by (see (5.1.17))

$$[\mathbf{D}_{\underline{a}}\,, \mathbf{D}_{\underline{b}}] = t_{\underline{a}\underline{b}}{}^{\underline{c}}\mathbf{D}_{\underline{c}} + r_{\underline{a}\underline{b}}{}^{\boldsymbol{\iota}}M_{\boldsymbol{\iota}} \quad , \tag{5.6.31a}$$

where

$$t_{\underline{a}\underline{b}}{}^{\underline{c}} = c_{\underline{a}\underline{b}}{}^{\underline{c}} + \phi_{[\underline{a}}{}^{\boldsymbol{\iota}}(M_{\boldsymbol{\iota}})_{\underline{b}]}{}^{\underline{c}} \quad , \tag{5.6.31b}$$

$$[e_{\underline{a}}, e_{\underline{b}}] = c_{\underline{a}\underline{b}}{}^{\underline{c}} e_{\underline{c}} \quad , \tag{5.6.31c}$$

$$r_{\underline{a}\underline{b}}{}^{\boldsymbol{\iota}} = e_{[\underline{a}}\phi_{\underline{b}]}{}^{\boldsymbol{\iota}} - c_{\underline{a}\underline{b}}{}^{\underline{c}}\phi_{\underline{c}}{}^{\boldsymbol{\iota}} + \phi_{\underline{a}}{}^{\boldsymbol{\iota}_1}\phi_{\underline{b}}{}^{\boldsymbol{\iota}_2} f_{\boldsymbol{\iota}_1\boldsymbol{\iota}_2}{}^{\boldsymbol{\iota}} \quad , \tag{5.6.31d}$$

and $[M_{\boldsymbol{\iota}_1}, M_{\boldsymbol{\iota}_2}] = f_{\boldsymbol{\iota}_1\boldsymbol{\iota}_2}{}^{\boldsymbol{\iota}_3} M_{\boldsymbol{\iota}_3}$. We define a curvature for $\psi_{\underline{a}}{}^{\underline{\beta}}$ by

$$t_{\underline{a}\underline{b}}{}^{\underline{\gamma}} \equiv \mathbf{D}_{[\underline{a}}\psi_{\underline{b}]}{}^{\underline{\gamma}} - t_{\underline{a}\underline{b}}{}^{\underline{d}}\psi_{\underline{d}}{}^{\underline{\gamma}}$$

$$= e_{[\underline{a}}\psi_{\underline{b}]}{}^{\underline{\gamma}} - c_{\underline{a}\underline{b}}{}^{\underline{d}}\psi_{\underline{d}}{}^{\underline{\gamma}} - \psi_{[\underline{b}}{}^{\underline{\delta}}\phi_{\underline{a}]\underline{\delta}}{}^{\underline{\gamma}} \quad . \tag{5.6.32}$$

We now have all the $x$-space field strengths. From the fact that $e_{\underline{a}}{}^{\underline{m}}$, $\psi_{\underline{a}}{}^{\underline{\gamma}}$, and $\phi_{\underline{a}}{}^{\boldsymbol{\iota}}$ are gauge fields, $t_{\underline{a}\underline{b}}{}^{\underline{c}}$, $t_{\underline{a}\underline{b}}{}^{\underline{\gamma}}$, and $r_{\underline{a}\underline{b}}{}^{\boldsymbol{\iota}}$ are the appropriate field strengths. The field strength associated with $M_{\gamma\delta}$ (and $\bar{M}_{\dot{\gamma}\dot{\delta}}$), $r_{\underline{a}\underline{b},\gamma\delta} = -\overline{r_{\underline{a}\underline{b},\dot{\gamma}\dot{\delta}}}$ is the Riemann curvature tensor. With these definitions we can express the superspace torsion and curvatures *at* $\theta = 0$ as

$$T_{\underline{a}\underline{b}}{}^{C} = t_{\underline{a}\underline{b}}{}^{C} + \psi_{[\underline{a}}{}^{\delta} T_{\delta\underline{b}]}{}^{C} + \bar{\psi}_{[\underline{a}}{}^{\dot{\delta}} T_{\dot{\delta}\underline{b}]}{}^{C} + \psi_{[\underline{a}}{}^{\delta}\bar{\psi}_{\underline{b}]}{}^{\dot{\epsilon}} T_{\delta\dot{\epsilon}}{}^{C}$$

$$+ \psi_{\underline{a}}{}^{\delta}\psi_{\underline{b}}{}^{\epsilon} T_{\delta\epsilon}{}^{C} + \bar{\psi}_{\underline{a}}{}^{\dot{\delta}}\bar{\psi}_{\underline{b}}{}^{\dot{\epsilon}} T_{\dot{\delta}\dot{\epsilon}}{}^{C} \quad , \tag{5.6.33}$$

$$R_{\underline{a}\underline{b}}{}^{\boldsymbol{\iota}} = r_{\underline{a}\underline{b}}{}^{\boldsymbol{\iota}} + \psi_{[\underline{a}}{}^{\delta} R_{\delta\underline{b}]}{}^{\boldsymbol{\iota}} + \bar{\psi}_{[\underline{a}}{}^{\dot{\delta}} R_{\dot{\delta}\underline{b}]}{}^{\boldsymbol{\iota}}$$

$$+ \psi_{[\underline{a}}{}^{\delta}\bar{\psi}_{\underline{b}]}{}^{\dot{\epsilon}} R_{\delta\dot{\epsilon}}{}^{\boldsymbol{\iota}} + \psi_{\underline{a}}{}^{\delta}\psi_{\underline{b}}{}^{\epsilon} R_{\delta\epsilon}{}^{\boldsymbol{\iota}} + \bar{\psi}_{\underline{a}}{}^{\dot{\delta}}\bar{\psi}_{\underline{b}}{}^{\dot{\epsilon}} R_{\dot{\delta}\dot{\epsilon}}{}^{\boldsymbol{\iota}} \quad . \tag{5.6.34}$$

We have used

$$[\nabla_{\underline{a}}, \nabla_{\underline{b}}]| = [\nabla_{\underline{a}}|, \nabla_{\underline{b}}|] + \psi_{[\underline{a}}{}^{\gamma}\nabla_{\gamma}\nabla_{\underline{b}]}| + \bar{\psi}_{[\underline{a}}{}^{\dot{\gamma}}\bar{\nabla}_{\dot{\gamma}}\nabla_{\underline{b}]}| \tag{5.6.35}$$

and (5.6.15). We see these tensors differ from their $x$-space analogs (5.1.17,18) by additional gravitino terms. The superspace field strengths are covariant: Therefore the $\theta = 0$ projections of (5.6.33,34) are the supercovariant $x$-space field strengths.

For the gauge fields of internal symmetries, covariant field strengths are also necessary. These field strengths are defined by exactly the same formulae as the curvatures above.

### f. Supercovariant supergravity field strengths

We now use the explicit solution of the $n = -\frac{1}{3}$ Bianchi identities to obtain from (5.6.33,34) the component field strengths. The solution of the Bianchi identities contains all of the necessary information about the torsions and curvatures. For the torsions and curvatures with at least one lower spinorial index, we substitute from (5.2.81) into the left hand side of (5.6.33).

Considering first $T_{\underline{ab}}{}^{\gamma}$ we find

$$T_{\underline{ab}}{}^{\gamma} = t_{\underline{ab}}{}^{\gamma} + i(\,\psi_{\underline{a}\beta} G^{\gamma}{}_{\dot{\beta}} - \psi_{\underline{b}\alpha} G^{\gamma}{}_{\dot{\alpha}}\,) - i(\,\bar{\psi}_{\underline{a}\dot{\beta}} \delta_{\beta}{}^{\gamma} - \bar{\psi}_{\underline{b}\dot{\alpha}} \delta_{\alpha}{}^{\gamma}\,)\, R\ . \tag{5.6.36}$$

This equation is correct to $\theta$-independent order and thus a supercovariant gravitino field strength, $f_{\underline{ab}}{}^{\gamma}$, is defined by

$$f_{\underline{ab}}{}^{\gamma} = T_{\underline{ab}}{}^{\gamma}|\ \ . \tag{5.6.37}$$

For $T_{\underline{ab}}{}^{\underline{c}}$ we use (5.6.33) in a slightly different way. Along with the torsions with at least one lower spinorial index, we also substitute for $T_{\underline{ab}}{}^{\underline{c}}$ on the right side. This yields

$$t_{\underline{ab}}{}^{\underline{c}} + i\psi_{[\underline{a}}{}^{\gamma} \bar{\psi}_{\underline{b}]}{}^{\dot{\gamma}} = i(C_{\alpha\beta} \delta_{\dot{\alpha}}{}^{\dot{\gamma}} G^{\gamma}{}_{\dot{\beta}} - C_{\dot{\alpha}\dot{\beta}} \delta_{\alpha}{}^{\gamma} G_{\beta}{}^{\dot{\gamma}})\ \ . \tag{5.6.38}$$

Now we can take this result, use it to solve for the component spin-connection, and thus obtain a second order formalism. Before doing this it is convenient to observe that

$$i(\,C_{\alpha\beta} C_{\dot{\alpha}\dot{\gamma}} G_{\gamma\dot{\beta}} - C_{\dot{\alpha}\dot{\beta}} C_{\alpha\gamma} G_{\beta\dot{\gamma}}\,) = -\,\epsilon_{\underline{abcd}} G^{\underline{d}}\ , \tag{5.6.39}$$

so that $\phi_{\underline{abc}}$ can be expressed as

$$\phi_{\underline{abc}} = \phi(e)_{\underline{abc}} + i\,\frac{1}{2}\,(\,\psi_{[\underline{b}\alpha} \bar{\psi}_{\underline{c}]\dot{\alpha}} + \psi_{[\underline{a}\beta} \bar{\psi}_{\underline{c}]\dot{\beta}} - \psi_{[\underline{a}\gamma} \bar{\psi}_{\underline{b}]\dot{\gamma}}\,) - \frac{1}{2}\,\epsilon_{\underline{abcd}} A^{\underline{d}}\ \ . \tag{5.6.40}$$

where $\phi(e)_{\underline{abc}}$ is defined in (5.1.19).

Finally the supercovariantized Riemann curvature tensor is treated analogously to $T_{\underline{ab}}{}^{\gamma}$. We begin by substituting from (5.2.81) for the curvatures with at least one spinorial index.

$$R_{\underline{ab}\,\gamma\delta} = r_{\underline{ab}\,\gamma\delta} + \bar{R}\psi_{\underline{a}(\gamma} \psi_{\underline{b}\delta)} + i\{\,[\,\bar{\psi}_{\underline{a}\dot{\beta}} W_{\beta\gamma\delta}$$

$$- \frac{1}{2} \psi_{\underline{a}}{}^{\epsilon} ( C_{\epsilon\beta} \nabla_{(\gamma} G_{\delta)\dot{\beta}} + ( \overline{\nabla}_{\dot{\beta}} \overline{R} ) C_{\epsilon(\gamma} C_{\delta)\beta} ) ] - ( \underline{a} \longleftrightarrow \underline{b} ) \} . \qquad (5.6.41)$$

However, we must carry out one further step before we have an expression which can be evaluated in terms of component fields. We must eliminate $\nabla_{\gamma} G_{\delta\dot{\beta}}$, $\overline{\nabla}_{\dot{\beta}} \overline{R}$, and $W_{\beta\gamma\delta}$ from this expression. This can be done by considering the coefficients of $\nabla_{\gamma}$ on both sides of (5.2.81). On the left hand side we find $f_{\underline{ab}}{}^{\gamma}$, while on the right hand side $W_{\alpha\beta\gamma}$, $\overline{\nabla}_{\dot{\alpha}} G_{\gamma\dot{\beta}}$, and $\nabla_{\alpha} R$ appear. We can therefore solve for these quantities in terms of $f_{\underline{ab}}{}^{\gamma}$ which is expressed in terms of component fields in (5.6.36).

$$W_{\alpha\beta\gamma} = \frac{1}{12} f_{(\alpha\dot{\alpha},\beta}{}^{\dot{\alpha}}{}_{\gamma)} \ , \qquad (5.6.42a)$$

$$\nabla_{\alpha} G_{\underline{b}} = \ - \frac{1}{2} [ \overline{f}_{\alpha\dot{\alpha},\beta}{}^{\dot{\alpha}}{}_{,\dot{\beta}} - \frac{1}{3} C_{\alpha\beta} \overline{f}_{\gamma\dot{\beta},}{}^{\gamma\dot{\delta},}{}_{\dot{\delta}} ] \ , \qquad (5.6.42b)$$

$$\nabla_{\alpha} R = - \frac{1}{3} f_{\alpha\dot{\beta},\gamma}{}^{\dot{\beta},\gamma} \ . \qquad (5.6.42c)$$

These expressions can now be substituted into (5.6.41) which results in a well defined (at the component level) supercovariantized Riemann curvature tensor.

As a by-product of this process we have also derived the component supersymmetry transformation law of the auxiliary fields $A_{\underline{a}}$ and $S$ where $S \equiv R|$ and $A_{\underline{a}} \equiv G_{\underline{a}}|$. These are supercovariants and hence their supersymmetry variations are given by,

$$\delta_{\mathrm{Q}} A_{\underline{a}} = i K_{\mathrm{Q}} G_{\underline{a}}| = \ - \frac{1}{2} [ \epsilon^{\gamma} \overline{f}_{\gamma\dot{\beta},\alpha}{}^{\dot{\beta}}{}_{,\dot{\alpha}} + \frac{1}{3} \epsilon_{\alpha} \overline{f}_{\beta\dot{\alpha},}{}^{\beta}{}_{\dot{\delta}}{}^{\dot{\delta}} ] + h.\,c. \ , \qquad (5.6.43a)$$

$$\delta_{\mathrm{Q}} S = i K_{\mathrm{Q}} R| = - \frac{1}{3} \epsilon^{\alpha} f_{\alpha\dot{\beta},\gamma}{}^{\dot{\beta},\gamma} \ . \qquad (5.6.43b)$$

## g. Tensor calculus

The component rules for the manipulation of locally supersymmetric quantities are called the tensor calculus for supergravity theories. These rules give a component by component description of supersymmetric theories. Superfields on the other hand provide a concise description of these theories in much the same way that vector notation provides a more concise description of Maxwell's equations. Superfields can always be reduced to their component field content in the case of global supersymmetry and in this section we discuss the analogous procedure in the locally supersymmetric case.

As an example, let us consider for $N = 1$ supergravity a local scalar multiplet described by a covariantly chiral superfield $\eta$, $\overline{\nabla}_{\dot{\alpha}}\eta = 0$. The component fields of this multiplet are defined by projection

$$A \equiv \eta| \quad ,$$

$$\psi_\alpha \equiv \nabla_\alpha \eta| \quad ,$$

$$F \equiv \nabla^2 \eta| \quad . \tag{5.6.44}$$

The infinitesimal supersymmetry transformations of *all* quantities are obtained by commutation with $iK_Q(\epsilon)$. Thus, using (5.6.18)

$$\delta_{\rm Q} A = iK_Q(\epsilon)\eta| = -\,\epsilon^\alpha \nabla_\alpha \eta| = -\,\epsilon^\alpha \psi_\alpha \quad ,$$

$$\delta_{\rm Q}\psi_\alpha = -\,(\epsilon^\beta \nabla_\beta + \bar{\epsilon}^{\dot{\beta}}\overline{\nabla}_{\dot{\beta}})\nabla_\alpha \eta|$$

$$= -\,\big[\epsilon^\beta(\tfrac{1}{2}\{\nabla_\alpha\,,\nabla_\beta\} + C_{\alpha\beta}\nabla^2) + \bar{\epsilon}^{\dot{\beta}}\{\nabla_\alpha\,,\overline{\nabla}_{\dot{\beta}}\}\big]\eta| \quad . \tag{5.6.45}$$

At this point, no specific choice of auxiliary fields for supergravity has been made. The only constraints on the superspace torsions necessary are those which follow as consistency requirements for the existence of chiral superfields, i.e., the representation-preserving constraints. Using the solution to the Bianchi identities for the case of $N = 1$, $n = -\,\frac{1}{3}$ supergravity we obtain

$$\delta_{\rm Q}\psi_\alpha = \epsilon_\alpha F - i\bar{\epsilon}^{\dot{\beta}}(\nabla_{\alpha\dot{\beta}}\eta)| \; . \tag{5.6.46}$$

Using (5.6.6), we have

$$\delta_{\rm Q}\psi_\alpha = \epsilon_\alpha F - i\bar{\epsilon}^{\dot{\alpha}}(\mathbf{D}_{\alpha\dot{\alpha}}A + \psi_{\alpha\dot{\alpha}}{}^\gamma \psi_\gamma) \quad . \tag{5.6.47}$$

The last expression illustrates the concept of a *supercovariant derivative* at the component level. The combination $[\mathbf{D}_{\underline{a}}A + \psi_{\underline{a}}{}^\gamma\psi_\gamma]$, which generalizes the ordinary covariant derivative $\mathbf{D}_{\underline{a}}A$, transforms without a term proportional to $\mathbf{D}_{\underline{a}}\epsilon^\gamma$. Thus, this combination of fields is covariant with respect to a local component supersymmetry transformation. Finally, for the transformation law for the auxiliary field we find

$$\delta_Q F = -(\epsilon^\alpha \nabla_\alpha + \bar{\epsilon}^{\dot{\alpha}} \bar{\nabla}_{\dot{\alpha}}) \nabla^2 \eta|$$

$$= -\bar{\epsilon}^{\dot{\alpha}} [i\mathbf{D}^\beta{}_{\dot{\alpha}} \psi_\beta - A^\beta{}_{\dot{\alpha}} \psi_\beta - i\psi^\beta{}_{\dot{\alpha}\beta} F - \psi^\beta{}_{\dot{\alpha}}{}^{\dot{\beta}} (\mathbf{D}_{\beta\dot{\beta}} A + \psi_{\beta\dot{\beta}}{}^\gamma \psi_\gamma)] + \bar{S} \epsilon^\alpha \psi_\alpha$$

$$= -\bar{\epsilon}^{\dot{\alpha}} [i\underset{\sim}{\mathbf{D}}^\beta{}_{\dot{\alpha}} - A^\beta{}_{\dot{\alpha}}] \psi_\beta + \bar{S} \epsilon^\alpha \psi_\alpha \quad . \tag{5.6.48}$$

(Analogous transformations for a chiral scalar multiplet can be found for $n \neq -\frac{1}{3}$ by using the appropriate solution to the Bianchi identities.) On the second line above we have introduced the notation $\underset{\sim}{\mathbf{D}}_{\underline{a}} \psi_\beta$ for the supercovariant derivative of the spinor matter field.

We can also find the components of the product of two different multiplets in terms of the components of the original multiplets. Thus, for example, a product of two chiral scalar multiplets described by chiral superfields $\eta_1$, $\eta_2$ is the scalar multiplet described by the chiral superfield $\eta_3 = \eta_1 \eta_2$:

$$A_3 = \eta_1 \eta_2| = A_1 A_2 \quad ,$$

$$(\psi_3)_\alpha = \nabla_\alpha (\eta_1 \eta_2)| = ([\nabla_\alpha \eta_1] \eta_2 + \eta_2 [\nabla_\alpha \eta_2])| = (\psi_1)_\alpha A_2 + A_1 (\psi_2)_\alpha \quad ,$$

$$F_3 = \nabla^2 (\eta_1 \eta_2)| = ([\nabla^2 \eta_1] \eta_2 + [\nabla^\beta \eta_1][\nabla_\beta \eta_2] + \eta_1 [\nabla^2 \eta_2])|$$

$$= F_1 A_2 + (\psi_1)^\beta (\psi_2)_\beta + A_1 F_2 \quad . \tag{5.6.49}$$

The components of $\Phi_3$ transform according to (5.6.45,47,48). This multiplication law is just like in the global case (3.6.11).

Another possible product of two scalar multiplets is found by taking the product of a chiral superfield $\eta_1$ and an antichiral superfield $\bar{\eta}_2$; this gives the complex general scalar superfield $\Psi = \eta_1 \bar{\eta}_2$:

$$\Psi| = A_1 \bar{A}_2 \quad ,$$

$$\nabla_\alpha \Psi| = \psi_{1\alpha} \bar{A}_2 \quad ,$$

$$\nabla^2 \Psi| = F_1 \bar{A}_2 \quad ,$$

$$[\nabla_\alpha , \overline{\nabla}_{\dot{\alpha}}]\Psi| = 2\psi_{1\alpha}\overline{\psi}_{2\dot{\alpha}} \ ,$$

$$(\overline{\nabla}^2 + R)\nabla_\alpha\Psi| = \overline{F}_2\psi_{1\alpha} + i\overline{\psi}_2{}^{\dot{\beta}}(\underset{\sim}{\mathbf{D}}_{\alpha\dot{\beta}}A_1) \ ,$$

$$\nabla^\alpha(\overline{\nabla}^2 + R)\nabla_\alpha\Psi| = -\,\psi_{1\alpha}(\,i\underset{\sim}{\mathbf{D}}^{\alpha\dot{\beta}} + A^{\alpha\dot{\beta}})\overline{\psi}_{2\dot{\beta}} + \overline{\psi}_2{}^{\dot{\beta}}(-\,i\underset{\sim}{\mathbf{D}}_{\alpha\dot{\beta}} + 2A_{\alpha\dot{\beta}})\psi_1{}^\alpha$$

$$+\, 2\overline{F}_2F_1 - (\underset{\sim}{\mathbf{D}}^{\alpha\dot{\beta}}\overline{A}_2)(\underset{\sim}{\mathbf{D}}_{\alpha\dot{\beta}}A_1) \ . \qquad (5.6.50)$$

where we have used $(\overline{\nabla}^2 + R)\nabla_\alpha\eta = 0$ which can be obtained from (5.4.16). Note the appearance of the supercovariant derivative $\underset{\sim}{\mathbf{D}}_{\underline{a}}$.

We can also give the components of a chiral superfield made out of an antichiral one $\eta_1 = (\overline{\nabla}^2 + R)\overline{\eta}$. This is sometimes called the kinetic multiplet. Its components are:

$$A_1 = (\overline{\nabla}^2 + R)\overline{\eta}| = \overline{F} + S\overline{A} \ ,$$

$$(\psi_1)_\alpha = -\,(\,i\underset{\sim}{\mathbf{D}}_{\alpha\dot{\beta}} + A_{\alpha\dot{\beta}})\overline{\psi}^{\dot{\beta}} - \frac{1}{3}\,f_{\alpha\dot{\beta},\gamma}{}^{\dot{\beta},\gamma}\overline{A} \ ,$$

$$F_1 = (\,(\,\underset{\sim}{\mathbf{D}}^{\underline{a}} + i3A^{\underline{a}})\underset{\sim}{\mathbf{D}}_{\underline{a}} - \frac{1}{3}\,R_{\alpha\dot{\alpha},}{}^{\alpha}{}_{\dot{\beta}}{}^{\dot{\alpha}\dot{\beta}} - 4S\overline{S})\overline{A} - 4\overline{S}F - \frac{1}{3}\,\overline{f}_{\alpha\dot{\beta},}{}^{\alpha}{}_{\dot{\gamma}}{}^{,\dot{\gamma}}\overline{\psi}^{\dot{\beta}} \ . \qquad (5.6.51)$$

where we have made use of the result

$$\nabla^2 R + 2R\overline{R} = -\,\frac{1}{6}\,R_{\alpha\dot{\alpha},}{}^{\alpha}{}_{\dot{\beta}}{}^{\dot{\alpha}\dot{\beta}} \ . \qquad (5.6.52)$$

The $x$-space supercovariant curvature $R_{\alpha\dot{\alpha},}{}^{\alpha}{}_{\dot{\beta}}{}^{\dot{\alpha}\dot{\beta}}$ in (5.6.51) is given by (5.6.41). The computation of these results is straightforward but tedious. All of the above results have made extensive use of the commutator algebra in (5.2.82).

As a further example of component tensor calculus, we consider the vector multiplet. The local components are defined in the same way as in the global case (4.3.5), but with $\nabla_A$ replaced by $\underset{\sim}{\nabla}_A$, the supergravity and Yang-Mills covariant derivative. The field strengths and Bianchi identities for the vector multiplet take the forms

$$F^{YM}{}_{\alpha\beta} = F^{YM}{}_{\alpha\dot{\beta}} = 0,$$

$$F^{YM}{}_{\alpha,\underline{b}} = -\, iC_{\beta\alpha}\bar{W}_{\dot{\beta}}{}^{YM} \quad ,$$

$$F^{YM}{}_{\underline{a},\underline{b}} = C_{\dot{\alpha}\dot{\beta}}\tilde{f}_{\alpha\beta} + C_{\alpha\beta}\tilde{\bar{f}}_{\dot{\alpha}\dot{\beta}} \quad ,$$

$$\bar{\underset{\sim}{\nabla}}_{\dot{\alpha}} W_{\alpha}{}^{YM} = \underset{\sim}{\nabla}^{\alpha} W_{\alpha}{}^{YM} + \bar{\underset{\sim}{\nabla}}{}^{\dot{\alpha}} \bar{W}_{\dot{\alpha}}{}^{YM} = 0 \quad . \tag{5.6.53}$$

The quantity $\tilde{f}_{\alpha\beta}$ is a supercovariant field strength (see below). The local components of the multiplet are thus defined by

$$v_{\underline{a}} = \Gamma_{\underline{a}}| \ , \ \lambda_{\alpha} = W_{\alpha}{}^{YM}| \ , \ \mathrm{D}' = -\, i\,\frac{1}{2}\,\underset{\sim}{\nabla}^{\alpha} W_{\alpha}{}^{YM}| \quad . \tag{5.6.54}$$

The supersymmetry variations of the covariant components ,$\lambda_{\alpha}$ and $\mathrm{D}'$, are obtained as with the components of the chiral multiplet (see (5.6.46)).

$$\delta_{\alpha\mathrm{Q}}\lambda = -\, \epsilon^{\beta}\tilde{f}_{\alpha\beta} + i\epsilon_{\alpha}\mathrm{D}' \quad ,$$

$$\delta_{\mathrm{Q}}\mathrm{D}' = \frac{1}{2}\,(\, \epsilon^{\alpha}\underset{\sim}{\mathbf{D}}_{\alpha\dot{\beta}}\bar{\lambda}^{\dot{\beta}} - \bar{\epsilon}^{\dot{\beta}}\underset{\sim}{\mathbf{D}}_{\alpha\dot{\beta}}\lambda^{\alpha}\,) \quad , \tag{5.6.55}$$

where $\underset{\sim}{\mathbf{D}}_{\alpha\dot{\beta}}\bar{\lambda}^{\dot{\beta}}$ is the supercovariant derivative of $\lambda$,

$$\underset{\sim}{\mathbf{D}}_{\alpha\dot{\beta}}\bar{\lambda}^{\dot{\beta}} = \mathbf{D}_{\alpha\dot{\beta}}\bar{\lambda}^{\dot{\beta}} - \bar{\psi}_{\alpha\dot{\beta}\dot{\gamma}}(\,\tilde{\bar{f}}^{\dot{\beta}\dot{\gamma}} - iC^{\dot{\beta}\dot{\gamma}}\mathrm{D}'\,) \quad . \tag{5.6.56}$$

This follows from (5.6.13) and the Bianchi identities of the vector multiplet (4.2.90) which are valid in a curved superspace. The quantity $\tilde{f}_{\alpha\beta}$ (and its conjugate $\tilde{\bar{f}}_{\dot{\alpha}\dot{\beta}}$) can be calculated in the same way as (5.6.41) from ( 5.6.34,53)

$$\tilde{f}_{\alpha\beta} = \frac{1}{2}\,[\, f^{YM}{}_{\alpha\dot{\alpha},\beta}{}^{\dot{\alpha}} + i(\,\psi_{(\alpha\dot{\alpha},\beta)}\bar{\lambda}^{\dot{\alpha}}) + i(\,\bar{\psi}_{(\alpha\dot{\alpha},}{}^{\dot{\alpha}}\lambda_{\beta)})\,] \quad , \tag{5.6.57}$$

where $f^{YM}{}_{\underline{a},\underline{b}}$ is the ordinary $x$-space Yang-Mills field strength. For the transformation law of $v_{\underline{a}}$, we use (5.6.28). (Even though the derivation of that result was for the gauge fields for tangent space symmetries, it also applies to the gauge fields for internal symmetries.)

$$\delta_{\mathrm{Q}}v_{\underline{a}} = i(\,\epsilon_{\alpha}\bar{\lambda}_{\dot{\alpha}} + \bar{\epsilon}_{\dot{\alpha}}\lambda_{\alpha}) - i(\,\epsilon^{\beta}\bar{\psi}_{\underline{a}}{}^{\dot{\beta}} + \bar{\epsilon}^{\dot{\beta}}\psi_{\underline{a}}{}^{\beta})v_{\beta\dot{\beta}} \quad . \tag{5.6.58}$$

Just as for the gravitino tansformation law in (5.6.29b), the last two terms above are absent if we consider the transformation law of $v_{\underline{m}}$.

We have presented the above results for $N = 1$, $n = -\frac{1}{3}$ supergravity; they can be generalized to all $n$ by using the appropriate set of Bianchi identities (5.4.16-17).

In our discussion of global supermultiplets we found a large number of gauge multiplets where the component gauge field was not a spin-one field (for example the tensor multiplet). Since we gave a completely geometrical treatment of these multiplets using $p$-forms within global supersymmetry, their extension to the locally supersymmetric case (i.e., transformation laws, supercovariant field strengths, etc.) is obtained by the straightforward generalization of the methods which we used to treat the spin-one case. The only complication that can occur is that the existence of the unconstrained prepotential must be consistent with the set of constraints that describe the supergravity background. An example of an $N = 1$ multiplet for which the superfield extension to local supersymmetry is not known is the matter gravitino multiplet. This is not surprising since a second supersymmetry (i.e., $N = 2$ supersymmetry) is required for the consistency of the equations of motion for the matter gravitino.

### h. Component actions

Finally we give formulae to obtain component actions from the $N = 1$ $n = -\frac{1}{3}$ superspace actions

$$S_1 = \int d^4x \, d^2\theta \, \phi^3 \, \mathbb{L}_{chiral} \quad , \tag{5.6.59a}$$

$$S_2 = \int d^4x \, d^4\theta \, E^{-1} \, \mathbb{L}_{general} \quad . \tag{5.6.59b}$$

We first have

$$S_1 = \int d^4x \; \mathrm{e}^{-1} [\nabla^2 + i \bar{\psi}^{\alpha\dot{\alpha}}{}_{\dot{\alpha}} \nabla_\alpha + 3\bar{S} + \frac{1}{2} \bar{\psi}_{\alpha(\dot{\alpha}|}{}^{\dot{\alpha}} \bar{\psi}^{\alpha}{}_{|\dot{\beta})}{}^{\dot{\beta}} ] \mathbb{L}_{chiral}| \quad . \tag{5.6.60}$$

To derive (5.6.60) there are several steps. First, using (5.5.9) and choosing $\mathbb{L} = R^{-1}$, we see that $\phi^3 = \bar{D}^2 E^{-1} R^{-1}$. This is a density under $x$-space coordinate transformations. But in $x$-space, a density is $\mathrm{e}^{-1}$ multiplied possibly by a dimensionless $x$-space scalar.

No such dimensionless scalars can be constructed in the minimal theory. Therefore $\phi^3$ at lowest order in $\theta$ must be proportional to $\mathrm{e}^{-1}$. (It should be noted that there are no explicit factors of $\kappa$ anywhere except that multiplying the supergravity action.) This situation is not true for the nonminimal theories, in which it is possible to construct a dimensionless scalar from some of the additional auxiliary fields. This is precisely what happens for the F-type density for the nonminimal theory, and is responsible for the nonpolynomiality discussed in subsec. 5.5.f.3.

Once we know that the lowest component of $\phi^3$ is $\mathrm{e}^{-1}$, we derive (5.6.60) by multiplying $\mathrm{e}^{-1}$ by the highest component $F$ of a chiral superfield and performing a supersymmetry transformation. This generates a term proportional to $F$ times the gravitino, which we can cancel by adding to $\mathrm{e}^{-1}F$ a term proportional to $\bar{\psi}^{\alpha\dot{\alpha}}{}_{\dot{\alpha}}\psi_\alpha$. This new term generates supersymmetry variations proportional to $\underset{\sim}{\mathbf{D}}A$ times the gravitino. These can be canceled by adding a term proportional to $\bar{\psi}^2 A$ to the starting point. Finally, we determine the contribution of the $\bar{S}A$ term by canceling variations proportional to $\psi_\alpha \bar{S}$. By dimensional analysis, there can be no other contributions, and we have obtained the density formula of (5.6.60).

To find the corresponding expression for $S_2$, we use

$$S_2 = \int d^4x \, d^2\theta \; \phi^3 \, (\bar{\nabla}^2 + R) I\!\!L_{general} \tag{5.6.61}$$

and the formula (5.6.60) for the chiral case. The covariant derivatives act on the superfields in the Lagrangian and project out the components.

As a simple example, we compute the mass term for a chiral superfield $\eta$:

$$\begin{aligned} S &= \frac{1}{2} m \int d^4x \, d^2\theta \; \phi^3 \, \eta^2 \\ &= \frac{1}{2} m \int d^4x \; \mathrm{e}^{-1} [2FA + \psi^\alpha \psi_\alpha + i2\bar{\psi}^{\alpha\dot{\alpha}}{}_{\dot{\alpha}}\psi_\alpha \, A \\ &\qquad + (3\bar{S} + \frac{1}{2} \bar{\psi}_{\alpha(\dot{\alpha}|}{}^{\dot{\alpha}} \bar{\psi}^{\alpha}{}_{|\dot{\beta})}{}^{\dot{\beta}}) A^2] \ . \end{aligned} \tag{5.6.62}$$

As a second example we compute the $N = 1$, $n = -\frac{1}{3}$ component supergravity action and cosmological term.

From (5.5.4,34) we have

$$S_{SG} = \frac{1}{\kappa^2} \int d^4x \; \mathrm{e}^{-1} \, [ \, \frac{1}{2} \, r_{\alpha \dot{\alpha},}{}^{\alpha}{}_{\dot{\beta},}{}^{\dot{\alpha}\dot{\beta}} - \epsilon^{\underline{abcd}} \overline{\psi}_{\underline{a},\dot{\beta}} \mathbf{D}_{\underline{c}} \psi_{\underline{d},\beta} - 3|S|^2 \, ] \; . \tag{5.6.63}$$

This is the component form of the supergravity action with the improved spin connection. The axial vector auxiliary field is present implicitly in the first term since the spin connection, as defined in (5.6.40), depends on $A_{\underline{a}}$. If we separate out from $\phi(e)_{\underline{abc}}$ the contribution of $A_{\underline{a}}$ it appears *only* quadratically in the action. In particular, there is a cancellation among terms of the form $A_{\underline{a}} \psi_{\underline{b},\gamma} \overline{\psi}_{\underline{d},\epsilon}$ which come from the first two terms in the action.

For the cosmological term from (5.5.42) and using (5.6.60), we have

$$\begin{aligned} S_{cosmo} &= \lambda \kappa^{-2} \int d^4x \, d^2\theta \; \phi^3 + h.\,c. \\ &= \lambda \kappa^{-2} \int d^4x \; \mathrm{e}^{-1} [\, 3\overline{S} + \frac{1}{2} \, \overline{\psi}_{\alpha(\dot{\alpha}|}{}^{\dot{\alpha}} \overline{\psi}^{\alpha}{}_{|\dot{\beta})}{}^{\dot{\beta}} + h.\,c. ] \end{aligned} \tag{5.6.64}$$

The cosmological term contains at the component level an apparent mass term for the gravitino. However, in the deSitter background geometry the gravitino is actually massless, since it is still a gauge field.

In closing we make two observations: Although (5.6.61) was computed after the constraints were imposed on the covariant derivatives, in principle one can compute such an action formula without imposing *any* constraints at all. This follows because the transformation laws for the components of the totally unconstrained superspace are directly obtainable from (5.6.28) and, for matter multiplets, from equations analogous to (5.6.45-48). A large number of auxiliary fields defined as the $\theta = 0$ value of the various superspace torsions will enter such a construction. Among these occurs an auxiliary field which is a Lagrange multiplier that multiplies $\mathrm{e}^{-1}$. (The variation of this Lagrange multiplier will constrain the geometry of $x$-space.) Clearly this is unacceptable, and we have seen how, for $N = 1$ supergravity, this can be avoided. However, understanding the role of such fields may be necessary to understand $N > 4$ off-shell theories.

The second point is that we lack at present a direct method for computing density formulae analogous to (5.6.60). We can always compute such a formula by hand: We start with $\mathrm{e}^{-1} \times \nabla^{2N}\Psi$ (where $\Psi$ is an arbitrary superfield) and perform supersymmetry variations to obtain an entire density multiplet. What is lacking is a way to obtain this result without laborious calculation.

## 5.7. DeSitter supersymmetry

In sec. 3.2.f, we discussed the super-deSitter algebra (3.2.14). Here we describe how supersymmetric deSitter covariant derivatives can be obtained from supergravity covariant derivatives. We first discuss the nonsupersymmetric analog. Nonsupersymmetric deSitter covariant derivatives can be obtained from gravitational covariant derivatives by eliminating all field components except the (density) compensating field (i.e., the determinant of the metric or vierbein). This follows from the fact that in deSitter space the Weyl tensor vanishes, which says that there is no conformal (spin 2) part to the metric: It is "conformally flat". On the other hand, the scalar curvature tensor is a nonzero constant $r = 2\lambda^2$ (this is the gravity field equation).

We can write $e_{\underline{a}}{}^{\underline{m}} = \phi^{-1}\delta_{\underline{a}}{}^{\underline{m}}$ where $\phi$ is the compensator of (5.1.33). After setting the other components to zero, the action for deSitter gravity (Poincaré plus cosmological term) is just the action for a massless scalar field with a quartic self-interaction term. (The rest of gravity, the conformal part, is simply the locally conformal coupling of gravity to this scalar.) The equation of motion corresponding to the covariant equation $r = 2\lambda^2$,

$$\Box\phi = 2\lambda^2\phi^3 \quad , \tag{5.7.1}$$

has the solution, with appropriate boundary conditions,

$$\phi^{-1} = 1 - \lambda^2 x^2 \quad . \tag{5.7.2}$$

The deSitter covariant derivatives are now obtained from the gravity covariant derivatives of sec. 5.1 by substituting $e_{\underline{a}}{}^{\underline{m}} = \phi^{-1}\delta_{\underline{a}}{}^{\underline{m}}$, with $\phi^{-1}$ given by (5.7.2).

In the supersymmetric case, we start with the supergravity action and a cosmological term (5.5.16). We set $H$ to zero, and solve for the chiral density compensator $\phi$: In super-deSitter space $W_{\alpha\beta\gamma}$ vanishes (as does $G_{\underline{a}}$), while $R = \lambda$.

The action for the compensator is the massless Wess-Zumino action (again a conformal action, whose superconformal coupling to $H$ gives the deSitter supergravity action). The field equations in the chiral representation

$$\bar{D}^2\bar{\phi} = \lambda\phi^2 \tag{5.7.3}$$

have the solution

$$\phi^{-1} = 1 - \lambda\bar{\lambda}x^2 + \bar{\lambda}\theta^2 \quad . \tag{5.7.4}$$

The (real part of the) $\theta = 0$ component of $\phi$ is thus the gravity compensator of (5.7.1,2). The super-deSitter covariant derivatives are obtained by substituting this solution for $\phi$ (with $H = 0$) into the expressions for the supergravity covariant derivatives given in sec. 5.2.

The preceding discussion involved the $n = -\frac{1}{3}$ compensator $\phi$. For other $n$, we find strange pathologies: deSitter space cannot be described for $n \neq -\frac{1}{3}$ in a globally (deSitter) supersymmetric way. For $n = -\frac{1}{3}$, empty deSitter space is described by $R = \lambda$, $G_{\underline{a}} = W_{\alpha\beta\gamma} = 0$, but for nonminimal $n$ we would require $G_{\underline{a}} = W_{\alpha\beta\gamma} = 0$ with $T_\alpha \sim \lambda\theta_\alpha$. This follows from the fact that the commutators of covariant derivatives must take the following form to describe deSitter superspace

$$\{\nabla_\alpha , \nabla_\beta\} = -\, 2\lambda M_{\alpha\beta} \ \ , \ \ \{\nabla_\alpha , \bar{\nabla}_{\dot{\alpha}}\} = i\nabla_{\alpha\dot{\alpha}} \ \ , \ \ [\nabla_\alpha , \nabla_{\underline{b}}] = -\, i\,\lambda C_{\alpha\beta}\bar{\nabla}_{\dot{\beta}} \ \ ,$$

$$[\nabla_{\underline{a}} , \nabla_{\underline{b}}] = 2\lambda\bar{\lambda}(\, C_{\alpha\beta}\bar{M}_{\dot{\alpha}\dot{\beta}} + C_{\dot{\alpha}\dot{\beta}}M_{\alpha\beta}\,) \quad . \tag{5.7.5}$$

This requires *spontaneous breakdown* of $N = 1$ supersymmetry, since $T_\alpha$ is a tensor: $T_\alpha| = 0$ would imply $T_\alpha = 0$ if global (deSitter or other) supersymmetry were maintained. ($T_\alpha$ must be nonzero for $R$ to be nonzero in the nonminimal theory. See (5.2.80b)). For $n = 0$, $G_{\underline{a}} = W_{\alpha\beta\gamma} = 0$ already implies Minkowski space.

**Contents of 6. QUANTUM GLOBAL SUPERFIELDS**

## 6. QUANTUM GLOBAL SUPERFIELDS

### 6.1. Introduction to supergraphs

As we have seen in previous chapters, at the component level supersymmetric models are described by ordinary field theory Lagrangians, and their quantization and renormalization uses conventional methods. Evidently the quantum theory should be renormalized in a manner that preserves supersymmetry. Unless a manifestly supersymmetric regularization method is used, this requires applying the Ward-Takahashi identities of supersymmetry at each order of perturbation theory.

Supersymmetric models are in general less divergent than naive component power counting indicates, and this can be traced to the equality of numbers of bosonic and fermionic degrees of freedom, together with relations between coupling constants that are imposed by supersymmetry. We find that the vacuum energy (or, when (super)gravity is present, the cosmological term) receives no radiative corrections, and that, in renormalizable models, a common wave-function renormalization constant is sufficient to renormalize terms involving only scalar multiplet fields (the *no renormalization* theorem). A related result is a theorem that if the classical potential has a supersymmetric minimum (no spontaneous supersymmetry breaking), so does the effective potential to all orders of perturbation theory (no Coleman-Weinberg mechanism: see sec. 8.3.b).

Improved convergence due to supersymmetry is also evident in supergravity. For all $N$, the S-matrix of (extended) supergravity is finite at the first two loops; we argue in sec.7.7 that it is also finite at less than $N-1$ loops. In suitable supersymmetric gauges this finiteness also holds for the off-shell Green functions.

In supersymmetric theories the one-loop superconformal anomalies (trace of the energy-momentum tensor, $\gamma$-trace of the component supersymmetry current, and the divergence of the axial current) form a supersymmetric multiplet, the "supertrace", so that their coefficients are equal. There exist other anomalies as well. We show in sec. 7.10 that in nonminimal $N=1$ supergravity ($n \neq -\frac{1}{3}$), anomalies may be present in the Ward identities of local supersymmetry. Thus, in general, only minimal $N=1$ supergravity is consistent at the quantum level (but extended theories that have nonminimal $N=1$ supergravity as a submultiplet are consistent because of anomaly cancellation mechanisms).

Superfields greatly simplify classical calculations: Supersymmetric actions can be easily constructed, and the tensor calculus of supersymmetry becomes trivial. However, the greatest advantages of superfields appear at the quantum level. There are algebraic simplifications in supersymmetric Feynman graph ("supergraph") calculations for a number of reasons: (1) compactness of notation, (2) decrease in the number of indices (e.g., the vector field $A_{\underline{a}}$ is hidden inside the scalar superfield $V$), and (3) automatic cancellation of component graphs related by supersymmetry (which would require separate calculation in component formulations). Furthermore, the use of superfields leads to power-counting rules which explain many component results and can be used to derive additional finiteness predictions, especially when combined with supersymmetric background-field methods.

Renormalization is much simpler in the superfield formalism. Supersymmetry is manifest and, as we discuss later, *any* regularization method that preserves translational invariance in superspace will maintain it. For gauge theories we can use supersymmetric gauge-fixing terms. By contrast component Wess-Zumino gauge calculations explicitly break supersymmetry and have the disadvantage that the Ward-Takahashi identities for global supersymmetry cannot be directly applied due to their nonlinearity.

In this chapter and the next one we discuss the quantization of $N = 1$ superfield theories. We consider classical superfield actions $S(\Psi)$ and use functional methods to construct the generating functional $Z(J)$ and the effective action $\Gamma(\Psi)$. If $\Psi$ is a gauge field we quantize covariantly, introducing gauge-fixing terms, gauge averaging, and superfield Faddeev-Popov ghosts. We then derive Feynman rules for supergraphs using superspace propagators $\Delta(x, x', \theta, \theta')$. The methods are completely analogous to those for component fields, but some new features are present: We must deal with constrained (chiral) superfields, and we encounter not only $i\partial_{\underline{a}} = p_{\underline{a}}$ operators, but also spinor derivatives $D_\alpha$ acting on the arguments of propagators or external lines. We show how these operators are manipulated and how, for any graph, the $\theta$-integrals at each vertex can be done, leaving us with one overall $\theta$-integral for the whole graph (the effective action is local in $\theta$), and ordinary loop-momentum integrals. At all steps of the calculations manifest supersymmetry is maintained.

We discuss next the background field method for supersymmetric Yang-Mills theories. This is similar to that for component theories, with one significant difference: The quantum-background splitting is nonlinear, reflecting the nonlinearities of the gauge transformations of the superfield $V$. The method simplifies many calculations and can be used to study higher-loop finiteness questions.

For supergraphs the simplest regularization procedure is to use dimensional regularization of momentum integrals *after* the contribution from a graph has been reduced to a single $\theta$ integral. The resulting effective action, which is a (local in $\theta$) functional of the external superfields, is manifestly supersymmetric. However, this regularization method corresponds to (component) regularization by dimensional reduction, which is known to be inconsistent. The superfield results, although supersymmetric, may reflect this inconsistency by exhibiting ambiguities associated with the order in which some of the $\theta$-integrations have been carried out. We also discuss alternative regularization procedures. Besides giving power counting rules we do not discuss the details of the renormalization of superfield theories. We work with Wick-rotated time coordinates: $d^4x \to id^4x$, so $e^{-iS} \to e^S$. (The metric $\eta_{\underline{ab}}$ has signature $(-+++) \to (++++)$, so $\Box \to +\Box$, etc. Note that in our conventions, $i$ is opposite in sign from usual conventions: Thus positive-energy states are described by $e^{i\omega t}$ and propagators are $(p^2+m^2+i\epsilon)^{-1}$. We further warn the reader that the gauge coupling constant $g$ is $\sqrt{2}$ times the usual $g$ (see page 55).)

## 6.2. Gauge fixing and ghosts

The quantization of supersymmetric gauge theories is similar to that of ordinary gauge theories. There are two related aspects of the situation: (a) The action is invariant under gauge transformations and therefore the functional integration should be restricted to the subset of gauge inequivalent fields. (b) The kinetic operator is not invertible over the space of all field configurations so that the propagator, needed for doing perturbation theory, cannot be defined unless the set of fields is restricted. In component gauge theories, imposing an algebraic restriction explicitly in the functional integral leads to an axial gauge which breaks manifest Lorentz invariance. Alternatively, we can quantize covariantly using the Faddeev-Popov procedure: We introduce gauge fixing function(s), weighted gauges and Faddeev-Popov ghosts. In supersymmetric gauge theories the analog of the axial gauge is the Wess-Zumino gauge. In this gauge, quantization breaks manifest supersymmetry. In contrast, covariant superfield quantization maintains manifest supersymmetry.

### a. Ordinary Yang-Mills theory

For orientation we briefly recall the quantization method for ordinary Yang-Mills theory. The Yang-Mills gauge action is

$$S_{YM} = \frac{1}{g^2} tr \int d^4x \, [-\tfrac{1}{8} f^{\underline{ab}} f_{\underline{ab}}] \quad , \qquad f_{\underline{ab}} = \partial_{[\underline{a}} A_{\underline{b}]} - i[A_{\underline{a}}, A_{\underline{b}}] \quad , \tag{6.2.1}$$

with gauge invariance under the transformation

$$A'_{\underline{a}} \equiv A_{\underline{a}}{}^{\omega} = e^{i\omega}[A_{\underline{a}} + i\partial_{\underline{a}}]e^{-i\omega} \quad , \tag{6.2.2a}$$

or, infinitesimally,

$$\delta A_{\underline{a}} = \nabla_{\underline{a}}\omega = \partial_{\underline{a}}\omega + i[\omega, A_{\underline{a}}] \quad . \tag{6.2.2b}$$

Here $\omega$ is an element of the gauge algebra. Both $A$ and $\omega$ are matrices in the adjoint representation. We observe that the kinetic (quadratic) part of the Lagrangian can be written (after rescaling $A \to gA$) in the form $\frac{1}{2} A \Box \Pi^T A$ where $(\Pi^T)_{\underline{ab}} = \eta_{\underline{ab}} - \frac{1}{2} \partial_{\underline{a}} \partial_{\underline{b}} \Box^{-1}$ is a transverse projection operator (see (3.11.2)).

We start with the normalized functional integral

$$Z = N' \int I\!\!D A_{\underline{a}} \, e^{S_{inv}} \quad , \tag{6.2.3}$$

where we have included in $S_{inv}$ possible terms with sources coupled to gauge invariant operators. We define the gauge invariant integral over the group manifold

$$\Delta_F(A_{\underline{a}}) = \int I\!\!D\omega \, \delta[F(A_{\underline{a}}{}^{\omega}) - f(x)] \quad , \tag{6.2.4}$$

where $f(x)$ is an arbitrary field-independent function, and $F$ is a gauge-variant function such that $F = f$ for some value of $\omega$. It is important to verify that this is the case. We introduce a factor of 1 in the functional integral, in the form $\Delta_F{}^{-1}\Delta_F$:

$$Z = N' \int I\!\!D A_{\underline{a}} \, \Delta_F{}^{-1}(A_{\underline{a}}) \int I\!\!D\omega \, \delta[F(A_{\underline{a}}{}^{\omega}) - f] e^{S_{inv}}$$

$$= N' \int I\!\!D A_{\underline{a}} \, \Delta_F{}^{-1}(A_{\underline{a}}) \int I\!\!D\omega \, \delta[F(A_{\underline{a}}) - f] e^{S_{inv}} \quad , \tag{6.2.5}$$

where the last form follows from a change of variables that is a gauge transformation, and the gauge invariance of $\Delta_F$ and $S$. The $\omega$ integral now gives a constant infinite factor that we absorb into the normalization $N'$, leading to the form

$$Z = N' \int I\!\!D A_{\underline{a}} \, \Delta_F{}^{-1}(A_{\underline{a}}) \, \delta[F(A_{\underline{a}}) - f] e^{S_{inv}} \quad . \tag{6.2.6}$$

By construction $Z$ is independent of $F$ and $f$, and hence we can average over $f$ with an arbitrary (normalized) weighting factor. In particular, if we introduce a factor $1 = N'' \int I\!\!D f \, exp(-\frac{1}{g^2\alpha} tr \int d^4x \, f^2)$, the $\delta(F(A) - f)$ factor can be used to carry out the integration and leads to the form

$$Z = N' \int I\!\!D A_{\underline{a}} \, \Delta_F{}^{-1} \, e^{S_{inv} + S_{GF}} \quad ,$$

$$S_{GF} = -\frac{1}{g^2\alpha} tr \int d^4x \, [F(A_{\underline{a}})]^2 \quad . \tag{6.2.7}$$

where we have absorbed $N''$ into $N'$.

We can parametrize the gauge group by a gauge parameter $\omega(x)$ such that $F(A_a{}^{\omega}) = f(x)$ for $\omega = 0$. Then

$$\Delta_F(A_{\underline{a}}) = \int I\!\!D\omega\ \delta[F(A_{\underline{a}}{}^{\omega}) - f] = \int I\!\!D\omega\ [\frac{\delta F}{\delta\omega}]^{-1}\ \delta(\omega)$$

$$= \int I\!\!D\omega\ \delta[\frac{\delta F}{\delta\omega}\,\omega] = \int I\!\!D\omega\ I\!\!D\omega'\ e^{\int \omega' \frac{\delta F}{\delta\omega}\omega\, dx} \quad , \tag{6.2.8}$$

where we have written an integral representation for the functional $\delta$-function. In the second line of (6.2.8), and in the equations below, $\frac{\delta F}{\delta\omega}$ *is evaluated at* $\omega = 0$. To obtain ${\Delta_F}^{-1}$ we replace $\omega$ and $\omega'$ by real anticommuting (Faddeev-Popov ghost) fields $c(x)$ and $c'(x)$ (see sec. 3.7). Finally, we can choose for the gauge fixing function the form $F(A_a) = \frac{1}{2}\,\partial^{\underline{a}} A_{\underline{a}}$. Then $\frac{\delta F}{\delta\omega}\,\omega = \frac{1}{2}\,\partial^{\underline{a}}\nabla_{\underline{a}}\omega$, and we have

$$Z = N' \int I\!\!D A_{\underline{a}}\, I\!\!D c\, I\!\!D c'\ e^{S_{\mathit{eff}}} \quad ,$$

$$S_{\mathit{eff}} = \frac{1}{g^2}\,tr \int d^4x\ [L_{inv}(A_{\underline{a}}) - \frac{1}{\alpha}\,F(A)^2 + ic'\,\frac{\delta F}{\delta\omega}\,c]$$

$$= \frac{1}{g^2}\,tr \int d^4x\ [L_{inv}(A_{\underline{a}}) - \frac{1}{4\alpha}\,(\partial^{\underline{a}} A_{\underline{a}})^2 + ic'\,\frac{1}{2}\,\partial^{\underline{a}}\nabla_{\underline{a}} c] \quad . \tag{6.2.9}$$

(The $i$ is for hermiticity.) The gauge-fixing term can be written in the form $\frac{1}{2\alpha}\,A\Box\Pi^L A$ where $\Pi^L = 1 - \Pi^T$ is the longitudinal projection operator $(\Pi^L)_{\underline{ab}} = \frac{1}{2}\,\partial_{\underline{a}}\partial_{\underline{b}}\Box^{-1}$. The total kinetic operator becomes $\Box(1 + (\frac{1}{\alpha} - 1)\Pi^L)$, which is invertible: Minus its inverse (the propagator) is $-\Box^{-1}(1 + (\alpha - 1)\Pi^L)$. In the Fermi-Feynman gauge, $\alpha = 1$, the propagator is $-\Box^{-1}$.

The gauge-fixed Lagrangian, including ghosts, is invariant under the *global* BRST transformations:

$$\delta A_{\underline{a}} = i\xi\nabla_{\underline{a}} c \quad ,$$

$$\delta c' = \frac{2}{\alpha}\,\xi F = \frac{1}{\alpha}\,\xi\partial_{\underline{a}} A^{\underline{a}} \quad ,$$

$$\delta c = -\,\xi c^2 \quad , \tag{6.2.10}$$

with constant Grassmann parameter $\xi$. These transformations are nilpotent: $\delta^2$ on any field vanishes (when the antighost equations of motion are imposed). The Ward identities for this global invariance are the Slavnov-Taylor identities of the gauge theory.

**b. Supersymmetric Yang-Mills theory**

For supersymmetric Yang-Mills theory we quantize following the same procedure. We start with the functional integral for a gauge real scalar superfield $V = V^{A} T_{A}$, where $T_{A}$ are the generators of the gauge group:

$$Z = \int I\!\!DV \; e^{S_{inv}(V)} \quad . \tag{6.2.11}$$

Note that in supersymmetric theories the normalization factor of (6.2.3) $N' = 1$ (see sec. 3.8.b). The action is

$$\begin{aligned} S_{inv} &= \frac{1}{g^2}\, tr \int d^4x \, d^2\theta \; W^2 \\ &= -\frac{1}{2g^2}\, tr \int d^4x \, d^4\theta \; (e^{-V} D^{\alpha} e^{V}) \bar{D}^2 (e^{-V} D_{\alpha} e^{V}) \\ &= \frac{1}{2g^2}\, tr \int d^4x \, d^4\theta \; [V D^{\alpha} \bar{D}^2 D_{\alpha} V + higher - order \;\; terms \,] \, . \end{aligned} \tag{6.2.12}$$

It is invariant under the gauge transformations

$$e^{V'} = e^{i\bar{\Lambda}} \, e^{V} \, e^{-i\Lambda} \quad , \tag{6.2.13a}$$

or, for infinitesimal $\Lambda$ (see (4.2.28)),

$$\delta V = L_{\frac{1}{2}V} \, [-i(\bar{\Lambda} + \Lambda) + coth \, L_{\frac{1}{2}V} \, i(\bar{\Lambda} - \Lambda)] \quad , \quad L_X \, Y = [X, Y] \;\; . \tag{6.2.13b}$$

In the abelian case, this is $\delta V = i(\bar{\Lambda} - \Lambda)$. The kinetic operator is $\Box \Pi_{\frac{1}{2}}$ with the superspin $\frac{1}{2}$ projection operator $\Pi_{\frac{1}{2}} = -\Box^{-1} D^{\alpha} \bar{D}^2 D_{\alpha}$ , and is not invertible because it annihilates the chiral and antichiral superspin zero parts of $V$: $V_0 = \Pi_0 V = \Box^{-1}(D^2 \bar{D}^2 + \bar{D}^2 D^2) V$.

We must now choose gauge-fixing functions. Corresponding to the chiral gauge parameter $\Lambda$ we need a gauge-variant function that can be made to vanish by a suitable gauge transformation. Therefore it must have the same spin and superspin as the gauge parameter and hence should be chosen a chiral scalar. The gauge-variant quantity $F = \bar{D}^2 V$ is a suitable gauge-fixing function. For any chiral function $f(x, \theta)$, we verify that gauge transformations can be found to make $F = f$. For example, in the abelian case, under a gauge transformation $F(V) \to F(V^\Lambda) = \bar{D}^2 V + i\bar{D}^2\bar{\Lambda}$; if we choose $i\bar{\Lambda} = \Box^{-1} D^2 (f - \bar{D}^2 V)$, we find $F^\Lambda = f$.

We define the functional determinant

$$\Delta(V) = \int I\!D\Lambda\, I\!D\bar{\Lambda}\, \delta[F(V, \Lambda, \bar{\Lambda}) - f]\, \delta[\bar{F}(V, \Lambda, \bar{\Lambda}) - \bar{f}] \quad . \tag{6.2.14}$$

We first write (cf. (6.2.5))

$$Z = \int I\!DV\ \Delta^{-1}(V)\, \delta[\bar{D}^2 V - f]\, \delta[D^2 V - \bar{f}]\, e^{S_{inv}} \quad . \tag{6.2.15}$$

As in (6.2.7), we average over $f$ and $\bar{f}$ with a weighting factor $\int I\!Df I\!D\bar{f} exp(-\frac{1}{g^2\alpha} tr \int d^4x d^4\theta\, \bar{f} f)$, and obtain the form

$$Z = \int I\!DV\ \Delta^{-1}(V) e^{S_{inv} + S_{GF}} \quad , \tag{6.2.16}$$

where

$$S_{GF} = -\frac{1}{\alpha g^2} tr \int d^4x\, d^4\theta\, (D^2 V)\, (\bar{D}^2 V) \ . \tag{6.2.17}$$

We write

$$\Delta(V) = \int I\!D\Lambda\, I\!D\bar{\Lambda}\, I\!D\Lambda'\, I\!D\bar{\Lambda}'\, e^{\int d^4x\, d^2\theta\, \Lambda'\left(\frac{\delta F}{\delta \Lambda}\Lambda + \frac{\delta F}{\delta \bar{\Lambda}}\bar{\Lambda}\right) + \int d^4x\, d^2\bar{\theta}\, \bar{\Lambda}'\left(\frac{\delta \bar{F}}{\delta \Lambda}\Lambda + \frac{\delta \bar{F}}{\delta \bar{\Lambda}}\bar{\Lambda}\right)} \quad , \tag{6.2.18}$$

where we have replaced the $\delta$-functions involving (anti)chiral quantities by integral representations involving (anti)chiral parameters and integration measures. The variational derivatives of $F$, $\bar{F}$, are evaluated at $\Lambda = \bar{\Lambda} = 0$. In the functional integral, where $\Delta^{-1}(V)$ appears, we replace the parameters $\Lambda, \Lambda'$ by anticommuting chiral ghost fields $c, c'$. Finally, we find

$$Z = \int I\!DV\, I\!Dc\, I\!Dc'\, I\!D\bar{c}\, I\!D\bar{c}'\, e^{S_{inv} + S_{GF} + S_{FP}} \quad , \tag{6.2.19}$$

where

$$S_{FP} = i\,tr \int d^4x\, d^2\theta\; c'\overline{D}^2(\delta V) + i\,tr \int d^4x\, d^2\overline{\theta}\; \overline{c}' D^2(\delta V)$$

$$= tr \int d^4x\, d^4\theta\; (c' + \overline{c}') L_{\frac{1}{2}V} \left[ (c + \overline{c}) + coth\, L_{\frac{1}{2}V}\, (c - \overline{c}) \right] \ . \qquad (6.2.20)$$

Integrating by parts, we can write $(D^2V)\,(\overline{D}^2V) = \frac{1}{2}V(D^2\overline{D}^2 + \overline{D}^2D^2)V = \frac{1}{2}V\Box\Pi_0 V$. The quadratic part of the gauge field action has now the form

$$-\frac{1}{2}V\Box(\Pi_{\frac{1}{2}} + \alpha^{-1}\Pi_0)V = -\frac{1}{2}V\Box[1 + (\alpha^{-1} - 1)\Pi_0]V \quad , \qquad (6.2.21)$$

and the operator is invertible. To avoid $\Box^{-2}$ terms in the propagator and thus bad infrared behavior, we choose the supersymmetric Fermi-Feynman gauge $\alpha = 1$, which leads to a simple $\Box^{-1}$ propagator. (The $-$ sign in (6.2.21) leads to the usual kinetic term for the component gauge field: $-\int d^4\theta V\Box V \sim A_{\underline{a}}\Box A^{\underline{a}}$.)

The quadratic part of the ghost action has the form

$$S^{(2)}{}_{FP} = tr \int d^4x\, d^4\theta\; (c' + \overline{c}')(c - \overline{c}) = tr \int d^4x\, d^4\theta\; (\overline{c}'c - c'\overline{c}) \quad . \qquad (6.2.22)$$

The chiral and antichiral $c'c$ and $\overline{c}'\overline{c}$ terms vanish when integrated with $d^4\theta$ and have been dropped. (Such terms cannot be dropped in the presence of supergravity fields: see, for example, (5.5.16)).

The total action is invariant under superfield BRST transformations. These take the form

$$\delta V = \delta_\Lambda V|_{\Lambda = i\xi c} = \xi L_V[(c + \overline{c}) + coth\, L_{\frac{1}{2}V}\,(c - \overline{c})] \quad ,$$

$$\delta c' = \frac{1}{\alpha}\xi\overline{D}^2\overline{F} = \frac{1}{\alpha}\xi\overline{D}^2D^2V \quad , \qquad \delta\overline{c}' = \frac{1}{\alpha}\xi D^2F = \frac{1}{\alpha}\xi D^2\overline{D}^2V \quad ,$$

$$\delta c = -\,\xi c^2 \quad , \qquad \delta\overline{c} = -\,\xi\overline{c}^2 \quad , \qquad (6.2.23)$$

and the invariance can be used to derive the Slavnov-Taylor identities of the theory.

Before performing perturbation expansions, we rescale $V \rightarrow gV$. Then all quadratic terms are $O(g^0)$, cubic terms are $O(g)$, etc. We rescale back $gV \rightarrow V$ in the effective action $\Gamma$. Alternatively, we simply provide each graph with a factor $(g^2)^{L-1}$, where $L$ is the number of loops.

### c. Other gauge multiplets

We give two other examples of the gauge-fixing procedure: For a chiral superfield $\Phi$, the solution of the chirality constraint, $\Phi = \bar{D}^2\Psi$, gives the kinetic action

$$S_{inv} = \int d^4x\, d^4\theta\; \bar{\Psi} D^2 \bar{D}^2 \Psi \quad , \tag{6.2.24}$$

and introduces the gauge invariance

$$\delta\Psi = \bar{D}^{\dot{\alpha}} \bar{\omega}_{\dot{\alpha}} \quad , \qquad \delta\bar{\Psi} = D^{\alpha}\omega_{\alpha} \quad , \tag{6.2.25}$$

for an arbitrary spinor parameter $\omega_\alpha$ (see (4.5.1-4)). Suitable gauge fixing functions are the linear spinor superfields

$$F_{\alpha} = D_{\alpha}\Psi \quad , \qquad \bar{F}_{\dot{\alpha}} = \bar{D}_{\dot{\alpha}}\bar{\Psi} \quad . \tag{6.2.26}$$

To obtain a convenient gauge fixing term we average with $\bar{f}^{\dot{\alpha}} M_{\alpha\dot{\alpha}} f^{\alpha}$, where

$$M_{\alpha\dot{\alpha}} = D_{\alpha}\bar{D}_{\dot{\alpha}} + \frac{3}{4}\bar{D}_{\dot{\alpha}} D_{\alpha} \quad . \tag{6.2.27}$$

This leads to

$$S_{inv} + S_{GF} = \int d^4x\, d^4\theta\; \bar{\Psi} \Box \Psi \quad , \tag{6.2.28}$$

and a standard $p^{-2}$ propagator.

A second example is for the action of the chiral spinor superfield $\Phi_\alpha$ that describes the tensor multiplet (4.4.46):

$$S_{inv} = -\frac{1}{2}\int d^4x\, d^4\theta\; G^2 = -\frac{1}{8}\int d^4x\, d^4\theta\; (D^{\alpha}\Phi_{\alpha} + \bar{D}^{\dot{\alpha}}\bar{\Phi}_{\dot{\alpha}})^2 \quad , \tag{6.2.29}$$

with gauge invariance under

$$\delta\Phi_{\alpha} = i\bar{D}^2 D_{\alpha} K \quad , \qquad K = \bar{K} \quad . \tag{6.2.30}$$

A suitable gauge fixing function is

$$F = -\,i\,\frac{1}{2}(D^{\alpha}\Phi_{\alpha} - \bar{D}^{\dot{\alpha}}\bar{\Phi}_{\dot{\alpha}}) \quad , \tag{6.2.31}$$

where $F$ is linear. The gauge-fixed action

$$S_{inv} + S_{GF} = \frac{1}{2}\int d^4x\, d^4\theta\, [-G^2 + \frac{1}{\alpha}F^2]$$

$$= -\frac{1}{4}\int d^4x d^2\theta\, \Phi^\alpha \Box [\frac{1}{2}(1+\mathbf{K}) + \frac{1}{\alpha}\frac{1}{2}(1-\mathbf{K})]\Phi_\alpha + h.\,c.$$

$$= \frac{1}{2}\int d^4x\, d^4\theta\, [-\frac{1}{2}(1+\alpha)(\frac{1}{2}\Phi^\alpha D^2\Phi_\alpha + h.\,c.) + \frac{1}{2}(1-\alpha)\bar{\Phi}^{\dot{\alpha}} i\partial^\alpha{}_{\dot{\alpha}}\Phi_\alpha]$$

(6.2.32)

(with $\mathbf{K}$ as in sec 3.11) takes two convenient forms:

For $\alpha = 1$,

$$S_{inv} + S_{GF} = -\frac{1}{4}\int d^4x\, d^4\theta\, (\Phi^\alpha D^2\Phi_\alpha + \bar{\Phi}^{\dot{\alpha}}\bar{D}^2\bar{\Phi}_{\dot{\alpha}}) \;\; ; \qquad (6.2.33)$$

for $\alpha = -1$,

$$S_{inv} + S_{GF} = \frac{1}{2}\int d^4x\, d^4\theta\, \bar{\Phi}^{\dot{\alpha}} i\partial^\alpha{}_{\dot{\alpha}}\Phi_\alpha \qquad (6.2.34)$$

(cf. (3.8.36)).

## 6.3. Supergraph rules

Given an action $S(\Psi)$, we define the generating functional for Green functions

$$Z(J) = \int I\!D\Psi \, e^{S(\Psi)+\int J\Psi} \quad , \tag{6.3.1}$$

where $J$ is a source of the same type as the field $\Psi$ (general if $\Psi$ is general, chiral if $\Psi$ is chiral, etc.). The generating functional of *connected* Green functions is

$$W(J) = ln\, Z(J) \quad . \tag{6.3.2}$$

The expectation value of the field $\Psi$ or the "classical field" $\hat{\Psi}$ in the presence of the source is

$$\hat{\Psi}(J) = \frac{\delta W}{\delta J} \quad . \tag{6.3.3}$$

This relation can be inverted to give $J(\hat{\Psi})$. The effective action $\Gamma(\hat{\Psi})$, the generating functional of one particle irreducible graphs, is defined by a functional Legendre transform

$$\Gamma(\hat{\Psi}) = W[J(\hat{\Psi})] - \int J(\hat{\Psi})\hat{\Psi} \quad . \tag{6.3.4}$$

In this section we derive the Feynman rules for the pertubative expansion of the effective action. The derivation of the Feynman rules for unconstrained superfields presents few surprises. Instead of having $d^4x$ integrals we have $d^4xd^4\theta$ integrals. Propagators are obtained from the inverses of the kinetic operators, and vertices can be read directly from the interaction terms. However, for chiral superfields the Feynman rules reflect the chirality constraints.

### a. Derivation of Feynman rules

We begin by deriving the rules for the real scalar gauge superfield. The gauge fixed action (in the Fermi-Feynman gauge, $\alpha = 1$) reads

$$S_V = tr \int d^4x\, d^4\theta\, [-\frac{1}{2} V\Box V + \frac{1}{2}[V, (D^\alpha V)](\bar{D}^2 D_\alpha V) + \cdots] \quad . \tag{6.3.5}$$

The Feynman rules can be read directly from this expression: The propagator is minus the inverse of the kinetic operator, $\Box^{-1}\delta^4(x-x')\delta^4(\theta-\theta')$ or, in momentum space, $-p^{-2}\delta^4(\theta-\theta')$. Since the spinor derivatives $D$ contain explicit $\theta$'s, we do *not* Fourier transform with respect to the $\theta$ variables. (If one does Fourier transform with respect to $\theta$, there is little change in the Feynman rules.) Vertices can be read from the interaction terms. Thus, the cubic term $\frac{1}{2}tr[V,(D^\alpha V)](\bar{D}^2 D_\alpha V)$ leads to a three-point vertex with factors of $D^\alpha$ and $\bar{D}^2 D_\alpha$ acting on two of the lines, and a group theory factor. In addition we integrate over $x$'s and $\theta$'s at each vertex or, equivalently, over loop momenta and over $\theta$'s at each vertex.

These rules can also be obtained by starting with the functional integral:

$$
\begin{aligned}
Z(J) &= \int I\!\!DV\ e^{\int [-\frac{1}{2}V\Box V + I\!\!L_{int}(V) + JV]} \\
&= e^{\int I\!\!L_{int}\left(\frac{\delta}{\delta J}\right)} \int I\!\!DV\ e^{\int [-\frac{1}{2}V\Box V + JV]} \\
&= e^{\int I\!\!L_{int}\left(\frac{\delta}{\delta J}\right)}\ e^{\frac{1}{2}\int J\Box^{-1}J} \quad ,
\end{aligned}
\tag{6.3.6}
$$

where in the last step we have performed the Gaussian integral over $V$. The Feynman rules can be obtained using $\frac{\delta J(x,\theta)}{\delta J(x',\theta')} = \delta^4(x-x')\delta^4(\theta-\theta')$ and expanding the exponentials in power series. Thus, we obtain factors of $I\!\!L_{int}(\frac{\delta}{\delta J})$ corresponding to vertices, and the $\frac{\delta}{\delta J}$ operators, when acting on the factors of $J\Box^{-1}J$ remove the $J$'s and produce propagators $\Box^{-1}$ connecting the vertices. The result is exactly as for ordinary field theory, with the additional feature of $d^4\theta$ integrals at each vertex, and additional $\delta^4(\theta-\theta')$ factors in each propagator.

Chiral scalar superfields usually have a *kinetic* action (with chiral sources $j$, $\bar{j}$) of the form

$$
\begin{aligned}
S^{(2)} = &\int d^4x\, d^4\theta\ \bar{\Phi}\Phi - \frac{1}{2}\int d^4x\, d^2\theta\ m\Phi^2 - \frac{1}{2}\int d^4x\, d^2\bar{\theta}\ m\bar{\Phi}^2 \\
&+ \int d^4x\, d^2\theta\ j\Phi + \int d^4x\, d^2\bar{\theta}\ \bar{j}\bar{\Phi} \quad .
\end{aligned}
\tag{6.3.7}
$$

To perform the Gaussian integration, we rewrite chiral integrals as integrals over full superspace. A chiral integral

$$I_c = \int d^4x\, d^2\theta\; FG \quad , \tag{6.3.8}$$

where $F$ and $G$ are *arbitrary* chiral expressions, can be rewritten as

$$I_c = \int d^4x\, d^4\theta\; F\Box^{-1}D^2 G \quad , \tag{6.3.9}$$

using $\Box^{-1}\bar{D}^2 D^2\, G = G$ (3.4.10) and $\int d^4x\, d^4\theta \;=\int d^4x\, d^2\theta\; \bar{D}^2$. The Gaussian integral can be rewritten as

$$e^{W_0(j)} \equiv \int I\!\!D\Phi\, I\!\!D\bar{\Phi}\; exp\int d^4x\, d^4\theta\; [\tfrac{1}{2}\,(\Phi\;\; \bar{\Phi})\mathbf{O}\begin{pmatrix} \Phi \\ \bar{\Phi} \end{pmatrix} + (\Phi\;\; \bar{\Phi})\begin{pmatrix} \Box^{-1}D^2 j \\ \Box^{-1}\bar{D}^2\bar{j} \end{pmatrix}] \tag{6.3.10}$$

where

$$\mathbf{O} = \begin{pmatrix} -\dfrac{mD^2}{\Box} & 1 \\ 1 & -\dfrac{m\bar{D}^2}{\Box} \end{pmatrix} \quad . \tag{6.3.11}$$

The inverse of $\mathbf{O}$ is

$$\mathbf{O}^{-1} = \begin{pmatrix} \dfrac{m\bar{D}^2}{\Box - m^2} & 1 + \dfrac{m^2\bar{D}^2D^2}{\Box(\Box - m^2)} \\ 1 + \dfrac{m^2D^2\bar{D}^2}{\Box(\Box - m^2)} & \dfrac{mD^2}{\Box - m^2} \end{pmatrix} \quad . \tag{6.3.12}$$

Performing the integral we obtain

$$W_0(j) = \int d^4x\, d^4\theta\; [-\,\bar{j}\,\frac{1}{\Box - m^2}\,j - \tfrac{1}{2}\,(\,j\,\frac{mD^2}{\Box(\Box - m^2)}\,j + h.\,c.)] \quad . \tag{6.3.13}$$

For a general interaction Lagrangian $I\!\!L_{int}(\Phi, \bar{\Phi})$ we can write

$$Z(j) = e^{\int d^4x\, d^4\theta\; I\!\!L_{int}\left(\frac{\delta}{\delta j}, \frac{\delta}{\delta \bar{j}}\right)}\; e^{W_0(j)} \quad , \tag{6.3.14}$$

and the Feynman rules can be obtained from this expression. Since $\dfrac{\delta j(x,\theta)}{\delta j(x',\theta')} = \bar{D}^2\delta^4(\theta-\theta')\delta^4(x-x')$ (3.8.10), there is an operator $\bar{D}^2$ acting on each chiral field line leaving a vertex. Similarly, there is an operator $D^2$ acting on each antichiral line leaving a vertex. However, at a purely chiral vertex, e.g. $\int d^2\theta\Phi^n$, we use one of these factors to convert the $d^2\theta$ integral to a $d^4\theta$ integral. Therefore at such vertices we omit one factor of $\bar{D}^2$.

We now summarize the Feynman rules for interacting gauge and chiral superfields.

(a) Propagators:

$$VV: \qquad -\frac{1}{p^2}\,\delta^4(\theta-\theta') \quad , \tag{6.3.15a}$$

$$\bar{\Phi}\Phi: \qquad \frac{1}{p^2+m^2}\,\delta^4(\theta-\theta') \quad , \tag{6.3.15b}$$

$$\Phi\Phi: \qquad -\frac{mD^2}{p^2(p^2+m^2)}\,\delta^4(\theta-\theta') \quad , \tag{6.3.15c}$$

$$\bar{\Phi}\bar{\Phi}: \qquad -\frac{m\bar{D}^2}{p^2(p^2+m^2)}\,\delta^4(\theta-\theta') \quad . \tag{6.3.15d}$$

In the massive case, the $p^{-2}$ factors in the $\Phi\,\Phi$ and $\bar{\Phi}\,\bar{\Phi}$ propagators are always canceled by numerator factors (e.g., for $\Phi\Phi$ the vertices give $\bar{D}^2$ factors and, as we discuss later, we obtain $\bar{D}^2 D^2 \bar{D}^2 = -\,p^2 \bar{D}^2$). In the massless case these propagators are absent.

(b) Vertices: These are read directly from the interaction Lagrangian, with the additional feature that for each chiral or antichiral line leaving a vertex there is a factor $\bar{D}^2$ or $D^2$ acting on the corresponding propagator, and the rule that at purely chiral or antichiral vertices we omit one $\bar{D}^2$ or $D^2$ factor from among the ones acting on the propagators.

(c) We integrate over $d^4\theta$ at each vertex, and in momentum space we have loop-momentum integrals $\int d^4p(2\pi)^{-4}$ for each loop, and an overall factor $(2\pi)^4\delta(\sum k_{ext})$.

(d) To obtain the effective action $\Gamma$, we compute one-particle-irreducible graphs. For each external line with outgoing momentum $k_i$, we multiply by a factor $\int d^4k_i(2\pi)^{-4}\,\Psi(k_i)$ where $\Psi$ stands for any of the fields in the effective action. For each external chiral or antichiral line, we have a $\Phi$ or $\bar{\Phi}$ factor, but no $\bar{D}^2$ or $D^2$ factors.

(e) Finally, there may be symmetry factors associated with certain graphs.

An alternative derivation of the Feynman rules for chiral superfields can be obtained by solving the chirality constraints in terms of an unconstrained field (see sec. 4.5a):

$$\Phi = \bar{D}^2\Psi \quad , \qquad \bar{\Phi} = D^2\bar{\Psi} \quad , \tag{6.3.16}$$

where $\Psi$ is a general, complex scalar superfield. The action, including source terms, becomes

$$S = \int d^4x\, d^4\theta\, [(D^2\bar{\Psi})(\bar{D}^2\Psi) + I\!L_{int}(\bar{D}^2\Psi\,, D^2\bar{\Psi})]$$

$$+ \int d^4x\, d^2\theta\, (\bar{D}^2\Psi)(-\frac{1}{2}m\bar{D}^2\Psi + j) + h.\,c. \quad . \tag{6.3.17}$$

Chiral integrals can be rewritten as full integrals by using up a $\bar{D}^2$ factor. We recall that in terms of $\Psi$ we have an abelian gauge invariance, $\Psi \rightarrow \Psi + \bar{D}^{\dot{\alpha}}\bar{\omega}_{\dot{\alpha}}$ (4.5.4). Consequently, the kinetic operator appearing in the action, $\bar{\Psi}D^2\bar{D}^2\Psi$, is not invertible. As discussed in sec. 6.2, we can fix the gauge and arrive at an invertible quadratic action (the ghosts decouple)

$$S^{(2)} = \int d^4x\, d^4\theta\, \frac{1}{2}\begin{pmatrix}\Psi & \bar{\Psi}\end{pmatrix}\begin{pmatrix}-mD^2 & \Box \\ \Box & -m\bar{D}^2\end{pmatrix}\begin{pmatrix}\Psi \\ \bar{\Psi}\end{pmatrix} \quad . \tag{6.3.18}$$

The Feynman rules are now the naive ones and are identical to the ones we have obtained before (after using the $D^2$, $\bar{D}^2$ factors at the vertices to simplify the propagators, obtained from (6.3.12)). In particular, from $I\!L_{int}(\bar{D}^2\Psi, D^2\bar{\Psi})$, we again find factors of $\bar{D}^2$, $D^2$ acting on the propagators, except that one such factor is missing at purely (anti)chiral vertices, since we convert everywhere to full $d^4\theta$ integrals.

It is simple to obtain the supergraph rules for the tensor multiplet, with gauge-invariant action

$$S = \int d^4x d^4\theta\, f(G) \quad , \; f(G) = -\frac{1}{2}G^2 + \cdots , \tag{6.3.19}$$

with $\Phi\Phi$ propagator $-2p^{-4}\delta_\alpha{}^\beta D^2\delta^4(\theta-\theta')$ (and the hermitian conjugate for $\bar{\Phi}\bar{\Phi}$) from (6.2.33). However, there is a much simpler form of the rules which resembles the rules for the scalar multiplet (to which the tensor multiplet is on-shell equivalent by a duality transformation: see sec. 4.4.c.2). We first note that the vertex at either end of a $\Phi\Phi$ propagator has a $D$ at the vertex (from $f(G)$ with $G = D\Phi + \bar{D}\bar{\Phi}$), and next to it a $\bar{D}^2$ (as occurs at the end of any chiral propagator, rule (b) above; this kills the $\bar{D}\bar{\Phi}$ part of the vertex). Also note that the spinor index at the vertex contracts directly with the corresponding spinor index of the propagator (because the vertex is a function of only $G = \frac{1}{2} D_\alpha \Phi^\alpha + h.\,c.$). Contracting these spinor indices, and integrating by parts all $D$'s from the vertices onto the propagators, we obtain the same expression for the $\Phi\Phi$ and $\bar{\Phi}\bar{\Phi}$ propagators (with the same vertices), which can now be added together (i.e., the total contribution from graphs with both types of propagators is the same as that from only one type, but with an overall factor of 2 for each propagator). The rules are thus cast into the following form: All vertices are now simply *constants,* read from the expansion of $f(G)$ in $G$. There is only one type of propagator, with no spinor indices, which is

$$- \frac{1}{p^2} D^\alpha \bar{D}^2 D_\alpha \delta^4(\theta - \theta') = -\, \Pi_{\frac{1}{2}} \delta^4(\theta - \theta') \ . \qquad (6.3.20)$$

(The algebra from the various contributing factors is $D\bar{D}^2 D^2 \bar{D}^2 D = D\bar{D}^2 D \Box$.) Each external line gets a factor of $G$. If we were to perform the same rearrangement of vertex factors for the supergraphs of the dual scalar-multiplet theory, we would obtain $\Pi_0$ instead of $\Pi_{\frac{1}{2}}$, the external line factors would be $\Phi + \bar{\Phi}$, and the constants at the vertices would be obtained from $\tilde{f}(\Phi + \bar{\Phi})$ in terms of the function $\tilde{f}$ dual to $f$ (see again sec. 4.4.c.2). The on-shell equivalence then follows from the fact that the combinatorics resulting from using $\Pi_{\frac{1}{2}} = 1 - \Pi_0$, with a propagator 1 collapsing to a point (in $\theta$ *and* $x$), performs the duality, where for the external lines $G = \Phi + \bar{\Phi}$ on shell.

## b. A sample calculation

We now give an example in a theory of a massless chiral superfield $\Phi$ interacting with a gauge superfield $V$. We compute the one-loop contribution from the chiral superfield to the $V$ two-point function. The relevant interaction is obtained from $\bar{\Phi} e^V \Phi = \bar{\Phi}\Phi + \bar{\Phi} V \Phi + \cdots$. We find a contribution to the effective action, according to our rules and Fig. 6.3.1,

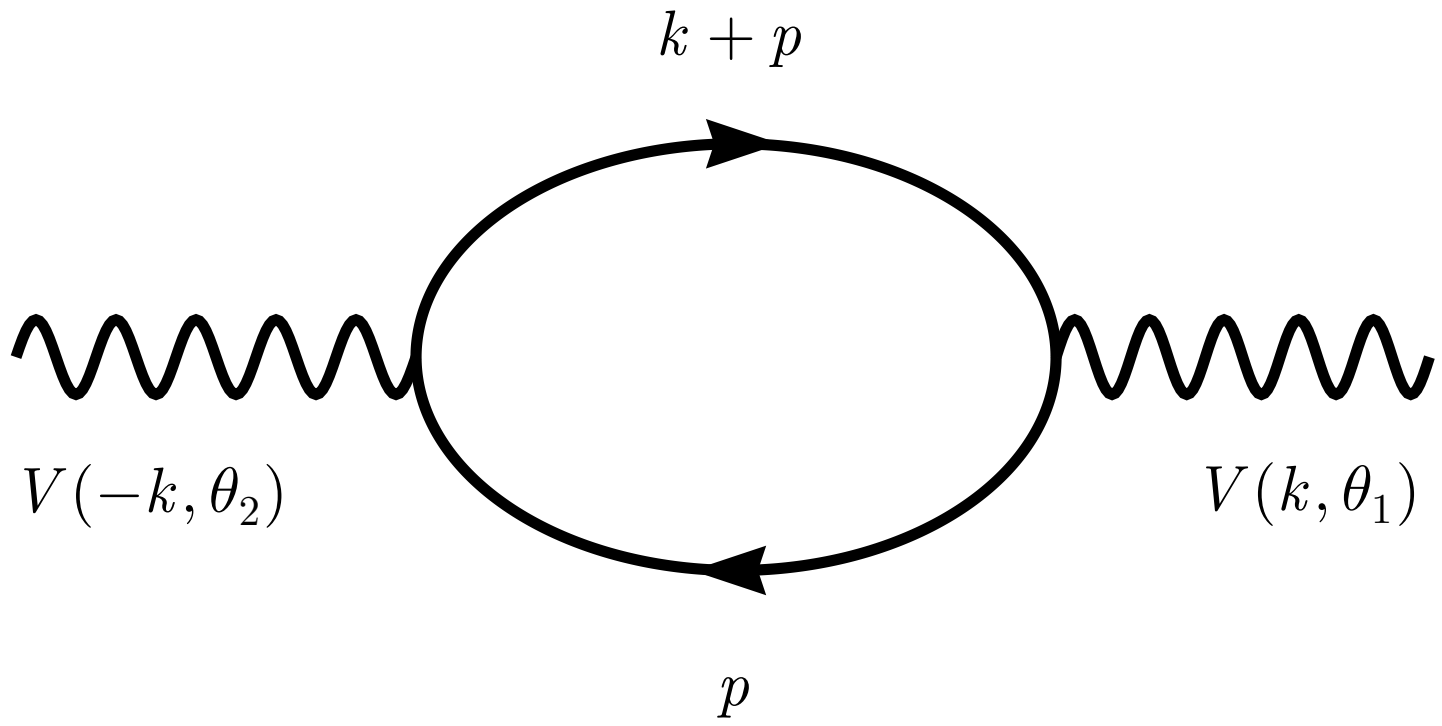

*Fig. 6.3.1*

$$\frac{1}{2}\int \frac{d^4k}{(2\pi)^4}\, d^4\theta_1\, d^4\theta_2\; V(-k,\theta_2)V(+k,\theta_1)$$

$$\cdot\int \frac{d^4p}{(2\pi)^4}\,\frac{{D_1}^2\delta^4(\theta_1-\theta_2)\overleftarrow{\bar{D}}{}_2{}^2}{p^2}\,\frac{{D_2}^2\delta^4(\theta_2-\theta_1)\overleftarrow{\bar{D}}{}_1{}^2}{(p+k)^2} \quad . \tag{6.3.21}$$

Note that in the above expression

$$D_{1\alpha} = \frac{\partial}{\partial{\theta_1}^\alpha} + \frac{1}{2}\bar{\theta}_1{}^{\dot\alpha}\, p_{\alpha\dot\alpha} \quad , \qquad \bar{D}_{1\dot\alpha} = \frac{\partial}{\partial\bar{\theta}_1{}^{\dot\alpha}} - \frac{1}{2}{\theta_1}^\alpha(k+p)_{\alpha\dot\alpha} \quad ,$$

$$D_{2\alpha} = \frac{\partial}{\partial{\theta_2}^\alpha} + \frac{1}{2}\bar{\theta}_2{}^{\dot\alpha}\,(k+p)_{\alpha\dot\alpha} \quad , \qquad \bar{D}_{2\dot\alpha} = \frac{\partial}{\partial\bar{\theta}_2{}^{\dot\alpha}} - \frac{1}{2}{\theta_2}^\alpha p_{\alpha\dot\alpha} \quad . \tag{6.3.22}$$

Although we do not indicate the momentum dependence explicitly, it is implicit that the momentum is that *leaving* the vertex through the propagator on which the operators act. (From the $\bar{\Phi}V^2\Phi$ interaction term we also obtain a tadpole-type diagram; its contribution cancels a similar contribution from the diagram we are considering, or vanishes if we use dimensional regularization).

The $D$'s can be manipulated like ordinary derivatives. They obey a Leibnitz rule, and a "transfer" rule

$$\delta^4(\theta_1-\theta_2)\overleftarrow{\bar{D}}_{\dot\alpha 2}(p) = -\,\overrightarrow{\bar{D}}_{\dot\alpha 1}(-p)\delta^4(\theta_1-\theta_2) \quad , \tag{6.3.23}$$

which can be checked by examining the explicit form of the operators. Another example of the transfer rule is

$$D_1{}^2\delta^4(\theta_1-\theta_2)\overleftarrow{\bar{D}}_2{}^2 = D_1{}^2\bar{D}_1{}^2\delta^4(\theta_1-\theta_2) \neq \bar{D}_1{}^2 D_1{}^2\delta^4(\theta_1-\theta_2)\ . \tag{6.3.24}$$

Inside integrals the $D$'s can be integrated by parts (see sec. 3.7). Thus

$$\int d^4\theta[D_\alpha(p)f(\theta,p)]g(\theta,-p) = -\int d^4\theta f(\theta,p)D_\alpha(-p)g(\theta,-p)\ \ . \tag{6.3.25}$$

This can be most easily understood in $x$-space. Since $D_\alpha = \partial_\alpha + \frac{1}{2}i\bar{\theta}^{\dot\alpha}\partial_{\alpha\dot\alpha}$ we are doing integration by parts in $\partial_\alpha$ and in $\partial_{\alpha\dot\alpha}$.

Armed with these facts we return to the evaluation of the expression in (6.3.21). We concentrate on the $\theta$ dependence and write the relevant part as

$$\int d^4\theta_1\, d^4\theta_2\ V(-k,\theta_2)[D_1{}^2\bar{D}_1{}^2\delta_{12}][\bar{D}_1{}^2 D_1{}^2\delta_{12}]V(k,\theta_1)\ \ . \tag{6.3.26}$$

We have abbreviated $\delta^4(\theta_1-\theta_2)=\delta_{12}$. We now integrate by parts and find first of all

$$\begin{aligned}[D^2\bar{D}^2\delta][\bar{D}^2 D^2\delta]V &= \bar{D}^2\delta D^2[(\bar{D}^2 D^2\delta)V] \\ &= \delta\bar{D}^2[(D^2\bar{D}^2D^2\delta)V + (D^\alpha\bar{D}^2D^2\delta)D_\alpha V + (\bar{D}^2D^2\delta)D^2V] \\ &= \delta\bar{D}^2[-p^2(D^2\delta)V + p^{\alpha\dot\alpha}(\bar{D}_{\dot\alpha}D^2\delta)D_\alpha V + (\bar{D}^2D^2\delta)D^2V]\ \ ,\end{aligned} \tag{6.3.27}$$

where we have used $(D)^3=0$ and the anticommutation relations $\{D_\alpha,\bar{D}_{\dot\alpha}\}=p_{\alpha\dot\alpha}$ when acting on the propagator with momentum $p$.

Before proceeding we make the following important observation: Since $\delta^4(\theta)=\theta^2\bar{\theta}^2$, multiplying two identical $\delta$-functions together, or multiplying one by $\theta$ gives zero. We have therefore the following relations:

$$\delta_{21}\,\delta_{21} = \delta_{21}\,\delta_{12} = 0\ \ ,$$

$$\delta_{21}\ D^\alpha\delta_{21} = 0\ \ ,$$

$$\delta_{21}\ D^2\delta_{21} = 0\ \ ,$$

$$\delta_{21}\ D^\alpha\bar{D}^{\dot\alpha}\delta_{21} = 0\ \ ,$$

$$\delta_{21}\ D^\alpha\bar{D}^2\delta_{21} = 0\ \ ,$$

$$\delta_{21}\ D^2\bar{D}^2\delta_{21} = \delta_{21}\ \bar{D}^2D^2\delta_{21} = \delta_{21}\ \frac{1}{2}D^\alpha\bar{D}^2D_\alpha\delta_{21} = \delta_{21} \quad ,$$

$$\delta_{21}\ D^\alpha\bar{D}^2D^\beta\delta_{21} = C^{\beta\alpha}\delta_{21} \quad . \tag{6.3.28}$$

In these relations, we obtain a nonzero result only if all the $\theta$'s in the second $\delta$-function are removed by differentiation. Hence two $D$'s and two $\bar{D}$'s are needed and only their momentum independent parts contribute. Expressions of this kind, but with more $D$'s, can be reduced to one of the above forms by using the anticommutation relations. In the expressions with four $D$'s the order is irrelevant (except for producing some minus signs).

Returning to our calculation (6.3.27), and letting $\bar{D}^2$ act on the factors to its right, we see that out of the a priori possible six terms, only three survive:

$$\delta[-p^2(\bar{D}^2D^2\delta)V - p^{\alpha\dot{\alpha}}(\bar{D}^{\dot{\beta}}\bar{D}_{\dot{\alpha}}D^2\delta)\bar{D}_{\dot{\beta}}D_\alpha V + (\bar{D}^2D^2\delta)(\bar{D}^2D^2V)] \quad . \tag{6.3.29}$$

Finally, using $\bar{D}^{\dot{\alpha}}\bar{D}_{\dot{\beta}} = \delta_{\dot{\beta}}{}^{\dot{\alpha}}\bar{D}^2$ we find

$$\delta^4(\theta_1-\theta_2)[-p^2 - p^{\alpha\dot{\alpha}}\bar{D}_{\dot{\alpha}}D_\alpha + \bar{D}^2D^2]V(k,\theta_1) \quad . \tag{6.3.30}$$

Inserting this result into the original integral (6.3.21), we use the remaining $\delta$-function to do the $\theta_2$ integral, and finally obtain

$$\tfrac{1}{2}\int\frac{d^4k}{(2\pi)^4}\,d^4\theta\ V(-k,\theta)\Big[\int\frac{d^4p}{(2\pi)^4}\ \frac{-p^2 - p^{\alpha\dot{\alpha}}\bar{D}_{\dot{\alpha}}D_\alpha + \bar{D}^2D^2}{p^2(k+p)^2}\Big]V(k,\theta). \tag{6.3.31}$$

The result consists of an ordinary loop momentum integral, with usual propagators and some momentum factors in the numerator, and operators $D$,$\bar{D}$, acting on the external superfields (i.e., $D$ and $\bar{D}$ depend on $k$, not $p$). The $p^2$ term is canceled by the tadpole diagram mentioned above, (or gives zero in dimensional regularization) so that the final contribution to the one-loop self-energy is logarithmically divergent. This is a consequence of gauge invariance. The remaining terms in the numerator, in a gauge-invariant regularization (such as dimensional), combine to form $\frac{1}{2}D^\alpha\bar{D}^2D_\alpha$, giving a result proportional to $W^2$.

## c. The effective action

In the example above, the $\delta$-function has reduced the expression to one which involves a single $\theta$. This is a general and important result of superfield perturbation theory. The effective action is a sum of terms involving products of fields evaluated at different points, and a Green function which is a nonlocal function of its arguments:

$$\Gamma = \sum_n \int d^4x_1 \cdots d^4x_n \, d^4\theta_1 \cdots d^4\theta_n \;\; G(x_1 \cdots x_n; \theta_1 \cdots \theta_n)\, \Phi(x_1, \theta_1) \cdots D^\alpha V(x_i, \theta_i) \cdots \quad . \tag{6.3.32}$$

It turns out, however, that by manipulation of the contributions from any graph, we can reduce it to an expression that is local in $\theta$, i.e.

$$\Gamma = \sum_n \int d^4x_1 \cdots d^4x_n \, d^4\theta \;\; \tilde{G}(x_1 \cdots x_n)\, \Phi(x_1, \theta) \cdots D^\alpha V(x_i, \theta) \cdots \quad . \tag{6.3.33}$$

We do this as follows: Consider an arbitrary $L$-loop contribution to the effective action. It consists of propagators, with factors $\delta^4(\theta_i - \theta_{i+1})$ and $D$ operators acting on them, external superfield factors, and $d^4\theta_i$ integrals. We choose any propagator from a particular vertex $v$ to another vertex $v'$, and integrate by parts to remove all the $D$'s from its $\delta$-function. The original contribution now becomes a sum of terms. If there are other propagators, each of which connects $v$ and $v'$, we use the relations (6.3.28): The terms vanish unless *each* of the other $\delta$-functions has exactly two $D$'s and two $\bar{D}$'s acting on it, in which case they can be replaced by 1. We now use the free $\delta$-function to do the $\theta$-integral at $v'$ and shrink all the propagators between the two vertices to a point in $\theta$-space. We repeat the procedure, choosing a propagator leading to a new vertex $v''$, until we have removed all $\delta$-functions and performed all $\theta$-integrals except the original one at $v$. Whenever we have more than two $D$'s and two $\bar{D}$'s on a line we use the anticommutation relations to replace $D$,$\bar{D}$ pairs by momenta. We are left with a sum of terms, all with a single $\theta$ integral, and various factors of loop-momenta coming from the anticommutators of $D$'s, as well as $D$ factors acting on the external superfields, coming from the integration by parts.

In the course of evaluating Feynman diagrams, we may encounter loop-momentum ultraviolet divergences, and a suitable regularization procedure is needed to handle them. We discuss regularization issues later on. For the time being we assume that there exists a procedure that allows us to carry out the manipulations we have described above inside momentum integrals.

The expression for the effective action in (6.3.33) reveals one important fact: We have ended up with a $d^4\theta$ integral, even though in the original classical action we may have had $d^2\theta$ integrals. This is a consequence of our Feynman rules: All our vertices carry $d^4\theta$ integrals, and nowhere in our manipulations does a $d^2\theta$ appear. In particular, if the original action had purely chiral $d^2\theta$ mass or cubic interaction terms $\Phi^2$ or $\Phi^3$, radiative corrections do not induce finite or infinite modifications of these terms. This is the *no-renormalization theorem* for chiral superfields. Masses and coupling constants *are* renormalized, but only as a consequence of wave function renormalization. (Any $d^4\theta$ integral can be written as a $d^2\theta$ integral and a $\bar{D}^2$ operator acting on the integrand; however, this will not produce the above terms.) This theorem is valid in perturbation theory. So far no one has succeeded in giving examples, in four dimensions, where it might fail nonperturbatively, but a proof of its general validity does not exist. Even within perturbation theory there exists the possibility of a pathological infrared-type behavior which might invalidate it. For example, if in the course of evaluating the effective action a term $\int d^4\theta \Phi^2 \frac{D^2}{\Box} \Phi$ were produced, the $\bar{D}^2$ operator which comes from converting the integral to chiral form, when acting on the chiral field, would give $\bar{D}^2 D^2 \Box^{-1} \Phi = \Phi$ and we would end up with a contribution to the chiral cubic vertex. Whether such pathological behavior can be obtained in any calculation with a sensible infrared regularization is doubtful.

## d. Divergences

We now discuss the divergence structure of the effective action. There are two issues involved: We must determine which terms in the effective action are divergent (power counting), and which terms in the classical action lead only to divergences that can be absorbed in a renormalization of the parameters (renormalizable interactions). We restrict our discussion to interacting gauge and chiral scalar superfields (with no negative-dimension coupling constants).

The possible divergences of the effective action can be understood by straight power counting (see sec. 6.6) or simply by a dimensional argument: The divergent parts of graphs that contain no subdivergences give rise to local terms in the effective action of

the form

$$\Gamma_{\infty} = \int d^4x \, d^4\theta \; I\!P(\Phi, \overline{\Phi}, V, D_\alpha \Phi, \dots) \quad , \tag{6.3.34}$$

where $I\!P$ is a polynomial in the fields and their derivatives. Since the effective action must be dimensionless, and $d^4\theta$ has dimension 2, $I\!P$ must also have dimension 2. $\Phi$ has dimension 1, $D_\alpha$ has dimension $\frac{1}{2}$, and $V$ is dimensionless. Therefore, graphs with more than two external $\Phi$'s are convergent. A $\Phi\Phi$ or $\overline{\Phi}\overline{\Phi}$ propagator produces a numerator factor of $m$ which contributes to the dimension of $I\!P$ and therefore reduces the degree of divergence. If $I\!P$ is made up of only chiral superfields the $\theta$ integration will give zero unless some $D$'s (at least two of them, to contract indices ) are present to make the integrand nonchiral, and again the $D$'s contribute to the dimension of $I\!P$ reducing the number of fields that can appear. Finally, *if* gauge invariance requires $V$ to appear through its field strength $W_\alpha = i\bar{D}^2 D_\alpha V$ (or its nonabelian generalization), this limits the possible divergences involving $V$ fields. (This is an oversimplification: We must use the full machinery of Slavnov-Taylor identities, at least in the nonabelian case, or use the background-field method (see sec. 6.5) to analyze the divergences involving gauge superfields.)

The net result of the analysis is to establish that the only local divergent terms contain at most one $\Phi$ and one $\overline{\Phi}$ and, while they contain an arbitrary number of $V$ factors, these enter in a manner which is controlled by the Slavnov-Taylor identities. For a renormalizable theory of chiral scalar multiplets interacting with a vector multiplet (we omit the ghost terms) the renormalized classical action has the form

$$\int d^4x \, d^4\theta \; [\overline{\Phi}_R e^{g_R V_R} \Phi_R + tr\nu_R V_R - tr{\alpha_R}^{-1}(\bar{D}^2 V_R)(D^2 V_R)]$$

$$+ \frac{1}{{g_R}^2} \int d^4x \, d^2\theta \; tr{W_R}^2 + [\int d^4x \, d^2\theta \; I\!P(\Phi_R) + h.c.] \quad , \tag{6.3.35}$$

where the subscript $R$ labels renormalized quantities. (In the exponential we have written explicitly the gauge coupling constant $g$ that we normally absorb into $V$.)

Since $V$ is dimensionless, $V_R$ is in general a nonlinear function of $V$, i.e., the wave-function renormalization factor may be a function of $V$: We can have functional renormalizations $V_R = f(V)$, where each coefficient in the Taylor expansion of $f$ is a renormalization constant. Since all such renormalizations are proportional to the field equations ($\delta S = \int \frac{\delta S}{\delta V}\,\delta V$), they vanish on shell. (Such noncovariant renormalizations are avoided in the background field gauges that we discuss below.)

Ghosts are described by chiral superfields which follow the same rules. The divergences of the theory are all logarithmic, except that of the Fayet-Iliopoulos term, which is quadratic. (However, as we shall discuss in sec. 6.5, this term is not produced by radiative corrections.)

In general, renormalizable interactions are associated with dimensionless (or positive dimension) coupling constants. For Feynman graphs, since at each vertex we have a $d^4\theta$ integral with dimension 2 and a $d^4x$ integral with dimension $-4$, we may allow up to the equivalent of four $D$'s at each vertex. This is indeed the case with the gauge field self-couplings, and also the usual vertices involving chiral superfields, where the $D$ factors come from our Feynman rules. On the other hand a term such as $\bar{\Phi}^2\Phi$ , or $\Phi^4$, would lead to an excess of $D$'s at the vertices and a nonrenormalizable theory.

## e. D-algebra

In the next section we give a number of examples of evaluation of supergraphs. As preparation we discuss several simplifications that we use in performing the manipulation of the $D$'s and the $\theta$ integration. The numerous integrations by parts that have to be performed can lead to long intermediate expressions, and a lot of effort (and paper) can be saved by doing the manipulations directly on the graphs.

We draw the supergraph and indicate on it, adjacent to the vertices, the $D$ factors acting on the propagators in the order in which they act. We ignore signs having to do with the ordering of the $D$'s: These will be determined later. Thus, an expression such as $\bar{D}^2 D_\alpha \delta^4(\theta-\theta')\overleftarrow{D}_\beta \overleftarrow{\bar{D}}_{\dot\beta} \overleftarrow{D}_\gamma$ , with the last three $D$'s acting backwards on the $\theta'$ argument of the $\delta$-function (and thus in the order $D_\beta$ first, then $\bar{D}_{\dot\beta}$ next, etc.), would be represented on the graph as shown in fig. 6.3.2:

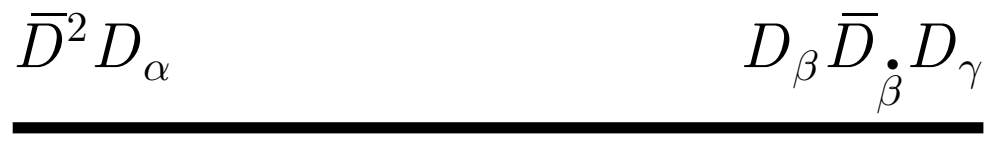

*Fig. 6.3.2*

The transfer rule can be implemented by "sliding" the $D$'s to the left, keeping the order. This corresponds to writing $-\bar{D}^2 D_\alpha D_\beta \bar{D}_{\dot\beta} D_\gamma \delta^4(\theta - \theta')$ with $D_\gamma$ acting first on the $\theta$ argument of the $\delta$-function. We must keep track of the $-$ sign coming from transferring an odd number of $D$'s. (Note the order in these expressions, e.g., $D_1{}^\alpha \delta_{12} \overleftarrow{D}_2{}^\beta = D_1{}^\alpha D_2{}^\beta \delta_{12} = -D_1{}^\alpha D_1{}^\beta \delta_{12}$.)

We use the commutation relations to replace the rightmost $\bar{D}, D$ by $p_{\dot\beta\gamma}$ . (Since $p$ is hermitian $p_{\dot\beta\gamma} = p_{\gamma\dot\beta}$ but we maintain the distinction to keep track of the order in which the $D$'s appeared.) When we encounter expressions such as $D^2 \bar{D}^2 D^2$ we replace them with $-p^2 D^2$.

The integration by parts can also be carried out directly on the graphs. For example, we show in fig. 6.3.3 the integration by parts on a vertex coming from the $\bar{\Phi} V \Phi$ interaction:

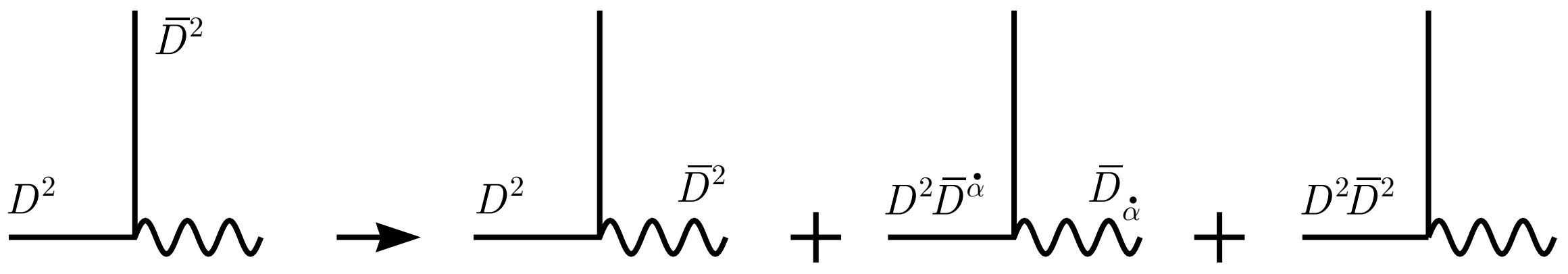

*Fig. 6.3.3*

Starting from a given graph in general we obtain several, because integration by parts gives several contributions. We remove the $D$'s from any given line and use the $\delta$-function to do one of the $\theta$ integrals, thus contracting the line to a point in $\theta$ space. We need not indicate explicitly this contraction: A line without any operators on it is understood to be contracted. Whenever several lines connect the same pair of vertices, if all the lines (other than the one we have cleared of $D$'s) have exactly two $D$'s and two $\bar{D}$'s each, we use (6.3.28) to replace them by 1. If any line has fewer $D$'s or $\bar{D}$'s, the contribution vanishes. If any line has more $D$'s or $\bar{D}$'s, we use the anticommutation relations to reduce their number. In the end we have a sum of graphs, with momentum factors from the anticommutators, and $D$'s acting on the external lines only.

To make this procedure clear, we redo the example considered above (fig. 6.3.1, (6.3.21-31)), working directly on the graph, as shown in fig. 6.3.4:

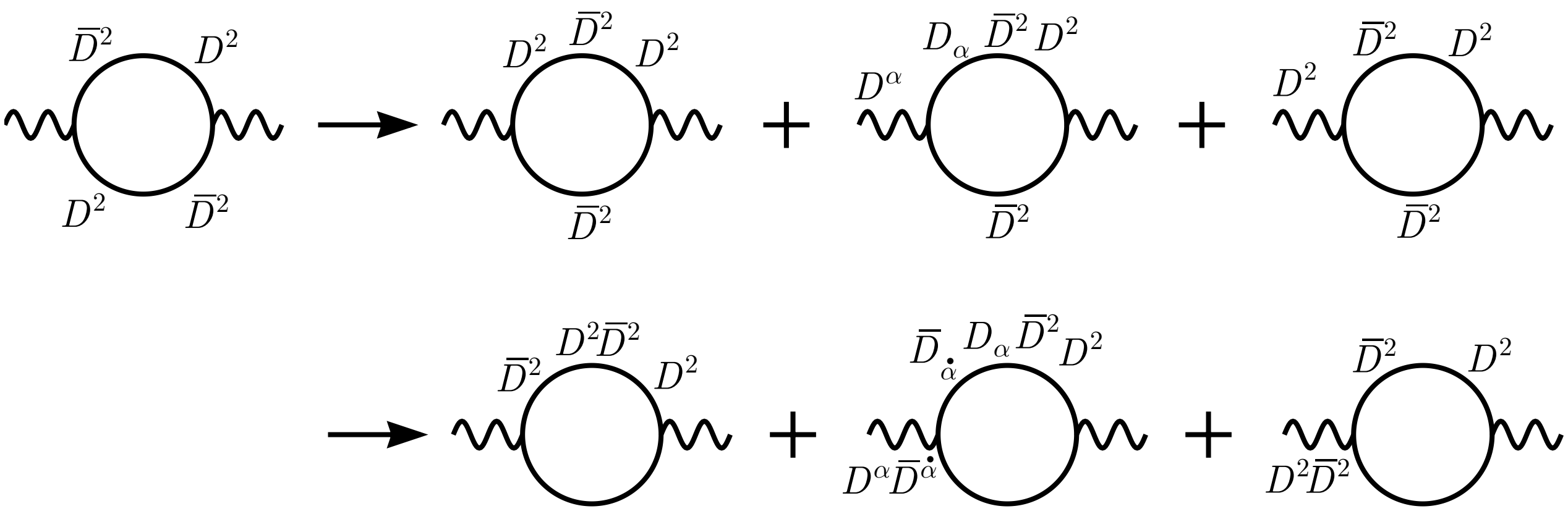

*Fig. 6.3.4*

In the last step we have only indicated nonzero contributions (the others vanish trivially because of (6.3.28).)

In the course of the manipulations on the graphs, we must keep track of $-$ signs coming from transfers and from integration by parts. However, we need not keep track of $-$ signs that come from passing a $D$ past another $D$, nor from signs that come from raising or lowering indices. (On the graphs, we do not indicate the relative order of $D$'s on different lines, nor which indices are up or down). These signs can be determined at the end of the computation in the following manner: On the original graph, we have factors such as $D^2\bar{D}^2 = \frac{1}{4} D^\alpha D_\alpha \bar{D}^{\dot\alpha}\bar{D}_{\dot\alpha}$ , and also, from a vertex such as $V(D^\alpha V)(\bar{D}^2 D_\alpha V)$, *adjacent* factors $D^\alpha$ and $\frac{1}{2}\bar{D}^{\dot\alpha}\bar{D}_{\dot\alpha}D_\alpha$ where we determine the initial sign by requiring that in any contracted pair, the first $D$ or $\bar{D}$ has the upper index. These various factors may end up in a different order in the final expression, e.g., $D^\alpha \cdots \bar{D}^{\dot\alpha} \cdots D_\alpha \cdots \bar{D}_{\dot\alpha}$, possibly acting on different superfields; however, we still write a contracted pair with the first index raised. To determine the final overall sign, we count the number of transpositions needed in the final expression to bring the $D$'s back to the original order. This is true even if some of the $D$'s have been replaced by momenta: An expression such as $p_{\beta\dot\gamma}$ will correctly keep track of the transpositions. We do *not* count transpositions of contracted pairs, since the convention for raising and lowering indices cancels such signs: $X^\alpha Y_\alpha = +Y^\alpha X_\alpha$. A quick way to count the transpositions is to draw lines connecting all contracted pairs with lines, and count the number of intersections: an odd number means an odd number of transpositions, and hence a $-$ sign, whereas an even number means no $-$ sign.

There are many other tricks that one can use to simplify the manipulations. We give the following "twingling" rule which is often useful:

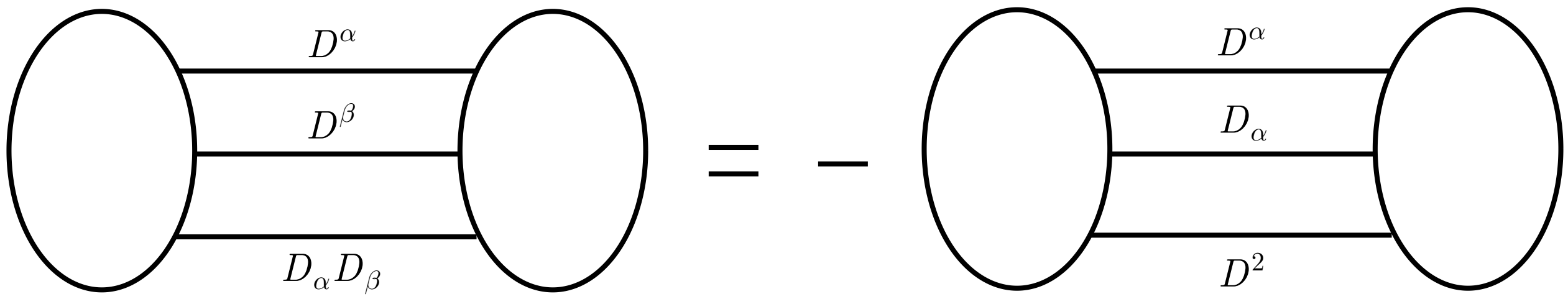

*Fig. 6.3.5*

We are now ready to consider further examples of graph evaluation.

## 6.4. Examples

In this section we give a number of examples of supergraph calculations. Most of the examples were encountered in various calculations that have been performed. For more complicated ones we refer the reader to the calculations of the 3-loop $\beta$-function in $N = 4$ Yang-Mills, and the 3- and 4-loop $\beta$-function in the Wess-Zumino model.

We do our manipulations directly on the graphs until only an ordinary momentum integral remains. For notational convenience we sometimes indicate a factor $p^2$ multiplying a propagator with the same momentum by a $\Box$ drawn on the corresponding line. In the case of a line with no $D$'s acting on it, we sometimes leave it in the graph, while at other times, when we draw $\theta$-space graphs, we contract it out. To establish the procedure we begin with some simple examples.

For the massive Wess-Zumino model we consider first some self energy graphs:

(1)

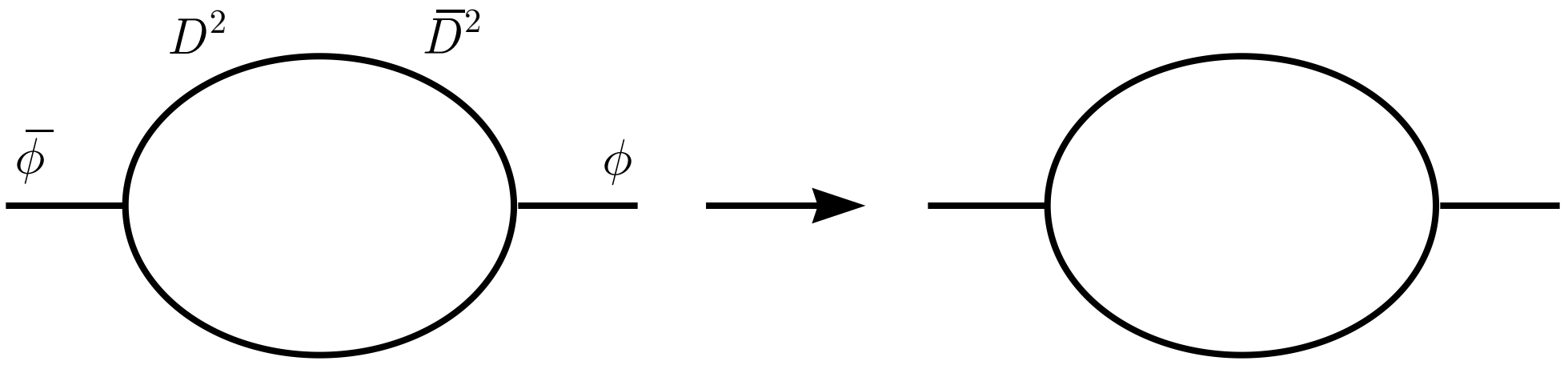

*Fig. 6.4.1*

$$\rightarrow \int d^4\theta\, \overline{\Phi}(-p,\theta)\Phi(p,\theta) \int \frac{d^4k}{(2\pi)^4} \frac{1}{(k^2+m^2)[(k+p)^2+m^2]} \quad . \qquad (6.4.1)$$

(2)

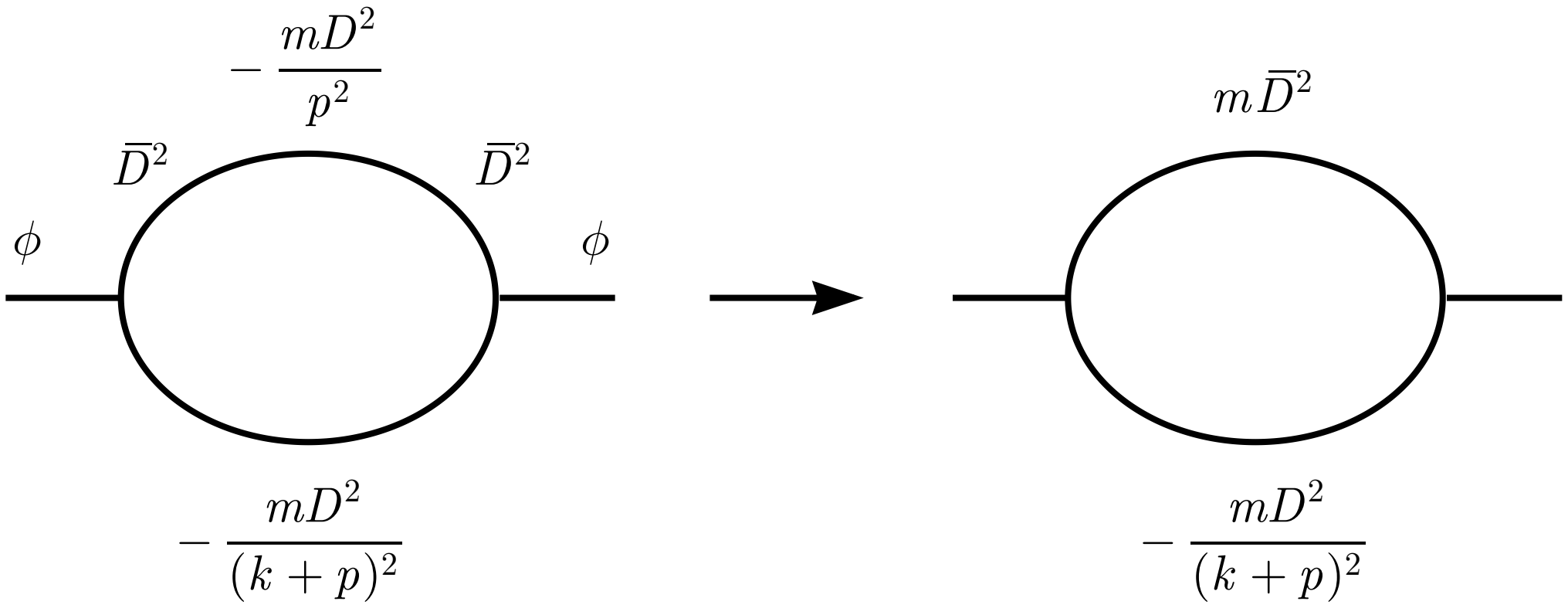

*Fig. 6.4.2*

We have used $\bar{D}^2 D^2 \bar{D}^2 = -p^2 \bar{D}^2$. At a chiral vertex the $\bar{D}^2$ factors can be put on either line by integration by parts. The result is $\int d^4x d^4\theta\, \Phi\Phi = 0$ because the integrand is chiral. We consider next a triangle diagram with $\Phi^3$ and $\bar{\Phi}^3$ vertices:

(3)

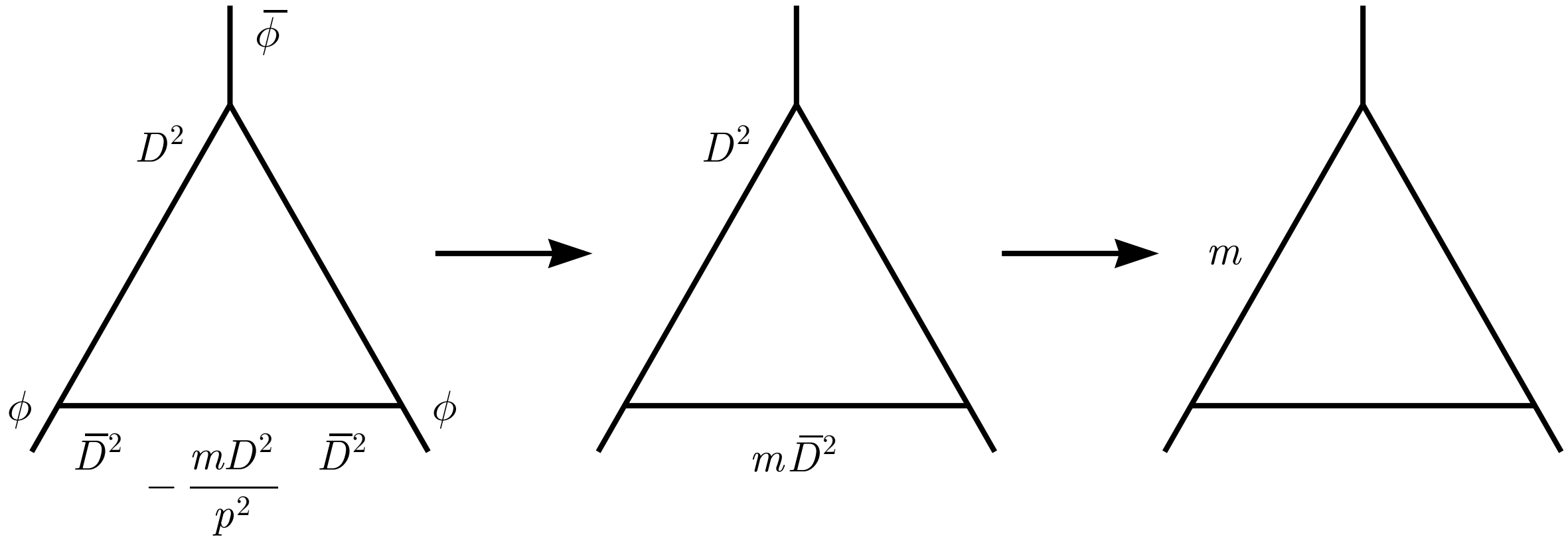

*Fig. 6.4.3*

Thus, the one-loop contributions to this three-point function are zero in the massless case and nonlocal in the massive case.

(4) For $m = 0$, with $\Phi^3$ and $\bar{\Phi}^3$ vertices,

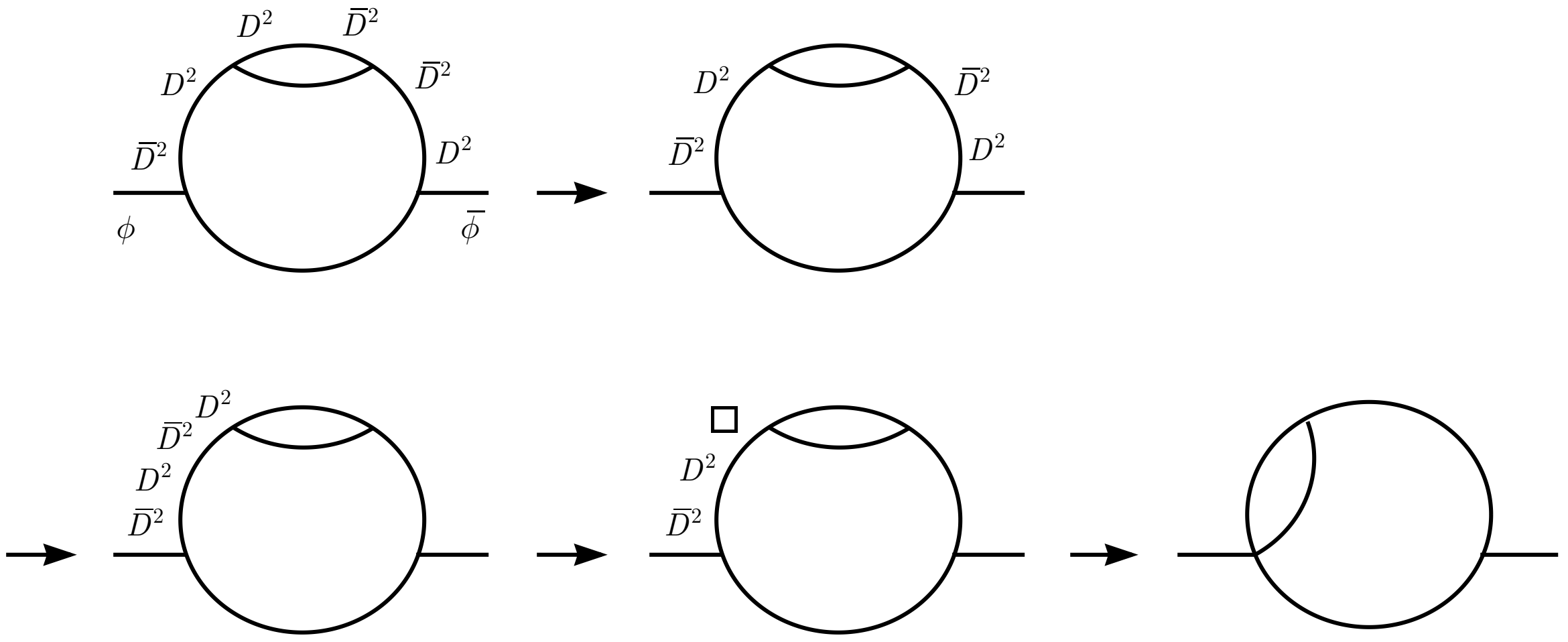

*Fig. 6.4.4*

In the second graph, the small loop is contracted to a point in $\theta$-space, after which the $\bar{D}^2 D^2$ operators can be transferred across it and all act on the same line. The final graph shows the actual momentum-space diagram one would have to evaluate. A propagator has been canceled by $\Box$.

(5) Again in the massless case,

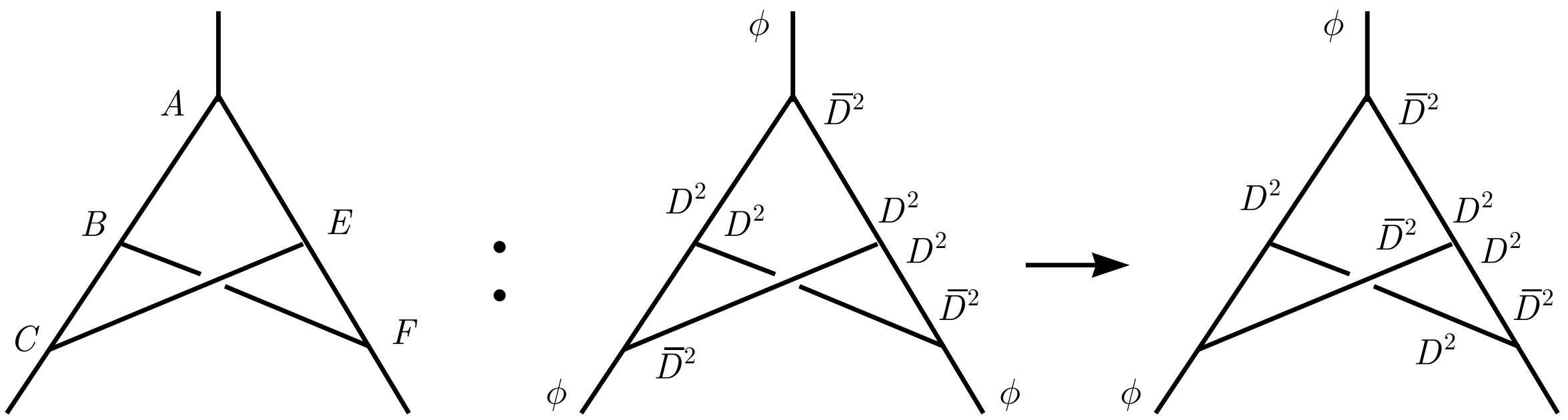

*Fig. 6.4.5a*

We have placed the $D$'s and $\bar{D}$'s on certain two of the three lines for convenience, and have labeled the vertices. In the second graph we have transferred $\bar{D}^2$, $D^2$ from C,B, to E,F, respectively. We now integrate by parts the $\bar{D}^2$ factor at E. This will generate three terms.

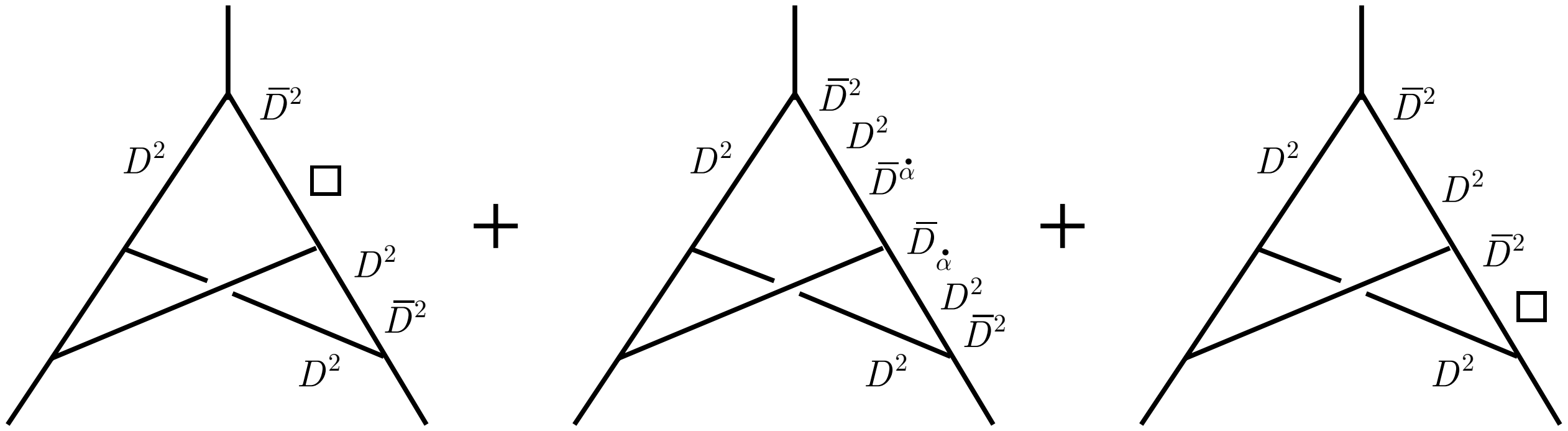

*Fig. 6.4.5b*

By examining the $\theta$-space loops AECBA and EFBCE, it is clear that in each graph the $D^2$'s on AB and BF, respectively, must give a single term when integrated by parts at their respective vertices (A and F). We obtain

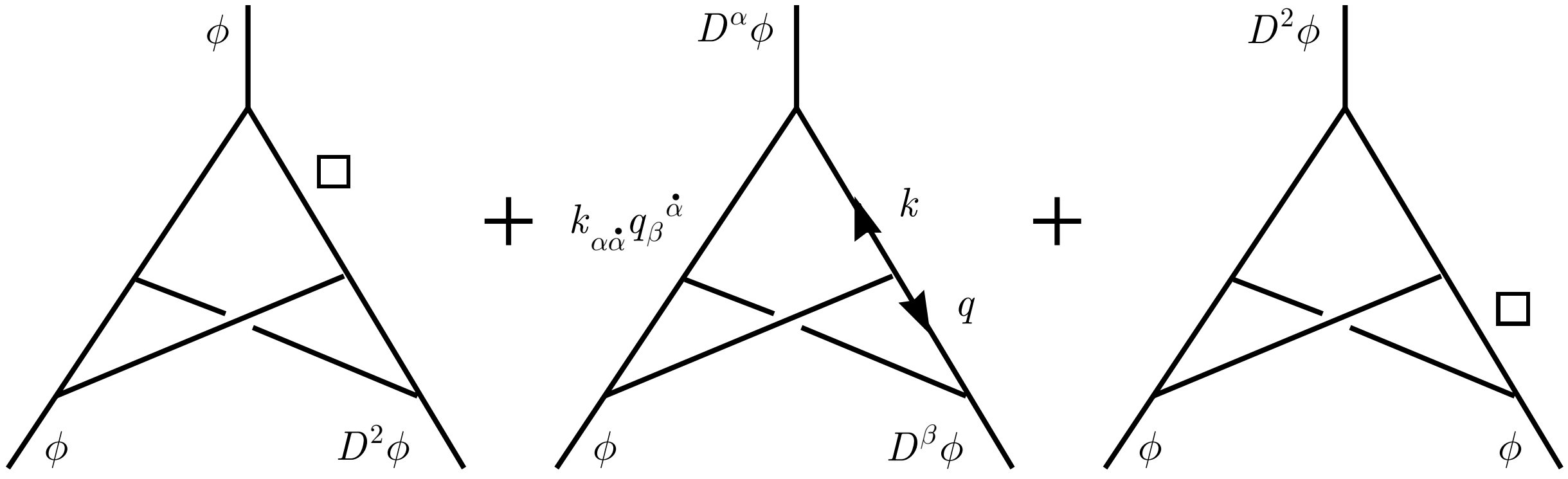

*Fig. 6.4.5c*

We have used

$$\delta \bar{D}^{\dot\alpha} D^2 \bar{D}^2 D^\alpha \delta = k^{\beta\dot\alpha} \delta D_\beta \bar{D}^2 D^\alpha \delta = -\, k^{\underline{a}} \delta \quad . \qquad (6.4.2)$$

Our next example has $\bar{\Phi} e^V \Phi$ interactions:

(6)

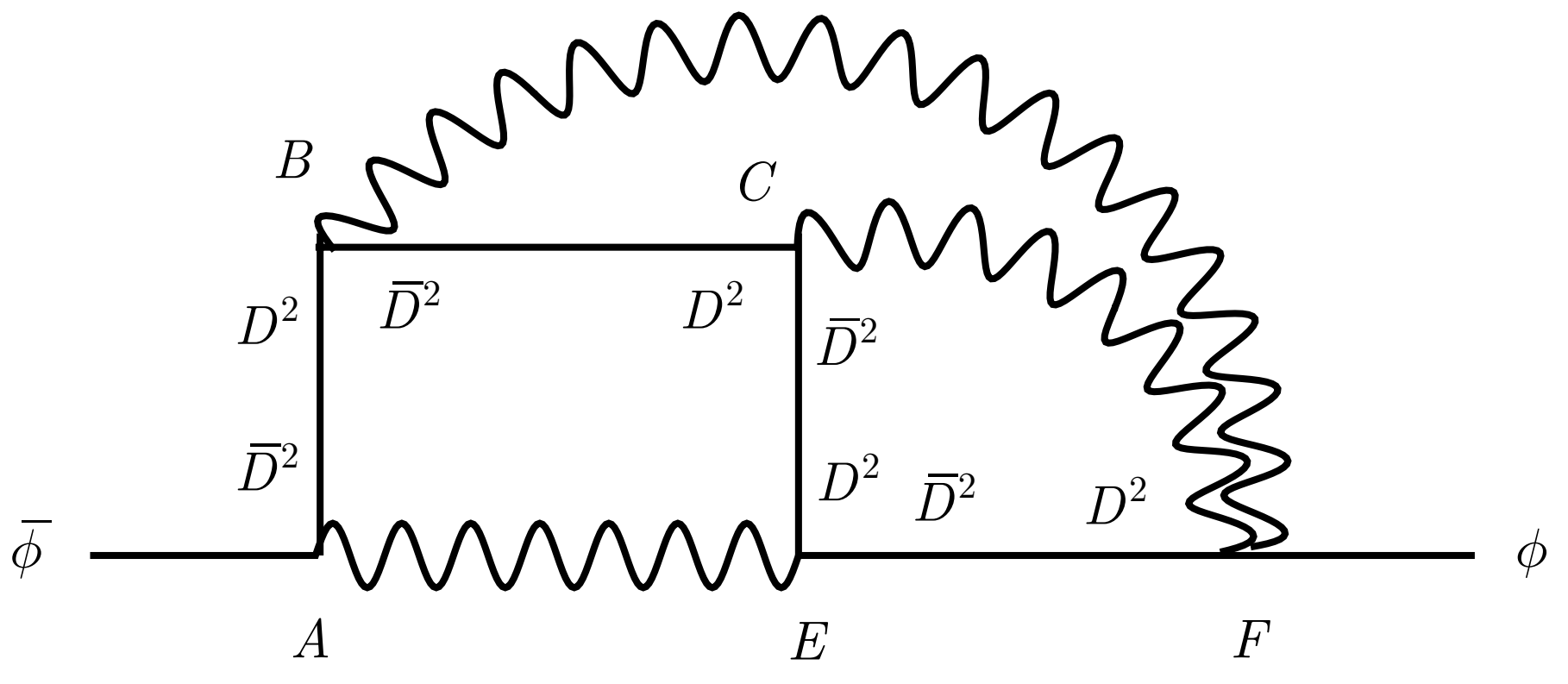

*Fig. 6.4.6a*

In the loop BCF we have just a $D^2\bar{D}^2$ factor, so we contract it to a point. Similarly, we contract the AE line to a point. For clarity we draw a $\theta$-space graph where $q$ and $h$ are the momenta of the AB and EF lines:

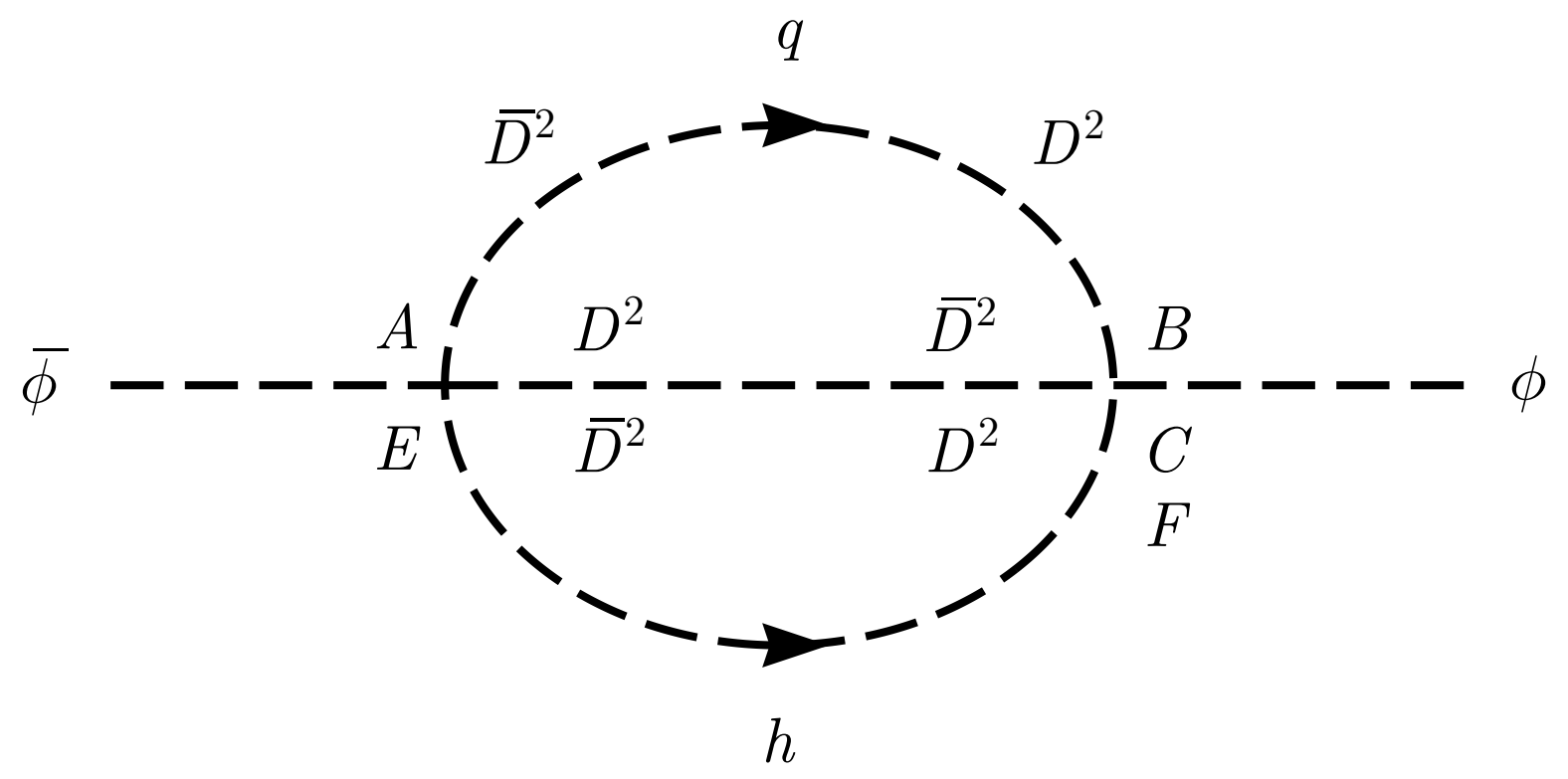

*Fig. 6.4.6b*

We integrate the $D^2$ factor off the middle line. It cannot go on $\bar{\Phi}$ so it must go on either the top or bottom line, or split. Because of (6.3.28) the $\bar{D}^2$ factor must follow it. This is the same as computing $\bar{D}^2 D^2[\eta(q)\eta(h)]$ where the $\eta$'s are chiral. The result is simply $-(q+h)^2\eta(q)\eta(h)$. Therefore, the $D$ manipulation is finished and we obtain $-\bar{\Phi}(-p,\theta)\Phi(p,\theta)(q+k)^2$ multiplying a standard momentum integral with the propagators of the original diagram.

We give now an example in nonabelian Yang-Mills theory:

(7)

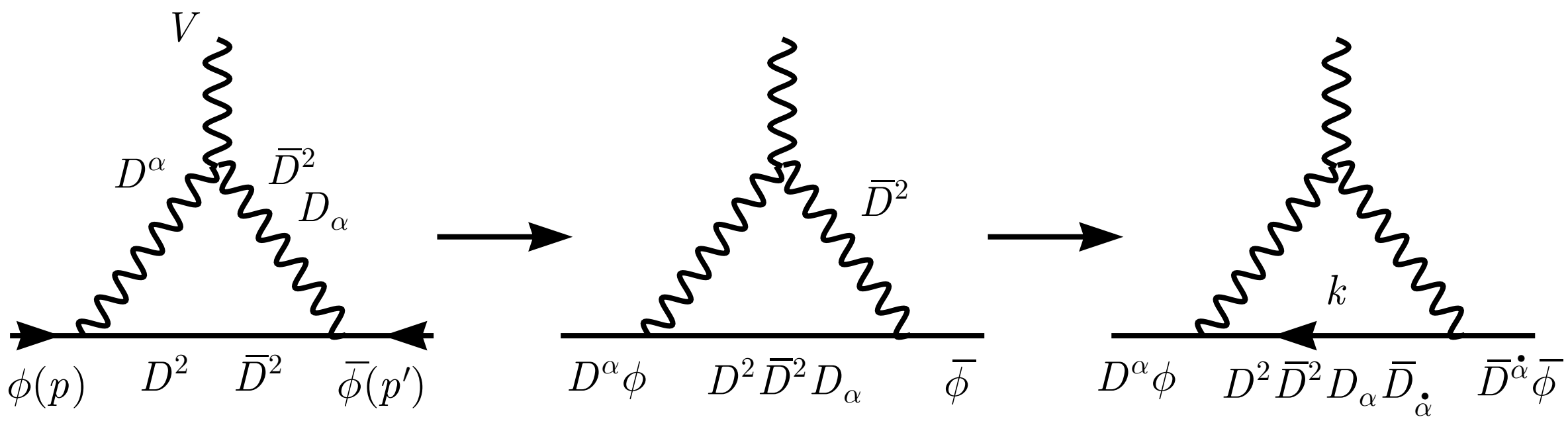

*Fig. 6.4.7*

$$\to V(-p-p')\,D^\alpha\Phi(p)\,\bar{D}^{\dot{\alpha}}\bar{\Phi}(p')\int\frac{d^4k}{(2\pi)^4}\,\frac{k_{\alpha\dot{\alpha}}}{k^2(k-p)^2(k+p')^2}\quad . \tag{6.4.3}$$

(8) $N=4$ Yang-Mills theory.

In sec. 4.6.b we have given the classical action of this theory in terms of $N=1$ superfields. Here we discuss some of its quantum properties.

The theory is described by superfields $V\,,\Phi^i$ $(i=1,2,3)$, all in the adjoint representation of an arbitrary group, and with interactions governed by a common coupling constant $g$. It is classically scale invariant, and both component and superfield calculations have established that its $\beta$-function vanishes to three loops, so that, perturbatively, the scale invariance survives quantization. Proofs exist that extend this conclusion to all orders of perturbation theory. Here we discuss some of the explicit supergraph calculations for establishing $\beta(g)=0$, and leave the general arguments to sec. 7.7.

We add to the classical action (4.6.38) the gauge fixing and ghost terms (6.2.17,20-22) with gauge parameter $\alpha=1+O(g^2)$. To $O(g^0)$ this choice gives the Fermi-Feynman gauge and a propagator $\Box^{-1}$ that avoids serious infrared problems. However, the transverse part of the self-energy receives (local) radiative corrections, whereas the longitudinal part does not (as follows from the Ward identities). To stay in the Fermi-Feynman gauge, we must maintain the equality of the longitudinal and transverse parts, and we do this by adjusting the $O(g^2)$ parts of $\alpha$ in each order of perturbation theory (actually, the radiative corrections vanish at one loop, and only arise at $O(g^4)$).

At the classical level the $O(4)$ invariance of the theory requires the equality of the gauge and $\Phi_1\Phi_2\Phi_3$ coupling constants. Although the gauge fixing procedure breaks the $O(4)$ symmetry (this could be avoided if we had an $N = 4$ superfield formalism), gauge invariance should insure that the coupling constants receive a common renormalization. Therefore, the theory has only one $\beta$-function, which we can compute, for example, by comparing the renormalization of the $C_{ijk}\Phi^i\Phi^j\Phi^k$ vertex function and the $\bar{\Phi}_i\Phi^i$ wave function renormalization. However, the vertex being chiral, receives no radiative corrections (see sec. 6.3), so that to establish $\beta(g) = 0$ it is sufficient to show that the $\bar{\Phi}_i\Phi^i$ self-energy is finite. (We observe that if $O(4)$ invariance of the quantum effective action were not spoiled by the gauge fixing procedure, finiteness to all orders would follow immediately: The finiteness of the $C_{ijk}\Phi^i\Phi^j\Phi^k$ vertex would imply the finiteness of all other local terms in the effective action. In principle, the desired result should still follow from the $O(4)$ Ward identities, but in practice the nonlinearity of the transformations (4.6.39,40) makes them difficult to apply.)

To low orders in $V$ (sufficient for the three-loop calculation) the action is

$$S = tr \int d^4x\, d^4\theta\, \{\bar{\Phi}_i\Phi^i - \frac{1}{2}V\Box V + \bar{c}'c - c'\bar{c}$$

$$+ g[\bar{\Phi}_i, V]\Phi^i + \frac{1}{2}gV\{D^\alpha V, \bar{D}^2 D_\alpha V\} + \frac{1}{2}g(c' + \bar{c}')[V, c + \bar{c}]$$

$$+ \frac{1}{2}g^2[[\bar{\Phi}_i, V], V]\Phi^i + \frac{1}{8}g^2[V, D^\alpha V]\bar{D}^2[V, D_\alpha V]$$

$$+ \frac{1}{6}g^2(\bar{D}^2 D^\alpha V)[V, [V, D_\alpha V]] + \frac{1}{12}g^2(c' + \bar{c}')[V, [V, c - \bar{c}]]$$

$$\frac{1}{3!}g^3[[[\bar{\Phi}_i, V], V], V]\Phi^i - \frac{1}{2}(\frac{1}{\alpha} - 1)V\Pi_0\Box V + \cdots\}$$

$$+ tr\{\int d^4x\, d^2\theta\, ig\,\frac{1}{3!}C_{ijk}\Phi^i[\Phi^j, \Phi^k] + h.c.\} \tag{6.4.4}$$

with

$$V = V^{\text{A}}T_{\text{A}} \ , \quad \Phi^i = \Phi^{i\text{A}}T_{\text{A}} \ , \quad c = c^{\text{A}}T_{\text{A}} \ ,$$

$$[T_{\text{A}}, T_{\text{B}}] = i\, f_{\text{AB}}{}^{\text{C}}T_{\text{C}} \ , \quad f_{\text{AB}}{}^{\text{C}} f_{\text{DC}}{}^{\text{B}} = -\, k\delta_{\text{AD}} \ . \tag{6.4.5}$$

Using the Feynman rules it is trivial to see that at the one-loop level, in the Fermi-Feynman gauge defined above, the $\overline{\Phi}\Phi$ self energy is identically zero. The contributions from the two graphs below cancel:

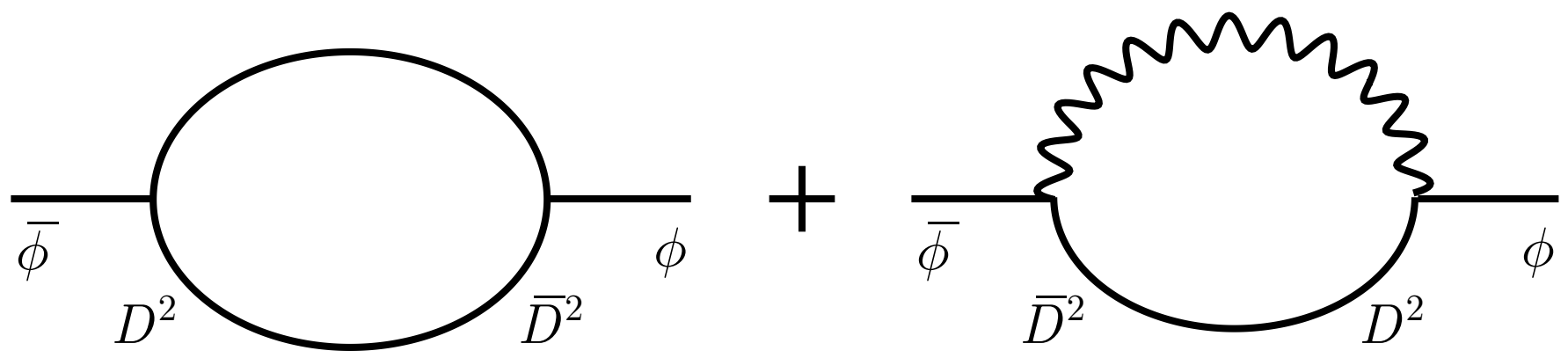

*Fig. 6.4.8a*

It is easy to verify that the one-loop corrections to the ghost and vector self-energies also completely vanish. For the former, this is true in any theory, but for the latter it is due to the multiplicity (3) of the chiral multiplets, which leads to cancellations among the three graphs below:

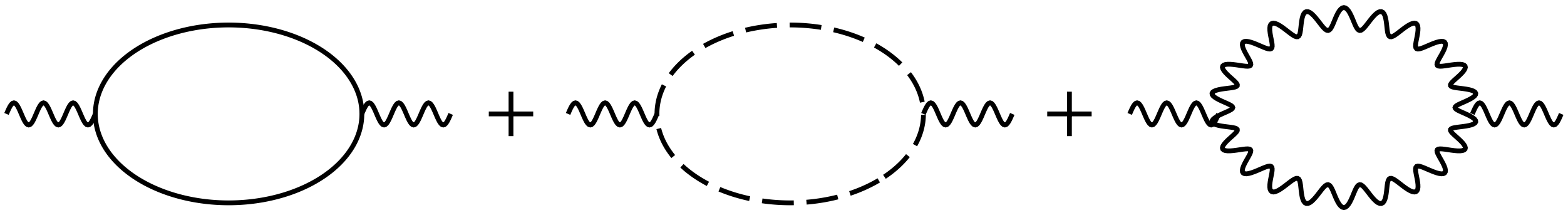

*Fig. 6.4.8b*

This result is trivially true in the background field method: (see sec. 6.5). Then the field $V$ does not contribute and the three chiral fields exactly cancel contributions from *three* chiral ghosts. This also occurs for the $V$ three-point function, and is sufficient to establish in an independent way that $\beta(\text{one-loop}) = 0$. We refer the reader to the literature for other one- and higher-loop calculations and summarize the supergraph results: (a) One-loop three-point functions are finite. The $V$-field four- and higher-point functions have a divergence that can be removed by a nonlinear $V$-field renormalization. (The divergence never arises in the background field method.) (b) At the two-loop level the ghost self-energies are still zero, whereas those for $V$ and $\Phi_i$ are only finite. (Consequently, the higher-order contributions to the gauge parameter are $O(g^4)$.) (c) At the three-loop level the $\Phi^i$ self-energy is finite, thus ensuring the vanishing of the $\beta$-function. We present arguments for proving the results to all orders in sec. 7.7.

The vanishing of the $\beta$-function to all orders of perturbation theory leads to the conclusion that $N = 4$ Yang-Mills is a finite four-dimensional field theory (up to gauge artifacts; e.g., except in supersymmetric background or light-cone gauges, divergences are present, but only in gauge-dependent quantities).

Another theory with interesting finiteness properties is $N = 2$ Yang-Mills theory. It has one-loop divergences, but is finite to all higher orders of perturbation theory (as explicitly verified at two and three loops), making it superrenormalizable. By coupling an appropriate number of $N = 2$ hypermultiplets to $N = 2$ Yang-Mills theory, one can arrange for the one-loop divergences to cancel, and thus construct a completely finite theory (in perturbation theory). (In the special case of one adjoint-representation hypermultiplet, we obtain $N = 4$ Yang-Mills theory.)

## 6.5. The background field method

### a. Ordinary Yang-Mills

The background field method is extremely useful in supersymmetric Yang-Mills theory, and essential in the quantum theory of supergravity. In this section we review the background field method for ordinary Yang-Mills, and then extend it to the supersymmetric case. The extension involves some subtleties, primarily because of the nonlinearity of the gauge transformations.

In gauge theories we start with the gauge-fixed functional integral (6.2.9) and introduce sources coupled to the fields, defining

$$Z(J) = \int I\!DA\, I\!Dc\, I\!Dc'\, e^{S_{eff} + JA} \quad , \qquad JA \equiv \int d^4x\; J^{\underline{a}} A_{\underline{a}} \quad . \tag{6.5.1}$$

We introduce $W(J) = \ln Z(J)$ and define the effective action by a Legendre transform

$$\Gamma(\hat{A}) = W(J) - J\hat{A} \quad , \qquad \hat{A}_{\underline{a}} = \frac{\delta W}{\delta J^{\underline{a}}} \quad . \tag{6.5.2}$$

This quantity is not gauge invariant in general. Physical quantities computed from it are gauge invariant, and the Green functions satisfy Slavnov-Taylor identities that express the underlying gauge invariance, but manifest gauge invariance is lost because of the gauge fixing procedure. On the other hand, the effective action computed in the background field method is manifestly gauge invariant. It is equivalent to the usual one, but is more convenient to handle.

In the background field quantization of Yang-Mills theories we follow a procedure similar to that of sec. 6.2a. We start with the gauge-invariant Lagrangian $I\!\!L_{inv}(A_{\underline{a}})$ (other fields may be present but we do not indicate them explicitly) and *split* the field into a background and quantum part: $I\!\!L_{inv}(\mathbf{A}_{\underline{a}} + A_{\underline{a}})$. The action is invariant under two kinds of transformations that give the same $\delta(\mathbf{A}_{\underline{a}} + A_{\underline{a}})$:

Quantum:

$$\delta \mathbf{A}_{\underline{a}} = 0 \quad , \qquad \delta A_{\underline{a}} = \boldsymbol{\nabla}_{\underline{a}} \lambda + i[\lambda, A_{\underline{a}}] \quad , \tag{6.5.3}$$

$$\boldsymbol{\nabla}_{\underline{a}} \lambda = \partial_{\underline{a}} \lambda + i[\lambda, \mathbf{A}_{\underline{a}}] \quad ,$$

Background:

$$\delta \mathbf{A}_{\underline{a}} = \boldsymbol{\nabla}_{\underline{a}} \lambda \quad , \quad \delta A_{\underline{a}} = i[\lambda, A_{\underline{a}}] \quad . \tag{6.5.4}$$

We consider now the functional

$$\mathbf{Z}(\mathbf{A}) = \int I\!\!D A_{\underline{a}}\, e^{\int I\!\!L_{inv}(A_{\underline{a}} + \mathbf{A}_{\underline{a}})} \quad , \tag{6.5.5}$$

and quantize as before to fix the *quantum* gauge invariance except that, to maintain manifest invariance with respect to the *background* gauge transformations, we choose the gauge-fixing function so that it transforms *covariantly* under these transformations. This requires in particular that we covariantize the derivatives that appear there with respect to the background field: $\partial^{\underline{a}} A_{\underline{a}} \to \boldsymbol{\nabla}^{\underline{a}} A_{\underline{a}} = \partial^{\underline{a}} A_{\underline{a}} - i[\mathbf{A}^{\underline{a}}, A_{\underline{a}}]$. The remainder of the quantization procedure is the same. We require the Faddeev-Popov ghosts to transform covariantly under background gauge transformations, and we choose the weighting function $exp(-\frac{1}{\alpha g^2} tr \int f^2)$ to be invariant. We thus obtain the following expression :

$$\mathbf{Z}(\mathbf{A}) = \int I\!\!D A_{\underline{a}}\, I\!\!D c\, I\!\!D c'\, e^{S_{inv}(A + \mathbf{A}) - \int \frac{1}{4\alpha g^2} tr(\boldsymbol{\nabla} \cdot A)^2 + S_{FP}} \quad . \tag{6.5.6}$$

$\mathbf{Z}$ is manifestly invariant under background gauge transformations but its significance is not obvious. To elucidate its meaning we consider an object defined exactly like $\mathbf{Z}$ except that we also couple the quantum field to a source:

$$\tilde{Z}(J, \mathbf{A}) = \int I\!\!D A_{\underline{a}}\, I\!\!D c\, I\!\!D c'\, e^{S_{eff}(A_{\underline{a}}, \mathbf{A}_{\underline{a}}) + JA} \quad . \tag{6.5.7}$$

We can now pass from $\tilde{Z}(J, \mathbf{A})$ to $\tilde{\Gamma}(\tilde{A}, \mathbf{A})$ by a Legendre transformation in the presence of the fixed field $\mathbf{A}$. On the other hand, returning to $\tilde{Z}$ itself, we can make a change of variables $A -> A - \mathbf{A}$ which gives

$$\tilde{Z}(J, \mathbf{A}) = e^{-J\mathbf{A}} \int I\!\!D A_{\underline{a}}\, I\!\!D c\, I\!\!D c'\, e^{\int [\, I\!\!L_{inv}(A_{\underline{a}}) + I\!\!L_{GF} + I\!\!L_{FP} + JA]}$$

$$= e^{-J\mathbf{A}} Z(J, \mathbf{A}) \quad . \tag{6.5.8}$$

It contains the usual $I\!\!L_{inv}(A_{\underline{a}})$ but unusual gauge fixing and ghost terms that have additional dependence on $\mathbf{A}_{\underline{a}}$. Therefore, $Z(J, \mathbf{A})$ is the usual generating functional but with **A**-dependent gauge fixing and ghost terms. Now we have

$$\widetilde{W}(J,\mathbf{A}) = ln\,\tilde{Z}(J,\mathbf{A}) = ln\,Z(J,\mathbf{A}) - J\mathbf{A} = W(J,\mathbf{A}) - J\mathbf{A} \quad , \tag{6.5.9}$$

and

$$\tilde{\Gamma}(\tilde{A},\mathbf{A}) = \widetilde{W}(J,\mathbf{A}) - J\tilde{A} = [W(J,\mathbf{A}) - J\mathbf{A}] - J\tilde{A}$$

$$= W(J,\mathbf{A}) - J(\tilde{A}+\mathbf{A}) \quad , \qquad \tilde{A} = \frac{\delta\widetilde{W}}{\delta J} = \frac{\delta W}{\delta J} - \mathbf{A} \quad . \tag{6.5.10}$$

Therefore

$$\tilde{\Gamma}(\tilde{A},\mathbf{A}) = \Gamma(\tilde{A}+\mathbf{A}\,,\mathbf{A}) \quad , \tag{6.5.11}$$

is the usual effective action $\Gamma(\hat{A})$, evaluated in an unusual **A**-dependent gauge, and at $\hat{A} = \tilde{A} + \mathbf{A}$. In particular, if in the evaluation of $\tilde{\Gamma}(\tilde{A},\mathbf{A})$ we restrict ourselves to graphs with no external $\tilde{A}$ lines ("vacuum" graphs), i.e., set $\tilde{A} = 0$, we will obtain $\Gamma(\mathbf{A})$, the usual effective action. But these "$\tilde{A}$-vacuum" graphs are simply the one-particle-irreducible subset of the graphs obtained from $\tilde{Z}(0,\mathbf{A}) = \mathbf{Z}(\mathbf{A})$. (Actually, this is an oversimplification. What one obtains is not exactly the effective action, because **A** lines from the gauge-fixing term give additional contributions. However, because of gauge invariance, it can be shown that these have no effect on S-matrix elements so that the identification, though strictly speaking not correct, can be used when computing physical quantities.)

Our conclusion is that the effective action is obtained from $\mathbf{Z}(\mathbf{A})$ by evaluating in perturbation theory one-particle-irreducible graphs with only internal $A_{\underline{a}}$ lines and external $\mathbf{A}_{\underline{a}}$ lines (as well as ghost, and other non-gauge field lines). In particular, if we expand $I\!\!L_{inv}(\mathbf{A}+A) = I\!\!L(\mathbf{A}) + I\!\!L'(\mathbf{A})A + I\!\!L''(\mathbf{A})A^2 + \cdots$, the first term does not contribute to loop graphs (it is the classical contribution to $\Gamma$), and the second can be dropped because it does not contribute to one-particle-irreducible graphs with no external $A$ lines. The $A^2$ term gives the complete contribution (from the gauge field) to one-loop graphs. For higher-loop graphs, internal vertices are read from the higher order expansion, and all the terms contribute to vertices that involve the external **A** lines.

There is an additional feature of the background-field quantization that is not usually encountered in the Yang-Mills case but that is important. This is the appearance of the Nielsen-Kallosh ghost. In the gauge-averaging procedure we used the simplest exponential factor to produce the gauge-fixing term in the effective Lagrangian. However, a more complicated averaging function could be used, e.g., $exp \int (fMf)$ where $M$ is any operator (matrix). To properly normalize the averaging procedure, we must divide by $det\, M$. If $M$ is field independent, this is a trivial factor. However, if $M$ is a function of the background field, we normalize the gauge averaging by introducing into the functional integral a factor

$$\int I\!\!D f\, I\!\!D b\; e^{fMf}\; e^{bMb} \quad , \tag{6.5.12}$$

where $b$ is a *ghost* field, with opposite statistics to $f$. When we carry out the $f$ integration using the $\delta$-function of sec. 6.2.a, we are left with the $b$ field. Thus, the final form is

$$\mathbf{Z}(\mathbf{A}_{\underline{a}}) = \int I\!\!D A_{\underline{a}}\, I\!\!D c\, I\!\!D c'\, I\!\!D b\; e^{\int [I\!\!L_{inv}(A+\mathbf{A}) + I\!\!L_{GF}(A,\mathbf{A}) + I\!\!L_{FP}(c,c',A,\mathbf{A}) + I\!\!L_{NK}(b,\mathbf{A})]} \tag{6.5.13}$$

where $I\!\!L_{NK} = bMb$. If $M$ is independent of the background field, the additional ghost gives trivial contributions and can be dropped; otherwise, since the ghost field $b$ has no interactions with other quantum fields and since it enters quadratically, it only contributes at the one-loop level.

To motivate the procedure we use in the background field quantization of supersymmetric Yang-Mills theory, we point out two aspects of the background-quantum splitting $A_{\underline{a}} \to \mathbf{A}_{\underline{a}} + A_{\underline{a}}$ of ordinary Yang-Mills. This splitting has the virtue that the transformations $\delta(\mathbf{A}_{\underline{a}} + A_{\underline{a}}) = (\partial_{\underline{a}}\omega - i[\mathbf{A}_{\underline{a}}, \omega]) + (-i[A_{\underline{a}}, \omega])$, which leave the action invariant, can be interpreted as ordinary gauge transformations of the background field accompanied by covariant gauge rotations (linear and homogeneous) of the quantum field. Furthermore, in an expansion

$$I\!\!L(A + \mathbf{A}) = \sum [I\!\!L^{(n)}(\mathbf{A})](A)^n \quad , \tag{6.5.14}$$

each term in the power series is separately invariant under these transformations, since $A_{\underline{a}}$ transforms linearly and homogeneously, as do the functional derivatives of $I\!\!L(\mathbf{A})$.

Thus, if we truncate the series, as we do in a perturbative loop-by-loop evaluation of the effective action, we maintain the background gauge invariance. *This would not be true if the transformation of the quantum field were nonlinear.*

## b. Supersymmetric Yang-Mills

In supersymmetric Yang-Mills theory the classical action is invariant under *nonlinear* gauge transformations $e^V \to e^{V'} = e^{i\bar{\Lambda}} e^V e^{-i\Lambda}$, and the splitting $V \to V + \mathbf{V}$ is unsuitable. To motivate the subsequent procedure, we first reexamine the background-quantum splitting of ordinary Yang-Mills theory from a different point of view. We start with the original gauge and matter action, invariant under the local transformations $\delta A_{\underline{a}} = \partial_{\underline{a}}\omega - i[A_{\underline{a}}, \omega]$ and, for some matter field, $\delta\psi = i[\omega, \psi]$. Under *global* transformations with constant $\omega$ we still have invariance, with the gauge fields rotating like the matter fields. For local $\omega$, we can introduce a new invariance by keeping the covariant transformations $i[\omega, A_{\underline{a}}]$ and $i[\omega, \psi]$ for all fields, and introducing a separate gauge field, the background field $\mathbf{A}_{\underline{a}}$ to covariantize the derivatives. Since in the original action all derivatives entered in the form $\nabla_{\underline{a}} = \partial_{\underline{a}} - i[A_{\underline{a}},\ \ ]$, this covariantization amounts to the replacement

$$\nabla_{\underline{a}} = \partial_{\underline{a}} - i[A_{\underline{a}},\ \ ] \to \boldsymbol{\nabla}_{\underline{a}} - i[A_{\underline{a}},\ \ ] = \partial_{\underline{a}} - i[\mathbf{A}_{\underline{a}} + A_{\underline{a}},\ \ ] \quad , \tag{6.5.15}$$

which is equivalent to the ordinary quantum-background splitting. We now have two invariances: the original one where the background field is inert, and the new one, under which all the fields transform. *We obtain a linear splitting $A \to A + \mathbf{A}$ because the gauge field enters linearly in the covariant derivative.* In supersymmetric Yang-Mills theory this is so in the abelian case, but not in the nonabelian case. However, the philosophy is the same. We start with the locally invariant gauge theory, observe that it is invariant for global transformations (with $\Lambda = \lambda = \bar{\Lambda}$ a constant matrix, not a superfield), under which the gauge fields transform covariantly (linearly and homogeneously, since $e^V \to e^{i\lambda} e^V e^{-i\lambda}$ implies $V \to e^{i\lambda} V e^{-i\lambda}$) and now gauge this transformation by covariantizing with the aid of a background field. This amounts to the replacement $D_A \to \boldsymbol{\nabla}_A$ where $\boldsymbol{\nabla}_A$ is a background covariant derivative. We also have to treat the covariantly chiral superfields properly.

We recall that supersymmetric Yang-Mills theory can be formulated in terms of constrained covariant derivatives. The reason for solving the constraints and introducing the gauge prepotentials is that only these unconstrained objects are suitable for quantization. In solving the constraints we have the choice of working in the vector representation or in the chiral representation. The latter is more convenient for quantization, expressing the theory in terms of the real superfield $V$, rather than superfields $\Omega$, $\overline{\Omega}$ with a redundant gauge invariance. We want to maintain this advantage in the background field method and work with a quantum $V$. On the other hand, when we introduce the background covariant derivatives, it is useful to think of them in the vector representation. In fact it is possible to express all our results in terms of the (constrained) background covariant derivatives themselves, without ever introducing explicitly the background gauge superfields, i.e. without solving the constraints, and in that case the only representation that is available is the vector representation. The advantage of working with the background derivatives directly is that background covariance is manifest and we obtain significant simplifications and improvement in the power counting rules for Feynman graphs.

We also express covariantly chiral superfields in terms of the quantum field $V$ and background-covariantly chiral superfields. The latter therefore depend implicitly on the background fields and would seem not to be suitable for quantization. However, this is not always the case: At more than one loop, and even at one loop for real representations of the gauge group, we formulate covariant Feynman rules directly for covariantly chiral superfields that lead to considerable improvement over the ordinary ones.

Starting with the ordinary covariant derivatives we perform the splitting by writing them, in the *quantum-chiral but background-vector* representation, as

$$\nabla_\alpha = e^{-V}\boldsymbol{\nabla}_\alpha e^V \quad , \qquad \overline{\nabla}_{\dot\alpha} = \overline{\boldsymbol{\nabla}}_{\dot\alpha} \quad , \qquad \nabla_{\underline{a}} = -\, i\{\nabla_\alpha, \overline{\nabla}_{\dot\alpha}\} \quad ; \tag{6.5.16}$$

where $\boldsymbol{\nabla}_\alpha$ and $\overline{\boldsymbol{\nabla}}_{\dot\alpha}$ are background covariant derivatives satisfying the usual constraints. The $\nabla$'s transform covariantly under two sets of transformations:

(a) Quantum:

$$e^V \to e^{i\overline{\Lambda}}\, e^V\, e^{-i\Lambda} \quad ,$$

$$\boldsymbol{\nabla}_A \to \boldsymbol{\nabla}_A \quad , \tag{6.5.17}$$

with background covariantly chiral parameters $\boldsymbol{\nabla}_\alpha\overline{\Lambda} = \overline{\boldsymbol{\nabla}}_{\dot\alpha}\Lambda = 0$ , i.e.,

$$\nabla_A \to e^{i\Lambda}\nabla_A\, e^{-i\Lambda} \quad . \tag{6.5.18}$$

(b) Background:

$$e^V \to e^{iK}\, e^V\, e^{-iK} \quad ,$$

$$\boldsymbol{\nabla}_A \to e^{iK}\boldsymbol{\nabla}_A e^{-iK} \quad , \tag{6.5.19}$$

with a real parameter $K = \bar{K}$ , i.e.,

$$\nabla_A \to e^{iK}\nabla_A e^{-iK} \quad . \tag{6.5.20}$$

The background field transformations of $V$ can be rewritten as

$$V \to e^{iK} V\, e^{-iK} \quad , \tag{6.5.21}$$

i.e., $V$ transforms covariantly.

* * *

While this procedure has given us a correct quantum-background splitting, in contrast to the component Yang-Mills case it results in different transformations of $\nabla_A$ under quantum and background transformations. However, the transformation of the unsplit gauge field *is* the same. To understand the splitting of the gauge field we solve the constraints on the background covariant derivatives:

$$\boldsymbol{\nabla}_\alpha = e^{-\boldsymbol{\Omega}} D_\alpha e^{\boldsymbol{\Omega}} \quad , \quad \bar{\boldsymbol{\nabla}}_{\dot\alpha} = e^{\bar{\boldsymbol{\Omega}}} \bar{D}_{\dot\alpha} e^{-\bar{\boldsymbol{\Omega}}} \quad . \tag{6.5.22}$$

Hence the splitting of the full derivatives is

$$\nabla_\alpha = e^{-V} e^{-\boldsymbol{\Omega}} D_\alpha e^{\boldsymbol{\Omega}} e^V \quad , \quad \bar{\nabla}_{\dot\alpha} = e^{\bar{\boldsymbol{\Omega}}} \bar{D}_{\dot\alpha} e^{-\bar{\boldsymbol{\Omega}}} \quad . \tag{6.5.23}$$

We transform to a background chiral representation by pre- and post-multiplying all quantities by $e^{-\bar{\boldsymbol{\Omega}}}$ and $e^{\bar{\boldsymbol{\Omega}}}$, respectively. Then

$$\nabla_\alpha \to e^{-\bar{\boldsymbol{\Omega}}} e^{-V} e^{-\boldsymbol{\Omega}} D_\alpha e^{\boldsymbol{\Omega}} e^V e^{\bar{\boldsymbol{\Omega}}} \quad , \quad \bar{\nabla}_{\dot\alpha} \to \bar{D}_{\dot\alpha} \quad , \tag{6.5.24}$$

and the splitting is equivalent to replacing $e^V$ by

$$e^{V(split)} = e^{\boldsymbol{\Omega}} e^V e^{\bar{\boldsymbol{\Omega}}} \quad . \tag{6.5.25}$$

In other words, we split the full $V$ into a quantum $V$ and background $\boldsymbol{\Omega}$ and $\bar{\boldsymbol{\Omega}}$ in a particular, nonlinear fashion. (In the abelian case this reduces to $V \to V + \mathbf{\Omega} + \overline{\mathbf{\Omega}}$ which is just the ordinary splitting since by (4.2.72) $\mathbf{\Omega} + \overline{\mathbf{\Omega}} = \mathbf{V}$.)

The usual chiral representation transformations of (6.5.25)

$$(e^{\mathbf{\Omega}} e^V e^{\overline{\mathbf{\Omega}}})' = e^{i\overline{\Lambda}_0}(e^{\mathbf{\Omega}} e^V e^{\overline{\mathbf{\Omega}}}) e^{-i\Lambda_0} \quad , \tag{6.5.26}$$

(where $\Lambda_0$ is ordinary chiral, $\bar{D}_{\dot{\alpha}}\Lambda_0 = 0$), can be written in two ways:

(a)

$$(e^{\mathbf{\Omega}} e^V e^{\overline{\mathbf{\Omega}}})' = e^{\mathbf{\Omega}}[(e^{-\mathbf{\Omega}} e^{i\overline{\Lambda}_0} e^{\mathbf{\Omega}}) e^V (e^{\overline{\mathbf{\Omega}}} e^{-i\Lambda_0} e^{-\overline{\mathbf{\Omega}}})] e^{\overline{\mathbf{\Omega}}}$$

$$= e^{\mathbf{\Omega}}(e^{i\overline{\Lambda}} e^V e^{-i\Lambda}) e^{\overline{\mathbf{\Omega}}} \quad , \tag{6.5.27a}$$

i.e., the quantum transformations (6.5.17), *with background covariantly chiral* $\Lambda$, or

(b)

$$(e^{\mathbf{\Omega}} e^V e^{\overline{\mathbf{\Omega}}})' = (e^{i\overline{\Lambda}_0} e^{\mathbf{\Omega}} e^{-iK})\,(e^{iK} e^V e^{-iK})\,(e^{iK} e^{\overline{\mathbf{\Omega}}} e^{-i\Lambda_0}) \quad , \tag{6.5.27b}$$

i.e., the background transformations (6.5.19) (cf. (4.2.70-71); recall that the $\Lambda_0$ part of the transformation of $\mathbf{\Omega}$ does not affect the transformation of the background covariant derivatives). This is very similar to the situation in component Yang-Mills.

The gauge Lagrangian has the form

$$tr W^2 = -\,tr(\tfrac{1}{2}\,[\overline{\nabla}^{\dot{\alpha}}, \{\overline{\nabla}_{\dot{\alpha}}, \nabla_\alpha\}])^2 \quad . \tag{6.5.28}$$

When we substitute (6.5.16) into (6.5.28), we obtain a splitting of the action into explicit quantum $V$'s and background covariant derivatives. Since the $\nabla$'s transform covariantly, the Lagrangian will be invariant under both background and quantum transformations. Furthermore, since $V$ transforms homogeneously, expanding the Lagrangian in powers of $V$ will maintain the background invariance term-by-term, which is one of the required properties of a good splitting.

When covariantly chiral superfields $\Phi$, $\overline{\nabla}_{\dot{\alpha}}\Phi = 0$ are present, we first express them in terms of background covariantly chiral superfields by $\Phi = \Phi$, $\tilde{\overline{\Phi}} = \overline{\Phi} e^V$ (in the *quantum* chiral representation) $\overline{\mathbf{\nabla}}_{\dot{\alpha}}\Phi = \mathbf{\nabla}_\alpha \overline{\Phi} = 0$, and then *linearly* split them into a sum of background and quantum fields. The quantum fields transform under

(a)Quantum transformations:

$$\Phi' = e^{i\Lambda}\Phi \quad ,$$

$$\bar{\Phi}' = \bar{\Phi}e^{-i\bar{\Lambda}} \quad . \tag{6.5.29}$$

(b)Background transformations:

$$\Phi' = e^{iK}\Phi \quad ,$$

$$\bar{\Phi}' = \bar{\Phi}e^{-iK} \quad . \tag{6.5.30}$$

The chiral field action is invariant under both quantum and background transformations.

We examine now the background field quantization. We proceed as in the conventional approach, but compute

$$\mathbf{Z} = \int I\!\!D V \, I\!\!D c I\!\!D c' I\!\!D \bar{c} I\!\!D \bar{c}' \, \delta(\bar{\boldsymbol{\nabla}}^2 V - f)\delta(\boldsymbol{\nabla}^2 V - \bar{f})e^{S_{inv}+S_{FP}} \quad . \tag{6.5.31}$$

We have chosen background-covariantly chiral gauge fixing functions, and this means that the Faddeev-Popov ghosts, introduced as in sec. 6.2, are also background covariantly chiral. Finally, we gauge average with

$$\int I\!\!D f \, I\!\!D \bar{f} \, I\!\!D b \, I\!\!D \bar{b} \, e^{-\int d^4x \, d^4\theta \, [\bar{f}f + \bar{b}b]} \quad , \tag{6.5.32}$$

where the background covariantly chiral Nielsen-Kallosh ghosts $b$, $\bar{b}$ have been introduced to normalize to 1 the averaging over $f$, $\bar{f}$. This leads to the final form

$$\mathbf{Z} = \int I\!\!D V \, I\!\!D c \, I\!\!D c' \, I\!\!D \bar{c} \, I\!\!D \bar{c}' \, I\!\!D b \, I\!\!D \bar{b} \, e^{S_{eff}} \quad ,$$

$$S_{eff} = S_{inv} + S_{GF} + S_{FP} + \int \bar{b}b \quad ; \tag{6.5.33}$$

which, except for the Nielsen-Kallosh ghosts, is like (6.2.19), but with background covariant derivatives and covariantly chiral superfields. The N.-K. ghosts interact with the background field, and only give one-loop contributions. If we couple external sources to the quantum fields, e.g., $\int d^4x d^4\theta JV$, the generating functional $\mathbf{Z}(J, \boldsymbol{\Omega})$ will still be invariant under background transformations, provided we require the sources to transform covariantly, $\delta J = i[K, J]$.

What we must do now is argue that the background field functional, obtained by setting sources to zero and computing one-particle-irreducible graphs with only internal quantum lines and external background lines, is equivalent to the usual effective action, except for being computed in a different gauge. This is less direct than in the ordinary case because the splitting is highly nonlinear. We present the following argument:

The splitting (6.5.25) is

$$V \to V + \boldsymbol{\Omega} + \overline{\boldsymbol{\Omega}} + \textit{nonlinear terms} \quad . \tag{6.5.34}$$

In a gauge for the background fields where $\boldsymbol{\Omega} = \overline{\boldsymbol{\Omega}} = \frac{1}{2}\boldsymbol{V}$ we write this as $V \to f(V, \mathbf{V})$ where $f(0, \mathbf{V}) = \mathbf{V}$ and $f(V, 0) = V$. If now in the original functional integral we add a source term $\int d^4x d^4\theta J[f(V, \mathbf{V}) - f(0, \mathbf{V})]$ to define a $\tilde{Z}(J, \mathbf{V})$ we will have a $JV$ coupling, and coupling to higher order terms in $V$ and $\mathbf{V}$ (which are irrelevant when computing the S-matrix), but no linear coupling $J\mathbf{V}$. When we set $\mathbf{V} = 0$ we obtain the conventional $Z(J)$. As in (6.5.8) we make a change of variables of integration which involves the *inverse* of the function $f$. Under this transformation the invariant gauge action goes to its usual form in terms of $V$, the gauge fixing and ghost terms change in a complicated, but physically irrelevant manner, and the coupling to the source becomes simply $JV - J\mathbf{V}$. Furthermore, the Jacobian of the transformation is 1 (see sec. 3.8.b). We now have the same form as in ordinary Yang-Mills theory, and we conclude that the background field functional computed by setting $J = 0$, i.e., evaluating graphs with only internal quantum lines, does give the usual effective action as a function of the background field, albeit in an unconventional gauge. Therefore all physically relevant quantum corrections can be obtained from the background field functional. We now discuss how to evaluate it in perturbation theory.

### c. Covariant Feynman rules

We consider first contributions from only the quantum gauge field $V$. The effective Lagrangian is

$$-\frac{1}{2g^2}\, tr[(e^{-V}\boldsymbol{\nabla}^\alpha e^V)\overline{\boldsymbol{\nabla}}^2(e^{-V}\boldsymbol{\nabla}_\alpha e^V) + V(\boldsymbol{\nabla}^2\overline{\boldsymbol{\nabla}}^2 + \overline{\boldsymbol{\nabla}}^2\boldsymbol{\nabla}^2)V] \quad . \tag{6.5.35}$$

All the dependence on the background fields is through the connection coefficients and never through the gauge fields themselves.

The quadratic action has the form

$$-\frac{1}{2g^2}\, tr V[\,-\boldsymbol{\nabla}^\alpha \bar{\boldsymbol{\nabla}}^2 \boldsymbol{\nabla}_\alpha - i\mathbf{W}^\alpha \boldsymbol{\nabla}_\alpha + \frac{1}{2} i(\boldsymbol{\nabla}_\alpha \mathbf{W}^\alpha) + \boldsymbol{\nabla}^2 \bar{\boldsymbol{\nabla}}^2 + \bar{\boldsymbol{\nabla}}^2 \boldsymbol{\nabla}^2]V \quad . \tag{6.5.36}$$

Using the commutation relations

$$[\bar{\boldsymbol{\nabla}}_{\dot\alpha}, \boldsymbol{\nabla}_{\alpha\dot\beta}] = C_{\dot\alpha\dot\beta} \mathbf{W}_\alpha \quad , \tag{6.5.37}$$

this can be rewritten as

$$-\frac{1}{2g^2}\, tr V[\boldsymbol{\Box} - i\mathbf{W}^\alpha \boldsymbol{\nabla}_\alpha - i\bar{\mathbf{W}}^{\dot\alpha} \bar{\boldsymbol{\nabla}}_{\dot\alpha}]V \quad , \tag{6.5.38}$$

where $\boldsymbol{\Box} = \frac{1}{2}\boldsymbol{\nabla}^{\underline{a}}\boldsymbol{\nabla}_{\underline{a}}$ is the background covariant d'Alembertian and $\mathbf{W}_\alpha$ is the background field strength. Introducing connection coefficients (depending on the background fields) by $\boldsymbol{\nabla}_A = D_A - i\boldsymbol{\Gamma}_A$, we can separate out a free kinetic term, and interactions with the background:

$$-\frac{1}{2g^2}\, tr V[\Box_0 - i\boldsymbol{\Gamma}^{\underline{a}}\partial_{\underline{a}} - \frac{1}{2} i(\partial^{\underline{a}}\boldsymbol{\Gamma}_{\underline{a}}) - \frac{1}{2}\boldsymbol{\Gamma}^{\underline{a}}\boldsymbol{\Gamma}_{\underline{a}}$$

$$- i\mathbf{W}^\alpha(D_\alpha - i\boldsymbol{\Gamma}_\alpha) - i\bar{\mathbf{W}}^{\dot\alpha}(\bar{D}_{\dot\alpha} - i\bar{\boldsymbol{\Gamma}}_{\dot\alpha})]V \quad . \tag{6.5.39}$$

This expression is sufficient for doing one-loop calculations using conventional propagators for real scalar superfields and the usual $D$-manipulations. Since the interaction with the background fields is at most linear in $D$'s, and at least four $D$'s are needed in a loop, *the first nonvanishing one-loop contribution from $V$ is in the four-point function.* Self-interactions for computing higher-loop contributions can be obtained from the higher-order in $V$ terms in the Lagrangian (6.5.35).

We now turn to contributions from (fully) covariantly chiral physical superfields and background covariantly chiral ghost superfields. In principle we have to solve the chirality constraint $\bar{\boldsymbol{\nabla}}_{\dot\alpha}\Phi = 0$ (by writing $\Phi = e^{\bar\Omega}\Phi_0$ in terms of an ordinary chiral superfield), but this introduces explicit dependence on the background gauge prepotentials which we wish to avoid if possible. Instead, we reexamine the derivation of the Feynman rules for chiral superfields.

We consider the generating functional of the form

$$Z(j,\bar{j}) = \int I\!D\Phi\, I\!D\bar{\Phi}\; e^{S + (\int d^4x\, d^2\theta\; j\Phi + h.c.)} \quad , \tag{6.5.40}$$

where $\Phi$ and $j$ are covariantly chiral $\bar{\nabla}_{\dot{\alpha}}\Phi = \bar{\nabla}_{\dot{\alpha}}j = 0$. For the time being we need not specify whether these are full covariant derivatives or just background covariant derivatives. In principle we define $\int I\!D\Phi$ as the integral over the corresponding chiral-representation field $\Phi_0$ (antichiral for $\bar{\Phi}$ integration), but in practice we simply define it by the Gaussian integral

$$\int I\!D\Phi e^{\int d^4x d^2\theta \frac{1}{2}\Phi^2} = 1 \quad . \tag{6.5.41}$$

Additional fields may be present but we need not indicate them explicitly.

We *define* covariant functional differentiation by

$$\frac{\delta\Phi(z)}{\delta\Phi(z')} = \bar{\nabla}^2\delta^8(z-z') \quad . \tag{6.5.42}$$

This form can be derived from (3.8.3), or by writing $\Phi = \bar{\nabla}^2\Psi$, in terms of a general superfield, and covariantizing (3.8.13). Manifestly covariant rules for chiral superfields can now be found by a direct covariantization of the usual method. The covariantization of the identity $\bar{D}^2D^2\Phi = \Box_0\Phi$ (where $\Box_0$ denotes now the free d'Alembertian) becomes

$$\bar{\nabla}^2\nabla^2\Phi = \Box_+\Phi \quad ,$$

$$\Box_+ = \Box - iW^\alpha\nabla_\alpha - \frac{1}{2}i(\nabla^\alpha W_\alpha) \quad , \tag{6.5.43}$$

with the covariant $\Box$. We consider first the massless case.

We carry out the functional integration over $\Phi$ by separating out the interaction terms and doing the Gaussian integral, and obtain

$$Z = \Delta \cdot e^{S_{int}(\frac{\delta}{\delta j},\frac{\delta}{\delta \bar{j}})}\; e^{-\int d^4x\, d^4\theta\; \bar{j}\Box_+{}^{-1}j} \quad , \tag{6.5.44}$$

where $\Delta$ is the functional determinant

$$\Delta = \int I\!D\Phi\, I\!D\bar{\Phi}\; e^{S_0} \quad , \qquad S_0 = \int d^4x\, d^4\theta\; \bar{\Phi}\Phi \quad . \tag{6.5.45}$$

In general the above expression must still be integrated over other quantum fields that may be present.

Before we evaluate $\Delta$, which will give a separate, one-loop contribution to the effective action from the $\Phi$ field, we examine the rest of the contributions. The expression for $Z$ is identical to the one in (6.3.14), except for the presence of covariant derivatives and covariantly chiral sources, and the factor $\Delta$. The perturbation expansion takes the same form, except that from the functional differentiation we get factors of $\overline{\nabla}^2$ or $\nabla^2$ acting on chiral and antichiral lines. The propagators are given by $-\Box_+{}^{-1}$, but in a perturbative calculation we separate $\Box_+$ into a free part, which leads to $p^{-2}$ propagators, and the remainder, which gives additional interaction vertices. However, at no stage do we encounter explicit gauge fields, only connections and field strengths. (The explicit dependence on the quantum gauge fields will be needed only when we functionally integrate over them.)

We now evaluate $\Delta$. It gives the complete one-loop contribution of the chiral superfield to graphs with only external $V$ lines and could be evaluated by using standard Feynman rules, but this we wish to avoid. This turns out to be possible only for *real* representations of the Yang-Mills group. Of course, real representations are frequently the ones of interest: e.g., the Yang-Mills ghosts are in the adjoint representation, which is always real. We therefore consider first the case of real representations, and return later to the complications caused by complex representations. We are still considering the massless case.

The action $S_0$ leads to the equations of motion (in the presence of sources)

$$\mathbf{O}\begin{pmatrix} \Phi \\ \overline{\Phi} \end{pmatrix} + \begin{pmatrix} j \\ \bar{j} \end{pmatrix} = 0 \quad , \quad \mathbf{O} \equiv \begin{pmatrix} 0 & \overline{\nabla}^2 \\ \nabla^2 & 0 \end{pmatrix} \quad . \tag{6.5.46}$$

We define an action whose equations of motion are

$$\mathbf{O}^2\begin{pmatrix} \Phi \\ \overline{\Phi} \end{pmatrix} - \begin{pmatrix} j \\ \bar{j} \end{pmatrix} = 0 \quad , \quad \mathbf{O}^2 = \begin{pmatrix} \overline{\nabla}^2\nabla^2 & 0 \\ 0 & \nabla^2\overline{\nabla}^2 \end{pmatrix} \quad , \tag{6.5.47}$$

in terms of the square of $\mathbf{O}$. This action is given by $S'_0 + \overline{S}'_0$, where

$$S'_0 = \int d^4x\, d^2\theta\, \frac{1}{2}\Phi\Box_+\Phi = \int d^4x\, d^4\theta\, \frac{1}{2}\Phi\nabla^2\Phi \quad . \tag{6.5.48}$$

In terms of it we can write the functional integral

$$\Delta^2 = \int I\!D\Phi\, I\!D\bar{\Phi}\ e^{S'_0 + \bar{S}'_0} = |\int I\!D\Phi\ e^{S'_o}|^2 = (\int I\!D\Phi\ e^{S'_o})^2 \quad . \qquad (6.5.49)$$

We have used the fact that $S'_0$ and its hermitian conjugate contribute equally to $\Delta$, as can be seen, for example, by examining the resulting Feynman rules below. (This procedure is analogous to the "doubling" trick in QED, where the analogue of **O** is $\nabla_{\alpha\dot{\beta}}$ and of $\Box_+$ is $C_{\alpha\beta}\Box + f_{\alpha\beta}$, with $f_{\alpha\beta}$ the electromagnetic field strength. )

We now integrate $S'_0$ by separating out $\bar{D}^2D^2$ from $\bar{\nabla}^2\nabla^2$ and treating $(\bar{\nabla}^2\nabla^2 - \bar{D}^2D^2)$ as an interaction term. The result is

$$\Delta = e^{\int d^4x\, d^2\theta\ \frac{1}{2}\frac{\delta}{\delta j}[\bar{\nabla}^2\nabla^2 - \bar{D}^2D^2]\frac{\delta}{\delta j}}\ e^{-\int d^4x\, d^2\theta\ \frac{1}{2}j\Box_o{}^{-1}j} \quad . \qquad (6.5.50)$$

(Writing instead $\bar{\nabla}^2\nabla^2 = \bar{D}^2e^{-V}D^2e^V$ in the chiral representation gives the rules for the one-loop expression in the usual noncovariant formalism.) Therefore, a calculation of the one-loop contribution consists in evaluating graphs with propagators $p^{-2}\delta^4(\theta - \theta')$ and vertices $\bar{\nabla}^2\nabla^2 - \bar{D}^2D^2$ giving rise to a string

$$\cdots[\bar{\nabla}^2\nabla^2 - \bar{D}^2D^2]_i\delta^4(\theta_i - \theta_{i+1})[\bar{\nabla}^2\nabla^2 - \bar{D}^2D^2]_{i+1}\cdots \quad , \qquad (6.5.51)$$

with $\int d^4\theta_i$ integrals at each vertex and one loop-momentum integral. We carry out the evaluation *in the chiral representation,* so that $\bar{\nabla}_{\dot{\alpha}} = \bar{D}_{\dot{\alpha}}$ . We concentrate on the $i$ vertex, and from the next vertex we temporarily transfer the $\bar{D}^2 = \bar{\nabla}^2$ factor across the $\delta$-function. We now use the identity $(\bar{\nabla}^2\nabla^2 - \bar{D}^2D^2)\bar{D}^2 = (\Box_+ - \Box_0)\bar{D}^2$ (in the chiral representation). Having performed this maneuver we return the $\bar{D}^2$ to its original place, and proceed to manipulate the next vertex in the same way. This procedure can be carried out at all vertices but one, which retains its original form. The resulting rules for the evaluation of $\Delta$ are, with the usual propagator,

one vertex:

$$\bar{D}^2(\nabla^2 - D^2) \quad , \qquad (6.5.52)$$

other vertices:

$$\Box_+ - \Box_0 \quad , \qquad (6.5.53)$$

with the covariant derivatives in chiral representation. Now only one vertex contributes any $\bar{D}$'s and as a consequence the evaluation of one-loop graph contributions from chiral superfields is considerably simplified. Higher loops are given by the rest of the expression in (6.5.44).

Up to now we have not specified whether the $\nabla$'s are full or background covariant derivatives. If both quantum and background gauge fields are present, it is more convenient to carry out the above procedure at an early stage, before we write $\tilde{\overline{\Phi}} = \overline{\Phi} e^V$, i.e., work with fully covariantly chiral superfields (but not for the ghosts, which are only background covariantly chiral). The result of the calculation is expressible in terms of the full covariant derivatives, and only at that stage, having integrated out the chiral superfields, do we need to make the background quantum splitting on the gauge fields.

The "doubling" trick cannot be applied covariantly when the scalar multiplet is in a complex representation of the Yang-Mills group. If we write the covariantly chiral $\Phi$ in terms of ordinary chiral $\Phi_0$ ($\overline{D}_{\dot{\alpha}} \Phi_0 = 0$) in the vector representation

$$\Phi = e^{\overline{\Omega}} \Phi_0 \quad , \tag{6.5.54}$$

we have

$$\overline{\Phi} = e^{\overline{\Omega}^*} \overline{\Phi}_0 \quad , \tag{6.5.55}$$

and

$$\overline{\nabla}^2 \overline{\Phi} = e^{-\Omega^*} \overline{D}^2 e^{\Omega^*} e^{\overline{\Omega}^*} \overline{\Phi}_0 \quad , \tag{6.5.56}$$

is not in the same representation as $\Phi$ (does not satisfy the same chirality condition) except when the representation is real (in which case $\Omega^* = -\overline{\Omega}$). Therefore, the operator $\mathbf{O}$ in (6.5.46) cannot be squared, since it is not representation-preserving. As a result, we must use rules *at one loop* which are not expressed manifestly in terms of connections $\Gamma_A$, but involve explicit gauge fields.

In (6.5.45) we express $\Phi$ in terms of $\Phi_0$, and introduce ordinary chiral sources $j_0$ ($\overline{D}_{\dot{\alpha}} j_0 = 0$). We have instead of (6.5.46) the following equations of motion in the presence of external super-Yang-Mills:

$$\hat{\mathbf{O}} \begin{pmatrix} \Phi_0 \\ \overline{\Phi}_0 \end{pmatrix} + \begin{pmatrix} j_0 \\ \overline{j}_0 \end{pmatrix} = 0 \quad , \quad \hat{\mathbf{O}} = \begin{pmatrix} 0 & \overline{D}^2 e^{V^*} \\ D^2 e^V & 0 \end{pmatrix} \quad . \tag{6.5.57}$$

The noncovariant object $\hat{\mathbf{O}}$ can be squared, since it preserves ($\bar{D}_{\dot{\alpha}}$-)chirality:

$$\hat{\mathbf{O}}^2 = \begin{pmatrix} \bar{D}^2 e^{V*} D^2 e^V & 0 \\ 0 & D^2 e^V \bar{D}^2 e^{V*} \end{pmatrix} \quad . \tag{6.5.58}$$

The action $\hat{S}_0{}'$ obtained from $\hat{\mathbf{O}}^2$ again gives a contribution equal to that of its hermitian conjugate. As in (6.5.50) we separate a $D^2$ from $e^{V*} D^2 e^V$ and treat the rest as an interaction. The propagator is as before, but the vertex is now

$$\bar{D}^2 (e^{V*} D^2 e^V - D^2) \quad . \tag{6.5.59}$$

Note that, for *real* representations, $V\,{}^* = -\bar{V} = -V$, so this vertex is just $\bar{D}^2(\nabla^2 - D^2)$, and the rules of (6.5.52,53) can be obtained. In general, for a group containing factors for which $\Phi$ is in a real representation, we can write $V = V_1 + V_2$, where $V_1{}^* = -V_1$, but $V_2{}^* \neq -V_2$ ( $[V_1, V_2] = 0$), and write the vertex as

$$\bar{D}^2 (e^{V_2{}^*} \nabla_1{}^2 e^{V_2} - D^2) \quad , \quad \nabla_{1\alpha} = e^{-V_1} D_\alpha e^{V_1} = D_\alpha + \Gamma_{1\alpha} \quad . \tag{6.5.60}$$

Then $V_1$ appears in the rules only as $\Gamma_{1A}$ while $V_2$ appears explicitly.

The net result is that the effective action is expressed manifestly in terms of $\nabla_A$ for Yang-Mills factors that occur coupled only to real representations, and *always* for higher-loop contributions. However, at one loop, and only for Yang-Mills factors coupled to complex representations, the contribution must be calculated in a way where the covariance is not manifest.

Our methods can also be applied to massive chiral superfields. In that case the term $\bar{j}\Box_+{}^{-1} j$ of (6.5.44) is replaced with

$$\bar{j}\, \frac{1}{\Box_+ - m^2}\, j + \tfrac{1}{2} [ j\, \frac{m \nabla^2}{\Box_+ (\Box_+ - m^2)}\, j + h.\, c.] \quad , \tag{6.5.61}$$

a direct covariantization of the result (6.3.13). In performing the doubling trick, we use $\Delta(m) = \Delta(-m)$. We replace the kinetic operator

$$\mathbf{O}(m) = \begin{pmatrix} m & \bar{\nabla}^2 \\ \nabla^2 & m \end{pmatrix} \tag{6.5.62}$$

by

$$\mathbf{O}(-m)\mathbf{O}(m) = \begin{pmatrix} \overline{\nabla}^2\nabla^2 - m^2 & 0 \\ 0 & \nabla^2\overline{\nabla}^2 - m^2 \end{pmatrix} \quad . \tag{6.5.63}$$

After making the corresponding replacements in (6.5.50), we obtain $-(\Box_0 - m^2)^{-1}$ for the propagator, while the vertices are the same as before.

We summarize the procedure for evaluating the effective action in the background field formalism: One-loop graphs with only external gauge field lines are obtained from the quadratic Lagrangian for $V$ in (6.5.39), and by evaluating $\Delta$ for each chiral superfield. Higher loops are obtained with vertices involving interactions of the quantum fields, either from the higher-order expansion of the $V$ Lagrangian, or from the perturbative evaluation of (6.5.44). The rules for loops with (some) external chiral lines follow from (6.5.44) and are the usual ones but with covariant propagators and vertices.

**d. Examples**

We now present some results. We begin by investigating the radiative generation of a Fayet-Iliopoulos term (4.3.3) for an abelian gauge field, and consider first the one-loop tadpole graph with a chiral field inside (Fig. 6.5.1).

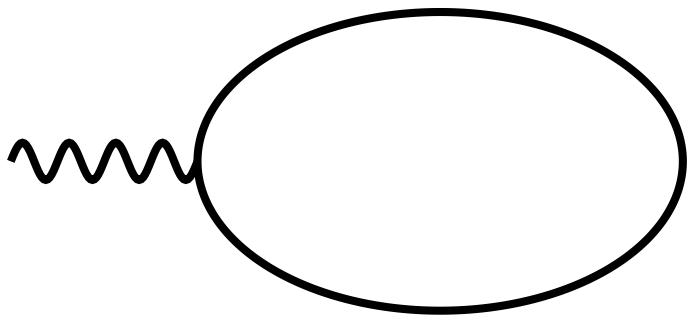

*Fig. 6.5.1*

If the chiral field is massless, we can drop it when using dimensional regularization. In the massive case, according to the usual rules, it would seem to contribute but gauge invariance requires that there be two chiral fields of opposite charges, and their contributions cancel. Therefore, gauge-invariant Pauli-Villars regulators cannot contribute to this graph either. As a result, the graph must be defined to vanish in the massless case in any gauge-invariant supersymmetric regularization procedure, dimensional or Pauli-Villars.

However, in the case of real representations, with the covariant rules, there is no need for such an argument. At the vertex we have a contribution (see (6.5.52); to linearized order we need not distinguish between full and background derivatives )

$$\bar{D}^2(\nabla^2 - D^2) = \bar{D}^2[-i\Gamma^\alpha D_\alpha - i(\frac{1}{2}D^\alpha\Gamma_\alpha)] \tag{6.5.64}$$

(to linearized order), and we do not have two $D$'s in the loop. Thus, for real representations, the graph vanishes just by $D$ algebra.

Actually, even this calculation is unnecessary, because we can give a simple proof that the Fayet-Iliopoulos term is never generated in perturbation theory for real representations: This term corresponds to a contribution to the effective action of the form $\int d^4x d^4\theta V$. However, according to our covariant Feynman rules, a $V$ never appears at a vertex, only connections and field strengths, so that no such term can be produced.

We next calculate the one-loop contribution to the $V$ self-energy from a *massive* chiral superfield in a real representation. If we use the ordinary non-background rules there are three graphs to compute (because we have massive $\Phi\Phi$ propagators), and they have to be combined to exhibit the gauge invariance of the final result. Also, there are some $D$ manipulations to be performed. Here, there is essentially nothing to do. We consider again the relevant graph, shown in Fig. 6.5.2.

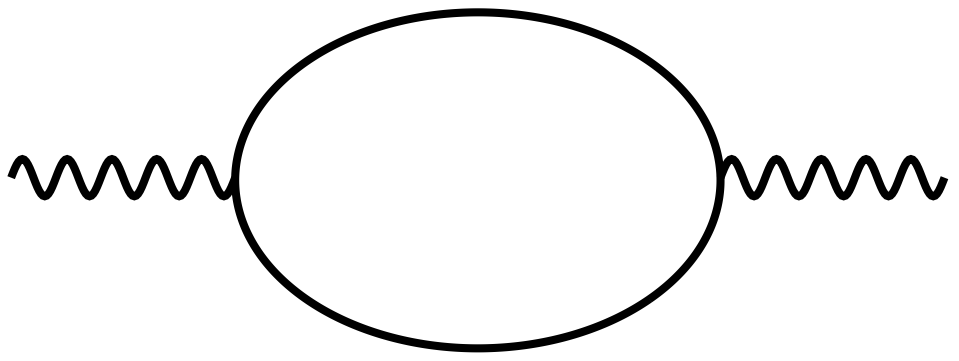

*Fig. 6.5.2*

One vertex is given by (6.5.64), while at the other vertex we have (again to linear order)

$$\Box_+ - \Box_0 = [-i\Gamma^{\underline{a}}\partial_{\underline{a}} - i(\frac{1}{2}\partial^{\underline{a}}\Gamma_{\underline{a}})] + [-iW^\alpha D_\alpha - i(\frac{1}{2}D^\alpha W_\alpha)] \quad . \tag{6.5.65}$$

Since we require two $D$'s and two $\bar{D}$'s in the loop, there is a unique term with contributions $-i\bar{D}^2\Gamma^\alpha D_\alpha$ from one vertex, and $-iW^\alpha D_\alpha$ from the other. The answer is

$$\frac{1}{4}k\, tr \int d^4\theta\ W^\alpha(p)\Gamma_\alpha(-p) \int \frac{d^4k}{(2\pi)^4} \frac{1}{(k^2+m^2)((k+p)^2+m^2)} \quad . \tag{6.5.66}$$

(The factor $k$ was defined in (6.4.5).) We observe that with the covariant rules there is no seagull-tadpole contribution. This would have to come from the nonlinear part of the one-vertex formula (6.5.64), but it does not have enough $D$'s to contribute.

We can obtain the $V$ self energy in nonabelian Yang-Mills theory without any calculation. According to the discussion following (6.5.39), if we look at graphs with two external lines, there are not enough $D$'s in the loop. The only source of $D$'s are the $W$-terms, and each factor of $W$ brings with it just one $D$. Thus the whole contribution to the self energy comes from the three chiral ghosts, and therefore we obtain an answer which is just $-3$ times that from the chiral field we considered above (with the fields now being background). This is a general feature: As already mentioned, $V$'s start contributing at one loop only beginning with the four-point function. To see the implications of this remark, we give now a computation of the one-loop, four-particle S-matrix in $N = 4$ Yang-Mills theory.

The one-loop contributions with external $V$ lines come from a $V$ loop, from the three chiral fields, or from the three chiral ghosts. Because of the statistics of the ghosts the chiral contributions cancel exactly. This is true for a graph with an arbitrary number of external vector lines. In particular, it implies that the two- and three-point functions are identically zero at the one-loop level. Therefore, we need only compute the $V$-loop contribution. We have just a box diagram, with factors $-i(\mathbf{W}^{\alpha} D_{\alpha} + \bar{\mathbf{W}}^{\dot{\alpha}} \bar{D}_{\dot{\alpha}})$ at each vertex, and we must keep terms with two $D$'s and two $\bar{D}$'s. The $D$-algebra is trivial, and we obtain for the four-$V$ amplitude

$$\Gamma = \frac{1}{2}\, tr \int \frac{d^4p_1 \ldots d^4p_4}{(2\pi)^{16}}\, d^4\theta\ (2\pi)^4 \delta(\Sigma p_i) G_o(p_1 \ldots p_{4)}$$

$$\times [\mathbf{W}^{\alpha}(p_1)\mathbf{W}_{\alpha}(p_2)\bar{\mathbf{W}}^{\dot{\alpha}}(p_3)\bar{\mathbf{W}}_{\dot{\alpha}}(p_4)$$

$$- \frac{1}{2}\mathbf{W}^{\alpha}(p_1)\bar{\mathbf{W}}^{\dot{\alpha}}(p_2)\mathbf{W}_{\alpha}(p_3)\bar{\mathbf{W}}_{\dot{\alpha}}(p_4)] \quad , \qquad (6.5.67)$$

where $G_0$ is the contribution from the four-point scalar box diagram

$$G_0 = \int \frac{d^4k}{(2\pi)^4} \frac{1}{k^2(k-p_1)^2(k-p_1-p_2)^2(k+p_4)^2} \quad . \qquad (6.5.68)$$

The trace is over internal symmetry indices, and all the superfields have the same $\theta$ argument.

This result is valid off-shell and is ultraviolet finite. On-shell it gives the one-loop S-matrix, but it is infrared divergent. (To obtain the S-matrix we drop the $p_i$ integrals and sum over $p_i$ permutations. The $W$'s give kinematical factors proportional to momenta and polarizations). The simplicity of the calculation is due in large part to the absence of chiral superfield contributions. In the particular gauge we are using there are no self-energy or triangle graphs to consider, and the whole S-matrix is given by the box graph. We will encounter a similar situation in supergravity (see sec. 7.8).

## 6.6. Regularization

### a. General

The perturbative renormalization of superfield theories is in principle no different from that of ordinary field theories. We need a procedure for regularizing divergent integrals, and a prescription for subtracting ultraviolet divergences. We must deal with renormalizable non-polynomial Lagrangians (e.g., supersymmetric Yang-Mills in a supersymmetric gauge), and use the supersymmetry Ward identities in the course of renormalization or, alternatively, use a regularization scheme that manifestly preserves supersymmetry. We do not have much to say about renormalization. For renormalizable models, we introduce renormalization constants in the classical action and use them to cancel, order by order in perturbation theory, the divergences we encounter.

As we have already mentioned, in supersymmetric theories there are fewer divergences present than in nonsupersymmetric ones. In general, the degree of divergence of any supergraph can be determined by the dimensional argument of sec. 6.3 or by the following power counting rules: In renormalizable theories all supersymmetric vertices have four $D$'s (either from the $D^2$ and $\bar{D}^2$ of chiral superfields, or the $D^\alpha$, $\bar{D}^2 D_\alpha$ of gauge superfields). In nonrenormalizable theories there are additional factors, but we first consider the renormalizable case. All vertices have a $d^4\theta$ factor.

We consider an $L$-loop graph with $V$ vertices, $P$ propagators of which $C$ are $\Phi\Phi$ or $\bar{\Phi}\bar{\Phi}$ massive chiral propagators, and $E$ external lines of which $E_c$ are chiral or antichiral. From the vertices there are $V$ factors of $D^2\bar{D}^2 \sim q^2$. The propagators produce $q^{-2}$ factors, but $\Phi\Phi$ or $\bar{\Phi}\bar{\Phi}$ propagators give an additional $D^2 q^{-2} \sim q^{-1}$ factor. Each loop produces a $d^4q \sim q^4$ and uses up a $D^2\bar{D}^2 \sim q^2$ factor from $\delta D^2\bar{D}^2\delta = \delta$. Each external chiral line accounts for one $\bar{D}^2 \sim q$ missing at the corresponding vertex. The *superficial* degree of divergence is (using $L - P + V = 1$)

$$\mathrm{D}_\infty = 4L - 2L - 2P + 2V - C - E_c = 2 - C - E_c$$

Therefore, for graphs with only external $V$'s the superficial degree of divergence is two (but gauge invariance improves this), and zero if there are two external chiral lines. Furthermore, if the external lines are all chiral an additional $D^2$ must come out of the loop: $\int d^4\theta\Phi^n = 0$ so one must have at least $\int d^4\theta\Phi^{n-1}D^2\Phi$ for a nonzero result. Therefore the convergence is improved and only graphs with one $\Phi$ and one $\overline{\Phi}$ line may be divergent. For renormalizable theories we obtain the results of sec. 6.3. In supergravity on the other hand, where at each vertex we have the equivalent of six factors of $D$, the degree of divergence of a graph is $2 - C - E_c + V$. This result can also be obtained by a dimensional argument (see sec. 7.7).

Regularization is an important part of any renormalization scheme. Although in principle any regularization may be used, in practice it is preferable to use a scheme that is computationally simple and maintains as many properties of the classical theory as possible. This simplifies the renormalization procedure, which must not only make the quantum theory finite by subtraction of divergences, but also must maintain unitarity by possible subtraction of additional finite quantities. For theories with (global or local) symmetries, such additional subtractions are determined by the requirement that renormalized Green functions satisfy Ward-Takahashi identities, and in the case of nonabelian gauge theories, Slavnov-Taylor identities. However, it is preferable to employ a regularization scheme that manifestly preserves all symmetries; this allows a renormalization scheme that requires the subtraction of only the divergent parts, so the application of Ward-Takahashi-Slavnov-Taylor identities is unnecessary.

**b. Dimensional reduction**

Dimensional regularization has proven to be the most practical method of regularization in component field theories because it has three properties: (1) It manifestly preserves (almost) all symmetries, thus bypassing the Ward-Takahashi or Slavnov-Taylor identities; (2) the regularized graphs are no harder to calculate than the unregularized ones and require only one regulator, the dimensionality of spacetime; (3) renormalization is a simple procedure, requiring only minimal subtraction. The prescription for dimensional regularization is: (1) Write the action in a form which is valid for any dimension D of spacetime; (2) calculate Feynman graphs formally in arbitrary spacetime dimensions, integrating over D components of each loop momentum, giving fields of any Lorentz representation the number of components appropriate to that value of D, and performing any algebraic manipulations that would be valid for finite integrals (i.e., performing the integral in dimensions D for which it is finite and analytically continuing in D); (3) renormalize by subtracting from divergent contributions (as D$\to$ 4) only their pole parts (proportional to $\frac{1}{D-4}$), and no additional finite parts, first in subdivergences and then for the superficial divergence (in amputated one-particle-irreducible graphs, i.e., the effective action).

This procedure has two drawbacks: (1) It must be supplemented by a prescription for handling symmetries which do not commute with parity, i.e., involving $\gamma_5$ or $\epsilon_{\underline{abcd}}$, and in particular for correctly obtaining chiral anomalies for those cases where they are present. (2) It does not maintain supersymmetry: The prescription for giving fields of any Lorentz representation the number of components appropriate to D dimensions does not keep Fermi and Bose degrees of freedom balanced. A modification of the prescription, which would continue a four-dimensional theory to a theory supersymmetric in D dimensions, is not possible either. For example, if D is increased past 10, a globally supersymmetric theory would have to be continued to a locally supersymmetric one. We would have spins $\geq 2$ because the number of supersymmetry generators increases with increasing D. We now describe a modification of dimensional regularization intended to preserve supersymmetry, and return later to the first difficulty.

Since the change in structure of supersymmetric theories as the number of supersymmetry generators ($4N$) is increased is not uniform, we consider keeping this number fixed. For regularizing ultraviolet divergences it is only necessary to continue to *lower* dimensions, and it is then possible to keep the number of supersymmetry generators fixed at their four-dimensional value. In general, our prescription for continuing to lower dimensions is to continue *only* the dimensionality of spacetime, but keep the range of all Lorentz indices the same, as if they were internal symmetry indices. As we reduce D, an $N$-extended supersymmetry can be reinterpreted as an $N'$-extended supersymmetry, $N' > N$. For example, $N = 1$ in D=4 dimensions can be regarded as $N = 2$ in D=3 dimensions. In this way, the number of bosonic and fermionic variables stay equal. Such a continuation is called “dimensional reduction”. Here we consider continuation only from D=4 to D<4.

Our rules for applying dimensional reduction to regularize component Feynman graphs are: (1) All indices on the fields, and corresponding matrices, coming from the action are treated as 4-dimensional indices; (2) as in ordinary dimensional regularization, all momentum integrals are integrated over D-component momenta, and all resulting Kronecker $\delta$'s are D-dimensional; (3) since D<4 always, any 4-dimensional Kronecker $\delta$ contracted with a momentum equals that momentum $\delta_{\underline{a}}{}^{\underline{b}} p_{\underline{b}} = p_{\underline{a}}$, and any 4-dimensional $\delta$ contracted with a D-dimensional one gives the D-dimensional one $\delta_{\underline{a}}{}^{\underline{b}} \hat{\delta}_{\underline{b}}{}^{\underline{c}} = \hat{\delta}_{\underline{a}}{}^{\underline{c}}$, where the " $\hat{}$ " indicates D-dimensional quantities. The first rule is necessary to preserve supersymmetry, since it keeps the number of components the same; the second rule preserves all the useful properties of dimensional regularization (e.g., gauge invariance); the last rule defines the regularization as dimensional reduction.

Unlike in ordinary dimensional regularization, both 4-dimensional and D-dimensional quantities occur. Therefore, when applied to components, dimensional reduction requires handling more types of fields: e.g., a 4-dimensional vector becomes a D-dimensional vector and 4-D scalars. This can cause difficulties in nonsupersymmetric theories, since a larger variety of divergences can occur, but in supersymmetric theories supersymmetry allows only divergences containing the full set of 4-dimensional fields. For example, $([A_{\underline{a}}, A_{\underline{b}}])^2 \rightarrow Z_A{}^2([\hat{A}_{\underline{a}}, \hat{A}_{\underline{b}}])^2 + Z_A Z_\phi([\hat{A}_{\underline{a}}, \phi_i])^2 + Z_\phi{}^2([\phi_i, \phi_j])^2$, but in supersymmetric theories the D-dimensional *extended* supersymmetry that results from reducing 4-dimensional supersymmetry ensures that $Z_A = Z_\phi$. (In theories with only scalars and spinors, the only difference from usual dimensional regularization is in the normalization of the spinor trace, and hence these problems do not arise.)

When applied to superfields, dimensional reduction is the unique form of dimensional regularization that allows the naive algebraic manipulation of the 4-dimensional spinor derivatives $D_\alpha$ in divergent as well as convergent supergraphs. This requirement leads to the following definition of regularization by dimensional reduction on supergraphs: (1) Perform all algebra as in D=4, obtaining a form where all $\theta$-integration has been performed i.e., the graph is expressed as an integral over a single $d^{4N}\theta$ of products of superfields of various momenta times an ordinary momentum integral *and is therefore manifestly supersymmetric;* (2) perform the remaining momentum integral in D-dimensions. In step (1), we use the 4-dimensional identity $p_{\alpha\dot{\beta}} p^{\beta\dot{\beta}} = \delta_\alpha{}^\beta p^2$ (recall $p^2 \equiv \frac{1}{2} p^{\underline{a}} p_{\underline{a}}$ (3.1.16,18)). No D-dimensional Kronecker deltas arise at this stage. In step (2), symmetric integrations generate D-dimensional Kronecker deltas, e.g.,

$$p_{\alpha\dot{\alpha}}\, p^{\beta\dot{\beta}} \rightarrow \frac{2}{\mathrm{D}} p^2\, \hat{\delta}_{\alpha\dot{\alpha}}{}^{\beta\dot{\beta}} \quad , \tag{6.6.1}$$

where $\hat{\delta}_{\alpha\dot{\alpha}}{}^{\beta\dot{\beta}} = \hat{\delta}_{\underline{a}}{}^{\underline{b}}$ has the properties

$$\hat{\delta}_{\alpha\dot{\alpha}}{}^{\beta\dot{\beta}}\delta_\beta{}^\alpha = \frac{\text{D}}{2}\delta_{\dot{\alpha}}{}^{\dot{\beta}} \quad , \quad \hat{\delta}_{\alpha\dot{\alpha}}{}^{\beta\dot{\beta}}\hat{\delta}_{\beta\dot{\beta}}{}^{\gamma\dot{\gamma}} = \hat{\delta}_{\alpha\dot{\alpha}}{}^{\gamma\dot{\gamma}} \quad , \tag{6.6.2}$$

and spinor indices are still manipulated as in 4 dimensions.

On the other hand, a dimensional regularization scheme, which like ordinary dimensional regularizations continued spinor indices (including the one on $\theta^\alpha$) to D-dimensional ones with $k = 2^{\frac{\text{D}}{2}-1}$ components, would have problems: e.g., $\int d^k\theta d^k\bar{\theta} = D^k\bar{D}^k$ would no longer have well defined statistics, and would introduce higher derivatives (for $k > 2$) into the action, requiring some nonminimal subtraction scheme (such as analytic regularization).

Unfortunately, although it preserves supersymmetry, regularization by dimensional reduction leads to ambiguities. For example, let us consider the expression

$$\hat{\delta}_{[\underline{a}}{}^{\underline{f}}p_{\underline{b}}q_{\underline{c}}r_{\underline{d}}s_{\underline{e}]} = 0 \tag{6.6.3}$$

which vanishes in D<5 because it is totally antisymmetric in 5 indices which take less than 5 values. (This also follows if we write the vector indices in terms of spinor indices and use 4-dimensional spinor manipulations.) If we now contract with $\hat{\delta}_{\underline{f}}{}^{\underline{a}}$ we obtain

$$(\text{D} - 4)p_{[\underline{a}}q_{\underline{b}}r_{\underline{c}}s_{\underline{d}]} = 0 \tag{6.6.4}$$

Since $p_{[\underline{a}}q_{\underline{b}}r_{\underline{c}}s_{\underline{d}]}$ does not vanish in D=4, and we must require it not to vanish in D≠4 to avoid generating arbitrary coefficients for such terms upon continuation, we have an inconsistency. This can also be viewed as an ambiguity: By evaluating a supergraph in two different ways, we may obtain results that differ by the left hand side of (6.6.4). If the supergraph is convergent, this ambiguity disappears in the limit D→4. However, if it is divergent, a finite difference between the two ways of evaluating the graph may result.

The same ambiguity is present in component theories (supersymmetric or otherwise) that have chiral anomalies, where the corresponding expression is

$$0 = \hat{\delta}_{\underline{f}}{}^{\underline{a}}[tr(\gamma_5\gamma^{\underline{f}}\gamma_{\underline{a}}\not{p}\not{q}\not{r}\not{s}) + tr(\gamma_5\gamma_{\underline{a}}\not{p}\not{q}\not{r}\not{s}\gamma^{\underline{f}})]$$

$$= (\text{D} - 4)tr(\gamma_5\not{p}\not{q}\not{r}\not{s}) \tag{6.6.5}$$

To derive this result, we have used the identities

$$\{\gamma_{\underline{a}}, \gamma_{\underline{b}}\} = 2\delta_{\underline{ab}} \quad , \quad \{\gamma_5, \gamma_{\underline{a}}\} = 0 \quad , \quad \hat{\delta}_{\underline{a}}{}^{\underline{a}} = \mathrm{D} \quad , \tag{6.6.6}$$

which are equivalent to the prescription (6.6.2) combined with the 4-dimensional spinor algebra.

This problem arises because we have required that our regularization respects local gauge invariance: When we consider theories with axial couplings, we must use a prescription such as (6.6.6) that respects chiral invariance. This makes it impossible to calculate (unambiguously) anomalies that should be there. Modifications of (6.6.6) exist that give the correct anomalies, but unfortunately, these also give spurious anomalies that must be eliminated by using Ward-Takahashi-Slavnov-Taylor identities, which is just what we were trying to avoid.

## c. Other methods

There do exist alternative schemes for supersymmetric regularization, at least for special systems. For theories that allow the introduction of mass terms, we can use supersymmetric Pauli-Villars regularization. This is the case, for example, in the Wess-Zumino model (or models with several chiral scalar superfields) where one can work with the regularized Lagrangian

$$I\!\!L_R = \int d^8z \, (\overline{\Phi}\Phi + \sum c_i \overline{\Phi}_i \Phi_i)$$

$$+ \int d^6z \, [\tfrac{1}{2}(m\Phi^2 + \sum M_i {\Phi^2}_i) + \tfrac{1}{3!}\lambda(\Phi + \sum \Phi_i)^3] + h.\,c. \quad , \tag{6.6.7}$$

or in models of chiral multiplets coupled to a Yang-Mills multiplet, *for regularizing chiral loops,*

$$I\!\!L_R = \int d^8z \, (\overline{\Phi} e^V \Phi + \sum c_i \overline{\Phi}_i e^V \Phi_i)$$

$$+ \tfrac{1}{2}\int d^6z \, (m\Phi^2 + \sum M_i {\Phi^2}_i) + h.\,c. \quad . \tag{6.6.8}$$

Here the $\Phi_i$ are chiral regulator fields, and the limit $M_i \to \infty$ is to be taken at the end of the calculations.

For supersymmetric Yang-Mills theories we can use higher derivative regularization. For example, the usual covariant action can be modified to read

$$tr \int d^4x\, d^2\theta\, \frac{1}{2} W^\alpha (1 + \xi^{-2}\Box_+) W_\alpha \quad , \tag{6.6.9}$$

and similar modifications can be made in the gauge fixing and ghost terms to produce propagators with $k^{-4}$ behavior for large $k$. As in ordinary Yang-Mills, all multiloop diagrams are superficially convergent. However this procedure must be supplemented by a different one-loop regularization.

Straightforward Pauli-Villars regularization cannot be used for Yang-Mills theories because it destroys gauge invariance. However, in the background field method it seems perfectly acceptable. In this method the effective action is manifestly covariant, and since the quantum fields transform covariantly (rotate like tensors), one can add a mass term, and therefore massive regulators, without destroying the gauge invariance. What is not entirely clear is that this can be done in general in a manifestly BRS invariant way, i.e., without destroying the unitarity of the S-matrix. But there seem to be no problems at the one-loop level, so that a combination of higher-derivative and one-loop Pauli-Villars regularization is a perfectly acceptable procedure for maintaining manifest supersymmetry in the background field formalism. A related procedure for one-loop graphs can be used even in a non-background formalism.

Another regularization procedure, which has been used for one-loop graphs, is point splitting. We first consider the nonsupersymmetric case. The regularization is applied to one-loop graphs by expressing them as traces of propagators in external fields, and separating the coincident end points of the propagator:

$$\int d^4x\, G(x\,, x) \rightarrow \int d^4x\, G(x\,, x+\epsilon) \quad , \tag{6.6.10}$$

where $\epsilon$ is an infinitesimal regulator. Writing the Green function $G$ as a functional average of fields,

$$G(x\,, y) = <\psi(x)\psi(y)> = \int I\!\!D\psi e^{S(\psi)} \psi(x)\psi(y) \quad , \tag{6.6.11}$$

we can express the point splitting in the following form in terms of an explicit operator:

$$G(x\,, x+\epsilon) = <\psi(x) e^{\epsilon\cdot\partial} \psi(x)> \quad . \qquad (6.6.12))$$

This procedure must be modified in order to preserve gauge invariances. However, for the form (6.6.12), gauge covariantization is trivial: We replace the partial derivative $\partial$ with a covariant one $\nabla = \partial - iA$:

$$< \psi(x) e^{\epsilon \cdot \nabla} \psi(x) > \quad . \tag{6.6.13}$$

This is equivalent to the form:

$$< \psi(x) I\!P \{ exp[-i \int_x^{x+\epsilon} dx' \cdot A(x')] \} \psi(x+\epsilon) > \quad , \tag{6.6.14}$$

where the line integral is along a straight line and $I\!P$ means path ordering. The equivalence can be proven, even for finite $\epsilon$, by writing the exponential in (6.6.13) as a product of exponentials of infinitesimals, and then reordering all the $\partial$'s to the right (which translates the $A$'s). In calculations, it is more convenient to have the manifestly covariant form (6.6.13) in terms of $\nabla$'s.

The supersymmetric generalization is straightforward: For $\nabla_{\underline{a}}$ use the superspace covariant derivative. (In principle one could also translate in $\theta$ with $\nabla_\alpha$, but the $\theta$ integration is already finite and doesn't need regularization.) The above equivalence to the path-ordered expression also holds in superspace.

While such regularization methods maintain supersymmetry, they are cumbersome. Some form of dimensional regularization is preferable, for all the reasons we gave earlier. As we have already discussed, at the supergraph level this amounts to doing first all the $D$-algebra in four dimensions, and then dimensionally continuing the momentum integrals. The results are manifestly supersymmetric, but presumably the inconsistencies we have discussed earlier will give rise to some ambiguities in the results.

## 6.7. Anomalies in Yang-Mills currents

As an example of our rules for chiral superfields and regularization methods, we calculate the supersymmetric version of the Adler-Bell-Jackiw anomaly. We consider a chiral multiplet coupled to both polar and axial vector gauge multiplets. The only physical component in the anomaly multiplet is the anomaly in the chiral symmetry current corresponding to phase transformations of the chiral superfields. This symmetry commutes with supersymmetry transformations and should be distinguished from R-symmetry, which does not commute with supersymmetry. The R-symmetry chiral anomaly appears in the multiplet of superconformal anomalies, which also includes the trace and supersymmetry current anomalies. We will discuss this in sec. 7.10.

We consider the action for scalar multiplets coupled to vector multiplets:

$$S = \int d^4x d^4\theta \, \bar{\Phi} e^V \Phi \quad . \tag{6.7.1}$$

For simplicity, we assume an even number of scalar multiplets, in pairs of opposite charge with respect to polar vector gauge fields. The two Weyl spinors in such a pair form a Dirac spinor, with the usual transformation under parity. The column vector $\Phi$ is thus in a real representation of the symmetries which the polar vectors gauge. We can also consider the coupling of axial vector gauge fields, with respect to which the two members of a pair have the same charge. The Dirac spinors of the pairs couple to these axial vectors with a $\gamma_5$. To indicate these two types of vectors, and the corresponding two types of vector multiplets, we separate $V$ into polar and axial parts:

$$V = V_+ + V_- \quad ; \quad V_+ = -\, V_+{}^* \quad , \quad V_- = \, V_-{}^* \quad . \tag{6.7.2}$$

The * refers to complex conjugation in the sense of (3.1.9) ($V^* = V^t$, since $V = V^\dagger$), but can refer to matrix complex conjugation if an appropriate representation is chosen. This is a special case of the situation discussed after (6.5.59). Since $\Phi$ is in a real representation of the group of $V_+$, the polar vector multiplet, we can use improved rules with respect to it, but must use the unimproved form (at one loop) for $V_-$, the axial vector multiplet.

We consider the one-loop graphs with two external polar vectors and one external axial vector. Depending on the group structure, the anomaly in the line of the axial vector may or may not cancel. If the axial anomaly is nonvanishing, axial gauge invariance is lost, and the axial vector cannot be considered physical. To be general, we consider the axial vector as merely a device for defining the appropriate axial current. We then evaluate the divergence of that current in terms of the polar vector fields appearing at the other two legs.

Varying the action with respect to any of the axial vector multiplets $V_-$, we obtain the axial current superfields

$$J_A = \bar{\Phi} T_A \Phi \quad , \tag{6.7.3}$$

where we have written $V_- = V_-{}^A T_A$. Gauge invariance of the action requires the on-shell conservation law $\bar{\nabla}^2 J_A = 0$ (as follows from substituting the transformation law $e^{V_-{}'} = e^{i\bar{\Lambda}} e^{V_-} e^{-i\Lambda}$ into the action and varying with respect to the chiral gauge parameter). Therefore we define the anomaly $A_A$ by

$$\bar{\nabla}^2 J_A = A_A \quad . \tag{6.7.4}$$

We will find that the anomaly $\bar{\nabla}^2 J$ is proportional to $W^2$. The component (axial) current is given by $j_{\alpha\dot{\alpha}} = \frac{1}{2}[\bar{\nabla}_{\dot{\alpha}}, \nabla_\alpha] J|$. Its divergence is therefore given by $\nabla^{\alpha\dot{\alpha}} j_{\alpha\dot{\alpha}} \sim [\nabla^2, \bar{\nabla}^2] J| \sim (\nabla^2 W^2 - \bar{\nabla}^2 \bar{W}^2)| \sim \epsilon^{\underline{abcd}} f_{\underline{ab}} f_{\underline{cd}}$, which is the familiar component result.

The anomaly can be calculated by evaluating the matrix element

$$\bar{\nabla}^2 < \bar{\Phi} T_A \Phi > \quad , \tag{6.7.5}$$

where $\Phi$ is now covariantly chiral with respect to $V_+$. In the calculations below we omit the group theory factor. We compute the matrix element $< \bar{\Phi}\Phi >$ and at the end we must take the trace of its product with $T_A$.

We will evaluate the anomaly by three methods: (1) the Adler-Rosenberg method, (2) with a Pauli-Villars regulator, and (3) with point-splitting regularization.

In the Adler-Rosenberg method we need only compute a triangle graph with one axial and two polar vectors at the vertices. Other, self-energy-type graphs, with one vector at one vertex and two at the other also contribute. However, their contribution merely covariantizes that from the triangle graph and therefore, by imposing gauge invariance, the full result can be extracted from this graph. We use the background-field formalism of sec. 6.5. At the axial vertex we have, from $< \bar{\Phi}\Phi >$ itself,

$$D^2\bar{D}^2 \ , \tag{6.7.6}$$

while at the other two we use the linearized expression (6.5.65) with *on-shell,* Landau-gauge polar vectors ($\partial^{\underline{a}}\Gamma_{\underline{a}} = D^\alpha W_\alpha = 0$):

$$-i(\Gamma^{\underline{a}}\partial_{\underline{a}} + W^\alpha D_\alpha) \ . \tag{6.7.7}$$

The supergraph is shown in fig. 6.7.1:

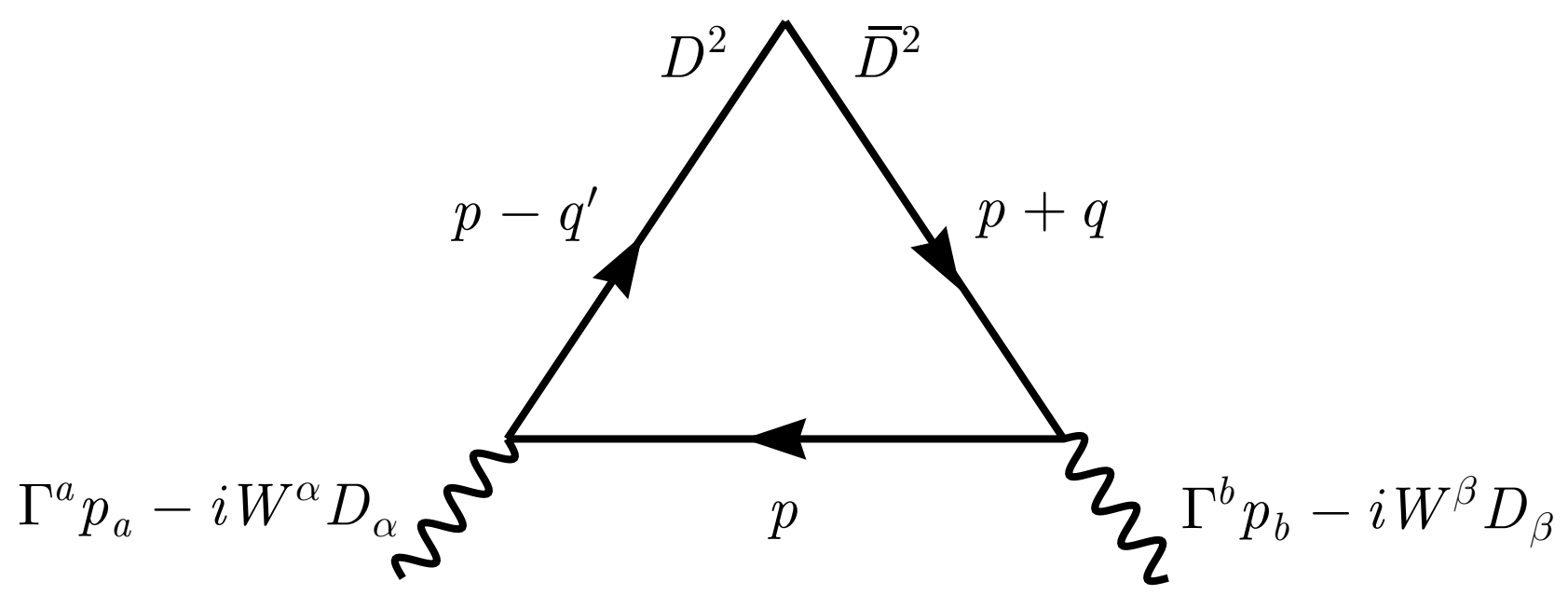

*Fig. 6.7.1*

It is easy to check, by integration by parts, that the $W^\alpha D_\alpha$, $W^\beta D_\beta$ terms do not contribute on shell. Therefore the $D$ manipulation is trivial and we must evaluate an ordinary graph, as in scalar QED, with 1 at one vertex, and $-i\Gamma^{\underline{a}}\partial_{\underline{a}}$ at the others. The Feynman integral is

$$\int \frac{d^4p}{(2\pi)^4} \frac{p_{\underline{a}}p_{\underline{b}}}{p^2(p-q)^2(p+q')^2} \Gamma^{\underline{a}}(q)\Gamma^{\underline{b}}(q') \ . \tag{6.7.8}$$

This directly gives the contribution to the matrix element $< \bar{\Phi}\Phi >$. (If considered as an ordinary triangle graph, with external vectors attached afterwards, the factor of 2 from functionally differentiating the two $\Gamma^{\underline{a}}$'s corresponds to this graph plus that with crossed vector lines.)

According to our supersymmetric dimensional regularization prescription the rest of the evaluation should be carried in D dimensions and the other graphs should be included. However, gauge invariance requires that $\Gamma_{\underline{a}}(q)$ enter the result in the form $F_{\underline{ab}} = -\, iq_{[\underline{a}}\Gamma_{\underline{b}]}$, and a term of this form can only be obtained from the triangle graph, by extracting from the integral the (finite) part proportional to $q'_{\underline{a}}q_{\underline{b}}$. After introducing Feynman parameters and shifting the loop momentum this part can be easily extracted, and we obtain for the complete contribution to $< \bar{\Phi}\Phi >$

$$-\frac{1}{4}\frac{1}{(4\pi)^2}\frac{1}{(q+q')^2}F^{\underline{ab}}(q)F_{\underline{ab}}(q') \qquad (6.7.9)$$

or, in $x$-space

$$< \bar{\Phi}\Phi > = \frac{1}{4}\frac{1}{(4\pi)^2}\frac{1}{\Box}(F^{\underline{ab}}F_{\underline{ab}}) \quad . \qquad (6.7.10)$$

Here

$$F_{\underline{ab}} = \partial_{[\underline{a}}\Gamma_{\underline{b}]} = \frac{1}{2}C_{\alpha\beta}\bar{\nabla}_{(\dot{\alpha}}\bar{W}_{\dot{\beta})} + \frac{1}{2}C_{\dot{\alpha}\dot{\beta}}\nabla_{(\alpha}W_{\beta)} \qquad (6.7.11)$$

and hence

$$F^{\underline{ab}}F_{\underline{ab}} = \frac{1}{2}(\nabla^{(\alpha}W^{\beta)})(\nabla_{(\alpha}W_{\beta)}) + h.c.$$

$$= -4\nabla^2 W^2 + h.c. \ , \qquad (6.7.12)$$

where we have used the field equations $\nabla_\alpha W^\alpha = \nabla^2 W^\alpha = 0$.

The anomaly is given by

$$\bar{\nabla}^2[\frac{1}{4}\frac{1}{(4\pi)^2}\frac{1}{\Box}F^{\underline{ab}}F_{\underline{ab}}] = -\frac{1}{(4\pi)^2}\frac{1}{\Box}\bar{\nabla}^2\nabla^2 W^2 = -\frac{1}{(4\pi)^2}W^2 \quad . \qquad (6.7.13)$$

This must be multiplied by the group generator $T_{\mathrm{A}}$ and a trace taken (with $W_\alpha = W_\alpha{}^{\mathrm{B}}T_{\mathrm{B}}$).

In the Pauli-Villars regularization method we compute

$$\lim_{m\to\infty}\bar{\nabla}^2(< \Phi\bar{\Phi} > - < \Phi_m\bar{\Phi}_m >) \qquad (6.7.14)$$

where $\Phi_m$ is a massive regulator field. In this regularized expression we can use the equations of motion

$$\bar{\nabla}^2\bar{\Phi} = 0 \ , \qquad \bar{\nabla}^2\bar{\Phi}_m = m\Phi_m \qquad (6.7.15)$$

so that the relevant quantity to compute is

$$-\lim_{m\to\infty} m < \Phi_m\Phi_m > \qquad (6.7.16)$$

Using (6.5.44,61) we have

$$< \Phi_m \Phi_m > = \frac{\delta}{\delta j}\frac{\delta}{\delta j} W(j)$$

$$= \overline{\nabla}^2 \frac{-m\nabla^2}{\Box_+(\Box_+ - m^2)} \overline{\nabla}^2 = \frac{-m\overline{\nabla}^2}{\Box_+ - m^2} \quad . \tag{6.7.17}$$

We must therefore compute a loop graph, with $\overline{\nabla}^2$ replaced by $\bar{D}^2$ (this is the only source of $\bar{D}$'s), and $(\Box_+ - m^2)^{-1}$ expanded in powers of the background field (we need at least two $D$'s):

$$\frac{1}{\Box_+ - m^2} \simeq \frac{1}{\Box_0 - m^2}(-iW^\alpha D_\alpha)\frac{1}{\Box_0 - m^2}(-iW^\beta D_\beta)\frac{1}{\Box_0 - m^2} + \cdots \tag{6.7.18}$$

The only nonzero contribution in the $m \to \infty$ limit comes from the term explicitly written. In momentum space it corresponds to a triangle graph with $\bar{D}^2$ at one vertex and $W^\alpha D_\alpha$, $W^\beta D_\beta$ at the other two. The anomaly is therefore given by

$$- \lim_{m\to\infty} m^2 \int \frac{d^4p}{(2\pi)^4} \frac{1}{[p^2+m^2][(p-q)^2+m^2][(p+q')^2+m^2]}(2W^2)$$

$$= - \frac{1}{(4\pi)^2} W^2 \tag{6.7.19}$$

as before.

In the point-splitting method we compute

$$\bar{D}^2 < \Phi e^{\epsilon\cdot\nabla}\overline{\Phi} > = < \Phi\overline{\nabla}^2 e^{\epsilon\cdot\nabla}\overline{\Phi} >$$

$$= < \Phi e^{\epsilon\cdot\nabla}(e^{-\epsilon\cdot\nabla}\overline{\nabla}e^{\epsilon\cdot\nabla})^2\overline{\Phi} > \quad . \tag{6.7.20}$$

Using the commutation relations (4.2.90) of the covariant derivatives, we find

$$e^{-\epsilon\cdot\nabla}\overline{\nabla}_{\dot\alpha}e^{\epsilon\cdot\nabla} = \overline{\nabla}_{\dot\alpha} - \epsilon^\alpha{}_{\dot\alpha}W_\alpha + \frac{1}{2}\epsilon^\alpha{}_{\dot\alpha}\epsilon^{\beta\dot\beta}(\nabla_{\beta\dot\beta}W_\alpha) + O(\epsilon^3) \quad . \tag{6.7.21}$$

We then expand the remaining $e^{\epsilon\cdot\nabla} = e^{\epsilon\cdot\partial}[1 - i\epsilon\cdot\Gamma + O(\epsilon^2)]$, express everything in terms of $< \Phi(x)e^{\epsilon\cdot\partial}\overline{\Phi}(x) > = < \Phi(x)\overline{\Phi}(x+\epsilon) >$, and take the limit $\epsilon \to 0$.

However, we can limit the number of terms we need consider by evaluating $<\Phi(x)\bar{\Phi}(x+\epsilon)>$ first. The only terms which are divergent in the limit $\epsilon \to 0$ (from power counting) are given by the tadpole and propagator graphs, as shown in fig. 6.7.2:

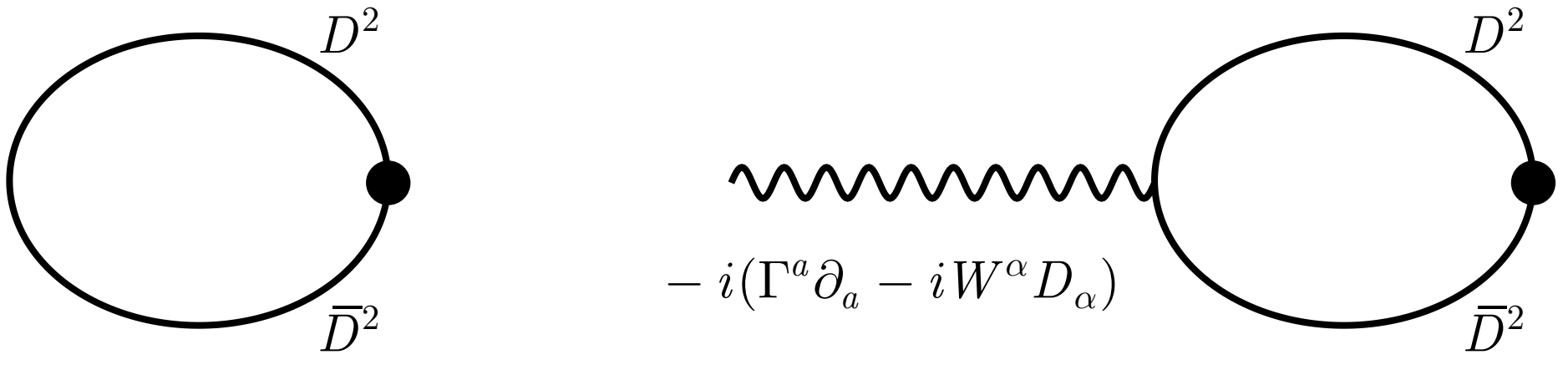

*Fig. 6.7.2*

(As before, we have two factors $\bar{D}^2$ and $D^2$ from $\Phi$ and $\bar{\Phi}$, and $\Box_+ - \Box_0 \simeq -i(\Gamma^{\underline{a}}\partial_{\underline{a}} + W^\alpha D_\alpha)$.) The $W^\alpha D_\alpha$ term does not contribute. The graphs are evaluated as

$$\int \frac{d^4p}{(2\pi)^4}\, e^{-i\epsilon\cdot p}\, \frac{1}{p^2} = \frac{1}{(4\pi)^2}\frac{1}{\epsilon^2} \quad ,$$

$$-\Gamma^{\underline{a}} \int \frac{d^4p}{(2\pi)^4}\, e^{-i\epsilon\cdot p}\, \frac{p_{\underline{a}}}{p^2(p+k)^2} = \frac{1}{(4\pi)^2}\frac{1}{\epsilon^2}\, i\epsilon\cdot\Gamma \qquad (6.7.22)$$

so that

$$<\Phi(x)\bar{\Phi}(x+\epsilon)> = \frac{1}{(4\pi)^2}\frac{1}{\epsilon^2}\,[1 + i\epsilon\cdot\Gamma + O(\epsilon^2)] \quad . \qquad (6.7.23)$$

We thus keep factors multiplying $<\Phi\bar{\Phi}>$ only to $O(\epsilon^2)$, and also drop factors $O(\epsilon^2)$ which have a $\bar{D}_{\dot{\alpha}}$ acting on $<\Phi\bar{\Phi}>$. The result is

$$\bar{D}^2 <\Phi e^{\epsilon\cdot\nabla}\bar{\Phi}> \simeq (-\epsilon^{\alpha\dot{\alpha}}W_\alpha\bar{D}_{\dot{\alpha}} - \frac{1}{2}\epsilon^{\alpha\dot{\alpha}}W_\alpha\epsilon_{\beta\dot{\alpha}}W^\beta) <\Phi e^{\epsilon\cdot\partial}\bar{\Phi}>$$

$$\simeq (-\epsilon^{\alpha\dot{\alpha}}W_\alpha\bar{D}_{\dot{\alpha}} + \epsilon^2W^2)\,\frac{1}{(4\pi)^2}\frac{1}{\epsilon^2}\,(1 + i\epsilon\cdot\Gamma)$$

$$\simeq \frac{1}{(4\pi)^2}\,(-i\,\frac{1}{\epsilon^2}\,\epsilon^{\alpha\dot{\alpha}}\epsilon^{\beta\dot{\beta}}W_\alpha\bar{D}_{\dot{\alpha}}\Gamma_{\beta\dot{\beta}} + W^2) \quad . \qquad (6.7.24)$$

(Note in particular that only the $e^{\epsilon\cdot\partial}$ part of the remaining $e^{\epsilon\cdot\nabla}$ contributes, and only up to $O(\epsilon)$ in $e^{-\epsilon\cdot\nabla}\overline{\nabla}_{\dot{\alpha}}e^{\epsilon\cdot\nabla}$.)

Using the chiral representation relation

$$\bar{D}_{\dot{\alpha}}\Gamma_{\beta\dot{\beta}} = iC_{\dot{\alpha}\dot{\beta}}W_{\beta} \quad , \tag{6.7.25}$$

we obtain the same result as by the previous two methods. (Note that in ordinary QED the calculation is slightly simpler because the point-split propagator goes only as $\epsilon^{-1}$.)

At higher loops, and also at one loop for real representations, our covariant Feynman rules apply. Consequently the triangle graph contribution to the effective action depends on the connections and field strengths and not on the gauge fields themselves and, by simple power counting, it is therefore superficially convergent. We draw two conclusions: There are no one-loop chiral anomalies for the Yang-Mills multiplet itself (the chiral ghosts are in a real representation of the group), and there are *no higher-loop chiral anomalies* for any multiplet: For the chiral current defined by (6.7.3) the Adler-Bardeen theorem holds.

The Adler-Bardeen theorem is not in conflict with the existence of higher-order contributions to the $\beta$-function. As we mentioned at the beginning of this section, the chiral current that is in the same multiplet with the energy-momentum tensor is not the one we have discussed here, but the R-symmetry axial current. It is a member of the *supercurrent* defined by coupling to supergravity, and in general its anomaly does receive higher-order radiative corrections as do the anomalies of its supersymmetric partners.

**Contents of 7. QUANTUM N=1 SUPERGRAVITY**

# 7. QUANTUM N=1 SUPERGRAVITY

## 7.1. Introduction

The quantization of superfield supergravity presents a number of new features and complications that we discuss in this chapter. Once the gauge-fixing procedure and ghost structure have been determined Feynman rules can be obtained. Useful as supergraphs are in global supersymmetry, their power is awesome when it comes to doing perturbation theory calculations in supergravity. Much of the simplicity of superfield calculations in supergravity, as compared with component calculations, occurs because, as in global supersymmetry, we deal with objects having fewer Lorentz indices. The supergravity superfield is a Lorentz vector, as compared to the Lorentz second-rank tensor and vector-spinor of component supergravity. Consequently the interaction Lagrangians have fewer terms, and the tensor algebra is much simpler. For example, the three-graviton vertex contains 171 terms, while the corresponding three-vertex in supergravity consists of only 27. As a result, it is possible to do calculations in superfield supergravity that have not even been attempted in ordinary quantum gravity or component supergravity.

The investigation of the divergence structure of quantum supergravity is also very much facilitated by the use of superfields. Many cancellations due to supersymmetry happen automatically, and it is much easier to list and understand the possible counterterms. In component calculations, with non-supersymmetric gauge-fixing terms for the graviton and gravitino, the corresponding cancellations do not occur automatically, and it is much more difficult to determine what infinities might be present or absent.

Background field methods play a crucial role here. Since the calculations are never very easy, and the algebra does get complicated, it is essential to keep some control of the gauge invariance of the theory, and this is best accomplished by working in the manifestly gauge-invariant background field formalism. In particular, we have the usual property that all divergences are gauge invariant: The formalism avoids the noncovariant divergences of gravitational theories quantized in nonbackground gauges. We shall see that, just as in Yang-Mills theory, the background field quantization has the further virtue of simplifying some of the vertices. It also allows us to work with only background covariant derivatives rather than prepotentials, and consequently leads to some improvement in the power counting rules for the theory.

The new feature of the quantization procedure is the appearance of large numbers and new types of ghosts, besides the Faddeev-Popov and Nielsen-Kallosh ghosts. They arise either because of certain constraints that the gauge-fixing functions satisfy, or are introduced to remove certain nonlocalities that have been produced by the gauge-fixing procedure. In addition, we find numerous ghosts-for-ghosts.

We discuss first the background field quantization procedure. Ordinary quantization Feynman rules can easily be obtained from the ones we derive by setting the background fields to zero, but in pure supergravity there is little advantage to using them: In general, $L$-loop calculations in ordinary field theory present about the same level of difficulty as $L+1$-loop calculations in the background field method. In the following sections we discuss the background-quantum splitting, which we pattern after the one in Yang-Mills theory, the number and kinds of ghosts one may encounter, and the choice of gauge-fixing function. Once we have the Lagrangian, we can discuss general properties of the effective action, derive supergraph rules, and do loop calculations.

We consider only $n = -\frac{1}{3}$ supergravity. In sec. 7.10.e we shall argue that $N = 1$, $n \neq -\frac{1}{3}$ theories are inconsistent at the quantum level due to anomalies in the Ward identities of local supersymmetry (except when part of an extended theory).

## 7.2. Background-quantum splitting

We divide the discussion in two parts: the splitting itself, and the expansion of the action. As in the Yang-Mills case, the splitting into quantum and background fields is nonlinear and done in terms of exponentials. The simplest way to understand it is as an expansion of the (constrained) covariant derivatives in terms of unconstrained quantum prepotentials (needed for quantization) and constrained background derivatives; the simplest way to obtain it is by re-solving the constraints, as in sec. 5.3., but using background covariant instead of flat superspace derivatives. Except for some small modifications explained below, the results can be written almost immediately.

The expansion of the action is algebraically lengthy, but straightforward. We give the part quadratic in the quantum fields, but the procedure can be extended for finding higher order terms.

### a. Formalism

The quantum-background splitting in supergravity follows a pattern very similar to that of Yang-Mills, and we simply repeat it here, referring the reader back to sec. 6.5 for motivation and an explanation of the procedure. We start with the conventional derivatives (with degauged $U(1)$; see sec. 5.3.b.8)

$$\nabla_A = E_A{}^M D_M + (\Phi_{A\beta}{}^\gamma M_\gamma{}^\beta + \Phi_{A\dot\beta}{}^{\dot\gamma} \bar M_{\dot\gamma}{}^{\dot\beta}) \quad ,$$

$$[\nabla_A, \nabla_B\} = T_{AB}{}^C \nabla_C + (R_{AB\gamma}{}^\delta M_\delta{}^\gamma + R_{AB\dot\gamma}{}^{\dot\delta} \bar M_{\dot\delta}{}^{\dot\gamma}) \quad ; \tag{7.2.1}$$

covariant under the (vector-representation) transformations:

$$\nabla_A{}' = e^{iK} \nabla_A e^{-iK} \ , \ \ K = \bar K \quad ;$$

$$K = K^M iD_M + (K_\alpha{}^\beta iM_\beta{}^\alpha + \bar K_{\dot\alpha}{}^{\dot\beta} i\bar M_{\dot\beta}{}^{\dot\alpha}) \quad . \tag{7.2.2}$$

The solution to the constraints expresses the derivatives in terms of the unconstrained prepotentials $H_{\alpha\dot\alpha}$ and $\phi$ and ordinary flat-space derivatives $D_M$, in the chiral representation, as in sec. 5.3. For the time being we use a chiral density compensator. We achieve our quantum-background splitting by substituting into the solution background covariant derivatives for the flat derivatives. The fields $H_{\alpha\dot\alpha}$ and $\phi$ are the quantum fields, while the background fields appear implicitly in the background derivatives.

We take the background covariant derivatives $\boldsymbol{\nabla}_A$ in the vector representation, $\overline{\boldsymbol{\nabla}}_{\dot\alpha} = (\boldsymbol{\nabla}_\alpha)^\dagger$, $i\boldsymbol{\nabla}_{\underline{a}} = (i\boldsymbol{\nabla}_{\underline{a}})^\dagger$, and require them to satisfy the same constraints as $\nabla_A$. This immediately allows us to solve the representation preserving constraints:

$$\{\nabla_\alpha , \nabla_\beta\} = T_{\alpha\beta}{}^\gamma \nabla_\gamma + R_{\alpha\beta}(M) \tag{7.2.3}$$

and its hermitian conjugate are obviously satisfied by (cf. sec. 5.3.b.2)

$$\nabla_{\dot\alpha} = \Psi(\overline{\boldsymbol{\nabla}}_{\dot\alpha} + \bar\omega_{\dot\alpha\beta}{}^\gamma M_\gamma{}^\beta + \bar\omega_{\dot\alpha\dot\beta}{}^{\dot\gamma} \bar M_{\dot\gamma}{}^{\dot\beta}) \quad , \tag{7.2.4a}$$

$$\nabla_\alpha = e^{-H}\overline\Psi(\boldsymbol{\nabla}_\alpha + \omega_{\alpha\beta}{}^\gamma M_\gamma{}^\beta + \omega_{\alpha\dot\beta}{}^{\dot\gamma} \bar M_{\dot\gamma}{}^{\dot\beta})e^H \quad , \tag{7.2.4b}$$

where $H = H^A i\boldsymbol{\nabla}_A$ introduces the quantum field $H^A$ which also appears in $\Psi$ and the quantum connection $\omega$. We have written the derivatives in a chiral representation with respect to $H^A$. Replacing $e^H = e^\Omega e^{\overline\Omega}$ and multiplying all quantities by $e^{\overline\Omega}$ from the left and $e^{-\overline\Omega}$ from the right would take us to a quantum vector representation. However, for quantization, it is simpler to work with $H$.

It is also convenient to define *background covariant* hatted objects as in (5.2.23), but from background covariant derivatives and $H$:

$$\widehat{\overline\nabla}_{\dot\alpha} = \overline{\boldsymbol{\nabla}}_{\dot\alpha} \quad , \quad \widehat\nabla_\alpha = e^{-H}\boldsymbol{\nabla}_\alpha e^H \quad , \quad \widehat\nabla_{\alpha\dot\alpha} = -\, i\{\widehat\nabla_\alpha, \widehat{\overline\nabla}_{\dot\alpha}\} \quad ; \tag{7.2.5}$$

$$\widehat\nabla_A = \hat E_A{}^B \boldsymbol{\nabla}_B + (\hat\Phi_{A\beta}{}^\gamma M_\gamma{}^\beta + \hat\Phi_{A\dot\beta}{}^{\dot\gamma}\bar M_{\dot\gamma}{}^{\dot\beta}) = -\,(-1)^A e^{-H}\overline{(\widehat\nabla_A)}e^H \quad ; \tag{7.2.6}$$

and define $\hat T_{AB}{}^C$ and $\hat R_{AB}$ in terms of them. We also define the superdeterminants, with their appropriate hermiticity conditions:

$$E = sdet\ E_A{}^M \quad , \quad \mathbf{E} = sdet\ \mathbf{E}_A{}^M \quad , \quad \hat E = sdet\ \hat E_A{}^B \quad ;$$

$$E^{-1} = (E^{-1})^\dagger e^{-\overleftarrow{H}} \quad , \quad \mathbf{E}^{-1} = (\mathbf{E}^{-1})^\dagger \quad , \quad \hat E^{-1}\mathbf{E}^{-1} = (\hat E^{-1})^\dagger \mathbf{E}^{-1} e^{-\overleftarrow{H}} \quad . \tag{7.2.7}$$

We have used the identity, for any function $f$,

$$f\mathbf{E}^{-1}\overleftarrow{H}\,\mathbf{E} = Hf + fi(-1)^A\boldsymbol{\nabla}_A H^A \tag{7.2.8}$$

(dropping the term $iH^A(\mathbf{E}^{-1}\overleftarrow{\boldsymbol{\nabla}}_A\mathbf{E}) = -\,iH^A(-1)^B\boldsymbol{T}_{AB}{}^B = 0$ (see (5.3.42)). The

background covariantization of the operator $e^{-\overleftarrow{H}}$ is thus $e^{-\mathbf{E}^{-1}\overleftarrow{H}\mathbf{E}} = \mathbf{E}^{-1}e^{-\overleftarrow{H}}\mathbf{E}$. It results in the hermiticity conditions (from (5.3.51b) and (7.2.9))

$$(1\cdot e^{-\mathbf{E}^{-1}\overleftarrow{H}\mathbf{E}})^{-1} = e^{-H}(1\cdot e^{\mathbf{E}^{-1}\overleftarrow{H}\mathbf{E}})^{\dagger} \quad , \quad \hat{E}^{-1} = (\hat{E}^{-1})^{\dagger}e^{-\mathbf{E}^{-1}\overleftarrow{H}\mathbf{E}} \quad . \tag{7.2.9}$$

We use a conventional constraint to determine the vector covariant derivative

$$\nabla_{\alpha\dot{\alpha}} = -\, i\{\nabla_{\alpha}, \overline{\nabla}_{\dot{\alpha}}\} \quad , \tag{7.2.10}$$

and conventional constraints $T_{\dot{\alpha}\dot{\beta}}{}^{\dot{\gamma}} = 0$ (or equivalently $T_{\dot{\alpha}\beta(\dot{\beta}}{}^{\beta\dot{\gamma})} = 0$) and $T_{\dot{\alpha}(\beta\dot{\beta}}{}^{\gamma)\dot{\beta}} = 0$ to determine the spinor connections:

$$\bar{\omega}_{\dot{\alpha}\dot{\beta}}{}^{\dot{\gamma}} = -\,\delta_{\dot{\alpha}}{}^{(\dot{\gamma}}\overline{\boldsymbol{\nabla}}_{\dot{\beta})} ln\Psi \equiv -\,(\omega_{\alpha\beta}{}^{\gamma})^{\dagger} \quad ,$$

$$\bar{\omega}_{\dot{\alpha}\beta}{}^{\gamma} = -\,\frac{1}{2}\hat{T}_{\dot{\alpha},(\beta\dot{\delta}}{}^{\gamma)\dot{\delta}} \equiv -\,(\omega_{\alpha\dot{\beta}}{}^{\dot{\gamma}})^{\dagger} \quad ; \tag{7.2.11}$$

(compare to (5.3.55) and (5.3.25) after degauging).

Finally, we impose the $n = -\,\frac{1}{3}$ conformal-breaking constraint, which determines $\Psi$ by a procedure similar to that of sec 5.3:

$$\Psi = \phi^{-1}(e^{-H}\bar{\phi})^{\frac{1}{2}}(1\cdot e^{-\mathbf{E}^{-1}\overleftarrow{H}\mathbf{E}})^{\frac{1}{6}}\hat{E}^{-\frac{1}{6}} \quad , \tag{7.2.12}$$

where $\phi$ is background covariantly chiral: $\overline{\boldsymbol{\nabla}}_{\dot{\alpha}}\phi = \boldsymbol{\nabla}_{\alpha}\bar{\phi} = 0$. For quantum calculations, we have to either express $\phi$ explicitly in terms of an ordinary chiral superfield and the background gauge field or, what is preferable in general, derive covariant Feynman rules that allow us to work with it directly.

The full derivatives $\nabla_A$ transform covariantly under two sets of transformations:

(a) Background transformations:

$$\boldsymbol{\nabla}_A{}' = e^{iK}\boldsymbol{\nabla}_A e^{-iK} \quad ,$$

$$H' = e^{iK}He^{-iK} \quad , \quad \phi' = e^{iK}\phi e^{-iK} \quad , \quad \omega'(M) = e^{iK}\omega(M)e^{-iK} \quad ,$$

$$\nabla_A{}' = e^{iK}\nabla_A e^{-iK} \quad ,$$

$$K = \bar{K} = K^A i\boldsymbol{\nabla}_A + (K_{\alpha}{}^{\beta}iM_{\beta}{}^{\alpha} + \bar{K}_{\dot{\alpha}}{}^{\dot{\beta}}i\bar{M}_{\dot{\beta}}{}^{\dot{\alpha}}) \quad . \tag{7.2.13}$$

These transformations follow from the requirement that the full derivative $\nabla_A$ transform covariantly and $\phi$ be background covariantly chiral: From (7.2.4a) or (7.2.6,12) it follows that $\Psi$ and $\omega$ transform covariantly, and therefore, from (7.2.4b) $H$ transforms covariantly.

(b) Quantum transformations:

$$\boldsymbol{\nabla}_A{}' = \boldsymbol{\nabla}_A \quad ; \tag{7.2.14a}$$

$$e^{H'} = e^{i\bar{\Lambda}} e^{H} e^{-i\Lambda} e^{X(\Lambda)} \quad ,$$

$$\phi' = e^{i\Lambda - \frac{1}{3}(\boldsymbol{\nabla}_{\underline{a}}\Lambda^{\underline{a}} - \boldsymbol{\nabla}_{\alpha}\Lambda^{\alpha} - i\mathbf{G}_{\underline{a}}\Lambda^{\underline{a}})} \phi \quad , \tag{7.2.14b}$$

$$\nabla_A{}' = L_A{}^B(\Lambda) e^{i\Lambda} \nabla_B e^{-i\Lambda} \quad ; \tag{7.2.14c}$$

where

$$\Lambda = \Lambda^A i\boldsymbol{\nabla}_A \neq \bar{\Lambda} \quad , \quad [\bar{\boldsymbol{\nabla}}_{\dot{\alpha}}\, , \Lambda]\eta = 0 \quad ; \tag{7.2.14d}$$

for any background covariantly chiral $\eta$ ($\bar{\boldsymbol{\nabla}}_{\dot{\alpha}}\eta = 0$). We have taken the standard chiral representation transformations (5.2.16,67) for the quantum superfields $H$ and $\phi$ except for certain modifications required because of the $\boldsymbol{\nabla}_A$'s in $H$ and $\Lambda$. Since

$$[H\, , \Lambda] = (H\Lambda^A - \Lambda H^A) i\boldsymbol{\nabla}_A - \Lambda^B H^A (\mathbf{T}_{AB}{}^C \boldsymbol{\nabla}_C + \mathbf{R}_{AB}(M)) \quad , \tag{7.2.15}$$

$e^{i\bar{\Lambda}} e^H e^{-i\Lambda}$ generates Lorentz transformation terms. We introduce $X(\Lambda) = X_\alpha{}^\beta M_\beta{}^\alpha + X_{\dot{\alpha}}{}^{\dot{\beta}} \bar{M}_{\dot{\beta}}{}^{\dot{\alpha}}$ to cancel them. Similarly, the usual transformation law of $\phi$ has the additional $\mathbf{G}_{\underline{a}}\Lambda^{\underline{a}}$ term because we require $\phi'$ to be background chiral and $\bar{\boldsymbol{\nabla}}_{\dot{\beta}}(\boldsymbol{\nabla}_{\underline{a}}\Lambda^{\underline{a}} - \boldsymbol{\nabla}_\alpha\Lambda^\alpha - i\mathbf{G}_{\underline{a}}\Lambda^{\underline{a}}) = 0$. The transformation (7.2.14c) of $\nabla_A$ follows from (7.2.14a,b). The $\Lambda$-dependent Lorentz transformation $L_A{}^B(\Lambda)$ corresponds to the $\omega_A{}^B$ term in (5.2.21) with additional contributions from $X(\Lambda)$.

In practice, we never need to compute $X$ explicitly. Using (7.2.15) and the Baker-Hausdorff theorem,

$$e^{i\bar{\Lambda}} e^H e^{-i\Lambda} = e^{H' + Y(M)} \tag{7.2.16}$$

for some Lorentz transformation $Y(M)$. Since $H'$ is a scalar operator,

$$[H' , Y(M)] = Y'(M) \tag{7.2.17a}$$

for some Lorentz rotation $Y'$ and hence

$$e^{H' + Y} = e^{H'} e^{-X} \quad , \tag{7.2.17b}$$

thus defining $X$. Introducing $X$ into (7.2.14b) is equivalent to the prescription of dropping at any stage of the calculation Lorentz terms *other* than those implicit in $\boldsymbol{\nabla}_A$. For convenience, we introduce the operation $< >$ that removes explicit Lorentz generators as follows: For any

$$A = A^A \boldsymbol{\nabla}_A + (A_\alpha{}^\beta M_\beta{}^\alpha + A_{\dot\alpha}{}^{\dot\beta} \bar{M}_{\dot\beta}{}^{\dot\alpha}) \quad , \tag{7.2.18a}$$

we define

$$< A > \equiv A^A \boldsymbol{\nabla}_A \ , \ < e^A > \equiv e^{<A>} \quad . \tag{7.2.18b}$$

The transformation law for $H$ can be rewritten

$$e^{H'} = < e^{i\bar\Lambda} e^H e^{-i\Lambda} > \quad . \tag{7.2.19}$$

The quantum-background splitting we have described is equivalent to the following splitting of the prepotential in terms of a quantum chiral $H$ and background vector $\boldsymbol{\Omega}$:

$$e^{H(split)} = < e^{\boldsymbol{\Omega}} e^H e^{\bar{\boldsymbol{\Omega}}} > \quad , \tag{7.2.20}$$

which is analogous to the Yang-Mills case (6.5.25). The usual chiral representation transformation law

$$< e^{\boldsymbol{\Omega}} e^H e^{\bar{\boldsymbol{\Omega}}} > ' = e^{i\bar\Lambda_0} < e^{\boldsymbol{\Omega}} e^H e^{\bar{\boldsymbol{\Omega}}} > e^{-i\Lambda_0} \tag{7.2.21}$$

can be rewritten as either background (7.2.13) or quantum (7.2.14) transformations analogous to those of (6.5.27).

As in the nonbackground case, the quantum transformations must preserve chirality (7.2.14d). Therefore, $\Lambda$ takes the following form, expressing it in terms of the unconstrained supergravity gauge parameter $L^\alpha$ (cf. (5.2.14)):

$$\Lambda_{\alpha\dot\alpha} = - i \bar{\boldsymbol{\nabla}}_{\dot\alpha} \phi^{-3} L_\alpha \ , \ \Lambda_\alpha = \bar{\boldsymbol{\nabla}}^2 \phi^{-3} L_\alpha \quad . \tag{7.2.22}$$

(We have made the redefinition $L_\alpha \to \phi^{-3} L_\alpha$ to simplify quantization, as will be explained in sec. 7.4.) Furthermore, we choose the (quantum) $\Lambda_{\dot\alpha}$-gauge $H^\alpha = H^{\dot\alpha} = 0$

(see (5.2.18)), which determines $\Lambda_{\dot\alpha}$ in terms of $L_\alpha$ :

$$\Lambda_{\dot\alpha} = e^{-H}\boldsymbol{\nabla}^2\bar\phi^{-3}\bar L_{\dot\alpha} + \mathbf{O}(\mathbf{R},\mathbf{G},\mathbf{W}) \quad . \tag{7.2.23}$$

This supersymmetric gauge choice does not introduce any ghosts.

We now take the classical supergravity action (5.2.48) in terms of full superfields and express it in terms of the quantum gauge fields and the background covariant derivatives, using (7.2.4-12):

$$S_C = -\frac{3}{\kappa^2}\int d^4x d^4\theta \; E^{-1} \quad ,$$

$$E^{-1} = \mathbf{E}^{-1}\hat E^{-\frac{1}{3}}(1\cdot e^{-\mathbf{E}^{-1}\overleftarrow{H}\mathbf{E}})^{\frac{1}{3}}\phi e^{-H}\bar\phi \quad , \tag{7.2.24}$$

This expression is the direct background covariantization of (5.2.72), including a factor of $\mathbf{E}^{-1}$ to make it a density. The quantum fields appear explicitly and in $\hat E$, while the background fields appear implicitly in covariant derivatives and $\mathbf{E}$.

## b. Expanding the action

Our next task is to expand the action in powers of the quantum fields. This is a tedious but healthy exercise and we outline the steps needed to get the quadratic part, which we need for discussing gauge-fixing, and for doing one-loop calculations. Cubic and higher-order terms are needed for higher-loop calculations, but we do not derive them here.

We must expand the exponentials and the determinant $\hat E^{-\frac{1}{3}}$ in (7.2.24) in powers of $H$. We first define $\Delta_A$ by

$$\hat E_A = <\widehat{\nabla}_A> = \hat E_A{}^B\boldsymbol{\nabla}_B \equiv (\delta_A{}^B + \Delta_A{}^B)\boldsymbol{\nabla}_B = \boldsymbol{\nabla}_A + \Delta_A \quad . \tag{7.2.25}$$

The expansion of $\hat E^{-\frac{1}{3}}$ is then, to quadratic order in $\Delta$ (and $H$):

$$\hat E^{-\frac{1}{3}} = [sdet(1+\Delta)]^{-\frac{1}{3}} = exp(-\frac{1}{3}\,str\,ln(1+\Delta))$$

$$= 1 - \frac{1}{3}\,str\Delta + \frac{1}{18}\,(str\Delta)^2 + \frac{1}{6}\,str(\Delta^2)$$

$$= 1 - \frac{1}{3}(-1)^A \Delta_A{}^A + \frac{1}{18}[(-1)^A \Delta_A{}^A]^2 + \frac{1}{6}(-1)^A \Delta_A{}^B \Delta_B{}^A \quad . \tag{7.2.26}$$

To find the explicit form of $\Delta_A{}^B$, we return to (7.2.5,6) and write $\Delta_A$ explicitly. Since $< \widehat{\nabla}_{\dot{\alpha}} > = \widehat{\nabla}_{\dot{\alpha}} = \bar{\boldsymbol{\nabla}}_{\dot{\alpha}}$,

$$\Delta_{\dot{\alpha}} = 0 \quad . \tag{7.2.27a}$$

From

$$< \widehat{\nabla}_\alpha > = < e^{-H} \boldsymbol{\nabla}_\alpha e^H > = \boldsymbol{\nabla}_\alpha + < [\boldsymbol{\nabla}_\alpha , H] > + \frac{1}{2} < [[\boldsymbol{\nabla}_\alpha , H], H] > + < O(H^3) > \tag{7.2.27b}$$

we obtain $\Delta_\alpha{}^B$ $(+\delta_\alpha{}^B)$ from the coefficient of $\boldsymbol{\nabla}_B$ on the right hand side of (7.2.27b). Since $H$ is a scalar operator, we can drop Lorentz rotation terms produced by the commutators at each stage, because they can produce only more Lorentz terms (see (7.2.17a)). We obtain $\Delta_{\underline{a}}{}^B$ from the right hand side of

$$< \widehat{\nabla}_{\underline{a}} > = -i < \{\widehat{\nabla}_\alpha , \widehat{\nabla}_{\dot{\alpha}}\} >$$

$$= \boldsymbol{\nabla}_{\underline{a}} - i < \{\bar{\boldsymbol{\nabla}}_{\dot{\alpha}}, [\boldsymbol{\nabla}_\alpha , H]\} > - i\frac{1}{2} < \{\bar{\boldsymbol{\nabla}}_{\dot{\alpha}}, [[\boldsymbol{\nabla}_\alpha , H], H]\} > \quad . \tag{7.2.27c}$$

Again we can drop Lorentz terms at intermediate stages of the calculation: They contribute only to $\Delta_{\underline{a}}{}^{\dot{\beta}}$, and since $\Delta_{\dot{\alpha}}{}^B = 0$ (to all orders), they do not contribute to the determinant $\hat{E}$. We first find $\Delta_\alpha{}^B$ to lowest order in $H$:

$$[\boldsymbol{\nabla}_\alpha , H] = [\boldsymbol{\nabla}_\alpha , H^{\underline{b}} i \boldsymbol{\nabla}_{\underline{b}}] = i[\boldsymbol{\nabla}_\alpha , H^{\underline{b}}]\boldsymbol{\nabla}_{\underline{b}} + iH^{\underline{b}}[\boldsymbol{\nabla}_\alpha , \boldsymbol{\nabla}_{\underline{b}}]$$

$$= i(\boldsymbol{\nabla}_\alpha H^{\underline{b}})\boldsymbol{\nabla}_{\underline{b}} - H_\alpha{}^{\dot{\beta}}(\bar{\mathbf{R}}\bar{\boldsymbol{\nabla}}_{\dot{\beta}} - \mathbf{G}^\gamma{}_{\dot{\beta}}\boldsymbol{\nabla}_\gamma) + \textit{Lorentz terms} \quad ,$$

and hence

$$\Delta_\alpha{}^\beta = -H_{\alpha\dot{\gamma}}\mathbf{G}^{\beta\dot{\gamma}} \quad , \quad \Delta_\alpha{}^{\underline{b}} = i\boldsymbol{\nabla}_\alpha H^{\underline{b}} \quad . \tag{7.2.28}$$

(Again we can ignore $\Delta_\alpha{}^{\dot{\beta}}$.)

Proceeding in this manner we then find, to the order in $H^{\alpha\dot{\alpha}}$ necessary for the quadratic action ($H \cdot \boldsymbol{\nabla} \equiv H^{\alpha\dot{\alpha}}\boldsymbol{\nabla}_{\alpha\dot{\alpha}}$):

$$\Delta_\alpha{}^\beta = -\,(1 - \frac{1}{2}\,iH\cdot\boldsymbol{\nabla})H_{\alpha\dot\gamma}\mathbf{G}^{\beta\dot\gamma}$$

$$-\,\frac{1}{2}\,(\boldsymbol{\nabla}_\alpha H^{\gamma\dot\delta})H_{\epsilon\dot\zeta}[\frac{1}{2}\,\delta_\gamma{}^\epsilon\bar{\boldsymbol{\nabla}}_{(\dot\delta}\mathbf{G}^{\beta\dot\zeta)} + \delta_{\dot\delta}{}^{\dot\zeta}(\mathbf{W}_\gamma{}^{\epsilon\beta} - \frac{1}{2}\,\delta_{(\gamma}{}^\beta\boldsymbol{\nabla}^{\epsilon)}\mathbf{R})]$$

$$+\,\frac{1}{2}\,H_{\alpha\dot\delta}\mathbf{G}^{\gamma\dot\delta}H_{\gamma\dot\epsilon}\mathbf{G}^{\beta\dot\epsilon} - \frac{1}{2}\,\delta_\alpha{}^\beta\mathbf{R}\bar{\mathbf{R}}H^2 \quad ,$$

$$\Delta_\alpha{}^{\underline{b}} = i\boldsymbol{\nabla}_\alpha H^{\underline{b}} \quad ,$$

$$\Delta_{\underline{a}}{}^\beta = -\,i(-\bar{\boldsymbol{\nabla}}_{\dot\alpha}H_{\alpha\dot\epsilon}\mathbf{G}^{\beta\dot\epsilon} + \mathbf{R}\boldsymbol{\nabla}_\alpha H^\beta{}_{\dot\alpha}) \quad ,$$

$$\Delta_{\underline{a}}{}^{\underline{b}} = \bar{\boldsymbol{\nabla}}_{\dot\alpha}\boldsymbol{\nabla}_\alpha H^{\underline{b}} + \delta_{\dot\alpha}{}^{\dot\beta}\Delta_\alpha{}^\beta \quad . \qquad (7.2.29)$$

We have used the fact that, for the part of the action quadratic in $H$, we need only the linear parts of $\Delta_\alpha{}^{\beta\dot\beta}$ and $\Delta_{\alpha\dot\alpha}{}^\beta$ , and we can also drop any total derivatives of quadratic terms (but not if we were to compute higher-order terms in the action). After substituting (7.2.29) into (7.2.26) we find, again dropping irrelevant terms,

$$\hat{E}^{-\frac{1}{3}} = 1 - \frac{1}{3}\big\{\bar{\boldsymbol{\nabla}}_{\dot\beta}\boldsymbol{\nabla}_\alpha H^{\alpha\dot\beta} - (1 - \frac{1}{2}\,iH\cdot\boldsymbol{\nabla})\mathbf{G}\cdot H$$

$$-\,\frac{1}{4}\,(\boldsymbol{\nabla}_\alpha H^{\gamma\dot\delta})H_{\epsilon\dot\zeta}[\delta_\gamma{}^\epsilon\bar{\boldsymbol{\nabla}}_{(\dot\delta}\mathbf{G}^{\alpha\dot\zeta)} + \delta_{\dot\delta}{}^{\dot\zeta}(2\mathbf{W}_\gamma{}^{\epsilon\alpha} - \delta_{(\gamma}{}^\alpha\boldsymbol{\nabla}^{\epsilon)}\mathbf{R})]$$

$$+\,\frac{1}{2}\,[(\mathbf{G}\cdot H)^2 - 2\mathbf{G}^2H^2] - \mathbf{R}\bar{\mathbf{R}}H^2\big\} + \frac{1}{18}\,(\bar{\boldsymbol{\nabla}}_{\dot\beta}\boldsymbol{\nabla}_\alpha H^{\alpha\dot\beta} - \mathbf{G}\cdot H)^2$$

$$+\,\frac{1}{6}\big\{(\bar{\boldsymbol{\nabla}}_{\dot\beta}\boldsymbol{\nabla}_\alpha H^{\gamma\dot\delta} - \delta_{\dot\beta}{}^{\dot\delta}H_{\alpha\dot\epsilon}\mathbf{G}^{\gamma\dot\epsilon})(\bar{\boldsymbol{\nabla}}_{\dot\delta}\boldsymbol{\nabla}_\gamma H^{\alpha\dot\beta} - \delta_{\dot\delta}{}^{\dot\beta}H_{\gamma\dot\zeta}\mathbf{G}^{\alpha\dot\zeta})$$

$$-2(\bar{\boldsymbol{\nabla}}_{\dot\delta}H_{\gamma\dot\epsilon}\mathbf{G}^{\beta\dot\epsilon} - \mathbf{R}\boldsymbol{\nabla}_\gamma H^\beta{}_{\dot\delta})\boldsymbol{\nabla}_\beta H^{\gamma\dot\delta} - [(\mathbf{G}\cdot H)^2 - 2\mathbf{G}^2H^2]\big\} \quad . \qquad (7.2.30)$$

We also have the relevant terms of ( using (7.2.10), and expanding $\phi = 1 + \chi$):

$$(1\cdot e^{-\mathbf{E}^{-1}\overleftarrow{H}\mathbf{E}})^{\frac{1}{3}} = 1 - \frac{1}{3}\,i\boldsymbol{\nabla}\cdot H + \frac{1}{9}\,(\boldsymbol{\nabla}\cdot H)^2 \quad ,$$

$$\phi e^{-H}\bar\phi = 1 + (\chi + \bar\chi) + \chi\bar\chi - iH\cdot\boldsymbol{\nabla}\bar\chi \quad . \qquad (7.2.31)$$

Finally, we obtain the quadratic part of the Lagrangian by multiplying together the various contributions, to obtain :

$$\mathbf{E}E^{-1} = 1 + (\chi + \bar{\chi} + \frac{1}{3}\mathbf{G}\cdot H) + \chi\bar{\chi} - \frac{1}{3}i(\chi - \bar{\chi})\boldsymbol{\nabla}\cdot H + \frac{1}{3}(\chi + \bar{\chi})\mathbf{G}\cdot H$$
$$+ \frac{1}{3}\mathbf{R}\bar{\mathbf{R}}H^2 + \frac{1}{18}(\mathbf{G}\cdot H)^2 + \frac{1}{12}(\boldsymbol{\nabla}\cdot H)^2 - \frac{1}{36}([\bar{\boldsymbol{\nabla}}_{\dot{\beta}}, \boldsymbol{\nabla}_\alpha]H^{\alpha\dot{\beta}})^2$$
$$- \frac{1}{18}(\mathbf{G}\cdot H)[\bar{\boldsymbol{\nabla}}_{\dot{\beta}}, \boldsymbol{\nabla}_\alpha]H^{\alpha\dot{\beta}} + \frac{1}{3}\mathbf{R}(\boldsymbol{\nabla}^\alpha H^{\beta\dot{\gamma}})(\boldsymbol{\nabla}_\beta H_{\alpha\dot{\gamma}})$$
$$+ \frac{1}{12}(\boldsymbol{\nabla}_\alpha H^{\gamma\dot{\delta}})H_{\epsilon\dot{\zeta}}[\delta_\gamma{}^\epsilon\bar{\boldsymbol{\nabla}}_{(\dot{\delta}}\mathbf{G}^{\alpha\dot{\zeta})} + \delta_{\dot{\delta}}{}^{\dot{\zeta}}(2\mathbf{W}_\gamma{}^{\epsilon\alpha} - \delta_{(\gamma}{}^\alpha\boldsymbol{\nabla}^{\epsilon)}\mathbf{R})]$$
$$- \frac{1}{6}H^{\alpha\dot{\beta}}(\boldsymbol{\nabla}_\alpha\bar{\boldsymbol{\nabla}}_{\dot{\beta}}\bar{\boldsymbol{\nabla}}_{\dot{\delta}}\boldsymbol{\nabla}_\gamma + \boldsymbol{\nabla}_\gamma\bar{\boldsymbol{\nabla}}_{\dot{\delta}}\bar{\boldsymbol{\nabla}}_{\dot{\beta}}\boldsymbol{\nabla}_\alpha)H^{\gamma\dot{\delta}} \quad . \qquad (7.2.32)$$

By using the identity (with $\Box = \frac{1}{2}\boldsymbol{\nabla}^{\alpha\dot{\alpha}}\boldsymbol{\nabla}_{\alpha\dot{\alpha}}$)

$$\boldsymbol{\nabla}_\alpha\bar{\boldsymbol{\nabla}}_{\dot{\beta}}\bar{\boldsymbol{\nabla}}_{\dot{\delta}}\boldsymbol{\nabla}_\gamma + \boldsymbol{\nabla}_\gamma\bar{\boldsymbol{\nabla}}_{\dot{\delta}}\bar{\boldsymbol{\nabla}}_{\dot{\beta}}\boldsymbol{\nabla}_\alpha$$
$$= C_{\alpha\gamma}C_{\dot{\beta}\dot{\delta}}\big(-\Box + \{\boldsymbol{\nabla}^2, \bar{\boldsymbol{\nabla}}^2\} - \frac{1}{2}[-\bar{\mathbf{R}}\bar{\boldsymbol{\nabla}}^{\dot{\epsilon}} + \mathbf{G}^{\zeta\dot{\epsilon}}\boldsymbol{\nabla}_\zeta + (\boldsymbol{\nabla}^\zeta\mathbf{G}^{\eta\dot{\epsilon}})M_{\zeta\eta} + \bar{\mathbf{W}}^{\dot{\epsilon}}{}_{\dot{\zeta}}{}^{\dot{\eta}}\bar{M}_{\dot{\eta}}{}^{\dot{\zeta}}, \bar{\boldsymbol{\nabla}}_{\dot{\epsilon}}]\big)$$
$$+ 2\mathbf{R}\bar{\mathbf{R}}M_{\alpha\gamma}\bar{M}_{\dot{\beta}\dot{\delta}} - (\boldsymbol{\nabla}_{(\alpha}\mathbf{R})\boldsymbol{\nabla}_{\gamma)}\bar{M}_{\dot{\beta}\dot{\delta}} \quad , \qquad (7.2.33)$$

we can rewrite (7.2.32) as

$$\mathbf{E}E^{-1} = 1 + \big\{\chi + \bar{\chi} + \frac{1}{3}H^{\alpha\dot{\alpha}}\mathbf{G}_{\alpha\dot{\alpha}}\big\} + \big\{\chi\bar{\chi} + \frac{1}{3}i(\bar{\chi} - \chi)\boldsymbol{\nabla}_{\alpha\dot{\alpha}}H^{\alpha\dot{\alpha}} + \frac{1}{6}H^{\alpha\dot{\alpha}}\Box H_{\alpha\dot{\alpha}}$$
$$+ \frac{1}{12}(\boldsymbol{\nabla}_{\alpha\dot{\alpha}}H^{\alpha\dot{\alpha}})^2 - \frac{1}{36}([\bar{\boldsymbol{\nabla}}_{\dot{\alpha}}, \boldsymbol{\nabla}_\alpha]H^{\alpha\dot{\alpha}})^2 - \frac{1}{3}[(\bar{\boldsymbol{\nabla}}^2 + \frac{3}{2}\mathbf{R})H^{\alpha\dot{\alpha}}][(\boldsymbol{\nabla}^2 + \frac{3}{2}\bar{\mathbf{R}})H_{\alpha\dot{\alpha}}]\big\}$$
$$+ \big\{\frac{1}{3}(\chi + \bar{\chi})H^{\alpha\dot{\alpha}}\mathbf{G}_{\alpha\dot{\alpha}} + \frac{1}{18}(H^{\alpha\dot{\alpha}}\mathbf{G}_{\alpha\dot{\alpha}})^2 + \frac{2}{3}\mathbf{R}\bar{\mathbf{R}}H^{\alpha\dot{\alpha}}H_{\alpha\dot{\alpha}}$$
$$+ \frac{1}{12}(\boldsymbol{\nabla}^2\mathbf{R} + \bar{\boldsymbol{\nabla}}^2\bar{\mathbf{R}})H^{\alpha\dot{\alpha}}H_{\alpha\dot{\alpha}} + \frac{1}{6}H^{\alpha\dot{\alpha}}(\mathbf{R}\boldsymbol{\nabla}^2 + \bar{\mathbf{R}}\bar{\boldsymbol{\nabla}}^2)H_{\alpha\dot{\alpha}}$$
$$- \frac{1}{12}H^{\alpha\dot{\alpha}}\mathbf{G}^{\beta\dot{\beta}}[\bar{\boldsymbol{\nabla}}_{\dot{\beta}}, \boldsymbol{\nabla}_\beta]H_{\alpha\dot{\alpha}} + \frac{1}{12}H^{\alpha\dot{\alpha}}[(\boldsymbol{\nabla}_{(\alpha}\mathbf{G}_{\beta)}{}^{\dot{\beta}})\bar{\boldsymbol{\nabla}}_{\dot{\beta}}H^\beta{}_{\dot{\alpha}} + (\bar{\boldsymbol{\nabla}}_{(\dot{\alpha}}\mathbf{G}^\beta{}_{\dot{\beta})})\boldsymbol{\nabla}_\beta H_\alpha{}^{\dot{\beta}}]$$
$$- \frac{1}{18}(H^{\alpha\dot{\alpha}}\mathbf{G}_{\alpha\dot{\alpha}})[\bar{\boldsymbol{\nabla}}_{\dot{\beta}}, \boldsymbol{\nabla}_\beta]H^{\beta\dot{\beta}} + \frac{1}{6}H^{\alpha\dot{\alpha}}(\mathbf{W}_\alpha{}^{\beta\gamma}\boldsymbol{\nabla}_\beta H_{\gamma\dot{\alpha}} + \bar{\mathbf{W}}_{\dot{\alpha}}{}^{\dot{\beta}\dot{\gamma}}\bar{\boldsymbol{\nabla}}_{\dot{\beta}}H_{\alpha\dot{\gamma}})\big\} \quad , \qquad (7.2.34)$$

where we have also used the Bianchi identities (5.4.16,17,18). The expression in the first set of braces is linear in the quantum fields. If sources are coupled to them, variation with respect to $H$ and $\chi$ gives $\mathbf{R} = J$ and $\mathbf{G}_{\alpha\dot\alpha} = J_{\alpha\dot\alpha}$, using

$$\int d^4x\, d^4\theta\, \mathbf{E}^{-1}\chi = \int d^4x\, d^2\theta\, e^{-\bar{\boldsymbol{\Omega}}}\boldsymbol{\phi}^3(\bar{\boldsymbol{\nabla}}^2 + \mathbf{R})\chi = \int d^4x\, d^2\theta\, e^{-\bar{\boldsymbol{\Omega}}}\boldsymbol{\phi}^3\mathbf{R}\,\chi \quad (7.2.35)$$

(see sec. 5.5.e; we are in the background vector representation). The expression in the second set of braces is the direct covariantization of the free Lagrangian, which is obtained by setting all background fields to zero:

$$-\frac{1}{3}\mathbb{L}^{(2)} = \bar\chi\chi + \frac{1}{3}i(\bar\chi - \chi)\partial_{\alpha\dot\alpha}H^{\alpha\dot\alpha} + \frac{1}{6}H^{\alpha\dot\alpha}[\Box - D^2\bar D^2 - \bar D^2 D^2]H_{\alpha\dot\alpha}$$

$$+\frac{1}{12}(\partial_{\alpha\dot\alpha}H^{\alpha\dot\alpha})^2 - \frac{1}{36}([\bar D_{\dot\alpha}, D_\alpha]H^{\alpha\dot\alpha})^2 \quad . \quad (7.2.36)$$

## 7.3. Ghosts

In the next section we shall discuss in detail the quantization of superfield supergravity. The procedure is not entirely straightforward and runs into a number of subtleties not normally encountered in simpler theories, so it is desirable to know ahead of time the general ghost structure of the quantized theory. This is the topic of the present section. We discuss the following subjects: (a) how the linearized ghost structure can be easily determined before performing the quantization; (b) the modifications to the Faddeev-Popov procedure necessary when using constrained gauge-fixing functions in the presence of background fields; (c) how to obtain only propagators that go as $p^{-2}$, and avoid infrared difficulties, while still keeping the action local, by the introduction of additional fields; and (d) the necessity for appropriate parametrization of the gauge transformations (for which (density) compensators are crucial) so that the Faddeev-Popov procedure is applicable, and so that we only use the types of superfields allowed in an arbitrary supergravity background.

### a. Ghost counting

We first give a simple rule for counting all the ghosts in any gauge theory. In ordinary gauges some of these ghosts may decouple, but in background field gauges all the ghosts couple to the background. We begin by deriving the rules for a general component-field gauge theory. To streamline notation, we drop all indices and indicate abnormal-statistics fields by primes. The general quadratic Lagrangian for any gauge field can be written in the form $A\Box^n \Pi A$, where $\Pi$ is a projection operator and $n$ is an integer (when $A$ is a tensor ) or half-integer (when $A$ is a spinor : $\Box^{\frac{1}{2}} \equiv \not{\partial}$ ). For physical fields $n = 1$ or $\frac{1}{2}$. ( The operators $\Box$, $\not{\partial}$, etc. may be covariant with respect to background fields, and may include "nonminimal" couplings to the background.) The gauge invariance is expressed as $\delta A = \partial\lambda$ , with $\Pi\partial\lambda = 0$. After gauge fixing, the Lagrangian becomes $A\Box^n A$ but, in order to cancel the $A = \partial\lambda$ mode, which did not occur in the original Lagrangian, we must introduce a ghost $B'$. Its Lagrangian is obtained through the substitution $A -> A' = \partial B'$ in the gauge-fixed Lagrangian. We thus obtain

$$I\!L = A\Box^n A + (\partial B')\Box^n(\partial B') = A\Box^n A + B'\Box^{n+1}\Pi' B' \tag{7.3.1}$$

where $\Pi'$ is a new projection operator. In the simplest cases $\Pi' = 1$ (e.g., if $A$ is the photon field) and we are through. More generally (e.g., if $A$ is an antisymmetric tensor gauge field) $B'$ has a gauge invariance, and we must continue the procedure until no gauge invariance remains. (This is a finite procedure, since each ghost has one less vector index than its predecessor).

The final Lagrangian thus has the form

$$I\!\!L = A\Box^n A + B'\Box^{n+1} B' + C\Box^{n+2} C + \cdots \ . \tag{7.3.2}$$

For tensor fields, a field with kinetic operator $\Box^m$ represents $m$ fields of that type with kinetic operator $\Box$; for spinors, $\Box^m$ represents $2m$ fields with a $\not{p}$. (If $\Box$ includes background interactions their contribution to the effective action is $ln\ det\ \Box^m = m\ ln\ det\ \Box = 2m\ ln\ det \not{p}$ .) Thus, for physical tensor fields $A$ $(n = 1)$, the number of successive fields goes as $1, -2, 3, -4$,..., while for physical spinor fields $(n = \frac{1}{2})$, they go as $1, -3, 5, -7$,..., where the minus signs indicate abnormal statistics. These numbers represent the net number of normal-statistics minus abnormal-statistics quantum fields in the linearized Lagrangian, all coupling to the background fields. Furthermore, as we will see below, all the fields in this counting decouple at higher loops (and at one loop when one is quantizing in ordinary gauges rather than background field gauges) except for the physical fields and (for nonabelian theories) the Faddeev-Popov ghosts of the physical fields. There may also be additional compensating fields coming in pairs of opposite statistics ("catalysts": see below) which cancel in this counting, and which also cancel in one-loop background field calculations (but may contribute for higher loops). Examples where this counting includes more than just the physical and Faddeev-Popov fields are: (1) the gravitino, which has $1, -3$ instead of $1, -2$, due to the appearance of the Nielsen-Kallosh ghost (see below); (2) $p$-forms, which have $1, -2, 3, \ldots, (-1)^p(p + 1)$ instead of $1, -2, 4, \ldots, (-1)^p 2^p$, due to Nielsen-Kallosh ghosts and "hidden" ghosts (see also below).

Generalization of the counting rules to superfields is straightforward, though each case has to be treated separately because of the greater variety of superfield gauge transformations. We first generalize to superfields the result of the previous paragraph for obtaining the number of superfields with standard kinetic term corresponding to one with a higher-derivative kinetic term. When $\Box^m$ is the kinetic operator for general unconstrained superfields, it is equivalent to $m$ general tensor superfields with kinetic operator $\Box$, or $2m$ general spinor superfields with kinetic operator $\not{\partial}$. (Note that, as discussed in sec. 3.8, $sdet\Box = 1$ unless $\Box$ has nontrivial $\theta$-dependence.) However, for chiral superfields the situation is slightly more subtle (see (3.8.28-36)):

$$\int d^4x\, d^2\theta\, \Phi\Box^m\Phi \longleftrightarrow \sum_{i=1}^{m}\int d^4x\, d^2\theta\, \Phi_i\Box\Phi_i \longleftrightarrow \sum_{i=1}^{2m}\int d^4x\, d^4\theta\, \bar{\Phi}_i\Phi_i \ , \tag{7.3.3a}$$

$$\int d^4x\, d^4\theta\, \bar{\Phi}\Box^m\Phi \longleftrightarrow \sum_{i=1}^{2m+1}\int d^4x\, d^4\theta\, \bar{\Phi}_i\Phi_i \ , \tag{7.3.3b}$$

$$\int d^4x\, d^4\theta\, \bar{\Phi}^{\dot{\alpha}}\Box^m i\partial^{\alpha}{}_{\dot{\alpha}}\Phi_\alpha \longleftrightarrow \sum_{i=1}^{m+1}[\frac{1}{2}\int d^4x\, d^2\theta\, \Phi^\alpha\Box\Phi_\alpha + h.\,c.] \ . \tag{7.3.3c}$$

The $\bar{\Phi}\Phi$ form of (7.3.3a) gives the result for chiral scalar superfields, whereas the $\Phi\Box\Phi$ form is applicable to chiral (undotted) spinor superfields. This latter result can also be related to (7.3.3c) by noting that (7.3.3a) for $m = 1$ and (7.3.3c) for $m = 0$ are merely different gauge choices for the gauge-fixed action for the tensor multiplet (cf. (6.2.32-34)).

We now consider some examples of superfield ghost counting. The simplest is the vector multiplet. The classical Lagrangian is $V\Box\Pi_{\frac{1}{2}}V$ , with $\delta V = i(\bar{\Lambda} - \Lambda)$ , where $\Lambda$ is chiral and $\Pi_{\frac{1}{2}}$ is the superspin $\frac{1}{2}$ projection operator. After gauge fixing (which removes $\Pi_{\frac{1}{2}}$) and the substitution $V -> i(\bar{\Phi}' - \Phi')$ with chiral ghost $\Phi'$ to cancel the gauge modes, we obtain

$$I\!\!L = V\Box V + \bar{\Phi}'\Box\Phi' \ . \tag{7.3.4}$$

(With $d^4\theta$ integration the $\Phi'\Box\Phi'$ and $\bar{\Phi}'\Box\bar{\Phi}'$ terms give zero, modulo nonminimal couplings which we incorporate into the definition of $\bar{\Phi}'\Box\Phi'$.) The ghost Lagrangian is equivalent to three of the usual terms $\bar{\Phi}'\Phi'$ (see (7.3.3b)). We thus expect three chiral ghosts, which agrees with our results from explicit quantization in sec. 6.5.

A second example is that of the tensor multiplet, with classical action $\int d^4x d^2\theta\, \phi^\alpha\Box\Pi_{\frac{1}{2}+}\phi_\alpha$ and gauge invariance $\delta\phi_\alpha = i\bar{D}^2D_\alpha K,\ K = \bar{K}$. After gauge fixing (which removes the projector $\Pi_{\frac{1}{2}+}$), substitution leads to a first generation ghost Lagrangian $V'\Box^2\Pi_{\frac{1}{2}}V'$, and its gauge invariance $\delta V' = i(\bar{\Lambda} - \Lambda)$ leads to a second generation ghost Lagrangian $\bar{\phi}\Box^2\phi$. (The gauge invariance of $V'$ is the same as the gauge invariance of the *variation* of $\phi_\alpha$: $\delta\phi_\alpha(K) = \delta\phi_\alpha(K + i\bar{\Lambda} - i\Lambda)$.) We end up with one $\phi^\alpha\Box\phi_\alpha$ term, two $V'\Box V'$ terms, and five $\bar{\phi}\phi$ terms.

A third example is that of the general spinor superfield $\psi_\alpha$, which together with a compensating chiral scalar describes the $(\frac{3}{2}, 1)$ multiplet (see sec. 4.6; we are using the second form of (4.6.42), but with the compensator $V$ gauged to zero). The Lagrangian has the form

$$I\!\!L = \frac{1}{2}\,\bar{\Psi}^{\dot{\alpha}} i\partial^\alpha{}_{\dot{\alpha}}\tilde{\Pi}\Psi_\alpha + h.\,c. - 2\bar{\Phi}\Phi + crossterms \tag{7.3.5}$$

where $\tilde{\Pi}$ is a sum of projection operators and $I\!\!L$ has the gauge invariance $\delta\Psi_\alpha = \Lambda_\alpha + iD_\alpha K$, $\delta\Phi = -\bar{D}^2 K$, with chiral $\Lambda$ and real $K$. In the gauge fixed Lagrangian $\tilde{\Pi}$ is absent and we have a chiral spinor ghost $\phi'_\alpha$ and a real scalar ghost $V'$ (corresponding to $\Lambda_\alpha$ and $K$, respectively) without any further gauge invariance (the variation $\delta\psi_\alpha$, $\delta\phi$, is *not* invariant under any changes of $\Lambda^\alpha$ or $K$):

$$I\!\!L = \bar{\Psi}^{\dot{\alpha}} i\partial^\alpha{}_{\dot{\alpha}}\Psi_\alpha + \bar{\Phi}\Phi + \bar{\phi}'^{\dot{\alpha}} i\partial^\alpha{}_{\dot{\alpha}}\phi'_\alpha + V'\Box V' \tag{7.3.6}$$

($\psi^\alpha\phi$ cross terms can be eliminated by a suitable choice of gauge fixing function).

The other form of the theory has the Lagrangian $\frac{1}{2}\,\bar{\Psi}^{\dot{\alpha}} i\partial^\alpha{}_{\dot{\alpha}}\tilde{\Pi}\Psi_\alpha + h.\,c. + \cdots$ with a different $\tilde{\Pi}$, and gauge invariance $\delta\Psi_\alpha = i\bar{D}^2 D_\alpha K_1 + iD_\alpha K_2$. After including ghosts, it becomes

$$I\!\!L = \bar{\Psi}^{\dot{\alpha}} i\partial^\alpha{}_{\dot{\alpha}}\Psi_\alpha + V'_2\Box V'_2 + V'_1\Box^2 V'_1 + \bar{\Phi}\Box^2\Phi \quad . \tag{7.3.7}$$

The chiral scalar field $\Phi$ is a second-generation ghost, arising from the invariance $\delta V'_1 = i(\bar{\Lambda} - \Lambda)$ (due to the invariance of $\delta\Psi_\alpha$ under $K_1 \to K_1 + i(\bar{\Lambda} - \Lambda)$). We thus obtain the equivalent of three real scalar ghosts and five chiral scalar second-generation (normal statistics) ghosts.

We consider now $n = -\frac{1}{3}$ supergravity itself, with kinetic Lagrangian $H\Box\tilde{\Pi}H + \bar{\chi}\chi$ ($+ H\chi$ cross terms), where $\tilde{\Pi}$ is a sum of projection operators. We have the (linearized) gauge invariance $\delta H_{\alpha\dot{\beta}} = D_\alpha\bar{L}_{\dot{\beta}} - \bar{D}_{\dot{\beta}}L_\alpha$, $\delta\chi = \bar{D}^2 D_\alpha L^\alpha$, with general spinor gauge parameter $L_\alpha$. After gauge fixing we have a Lagrangian $H\Box H + \bar{\chi}\chi + \bar{\psi}^{\dot{\alpha}\prime}\Box i\partial_{\alpha\dot{\alpha}}\tilde{\Pi}'\psi^{\alpha\prime}$, where the ghost term is obtained by substitution in the gauge-fixed Lagrangian with the gauge parameter $L_\alpha$ replaced by the ghost $\psi_\alpha{}'$. We have a new gauge invariance, $\delta\psi_\alpha{}' = \Lambda_\alpha$ where $\Lambda$ is chiral. (This reflects the invariance of the original gauge transformations under $\delta L_\alpha = \Lambda_\alpha$.) We thus introduce a second generation chiral ghost $\phi_\alpha$ and are finally led to the form

$$I\!L = H \Box H + \bar{\chi}\chi + \bar{\psi}' \Box \not{\partial} \psi' + \bar{\phi} \Box \not{\partial} \phi \quad . \tag{7.3.8}$$

We obtain the equivalent of three first-generation general spinor ghosts, with Lagrangian $\bar{\psi}\not{\partial}\psi$, and two second-generation chiral spinor ghosts. In the next section we will derive these results from the gauge fixing procedure, and give the results for a variant form of the $n = -\frac{1}{3}$ compensator which leads to a different set of ghosts.

**b. Hidden ghosts**

In addition to the Nielsen-Kallosh ghost, which emerges from a careful application of the gauge-averaging procedure (see (6.5.12,13)), there is a second subtlety that may occur, and which must be handled correctly in order to arrive at the correct set of ghosts. This has to do with the occurrence of gauge-fixing functions which satisfy constraints. In the nonsupersymmetric case the simplest example is given by the 2-form $A_{\underline{ab}}$ in a background gravitational field. The naive 't Hooft gauge averaging

$$\int I\!D f_{\underline{a}}\ \delta(\nabla^{\underline{b}} A_{\underline{ab}} - f_{\underline{a}}) exp(-\int d^4x\, g^{\frac{1}{2}} f^2) = exp[-\int d^4x\, g^{\frac{1}{2}} (\nabla^{\underline{b}} A_{\underline{ab}})^2] \tag{7.3.9}$$

would give an incorrect result, since the constraint in the $\delta$ functional implies $\nabla \cdot f = 0$, and introduces extraneous dependence on the external gravitational field. We would therefore like to put just the transverse part of $f$ in the $\delta$ functional, and in the gauge-averaging function as $f^2 \to \frac{1}{2} f^{\underline{a}} (\delta_{\underline{ab}} - \frac{1}{2} \nabla_{\underline{a}} \Box^{-1} \nabla_{\underline{b}}) f^{\underline{b}}$. However, since then only the transverse part of $f$ appears in the functional integral, the integrand has a gauge invariance $\delta f_{\underline{a}} = \nabla_{\underline{a}} \lambda$, so we must introduce appropriate gauge-fixing and (Faddeev-Popov and Nielsen-Kallosh) ghost terms. The intermediate steps vary depending on the choice of gauge-fixing function (e.g., $\nabla \cdot f$ vs. $\Box^{-1} \nabla \cdot f$), but the net result is that one obtains $-1$ additional scalar fields, (a hidden ghost) and thus the total set of fields consists of 1 2-form, $-2$ 1-forms, and $+3$ scalars (vs. the $+4$ expected from considering just the Faddeev-Popov ghosts of the vector ghosts of the 2-form), in agreement with our general counting argument given above. Similar arguments apply to higher-rank forms.

A simpler form of the argument for the necessity of these "hidden" ghosts can be given in supersymmetric theories. Consider again the chiral spinor gauge superfield $\Phi_\alpha$, with gauge invariance $\delta\Phi_\alpha = i\bar{D}^2 D_\alpha K$ and gauge-fixing function $F = \bar{F} = \frac{1}{2}i(D_\alpha\Phi^\alpha - \bar{D}_{\dot\alpha}\bar{\Phi}^{\dot\alpha})$. Because of the chirality of $\Phi$, $F$ satisfies the constraint $\bar{D}^2 F = 0$, so that $F$ is a linear superfield. The usual gauge-fixing procedure involves introducing in the functional integral $\delta(F - f)$; however, the linear nature of $F$ would imply that $f$ is also linear, an unfortunate feature since it is impossible to functionally integrate or differentiate with respect to linear superfields.

This difficulty can be avoided by "completing" $F$ to a general superfield, by the addition of chiral and antichiral pieces to it. We do this by replacing $F$ in the $\delta$ functional with the expression

$$\hat{F} = F + (D^2\Box^{-1}\eta + \bar{D}^2\Box^{-1}\bar{\eta}) \ . \tag{7.3.10}$$

and functionally integrating over $\eta$ as well. The chiral superfield $\eta$ is the hidden ghost. Now $\bar{D}^2\hat{F} = \eta$ is unconstrained, and so $f$ is also.

To understand the procedure we examine its component form. The $\delta$-function $\delta(\hat{F} - f)$ is a product of $\delta$-functions for the individual components of $\hat{F} - f$ (see (3.8.17a)). Since $F$ is the $\Pi_{\frac{1}{2}}$ part of $\hat{F}$ and the rest is the $\Pi_0$ part, the components of the two terms in (7.3.10) appear in different component $\delta$-functions. The $\bar{D}^2 f|$, $D_\alpha\bar{D}^2 f|$ and $D^2\bar{D}^2 f|$ components are set equal to components of $\eta$ without spacetime derivatives (which is why we included the $D^2\Box^{-1}$ factor), and without any $F = \frac{1}{2}i(D_\alpha\Phi^\alpha - \bar{D}_{\dot\alpha}\bar{\Phi}^{\dot\alpha})$ contributions. Averaging with $exp\int d^4x\, d^4\theta\, f^2$ produces an action for these $\eta$ components that does not contribute to the functional integral upon $\eta$ integration (trivial kinetic terms). The $f|$, $D_\alpha f|$ components produce standard gauge-fixing terms for the gauge components of $\Phi_\alpha$ which are absent in the Wess-Zumino gauge (namely $\chi_\alpha$ and $B$ (4.5.30)), and whose contribution is therefore canceled by corresponding ghosts. Finally, the $[\bar{D}_{\dot\alpha}, D_\alpha]f|$ component gives the gauge-fixing function $\partial^{\underline{b}}A_{\underline{ab}} + \partial_{\underline{a}}\Box^{-1}G$, where $G = Im\ D^2\eta|$ and $A_{\underline{ab}}$ is the antisymmetric tensor component of $\Phi_\alpha$. Averaging of this component of $f$ gives a term $(\partial^{\underline{b}}A_{\underline{ab}})^2 - G\Box^{-1}G$: the usual $A_{\underline{ab}}$ gauge-fixing term as in (7.3.9) plus the hidden component ghost (+1 scalar with $\Box^{-1}$, counting as $-1$ scalar with $\Box$), in agreement with the above discussion in the component theory.

A similar situation occurs in supergravity due to the appearance of the gauge-fixing function $F_\alpha = \bar{D}^{\dot\alpha} H_{\alpha\dot\alpha}$ satisfying $\bar{D}^2 F_\alpha = 0$. This will be discussed in more detail in the next section.

Note that these hidden ghosts couple only to background fields, and thus contribute only at one loop. It would be desirable to have a general derivation of these ghosts based on BRST invariance of the gauge-fixed action, from which Slavnov-Taylor identities could be derived. The appropriate BRST transformations would be those whose Slavnov-Taylor identities implied gauge-independence of the effective action. At present this approach has not been worked out.

## c. More compensators

Unwanted terms in the Lagrangian, such as those leading to $p^{-4}$ terms in the propagator or nonlocal vertices, can sometimes be canceled by introducing additional fields and gauge-fixing them conveniently. Since only the ghosts discussed in the preceding sections are needed to preserve unitarity, contributions of these "catalyst" fields must themselves be canceled by their own ghosts, and indeed this happens at the one-loop level. The catalysts may in general interact with the other quantum fields, and hence contribute at higher loops, whereas their ghosts don't. If one were to integrate out the catalysts, their higher-loop contributions would simply reproduce the unwanted terms that the catalysts eliminated in the first place.

Catalysts are just a type of (tensor) compensator. For example, the compensator in the Stueckelberg formalism (sec. 3.10.a) is introduced simply to improve ultraviolet behavior of the propagator, and decouples due to gauge invariance of the interaction term. In our case, these compensators improve infrared behavior, and do not decouple. Furthermore, catalysts generally are introduced by ghost fields, whereas previously we discussed compensators related to only classical fields.

As an example, consider the linearized Lagrangian $I\!L = A\Box[(1-\Pi)+\alpha\Pi]A$, with $\Pi^2 = \Pi$ and $\alpha \neq 0, 1$. If $\Pi$ were a differential operator, e.g., $\Box^{-1}\partial\partial$, the above Lagrangian would lead to $p^{-4}$ propagators. To obtain the simpler Lagrangian $I\!L = A\Box A$ one could make a field redefinition $A' = [(1-\Pi)+\alpha^{-\frac{1}{2}}\Pi]A$, but if $\Pi$ were nonlocal this would introduce nonlocalities in the interaction terms.

Instead we introduce a catalyst field $B$ in the Lagrangian, either with a trivial kinetic term or together with a ghost which cancels it. We then make shifts $A \to A + OB$, $B \to B + O'A$ that cancel the unwanted terms, and give no $AB$ cross terms. For instance, in the example above, if $\Pi$ is a matrix, we choose $B$ so that $\Pi B = B$ (and introduce also a ghost field with opposite statistics and $\Pi B' = B'$ ) and we add it to the Lagrangian to obtain $I\!\!L' = I\!\!L + (1-\alpha)B\Box B + B'\Box B'$. First making the shift $B \to B + \Pi A$, then the shift $A \to A - (1-\alpha)B$, we obtain $I\!\!L'' = A\Box A + (1-\alpha)\alpha B\Box B + B'\Box B'$. For background interactions $B$ and $B'$ will cancel at one loop, but clearly the $A$ shift can lead to quantum interactions of $B$ (but not $B'$). An equivalent procedure consists of making the substitution $A \to A + B$ in the original Lagrangian, and going through the gauge fixing procedure for the new gauge invariance that has been introduced, namely $\delta A = \lambda$, $\delta B = -\lambda$ ( $\Pi\lambda = \lambda$ ). In general, this is the simplest procedure.

As a superfield example, we consider a real scalar $V$ in the presence of an on-shell background supergravity field, with Lagrangian

$$I\!\!L_0 = V(-\nabla^\alpha \bar{\nabla}^2 \nabla_\alpha + a\{\nabla^2, \bar{\nabla}^2\})V \tag{7.3.11}$$

with $a \neq 0, 1$ . The superfield $V$ has $p^{-4}$ terms in its propagator and complicated vertices (coupling to the background gravitational field), but it can be shown that the result for the effective action is independent of $a$. To show this using catalyst fields we introduce them, for example, by making the shift

$$V \to V + (\eta + \bar{\eta}) \ , \ \bar{\nabla}_{\dot{\alpha}} \eta = 0 \ . \tag{7.3.12}$$

We have now the gauge invariance $\delta V = \Lambda + \bar{\Lambda}$, $\delta\eta = -\Lambda$, with a chiral parameter. We choose the gauge fixing function $F = \bar{\nabla}^2(V - \frac{a}{1-a}\bar{\eta})$ and gauge fixing term $2(1-a)F\bar{F}$, and are led to the Lagrangian

$$I\!\!L = V\Box V + 2\frac{a}{1-a}\eta\Box\bar{\eta} \ . \tag{7.3.13}$$

The $\eta$ Lagrangian is equivalent to that for three ordinary chiral fields $\eta_i$, $i = 1, 2, 3$. The gauge-fixing procedure also introduces three chiral ghosts, just as for the usual $V$ superfield (two Faddeev-Popov and a Nielsen- Kallosh ghost). They exactly cancel the three ordinary chiral fields at the one-loop level, and leave us with $V\Box V$.

If we had considered a system similar to the above, but where $V$ had quantum interactions, the $\eta$'s would also have such interactions. Then the effect of the $\eta$'s would be to reproduce, if integrated out, the nonlocalities that would have been introduced if, instead of following the above procedure, we had made a nonlocal redefinition of $V$ to cast the original Lagrangian in the $V\Box V$ form. In this case, the (off-shell) Green functions have genuine $a$-dependence.

The general procedure is the following: Consider the Lagrangian of an arbitrary superfield $\psi$ of the form

$$\bar{\psi}\Box^n(\Pi_0 + \sum_i c_i \Pi_i)\psi \quad , \tag{7.3.14}$$

where $\Pi_0 + \sum_i \Pi_i = 1$ and $\Pi_0$ is a particular superspin chosen for convenience, e.g., the highest superspin in $\psi$ or the superspin that occurs most frequently in interaction terms. If some of the constants $c_i$ are equal, we may combine the corresponding projection operators $\Pi_i$ into a single one (including $\Pi_0$, which is merely a $\Pi$ with $c_0 = 1$). Also, some $c_i$ may vanish, which implies a corresponding gauge invariance. We now introduce catalysts by the shifts

$$\psi \to \psi + \sum_i O_i \hat{\psi}_i \quad , \quad \Pi_i O_j = \delta_{ij} O_j \quad , \tag{7.3.15}$$

where $O_i$ are operators that may be nonlocal, but only to the extent that all nonlocalities in the interaction terms can eventually be removed. Then we fix the corresponding gauge invariances

$$\delta\psi = \sum_i O_i \lambda_i \quad , \qquad \delta\hat{\psi}_i = -\,\lambda_i \quad , \tag{7.3.16}$$

in such a way that the Lagrangian for $\psi$ becomes simply $\bar{\psi}\Box^n\psi$, and all crossterms between $\psi$ and $\hat{\psi}_i$ are canceled: The gauge-fixing functions

$$F_i = O_i{}^\dagger(\psi - \frac{c_i}{1 - c_i}\, O_i \hat{\psi}_i) \tag{7.3.17}$$

with gauge-fixing terms

$$(1 - c_i)\bar{F}_i \Box^n (O_i{}^\dagger O_i)^{-1} F_i \tag{7.3.18}$$

give the Lagrangian

$$\bar{\psi}\Box^n\psi + \sum_i \frac{c_i}{1-c_i}\hat{\bar{\psi}}_i O_i{}^\dagger \Box^n O_i \hat{\psi}_i \quad . \tag{7.3.19}$$

($O_i{}^\dagger O_i$ is invertible on $F_i$. Also, it can be shown that $O_i(O_i{}^\dagger O_i)^{-1}O_i{}^\dagger = \Pi_i$.) Note that this procedure includes fixing of the ordinary gauge invariance. We then add the Faddeev-Popov and Nielsen-Kallosh ghosts as

$$\sum_i [\bar{\psi}'_{2i}(O_i{}^\dagger O_i)\psi'_{1i} + \bar{\psi}'_{3i}\Box^n(O_i{}^\dagger O_i)^{-1}\psi'_{3i} + h.\,c.] \quad . \tag{7.3.20}$$

In the interacting case, $O_i$ must be chosen so that any background dependence comes out local, including extra terms which may result from manipulations of the background dependent $\Box$ and $O_i$. It may be necessary to choose $O_i$ such that $\hat{\psi}_i$ has its own gauge invariance independent of $\psi$, or to combine several $\Pi_i$ in such a way that the above Lagrangian for $\hat{\psi}$ also needs fixing, in which case the entire procedure must be repeated for those $\hat{\psi}$'s. However, the $\hat{\psi}$'s always have fewer components than their $\psi$'s (at least for $N = 0$ or 1 supersymmetry), so the series must eventually terminate.

## d. Choice of gauge parameters

In any gauge theory, some care is required to ensure that the Faddeev-Popov quantization procedure will lead to correct, unitary results. One way to check unitarity is to compare results with those obtained in a ghost-free (e.g., axial) gauge. In supersymmetric theories such a gauge is the Wess-Zumino gauge, and one way to insure unitarity is by making certain that one can pass smoothly from covariant gauges to the physical gauge without introducing any extra unphysical degrees of freedom. This will certainly be the case if the superfield gauge transformations are such that they allow the gauging to zero of the unphysical components by algebraic, non-derivative transformations ($\delta A = \lambda$ and not, e.g., $\Box\lambda$ or $\partial_{\underline{a}}\lambda^{\underline{a}}$). For example, in ordinary component Yang-Mills theory the gauge transformation can be written either as $\delta A_{\underline{a}} = \nabla_{\underline{a}}\lambda$ or as $\delta A_{\underline{a}} = \nabla_{\underline{a}}\Box\lambda'$. However, the latter choice would give a Faddeev-Popov ghost Lagrangian $c'\partial\cdot\nabla\Box c$ (instead of just $c'\partial\cdot\nabla c$), and the extra $\Box$ would give a nontrivial contribution which would destroy unitarity. It is possible to modify the Faddeev-Popov prescription to correctly handle the situation, but the simplest procedure is to choose the gauge parameters in such a way as to avoid the problem.

In the case of supergravity, it can be verified by the procedure just described (cf. 5.2.10) that the $L_\alpha$ parametrization is the only correct one. As we saw in sec. 5.2.c this parametrization can be used only if we also have the compensator(s) in the theory. Elimination of the compensator would introduce constraints on the gauge parameter, the solution of which would express $L_\alpha$ in terms of derivatives of other superfield parameters. However, as in the example above, this would introduce spurious extra ghosts in the naive Faddeev-Popov procedure and unitarity would be lost, unless the procedure were modified.

There is one more restriction which must be observed in choosing gauge parametrizations (and thus ghosts): In general, not all superfields which are representations of global supersymmetry are also representations of local supersymmetry. In particular, for $n = -\frac{1}{3}$ the only type of chiral superfields allowed are ones with only undotted spinor indices: The existence of a dotted chiral spinor would imply

$$0 = \{\overline{\nabla}_{\dot\alpha} , \overline{\nabla}_{\dot\beta}\}\Phi_{\dot\gamma} = -2RM_{\dot\alpha\dot\beta}\Phi_{\dot\gamma} = -RC_{\dot\gamma(\dot\alpha}\Phi_{\dot\beta)} \neq 0 \quad . \tag{7.3.21}$$

Generally, the choice of gauge parameters must be restricted to those which can exist in an arbitrary background.

## 7.4. Quantization

In this section we present the details of the quantization procedure for supergravity. This involves choosing gauge-fixing functions which allow all kinetic terms to take simple forms, and finding the resulting ghosts (Faddeev-Popov, Nielsen-Kallosh, and hidden). Such simplifications often require the use of appropriate compensators and/or catalysts. This procedure is first applied to the physical fields, then to the resulting ghosts, the ghosts' ghosts, etc. (The ghosts reduce in size at each step, so the procedure quickly terminates.) For now we work with on-shell background fields ($\mathbf{R}=\mathbf{G}=0$), so the part of the action quadratic in the quantum fields becomes (see (7.2.24,34)), in units $\kappa = 1$ (or making the usual rescaling $(H,\chi) \rightarrow \kappa(H,\chi)$),

$$S = \int d^4x\, d^4\theta\, \mathbf{E}^{-1}[-3\chi\bar{\chi} + i(\chi - \bar{\chi})\boldsymbol{\nabla}\cdot H - \frac{1}{2} H \cdot \Box H - \frac{1}{4}(\boldsymbol{\nabla}\cdot H)^2$$

$$+ \frac{1}{12}([\bar{\boldsymbol{\nabla}}_{\dot{\beta}}, \boldsymbol{\nabla}_{\alpha}]H^{\alpha\dot{\beta}})^2 + (\bar{\boldsymbol{\nabla}}^2 H)\cdot(\boldsymbol{\nabla}^2 H)$$

$$- \frac{1}{2} H^{\alpha\dot{\beta}}(\mathbf{W}_{\alpha}{}^{\gamma\delta}\boldsymbol{\nabla}_{\gamma}H_{\delta\dot{\beta}} + \bar{\mathbf{W}}_{\dot{\beta}}{}^{\dot{\gamma}\dot{\delta}}\bar{\boldsymbol{\nabla}}_{\dot{\gamma}}H_{\alpha\dot{\delta}})] \quad . \tag{7.4.1}$$

The quantization with on-shell background fields is sufficient for computing physical quantities (S-matrix elements) in pure $N=1$ and extended supergravity. We will discuss the general situation later.

We have the following (off-shell) gauge invariance under the quantum transformations (from (7.2.14) and (7.2.22,23)):

$$\delta H_{\alpha\dot{\beta}} = (\boldsymbol{\nabla}_{\alpha}\bar{L}_{\dot{\beta}} - \bar{\boldsymbol{\nabla}}_{\dot{\beta}}L_{\alpha}) + O(H) + O(\chi) \quad ,$$

$$\delta\phi \;\; = -(\bar{\boldsymbol{\nabla}}^2 + \mathbf{R})\phi^{-3}L^{\alpha}\boldsymbol{\nabla}_{\alpha}\phi - \frac{1}{3}[(\bar{\boldsymbol{\nabla}}^2 + \mathbf{R})\boldsymbol{\nabla}^{\alpha}\phi^{-3}L_{\alpha}]\phi \quad . \tag{7.4.2a}$$

Note that the second equation can be rewritten as

$$\delta\phi^3 = (\bar{\boldsymbol{\nabla}}^2 + \mathbf{R})\boldsymbol{\nabla}_{\alpha}L^{\alpha} \quad . \tag{7.4.2b}$$

(On shell we can set $\mathbf{R} = 0$.)

To cancel the $H\chi$ crossterms, we choose the following gauge-fixing function:

$$F_{\alpha} = \bar{\boldsymbol{\nabla}}^{\dot{\beta}}(H_{\alpha\dot{\beta}} + i\bar{a}\boldsymbol{\nabla}_{\alpha\dot{\beta}}\Box_{-}^{-1}\bar{\phi}^3) \tag{7.4.3}$$

(for some constant $a$ to be determined below). This is the most convenient gauge choice. It corresponds to a modification of the transverse gauge $\overline{\nabla}^{\dot\beta} H_{\alpha\dot\beta} = 0$ (see sec. 7.5.b). We have defined the chiral d'Alembertians

$$\Box_+ = \Box + \mathbf{W}^{\alpha}{}_{\beta}{}^{\gamma}\nabla_\alpha M_\gamma{}^\beta \quad , \quad \Box_- = \Box + \overline{\mathbf{W}}^{\dot\alpha}{}_{\dot\beta}{}^{\dot\gamma}\overline{\nabla}_{\dot\alpha}\overline{M}_{\dot\gamma}{}^{\dot\beta} \quad , \tag{7.4.4a}$$

on arbitrary chiral superfields (i.e., with any number of undotted spinor indices) by

$$\Box_+\phi_{\alpha\cdots\beta} = \overline{\nabla}^2\nabla^2\phi_{\alpha\cdots\beta} \quad , \quad \Box_-\bar\phi_{\dot\alpha\cdots\dot\beta} = \nabla^2\overline{\nabla}^2\bar\phi_{\dot\alpha\cdots\dot\beta} \quad . \tag{7.4.4b}$$

We have used $\bar\phi^3 = (1+\bar\chi)^3$ in (7.4.3) instead of just $\bar\chi$ so that the Faddeev-Popov procedure will contribute nonlocal terms to *only the kinetic terms* of the ghosts, and not to their quantum interactions, due to the form of (7.4.2b). (This is the reason for our introduction of $\phi^{-3}$ into the transformation laws (7.2.22).) Nonlocal kinetic terms can be made local by use of catalyst ghosts, so this is a harmless nonlocality, whereas nonlocal interaction terms would be a problem.

We fix the gauge by first completing the linear superfield gauge fixing function $F_\alpha$ to a general superfield, thereby introducing hidden ghosts $\phi_\alpha$, and subsequently averaging over gauges by using a weighting function that leads to some of the desired simplifications: We wish to cancel all $H$ terms in the Lagrangian which would contribute to propagators except for $H\Box H$, including the $H\chi$ cross terms. This gauge is chosen by introducing in the functional integral the factor

$$\int I\!D\zeta_\alpha\, I\!D\bar\zeta_{\dot\alpha}\, I\!D\phi_\alpha\, I\!D\bar\phi_{\dot\alpha}\, \delta(F_\alpha + \nabla^2\Box_+^{-1}\phi_\alpha - \zeta_\alpha)\delta(\bar F_{\dot\alpha} + \overline{\nabla}^2\Box_-^{-1}\bar\phi_{\dot\alpha} - \bar\zeta_{\dot\alpha})$$

$$\times exp\{\int d^4x\, d^4\theta\, \mathbf{E}^{-1}$$

$$\times[-\tfrac{1}{4}(\nabla^\alpha\zeta_\alpha - \overline{\nabla}^{\dot\alpha}\bar\zeta_{\dot\alpha})^2 - \tfrac{1}{12}(\nabla^\alpha\zeta_\alpha + \overline{\nabla}^{\dot\alpha}\bar\zeta_{\dot\alpha})^2 + (\nabla^\alpha\bar\zeta^{\dot\beta})(\overline{\nabla}_{\dot\beta}\zeta_\alpha)]\} \quad , \tag{7.4.5}$$

and carrying out the integrals over $\zeta_\alpha$. This gives the gauge-fixing terms and the hidden ghost action

$$S_{GF} = \int d^4x\, d^4\theta\, \mathbf{E}^{-1}\{[\tfrac{1}{4}(\nabla\cdot H)^2 - \tfrac{1}{12}([\overline{\nabla}_{\dot\beta}, \nabla_\alpha]H^{\alpha\dot\beta})^2 - (\overline{\nabla}^2 H)\cdot(\nabla^2 H)]$$

$$+ [ -\frac{5}{3} i(\boldsymbol{\nabla} \cdot H)(a\phi^3 - \bar{a}\bar{\phi}^3) - \frac{4}{3}(a^2\phi^6 + \bar{a}^2\bar{\phi}^6) + \frac{10}{3} a\bar{a}\phi^3\bar{\phi}^3]$$

$$+ \bar{\phi}^{\dot{\beta}} i\boldsymbol{\nabla}_{\alpha\dot{\beta}} \boldsymbol{\Box}_+{}^{-1}\phi^\alpha \Big\} \quad . \tag{7.4.6a}$$

Upon linearization (and using the on-shell condition $\mathbf{R} = 0$), the $\phi$ terms become $-5i(\boldsymbol{\nabla} \cdot H)(a\chi - \bar{a}\bar{\chi}) + 30a\bar{a}\chi\bar{\chi}$, so we choose

$$a = \frac{1}{5} \tag{7.4.6b}$$

to cancel the $H\chi$ crossterms in (7.4.1).

We can show, however, that the hidden ghost Lagrangian gives no contributions. (Note that it has no quantum interactions and contributes at most at the one-loop level.) We perform two successive "rotations" (with unit Jacobian)

$$(1) \qquad \phi_\alpha \to \phi_\alpha + ib\boldsymbol{\nabla}_{\alpha\dot{\beta}} \bar{\boldsymbol{\nabla}}^2 \boldsymbol{\Box}_-^{-1} \bar{\phi}^{\dot{\beta}} \quad , \quad \bar{\phi}_{\dot{\beta}} \to \bar{\phi}_{\dot{\beta}} \ ; \tag{7.4.7a}$$

$$(2) \qquad \phi_\alpha \to \phi_\alpha \quad , \quad \bar{\phi}_{\dot{\beta}} \to \bar{\phi}_{\dot{\beta}} + ic\boldsymbol{\nabla}_{\alpha\dot{\beta}} \boldsymbol{\nabla}^2 \boldsymbol{\Box}_+^{-1} \phi^\alpha \ ; \tag{7.4.7b}$$

choose $b$ and $c$ to cancel the $\phi_\alpha \bar{\phi}_{\dot{\alpha}}$ crossterms, and rewrite the hidden-ghost action in chiral form

$$S = \int d^2\theta \; e^{-\bar{\boldsymbol{\Omega}}} \boldsymbol{\phi}^3 \phi^\alpha \phi_\alpha + h.\,c. \quad , \tag{7.4.8a}$$

(recall that the background is in vector representation) or, with the field redefinition $\phi_\alpha \to \boldsymbol{\phi}^{-\frac{3}{2}} \phi_\alpha$,

$$S = \int d^2\theta \; e^{-\bar{\boldsymbol{\Omega}}} \phi^\alpha \phi_\alpha + h.\,c. = \int d^2\theta \; \hat{\phi}^\alpha \hat{\phi}_\alpha + h.\,c. \ , \tag{7.4.8b}$$

in terms of an ordinary chiral superfield $\hat{\phi}_\alpha = e^{-\bar{\boldsymbol{\Omega}}} \phi_\alpha$. Thus the hidden ghost decouples from the background field and gives no contribution to the effective action. (This is just the covariantization of (7.3.3c) for $m = -1$.)

The Faddeev-Popov ghosts $\psi, \psi'$ are obtained in standard fashion from the gauge-fixing functions. For the time being we write only the kinetic terms (arising from the $H$ and $\chi$ independent part of the gauge transformation; the remainder gives rise to ghost-quantum field interactions). After some algebra we obtain

$$-(\bar{\nabla}^{\dot{\beta}}\psi'^{\alpha} - \nabla^{\alpha}\bar{\psi}'^{\dot{\beta}})(\bar{\nabla}_{\dot{\beta}}\psi_{\alpha} - \nabla_{\alpha}\bar{\psi}_{\dot{\beta}})$$

$$-\frac{1}{5}\left[(\bar{\nabla}^2\nabla^{\alpha}\psi'_{\alpha})\Box_{-}{}^{-1}(\nabla^2\bar{\nabla}^{\dot{\beta}}\bar{\psi}_{\dot{\beta}}) + (\nabla^2\bar{\nabla}^{\dot{\beta}}\bar{\psi}'_{\dot{\beta}})\Box_{+}{}^{-1}(\bar{\nabla}^2\nabla^{\alpha}\psi_{\alpha})\right] \ . \qquad (7.4.9)$$

We now wish to simplify the ghost Lagrangian by putting it in the standard form $\bar{\psi}'^{\dot{\alpha}}\nabla_{\alpha\dot{\alpha}}\psi^{\alpha}$. We therefore introduce catalysts with the shifts

$$\psi_{\alpha} \rightarrow \psi_{\alpha} + \nabla_{\alpha}(V_1 + iV_2) \ , \qquad \psi'_{\alpha} \rightarrow \psi'_{\alpha} + \nabla_{\alpha}(V'_1 + iV'_2) \ . \qquad (7.4.10)$$

In addition to the invariance due to these shifts, the Lagrangian has also the invariance due to the fact that the fields appear only as $\bar{\nabla}_{\dot{\alpha}}\psi_{\alpha}$, $\bar{\nabla}_{\dot{\alpha}}\psi'_{\alpha}$. We thus have the gauge transformations

$$\delta\psi_{\alpha} = \Lambda_{\alpha} + \nabla_{\alpha}L \ , \qquad \delta(V_1 + iV_2) = -L \ , \qquad \bar{\nabla}_{\dot{\beta}}\Lambda_{\alpha} = 0 \ ;$$

$$\delta\psi'_{\alpha} = \Lambda'_{\alpha} + \nabla_{\alpha}L' \ , \qquad \delta(V'_1 + iV'_2) = -L' \ , \qquad \bar{\nabla}_{\dot{\beta}}\Lambda'_{\alpha} = 0 \ ; \qquad (7.4.11)$$

with the chiral spinor parameters $\Lambda_{\alpha}$, $\Lambda'_{\alpha}$. These parameters will introduce second generation chiral spinor ghosts $\phi_{\alpha}$, $\phi'_{\alpha}$, and we also have the real scalar ghosts $V''_{1,2,3,4}$ associated with the invariances parametrized by the complex $L$, $L'$. After gauge fixing and some changes of variables the kinetic Lagrangian can be put in standard form (see (7.4.14a)). The one-loop contribution of $V_i$, $V'_i$ cancels that of $V''_i$, but $V_i$ and $V'_i$ have quantum interactions (because $\psi_{\alpha}$ and $\psi'_{\alpha}$ do, and $V_i$ and $V'_i$ enter through the shifts in (7.4.10)).

The averaging in (7.4.5) has to be normalized by introducing a Nielsen-Kallosh ghost $\Psi_{3\alpha}$, to compensate the contributions from the $\zeta_{\alpha}$ fields. Its Lagrangian is

$$-\frac{1}{4}(\nabla^{\alpha}\psi_{3\alpha} - \bar{\nabla}^{\dot{\alpha}}\bar{\psi}_{3\dot{\alpha}})^2 - \frac{1}{12}(\nabla^{\alpha}\psi_{3\alpha} + \bar{\nabla}^{\dot{\alpha}}\bar{\psi}_{3\dot{\alpha}})^2 + (\nabla^{\alpha}\bar{\psi}_3{}^{\dot{\beta}})(\bar{\nabla}_{\dot{\beta}}\psi_{3\alpha}) \ . \qquad (7.4.12)$$

We can now apply the usual procedure (as described in sec. 7.3) of the catalysts to place the Lagrangian in the standard form: We shift $\psi_{3\alpha}$ by the representations whose coefficients in (7.4.12) are not 1, and then fix the gauge to make them 1 (see (7.3.11-13) for an example). In this case, we make the shift

$$\psi_{3\alpha} \rightarrow \psi_{3\alpha} + \bar{\nabla}^2\nabla_{\alpha}\psi_3 + \nabla_{\alpha}\phi_3 \quad , \qquad \bar{\nabla}_{\dot{\alpha}}\phi_3 = 0 \ . \qquad (7.4.13)$$

This allows us to fix the superspin 0 ($\phi_3$) and two of the (four) superspin $\frac{1}{2}$ ($\psi_3$) parts of $\psi_{3\alpha}$'s kinetic term. We then choose the most general gauge-fixing functions and weightings for the new invariances (corresponding to arbitrary variations of $\phi_3$ and $\psi_3$, and the corresponding variations of $\psi_{3\alpha}$), introduce the appropriate new ghosts, make shifts, etc. The net result is that all the catalysts cancel as in the example (7.3.11), leaving us with just the Nielsen-Kallosh ghost with conventional Lagrangian. (This ghost has no quantum interactions.)

In fact, the form of the (quantum-quadratic) Lagrangian was predictable for all fields except $H$, since by dimensional analysis and Lorentz invariance (and, when relevant, chirality) only it could have nonminimal terms not resulting from direct background covariantization (i.e., $\mathbf{W}_{\alpha\beta\gamma}$ terms).

The final result of the quantization procedure is the following: We write the whole effective Lagrangian as a sum of a quadratic part and the rest, with the quadratic part being

$$S = \int d^4x\, d^4\theta\, \mathbf{E}^{-1}[-\frac{1}{2}H^{\alpha\dot{\alpha}}\widehat{\Box}H_{\alpha\dot{\alpha}} - \frac{9}{5}\chi\bar{\chi}$$
$$+ (\bar{\psi}'^{\dot{\alpha}} i\nabla^{\alpha}{}_{\dot{\alpha}}\psi_{\alpha} + \psi'^{\alpha} i\nabla_{\alpha}{}^{\dot{\alpha}}\bar{\psi}_{\dot{\alpha}} + \bar{\psi}_3{}^{\dot{\alpha}} i\nabla^{\alpha}{}_{\dot{\alpha}}\psi_{3\alpha})$$
$$+ (3V'_1\Box V_1 + V'_2\Box V_2) + \sum_{i=1}^{4}\frac{1}{2}V''_i\Box V''_i + \sum_{i=1}^{2}(\frac{1}{2}\phi_i{}^{\alpha}\nabla^2\phi_{i\alpha} + h.c.)] \quad , \tag{7.4.14a}$$

where

$$\widehat{\Box} = \Box + \mathbf{W}^{\alpha}{}_{\beta}{}^{\gamma}\nabla_{\alpha}M_{\gamma}{}^{\beta} + \bar{\mathbf{W}}^{\dot{\alpha}}{}_{\dot{\beta}}{}^{\dot{\gamma}}\bar{\nabla}_{\dot{\alpha}}\bar{M}_{\dot{\gamma}}{}^{\dot{\beta}} \quad . \tag{7.4.14b}$$

In these formulae $\psi^{\alpha}$, $\psi'^{\alpha}$, $\psi_3{}^{\alpha}$, $V_i$ and $V'_i$ have abnormal statistics. This expression is sufficient for one-loop calculations. Note that at one-loop the contributions from the various $V$'s cancel due to statistics.

The higher-loop contributions come from quantum interaction terms originating in three places: (a) the higher order (cubic, quartic, etc.) terms in the expansion of the classical action (7.2.24); (b) the gauge-fixing term; (c) the higher order terms in the Faddeev-Popov Lagrangian. The latter has the symbolic form $(antighost)\delta_{ghost}(gauge\ fixing\ function)$, where the variation is the full nonlinear variation (7.2.19,22,23), with the gauge parameter $L_\alpha$ replaced by the ghost $\psi_\alpha$. Since we have made the shifts (7.4.10) for the ghosts, the fields $V_i$, $V'_i$ will also appear. Thus, the quantum vertices are obtained from the higher order terms (beyond quadratic) in the expansion of (see (7.2.24,7.4.6))

$$S_C + S_{GF} - \{\boldsymbol{\nabla}^\alpha[\bar{\psi}'^{\dot\alpha} + \bar{\boldsymbol{\nabla}}^{\dot\alpha}(V\,'_1 - iV\,'_2)] - \bar{\boldsymbol{\nabla}}^{\dot\alpha}[\psi'^\alpha + \boldsymbol{\nabla}^\alpha(V\,'_1 + iV\,'_2)]\}$$

$$\times\delta H_{\alpha\dot\alpha}(\psi_\alpha + \boldsymbol{\nabla}_\alpha(V_1 + iV_2)) \quad , \tag{7.4.15}$$

where $\delta H_{\alpha\dot\alpha}(L_\alpha)$ is the expression obtained by substituting (7.2.22,23) into (7.2.19). We have performed an integration by parts in the second term. We note that while both terms in the gauge fixing function (7.4.3) lead to interactions of the ghosts with the background fields, only the first term $\bar{\boldsymbol{\nabla}}^{\dot\alpha}H_{\alpha\dot\alpha}$ leads to (local) interactions between the ghosts and the quantum fields. This is the end of the quantization process.

We have discussed the quantization procedure in the formulation with the chiral compensator $\phi$. As discussed in sec. 5.2.d, another possible choice for compensator is a real scalar superfield $V$ introduced through a variant representation. The treatment of the corresponding formulation can be obtained by making the substitution (even off shell)

$$\phi^3 \to 1 + (\bar{\boldsymbol{\nabla}}^2 + \mathbf{R})V \quad , \qquad V = \bar{V} \quad . \tag{7.4.16}$$

$V$ has the transformation laws

Background:

$$V\,' = e^{iK}V \quad ,$$

Quantum:

$$V\,' = V + (\boldsymbol{\nabla}_\alpha L^\alpha + \bar{\boldsymbol{\nabla}}_{\dot\alpha}\bar{L}^{\dot\alpha}) \quad . \tag{7.4.17}$$

We use now the gauge fixing function

$$F_\alpha = \bar{\boldsymbol{\nabla}}^{\dot\alpha}H_{\alpha\dot\alpha} - \frac{1}{5}\boldsymbol{\nabla}_\alpha V \quad . \tag{7.4.18}$$

The shifts (7.4.10) are again made, and by a procedure similar to the one described above we obtain the following results: The higher-order terms in the action are again given by (7.4.15), with the substitution (7.4.16) (but now $S_{GF}$ does not contribute).

However, the quadratic terms are now

$$S = \int d^4x\, d^4\theta\, \mathbf{E}^{-1}[-\frac{1}{2} H^{\alpha\dot{\alpha}} \widehat{\Box} H_{\alpha\dot{\alpha}} - \frac{1}{10} V \Box V$$

$$+ (\bar{\psi}'^{\dot{\alpha}} i \nabla^{\alpha}{}_{\dot{\alpha}} \psi_{\alpha} + \psi'^{\alpha} i \nabla_{\alpha}{}^{\dot{\alpha}} \bar{\psi}_{\dot{\alpha}} + \bar{\psi}_3{}^{\dot{\alpha}} i \nabla^{\alpha}{}_{\dot{\alpha}} \psi_{3\alpha})$$

$$+ (3V'_1 \Box V_1 + V'_2 \Box V_2) + \sum_{i=1}^{7} \frac{1}{2} V''_i \Box V''_i + \sum_{i=1}^{7} \chi_i \bar{\chi}_i] \quad . \qquad (7.4.19)$$

Here $\psi_\alpha$, $\psi_\alpha{}'$, $\psi_{3\alpha}$, $V_i$, $V_i{}'$, and $\chi_i$ have abnormal statistics, and $\chi_i$ are chiral.

In general, the total field content is the following: (a) physical fields $H$ and $\chi$ (or $V$), which contribute at all loops; (b) the first-generation Faddeev-Popov ghosts $\psi_\alpha$ and $\psi'_\alpha$, which contribute at all loops; (c) the first-generation Nielsen-Kallosh ghost $\psi_{3\alpha}$ and all higher-generation ghosts, which contribute only at one loop; (d) the catalyst ghosts $V_i$ and $V_i{}'$, which contribute at only more than one loop (being canceled at the one-loop level by the contribution from the $V''$s). We will discuss in sec. 7.10 some of the differences between the formulations (7.4.14) and (7.4.19).

## 7.5. Supergravity supergraphs

### a. Feynman rules

In the next section we shall consider further the background field quantization and discuss its applications. In this section we consider ordinary quantization and discuss the Feynman rules for supergravity-matter systems. There is no need to go again through the gauge-fixing procedure. We simply take the results of the previous section and set the background fields to zero. Therefore the supergravity quantum action is given by (7.4.14,15), where $\mathbf{E} = 1$, all the derivatives are flat space derivatives, and all chiral fields ordinary chiral. Furthermore, the fields $\psi_{3\alpha}$, $V_i{}''$ and $\phi_{i\alpha}$ can be dropped since they have no interactions (but the Faddeev-Popov fields $\psi_\alpha$, $\psi'{}_\alpha$, and the catalysts $V_i$, $V'{}_i$ do). Equivalently, we can work with (7.4.15,19), dropping $\psi_{3\alpha}$, $V_i{}''$, and $\chi_i$. Matter actions, covariantized with respect to $H_{\alpha\dot{\alpha}}$ and $\phi$, can be added.

From the flat space form of (7.4.14a) we obtain ordinary propagators. In particular we have

$$HH\ propagator: \quad -\frac{\delta_\alpha{}^\beta \delta_{\dot{\alpha}}{}^{\dot{\beta}}}{p^2}\,\delta^4(\theta-\theta') \tag{7.5.1}$$

$$\psi\,\overline{\psi}\ propagator: \quad \frac{p_\alpha{}^{\dot{\alpha}}}{p^2}\,\delta^4(\theta-\theta') \quad , \tag{7.5.2}$$

and the usual propagators for $\chi$ and $V$. Vertices are obtained from the expansion of (7.4.15), as well as from matter actions. For example, consider the kinetic action of a scalar multiplet $\eta_{cov}$:

$$S = \int d^4x\, d^4\theta\ E^{-1}\eta_{cov}\overline{\eta}_{cov}$$

$$= \int d^4x\, d^4\theta\ (\hat{E})^{-\frac{1}{3}}(1\cdot e^{-\overleftarrow{H}})^{\frac{1}{3}}\eta\phi e^{-H}\overline{\eta}\overline{\phi} \quad . \tag{7.5.3}$$

We have expressed $\eta_{cov}$ in terms of a flat space chiral superfield $\eta$ and we are working in the chiral representation. We must expand now the various factors in powers of $H_{\alpha\dot{\alpha}}$ and $\chi = \phi - 1$. However, the expansions were carried out in (7.2.30-32). Replacing background covariant derivatives with flat space derivatives, we find the cubic

interactions

$$S^{(3)} = \int d^4x\, d^4\theta\, \{(\chi + \bar{\chi})\eta\bar{\eta} + \eta[-H^{\underline{a}} i\partial_{\underline{a}} - \frac{1}{3}(\bar{D}_{\dot{\alpha}} D_\alpha H^{\underline{a}}) - \frac{1}{3}(i\partial^{\underline{a}} H_{\underline{a}})]\bar{\eta}\}$$

$$= \int d^4x\, d^4\theta\, [(\chi + \bar{\chi})\eta\bar{\eta} + H^{\underline{a}}(\frac{1}{2}\bar{\eta} i\overleftrightarrow{\partial}_{\underline{a}}\eta - \frac{1}{6}[\bar{D}_{\dot{\alpha}}, D_\alpha]\bar{\eta}\eta)] \quad . \qquad (7.5.4)$$

The same expansions can be used to find the supergravity vertices, but with backgrounds set to zero the algebra is much simpler. Thus, from $\widehat{\nabla}_\alpha = e^{-H} D_\alpha e^H$, $\overline{\nabla}_{\dot{\alpha}} = \bar{D}_{\dot{\alpha}}$, we obtain

$$\Delta_{\dot{\alpha}}{}^B = \Delta_\alpha{}^\beta = \Delta_\alpha{}^{\dot{\beta}} = \Delta_{\underline{a}}{}^\beta = \Delta_{\underline{a}}{}^{\dot{\beta}} = 0 \quad ,$$

$$\Delta_{\underline{a}}{}^{\underline{b}} = -\, i\bar{D}_{\dot{\alpha}}\Delta_\alpha{}^{\underline{b}} \quad , \qquad (7.5.5)$$

where $\Delta_\alpha{}^{\underline{b}}$ is obtained as the coefficient of $\partial_{\underline{b}}$ in (7.2.27b):

$$\Delta_\alpha{}^{\underline{b}} = i(D_\alpha H^{\underline{b}}) - \frac{1}{2}[(D_\alpha H^{\underline{c}})\partial_{\underline{c}} H^{\underline{b}} - H^{\underline{c}}(\partial_{\underline{c}} D_\alpha H^{\underline{b}})] + \cdots . \qquad (7.5.6)$$

(To find the cubic interactions we do not need the third order term in $\Delta_\alpha{}^{\underline{b}}$ since it only contributes a total $(\bar{D}_{\dot{\alpha}})$ derivative.) Therefore

$$\hat{E}^{-\frac{1}{3}} = [det(\delta_{\underline{a}}{}^{\underline{b}} + \Delta_{\underline{a}}{}^{\underline{b}})]^{-\frac{1}{3}}$$

$$= 1 - \frac{1}{3}\Delta_{\underline{a}}{}^{\underline{a}} + \frac{1}{6}\Delta_{\underline{a}}{}^{\underline{b}}\Delta_{\underline{b}}{}^{\underline{a}} + \frac{1}{18}(\Delta_{\underline{a}}{}^{\underline{a}})^2 - \frac{1}{18}\Delta_{\underline{a}}{}^{\underline{a}}\Delta_{\underline{b}}{}^{\underline{c}}\Delta_{\underline{c}}{}^{\underline{b}}$$

$$- \frac{1}{9}\Delta_{\underline{a}}{}^{\underline{b}}\Delta_{\underline{b}}{}^{\underline{c}}\Delta_{\underline{c}}{}^{\underline{a}} - \frac{1}{162}(\Delta_{\underline{a}}{}^{\underline{a}})^3 \quad . \qquad (7.5.7)$$

We also expand (7.2.31) one order higher which gives, again dropping a term which only contributes a total derivative,

$$(1.\, e^{-\overleftarrow{H}})^{\frac{1}{3}} = 1 - \frac{1}{3}i(\partial\cdot H) - \frac{1}{18}(\partial\cdot H)^2 - \frac{1}{6}H\cdot\partial(\partial\cdot H) + \frac{1}{162}i(\partial\cdot H)^3 \quad ,(7.5.8a)$$

$$\phi e^{-H}\bar{\phi} = 1 + \chi + \bar{\chi} + \chi\bar{\chi} - iH\cdot\partial\bar{\chi} - \chi iH\cdot\partial\bar{\chi} - \frac{1}{2}H\cdot\partial(H\cdot\partial\bar{\chi}) \quad . \qquad (7.5.8b)$$

The cubic supergravity action is obtained from the product of (7.5.7) and (7.5.8).

We obtain the cubic ghost-antighost-quantum field vertices from (7.4.15) and (7.2.19,22,23). We need $\delta H_{\alpha\dot{\alpha}}$ to first order in $H_{\alpha\dot{\alpha}}$ and $\chi$:

$$\delta H_{\alpha\dot{\alpha}} = (D_\alpha \bar{L}_{\dot{\alpha}} - \bar{D}_{\dot{\alpha}} L_\alpha) + 3(\chi \bar{D}_{\dot{\alpha}} L_\alpha - \bar{\chi} D_\alpha \bar{L}_{\dot{\alpha}})$$

$$- \frac{1}{2}[-i(D^\beta \bar{L}^{\dot{\beta}} + \bar{D}^{\dot{\beta}} L^\beta)\partial_{\beta\dot{\beta}} H_{\underline{a}} + (\bar{D}^2 L^\beta) D_\beta H_{\underline{a}}$$

$$+ (D^2 \bar{L}^{\dot{\beta}})\bar{D}_{\dot{\beta}} H_{\underline{a}} + iH \cdot \partial(D_\alpha \bar{L}_{\dot{\alpha}} + \bar{D}_{\dot{\alpha}} L_\alpha)] \ . \tag{7.5.9}$$

The cubic ghost action is obtained by substituting $L_\alpha = \psi_\alpha + \nabla_\alpha(V_1 + iV_2)$ (cf. (7.4.15)). These vertices are sufficient for doing some one-loop calculations. However, as we have already mentioned, at least for on-shell fields, the background field method is much simpler. In this method, the above (covariantized) vertices would be needed only for two-loop calculations.

## b. The transverse gauge

The Feynman rules we have discussed above use the particular (weighted) gauge of (7.4.3), which is the most convenient for internal lines. However, when computing gauge invariant quantities, we can use any gauge for the external lines (this is true in both ordinary and background field methods). We discuss here the choice of a globally supersymmetric gauge that is convenient for most calculations.

The superfields $H_{\alpha\dot{\alpha}}$, $\phi$ contain several irreducible representations of supersymmetry. According to (3.9.40) the superspin content of $H_{\alpha\dot{\alpha}}$ is $(\frac{3}{2} \oplus 1 \oplus \frac{1}{2} \oplus \frac{1}{2} \oplus 0)$, while $\phi$ has superspin 0. According to (3.9.36,37), the spinor gauge parameters $L_\alpha \oplus L_{\dot{\alpha}}$ contain superspins $(1 \oplus \frac{1}{2} \oplus \frac{1}{2} \oplus \frac{1}{2} \oplus \frac{1}{2} \oplus 0)$. (The extra superspin $\frac{1}{2}$ representations in the gauge parameter correspond to second-generation ghosts.) Therefore, we should be able to find a gauge where we have eliminated all superspins but $\frac{3}{2}$ and 0. We have two choices, corresponding to eliminating the superspin 0 in $\phi$ (gauging $\phi$ to 1), or in $H_{\alpha\dot{\alpha}}$ (in which case $\phi$ must be kept). The second choice is much more useful, and can be achieved by imposing the *transverse gauge condition* $D^\alpha H_{\alpha\dot{\alpha}} = 0$. Note that this condition (and its complex conjugate) implies $\partial \cdot H = [D_\alpha, \bar{D}_{\dot{\alpha}}]H^{\alpha\dot{\alpha}} = D^2 H^{\alpha\dot{\alpha}} = 0$.

**c. Linearized expressions**

For some computations, we need the explicit expressions for the geometrical quantities in terms of the prepotentials $H_{\alpha\dot\alpha}$, $\phi$. Here we outline the procedure for obtaining the linearized expressions; higher orders can be obtained in similar fashion. We work in the chiral representation, and in the Lorentz gauge $N_\alpha{}^\mu = \delta_\alpha{}^\mu$. After we obtain the results we will consider other Lorentz gauges.

We begin with the linearized expressions of (5.2.78a) (cf. also (7.5.5))

$$\hat{E}_{\dot\mu} = \bar{D}_{\dot\mu}$$

$$\hat{E}_\mu = D_\mu + [D_\mu\, , H] = D_\mu + i(D_\mu H^{\underline{m}})\partial_{\underline{m}} \tag{7.5.10}$$

We set $\phi = 1 + \chi$ and, using for example (7.5.5-7), we have at the linearized level

$$\hat{E} = 1 + \bar{D}_{\dot\nu} D_\nu H^{\nu\dot\nu} \tag{7.5.11}$$

From the form of $\Psi$ in (5.2.78c) we obtain

$$\Psi = 1 + X \tag{7.5.12}$$

where

$$X \equiv \frac{1}{2}\bar\chi - \chi - \frac{1}{6}(2\bar{D}_{\dot\alpha} D_\alpha + D_\alpha \bar{D}_{\dot\alpha})H^{\alpha\dot\alpha} \quad . \tag{7.5.13}$$

In the particular Lorentz gauge we are using we need not distinguish between flat and curved indices. We obtain then

$$E_\alpha = \check{E}_\alpha = D_\alpha + \bar{X} D_\alpha + i(D_\alpha H^{\underline{b}})\partial_{\underline{b}}$$

$$E_{\dot\alpha} = \check{E}_{\dot\alpha} = \bar{D}_{\dot\alpha} + X\bar{D}_{\dot\alpha} \quad . \tag{7.5.14}$$

To find $E_{\underline{a}}$ we write (again using $N_\alpha{}^\mu = \delta_\alpha{}^\mu$)

$$E_{\underline{a}} = \check{E}_{\underline{a}} + i\,\frac{1}{2}\check{C}_{\alpha,\beta(\dot\alpha}{}^{\beta\dot\gamma)}\check{E}_{\dot\gamma} + i\,\frac{1}{2}\check{C}_{\dot\alpha,(\alpha\dot\beta}{}^{\gamma)\dot\beta}\check{E}_\gamma \tag{7.5.15}$$

where $\check{E}_{\underline{a}} \equiv -\,i\{\check{E}_\alpha, \check{E}_{\dot\alpha}\}$. Therefore

$$\check{E}_{\underline{a}} = \partial_{\underline{a}} - i(D_\alpha X)\bar{D}_{\dot\alpha} - i(\bar{D}_{\dot\alpha}\bar{X})D_\alpha$$

$$+ [\bar{D}_{\dot{\alpha}} D_{\alpha} H^{\underline{b}} + (\frac{1}{6}[D_{\delta}, \bar{D}_{\dot{\delta}}]H^{\underline{d}} - \frac{1}{2}(\chi + \bar{\chi}))\delta_{\underline{a}}{}^{\underline{b}}]\partial_{\underline{b}} \tag{7.5.16}$$

The linearized expressions for the $\check{C}$'s can be worked out from their definition. We find finally

$$E_{\underline{a}} = \partial_{\underline{a}} + i[\frac{1}{2}\bar{D}^2 D_{(\alpha} H^{\gamma)}{}_{\dot{\alpha}} - (\bar{D}_{\dot{\alpha}}\bar{X})\delta_{\alpha}{}^{\gamma}]D_{\gamma} + i[-\frac{1}{2}D^2\bar{D}_{(\dot{\alpha}} H_{\alpha}{}^{\dot{\gamma})} - (D_{\alpha}X)\delta_{\dot{\alpha}}{}^{\dot{\gamma}}]\bar{D}_{\dot{\gamma}}$$

$$+ [(\bar{D}_{\dot{\alpha}} D_{\alpha} H^{\underline{b}}) + (X + \bar{X})\delta_{\underline{a}}{}^{\underline{b}}]\partial_{\underline{b}} \quad . \tag{7.5.17}$$

To find the connections we evaluate first the $C_{AB}{}^{C}$, and use the torsion constraints. We find

$$\Phi_{\alpha\beta\gamma} = -\, C_{\alpha(\beta} D_{\gamma)}\bar{X} \quad ,$$

$$\Phi_{\alpha\dot{\beta}\dot{\gamma}} = \frac{1}{2}D^2\bar{D}_{(\dot{\beta}} H_{\alpha\dot{\gamma})} \quad , \quad \Phi_{\dot{\alpha}\beta\gamma} = -\frac{1}{2}\bar{D}^2 D_{(\beta} H_{\gamma)\dot{\alpha}} \quad ,$$

$$\Phi_{\underline{a}\beta\gamma} = i\,\frac{1}{2}D_{\alpha}\bar{D}^2 D_{(\beta} H_{\gamma)\dot{\alpha}} + iC_{\alpha(\beta}\bar{D}_{\dot{\alpha}} D_{\gamma)}\bar{X} \quad . \tag{7.5.18}$$

The independent field strengths are

$$R = \bar{D}^2(\bar{\chi} - i\,\frac{1}{3}\partial_{\underline{a}} H^{\underline{a}}) \quad ,$$

$$G_{\underline{a}} = -\frac{2}{3}D^{\beta}\bar{D}^2 D_{\beta} H_{\underline{a}} - \frac{1}{6}\epsilon_{\underline{abcd}}\partial^{\underline{b}}[D^{\gamma}, \bar{D}^{\dot{\gamma}}]H^{\underline{d}} - \frac{1}{3}\partial_{\underline{a}}\partial_{\underline{b}} H^{\underline{b}} + i\partial_{\underline{a}}(\chi - \bar{\chi}) \quad ,$$

$$W_{\alpha\beta\gamma} = \frac{1}{6}\bar{D}^2 D_{(\alpha} i\partial_{\beta\dot{\beta}} H_{\gamma)}{}^{\dot{\beta}} \quad . \tag{7.5.19}$$

The remaining field strengths can be read from the solution of the Bianchi identities (5.4.16).

As we have mentioned several times, it is sometimes useful to choose a Lorentz gauge $N_{\alpha}{}^{\mu} \neq \delta_{\alpha}{}^{\mu}$ in which $\Phi_{\dot{\alpha}\beta}{}^{\gamma} = 0$ so that, in the chiral representation, when acting on a field with undotted indices, $\nabla_{\dot{\alpha}}\eta_{\alpha\beta\gamma\cdots} = \Psi\bar{N}_{\dot{\alpha}}{}^{\dot{\mu}}\bar{D}_{\dot{\mu}}\eta_{\alpha\beta\gamma\cdots}$. (That such a gauge is possible follows from $R_{\dot{\alpha}\beta\gamma}{}^{\delta} = 0$, which implies that the above connection is pure gauge.) We reach this gauge by the Lorentz transformation

$$\delta\nabla_A = [L, \nabla_A] \quad , \quad L = \omega_{\alpha}{}^{\beta} M_{\beta}{}^{\alpha} + h.\,c. \tag{7.5.20}$$

so that, in particular,

$$\delta\Phi_{\alpha\dot{\beta}}{}^{\dot{\gamma}} = \omega_\alpha{}^\delta\Phi_{\delta\dot{\beta}}{}^{\dot{\gamma}} - E_\alpha\omega_{\dot{\beta}}{}^{\dot{\gamma}} \quad . \tag{7.5.21}$$

At the linearized level, setting $\Phi'_{\alpha\dot{\beta}}{}^{\dot{\gamma}} = \Phi_{\alpha\dot{\beta}}{}^{\dot{\gamma}} + \delta\Phi_{\alpha\dot{\beta}}{}^{\dot{\gamma}} = 0$, we find

$$\omega_{\dot{\beta}}{}^{\dot{\gamma}} = \frac{1}{2} D_\beta \bar{D}_{(\dot{\beta}} H^{\beta\dot{\gamma})} \tag{7.5.22}$$

and $\omega_\beta{}^\gamma = \overline{(\omega_{\dot{\beta}}{}^{\dot{\gamma}})}$. In this gauge $\bar{N}_{\dot{\alpha}}{}^{\dot{\mu}} = \delta_{\dot{\alpha}}{}^{\dot{\mu}} + \omega_{\dot{\alpha}}{}^{\dot{\mu}}$.

In this gauge, the various quantities of (7.5.19) are shifted according to their index structure. In particular, we find that now

$$\Phi_{\alpha\beta\gamma} = - C_{\alpha(\beta} D_{\gamma)} \bar{X} + \frac{1}{2} D_\alpha \bar{D}_{\dot{\beta}} D_{(\beta} H_{\gamma)}{}^{\dot{\beta}} \quad . \tag{7.5.23}$$

## d. Examples

In this subsection we assume that a regularization scheme exists that preserves local supersymmetry. Such a scheme will be discussed in sec. 7.9. We first compute a massless chiral loop contribution to the supergravity self-energy. The relevant interaction is given by (7.5.4), and the supergraph is given in Fig. 7.5.1.

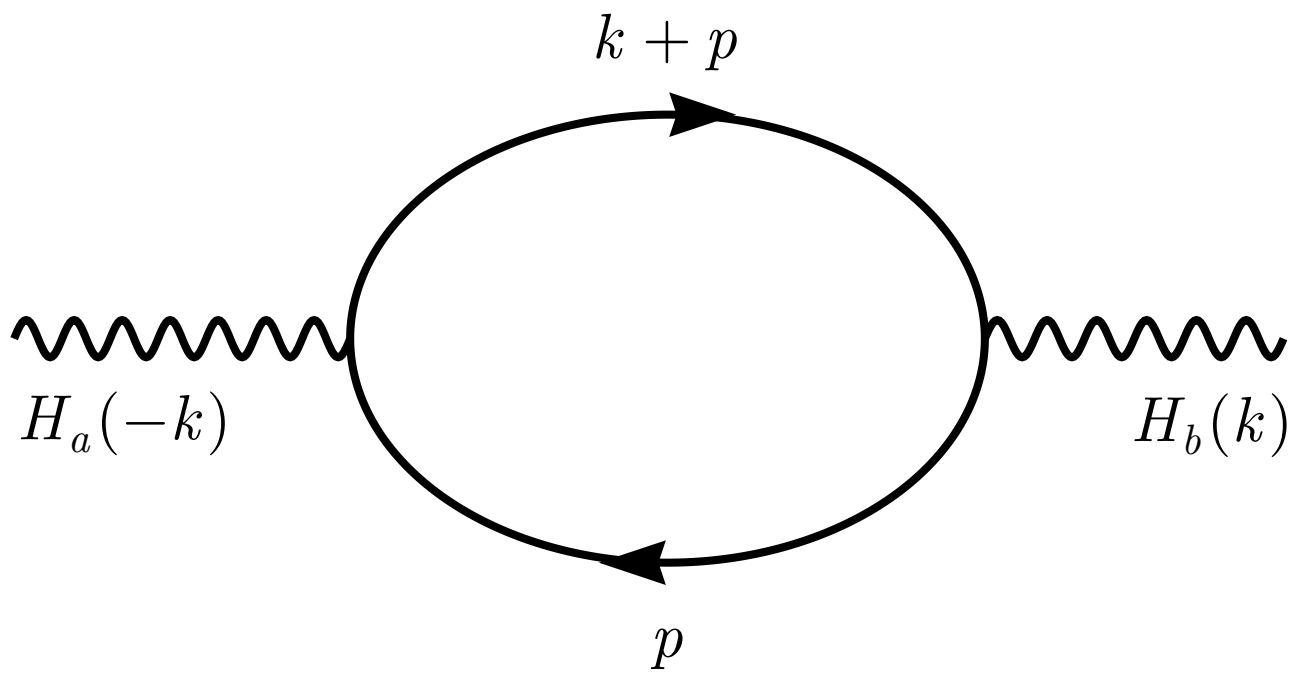

*Fig. 7.5.1*

We note the following simplifications: (a) The $(\chi + \bar{\chi})\eta\bar{\eta}$ vertex leads to only a tadpole contribution to the $\chi\chi$ or $\chi H$ self-energy diagram, and we set this to zero in dimensional regularization for massless $\eta$'s. Equivalently, we observe that in the original action (7.5.3) the compensator $\phi$ can be absorbed into $\eta$ by a field redefinition. (b) With suitable regularization the result should be gauge invariant, and we can work in the transverse gauge where two of the three terms in the $\eta\bar{\eta}H$ vertex do not contribute. The $-H^{\underline{a}}\eta i\partial_{\underline{a}}\bar{\eta}$ vertex is the same as in the Yang-Mills case, if we replace $V^{A}T_{A}$ by $H^{\underline{a}}i\partial_{\underline{a}}$. The result can then be read from (6.3.31) (with the additional momentum factors from $i\partial_{\underline{a}}$)

$$\frac{1}{2}\int\frac{d^4k}{(2\pi)^4}\,d^4\theta\,H^{\underline{a}}(-k,\theta)\int\frac{d^4p}{(2\pi)^4}\,\frac{-p^2-p^{\underline{c}}\bar{D}_{\dot{\gamma}}D_{\gamma}+\bar{D}^2D^2}{p^2(k+p)^2}\,p_{\underline{a}}(k+p)_{\underline{b}}\,H^{\underline{b}}(k,\theta)\quad . \qquad (7.5.24)$$

Using the gauge condition this can be reduced to

$$-\frac{1}{8(\mathrm{D}-1)}\int\frac{d^4k}{(2\pi)^4}\,d^4\theta\;H^{\underline{a}}(-k,\theta)\,k^2k^{\underline{c}}\bar{D}_{\dot{\gamma}}D_{\gamma}\,H_{\underline{a}}(k,\theta)\,I(k^2)\quad , \qquad (7.5.25)$$

where in dimensional regularization

$$I(k^2)=\int\frac{d^{\mathrm{D}}p}{(2\pi)^{\mathrm{D}}}\,\frac{1}{p^2(k+p)^2}=\frac{1}{(4\pi)^{\frac{1}{2}\mathrm{D}}}\,\frac{\Gamma(2-\frac{1}{2}\mathrm{D})[\Gamma(\frac{1}{2}\mathrm{D}-1)]^2}{\Gamma(\mathrm{D}-2)}\,(k^2)^{\frac{1}{2}\mathrm{D}-2}$$

$$=\frac{1}{(4\pi)^2}\,(\frac{1}{\epsilon}-\mathit{ln}\;k^2+\mathit{const.})\quad . \qquad (7.5.26)$$

When acting on $H_{\underline{a}}(k,\theta)$, again using the gauge condition, we can rewrite $k^{\underline{c}}\bar{D}_{\dot{\gamma}}D_{\gamma}=k^{\underline{c}}\frac{1}{2}\{\bar{D}_{\dot{\gamma}},D_{\gamma}\}=k^2$. The fully covariant result can be written as a contribution to the effective action of the form $[c_1\int d^2\theta(W_{\alpha\beta\gamma})^2+h.\,c.+c_2\int d^4\theta(G^2+2\bar{R}R)]I$. However, the coefficients $c_1$, $c_2$ cannot be determined from just a two-point calculation (except in the background field method: see sec. 7.8). On shell only the first term survives. Although the result is independent of $\phi$, the compensator reappears in the course of separating out the divergent part (see sec. 7.10).

As a second example we compute supergravity corrections to the chiral self-energy. The graphs are those of Fig. 7.5.2:

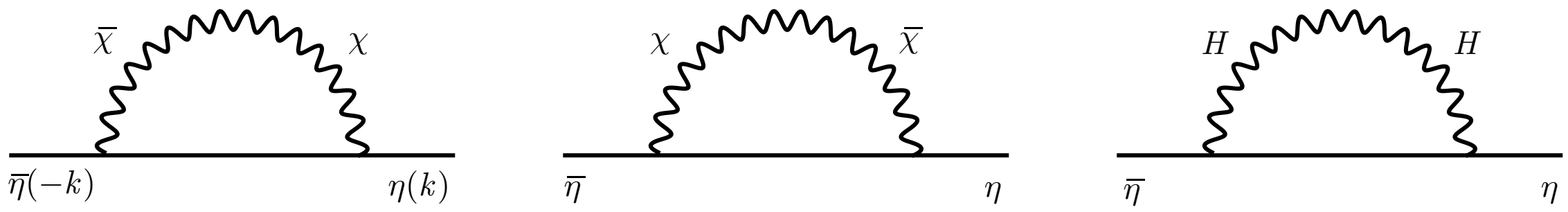

*Fig. 7.5.2*

The first graph gives no contribution (after $D$-algebra it is a tadpole) while the others add up to

$$\int \frac{d^4k}{(2\pi)^4} d^4\theta\, \bar\eta(-k,\theta) \int \frac{d^4p}{(2\pi)^4} \frac{1}{p^2(k+p)^2} \left[ -\frac{5}{9} \bar D^2 D^2 + \frac{1}{9} p\cdot k - \frac{2}{9}(p-k)^2 \right] \eta(k,\theta) \quad . \tag{7.5.27}$$

(The $-\frac{5}{9}$ for the $\chi\bar\chi$ propagator follows from its normalization in (7.4.14).) In the integral we can replace $p\cdot k = -\,k^2$ and $p^2 = 0$. Thus the total result vanishes.

## 7.6. Covariant Feynman rules

To the quantized supergravity action of sec. 7.4 we can add covariantized supersymmetric matter actions and consider general matter-supergravity systems. If the matter actions contain covariant derivatives, these must be split as in sec. 7.2. For constrained superfields we must first extract explicit quantum field dependence (e.g., $\phi \to \hat{\phi}$, $\bar{\phi} \to e^{-H}\hat{\bar{\phi}}$, where $\hat{\phi}$, $\hat{\bar{\phi}}$, are background covariantly chiral). In principle we can also split matter superfields into quantum and background parts and consider a general quantum system in a background of matter and supergravity. However, in general the procedure of sec. 7.4 is not applicable. We cannot impose the on-shell conditions $\mathbf{R} = \mathbf{G}_{\alpha\dot{\alpha}} = 0$. These conditions must be replaced by the equations $\mathbf{R} = J(matter)$, $\mathbf{G}_{\alpha\dot{\alpha}} = J_{\alpha\dot{\alpha}}(matter)$ and the quantization must be carried out with the supergravity fields off-shell. This is a straightforward but algebraically cumbersome procedure. Therefore in this section we will consider only pure on-shell supergravity backgrounds (no external matter). General systems can be handled by an extension of our quantization methods or by the ordinary (nonbackground) quantization of the preceding section.

Given the background field Lagrangian with quantum matter or supergravity fields, the Feynman rules can be derived in exactly the same way as for global supersymmetry. In general, for unconstrained quantum superfields, we can read the rules directly from the Lagrangian. We have two types of vertices: those arising from quantum self-interactions, and those containing also (or only) interactions with the background fields. The background fields appear only through field strengths and background covariant derivatives $\boldsymbol{\nabla}_A = \mathbf{E}_A{}^M D_M + \boldsymbol{\Phi}_A$ , with the flat superspace $D_M$. Therefore, we will encounter vertices with all quantum lines, or with a mixture of (at least two) quantum lines and background lines. For constrained, i.e., background covariantly chiral superfields, we must first of all solve the chirality constraints, i.e., write $\hat{\Phi} = e^{\bar{\boldsymbol{\Omega}}}\Phi_0$ in terms of an ordinary chiral superfield. This will introduce interactions involving explicitly the background potentials. We shall discuss below how to avoid this, but at any rate we end up with an action to which the methods of chapter 6 can be applied, with ordinary propagators and rules for calculation.

We observe that *at the one-loop level,* the contribution from the general spinors can also be obtained by squaring their kinetic operator and taking half of the resulting contribution to the effective action. We have

$$(i\nabla_\alpha{}^{\dot\gamma})(i\nabla^\beta{}_{\dot\gamma}) = \frac{1}{2}\delta_\alpha{}^\beta(i\nabla_\gamma{}^{\dot\gamma})(i\nabla^\gamma{}_{\dot\gamma}) + \frac{1}{2}[i\nabla_\alpha{}^{\dot\gamma}, i\nabla^\beta{}_{\dot\gamma}]$$

$$= \delta_\alpha{}^\beta \Box + \{\nabla_\alpha , \boldsymbol{W}^\beta{}_\gamma{}^\delta M_\delta{}^\gamma\} \quad , \tag{7.6.1}$$

where we have used (5.4.16). The spinor action in (7.4.14a) thus becomes

$$\int d^4x\, d^4\theta\, \mathbf{E}^{-1} \sum_{i=1}^{3} \frac{1}{2}\psi_i{}^\alpha(\delta_\alpha{}^\beta \Box + \{\nabla_\alpha , \boldsymbol{W}^\beta{}_\gamma{}^\delta M_\delta{}^\gamma\})\psi_{i\beta} + h.\,c.$$

$$= \int d^4x\, d^4\theta\, \mathbf{E}^{-1} \sum_i \frac{1}{2}\psi_i{}^\alpha \widehat{\Box}\psi_{i\alpha} + h.\,c. \quad , \tag{7.6.2}$$

and we observe that all the unconstrained superfields $(H,V,\psi)$ are described by similar actions, with the same operator $\widehat{\Box}$ given by (7.4.14b ). We shall discuss later applications of this result.

We now describe a modification of the Feynman rules for covariantly chiral superfields, analogous to the modification for the Yang-Mills case in sec. 6.5. The consequences of the modification are: It guarantees that the Feynman rules for chiral superfields will not introduce explicit background gauge potentials, but only the vielbein and connections, and it actually simplifies some of the $D$-algebra. We follow a procedure that is identical to that of sec. 6.5. We first *define* covariant functional differentiation for a general superfield $\Xi$ by

$$\frac{\delta\Xi(z)}{\delta\Xi(z')} \equiv E\delta^8(z - z') \quad , \tag{7.6.3}$$

which gives $(\frac{\delta}{\delta\Xi})\int d^8z\, E^{-1}L = \frac{\partial L}{\partial\Xi}$. We then define covariant functional differentiation for a covariantly chiral superfield $\eta$ (which could carry additional undotted spinor indices, but we do not indicate these explicitly) by

$$\frac{\delta\eta(z)}{\delta\eta(z')} \equiv (\overline{\nabla}^2 + R)E\delta^8(z - z') = \phi^{-3}\bar{D}^2\delta^8(z - z') \quad , \tag{7.6.4}$$

where the second form is obtained by using the identity (5.3.66b) and the chiral representation ($E_{\dot\alpha} = \Psi\bar{N}_{\dot\alpha}{}^{\dot\mu}\bar{D}_{\dot\mu}$) with the particular Lorentz gauge where $N_\alpha{}^\beta \neq \delta_\alpha{}^\beta$ such that

$$E_{\dot\alpha}{}^\mu = E_{\dot\alpha}{}^{\mu\dot\mu} = \Phi_{\dot\alpha\beta}{}^\gamma = 0 \quad . \tag{7.6.5}$$

In this representation covariantly chiral superfields are chiral in the usual sense: $\overline{\nabla}_{\dot{\alpha}}\eta_{...} = 0$ implies $\bar{D}_{\dot{\alpha}}\eta_{...} = 0$. Just as in the Yang-Mills case, at this point we need not be explicit as to whether the chiral covariance is with respect to full derivatives (containing both background and quantum fields) or just background fields, and the objects appearing in (7.6.4) can be functions of both, or just background fields (except when the chiral superfields are supergravity superfields, in which case the covariant derivatives can only be background). We can stay off-shell.

The covariantization of the usual expression $\bar{D}^2 D^2 \eta = \Box\eta$ becomes now

$$(\overline{\nabla}^2 + R)(\nabla^2 + \bar{R})\eta_{\alpha\cdots\beta} = \Box_+ \eta_{\alpha\cdots\beta} \quad ,$$

$$\Box_+ = \Box + W^{\alpha}{}_{\beta}{}^{\gamma}\nabla_{\alpha}M_{\gamma}{}^{\beta} + \frac{1}{2}\,i(\nabla^{\gamma\dot{\alpha}}G_{\beta\dot{\alpha}})M_{\gamma}{}^{\beta}$$

$$- \frac{1}{2}\,iG^{\alpha\dot{\alpha}}\nabla_{\alpha\dot{\alpha}} - R\nabla^2 - \frac{1}{2}\,(\nabla^{\alpha}R)\nabla_{\alpha} + R\bar{R} + (\overline{\nabla}^2\bar{R}) \quad , \qquad (7.6.6)$$

generalizing (7.4.4) off shell. We observe that on shell (recalling that chiral superfields can only have undotted indices) $\Box_+ = \widehat{\Box}$, where the latter quantity was defined in (7.4.14b), a result which we shall use later.

As in sec. 6.5c we start with the action

$$S = S_0 + S_{int}(\eta\, , \bar{\eta}) \quad , \quad S_0 = \int d^4x\, d^4\theta\ E^{-1}\bar{\eta}\eta \quad . \qquad (7.6.7)$$

$S_{int}$ also contains the other quantum fields but we have indicated explicitly only the dependence on $\eta$. We concentrate on the functional integral over $\eta$ which gives, using (7.6.3,4),

$$Z(J\, , \bar{J}) = \int I\!D\eta\, I\!D\bar{\eta}\ exp[S + (\int d^4x\, d^2\theta\ \phi^3 J\eta + h.\,c.)]$$

$$= \Delta \times [exp\ S_{int}(\frac{\delta}{\delta J}\, , \frac{\delta}{\delta \bar{J}})][exp(-\int d^4x\, d^4\theta\ E^{-1}\bar{J}\Box_+{}^{-1}J)] \quad , \qquad (7.6.8)$$

where $\Delta$ is the functional determinant

$$\Delta = \int I\!D\eta\, I\!D\bar{\eta}\, e^{S_0} \quad . \qquad (7.6.9)$$

In general the above expression for $Z$ depends on, and is to be integrated over, the other quantum fields. We are considering the massless case, but the results for the massive case can be obtained easily. Except for $\Delta$, the other factors, which contain quantum field self-interactions, contribute only beyond one loop, or to diagrams containing external chiral lines.

The determinant $\Delta$ gives the complete one-loop contribution from chiral superfields of diagrams with only external supergravity lines, and could be evaluated by using standard superfield Feynman rules, but we wish to avoid this. Instead, we shall use the "doubling" trick as in sec.6.5c. (In supergravity we are always dealing with real representations). We now have

$$\mathbf{O}\begin{pmatrix}\eta\\ \bar\eta\end{pmatrix}+\begin{pmatrix}J\\ \bar J\end{pmatrix}=0 \quad , \quad \mathbf{O}=\begin{pmatrix}0 & \bar\nabla^2+R\\ \nabla^2+\bar R & 0\end{pmatrix} \quad . \tag{7.6.10}$$

Its square,

$$\mathbf{O}^2\begin{pmatrix}\eta\\ \bar\eta\end{pmatrix}-\begin{pmatrix}J\\ \bar J\end{pmatrix}=0 \quad , \quad \mathbf{O}=\begin{pmatrix}(\bar\nabla^2+R)(\nabla^2+\bar R) & 0\\ 0 & (\nabla^2+\bar R)(\bar\nabla^2+R)\end{pmatrix} \quad , \tag{7.6.11}$$

corresponds to an action

$$S'_0=\int d^4x\, d^2\theta\, \phi^3\,\frac{1}{2}\,\eta\Box_+\eta=\int d^4x\, d^4\theta\ E^{-1}\,\frac{1}{2}\,\eta(\nabla^2+\bar R)\eta \quad , \tag{7.6.12}$$

and in terms of it we can write the functional integral

$$\Delta^2=\int I\!D\eta\, I\!D\bar\eta\ exp[S'_0(\eta)+h.\,c.]=(\int I\!D\eta\ e^{S'_0})^2 \quad . \tag{7.6.13}$$

We integrate $S'_0$ by separating out $\phi^3 D^2$ from $E^{-1}(\nabla^2+\bar R)$, treating $\frac{1}{2}\,\eta[E^{-1}(\nabla^2+\bar R)-\phi^3 D^2]\eta$ as an interaction term. The result is

$$\begin{aligned}\Delta&=\int I\!D\eta\ e^{S'_0}\\ &=\{exp\int d^4x\, d^2\theta\, \phi^3\,\frac{1}{2}\,\frac{\delta}{\delta J}\,[(\bar\nabla^2+R)(\nabla^2+\bar R)-\bar D^2 D^2]\,\frac{\delta}{\delta J}\,\}\\ &\quad\cdot[exp-\int d^4x\, d^2\theta\, \phi^3\,\frac{1}{2}\,J\Box_0{}^{-1}J]|_{J=0} \quad .\end{aligned} \tag{7.6.14}$$

(Note that writing instead $(\bar\nabla^2+R)(\nabla^2+\bar R)\rightarrow \bar D^2E^{-1}\phi^{-3}e^{-H}D^2E^{-1}\bar\phi^{-3}e^{H}$ would give the usual rules, except for the extra $\phi^3$'s from the definition (7.6.4)). Therefore, a calculation of the one-loop contribution of $\eta$ to the effective action (i.e., $ln\ \Delta$) consists in evaluating graphs with propagators $p^{-2}\delta^4(\theta - \theta')$ and vertices $[(\overline{\nabla}^2 \cdots]$ giving rise to a string

$$\cdots [(\overline{\nabla}^2 + R)(\nabla^2 + \bar{R}) - \bar{D}^2 D^2]_i \, \delta^4(\theta_i - \theta_{i+1}) \, [(\overline{\nabla}^2 + R)(\nabla^2 + \bar{R}) - \bar{D}^2 D^2]_{i+1} \cdots \tag{7.6.15}$$

with $\int d^4\theta_i$ integrals at each vertex. We concentrate on a given vertex and at the next one we rewrite $(\overline{\nabla}^2 + R) = \bar{D}^2\phi^{-3}E^{-1}$ . We temporarily transfer the $\bar{D}^2$ factor across the $\delta$-function and use the identity

$$[(\overline{\nabla}^2 + R)(\nabla^2 + \bar{R}) - \bar{D}^2 D^2]\bar{D}^2 = (\Box_+ - \Box_0)\bar{D}^2 \quad . \tag{7.6.16}$$

We further simplify the expression by using the anticommutation relations to move the $\bar{D}$'s in $\Box_+$ to the right until they are annihilated by the $\bar{D}^2$. The resulting expression, which we call $\widehat{\Box}_+$, contains no $\bar{D}$'s. We now return the $\bar{D}^2$ factor to its original place, reexpress the vertex in its original form, and proceed to manipulate it in the same way. We can continue around the loop and treat in this way all vertices but the last, and we are led to the following rules:

$$\textit{one vertex}: \quad \bar{D}^2[\phi^{-3}E^{-1}(\nabla^2 + \bar{R}) - D^2] \quad ,$$

$$\textit{other vertices}: \quad \widehat{\Box}_+ - \Box_0 \quad . \tag{7.6.17}$$

The massive case is obtained simply by adding a mass term in the denominator of the propagator.

These rules lead to a simpler evaluation of the one-loop contribution, since there are no $\bar{D}$'s in the loop except the one $\bar{D}^2$, but more importantly the contribution is manifestly expressible only in terms of objects which appear in the covariant derivatives, and not the gauge prepotentials. This is evidently true of the higher-loop contributions as well. From $S_{int}$ and the definition of the covariant functional derivative in (7.6.4), the expression (7.6.8) leads to higher-loop Feynman rules which do not explicitly depend on the background prepotentials. We obtain propagators $\widehat{\Box}_+{}^{-1}$ for chiral lines, where the full $\widehat{\Box}_+$ can be expressed in terms of the quantum $H_{\alpha\dot{\alpha}}$, $\phi$, and the background covariant derivatives. From $S_{int}(\frac{\delta}{\delta J}, \frac{\delta}{\delta \bar{J}})$ we obtain vertices with factors $(\overline{\nabla}^2 + R)E$ or $(\nabla^2 + \bar{R})E$ operating on each chiral or antichiral line leaving the vertices. (These generalize the ordinary flat space rules.) Again we can express these quantities in terms of the quantum $H_{\alpha\dot{\alpha}}$ and $\phi$, and the background covariant derivatives. For an actual momentum space calculation these have to be further expressed in terms of ordinary derivatives and background vielbein and connections. We now have the result that for all superfields, when calculations are carried out in the background field method, the contributions to the effective action from individual graphs do not involve the background supergravity prepotentials themselves, but only vielbein and connections, which depend on *(multi)derivatives* of the prepotentials. Consequently there is some improvement in the power counting rules for potentially divergent graphs.

## 7.7. General properties of the effective action

We analyze in this section the general form of the effective action (background field functional) $\Gamma$, as constrained by the requirement of background field invariance. This analysis is particularly important for determining the divergence structure of supergravity. Divergences, which could be canceled by counterterms in the Lagrangian, correspond to *local* terms in the effective action, and their form is limited by gauge invariance and dimensionality. In some cases we obtain stronger results by restricting ourselves to on-shell background fields. The on-shell restriction is not serious: The theory is not perturbatively renormalizable, Green's functions are gauge-dependent and divergent, and at best we can hope that gauge-independent, on-shell quantities (e.g., the S-matrix) are finite. Therefore, only the divergences which do not vanish on-shell are significant (divergences which are proportional to the field equations can be removed by a field redefinition which does not affect the S-matrix).

We will discuss first the situation in $N = 1$ supergravity. The discussion is applicable then to extended supergravity expressed in terms of $N = 1$ superfields. However, stronger statements can be made if the extended theories can be expressed in terms of extended superfields. Since the discussion does not depend on details of the extended superfield constructions, but only on properties that generalize our $N = 1$ background quantization methods, we devote a subsection to this case.

### a. N=1

Our background fields are supergravity fields, while the quantum fields can be supergravity or matter superfields or both. Since in the background field formalism the effective action is gauge invariant, it can be constructed from the field strengths $\mathbf{R}$, $\mathbf{G}_{\alpha\dot{\alpha}}$ ,$\mathbf{W}_{\alpha\beta\gamma}$ , and covariant derivatives (with an overall factor of $\mathbf{E}^{-1}$, or $\boldsymbol{\phi}^3$ for chiral integrands). Furthermore, on shell $\mathbf{R} = \mathbf{G}_{\alpha\dot{\alpha}} = 0$. (We consider only vanishing cosmological term: Otherwise, $\mathbf{R}$ is a nonvanishing dimensional constant, and the dimensional analysis is changed.) Thus, only the chiral field strength $\mathbf{W}_{\alpha\beta\gamma}$ (and its complex conjugate) and covariant derivatives can appear. We also have the on-shell conditions $\boldsymbol{\nabla}^\alpha \mathbf{W}_{\alpha\beta\gamma} = \boldsymbol{\nabla}^{\alpha\dot{\alpha}} \mathbf{W}_{\alpha\beta\gamma} = 0$ as well as the corresponding equations for $\bar{\mathbf{W}}$. In determining the form of the effective action we can also use the following facts: (a) The effective action is dimensionless. The dimensions of the various quantities which can appear are

$[d^4x] = -4$, $[d^2\theta] = 1$, $[\mathbf{W}] = \frac{3}{2}$, $[\boldsymbol{\nabla}_\alpha] = \frac{1}{2}$. In addition, with only supergravity interactions, for an $L$-loop contribution we have a factor $\kappa^{2(L-1)}$ with dimension $-2(L-1)$. (b) Functions $G(x_1, \ldots)$ arise from loop integrals after all the $D$-algebra has been carried out and, if they have odd dimension, must contain an odd number of space-time derivatives (momentum factors) which have an index structure $\partial_{\alpha\dot{\beta}}$. (c) Integrals with the chiral measure $d^2\theta$ must have chiral integrands, i.e., factors of $\mathbf{W}$ or $\bar{\boldsymbol{\nabla}}^2(\bar{\mathbf{W}}^2)$ ($\bar{\boldsymbol{\nabla}}^2\bar{\mathbf{W}} = 0$ on shell), etc. (but in the latter case they can be rewritten as full integrals anyway). (d) Dotted and undotted indices must be separately saturated.

Another important feature is the fact that all the $d^4\theta$ terms in $\Gamma$ have an equal number of (spinor-) undifferentiated $\mathbf{W}$'s and $\bar{\mathbf{W}}$'s. This is a consequence of the global chiral R-invariance of the theory (cf. (5.3.10); the $Y$ transformations are global invariances; in terms of prepotentials, they are simply phase transformations of $\phi$, leaving $H$ invariant). We should also remark that, a priori, as discussed in the previous section, perturbation theory does not lead to a form involving only the $\mathbf{W}$'s and their covariant derivatives, but rather the quantities which appear in the background covariant derivatives, i.e., background vielbein and connection coefficients, with a $d^4\theta$ integral. However, because of background invariance, these quantities must arrange themselves into a form that is manifestly covariant or contains one noncovariant factor times a covariant object that satisfies a Bianchi identity. (The noncovariant term, when varied, produces a derivative which, when integrated by parts, gives zero upon use of the Bianchi identity.) Thus, the term $\int d^4x d^2\theta\, \boldsymbol{\phi}^3 \mathbf{W}^2$ really arises from an expression (in the gauge (7.6.5)) $\int d^4x d^4\theta\, \mathbf{E}^{-1}\boldsymbol{\Phi}^{\alpha\beta\gamma}\mathbf{W}_{\alpha\beta\gamma}$, which can then be rewritten as a chiral integral.

We first discuss all *local* terms in $\Gamma$, i.e., all possible on-shell local divergences of the theory. A generic local term will have the structure

$$(\boldsymbol{\nabla}^{\underline{a}})^l (\mathbf{W}\bar{\mathbf{W}})^m (\boldsymbol{\nabla}_\beta \mathbf{W})^n (\bar{\boldsymbol{\nabla}}_{\dot{\gamma}}\bar{\mathbf{W}})^r \quad , \tag{7.7.1}$$

with $\boldsymbol{\nabla}_\alpha \mathbf{W} \equiv \boldsymbol{\nabla}_{(\alpha}\mathbf{W}_{\beta\gamma\delta)}$ and the indices contracted in various ways, and the space-time derivatives distributed in various ways. We have used the invariance under R-transformations to write only terms with equal powers of $\mathbf{W}$ and $\bar{\mathbf{W}}$. We note that unless some space-time derivatives act on them, $\mathbf{W}$ and $\bar{\mathbf{W}}$ cannot be raised to a power higher than 4, because they are symmetric in their three spinor indices and hence contain only four independent Lorentz components. The dimensionality of the above term is $l + 3m + 2n + 2r$. If it appears at $L$ loops it is multiplied by $(\kappa^2)^{L-1}$ with dimension $-2(L-1)$. The overall dimension of the action must be zero: $d = l + 3m + 2n + 2r - 2(L-1) - 2 = 0$. The only purely chiral term, involving just $\mathbf{W}$, is the quadratic expression

$$\int d^4x d^2\theta \boldsymbol{\phi}^3 \mathbf{W}^{\alpha\beta\gamma}\mathbf{W}_{\alpha\beta\gamma} + h.c. = -\int d^4x\ e^{-1} w^{\alpha\beta\gamma\delta} w_{\alpha\beta\gamma\delta} + h.c. \tag{7.7.2}$$

where $w$ is the Weyl tensor. On dimensional grounds it can only appear at the one-loop level (with no $\kappa$ factor, as follows from our discussion above), and the integrand is a total derivative on shell. It is, in fact, on dimensional grounds, the only local term (i.e., possible divergence) which can occur at the one-loop level, on shell. However, due to the Gauss-Bonnet theorem, it is just a topological constant, and vanishes in topologically trivial spaces. (For a further discussion, see sec. 7.10.)

At the two-loop level no local terms are possible. We have a factor $\kappa^2$ of dimension $-2$, and it is easy to check that there is no way, from among the generic expression above, to find either a chiral expression (to be integrated with $d^4xd^2\theta$), or a general expression (to be integrated with $d^4xd^4\theta$), which can lead to a term with dimension zero. Thus, at the two-loop level no on-shell divergences can arise in supergravity.

At higher loops the number and variety of local terms increases. For example, at three loops, the combination $\mathbf{W}^2\bar{\mathbf{W}}^2$ is a possible local term and therefore a potentially divergent one (the only on-shell one, in fact). At any given loop there are of course limitations due to dimensionality, chirality, and index contraction. In particular it is easy to verify that, on dimensional grounds and in order to saturate indices, all higher loop terms must have factors of both $\mathbf{W}$ and $\bar{\mathbf{W}}$ and therefore vanish when either $\bar{\mathbf{W}} = 0$ or $\mathbf{W} = 0$. This situation describes background field configurations which are self-dual or antiself-dual. We conclude that such configurations receive no radiative corrections.

The nonlocal part of the effective action has a structure as in (7.7.1), including however a nonlocal function $G(x_1, x_2, \ldots.)$ and with fields evaluated at different points $(x_1, \theta)$, $(x_2, \theta)$, ... . We find, using superspace perturbation theory and dimensional analysis that, if no massive fields are present, the on-shell effective action has the form

$$\Gamma \sim \int d^4x_1 d^4x_2 d^2\theta\, \boldsymbol{\phi}^3 \mathbf{W}^{\alpha\beta\gamma}(x_1,\theta) G(x_1,x_2) \mathbf{W}_{\alpha\beta\gamma}(x_2,\theta)\ +\ h.c.$$

$$+ \int d^4x_1 \cdots d^4x_4 d^4\theta \mathbf{W}^{\alpha\beta\gamma}(x_1,\theta)\mathbf{W}_{\alpha\beta\gamma}(x_2,\theta)\bar{\mathbf{W}}^{\dot\alpha\dot\beta\dot\gamma}(x_3,\theta)\bar{\mathbf{W}}_{\dot\alpha\dot\beta\dot\gamma}(x_4,\theta)$$

$$\times G(x_1 \cdots x_4) + \cdots , \tag{7.7.3}$$

where the terms not explicitly written contain more than four $\mathbf{W}$'s and their covariant derivatives. The nonlocal functions $G(x_1, x_2, \dots)$ can be thought of as polynomials in the space-time covariant derivatives acting on the fields, and functions of the covariant d'Alembertian (e.g., its inverse), corresponding to the result of doing various loop integrals in momentum space. The important point is that other, a priori possible terms with two, three, or four $\mathbf{W}$'s are not present. (For example, $d^4\theta\mathbf{W}\mathbf{W}\bar{\mathbf{W}}$ cannot have its indices saturated even if derivatives are included while $d^2\theta(\mathbf{W})^4$ has the wrong dimension, etc.) In secs. 7.8,10 we shall discuss in more detail the form of the $G$ functions.

## b. General N

We shall assume in this section that unconstrained superfield formalisms exist for all supersymmetric systems of interest. Such formalisms have not yet been developed except for $N = 2$, and there are indications that if they exist they have an unfamiliar form. We shall only assume that there exist constraints on the covariant derivatives that allow them, and the action, to be expressed in terms of ordinary derivatives and unconstrained prepotentials. We can then mimic the $N = 1$ background-quantum splitting for general $N$. We replace the unconstrained prepotentials by quantum prepotentials, and the ordinary derivatives by background covariant derivatives. Furthermore, if covariantly constrained (e.g., chiral) superfields are present, we can derive covariant rules for them as we did in $N = 1$; the procedure is general. We will not restrict ourselves to on-shell backgrounds.

By an extension of our fully covariant background field method of sec. 7.6, we obtain improved power-counting rules for discussing local divergences. These rules simply follow from the fact that *all quantum terms in the effective action are automatically expressed directly in terms of the constrained background covariant derivatives* (and their field strengths) and an explicit expansion in terms of unconstrained background prepotentials is unnecessary. (One might also expect to need the superspace generalization of antisymmetric tensor gauge fields, e.g., the three-form of D = 11 supergravity, but the background-quantum split action can always be written in a form where such background fields appear only as their field strengths, and these field strengths already appear among the field strengths of the background covariant derivatives.) This implies that all divergent terms must be expressible as *local functions of the covariant derivatives.* Thus, for supersymmetric Yang-Mills, where the same ideas apply, all counterterms must be local functions of $\Gamma_{\underline{\alpha}}$, and for supergravity of $\Gamma_{\underline{\alpha}}$ and also $E_{\underline{\alpha}}{}^M$. (Conventional constraints determine all of $\nabla_A$ from $\Gamma_{\underline{\alpha}}$and $E_{\underline{\alpha}}{}^M$.) Furthermore, because in the derivation of the Feynman rules vertices are always *integrated over full superspace,* they will carry a full $\int d^4x d^{4N}\theta$ for $N$-extended supersymmetry.

For the case of extended supersymmetry, treated with extended superfields, there is a technical difficulty in the background field method because of the appearance of an infinite number of generations of ghost superfields with progressively increasing superspin. For example, $N = 2$ Yang-Mills theory is described by a real isovector superfield $V_a{}^b$ with gauge invariance $\delta V_a{}^b = D_{c\alpha}\chi_{(a}{}^{bc)\alpha} + \bar{D}^c{}_{\dot{\alpha}}\bar{\chi}_{(ac}{}^{b)\dot{\alpha}}$. This transformation implies that the corresponding ghost has a gauge invariance $\delta\psi^{(abc)\alpha} = D_{d\beta}\chi^{(abcd)(\alpha\beta)}$, which in turn implies a ghost with invariance $\delta\psi^{(abcd)(\alpha\beta)} = D_{e\gamma}\chi^{(abcde)(\alpha\beta\gamma)}$, etc. The gauge superfields unavoidably contain fields of spin higher than those (physical and auxiliary) occurring in the gauge-invariant action. To gauge these away the gauge superparameters (and therefore the corresponding ghosts) must contain higher spins than the gauge superfields. However, only a finite number of ghosts (i.e., the usual Faddeev-Popov ghosts, plus perhaps certain catalyst ghosts) contribute at more than one loop. Therefore, the higher-loop contributions to the effective action can be calculated in a manifestly background covariant form and will obey the power-counting rules that we derive below, whereas the one-loop contribution may have to be treated separately. (For example, we could choose background noncovariant gauges for some of the ghosts which contribute only at one loop in such a way that all but a finite number of these ghosts decouple. The effect of such a choice would be to produce a one-loop effective action which is noncovariant, but this would have no effect on physical quantities). We discuss now the implications of these remarks and the improved power-counting rules to which they lead. For completeness we discuss first the situation in global theories.

The first example of the improved power counting was already given in sec. 6.5 for the Fayet-Iliopoulos D-term in $N = 1$ Yang-Mills theory. Background covariance immediately implies the vanishing of such a term beyond one loop. For $N > 1$ we obtain

stronger results:

Since Yang-Mills theory is renormalizable, the only allowed divergence in the background field method is proportional to the classical action. However, beyond one loop it must have the form $\int d^4x d^{4N}\theta\, \Gamma^{\underline{\alpha}}\Omega\Gamma_{\underline{\alpha}}$ at the lineared level because of the covariant Feynman rules. Here $\Gamma_{\underline{\alpha}}$ has dimension $\frac{1}{2}$ and $\Omega$ is a local operator (nonnegative dimension). Since the action is dimensionless, we obtain the inequality $-4 + 2N + \frac{1}{2} + \frac{1}{2} \leq 0$, which implies that only $N = 0$ or 1 can have divergences beyond one loop. Thus $N = 2$ and $N = 4$ supersymmetric Yang-Mills theory must be finite beyond one loop. Furthermore, we know from explicit one-loop calculations using $N = 1$ superfields (see sec. 6.4) that $N = 4$ is one-loop finite as well. On the other hand, $N = 2$ does have one-loop divergences. (Also, as for $N = 1$, loop corrections to the $N = 2$ Fayet-Iliopoulos term vanish.) We emphasize that we had to make a separate one-loop argument because of the problem with infinite numbers of ghosts.

We can apply similar arguments to $N$-extended supergravity. The local (divergent) part of the effective action consists of the integral of $E^{-1}$ times a (covariant) product of factors of vielbein and connections. At $L$ loops the lowest dimensional such term is

$$\Gamma_{loc} \sim \kappa^{2(L-1)} \int d^4x d^{4N}\theta\, E^{-1} \quad , \tag{7.7.4}$$

multiplied perhaps by some function of a dimensionless scalar field strength for $N \geq 4$; such a function may however be forbidden by global on-shell invariance (other additional factors would have positive dimension). Requiring this expression, possibly multiplied by a polynomial in the fields (with nonnegative dimension), to be dimensionless, we obtain the inequality $-2(L-1) - 4 + 2N \leq 0$, which implies $L \geq N - 1$. (Similar arguments for Yang-Mills give the improved $-4 + 2N + 2 \leq 0$ instead of the above $-4 + 2N + 1 \leq 0$.) Thus, from these arguments alone, we find that in $N$-extended supergravity the effective action can have local terms, and therefore possible divergences, only at $N - 1$ loops and beyond. (This is so even though possible lower-loop invariants can be constructed. The important point is that our Feynman rules imply integration over full superspace with integrands that involve covariant objects.) Note that the divergences excluded by these rules are absent *both on and off shell.*

A similar analysis in higher dimensions gives the result that higher-loop divergences are absent is supersymmetric Yang-Mills theory for $L < 2\,\frac{N-1}{\mathrm{D}-4}$, and in supergravity for $L < 2\,\frac{N-1}{\mathrm{D}-2}$ (for $L$ loops in D-dimensions, where $N$ refers to the four-dimensional value, i.e. the number of anticommuting coordinates is $4N$). For lower dimensions (super-)Yang-Mills is renormalizable anyway; for supergravity the above inequality holds for $\mathrm{D} = 3$ while for $\mathrm{D} = 2$ we find higher-loop finiteness for $N > 1$.

Our background field approach leads to a further result which is not apparent in ordinary quantization or nonsupersymmetric gauge fixing: At the one-loop level, in $N = 1$ language and using the background field formalism, the only contributions to the (on-shell, "topological") divergences are proportional to $(W_{\alpha\beta\gamma})^2$ and come from chiral superfields. To understand this we observe that the divergence is just a covariantization of the divergence in the two-point function, and its coefficient can be determined by calculating a self-energy diagram. However, in our gauge, examination of the quadratic action in (7.4.14) (which gives the general form for any type of superfield in an on-shell supergravity background), reveals that only chiral superfield vertices have enough $D$'s and $\bar{D}$'s to give nonzero contributions. Therefore *in a theory with a net zero number (physical minus ghost) of chiral superfields* (any $N \geq 3$ theory with appropriate choice of auxiliary fields (compensating multiplets)) *there are no (topological) one-loop divergences.* At the two-loop level *no* supergravity theory has on-shell divergences.

We summarize our results in Table 7.7.1, which lists all cases where divergences must be absent in pure supersymmetric gauge theories. The results can be classified into three types: (A) absence of divergences due to one-loop cancellations in $N \geq 3$ supersymmetry of contributions of $N = 1$ chiral superfields; (B) absence of two-loop supergravity counterterms because invariants of appropriate dimension do not exist; (C) absence of divergences at higher loops which is established by our arguments above.

The absence of higher-loop divergences cannot be established rigorously until the corresponding supergraph rules are explicitly constructed. Possible difficulties with carrying out the program are infrared problems due to large negative powers of momenta in the superfield propagators, and the explicit construction of the classical action (whose form may surprise us, if the properties of extended superspace are not a simple extension of those for $N = 1$ superspace). However, we emphasize that once the action has been written, the power counting rules and our conclusions immediately follow. We note that in the $N = 4$ Yang-Mills case the finiteness can already be proven when the theory is written in terms of $N = 2$ superfields, i.e., $N = 2$ Yang-Mills coupled to an $N = 2$ scalar multiplet; the $N = 2$ power counting rules can then be applied.

| | $N$ | loops 1 | 2 | 3 | 4 | 5 | 6 | $\geq 7$ |
|---|---|---|---|---|---|---|---|---|
| Yang-Mills | 0 | | | | | | | |
| | 1 | | | | | | | |
| | 2 | | C | C | C | C | C | C |
| | 4 | A | C | C | C | C | C | C |
| supergravity | 0 | | | | | | | |
| | 1 | | B | | | | | |
| | 2 | | B | | | | | |
| | 3 | A | B | | | | | |
| | 4 | A | B,C | | | | | |
| | 5 | A | B,C | C | | | | |
| | 6 | A | B,C | C | C | | | |
| | 8 | A | B,C | C | C | C | C | |

*Table 7.7.1. Absence of divergences in supersymmetric theories*

## 7.8. Examples

In this section we shall give some examples and applications of our covariant formalism for computing supergraphs in supergravity. We restrict ourselves to one-loop calculations. Higher-loop calculations are possible, but the algebra is complicated if there are internal supergravity superfields. We shall consider first some one-loop calculations with matter fields inside the loop and background supergravity superfields. The algebra simplifies considerably in the on-shell situation.

We begin by finding one-loop chiral-field contributions to the (covariantized) on-shell two-point function, corresponding to the first term in the on-shell effective action (7.7.3). In contrast to the calculation of sec. 7.5.d, the separate coefficients of the $W^2$ and $G^2 + 2\bar{R}R$ terms in the effective action can be determined from the two-point function alone when the covariant rules are used (although here we find only the former term since the latter term vanishes in our on-shell calculation). However, we must use dimensional regularization to keep track of terms that are total derivatives *only* in four dimensions, since in a four-dimensional momentum-space Feynman-graph calculation they vanish by momentum conservation. We shall use the on-shell conditions on the background superfields, but keep the external momentum $k$ off shell ($k^2 \neq 0$) in the loop integral. Also, we shall write our expressions in four dimensions. However, the calculation should be carried out in D dimensions, both to avoid ultraviolet divergences, and to circumvent the fact that, when D=4, the linearized result is a total divergence.

We consider a chiral superfield $\eta_{\alpha\beta\cdots\dot{\gamma}\dot{\delta}\cdots}$ with $2A$ undotted and $2B$ dotted indices, and action $\frac{1}{2}\int d^4x\, d^2\theta\, \boldsymbol{\phi}^3\eta\boldsymbol{\Box}_+\eta + h.c.$. (If $B \neq 0$ such fields can exist only in on-shell backgrounds.) From (7.6.17), and in the Lorentz gauge $\boldsymbol{\Phi}_{\dot{\alpha}\beta}{}^{\gamma} = 0$, the linearized vertices are (on shell $\mathbf{E}^{-1} = \boldsymbol{\phi} = 1$)

$$\textit{One vertex}: \; \bar{D}^2(\boldsymbol{\nabla}^2 - D^2) \simeq \bar{D}^2[\mathbf{E}^{\alpha\underline{a}}\partial_{\underline{a}}D_\alpha + \frac{1}{2}(\partial_{\underline{a}}\mathbf{E}^{\alpha\underline{a}})D_\alpha + \boldsymbol{\Phi}^{\alpha}{}_{\beta}{}^{\gamma}M_\gamma{}^\beta D_\alpha] \;, \tag{7.8.1a}$$

$$\textit{Other vertex}: \; \widehat{\boldsymbol{\Box}}_+ - \boldsymbol{\Box}_0 \simeq \mathbf{E}^{\underline{a}\alpha}\partial_{\underline{a}}D_\alpha + \frac{1}{2}(\partial_{\underline{a}}\mathbf{E}^{\underline{a}\alpha})D_\alpha + \mathbf{W}^{\alpha}{}_{\beta}{}^{\gamma}M_\gamma{}^\beta D_\alpha \;. \tag{7.8.1b}$$

The propagator is $p^{-2}\delta_\alpha{}^{\alpha'}\cdots\delta_{\dot{\gamma}}{}^{\dot{\gamma}'}\cdots\delta^4(\theta-\theta')$.

The $D$-algebra is trivial. The $M_\gamma{}^\beta$ terms give a contribution proportional to the number of undotted indices, and we have a factor from a trace over all spinor indices.

We find a contribution to the effective action

$$\Gamma_2 = (-2)^{2(A+B)} \int \frac{d^4k}{(2\pi)^4} \frac{d^4p}{(2\pi)^4} \frac{1}{p^2(k+p)^2}$$

$$\times \frac{1}{4} \int d^2\theta \, [(p+\frac{1}{2}k)^{\underline{a}}(p+\frac{1}{2}k)^{\underline{b}} \mathbf{E}_{\underline{a}}{}^{\alpha}(-k)\mathbf{E}_{\underline{b}\alpha}(k) + A\mathbf{W}^{\alpha\beta\gamma}(-k)\mathbf{W}_{\alpha\beta\gamma}(k)] + h.\,c. \tag{7.8.2}$$

We have used the linearized, on-shell relations $\mathbf{W}_{\alpha\beta\gamma} = \bar{D}^2\mathbf{\Phi}_{\alpha\beta\gamma}$, $\mathbf{E}_{\underline{a}\alpha} = \bar{D}^2\mathbf{E}_{\alpha\underline{a}}$, and converted the $d^4\theta$ integral to a $d^2\theta$ integral. Finally, doing the momentum integral and using the linearized relation

$$\partial_{\underline{a}}\mathbf{E}_{\underline{b}\gamma} - \partial_{\underline{b}}\mathbf{E}_{\underline{a}\gamma} = C_{\dot{\alpha}\dot{\beta}}\mathbf{W}_{\alpha\beta\gamma} \ , \tag{7.8.3}$$

we obtain the result

$$\Gamma_2 = (-2)^{2(A+B)}(\frac{1}{12} - A)\,\frac{1}{2} \int d^4x \, d^2\theta \, \frac{1}{2}\,\mathbf{W}_{\alpha\beta\gamma}(x,\theta)I(-\Box)\mathbf{W}^{\alpha\beta\gamma} + h.\,c. \tag{7.8.4a}$$

with the logarithmically divergent integral $I$ of (7.5.26). After covariantization $\Gamma_2$ also contains some contributions from graphs with 3, 4, etc. external lines. The chiral integral above, after covariantization, also contains a $\boldsymbol{\phi}^3$ factor.

Separating out the divergent part, we have

$$\Gamma_2 = k_1 \frac{1}{(4\pi)^2}\,\frac{1}{2} \int d^4x \, d^2\theta \, \boldsymbol{\phi}^3 \, \frac{1}{2}\,\mathbf{W}_{\alpha\beta\gamma}[\frac{1}{\epsilon} - ln\,\frac{\Box}{\mu^2}]\mathbf{W}^{\alpha\beta\gamma} + h.\,c. \tag{7.8.4b}$$

where $\mu$ is a renormalization mass and

$$k_1 = (-2)^{2(A+B)}(\frac{1}{12} - A) \tag{7.8.5}$$

For a chiral scalar $\eta$, $k_1 = \frac{1}{24}$. (We have included a factor of $\frac{1}{2}$ to cancel the 2 due to our using the action $\int \eta\Box\eta + h.\,c.$ instead of $\int \bar{\eta}\eta$.) For a chiral spinor $\eta_\alpha$, $k_1 = \frac{20}{24}$. If the chiral spinor superfield is the gauge field of the tensor multiplet, there will be an additional contribution $\Delta k_1 = \frac{5}{24}$ from the five second generation chiral scalar ghosts discussed in sec. 7.3.a. (The $V$ ghosts do not contribute: see below.)

The next calculation we could imagine performing is that of a triangle diagram. However, since no $\mathbf{WWW}$ or $\mathbf{WW\bar{W}}$ term is present in the effective action (cf. our discussion following (7.7.3)), the contribution from such a diagram must be completely contained in the third order (in $H_{\alpha\dot{\alpha}}$ or $\phi$) terms in the expansion of $\Gamma_2$.

We observe that, once the self-energy contribution from a chiral superfield has been computed, that from a vector multiplet $V$ is trivial, because the whole contribution comes from the three chiral ghosts. Indeed, the $V$-background interactions are extracted from the covariant $V\widehat{\Box}V$ quadratic action. However, just as in the background Yang-Mills calculation, each $D$ or $\bar{D}$ contained in the $\widehat{\Box}$ operator of (7.4.14b) brings with it one factor of the external field, and for graphs with less than four external lines we do not have enough $D$'s. This is an important feature of the background-field method: In off-shell Yang-Mills or on-shell supergravity background, *general (nonchiral) superfields do not contribute to one-loop two- and three-point functions; only their chiral ghosts do.* Thus, in this case the contribution to the supergravity self-energy from a vector multiplet is $-3$ times that from a physical chiral scalar superfield (3 ghosts with wrong statistics).

The calculation of the on-shell one-loop self-energy contributions in self-interacting (quantum and background) supergravity is now trivial. No new calculations need to be performed because, from the action in (7.4.14a) or (7.4.19), we see again that only chiral superfields contribute. In the form with $V$ compensators (5.2.75a), the result is simply $-7$ times that from a physical chiral field (7 chiral scalar ghosts with wrong statistics): $k_1 = -\frac{7}{24}$. In the form with a chiral compensator, we have the contribution from the physical $\chi$ field, and contributions from two chiral spinors with the action $\frac{1}{2}\int d^4x d^4\theta \phi^\alpha \nabla^2 \phi_\alpha + h.c.$. The final answer is 41 times the contribution from a physical chiral scalar: $k_1 = \frac{41}{24}$. Finally, if we used the spinor compensator (5.2.75b) we would obtain $k_1 = -\frac{55}{24}$.

The calculation of a chiral- or general-field box diagram contributing to the $\mathbf{W}\mathbf{W}\bar{\mathbf{W}}\bar{\mathbf{W}}$ term in (7.7.4) is in principle no more difficult. For a chiral field we have one vertex (7.8.1a) and three vertices (7.8.1b) with one $\bar{D}^2$ and four $D$'s in the loop, two of which have to be integrated by parts onto external lines. For a real field we have one factor of $D$ or $\bar{D}$ at each vertex (coming from the linearization of $\widehat{\Box}$), so the $D$-algebra is trivial. What is left then is a Feynman integral for a box diagram, with some momentum factors in the numerator.

Our next example is that of the calculation of some one-loop, four-particle S-matrix elements, similar to the calculation we carried out for $N = 4$ Yang-Mills. There we saw that there were cancellations and the whole contribution came from the $V$-loop. The corresponding situation that we will find here is that a similar cancellation between ghosts, physical fields, and certain other contributions takes place in $N = 8$ supergravity so that in that case the result is essentially identical to the Yang-Mills case, and can be obtained without further calculation. We now discuss the situation in detail.

For $N$-extended supergravity described by $N = 1$ superfields, to calculate contributions to background $N = 1$ supergravity, one simply adds contributions from superfields representing all the $N = 1$ multiplets. The superfields which enter in addition to $H$ are of the same type as above, namely $V$'s, $\psi$'s, and $\chi$'s, and their Lagrangians have the same form (up to choices of compensating fields and duality transformations, which may change the values of contributions to topological invariants and the corresponding superconformal anomalies (see sec. 7.10) but do not affect the S-matrix). In Table 7.8.1 we give the number of fields of each type (the minus signs indicate abnormal statistics) for each value of $N$.

| $N$ | $\chi$ | $V$ | $\psi_\alpha$ | $H_{\alpha\dot\alpha}$ |
|---|---|---|---|---|
| 1 | -7 | 4 | -3 | 1 |
| 2 | -2 | 1 | -2 | 1 |
| 3 | 0 | -1 | -1 | 1 |
| 4 | 0 | -2 | 0 | 1 |
| 5 | 0 | -2 | 1 | 1 |
| 6 | 0 | 0 | 2 | 1 |
| 8 | 0 | 4 | 4 | 1 |

*Table 7.8.1. Number of fields of each type contributing to the one-loop effective action*

We use the form of $N = 1$ supergravity with $V$ compensators. The $(\frac{3}{2}, 1)$ multiplet is described by a general spinor superfield (see sec. 4.5.e) and its quadratic Lagrangian, including ghosts (which is all that is needed) is given by the background covariantization of (7.3.7).

The four-particle S-matrix is obtained from the diagrams in Fig. 7.8.1.

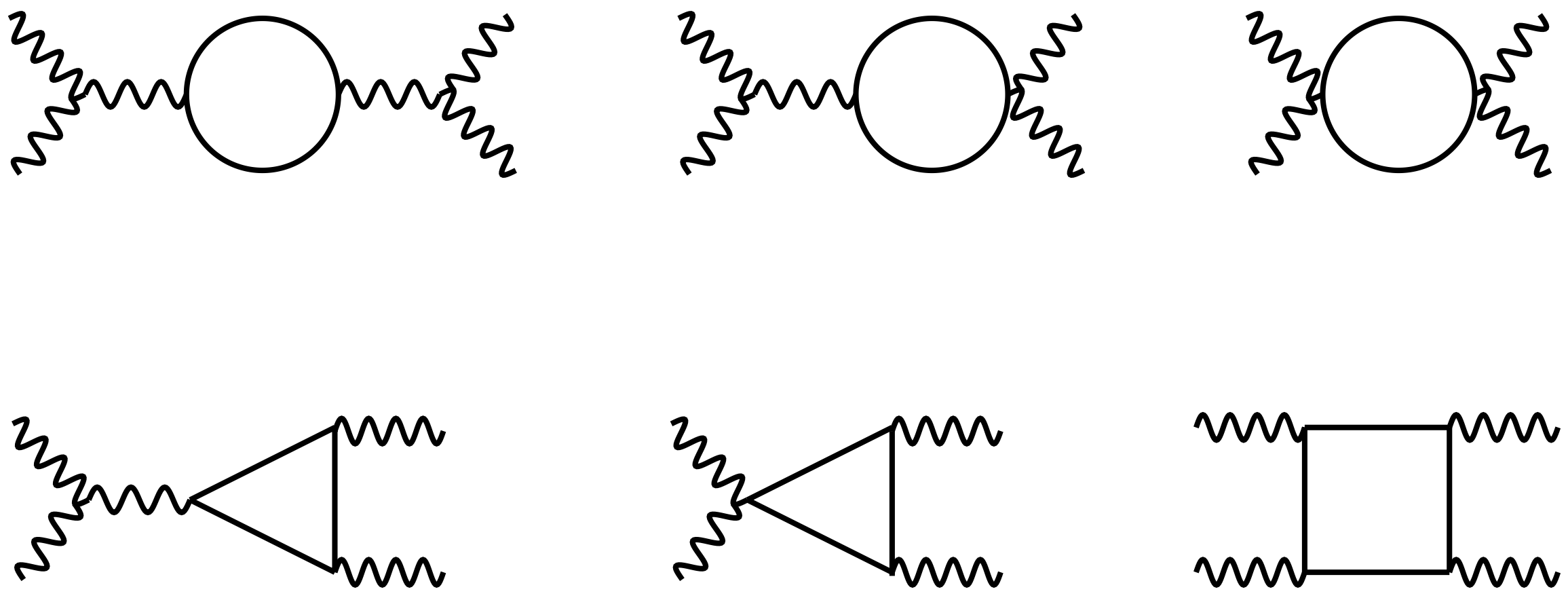

*Fig. 7.8.1*

The external wavy lines correspond to $N = 1$ supergravity fields, while the solid line loop corresponds to various $N = 1$ multiplets. All but the last diagram correspond to the first term $\Gamma_2$ of (7.7.3). However, by covariance the S-matrix must contain four factors of $\mathbf{W}$, and by dimensionality the complete contribution must come from the second term $\Gamma_4$ of (7.7.3), and therefore can be obtained from the box diagram. (Only the box diagram has enough denominator factors to balance the dimensions of four factors of $\mathbf{W}$.)

We write the contribution from the relevant part of $\Gamma_4$ as

$$\Gamma = \frac{1}{8}\int \frac{d^4p_1 \cdots d^4p_4}{(2\pi)^{16}}\, d^4\theta\, \delta(\sum(p_i))$$

$$\times[\mathbf{W}^{\alpha\beta\gamma}(p_1)\,\mathbf{W}_{\alpha\beta\gamma}(p_2)\,\bar{\mathbf{W}}^{\dot\alpha\dot\beta\dot\gamma}(p_3)\,\bar{\mathbf{W}}_{\dot\alpha\dot\beta\dot\gamma}(p_4)(C_4 G_4 + C_2 G_2 + C_0 G_0)(p_i)$$

$$-\frac{1}{2}\mathbf{W}^{\alpha\beta\gamma}(p_1)\,\bar{\mathbf{W}}^{\dot\alpha\dot\beta\dot\gamma}(p_2)\,\mathbf{W}_{\alpha\beta\gamma}(p_3)\,\bar{\mathbf{W}}_{\dot\alpha\dot\beta\dot\gamma}(p_4)(C_4 G'_4 + C_2 G'_2 + C_0 G_0)(p_i)] . \quad (7.8.6)$$

Here $G_0$ is the Feynman integral for a scalar box diagram (6.5.68), while $G_2(G'_2)$ , $G_4(G'_4)$ are similar contributions *extracted* from box diagrams with two and four momentum factors in the numerator. Unlike $N = 4$ Yang-Mills, the above expression is valid only on-shell, whereas off-shell the effective action diverges. (Because of covariantization the expression above contains terms with more than four fields, but it does not give the complete contribution to more-than-four-particle amplitudes.)

The coefficients $C_i$ are different for each value of $N$. To determine $C_i$ we must compute contributions from each of the superfields listed in Table 7.8.1. The only difficult calculation is of the contribution from the chiral superfields $\chi_i$. We shall therefore restrict ourselves here to discussing results for $N \geq 3$ where no chiral superfields appear.

As we discussed in sec. 7.6, it is useful to square the kinetic operator for the spinors (and take one half of the corresponding contribution to the effective action) in order to make it similar to the other kinetic terms. The kinetic operator for $V$, $\psi^\alpha$, and $H^{\alpha\dot{\alpha}}$ then takes the *universal* form

$$\widehat{\Box} = \Box + \mathbf{W}^\alpha{}_\beta{}^\gamma \, \boldsymbol{\nabla}_\alpha \, M_\gamma{}^\beta + \bar{\mathbf{W}}^{\dot{\alpha}}{}_{\dot{\beta}}{}^{\dot{\gamma}} \, \bar{\boldsymbol{\nabla}}_{\dot{\alpha}} \, \bar{M}_{\dot{\gamma}}{}^{\dot{\beta}} \quad , \tag{7.8.7}$$

The relative coefficient of the $\Box$ and $\mathbf{W}$ terms is *independent* of the choice of field. Therefore, in performing one-loop calculations we need only keep track of the index structure. In particular, there is always a factor from a trace over the Lorentz index: 1 for $V$, $-1$ for $\psi^\alpha$, $-1$ for $\bar{\psi}^{\dot{\alpha}}$ and 4 for $H^{\alpha\dot{\alpha}}$. ($\psi^\alpha$ and $\bar{\psi}^{\dot{\alpha}}$ each count as $-1 \cdot \frac{1}{2} \cdot 2 = -1$ due to a $-1$ for Fermi statistics, a $\frac{1}{2}$ to cancel the effect of having squared the kinetic operator $\nabla$, and 2 for the trace over $\alpha$.)

Looking at the kinetic operator and again requiring that each loop contain at least two $D_\alpha$'s and two $\bar{D}_{\dot{\alpha}}$'s, we discover two sources for such terms: The explicit $\mathbf{W}^\alpha{}_\beta{}^\gamma \, \boldsymbol{\nabla}_\alpha \, M_\gamma{}^\beta$ (to this order we can replace $\boldsymbol{\nabla}_\gamma$ by flat superspace $D_\gamma$) and those contained in the covariant d'Alembertian:

$$\Box = \frac{1}{2} \boldsymbol{\nabla}^{\underline{a}} \, \boldsymbol{\nabla}_{\underline{a}}$$

$$= \frac{1}{2} \left( \mathbf{E}^{\underline{a}\underline{m}} \partial_{\underline{m}} + \mathbf{E}^{\underline{a}\mu} D_\mu + \mathbf{E}^{\underline{a}\dot{\mu}} \bar{D}_{\dot{\mu}} + \boldsymbol{\Phi}^{\underline{a}}(M) \right) \left( \mathbf{E}_{\underline{a}}{}^{\underline{n}} \partial_{\underline{n}} + \mathbf{E}_{\underline{a}}{}^\nu D_\nu + \mathbf{E}_{\underline{a}}{}^{\dot{\nu}} \bar{D}_{\dot{\nu}} + \boldsymbol{\Phi}_{\underline{a}}(M) \right) \quad , \tag{7.8.8}$$

where $\mathbf{E}_{\underline{a}}{}^{\underline{m}} - \delta_{\underline{a}}{}^{\underline{m}}$ and all other quantities contain at least one factor of the external fields. (The connection terms $\Phi$ can be dropped: They cannot contribute to any graph with at most four external lines because they do not bring with them any $D$'s.)

We now make the following observations:

(1) Since connection terms can be dropped, $\Box$ acts in the same way on all fields.

(2) Since we need two $D$'s and two $\bar{D}$'s, and each one brings with it an $\mathbf{E}$ or a $\mathbf{W}$ and an $\bar{\mathbf{E}}$ or a $\bar{\mathbf{W}}$, our result will contain four such factors, two barred and two unbarred. The $\mathbf{E}$ vertices have the form $\mathbf{E}^{\underline{a}\beta}\partial_{\underline{a}}D_\beta$ or $\mathbf{E}^{\underline{a}\dot{\beta}}\partial_{\underline{a}}\bar{D}_{\dot{\beta}}$.

(3) Since $\partial_{[\underline{a}}\mathbf{E}_{\underline{b}]\gamma} \sim \mathbf{W}_{\alpha\beta\gamma}$, and the vector indices on the two $\mathbf{E}$'s in a term with only two such factors must be contracted (and therefore also the spinor indices) in order to produce a covariant contribution, there are $\mathbf{E}^2\,\bar{\mathbf{W}}^2$ and $\bar{\mathbf{E}}^2\,\mathbf{W}^2$ terms but no $\mathbf{E}\,\bar{\mathbf{E}}\,\mathbf{W}\,\bar{\mathbf{W}}$ terms. (Similarly there are no $\mathbf{E}^2\,\bar{\mathbf{E}}\,\bar{\mathbf{W}}$ terms or $\mathbf{E}\,\mathbf{W}\,\bar{\mathbf{W}}^2$ terms.)

(4) Due to the algebra of the Lorentz generators, the $\mathbf{W}\,M\,\mathbf{W}\,M$ factors in either the $\bar{\mathbf{E}}^2\,\mathbf{W}^2$ or $\mathbf{W}^2\,\bar{\mathbf{W}}^2$ produce an extra numerical factor of $a$ (and $b$ from $\bar{\mathbf{W}}\,\bar{M}\,\bar{\mathbf{W}}\,\bar{M}$ ) related to the number of spinor indices.

Therefore, there are three types of terms to consider: (1) $(\mathbf{E}\cdot\partial\,)^2\,(\bar{\mathbf{E}}\cdot\partial\,)^2$ , (2) $(\mathbf{E}\cdot\partial\,)^2\,(\bar{\mathbf{W}}\;\bar{M}\,)^2$ (and $h.\,c.$), (3) $(\mathbf{W}\,M)^2\,(\bar{\mathbf{W}}\,\bar{M})^2$. Each term takes the same form for all $N$, but with a coefficient determined by summing over $V$, $\psi$, $\bar{\psi}$, and $H$ the product *(number of such fields) · (Lorentz trace factor) · (M-factor).* The number of fields is given in Table 7.8.1, the trace factor was discussed earlier, and the "$M$-factor" is, for each of the three types of terms, respectively: (1) 1, (2) $b$ ($a$ for the h.c.), (3) $a\cdot b$. The values of the overall numerical coefficient are presented in Table 7.8.2. These coefficients are labeled (1) $C_4$, (2) $C_2$, (3) $C_0$, and appear in eq.(7.8.6). (Note that only $H$ contributes to the last column, since only $H$ has both a dotted and an undotted index, and so has both $\mathbf{W}$ and $\bar{\mathbf{W}}$ terms in its kinetic operator.) Our result is thus that: (1) $N = 1, 2$ contain *all* types of terms, including contributions from chiral superfields, which we have not discussed; (2) $N = 3, 4$ receive no contributions from chiral superfields; (3) $N = 5, 6$ lack *also* the $\mathbf{E}^2\,\bar{\mathbf{E}}^2$ type term; (4) $N = 8$ receives a contribution from *only* the $\mathbf{W}^2\,\bar{\mathbf{W}}^2$ term, in analogy to the $N = 4$ supersymmetric Yang-Mills calculation.

| $N$ | $C_4$ | $C_2$ | $C_0$ |
|---|---|---|---|
| 3 | 5 | 5 | 4 |
| 4 | 2 | 4 | 4 |
| 5 | 0 | 3 | 4 |
| 6 | 0 | 2 | 4 |
| 8 | 0 | 0 | 4 |

*Table 7.8.2. Multiplicity of contributions of each type to the one-loop effective action*

The calculation of the effective action proceeds as follows: For $N = 8$ supergravity there is nothing more to do. One has a box graph, with two factors of $\mathbf{W}$ and two factors of $\bar{\mathbf{W}}$ and a scalar loop integral to perform. We obtain the $G_0$ terms in eq.(7.8.6) with a factor $C_0 = 4$, while $C_4 = C_2 = 0$. For $N = 5, 6$ one has a box graph with two vertices of the form $\mathbf{E} \cdot \partial$ and two with $\mathbf{W}$'s, as well as a triangle graph with one vertex containing two $\mathbf{E}$-factors. The loop integral for the box graph contains now two loop-momentum factors, but gauge (local supersymmetry) invariance can be used to split off a part which gives $\partial_{[\underline{a}}\mathbf{E}_{\underline{b}]}$ so as to produce $\mathbf{W}$'s, while the rest must cancel the triangle graph contribution. (They actually may contribute to $\Gamma_2$, which is zero by momentum conservation.) Finally, for $N = 3, 4$ one has a box graph with one $\mathbf{E} \cdot \partial$ factor at each vertex, a triangle graph with one vertex containing two $\mathbf{E} \cdot \partial$ factors, and also a self-energy type graph with both vertices containing two $\mathbf{E} \cdot \partial$ factors. Again gauge invariance can be used to extract the complete contribution from the box graph, the remainder adding up to zero.

The S-matrix can be obtained from the effective action by dropping the $p_i$ integrals and taking a sum of $G$ terms over permutations of the Mandelstam invariants $s = (p_1 + p_2)^2$, $t = (p_1 + p_4)^2$, $u = (p_1 + p_3)^2$. (In the Yang-Mills case this also involves interchange of internal symmetry indices). Thus, for $N = 8$ supergravity we have

$$S(s, t, u) = (2\pi)^4 \delta\left(\sum (p_i)\right) \int d^4\theta$$

$$\times \mathbf{W}^{\alpha\beta\gamma}(p_1, \theta)\, \mathbf{W}_{\alpha\beta\gamma}(p_2, \theta)\, \bar{\mathbf{W}}^{\dot{\alpha}\dot{\beta}\dot{\gamma}}(p_3, \theta)\, \bar{\mathbf{W}}_{\dot{\alpha}\dot{\beta}\dot{\gamma}}(p_4, \theta)$$

$$\times [\, G_0(s, t, u) + G_0(s, u, t) + G_0(u, t, s)\,] \ . \qquad (7.8.9)$$

The $\theta$-integration splits up the product of the superfields $\mathbf{W}$ into a sum of terms involving products of Weyl tensors and gravitino field strengths which can be replaced with momenta and polarization vectors for the various processes. The actual value of $G_0$ is

$$G_0(s,t,u) = \frac{\pi^{2-\epsilon}}{(2\pi)^4} \cdot \frac{(\Gamma(-\epsilon))^2\Gamma(\epsilon)}{\Gamma(-2\epsilon)}$$

$$\times\big[s^{-2-\epsilon} F(1,1,1-\epsilon,-\frac{u}{s}) + t^{-2-\epsilon} F(1,1,1-\epsilon,-\frac{u}{t})\big], \tag{7.8.10}$$

where $\epsilon = 2 - \frac{\text{D}}{2}$ and $F$ is a hypergeometric function. It is ultraviolet finite but infrared divergent for both Yang-Mills and supergravity, but in the latter case the divergence is milder because of cancellations in the $s, t, u$ permutations. The expressions for $G_2$ , $G'_2$ , $G_4$, and $G'_4$ are somewhat more complicated and will not be given here.

So far our results are with only $N = 1$ supergravity external particles (gravitons and gravitini). However, the S-matrix can be extended immediately to the other particles of an $N > 1$ multiplet either by direct global supersymmetry transformations on the S-matrix or by realizing that the $\int d^4\theta\, (W^{\alpha\beta\gamma})^2\, (\bar{W}_{\dot\alpha\dot\beta\dot\gamma})^2$ can be extended to a similar expression involving products of four on-shell field strengths for extended supergravity. We also note that the $\mathbf{W}^2\, \bar{\mathbf{W}}^2$ form of the result implies the helicity conservation properties of the supersymmetric S-matrix.

The remarkable simplicity of the calculations and results is due in part to the *surprising* decrease in number of diagrams one has to consider as one proceeds from $N = 1$ to $N = 8$. In particular, the absence of chiral superfields for $N \geq 3$ produces the crucial simplification, and the ensuing cancellations between various fields culminates in the absolute triviality of the calculation for $N = 8$ supergravity. At the other extreme, the $N = 0$ theory (ordinary Einstein gravity) would seem to require a major computer calculation.

## 7.9. Locally supersymmetric dimensional regularization

If the only interesting supergravity theories are those that are finite, the construction of a regularization which manifestly preserves local supersymmetry is somewhat of an academic exercise. Nevertheless, we shall discuss the procedure here for completeness, and because the general method is useful for discussion of dimensional reduction to integral dimensions.

It is possible to extend the supersymmetric dimensional regularization method of sec. 6.6 for application to supergravity. Some modifications are required because, unlike matter (scalar or vector) multiplets, supergravity is no longer on-shell irreducible after dimensional reduction. The dimensional reduction must therefore be performed in a manner that picks out the irreducible part. In superfield language, the difference occurs because ($N=1$) matter multiplets are described by scalar superfields, whereas supergravity is described by a vector superfield. Upon naive dimensional reduction to D-dimensions, this vector superfield reduces to a superfield which is a D-dimensional vector, describing pure supergravity, plus 4-D scalar superfields, describing vector multiplets. The dimensional reduction must therefore be redefined so that only the D-dimensional-vector superfield appears. We therefore need a superfield formulation that describes pure supergravity for arbitrary D<4. For simplicity we will describe the construction for $N=1$, but the method can be easily generalized to any four-dimensional superfield theory. For integral dimensions, the $N=1$ theory reduces to pure $N=2$ supergravity for D=3 or 2, and pure $N=4$ supergravity for D=1.

We begin by constructing covariant derivatives. Since upon dimensional reduction the Lorentz group $SO(3,1)$ is broken down to $SO(\mathrm{D}-1,1)\otimes SO(4-\mathrm{D})$, our covariant derivatives take the form

$$\nabla_A = E_A + \frac{1}{2}\Phi_{A\underline{b}}{}^{\underline{c}} M_{\underline{c}}{}^{\underline{b}} + \frac{1}{2}\Phi_{Ab}{}^{c} M_{c}{}^{b} \quad , \tag{7.9.1}$$

where the supervector index $A=(\alpha,\dot{\alpha},\underline{a})$, and we have reduced a 4-dimensional vector index "$\alpha\dot{\alpha}$" into a D-dimensional vector index "$\underline{a}$" plus an internal symmetry ($SO(4-\mathrm{D})$) index "$a$". The field strengths are defined as usual:

$$[\nabla_A , \nabla_B\} = T_{AB}{}^{C}\nabla_C + \frac{1}{2} R_{AB\underline{c}}{}^{\underline{d}} M_{\underline{d}}{}^{\underline{c}} + \frac{1}{2} R_{ABc}{}^{d} M_{d}{}^{c} \quad . \tag{7.9.2}$$

The derivatives contained in $E_A = E_A{}^M D_M$ range over D (commuting) + 4

(anticommuting) coordinates. The fact that the spacetime coordinates have been reduced from 4 to D automatically takes care of reducing the fundamental superfield $H^{\underline{a}}$ to a D-component vector, as discussed above ($H = H^{\underline{a}} i\partial_{\underline{a}}$). However, as compared to the usual 4-dimensional covariant derivatives, we have less gauge freedom due to the absence of a Lorentz generator of the mixed type $M_{a\underline{b}}$, so that an additional constraint is needed to account for this lost invariance. Specifically, this means that the object $N_\alpha{}^\beta$ which appears upon solving the constraints, and which is gauged away by local Lorentz transformations in D=4, must have the components which are not gauged away in D<4 constrained away. We therefore impose the following set of constraints for arbitrary D$\leq$ 4 (for simplicity, we choose the case $n = -\frac{1}{3}$):

$$\textit{Conventional}: \quad T_{\alpha\beta}{}^\gamma = T_{\alpha[\underline{b}}{}^{\underline{c}]} = T_{\alpha\dot\beta}{}^{\dot\gamma} = \sigma_{\underline{a}}{}^{\alpha\dot\beta} R_{\alpha\dot\beta\underline{c}}{}^{\underline{d}} = \sigma_{\underline{a}}{}^{\alpha\dot\beta} R_{\alpha\dot\beta c}{}^{d} = 0 \quad ,$$

$$\sigma_{\underline{a}}{}^{\alpha\dot\beta} T_{\alpha\dot\beta}{}^{\underline{c}} = i\delta_{\underline{a}}{}^{\underline{c}} \quad ; \tag{7.9.3a}$$

$$\textit{Chirality preserving}: \quad T_{\alpha\beta}{}^{\underline{c}} = T_{\alpha\beta}{}^{\dot\gamma} = 0 \quad ; \tag{7.9.3b}$$

$$\textit{Conformal breaking}: \quad T_{\alpha\underline{b}}{}^{\underline{b}} - T_{\alpha\dot\beta}{}^{\dot\beta} = 0 \quad ; \tag{7.9.3c}$$

$$\textit{Additional conventional}: \quad \sigma_a{}^{\alpha\dot\beta} T_{\alpha\dot\beta}{}^{\underline{c}} = 0 \quad ; \tag{7.9.3d}$$

where the additional conventional constraint is the one that constrains the extra components of $N_\alpha{}^\beta$. ($\sigma_{\underline{a}}{}^{\alpha\dot\beta}$ is the D-dimensional Pauli matrix, which projects out the D-component vector index $\underline{a}$ from the 4-dimensional $\alpha\dot\beta$; similarly, $\sigma_a{}^{\alpha\dot\beta}$ projects out the 4-D-component index $a$. Our normalization here is $\sigma_{\underline{a}}{}^{\alpha\dot\beta}\sigma^{\underline{b}}{}_{\alpha\dot\beta} = \delta_{\underline{a}}{}^{\underline{b}}$, $\sigma_a{}^{\alpha\dot\beta}\sigma^b{}_{\alpha\dot\beta} = \delta_a{}^b$, $\sigma_{\underline{a}}{}^{\alpha\dot\beta}\sigma^{\underline{a}}{}_{\gamma\dot\delta} + \sigma_a{}^{\alpha\dot\beta}\sigma^a{}_{\gamma\dot\delta} = \delta_\gamma{}^\alpha \delta_{\dot\delta}{}^{\dot\beta}$.) The conventional constraints as written are somewhat redundant, but it can be shown that, in conjunction with the remaining constraints, they serve to determine $\nabla_A$ in terms of $E_\alpha$, as usual. The chirality-preserving and conformal-breaking constraints have a solution similar to that of D=4, except that $H^M$ in $H = H^M iD_M$ is now a D+4 component supervector. The solution to the constraints is (cf. sec. 5.3):

$$E_\alpha = \bar\Psi N_\alpha{}^\beta \hat{E}_\beta \quad ,$$

$$E_{\underline{a}} = \Psi\overline{\Psi}(N_{\underline{a}}{}^{\underline{b}} - N_{\underline{a}}{}^{c} A_c{}^{\underline{b}})\hat{E}_{\underline{b}} + (f_{\underline{a}}{}^{\alpha}\hat{E}_{\alpha} + \bar{f}_{\underline{a}}{}^{\dot{\alpha}}\hat{E}_{\dot{\alpha}}) \quad ,$$

$$\hat{E}_{\alpha} = e^{-\Omega}D_{\alpha}e^{\Omega} \quad , \quad \hat{E}_{\underline{a}} = -i\sigma_{\underline{a}}{}^{\alpha\dot{\alpha}}\{\hat{E}_{\alpha}, \hat{E}_{\dot{\alpha}}\} \quad , \quad \Omega = \Omega^M iD_M \quad ,$$

$$[\hat{E}_A, \hat{E}_B\} = \hat{C}_{AB}{}^C\hat{E}_C \quad , \quad A_a{}^{\underline{b}} = i\sigma_a{}^{\alpha\dot{\alpha}}\hat{C}_{\alpha\dot{\alpha}}{}^{\underline{b}} \quad ,$$

$$N_{\alpha\dot{\alpha}}{}^{\beta\dot{\beta}} = N_{\alpha}{}^{\beta}\bar{N}_{\dot{\alpha}}{}^{\dot{\beta}} \quad , \quad det\ N_{\alpha}{}^{\beta} = 1 \quad , \quad \hat{N} = det(N_{\underline{a}}{}^{\underline{b}} - N_{\underline{a}}{}^{c}A_c{}^{\underline{b}}) \quad ,$$

$$\overline{\Psi} = \phi^{\frac{1}{2}}\bar{\phi}^{-\frac{\mathrm{D}}{2(\mathrm{D}-2)}}[(1\cdot e^{-\overleftarrow{\Omega}})^{\mathrm{D}}(1\cdot e^{\overleftarrow{\bar{\Omega}}})^{-(\mathrm{D}-2)}\hat{E}^2\hat{N}^2]^{-\frac{1}{4(\mathrm{D}-1)}} \quad , \quad \hat{E}_{\dot{\alpha}}\phi = 0 \quad . \tag{7.9.4}$$

The matrix $N_{\alpha\dot{\alpha}}{}^{\beta\dot{\beta}}$ is determined by (7.9.3d) to take the form

$$N_{\alpha\dot{\alpha}}{}^{\beta\dot{\beta}} = \begin{pmatrix} N_{\underline{a}}{}^{\underline{b}} & N_{\underline{a}}{}^{b} \\ N_{a}{}^{\underline{b}} & N_{a}{}^{b} \end{pmatrix} = \begin{pmatrix} (1+A^TA)^{-\frac{1}{2}} & -(1+A^TA)^{-\frac{1}{2}}A^T \\ (1+AA^T)^{-\frac{1}{2}}A & (1+AA^T)^{-\frac{1}{2}} \end{pmatrix} \tag{7.9.5}$$

in an appropriate Lorentz ⊗ internal gauge, where double spinor indices are converted into vector indices and back again with $\sigma$'s (of the appropriate type), and the last equation is written in matrix notation with $A = A_a{}^{\underline{b}}$. $N_{\alpha}{}^{\beta}$ is determined from this expression for $N_{\alpha\dot{\alpha}}{}^{\beta\dot{\beta}}$ by using the relation

$$N_{\alpha\dot{\alpha}}{}^{\beta\dot{\beta}} = (e^X)_{\alpha\dot{\alpha}}{}^{\beta\dot{\beta}} \quad , \quad X_{\alpha\dot{\alpha}}{}^{\beta\dot{\beta}} = \delta_{\dot{\alpha}}{}^{\dot{\beta}}Y_{\alpha}{}^{\beta} + \delta_{\alpha}{}^{\beta}Y_{\dot{\alpha}}{}^{\dot{\beta}}$$

$$\rightarrow N_{\alpha}{}^{\beta} = (e^Y)_{\alpha}{}^{\beta} \quad . \tag{7.9.6}$$

Since $N_{\alpha\dot{\alpha}}{}^{\beta\dot{\beta}}$ is orthogonal, $X$ is antisymmetric, and therefore can be expressed in terms of the traceless $Y$. We have not given the explicit expression for $f_{\underline{a}}{}^{\alpha}$ in (7.9.4), nor the solutions for the connections, but they can be obtained as in D=4 without further complications, and will not be needed here.

The supergravity action is

$$S = -2\frac{\mathrm{D}-1}{\mathrm{D}-2}\kappa^{-2}\int d^{\mathrm{D}}x d^4\theta\, E^{-1}$$

$$= -\,2\,\frac{\mathrm{D}-1}{\mathrm{D}-2}\,\kappa^{-2}\int d^{\mathrm{D}}x d^4\theta\,\hat{E}^{-\frac{1}{(\mathrm{D}-1)}}[det(\delta_{\underline{a}}{}^{\underline{b}}+A^c{}_{\underline{a}}A_c{}^{\underline{b}})]^{-\frac{1}{2(\mathrm{D}-1)}}[(1\cdot e^{-\overleftarrow{\Omega}})(1\cdot e^{\overleftarrow{\bar{\Omega}}})]^{\frac{(\mathrm{D}-2)}{2(\mathrm{D}-1)}}\phi\bar{\phi} \quad . \tag{7.9.7}$$

Projection operator methods can be used to show that the linearized action contains only the usual superspins $\frac{3}{2}\oplus 0$. Coupling to matter can now be performed as in D=4, with chiral Lagrangians integrated by $\int d^{\mathrm{D}}x d^2\theta\,\phi^{\frac{2(\mathrm{D}-1)}{(\mathrm{D}-2)}}$. Supergraph calculations can be performed with the usual four-dimensional $D$-algebra. We do momentum integration as in conventional dimensional regularization, and minimally subtract the divergent part using $\frac{1}{\epsilon}$ times a local, covariant, D-dimensional counterterm constructed from the D-dimensional covariants.

The same inconsistencies that occurred in globally supersymmetric dimensional regularization of course remain in the local case. Nevertheless, as in the global case, in actual computations the inconsistencies seem to disappear after taking D$\to$ 4. After minimal subtraction, the remaining finite quantity satisfies the 4-dimensional local supersymmetry Ward-Takahashi identities (after taking D$\to$ 4). Furthermore, the method is perfectly consistent for reduction to integral dimensions, and can be used for describing extended supergravity in lower dimensions. However, we observe that the above superfield description in nonintegral dimensions defies understanding in terms of components. (E.g., since the "D-bein" in $H^{\underline{a}}$ has no $e_a{}^b$ part, what field gauges the internal $M_a{}^b$ symmetry?)

Since the subtraction procedure preserves local scale invariance when the compensator $\phi$ is included, the renormalized effective action will be superconformally invariant. However, D= 4 superconformal anomalies are in general present precisely because the renormalized effective action depends on $\phi$. We discuss this in the next section.

## 7.10 Anomalies

### a. Introduction

Anomalies in local conservation laws are harmless as long as no fields couple to the corresponding current. In divergent component theories there is always at least one such anomaly: the scale anomaly. This anomaly, which can be expressed as an additional contribution to the trace of the energy-momentum tensor, occurs because a new mass scale is introduced at the quantum level, the renormalization mass parameter. For example, a theory that is classically conformally invariant, and thus has a classical energy-momentum tensor with vanishing trace, gets quantum contributions to the trace. When Einstein gravity is coupled to the quantum system, this anomaly is harmless, as general coordinate invariance merely requires conservation of the energy-momentum tensor (i.e., the vanishing of its covariant divergence). However, it would be harmful in conformal gravity, since local (Weyl) scale invariance does require vanishing of the trace.

Quantum corrections to the energy-momentum tensor are most conveniently defined by coupling the quantum system to background gravity and defining

$$T^{\underline{m}\underline{n}} = \frac{\Delta\Gamma_R}{\Delta g_{\underline{m}\underline{n}}} \tag{7.10.1}$$

where $\Gamma_R$ is the renormalized effective action and $\Delta$ is a suitably defined variation (see below). Its trace is given by

$$T_{\underline{m}}{}^{\underline{m}} = g_{\underline{m}\underline{n}} \frac{\Delta\Gamma_R}{\Delta g_{\underline{m}\underline{n}}} \tag{7.10.2}$$

Alternatively, it can be obtained by first introducing a compensating scalar into the theory (5.1.34). For conformal theories the compensator decouples from the classical action, but in general it enters in the renormalized action where it couples to the trace. Therefore, varying $\Gamma_R$ with respect to the compensator determines $T_{\underline{m}}{}^{\underline{m}}$.

In supersymmetry, the energy-momentum tensor is the $\theta\bar{\theta}$ component of a superfield, the *supercurrent* $J_{\alpha\dot{\alpha}}$. (More precisely, $\frac{1}{8}[\bar{D}_{\dot{\alpha}}, D_\alpha]J_{\beta\dot{\beta}}| + \underline{a} \longleftrightarrow \underline{b} = T_{\underline{a}\underline{b}} - \frac{1}{3}\eta_{\underline{a}\underline{b}}T_{\underline{c}}{}^{\underline{c}}$.) The trace of the energy-momentum tensor is a component of a related superfield, the *supertrace* $J$ ($\frac{1}{2}(D^2 J| + h.c.) = \frac{1}{3}T_{\underline{a}}{}^{\underline{a}}$). Just as the energy-momentum tensor can be defined from the coupling to gravity, the supercurrent $J_{\alpha\dot{\alpha}}$ can be defined from the coupling to the supergravity superfield $H^{\alpha\dot{\alpha}}$:

$$J_{\alpha\dot{\alpha}} = \frac{\Delta\Gamma}{\Delta H^{\alpha\dot{\alpha}}} \quad . \tag{7.10.3}$$

As we will discuss below, the supertrace can be defined by functional differentiation with respect to the compensating superfield. In classical locally superscale invariant theories the compensator decouples and therefore the supertrace vanishes. In general, its presence is a measure of the breaking of local superscale invariance.

The supercurrent also contains the supersymmetry current (at linear order in $\theta$) and the R-symmetry axial current (at $\theta = 0$); the supertrace also contains the $\gamma$-trace of the supersymmetry current (at linear order in $\theta$) and the divergence of the axial current (the imaginary part of the $\theta^2$ component). Thus, in a supersymmetric theory where scale invariance is broken, the axial current has a chiral anomaly and the supersymmetry current has an $S$-supersymmetry anomaly, and the coefficients of all three anomalies are equal. However, just as translational invariance is not violated (the trace of the energy-momentum tensor is anomalous, not its divergence), neither is ordinary $Q$-supersymmetry (the $\gamma$-trace of the supersymmetry current is anomalous, not its divergence).

In locally supersymmetric theories, in addition to the superconformal anomalies described by the supertrace, there may exist anomalies in the Ward identities of local (Poincaré) supersymmetry. For the cases that have been studied they do not occur in $N = 1$ theory for $n = -\frac{1}{3}$ (the minimal set of auxiliary fields) because we can regularize in a manner consistent with local supersymmetry; they do occur in general for nonminimal (and new minimal) $N = 1$, $n \neq -\frac{1}{3}$ theories. We will use the existence of superconformal anomalies to infer the existence of these "auxiliary-field" anomalies and conclude that in general only $n = -\frac{1}{3}$ theory is quantum consistent.

**b. Conformal anomalies**

We first review one-loop "on-shell" scale anomalies in component theories. We are interested in quantum corrections to matrix elements of the energy-momentum tensor between the vacuum and a state containing two or more gravitons. (We could consider other external particles, and also "off-shell" anomalies, but when gravity is quantized only the on-shell ones are unambiguous.) Equivalently, we compute the one-loop effective action for a system in a gravitational background, functionally differentiate with respect to $g_{\underline{m}\underline{n}}$ and *then* set the background field on shell. At the level of Feynman diagrams, because of covariance, we need only consider the two-point function (graviton propagator correction), which determines all the one-loop divergences, and the three-point function ("triangle graph"), which determines the trace of the energy-momentum tensor. In fact, if the classical theory for the field in the loop is conformally invariant, *all* the relevant information can be extracted from the two-point function.

In classical theories conformal invariance is broken in two ways: (a) by mass terms that break it softly and whose effect can be separated out, as they can be for the divergence of the axial current, and: (b) by hard terms, e.g., derivatives of fields, as for Yang-Mills in D$\neq$ 4 dimensions, and for antisymmetric tensor fields. In the subsequent discussion, when we refer to nonconformal theories we mean classical theories where the breaking is hard.

We consider first a classically conformally invariant theory so that the trace of the classical energy-momentum tensor is zero. When coupled to gravity, the classical theory is locally scale invariant. By introducing appropriate D-dependence the theory can be dimensionally regularized so that this invariance is preserved in the regularized effective action near D$=$ 4. (The invariance is broken only to order $(\mathrm{D}-4)^2$, and thus has no effect even in $(\mathrm{D}-4)^{-1}$ divergent terms.) However, the coefficient of the $(\mathrm{D}-4)^{-1}$ factor is not separately locally scale invariant except at D $=4$, and there is no local finite term that can be added to it to make it so. Therefore, the renormalized effective action, defined by subtracting this D-dimensional, local, covariant divergent term from the regularized effective action (i.e., by adding a counterterm $S_\infty$) is not locally scale invariant. Consequently, when we compute the trace of $T^{\underline{m}\underline{n}}$ defined in terms of the renormalized effective action we find a nonzero result. Since the regularized, unrenormalized effective action $\Gamma_U$ was scale invariant, the scale anomaly of the renormalized effective action $\Gamma_R$ is just the trace computed from the D-dimensional counterterm $S_\infty$.

We have defined $T^{\underline{m}\underline{n}}$ in (7.10.1). The variation $\Delta$ is defined in terms of $\delta$ by

$$\frac{\Delta f}{\Delta g_{\underline{m}\underline{n}}} = g^{-\frac{1}{2}}\frac{\delta f}{\delta g_{\underline{m}\underline{n}}} \tag{7.10.4}$$

($g = det(g_{\underline{m}\underline{n}})$), or directly by

$$\frac{\Delta g_{\underline{mn}}(x)}{\Delta g_{\underline{pq}}(x')} = \frac{1}{2}\,\delta_{(\underline{m}}{}^{\underline{p}}\delta_{\underline{n})}{}^{\underline{q}} g^{-\frac{1}{2}}\delta^4(x-x') \quad . \tag{7.10.5}$$

The local scale (trace) anomaly is then

$$g_{\underline{mn}}T^{\underline{mn}} = g_{\underline{mn}}\,\frac{\Delta\Gamma_R}{\Delta g_{\underline{mn}}} = g_{\underline{mn}}\,\frac{\Delta S_\infty}{\Delta g_{\underline{mn}}} \quad . \tag{7.10.6}$$

*The last equality holds only because we are considering classically conformal theories.* Otherwise, the former two expressions, *the total trace,* do not equal the last expression, *the trace anomaly.* (In general, the *anomaly* is understood to be a contribution to the trace due to the divergences of the theory.)

Since in classically conformal theories $\Gamma_U$ is locally scale invariant near $\mathrm{D}=4$, $S_\infty$ takes the form of $(\mathrm{D}-4)^{-1}$ times a local (general coordinate) invariant that is the dimensional continuation of a 4-dimensional object that is locally scale invariant. From dimensional and covariance considerations we find two independent four-dimensional objects of this form: In terms of the irreducible parts of the curvature of (5.1.21), they are the Euler number

$$\chi = \frac{1}{(4\pi)^2}\,\frac{1}{2}\int d^4x\, g^{\frac{1}{2}}[\tfrac{1}{2}\,(w^{\alpha\beta\gamma\delta}w_{\alpha\beta\gamma\delta} + h.\,c.\,) - r^{\alpha\beta\dot{\alpha}\dot{\beta}}r_{\alpha\beta\dot{\alpha}\dot{\beta}} + 3r^2] \quad , \tag{7.10.7}$$

a topological invariant whose functional variation vanishes and which itself vanishes in topologically trivial spacetime for $\mathrm{D}=4$, and the integral of just the Weyl tensor $(w^2+\overline{w}^2)$. The difference between the two vanishes on shell ($r_{\alpha\beta\dot{\alpha}\dot{\beta}} = r = 0$).

In quantum gravity the coefficient of any term that vanishes on shell is in general gauge-dependent, and in fact can be made to vanish by an appropriate gauge choice, or can be eliminated by a local field redefinition of the metric (since such redefinitions of the action are proportional to the field equations). Therefore, we consider only the on-shell part of the trace anomaly, which we write in terms of $w^2 = \frac{1}{2}\,w^{\alpha\beta\gamma\delta}w_{\alpha\beta\gamma\delta}$. We note that $w^2$ has the simple scaling property for *arbitrary* D

$$(g_{\underline{mn}}\,\frac{\Delta}{\Delta g_{\underline{mn}}})(x)(g^{\frac{1}{2}}w^2)(x') = \frac{1}{2}\,(\mathrm{D}-4)(g^{\frac{1}{2}}w^2)\delta^4(x-x') \quad . \tag{7.10.8}$$

In D-dimensions the relevant part of $\Gamma_U$ is given by a covariantization of a graviton self-energy graph and has the form

$$\Gamma_U \sim \frac{1}{D-4} \int d^D x\, g^{\frac{1}{2}} \frac{1}{2} w(\frac{``\Box''}{\mu^2})^{\frac{D}{2}-2} w + h.c. \quad , \tag{7.10.9}$$

where $``\Box''$ means $\Box$ + curvature terms necessary to make $\Gamma_U$ locally scale invariant in arbitrary D, and $\mu$ is a renormalization mass. ($\Gamma_U$ also contains finite locally scale invariant terms, and divergent terms that vanish on shell.) We then have

$$S_\infty \sim -\frac{1}{D-4} \int d^D x\, g^{\frac{1}{2}} w^2 + h.c. \quad , \tag{7.10.10}$$

$$\Gamma_R \sim \int d^4 x\, g^{\frac{1}{2}} \frac{1}{4} w\, ln(\frac{\Box}{\mu^2}) w + h.c. \quad . \tag{7.10.11}$$

By integration by parts (dropping finite terms proportional to the Euler number, which can be considered part of the corresponding infinite term (7.10.10)), $\Gamma_R$ can be rewritten at D = 4

$$\Gamma_R \sim \int d^4 x\, g^{\frac{1}{2}} \{ \frac{1}{2} r^{\alpha\beta\dot{\alpha}\dot{\beta}}\, ln(\frac{\Box}{\mu^2}) r_{\alpha\beta\dot{\alpha}\dot{\beta}} - \frac{3}{2} r\, ln(\frac{\Box}{\mu^2}) r$$

$$+ (w^2 + \bar{w}^2)\, ln[1 - \frac{1}{\Box + r} r] \} \tag{7.10.12}$$

plus more finite terms that are locally scale invariant, and terms of third or higher order in $r$ and $r_{\alpha\beta\dot{\alpha}\dot{\beta}}$. Since $[1 - (\Box + r)^{-1} r]$ satisfies the scale covariant equation $(\Box + r)\phi = 0$ (with our conventions of sec. 5.1, $\Box + r$ is the kinetic operator of a locally scale-covariant scalar), it can be shown that

$$(g_{\underline{mn}} \frac{\Delta}{\Delta g_{\underline{mn}}})(x)\, ln[1 - \frac{1}{\Box + r} r](x') = -\frac{1}{2} \delta^4(x - x') \quad , \tag{7.10.13}$$

Therefore, using (7.10.8), we see that (7.10.12) gives the same (on-shell) trace (from the last term) as $\Gamma_R$ in (7.10.11) (or as $S_\infty$ in (7.10.10)):

$$g_{\underline{mn}} T^{\underline{mn}} \sim -\frac{1}{2} (w^2 + \bar{w}^2) \quad . \tag{7.10.14}$$

We find (7.10.12) a more convenient form of representing $\Gamma_R$. The first two terms are covariantized self-energy contributions unambiguously expressed in terms of the curvature scalar and Ricci tensor, and are of no interest for on-shell traces since their variation vanishes on shell. The last term, when the *ln* is expanded in powers of $r$, has the

form

$$\int d^4x\, g^{\frac{1}{2}} w^2 \frac{1}{\Box} r + \cdots = \int d^4x\, g^{\frac{1}{2}} r \frac{1}{\Box} w^2 + \cdots , \qquad (7.10.15)$$

so that it receives contributions only from diagrams with at least three external lines. This term is a covariantized triangle graph contribution and could be computed using, for example, the Adler-Rosenberg method with $r$ at the "top" vertex (cf. also (6.7.10-13). The trace operation $g\frac{\delta}{\delta g}$ acting on $r$ in (7.10.15) is analogous to the $\overline{\nabla}^2$ in (6.7.13)).

In the more general case when the quantized theory is not classically conformally invariant, or gravity is also quantized so that $g\frac{\delta}{\delta g}\Gamma_U \neq 0$, local scale invariance cannot be used to determine $g_{\underline{m}\underline{n}}T^{\underline{m}\underline{n}}$ from $S_\infty$. It is then necessary to calculate the total trace from $\Gamma_R$ directly from a triangle graph. (In the case of quantum gravity we use a background field gauge to maintain covariance of $\Gamma_R$.) The general form of the unrenormalized effective action *near* D = 4 is

$$\Gamma_U = k_1 \frac{1}{\epsilon}\chi_{\rm D} + \frac{1}{(4\pi)^2}\int d^4x\, g^{\frac{1}{2}}[k_2 \frac{1}{2} r(\frac{1}{\epsilon} - ln(\frac{``\Box''}{\mu^2}))r$$

$$+ k_3 \frac{1}{2} r^{\alpha\beta\dot\alpha\dot\beta}(\frac{1}{\epsilon} - ln(\frac{``\Box''}{\mu^2}))r_{\alpha\beta\dot\alpha\dot\beta}$$

$$- k_4(w^2 + \overline{w}^2) ln(1 - \frac{1}{\Box + r} r)] \qquad (7.10.16)$$

where $\epsilon = 2 - \frac{\rm D}{2}$ and $\chi_{\rm D}$ is the dimensional continuation of the Euler number of (7.10.7):

$$\chi_{\rm D} = \frac{1}{(4\pi)^{\frac{1}{2}{\rm D}}}\frac{1}{2}\int d^{\rm D}x\, g^{\frac{1}{2}}[\frac{1}{2}(w^{\alpha\beta\gamma\delta}w_{\alpha\beta\gamma\delta} + h.c.) - r^{\alpha\beta\dot\alpha\dot\beta}r_{\alpha\beta\dot\alpha\dot\beta} + 3r^2] \ . \qquad (7.10.17)$$

$\Gamma_R$ is obtained by subtracting out the $\epsilon^{-1}$ terms.

The relevant term for the on-shell trace is again the last one in (7.10.16), although now its coefficient is not related to those of the preceding terms. The on-shell trace computed from $\Gamma_R$ is not equal to the trace computed from $S_\infty$. It receives additional contributions from the classically nonconformal part of the theory. As mentioned above, we will refer to the part attributable to $S_\infty$ as the trace anomaly, while calling the entire contribution from $\Gamma_R$ the total trace.

The number $k_1$ has been computed in a variety of ways, and determines the on-shell trace anomaly of fields in the loop. This quantity is usually written

$$(T_{\underline{m}}{}^{\underline{m}})_{anomalous} = k_1 \frac{1}{(4\pi)^2}\frac{1}{4}[w_{\alpha\beta\gamma\delta}w^{\alpha\beta\gamma\delta} + h.\,c.] \qquad (7.10.18)$$

with $360k_1 = 4, 7, -52, -233, 848, 364,$ for a scalar, Majorana spinor, vector, Rarita-Schwinger field, graviton, and second-rank antisymmetric tensor gauge field, respectively, including their ghosts. We note that although the first and last fields both describe the same spin zero particle (if the scalar has no improvement term), their trace anomalies are different. On the other hand, it can be argued that they have the same total trace, which is the physically relevant quantity, determined by $\Gamma_R$. (For the *improved* scalar field $(T_{\underline{m}}{}^{\underline{m}})_{tot} = (T_{\underline{m}}{}^{\underline{m}})_{anom}$, but for the antisymmetric tensor or unimproved scalar they are different: The latter theories are not classically conformally invariant.) In like fashion, third- and fourth-rank antisymmetric tensor fields, which have no physical degrees of freedom, have zero total trace (in fact, zero renormalized effective action), although because of the quantization procedure, they have a nonzero divergent contribution to $\Gamma_U$ and therefore a nonzero trace anomaly.

A useful method for making scale-breaking properties manifest is to introduce a compensating scalar as in (5.1.34). We then have

$$(g_{\underline{mn}}\frac{\Delta}{\Delta g_{\underline{mn}}} f)(\phi^2 g_{\underline{pq}}) = \frac{1}{2}\phi^{-3}\frac{\delta}{\delta\phi} f \equiv \frac{1}{2}\frac{\Delta}{\Delta\phi} f \quad , \qquad (7.10.19)$$

so the existence of a nonzero trace is equivalent to having dependence on $\phi$. For example, (7.10.10,11) becomes

$$S_\infty \sim -\frac{1}{\mathrm{D}-4}\int d^{\mathrm{D}}x\, g^{\frac{1}{2}}\phi^{\frac{2\mathrm{D}}{\mathrm{D}-2}}w^2 + h.\,c. \ , \qquad (7.10.20)$$

$$\Gamma_R \sim \int d^4x\, g^{\frac{1}{2}}\frac{1}{4}w\, ln(\frac{``\Box''}{\phi^2})w + h.\,c. \ ; \qquad (7.10.21)$$

(note that we have absorbed $\mu$ into $\phi$) and the last term in (7.10.12) becomes

$$\int d^4x\, g^{\frac{1}{2}}(w^2 + \overline{w}^2)\{ln[1 - \frac{1}{\Box + r}r] - ln\,\phi\} \quad . \qquad (7.10.22)$$

If the classical theory is conformally invariant, $\phi$ decouples from the classical action, and thus does not appear in the Feynman rules or in $\Gamma_U$: Its only appearance in $\Gamma_R$ is through $S_\infty$. It must be introduced in $S_\infty$ to make this term scale invariant in D-dimensions, and to compensate for this it must also appear in $\Gamma_R$, since $\Gamma_U$ is independent of $\phi$. On the other hand, if the classical theory is not conformally invariant, $\phi$ is present in $\Gamma_U$, and will enter in $\Gamma_R$ in a manner which is not related to the way it enters in $S_\infty$.

### c. Classical supercurrents

In this subsection we derive the classical supercurrents for various multiplets. These are the superfields that contain the superconformal component currents. They can be obtained in principle from the classical actions by means of Noether's theorem, or can be calculated as the variational derivatives of the covariantized actions with respect to the supergravity prepotentials. In general we do not immediately obtain the same results, unless we perform some field redefinitions. These redefinitions have no physical effect since they only change the currents by terms proportional to the field equations. We consider minimal supergravity with the chiral compensator.

The action for a scalar multiplet in the presence of (background) supergravity is (in the chiral representation)

$$S = \int d^4x\, d^4\theta\; E^{-1}\eta e^{-H}\bar{\eta} + [\int d^4x\, d^2\theta\; \phi^3(\frac{1}{2}m\eta^2 + \frac{1}{6}\lambda\eta^3) + h.\,c.] \quad . \tag{7.10.23}$$

If we make the field redefinition $\eta \to \phi^{-1}\eta$ and use the linearized equation (see (7.5.4))

$$E^{-1}\phi^{-1}(e^{-H}\bar{\phi})^{-1} = 1 - \frac{1}{3}\bar{D}_{\dot{\alpha}}D_\alpha H^{\alpha\dot{\alpha}} - \frac{1}{3}i\partial_{\underline{a}}H^{\underline{a}} \tag{7.10.24}$$

we obtain the *supercurrent*

$$J_{\alpha\dot{\alpha}} \equiv \bar{J}_{\alpha\dot{\alpha}} = \frac{\delta S}{\delta H^{\alpha\dot{\alpha}}} = -\frac{1}{6}[\bar{D}_{\dot{\alpha}}, D_\alpha]\bar{\eta}\eta + \frac{1}{2}\bar{\eta}i\overleftrightarrow{\partial}_{\alpha\dot{\alpha}}\eta$$

$$= -\frac{1}{3}(\bar{D}_{\dot{\alpha}}\bar{\eta})(D_\alpha\eta) + \frac{1}{3}\bar{\eta}i\overleftrightarrow{\partial}_{\alpha\dot{\alpha}}\eta \quad . \tag{7.10.25}$$

The $\theta$-independent component of $J_{\alpha\dot{\alpha}}$ is the (R-transformation) axial current

$J_{\alpha\dot{\alpha}}| = \frac{1}{3}\bar{A}i\overleftrightarrow{\partial}_{\alpha\dot{\alpha}}A - \frac{1}{3}\bar{\psi}_{\dot{\alpha}}\psi_\alpha$, the linear $\theta$-component is the supersymmetry current, and at the $\theta\bar{\theta}$ level we find the (improved) energy-momentum tensor.

We define the *supertrace*

$$J \equiv \frac{\delta S}{\delta \phi^3} = \frac{1}{6}m\eta^2 \quad , \quad \bar{D}_{\dot{\alpha}}J = 0 \quad . \tag{7.10.26}$$

We can verify, using the equations of motion, the conservation equation

$$\bar{D}^{\dot{\alpha}}J_{\alpha\dot{\alpha}} = D_\alpha J \quad . \tag{7.10.27}$$

Quite generally, this equation is a direct consequence of the invariance of the action under $L_\alpha$-transformations (5.2.7,7.4.2b), $\delta H_{\alpha\dot{\alpha}} = D_\alpha \bar{L}_{\dot{\alpha}} - \bar{D}_{\dot{\alpha}}L_\alpha$, $\delta\phi^3 = \bar{D}^2 D_\alpha L^\alpha$:

$$\delta_L S = \frac{\delta S}{\delta H^{\alpha\dot{\alpha}}}\,\delta_L H^{\alpha\dot{\alpha}} + (\frac{\delta S}{\delta \phi^3}\,\delta_L\phi^3 + h.\,c.\,) = 0 \quad . \tag{7.10.28}$$

If the classical theory is conformally invariant the covariantized action is superscale invariant (independent of $\phi$, possibly after field redefinitions, e.g., in the case above if $m = 0$), the supertrace vanishes, and

$$\bar{D}^{\dot{\alpha}}J_{\alpha\dot{\alpha}} = 0 \quad . \tag{7.10.29}$$

This equation expresses the conservation of the axial current, and the vanishing of the supersymmetry current $\gamma$-trace, and of the energy-momentum tensor trace.

For the vector multiplet the flat superspace component currents are contained in the supercurrent $J_{\alpha\dot{\alpha}} = \bar{W}_{\dot{\alpha}}W_\alpha$, where $W_\alpha$ is the flat superfield strength. However, to obtain this expression from coupling to supergravity requires some field redefinitions which we now describe. For simplicity we consider the abelian case.

In the supergravity chiral representation we have the reality condition $V^\dagger = e^H V$. Introducing $V' = e^{\frac{1}{2}H}V$ we have now $V'^\dagger = V'$ and

$$W_\alpha = i(\bar{\nabla}^2 + R)\nabla_\alpha(e^{-\frac{1}{2}H}V) = i\bar{D}^2\Psi^2\bar{\Psi}N_\alpha{}^\beta e^{-H}D_\beta e^{\frac{1}{2}H}V$$

$$= i\phi^{-\frac{3}{2}}\bar{D}^2\hat{E}^{-\frac{1}{2}}N_\alpha{}^\beta e^{-H}D_\beta e^{\frac{1}{2}H}V \quad . \tag{7.10.30}$$

It is convenient to calculate in the gauge (7.6.5) where $N_\alpha{}^\beta \neq \delta_\alpha{}^\beta$ so that spinor chiral fields are chiral in the usual sense. In this gauge, at the linearized level (see sec. 7.5.c)

$$\hat{E}^{-\frac{1}{2}} N_\alpha{}^\beta = \delta_\alpha{}^\beta - \bar{D}_{\dot\gamma} D_\alpha H^{\beta\dot\gamma} \quad . \tag{7.10.31}$$

However, if we calculate the supercurrent by

$$J_{\alpha\dot\alpha} = \frac{\delta}{\delta H^{\alpha\dot\alpha}} \int d^4x\, d^2\theta\ \phi^3\, \frac{1}{2}\, W^\alpha W_\alpha \tag{7.10.32}$$

it will not be (Yang-Mills) gauge invariant because the gauge transformation of $V$ (or $V'$) depends on $H_{\alpha\dot\alpha}$:

$$\delta V' = i(e^{-\frac{1}{2}H}\bar\Lambda - e^{\frac{1}{2}H}\Lambda) \quad , \quad \bar{D}_{\dot\alpha}\Lambda = 0 \quad . \tag{7.10.33}$$

We remedy this by making a further field redefinition

$$V' = (cosh\, \frac{1}{2}\, H + \frac{sinh\, \frac{1}{2}\, H}{\frac{1}{2}\, H}\, \frac{1}{2}\, H^{\alpha\dot\alpha}[\bar{D}_{\dot\alpha}, D_\alpha])V_0 \quad . \tag{7.10.34}$$

Thus $V_0$ has the $H$-independent transformation law $\delta V_0 = i(\bar\Lambda - \Lambda)$, which indicates that the component vector field in $V_0$ has a curved vector index, in contrast to the flat index on that in $V$. At the linearized level, we find

$$\phi^{\frac{3}{2}} W_\alpha = W_{0\alpha} - \bar{D}^2 H_{\alpha\dot\alpha} \bar{W}_0{}^{\dot\alpha} \quad , \tag{7.10.35}$$

where

$$W_{0\alpha} = i\bar{D}^2 D_\alpha V_0 \tag{7.10.36}$$

is the gauge-invariant field strength of $V_0$ (containing the component field strength with curved indices). From the action (7.10.32) we find then

$$J_{\alpha\dot\alpha} = \bar{W}_{0\dot\alpha} W_{0\alpha} \quad , \quad J = 0 \ ;$$

$$\bar{D}^{\dot\alpha} J_{\alpha\dot\alpha} = 0 \quad . \tag{7.10.37}$$

If the (covariantized) supersymmetric gauge-fixing term (6.2.17) is present, we have additional contributions (for $\alpha = 1$)

$$J_{\alpha\dot\alpha}{}^{GF} = -\, \frac{1}{6}\, [\bar{D}_{\dot\alpha}\, , D_\alpha][V_0\{\bar{D}^2\, , D^2\}V_0 + (\bar{D}^2 V_0)(D^2 V_0)]$$

$$+\, \frac{1}{2}\, (\bar{D}^2 V_0) i\overleftrightarrow{\partial}_{\alpha\dot\alpha}(D^2 V_0) - V_0 i\partial_{\underline{a}}[\bar{D}^2\, , D^2]V_0$$

$$- \frac{1}{2} ([\bar{D}_{\dot{\alpha}} , D_\alpha] V_0) \{ \bar{D}^2 , D^2 \} V_0 \ , \qquad (7.10.38a)$$

$$J^{GF} = \frac{1}{3} \bar{D}^2 (V_0 [\bar{D}^2, D^2] V_0) \ . \qquad (7.10.38b)$$

For the tensor multiplet (chiral spinor superfield) there are analogous complications due to the transformation law

$$\delta \eta_\alpha = i (\bar{\nabla}^2 + R) \nabla_\alpha (e^{-\frac{1}{2}H} K') \ . \qquad (7.10.39)$$

We have introduced $K'$ by analogy to $V'$. However, if we redefine $K'$ in terms of $K_0$, and $\eta_\alpha$ in terms of $\eta_{0\alpha}$, by analogy to (7.10.34,35), we find that the covariant field strength

$$G' = \frac{1}{2} e^{\frac{1}{2}H} \nabla_\alpha \eta^\alpha + h.\,c. = - \frac{1}{2} e^{\frac{1}{2}H} E (E^{-1} \eta^\alpha \overleftarrow{E}_\alpha) + h.\,c.$$

$$= - \frac{1}{2} e^{\frac{1}{2}H} E (\phi^{\frac{3}{2}} \eta^\alpha \hat{E}^{-\frac{1}{2}} N_\alpha{}^\beta e^{\overleftarrow{H}} \overleftarrow{D}_\beta e^{-\overleftarrow{H}} ) + h.\,c. \qquad (7.10.40)$$

can be expressed as

$$\phi \bar{\phi} G' = G_0 - \frac{1}{3} ([\bar{D}_{\dot{\alpha}}, D_\alpha] H^{\alpha\dot{\alpha}}) G_0 - \frac{1}{2} H^{\alpha\dot{\alpha}} [\bar{D}_{\dot{\alpha}}, D_\alpha] G_0$$

$$+ [(D_\alpha H^{\alpha\dot{\alpha}}) \bar{D}_{\dot{\alpha}} - (\bar{D}_{\dot{\alpha}} H^{\alpha\dot{\alpha}}) D_\alpha] G_0 \ , \qquad (7.10.41)$$

where $G_0 = \frac{1}{2} D_\alpha \eta^{0\alpha} + h.\,c.$.

From the action

$$S = - \frac{1}{2} \int d^4x \, d^4\theta \ E^{-1} e^{-\frac{1}{2}H} G'^2 \qquad (7.10.42)$$

we obtain

$$J_{\alpha\dot{\alpha}} = - \frac{1}{12} [\bar{D}_{\dot{\alpha}}, D_\alpha] G_0{}^2 + \frac{1}{2} G_0 [\bar{D}_{\dot{\alpha}}, D_\alpha] G_0$$

$$= - \frac{1}{3} (\bar{D}_{\dot{\alpha}} G_0)(D_\alpha G_0) + \frac{1}{3} G_0 [\bar{D}_{\dot{\alpha}} , D_\alpha] G_0 \ , \qquad (7.10.43a)$$

$$J = \frac{1}{6} \bar{D}^2 G_0{}^2 \ . \qquad (7.10.43b)$$

We note that the substitution $G_0 \to \eta + \bar{\eta}$ (cf. sec 4.4.c.2) gives the supercurrent $J_{\alpha\dot{\alpha}}$ for the nonconformal scalar multiplet with Lagrangian $\frac{1}{2}(\eta + \bar{\eta})^2$. This Lagrangian for the scalar multiplet, identical in flat space to the usual one, gives disimprovement terms to $J_{\alpha\dot{\alpha}}$ and $J$ because the extra terms $\frac{1}{2}(\eta^2 + \bar{\eta}^2)$ lead to nonminimal couplings $\int d^4x\, d^2\theta\, \phi^3 R\, \frac{1}{2}\eta^2 + h.\,c.$. On the other hand, the improved tensor multiplet (4.4.46) with action $-\int d^4x\, d^4\theta\, G\, ln\, G$ does have $J = 0$.

From the gauge fixing term

$$S_{GF} = -\frac{1}{2}\int d^4x\, d^4\theta\, E^{-1}(\frac{1}{2}\nabla_\alpha \eta^\alpha - h.\,c.)^2 \tag{7.10.44}$$

we obtain additional contributions. The combined current from (7.10.42,44) can be written

$$J_{\alpha\dot{\alpha}} = \frac{1}{6}(D^2\eta_\alpha) i\partial_{\beta\dot{\alpha}}\eta^\beta - \frac{1}{6}(D_\gamma\eta^\gamma) i\partial_{\beta\dot{\alpha}} D_{(\alpha}\eta^{\beta)}$$

$$-\frac{1}{2}\eta_\alpha i\partial_{\beta\dot{\alpha}} D^2\eta^\beta + \frac{1}{2}\eta_\alpha \Box \bar{\eta}_{\dot{\alpha}} + h.\,c. \quad , \tag{7.10.45a}$$

$$J = \frac{1}{12}\bar{D}^2[(D_\alpha\eta^\alpha)^2 + h.\,c.] \tag{7.10.45b}$$

As mentioned earlier, the field redefinitions we have performed change the form of the supercurrents, but only by adding terms proportional to the field equations. Such terms have no physical consequences.

The supercurrent for the supergravity multiplet itself can be obtained from the background-quantum splitting of sec. 7.2, by functional differentiation with respect to the background field. We will not give it here.

### d. Superconformal anomalies

The discussion of sec. 7.10.b can be taken over directly to the $N = 1$ supersymmetric case. We consider quantum corrections to the supercurrent $J_{\alpha\dot{\alpha}}$, and in particular to its supertrace $J$. For classically conformally invariant systems the supertrace can be obtained from the one-loop counterterm, and we will generally refer to this contribution as the *superanomaly.* If the classical theory is not conformally invariant the supertrace, computed from the renormalized effective action, does not equal the superanomaly. We discuss in this section the minimal $n = -\frac{1}{3}$ theory with a chiral compensator.

We define the renormalized currents

$$J_{\alpha\dot{\alpha}} = \frac{\Delta\Gamma_R}{\Delta H^{\alpha\dot{\alpha}}} \tag{7.10.46}$$

$$J = \frac{\delta\Gamma_R}{\delta\phi^3} \quad . \tag{7.10.47}$$

(In the version of the theory with variant multiplet compensators we have $\frac{\Delta\Gamma_R}{\Delta V} = J + \bar{J}$ or $\frac{\Delta\Gamma_R}{\Delta\psi^\alpha} = \nabla_\alpha J$.) We will assume for the time being that the minimal theory has no local supersymmetry anomalies. Invariance of the effective action under local supersymmetry transformations gives then

$$\bar{\nabla}^{\dot{\alpha}} J_{\alpha\dot{\alpha}} = \nabla_\alpha J \quad . \tag{7.10.48}$$

The supertrace is zero only if $\Gamma_R$ is independent of $\phi$. (The $\Delta$ operation is defined in (5.5.44).)

The superanomaly is given by

$$J^{an} = \frac{\delta S_\infty}{\delta\phi^3} \tag{7.10.49}$$

$J = J^{an}$ only if the classical theory is superconformal.

The relevant one-loop expressions corresponding to (7.10.9,10,17) are

$$\Gamma_U \sim \frac{1}{\mathrm{D}-4} \int d^{\mathrm{D}}x d^2\theta\, \phi^{\frac{2(\mathrm{D}-1)}{\mathrm{D}-2}} W_{\alpha\beta\gamma} \left(\frac{\Box_+}{\mu^2}\right)^{\frac{1}{2}\mathrm{D}-2} W^{\alpha\beta\gamma} + h.\,c. \ , \tag{7.10.50a}$$

$$S_\infty \sim -\frac{2}{\mathrm{D}-4} \int d^{\mathrm{D}}x d^2\theta\, \phi^{\frac{2(\mathrm{D}-1)}{\mathrm{D}-2}} W^2 + h.\,c. = -\frac{2}{\mathrm{D}-4} \int d^{\mathrm{D}}x\ g^{\frac{1}{2}}(w^2 + \bar{w}^2) \ , \tag{7.10.50b}$$

$$\chi_{\mathrm{D}} = \frac{1}{(4\pi)^{\frac{1}{2}\mathrm{D}}} \left[\frac{1}{2} \int d^{\mathrm{D}}x d^2\theta\, \phi^{\frac{2(\mathrm{D}-1)}{\mathrm{D}-2}} W^2 + h.\,c. + \int d^{\mathrm{D}}x d^4\theta\ E^{-1}(G^2 + 2\bar{R}R)\right] \ , \tag{7.10.51}$$

where $W^2 = \frac{1}{2} W_{\alpha\beta\gamma} W^{\alpha\beta\gamma}$, $\Box$ is a supercovariantized d'Alembertian, and $\chi_{\mathrm{D}}$ is the supersymmetric form of the Euler number of (7.10.7). The expression corresponding to (7.10.16) is

$$\Gamma_U = k_1 \frac{1}{\epsilon} \chi_D + \frac{1}{(4\pi)^2} \int d^4x\, d^4\theta\, E^{-1} [(k_2 - k_3) \frac{1}{2} \bar{R} (\frac{1}{\epsilon} - ln \frac{\Box_+}{\mu^2}) R$$

$$- k_3 \frac{1}{2} G^{\alpha\dot{\alpha}} (\frac{1}{\epsilon} - ln \frac{\Box}{\mu^2}) G_{\alpha\dot{\alpha}}]$$

$$- k_4 \frac{1}{(4\pi)^2} \{ \int d^4x\, d^2\theta\, \phi^3 W^2\, ln[1 - (\bar{\nabla}^2 + R) \frac{1}{\Box_-} \bar{R}] + h.c. \} \ , \tag{7.10.52}$$

and represents an unambiguous way of organizing the off-shell covariantized contributions from supergraphs with two or three external lines. Other terms, with more factors of $W_{\alpha\beta\gamma}$, do not contribute to on-shell supertraces. The d'Alembertian $\Box_-$ was defined in (7.4.4).

If the classical theory is superconformal $k_1 = k_4$. Otherwise, they have to be computed separately, e.g., from a self-energy and from a triangle supergraph, respectively. For example, the last term in (7.10.52) can be expanded as

$$k_4 \frac{1}{(4\pi)^2} \int d^4x\, d^4\theta\, E^{-1} W^2 \frac{1}{\Box_-} \bar{R} + h.c. + \cdots \ , \tag{7.10.53}$$

and gives, at the linearized level,

$$J_{\alpha\dot{\alpha}} = \frac{1}{3} k_4 \frac{1}{(4\pi)^2} i\partial_{\alpha\dot{\alpha}} \frac{1}{\Box} (\bar{D}^2 \bar{W}^2 - D^2 W^2) \ . \tag{7.10.54}$$

This corresponds to the contribution from a triangle graph with two legs on shell. Its form is uniquely determined by covariance and power counting, and the actual value of $k_4$ can be determined, for example, by the Adler-Rosenberg method.

The supertrace and superanomaly are given by

$$J = \frac{1}{3} k_4 \frac{1}{(4\pi)^2} W^2 \ ,$$

$$J^{an} = \frac{1}{3} k_1 \frac{1}{(4\pi)^2} W^2 \ . \tag{7.10.55}$$

The superanomaly can be read from the results contained in (7.8.5) which give the on-shell value of the first term in (7.10.52). For a scalar multiplet $k_1 = \frac{1}{24}$, while for a tensor multiplet, including ghosts, it is $k_1 = \frac{25}{24}$. In a background covariant gauge for the vector multiplet the contribution to $S_\infty$ comes entirely from the three chiral ghosts since, as discussed in sec. 7.8, general superfields do not contribute to the two-point function. Thus, for the Yang-Mills multiplet we have $k_1 = -\frac{3}{24}$. For the gravitino matter multiplet, with the effective Lagrangian of (7.3.6) or (7.3.7) we find, by adding contributions from the chiral ghosts, $k_1 = -\frac{19}{24}$ or $\frac{5}{24}$, respectively, for the two different sets of compensators. Finally, for the supergravity multiplet we obtain the values $k_1 = \frac{41}{24}\,, -\frac{7}{24}\,, -\frac{55}{24}$ depending on whether we use a $\phi$, $V$, or $\psi_\alpha$ compensator. These numbers can also be obtained from a component analysis of the theories, using the values $k_1$ of (7.10.18) for the component trace anomaly. (Changing from one compensator to another corresponds to replacing some of the spin zero auxiliary fields with divergences of vector auxiliary fields.)

For the scalar multiplet, which is classically superconformally invariant, $k_4 = k_1 = \frac{1}{24}$, and for the same reason, for the vector multiplet $k_4 = -\frac{3}{24}$. Since the tensor multiplet is physically equivalent to the scalar multiplet, it has the same value $k_4 = \frac{1}{24}$ ($\neq k_1$ since the classical theory is not superconformal). This result has been checked by an explicit calculation.

For the supergravity multiplet the explicit calculations have not been completely carried out. If we conjecture that the contributions to the supertrace again come completely from the chiral fields in the quantum action, we can determine the coefficients $k_4$. Since we are discussing the supertrace, chiral spinors are equivalent to chiral scalars or, what amounts to the same thing, the result is independent of the type of compensator we use. This gives the value $k_4 = -\frac{7}{24}$ for the supergravity multiplet and, by a similar reasoning, $k_4 = \frac{5}{24}$ for the $(\frac{3}{2}\,,1)$ gravitino matter multiplet. (For example, in the $(2,\frac{3}{2})$ multiplet, replacing the chiral compensator with a $V$ compensator replaces two $\phi_\alpha$'s and two $\bar{\phi}_{\dot\alpha}$'s with eight $\chi$'s with opposite statistics. For the $(\frac{3}{2}\,,1)$ multiplet the equivalent of one $\phi_\alpha$ and one $\bar{\phi}_{\dot\alpha}$ in (7.3.6) is four more $\chi$'s, as in (7.3.7).)

For the scalar and vector multiplets, the supertrace results are also consistent with the calculated values of the component axial current anomalies (provided we assign the correct R-weights $\frac{1}{3}$, $-1$ for the fermions of the scalar and vector multiplet, respectively). However, the conventionally quoted value for the gravitino axial anomaly $(\partial_{\underline{m}} j^{5\underline{m}} = \frac{21}{24}(4\pi)^{-2} r * r)$ does not match the energy-momentum trace for either the $(2, \frac{3}{2})$ or $(\frac{3}{2}, 1)$ multiplet. This is a consequence of the fact that the component anomaly was calculated for a classically *conserved* gravitino axial current, whereas the component current contained in $J_{\underline{a}}$ is not classically conserved: It contains additional terms which give nonvanishing contributions to $\partial_{\underline{m}} j^{5\underline{m}}$. (Its energy-momentum partner is not traceless: e.g., the trace of the *quadratic* part of the Einstein tensor, representing the energy-momentum tensor of the graviton field, is classically nonvanishing even on shell. This is due to the conformal noninvariance of Einstein gravity.)

| $N$ | $s_{\max}$ | total trace ($k_4$) | |
|---|---|---|---|
| 0 | 0 | 8/360* | |
| | 1/2 | 7/360 | |
| | 1 | -52/360 | $= (-1)^{2s_{\max}+1}(15{s_{\max}}^2 - 2)/90$ |
| | 3/2 | 127/360 | |
| | 2 | -232/360 | |
| 1 | 1/2 | 1/24 | |
| | 1 | -3/24 | $= (-1)^{2s_{\max}+1}(4s_{\max} + 1)/24$ |
| | 3/2 | 5/24 | |
| | 2 | -7/24 | |
| 2 | 1/2 | 1/12 | |
| | 1 | -1/12 | $= (-1)^{2s_{\max}+1}/12$ |
| | 3/2 | 1/12 | |
| | 2 | -1/12 | |
| ≥3 | all | 0 | |

*Table 7.10.1. Values of the total trace coefficients (*Complex conformal scalar)*

The values of the $k_4$ coefficients calculated on the basis of our conjecture are presented in Table 7.10.1, which gives the supertrace in $N = 0$, $N = 1$, and extended supersymmetry. That $k_4 = 0$ for $N \geq 3$ reflects again the absence of a net number of chiral superfields.

The verification of our statements awaits an explicit calculation of the relevant triangle supergravity supergraph, and a better understanding of some of the component calculations. If our conjecture is correct, it is rather curious, and not understood, that the supergravity theory with the $V$ compensator behaves as if it were superconformal $(k_1 = k_4)$ or, equivalently, that the superanomaly and supertrace differ only if chiral spinors are present.

## e. Local supersymmetry anomalies

We can use the existence of superconformal anomalies to infer the existence of anomalies in the Ward identities of local supersymmetry for $n \neq -\frac{1}{3}$. We demonstrate this explicitly for the case of quantum matter multiplets coupled to background supergravity, but expect similar results when supergravity itself is quantized. We first consider $N = 1$ supergravity.

At the linearized level, the Ward identities reflect the invariance of the effective action under the (linearized) local supersymmetry transformations $(L^{\alpha\beta} = L^{\beta\alpha})$

$$\delta H_{\alpha\dot{\alpha}} = D_\alpha \bar{L}_{\dot{\alpha}} - \bar{D}_{\dot{\alpha}} L_\alpha \ , \tag{7.10.56}$$

$$n = -\frac{1}{3}: \quad \delta\phi^3 = \bar{D}^2 D_\alpha L^\alpha \ ,$$

$$n = 0: \quad \delta\phi^\alpha = -2\bar{D}^2 L^\alpha + i\bar{D}^2 D^\alpha K \ ,$$

$$n \neq -\frac{1}{3}, 0: \quad \delta H^\alpha = i(-\frac{1}{3}\bar{D}^2 L^\alpha + \frac{1}{3}\bar{D}_{\dot{\alpha}} D^\alpha \bar{L}^{\dot{\alpha}} + \frac{1}{2}\frac{n+1}{3n+1} D^\alpha \bar{D}_{\dot{\alpha}} \bar{L}^{\dot{\alpha}} + D_\beta L^{\alpha\beta}) \ . \tag{7.10.57}$$

We have used the gauge $H^\alpha = 0$ for $n = -\frac{1}{3}$; the gauge $H^\alpha = -i\bar{D}_{\dot{\alpha}} H^{\alpha\dot{\alpha}}$ $(\rightarrow 1 \cdot \overleftarrow{H} = 0,$ $1 \cdot e^{-\overleftarrow{H}} = 1$; see sec. 5.2.b) for $n = 0$, so that $E^{-1}$ can be linearized as $1 + \frac{1}{2}(D_\alpha \phi^\alpha + \bar{D}_{\dot{\alpha}} \bar{\phi}^{\dot{\alpha}})$; and the gauge $\Upsilon = 1$ for other $n$. For $n \neq -\frac{1}{3}, 0$ we have made the shift $H^\alpha \rightarrow H^\alpha - \frac{1}{3} i\bar{D}_{\dot{\alpha}} H^{\alpha\dot{\alpha}}$ so that $J_{\alpha\dot{\alpha}}$ is the superconformal current (coupling to conformal supergravity's axial vector, and not the other auxiliary axial vector). We have

$$0 = \delta\Gamma_R = \int d^4x d^4\theta\, (\delta H^{\alpha\dot{\alpha}}) J_{\alpha\dot{\alpha}} + \begin{cases} \int d^4x d^2\theta\, (\delta\phi^3) J + h.\,c. \\ \int d^4x d^2\theta\, (\delta\phi^\alpha)\chi_\alpha + h.\,c. \\ \int d^4x d^4\theta\, (\delta i H^\alpha)\lambda_\alpha + h.\,c. \ , \end{cases} \tag{7.10.58}$$

where

$$J_{\alpha\dot{\alpha}} \equiv \frac{\delta\Gamma_R}{\delta H^{\alpha\dot{\alpha}}} \quad ,$$

$$J \equiv \frac{\delta\Gamma_R}{\delta\phi^3} \quad , \quad \chi_\alpha \equiv \frac{\delta\Gamma_R}{\delta\phi^\alpha} \quad , \quad \lambda_\alpha \equiv \frac{\delta\Gamma_R}{\delta(iH^\alpha)} \quad . \tag{7.10.59}$$

If we require that (7.10.58) be satisfied, we obtain the (linearized) conservation laws

$$n = -\frac{1}{3}: \quad \bar{D}^{\dot{\alpha}} J_{\alpha\dot{\alpha}} = D_\alpha J \quad , \qquad \bar{D}_{\dot{\alpha}} J = 0 \quad ;$$

$$n = 0: \quad \bar{D}^{\dot{\alpha}} J_{\alpha\dot{\alpha}} = -2\chi_\alpha \quad , \qquad \bar{D}_{\dot{\alpha}}\chi_\alpha = D_\alpha\chi^\alpha - \bar{D}_{\dot{\alpha}}\bar{\chi}^{\dot{\alpha}} = 0 \quad ;$$

$$other\ n: \quad \bar{D}^{\dot{\alpha}} J_{\alpha\dot{\alpha}} = \frac{1}{3}\bar{D}^2\lambda_\alpha + \frac{1}{3}\bar{D}^{\dot{\alpha}} D_\alpha \bar{\lambda}_{\dot{\alpha}} + \frac{1}{2}\frac{n+1}{3n+1} D_\alpha \bar{D}^{\dot{\alpha}}\bar{\lambda}_{\dot{\alpha}} \quad ,$$

$$D_{(\alpha}\lambda_{\beta)} = 0 \quad . \tag{7.10.60}$$

The invariances used to derive these conservation laws are those of Poincaré supergravity, and their violation would imply that the multiplet contributing to $\Gamma_R$ cannot be coupled consistently to the corresponding form of supergravity. On the other hand, the violation of the superconformal conservation law $\bar{D}^{\dot{\alpha}} J_{\alpha\dot{\alpha}} = 0$ implies only that the multiplet cannot be coupled consistently to conformal supergravity.

We evaluate matrix elements of the conservation equations (7.10.60) between the vacuum and an on-shell supergravity state. In particular, if we consider one-loop "triangle" graphs we know the precise form of the left-hand side. As discussed in the previous subsection, power counting and covariance determines uniquely the matrix element of the supercurrent:

$$< \Psi(\mathbf{H})|J_{\alpha\dot{\alpha}}|0 > \sim i\partial_{\alpha\dot{\alpha}}\frac{1}{\Box}[D^2(W_{\alpha\beta\gamma})^2 - \bar{D}^2(\bar{W}_{\dot{\alpha}\dot{\beta}\dot{\gamma}})^2] \quad . \tag{7.10.61}$$

Then we have for the matrix element of $\bar{D}^{\dot{\alpha}}J_{\alpha\dot{\alpha}}$

$$< \bar{D}^{\dot{\alpha}}J_{\alpha\dot{\alpha}} > \sim D_\alpha W^2 \quad . \tag{7.10.62}$$

It is not zero (except when the supertrace vanishes), and is independent of the form of the compensator.

We now examine the matrix element of the right hand side of (7.10.60). We begin by considering contributions to the one-loop effective action from a classically superconformal multiplet. Its (locally supersymmetric, covariantized) action is independent of the compensator. However, as discussed in the previous subsection, the compensator enters the (one-loop) renormalized effective action after the divergences have been subtracted out. We can ask now if the form (7.10.62) is compatible with the right hand side of (7.10.60). Since the compensator enters $\Gamma_R$ only because we have subtracted out the covariant, *local,* counterterm $S_\infty(H, \mathit{compensator})$, the corresponding current must also be local. For $n = -\frac{1}{3}$ a solution of (7.10.60) is $J \sim W^2$, but *for* $n \neq -\frac{1}{3}$ *there exists no local* $\chi_\alpha$ *or* $\lambda_\alpha$ *that satisfies the conservation equation.* We conclude that any superconformal $N = 1$ multiplet that has a nonzero one-loop supertrace gives a contribution to $\Gamma_R$ that violates the Poincaré supergravity conservation laws for $n \neq -\frac{1}{3}$, i.e. has a *local supersymmetry anomaly.* Therefore, in general, superconformal multiplets can be coupled consistently only to $n = -\frac{1}{3}$ supergravity. (The analysis above is inconclusive, however, if the supertrace vanishes, e.g. for a system of one vector and three scalar multiplets, which has no one-loop divergence or supertrace.)

In the case where the classical theory is nonsuperconformal, the compensators may couple to nonlocal terms in the effective action. Thus, for $n \neq -\frac{1}{3}, 0$,

$$< \lambda_\alpha > \sim D_\alpha \frac{1}{\Box}\bar{D}^2\bar{W}^2 \tag{7.10.63}$$

can satisfy (7.10.60) and we cannot conclude, without further analysis, that Poincaré supergravity anomalies are present. However, we can still conclude that an anomaly is present for $n = 0$ since (7.10.60) implies $\bar{D}^2 J_{\alpha\dot{\alpha}} = \partial^{\alpha\dot{\alpha}}J_{\alpha\dot{\alpha}} = 0$, whereas $\bar{D}^{\dot{\alpha}}J_{\alpha\dot{\alpha}} \sim D_\alpha W^2$ implies $\bar{D}^2 J_{\alpha\dot{\alpha}} \sim i\partial_{\alpha\dot{\alpha}}W^2$ and $\partial^{\alpha\dot{\alpha}}J_{\alpha\dot{\alpha}} \sim i(D^2W^2 - \bar{D}^2\bar{W}^2)$, neither of which vanish even on shell. This occurs because the supertrace is an irreducible multiplet of superspin 0 ($W^2$ is a chiral scalar), whereas the compensator multiplet for $n = 0$ has superspin $\frac{1}{2}$.

For nonminimal supergravity $(n \neq -\frac{1}{3}, 0)$ we will see below that anomalies are absent only under very special circumstances. In general, their presence is related to the nonexistence of a chiral measure. An interesting way to understand the origin of the anomaly is to use the fact that (in appropriate supersymmetric gauges) only (physical or ghost) chiral superfields contribute to the divergences and require regularization. In particular, we can ask if Pauli-Villars regularization is possible for chiral scalar superfields with the various nonconformal couplings of sec. 5.5. Since only $n = -\frac{1}{3}$ has a chiral measure that allows mass terms for chiral superfields with *conformal kinetic terms,* it is the only $n$ that allows Pauli-Villars regularization for those superfields. (In other regularization schemes, the same difficulty with chiral measures shows up in other ways: e.g., in dimensional regularization, finding analogs to the chiral integrands in the last term of (7.10.52).) In fact, we will show below that the only quantum-consistent couplings to supergravity are those which: (1) allow Pauli-Villars regularization, (2) have vanishing supertrace, or (3) have couplings that correspond to extended supersymmetry. For $n = -\frac{1}{3}$ all chiral superfields can have mass terms, so all couplings are possible. For other $n$ coupling to the vector multiplet alone is impossible (it is classically superconformal and has classically superconformal chiral ghosts), coupling to a scalar multiplet alone is possible only for the nonconformal coupling that allows mass (but not self-interaction) terms, and coupling to the combination of the two requires a cancellation that occurs in extended multiplets (and probably nowhere else, if the cancellation is to be exactly maintained at higher loops).

To discuss the situation quantitatively, we perform an explicit verification of the conservation law (7.10.60) for contributions from chiral scalars with nonsuperconformal couplings. From the actions of sec. 5.5 (with the definitions in (7.10.59)) we find the classical currents

$$J_{\alpha\dot{\alpha}} = \begin{cases} -\frac{1}{6}[\bar{D}_{\dot{\alpha}}, D_\alpha]\bar{\eta}\eta + \frac{1}{2}\bar{\eta}i\overleftrightarrow{\partial}_{\alpha\dot{\alpha}}\eta & \text{for } n \neq 0 \quad , \\ -\frac{2\tilde{n}+1}{2}[\bar{D}_{\dot{\alpha}}, D_\alpha]\bar{\eta}\eta + \frac{1}{2}\bar{\eta}i\overleftrightarrow{\partial}_{\alpha\dot{\alpha}}\eta & \text{for } n = 0 \quad , \end{cases}$$

$$\begin{cases} J = \frac{1}{3}\bar{D}^2\bar{\eta}\eta \quad , \quad n = -\frac{1}{3} \\ \chi_\alpha = \frac{3\tilde{n}+1}{2}\bar{D}^2 D_\alpha\bar{\eta}\eta \quad , \quad n = 0 \\ \lambda_\alpha = \frac{3\tilde{n}+1}{2} D_\alpha\bar{\eta}\eta \quad , \quad other\ n \quad . \end{cases} \tag{7.10.64}$$

We now imagine computing *renormalized* one-loop matrix elements of these currents between the vacuum and an on-shell background supergravity state. The matrix element of $J_{\alpha\dot{\alpha}}$ must have the form (7.10.61) and, in particular, its $\theta = 0$ component has the form $i\partial_{\alpha\dot{\alpha}}\Box^{-1}(w^2 - \bar{w}^2)$. We observe that $\bar{\eta} i\overleftrightarrow{\partial}_{\alpha\dot{\alpha}}\eta$ gives no contribution to this component and therefore no contribution at all, since any covariant superfield that vanishes at $\theta = 0$ vanishes identically. (The "top" vertex of the graph contains only crossterms $\mathrm{A}\overleftrightarrow{\partial}\mathrm{B}$ of $\eta| = \mathrm{A} + i\mathrm{B}$, whereas the gravitational couplings are proportional to AA and BB.) Therefore, to compute matrix elements of *any* of the currents in (7.10.63) it is sufficient to compute matrix elements $< \Psi(\mathbf{H})|\bar{\eta}\eta|0 >$ for two-particle on-shell graviton states $\Psi(\mathbf{H})$, and then apply appropriate operators (e.g., $< J_{\alpha\dot{\alpha}} > \sim < [\bar{D}_{\dot{\alpha}}, D_\alpha]\bar{\eta}\eta > = [\bar{D}_{\dot{\alpha}}, D_\alpha] < \bar{\eta}\eta >$, etc.).

By power counting and covariance arguments, the renormalized matrix element has the unique form

$$< \bar{\eta}\eta > = c\,\frac{1}{\Box}(D^2W^2 + \bar{D}^2\bar{W}^2) \tag{7.10.65}$$

where c is a numerical factor. We now substitute the corresponding expressions of (7.10.64) into the conservation laws (7.3.59). Since

$$< \chi_\alpha > = \frac{1}{2}(3\tilde{n} + 1)\bar{D}^2 D_\alpha < \bar{\eta}\eta > = 0 \tag{7.10.66}$$

always, we find that the conservation laws are *never* satisfied for $n = 0$ (unless $c = 0$). For $n \neq 0$ substituting (7.10.64) into (7.10.60) gives

$$-\frac{1}{2} D_\alpha\bar{D}^2 < \bar{\eta}\eta > = \frac{3\tilde{n} + 1}{3n + 1} D_\alpha\bar{D}^2 < \bar{\eta}\eta > \quad , \tag{7.10.67a}$$

which is satisfied only for

$$\tilde{n} = -\frac{1}{2}(n + 1) \quad . \tag{7.10.67b}$$

For $n = -\frac{1}{3}$, this is the only value of $\tilde{n}$ defined, even classically (see sec. 5.5.f.2). For other $n$, this value is exactly the one that allows a mass term.

To investigate anomaly canceling mechanisms, we consider contributions from one vector multiplet and $l$ identical scalar multiplets with arbitrary weight $\tilde{n}$. The contribution of the vector multiplet to the nonlocal part of $<\chi_\alpha>$ and $<\lambda_\alpha>$ must vanish because of classical superconformal invariance (furthermore, the ghosts must have $\tilde{n}_{ghost} = -\frac{1}{3}$). The contribution to $<\bar{D}^{\dot{\alpha}} J_{\alpha\dot{\alpha}}>$ is $-3$ times that of a physical scalar multiplet. The $l$ scalar multiplets contribute to both the left and right hand sides of (7.10.60). The conservation law now becomes

$$-\frac{1}{2} D_\alpha \bar{D}^2 <\bar{\eta}\eta> (l-3) = \frac{3\tilde{n}+1}{3n+1} D_\alpha \bar{D}^2 <\bar{\eta}\eta> l \ , \qquad (7.10.68a)$$

which gives the condition

$$\tilde{n}_l = \frac{1}{2}(\frac{3}{l} - 1)n + \frac{1}{2}(\frac{1}{l} - 1) \ . \qquad (7.10.68b)$$

In particular, for $l = 3$, which corresponds to the $N = 4$ vector multiplet, where divergences cancel, the superconformal coupling $\tilde{n} = -\frac{1}{3}$ is required. For $l = 1$, the $N = 2$ vector multiplet, we find $\tilde{n}_l = n$. Recall that for $n = 0$ the conservation laws require the supertrace $\bar{D}^{\dot{\alpha}} J_{\alpha\dot{\alpha}}$ to vanish *identically* even though the theory may still have divergences (i.e., the superanomaly may be nonzero). We thus have $-\frac{2\tilde{n}+1}{2}[\bar{D}_{\dot{\alpha}}, D_\alpha] <\bar{\eta}\eta> l - \frac{1}{6}[\bar{D}_{\dot{\alpha}}, D_\alpha] <\bar{\eta}\eta> (-3) = 0$, agreeing with (7.10.68) for $n = 0$.

We have thus found that for $N = 1$ only $n = -\frac{1}{3}$ is generally quantum consistent, while for other $n$ only very special nonsuperconformal couplings are allowed. These arguments can be applied to extended supergravity. In particular, the standard $N = 2$ theory, which (in terms of $N = 2$ superfields) has an isovector compensator ${V_a}^b$, is quantum consistent, basically because it has chiral measure. Thus an $N = 2$ vector multiplet will give contributions to the effective action which are anomaly-free. When analyzed in terms of $N = 1$ superfields, $N = 2$ supergravity decomposes into a $(\frac{3}{2}, 1)$ multiplet coupled to $N = 1$, $n = -1$ supergravity. The $N = 2$ vector multiplet decomposes into a $N = 1$ vector multiplet and a nonconformal scalar multiplet, but with $\tilde{n} = n = -1$ which is consistent with the no anomaly condition we derived above. It is likely that this extended supersymmetry is necessary for $n \neq -\frac{1}{3}$ for this cancellation of the anomalies in the supergravitational conservation law to occur at higher loops.

## f. Not the Adler-Bardeen theorem

In sec. 6.7 we considered the anomaly in the (axial) Yang-Mills current and, on the basis of the covariant rules, concluded that it (and its component axial current) satisfies the Adler-Bardeen theorem. On the other hand, the supertrace (anomaly) in general receives higher-order corrections (the $\beta$-function is not zero), and therefore the component R-current does not satisfy the Adler-Bardeen theorem. (We are considering here matrix elements of the current between the vacuum and on-shell Yang-Mills states, rather than supergravity states.) Although the currents look the same classically (for a scalar multiplet in an external vector multiplet or supergravity background the $A\overleftrightarrow{\partial}_{\underline{a}}\bar{A}$ term does not contribute), the difference arises because of different renormalization prescriptions.

In the first case, when the axial-vector gauge superfield is external (otherwise, in the presence of one-loop anomalies the quantum theory makes no sense), it is possible to renormalize the higher-loop effective action $\Gamma(V_+, V_-(ext))$ and define $J^{renorm}$ so that it is not anomalous. On the other hand, if $V_-(ext)$ is replaced with $H_{\alpha\dot{\alpha}}(ext)$, the higher-loop effective action $\Gamma(V_+, H_{\alpha\dot{\alpha}}(ext))$ is usually renormalized in a manner which is consistent with Poincaré supergravity gauge invariance. In that case, we do not have the freedom to redefine $J_{\alpha\dot{\alpha}}{}^{renorm}$ so as to remove its higher-loop supertrace (anomaly). If we give up super-Poincaré invariance, we can renormalize so that $\partial^{\alpha\dot{\alpha}} J_{\alpha\dot{\alpha}}{}^{renorm} = 0$ at higher loops. However, *this* $J_{\alpha\dot{\alpha}}{}^{renorm}$ will not contain a conserved (symmetric) energy-momentum tensor at the $\theta\bar{\theta}$ level. Therefore, the renormalized chiral R-current which is in the same multiplet with the renormalized conserved energy-momentum tensor does not satisfy the Adler-Bardeen theorem.

**Contents of 8. BREAKDOWN**

# 8. BREAKDOWN

## 8.1. Introduction

The most striking feature of the relation between supersymmetry and the observed world is the absence of any experimental evidence for the former in the latter. The particles we see do not fall into supersymmetric multiplets, nor do they show even an approximate mass equality that would indicate they were in multiplets before symmetry breaking. Thus if supersymmetry is an underlying symmetry of the physical world, it must be badly broken, or otherwise hidden from direct experimental verification.

At a fundamental level, it is difficult to accept the idea of a global supersymmetry without believing that there exists an underlying local supersymmetry: Since we believe that gravity must be quantized, and since even global supersymmetry implies that the graviton requires a spin $\frac{3}{2}$ gravitino partner, then the gravitino must be the gauge particle of local supersymmetry, however badly broken global supersymmetry may be. Then, as in any gauge theory, the supersymmetry breaking must be spontaneous (i.e., by the vacuum) and not explicit (i.e., in the action itself). If we believe in local supersymmetry with symmetry breaking, we must understand mechanisms for this breaking. It can be through the Higgs mechanism, or due to cosmological factors such as boundary conditions or high temperature effects in the early universe, or nonperturbative dynamical effects, or via dimensional compactification. It is also reasonable to believe that the breaking happens at a large energy scale. If this is so, we may hope that the dynamical effects of the supergravity fields can be ignored at a lower energy scale, and that the effective low energy theory is a broken globally supersymmetric theory. We can start with an exact globally supersymmetric theory, at some scale where supergravity fields have decoupled, and investigate its spontaneous breaking ab initio, or we can put the breaking in by hand, as an explicit manifestation of the original local breaking. (In general, if we start with a locally supersymmetric theory that exhibits symmetry breaking and set gravitational fields and couplings to zero, soft breaking terms are induced).

Unlike other symmetries, there are some interesting and unexpected restrictions on the possible breaking of global supersymmetry. Some of these have their origin in the supersymmetry algebra itself, while others are most easily obtainable in the context of superfield perturbation theory. The first restriction, which follows from the algebra, is

the following theorem:

> If supersymmetry is not spontaneously broken, i.e., if the vacuum is invariant under supersymmetry transformations, then its energy is zero; conversely, if there exists a state for which the expectation value of the Hamiltonian is zero, supersymmetry is not spontaneously broken. Furthermore, if supersymmetry is spontaneously broken without an attendant modification of the supersymmetry algebra, then the vacuum energy is positive.

As discussed in sec. 3.2, this result follows directly from the commutation relations $\{Q, \bar{Q}\} = P$, which give in particular

$$E_{vac} = -\frac{1}{2N}\,\delta^{\alpha\dot{\beta}} < 0|\{Q_{a\alpha}, \bar{Q}^{a}{}_{\dot{\beta}}\}|0> = \frac{1}{2N}\sum_{a\alpha} < 0||Q_{a\alpha}|^2|0> \tag{8.1.1}$$

Thus:

> If all the components of the supersymmetry charge (generators) annihilate the vacuum, its energy is zero. If any one of them does not annihilate the vacuum, then its energy is positive.

We emphasize that this theorem assumes that the supersymmetry algebra is not changed. With an appropriate interpretation of the total energy and charge, the theorem also holds in supergravity. On the other hand, explicit breaking does change the algebra and then negative or zero energy is possible.

The second important result, proved for a fairly large class of renormalizable models, is that in spontaneously broken global theories, there are mass sum rules relating fermion and boson masses, which take the form

$$\sum_{states} {m_B}^2 - \sum_{states} {m_F}^2 = 0 \ . \tag{8.1.2}$$

These sum rules are extremely restrictive, and make the construction of realistic models difficult; however, for locally supersymmetric and explicitly (softly) broken globally supersymmetric theories, the generalizations of this formula are phenomenologically acceptable.

A third result is the following theorem, which can be proven in perturbation theory *for four-dimensional theories:*

> If supersymmetry is not spontaneously broken at the tree level, then it is not broken by radiative corrections. A Coleman-Weinberg mechanism is not possible.

This theorem is not valid in two-dimensional supersymmetry.

If supersymmetry is spontaneously broken, a massless Goldstone fermion must be present. Therefore, if one can prove that no massless fermion states can exist, supersymmetry cannot be broken spontaneously. Using this fact, Witten has given certain criteria (index theorems) that allow one to rule out in a simple manner, in certain cases, the possibility of spontaneous supersymmetry breaking. In a locally supersymmetric theory, the Goldstone fermion is absorbed by the spin $\frac{3}{2}$ gravitino via a conventional Higgs mechanism. Index theorems have not been investigated in supergravity.

Global supersymmetry breaking is most easily discussed in superfield language as a breaking of $Q$-translational invariance in superspace. This can happen either because the vacuum is not $Q$-translationally invariant (spontaneous breaking), or because one has explicit $\theta$-dependence in the effective action (either at the tree level or nonperturbatively, for example via instanton effects, which could introduce such explicit dependence). If the breaking is spontaneous, it means in general that some superfield has a nonzero vacuum expectation value (if Lorentz and internal symmetry invariance are not to be broken, it has to be a neutral scalar superfield). Furthermore, the nonzero expectation value must reside in other than the $\theta$-independent component of the field, so that some explicit $\theta$-dependence is introduced. For $N = 1$ matter superfields, it means that one of the auxiliary fields must have a nonzero vacuum expectation value. (Unless gauge invariance is broken, vacuum expectation values for the gauge components cannot be physically relevant.)

Supersymmetry breaking in a local context and the superHiggs mechanism can also be described directly in superspace. All the standard methods, such as the theory of nonlinear realizations, can be applied and all the standard results, such as the conversion of the Goldstino into helicity modes of a massive gravitino and the existence of U-gauge, can be generalized to the superfield discussion of spontaneously broken supersymmetry; the resulting formalism is considerably simpler than a component approach. However, some issues (at the present time) can be settled only by considering components directly, e.g., what are component field masses, what are the conditions for spontaneous breaking to occur, what is the Witten index, etc. Therefore, although most of the material in this chapter is at the superfield level, we cannot avoid some component calculations, and we also omit some topics that have not, as yet, received an adequate superspace treatment.

We first discuss soft explicit breaking of global supersymmetry (sec. 8.2). Our criterion for softness is the analog of Symanzik's criterion in ordinary field theory: In renormalizable globally supersymmetric theories, the only relevant divergences are logarithmic. We ask what nonsupersymmetric terms can be added to the classical action without spoiling the delicate cancellations between boson and fermion contributions that are responsible for the absence of quadratic divergences. Since we can cast the problem in superfield language, we are able to take advantage of the superfield power counting rules of chapter 6.

We next treat spontaneous breaking of global supersymmetry for both renormalizable and nonrenormalizable theories (sec. 8.3). (Nonrenormalizable theories are relevant to our discussion of breaking in the context of local supersymmetry.) We do not discuss Witten's index theorem, or breaking of supersymmetry by instantons; as noted above, with our present techniques these issues can be handled only at the component level. We do, however, give a superspace derivation of the supertrace mass formulae (sec. 8.4). This derivation is much simpler than the component calculation (which we also give, partly for comparison, but also because it provides some extra information, e.g., the masses of the individual components).

Finally, we discuss the superHiggs effect. We show how the Goldstino can be described by a nonlinear (superfield) realization of supersymmetry, and how standard "radial" and "angle" variables can be introduced in models with spontaneously broken supersymmetry (sec. 8.5). We exhibit the superHiggs mechanism (sec. 8.6) and give a detailed discussion of the case of arbitrary supersymmetric "matter" systems coupled to supergravity (sec. 8.7).

## 8.2. Explicit breaking of global supersymmetry

One of the important features of supersymmetric theories is the perturbative no-renormalization theorem (sec. 6.3.c): The superspace potential P(Φ) for chiral superfields receives no radiative corrections, so that scalar multiplet masses and coupling constants are not renormalized (aside from the effect of wave function renormalizations). Furthermore, for renormalizable models, only logarithmic divergences are present (as discussed in sec. 6.5, quadratically divergent D′-terms are not generated if gauge invariant regularization is used). When supersymmetry is explicitly broken this is no longer the case, and, in general, quadratically divergent corrections can be induced. Equivalently, the parameters of an effective low energy theory can depend quadratically on masses associated with the theory defined at high energies, and some of the "naturalness" of supersymmetric theories is destroyed. However, there exists a set of supersymmetry breaking terms whose effect is *soft* : When added to a supersymmetric Lagrangian, any new divergences that these terms generate are logarithmic. More precisely, if we introduce counterterms in the classical Lagrangian to cancel the new divergences, after renormalization their dependence on the renormalization mass (or high energy cutoff) is only logarithmic. In this section, we describe the set of soft breaking terms, and the additional terms that they induce.

Breaking supersymmetry is breaking $Q$-translational invariance. This is done by introducing explicit $\theta$-dependence into the Lagrangian. Equivalently, we can introduce a superfield $\Psi(x,\theta)$ with a fixed $\theta$-dependent value. This suggests the following procedure: Given a supersymmetric action, we generate new terms by coupling, in a manifestly supersymmetric fashion, some external ("spurion") superfield(s) to the quantum fields. Supersymmetry breaking is achieved by giving these fields suitable ($\theta$-dependent) fixed values. At the component level, this introduces some nonsupersymmetric terms. Soft breaking is achieved by only allowing new couplings that are consistent with the (power counting) renormalizability criteria of superfield perturbation theory, so that no divergences worse than logarithmic are introduced. The induced infinities correspond to conventional divergent terms in the effective action involving products of the quantum and spurion fields. When the spurion fields are given their fixed values we can determine the corresponding new component infinities. Generally, we will find that in a component language symmetry breaking terms of dimension two are soft, but terms of dimension three are not (with some exceptions). These terms correspond to splitting the masses of particles in multiplets by hand, or adding some new, nonsupersymmetric interactions of a very special form. We will find that there are essentially five distinct types of soft breaking terms that can occur singly, or in combinations. In general, one such term induces the others, so that we should discuss them all at the same time. However, since their physical significance is different, we prefer to treat them one at a time.

We consider conventional renormalizable Lagrangians (cf. sec. 4.3) of the form

$$S = \int d^4x\, d^4\theta\, [\bar{\Phi}_i e^V \Phi^i + \nu tr V] + \int d^4x\, d^2\theta\, [\frac{1}{2} W^\alpha W_\alpha + \mathrm{P}(\Phi^i)] + h.c. \tag{8.2.1}$$

where P is a polynomial of degree three or less. By power counting we know that the only divergences of the theory correspond to terms in the effective action of the form

$$\int d^4x\, d^4\theta\, \bar{\Phi}\Phi \qquad \int d^4x\, d^4\theta\, \bar{\Phi} V^n \Phi$$

$$\int d^4x\, d^4\theta\, V(D)^2(\bar{D})^2 V^n \qquad \int d^4x\, d^4\theta\, V \tag{8.2.2}$$

where the $D$-derivatives are suitably distributed and terms with $n > 1$ are related to terms with $n = 1$ by gauge invariance (we include ghosts among the chiral fields in (8.2.2)). We break supersymmetry softly by coupling additional external superfields in a manner consistent with the power counting criteria (see sec. 6.3): No more than four $D$'s acting on the internal lines should appear at any vertex where the external spurion field is inserted. In addition to the original divergences of the theory, we may generate new ones, involving the spurion fields as well, and they are the ones that interest us. In this section we do not consider divergences involving spurion fields only, which correspond to insertions into vacuum diagrams and contribute only to the vacuum energy (cosmological constant); see, however, sec. 8.4.

Since the spurion fields can never introduce any additional spinor derivatives into a loop, if in any soft breaking term the spurion field is set to 1 the resulting term must be either a conventional renormalizable supersymmetric term or a total (spinor) derivative. The possible additional couplings that introduce explicit $\theta$-dependence into the action correspond to multiplying a spurion factor into $\bar{\Phi}\Phi$, $\Phi^2$, $W^2$, $\Phi^3$, or $D^\alpha(\Phi W_\alpha)$. In detail, we have:

(a)

$$S_{break} = \int d^4x\, d^4\theta\; U\overline{\Phi}\Phi \sim \int d^4x\; \mu^2 A\overline{A} \tag{8.2.3}$$

where $U = \mu^2\theta^2\overline{\theta}^2$ is a neutral dimension-zero fixed general scalar superfield. At the classical level, when added to (8.2.1), such a term breaks the equality of the boson and fermion masses of a scalar multiplet by adding $-\mu^2$ to the masses of $\mathrm{A} = 2^{-\frac{1}{2}} Re\, A$ and $\mathrm{B} = 2^{-\frac{1}{2}} Im\, A$. To investigate the divergences it introduces, we consider loops with ordinary vertices and external $U$ vertices. We look for local terms in the effective action, involving a $d^4\theta$ integral and factors of $U$ and the quantum fields, of dimension no greater than 2. (This is our standard power counting of sec. 6.3.) Since $U$ is dimensionless, such terms are: $U\Phi\overline{\Phi}$, corresponding to a logarithmic renormalization of (8.2.3), i.e., of $\mu^2$; $U(\Phi + \overline{\Phi})$ (but only if some chiral field is massive); and $UD^\alpha\overline{D}^2D_\alpha V$ (but only if the gauge group has a $U(1)$ factor; the $D$-factors are required by gauge invariance). Therefore, the action may receive additional logarithmically divergent corrections:

$$\Delta\Gamma \sim \int d^4x\, d^4\theta\; U\Phi + \int d^4x\, d^4\theta\; UD^\alpha W_\alpha + h.\,c.$$

$$\sim \int d^4x\; [\mu^2 m\mathrm{A} + \mu^2 \mathrm{D}'] \tag{8.2.4}$$

(b)

$$S_{break} = \int d^4x\, d^2\theta\; \chi\Phi^2 + h.\,c. \sim \int d^4x\; \mu^2(\mathrm{A}^2 - \mathrm{B}^2) \tag{8.2.5}$$

where $\chi = \mu^2\theta^2$ is a neutral dimension-one chiral superfield. This addition corresponds to another way of splitting the masses of scalars and pseudoscalars away from the mass of a spinor in a chiral multiplet. New, logarithmically divergent terms are given by

$$\Delta\Gamma \sim \int d^4x\, d^4\theta\; \overline{\chi}\Phi + h.\,c. \sim \int d^4x\; \mathrm{F} \tag{8.2.6}$$

where $\mathrm{F} = Re\, F$. Since $\chi$ is neutral under whatever internal symmetry groups may be present, no infinities involving gauge fields (as might arise from a $\overline{\chi}e^V\chi$ term) can be induced.

(c)

$$S_{break} = \frac{1}{2} \int d^4x\, d^2\theta\; \eta W^\alpha W_\alpha + h.c. \sim \frac{1}{2} \int d^4x\; \mu\lambda^\alpha\lambda_\alpha + h.c. \tag{8.2.7}$$

where $\eta = \mu\theta^2$ is a neutral, dimension-zero chiral superfield. Again this involves vertices with only four $D$'s and is therefore soft, and provides a mechanism for giving masses to fermions in gauge multiplets. The following divergent terms may be generated in addition to corrections to $S_{break}$ itself:

$$\Delta\Gamma \sim \int d^4x\, d^4\theta\; \bar{\eta}\Phi + h.c. \sim \int d^4x\; \mathrm{F}$$

$$\Delta\Gamma \sim \int d^4x\, d^4\theta\; (\eta\bar{\eta}\Phi + h.c.) \sim \int d^4x\; \mathrm{A}$$

$$\Delta\Gamma \sim \int d^4x\, d^4\theta\; \eta\bar{\Phi}\Phi + h.c. \sim \int d^4x\; [\mathrm{FA} - \mathrm{GB}]$$

$$\Delta\Gamma \sim \int d^4x\, d^4\theta\; \bar{\eta}\eta\bar{\Phi}\Phi \sim \int d^4x\; [\mathrm{A}^2 + \mathrm{B}^2] \tag{8.2.8}$$

For a given theory, not all of these terms need appear; for example, the third term will only be generated at the two loop level, and only if a massive chiral superfield is present.

(d)

$$S_{break} = \int d^4x\, d^2\theta\; \eta\Phi^3 + h.c. \sim \int d^4x\; \mu\, Re\,(A^3 - 3AB^2) \tag{8.2.9}$$

with $\eta$ as in (c). Unlike the previous cases, this introduces an allowed nonsupersymmetric interaction term. In general we induce the same divergences as in case (c).

The breaking term $\int d^4\theta(\eta + \bar{\eta})\bar{\Phi}\Phi$ can be reduced by a field redefinition $\Phi \rightarrow (1 + \eta)\Phi$ to the previous cases.

Another possibility, which gives a gauge invariant mass mixing between the fermions of a gauge multiplet and of a scalar multiplet in the adjoint representation, is

(e)

$$S_{break} = \int d^4x\, d^4\theta\; D^\alpha U\, \Phi\, W_\alpha + h.c.$$

$$= \int d^4x\, d^4\theta\, \bar{D}^2 D^\alpha U\, \Phi\, D_\alpha V + h.\,c. \sim \int d^4x\, \mu Re\, [\psi^\alpha \lambda_\alpha + A\mathrm{D}'] \tag{8.2.10}$$

with a dimension $-1$ field $U = \mu\theta^2\bar{\theta}^2$, or, equivalently, with a dimension $\frac{1}{2}$ chiral spinor superfield $\chi^\alpha = \bar{D}^2 D^\alpha U \sim \mu\theta^\alpha$. Logarithmic corrections are induced for $S_{break}$ itself and for:

$$\Delta\Gamma \sim \int d^4x\, d^4\theta\, \chi^\alpha \chi_\alpha \bar{\Phi} + h.\,c. \sim \int d^4x\, \mu^2 \mathrm{F} \ . \tag{8.2.11}$$

The above possibilities for soft breaking are flexible enough to cover all interesting physical situations without introducing a large number of arbitrary parameters. (With several multiplets, because cancellations are possible, other types of terms can be soft, e.g., $\int d^4x\, d^4\theta\, D^\alpha U \Phi_1 \overleftrightarrow{D}_\alpha \Phi_2 \sim \int d^4x\, (F_1 A_2 - F_2 A_1)$.)

It is also interesting to examine some cases of breaking that are not soft. We mention two:

(a')

$$S_{break} = \int d^4x\, d^4\theta\, U (D^\alpha \Phi)(D_\alpha \Phi) + h.\,c. \tag{8.2.12}$$

with $U$ as in (e), shifts the mass of the spinor in a scalar multiplet. But it leads to vertices with six $D$'s, as does

(b')

$$S_{break} = \int d^4x\, d^4\theta\, U (\Phi + \bar{\Phi})^3 \sim \int d^4x\, \mu \mathrm{A}^3 \ . \tag{8.2.13}$$

Both will produce quadratically divergent terms, for example

$$\Delta\Gamma \sim \int d^4x\, d^4\theta\, U\Phi + h.\,c. \sim \int d^4x\, \mathrm{A} \tag{8.2.14}$$

We can understand the difference between cases (a) and (b) on one hand, and (a') on the other as follows: They both lead to fermion-boson mass splittings for the scalar multiplet. However, the former, in addition to splitting masses, also affects some of the component interaction terms, and it is the delicate balance of mass terms and interactions that keeps the divergences under control. On the other hand, there is no difficulty giving mass to the fermion of a vector multiplet, or introducing mass mixing between the fermions of the two multiplets.

It is a useful and simple exercise to check some of the above conclusions by examining supergraphs involving the spurion fields. We note that some of the induced terms we have listed may be missing because of group theory restrictions, or, in some cases, because of the absence of masses, e.g., the third term in case (c). In certain cases possible terms are missing because the corresponding graphs require $\Phi\Phi$ or $\bar{\Phi}\bar{\Phi}$ propagators and these bring with them numerator mass factors that reduce the degree of divergence of the diagrams. For example, in cases (c) and (d) a term $\bar{\eta}\Phi^2$ cannot be produced because the corresponding diagrams must contain two mass factors and hence are convergent.

## 8.3. Spontaneous breaking of global supersymmetry

If global supersymmetry is spontaneously broken, a massless (Goldstone) fermion must be present. This can be established by the usual reasoning that proves the Goldstone theorem: If the supersymmetry charge does not annihilate the vacuum, there exist operators whose (anti)commutator with the supersymmetry charge has nonzero vacuum expectation value, and, in particular, we can write

$$< 0|\{Q_\alpha, S_{\underline{a}\beta}\}|0 > = \int d^4x \, \frac{\partial}{\partial x^{\underline{b}}} < 0|T(S^{\underline{b}}{}_\alpha(x) S_{\underline{a}\beta}(0))|0 > \tag{8.3.1}$$

where $S_{\underline{a}\alpha}$ is the supersymmetry current, satisfying $\partial^{\underline{a}} S_{\underline{a}\alpha} = 0$, and $Q_\alpha = \int d^3x S_{\underline{0}\alpha}$. The left hand side not being zero (it is actually proportional to the vacuum energy density (8.1.1)) implies that the right hand side receives a contribution from a surface term; this is the case only if the matrix element vanishes at infinity not faster than $|x|^{-3}$, which is possible only if a massless fermion intermediate state is present.

The spontaneous breaking of supersymmetry in globally supersymmetric theories can be investigated by examining the effective potential at its minimum, where it equals the vacuum energy. The effective potential $U$ is obtained from minus the effective action by setting all momenta and all component fields that are not scalars to zero. We must then minimize $U$ with respect to all the remaining component fields and ask if it vanishes at the minimum.

### a. Renormalizable theories

#### a.1. Classical effects

We first consider a system with only chiral scalar superfields, and a renormalizable classical action given by (4.1.11):

$$S = \int d^4x \, d^4\theta \, \overline{\Phi}_i \Phi^i + \int d^4x \, d^2\theta \, \mathrm{P}(\Phi^i) + h.\,c. \tag{8.3.2}$$

where P is a polynomial of degree no higher than three. To investigate the classical vacuum we set all momenta and fermion fields to zero; we then have effectively $\Phi = A - \theta^2 F$, and since each $\theta$ integration requires a $\theta^2\overline{\theta}^2$ factor, we obtain the classical potential

$$U = -\,\overline{F}_i F^i - [F^i \mathrm{P}_i(A) + h.\,c.] \tag{8.3.3}$$

where $\mathrm{P}_i = \dfrac{\partial \mathrm{P}}{\partial A^i}$ as in (4.1.13). The classical vacuum is described by the constant ($x$-independent) classical (expectation) values of the scalar fields obtained by solving the classical field equations for constant fields, i.e., by extremizing the classical potential. Extremizing with respect to the $F^i$ first, we find $\overline{F}_i = -\mathrm{P}_i(A)$; substituting into $U$ we obtain

$$U = \sum_i |\mathrm{P}_i(A)|^2 \quad . \tag{8.3.4}$$

We require

$$\frac{\partial U}{\partial \overline{A}_i} = \overline{P}^{ij}(\overline{A}) P_i(A) = 0 \quad . \tag{8.3.5}$$

The potential will vanish at the extremum and supersymmetry will not be broken only if the simultaneous equations $\mathrm{P}_i(A) = 0$ have a solution. This requirement is equivalent to that of requiring that all the $F$'s have zero vacuum expectation value. We can work directly in superspace, by defining $\mathrm{P}(\Phi)$ as the *superspace potential.* The condition for supersymmetry not to be broken is formally that the superspace potential have an extremum with respect to the superfields: $\dfrac{\partial \mathrm{P}}{\partial \Phi^i} = 0$.

We consider two examples:

(a) The Wess-Zumino model, with action

$$\int d^4x\, d^4\theta\; \overline{\Phi}\Phi + \int d^4x\, d^2\theta\; [a\Phi + \frac{1}{2} m\Phi^2 + \frac{1}{6}\lambda\Phi^3] + h.\,c. \tag{8.3.6}$$

The superspace potential, when differentiated with respect to $\Phi$ gives $a + m\Phi + \frac{1}{2}\lambda\Phi^2$ and setting this to zero always gives us a solution. Hence the vacuum energy is zero and there is no supersymmetry breaking. This is the case even if we consider an arbitrary (nonrenormalizable) polynomial potential.

(b) On the other hand we can consider the O'Raiferteaigh model, given by

$$\int d^4x\, d^4\theta\; [\overline{\Phi}_0\Phi_0 + \overline{\Phi}_1\Phi_1 + \overline{\Phi}_2\Phi_2] + \int d^4x\, d^2\theta\; [\Phi_0{\Phi_1}^2 + m\Phi_1\Phi_2 + \xi\Phi_0] + h.\,c. \tag{8.3.7}$$

for which we obtain the equations

$$\Phi_1{}^2 + \xi = 0$$

$$2\Phi_0\Phi_1 + m\Phi_2 = 0$$

$$m\Phi_1 = 0 \quad , \tag{8.3.8}$$

which have no solution. In this case supersymmetry is broken at the classical level.

In the general case, the situation depends on the topological structure of P. In particular the presence of an extremum, and its stability under variations of parameters in P, can be studied in rigorous fashion and gives rise to index theorems to determine whether supersymmetry can or cannot be spontaneously broken. We remark that supersymmetry breaking in the sense above implies that the fermion mass matrix, which is given by the matrix of second derivatives $\mathrm{P}_{ij}$ evaluated at the minimum of $U$ (see (4.1.12)), has a zero eigenvalue corresponding to the zero mass of the Goldstone fermion. Indeed, if (8.3.5) is satisfied with $\mathrm{P}_j \neq 0$, $\mathrm{P}_{ij}$ must be singular.

Including gauge invariant interactions with a real gauge superfield does not fundamentally change the discussion. Gauge invariant terms that can be added to (8.3.2) have the general form $\int d^4\theta[\bar{\Phi}e^V\Phi + \nu V]$ (the last term only if $V$ is a $U(1)$ gauge field), and the only component of $V$ that can have a nonzero expectation value is the auxiliary field $\mathrm{D}'$; this leads to additional terms in the classical potential of the form $-\frac{1}{2}(\mathrm{D}')^2 - \nu\mathrm{D}' - \bar{A}A\mathrm{D}'$. Extremizing with respect to $\mathrm{D}'$ and then eliminating it gives an additional contribution $\frac{1}{2}|\nu + \bar{A}A|^2$ to the classical potential, which must be separately zero for supersymmetry not to be broken. The expression for the classical potential can be read from (4.3.7). We note that if $\nu \neq 0$, if it can be arranged for $\nu + \bar{A}A$ to equal zero, then some (charged) scalar field must acquire a (nonvanishing) vacuum expectation value, and the gauge group will be spontaneously broken. Thus, to have both gauge invariance and supersymmetry, it is necessary that $\nu = 0$.

We observe that if some expectation value of an auxiliary field is nonzero, i.e., $f =< F >$ or $d =< \mathrm{D}' >$, the supersymmetry transformation of the spinor field of the multiplet becomes (see (3.6.5,6))

$$\delta\psi_\alpha = f\epsilon_\alpha + \cdots$$

or

$$\delta\lambda_\alpha = id\epsilon_\alpha + \cdots \quad , \tag{8.3.9}$$

which is typical behavior for a spontaneously broken symmetry. The spinor field describes the Goldstino, and $f$ or $d$ sets the scale of supersymmetry breaking.

## a.2. Loop corrections

We now establish the following result: In four dimensions, if supersymmetry is not spontaneously broken at the classical level, it is not broken by radiative corrections. This theorem can be proven most readily by using results of superfield perturbation theory, and it might be violated by nonperturbative effects, although no example is known in four dimensions. We first consider the situation with only chiral superfields.

A basic feature of perturbation theory is that the effective action is obtained with a $d^4\theta$ integral. If we consider classical constant fields of the form $\Phi = A - \theta^2 F$ , $D_\alpha$ acting on them is simply $\frac{\partial}{\partial\theta^\alpha}$, so that the derivatives do not introduce any $\theta$ factors; consequently, in the $d^4\theta$ integration, we must get $\theta$ and $\bar{\theta}$ factors from the $\Phi$'s, and these are accompanied by an $F$ and an $\bar{F}$ factor. Therefore, adding the classical potential to the quantum corrections, we have a total potential of the form

$$U_{eff} = -\sum_i [\bar{F}_i F^i + F^i \mathrm{P}_i(A) + \bar{F}_i \bar{\mathrm{P}}^i(\bar{A})] + \sum_{ij} \bar{F}_i F^j G^i{}_j(A, \bar{A}, F, \bar{F}) \tag{8.3.10}$$

Differentiating with respect to $A^i$ and $\bar{F}_i$ we obtain

$$-\frac{\partial U}{\partial A^i} = F^j \mathrm{P}_{ij} + \bar{F}_j F^k \frac{\partial}{\partial A^i} G_k{}^j \tag{8.3.11a}$$

$$-\frac{\partial U}{\partial \bar{F}_i} = F^i + \bar{\mathrm{P}}^i + F^j G_j{}^i + \bar{F}_j F^k \frac{\partial}{\partial \bar{F}_i} G_k{}^j \tag{8.3.11b}$$

Now, if at the classical level there exist values of the $A$'s such that $\mathrm{P}_i = 0$, so that $F^i = 0$ satisfy the extremum equations and make the classical $U$ vanish, it is clear from the above form that this result is not changed by the quantum corrections since the additional terms also vanishes for $F^i = 0$. The important ingredient is that the *quantum corrections are bilinear in the auxiliary fields.*

Therefore the classical minimum of the classical potential is still an extremum of the quantum corrected potential, and it is still such that the total potential vanishes there. Furthermore, if the supersymmetry algebra still holds, it must be an absolute minimum (no negative energy) and therefore supersymmetry cannot be broken. Conversely, if supersymmetry is broken at the classical level (some $\mathrm{P}_i \neq 0$ for any $A$'s), the above equation has no $F^i = 0$ solutions, and hence radiative corrections cannot restore the symmetry.

We remark that in lower dimensions the situation is slightly different. There the superspace integrations are $d^2\theta$, while superfields still have the form $A - \theta^2 F$. Therefore, the quantum corrections to the effective potential can have terms of the form $FG(A, F)$ with a single $F$. When taking derivatives with respect to $F$, the factor in front can disappear, and we find that the classical extremum no longer need be an extremum of the quantum potential.

In the presence of gauge superfields we can have additional contributions to the effective potential. Terms proportional to $\mathrm{D}'^2$ or $\mathrm{D}'F$ are quadratic in auxiliary fields and do not change our conclusions: If $F = \mathrm{D}' = 0$ are solutions of the classical equations, they will also be solutions of the quantum corrected equations. However, it is possible to generate terms of the form $\mathrm{D}'f(A, \bar{A})$, and such terms, no longer quadratic in the auxiliary fields, could change our conclusions (recall that a pure $\mathrm{D}'$ term is not generated). Nevertheless, as long as gauge invariance and supersymmetry are unbroken at the tree level (which implies that the theory does not have a Fayet-Iliopoulos term), even a term linear in $\mathrm{D}'$ is harmless. This is because such a term arises only from the covariantization of terms in the effective action of the form $\int d^4x\, d^4\theta\; g(\bar{\Phi}, \Phi) \rightarrow \int d^4x\, d^4\theta\; g(\bar{\Phi}e^V, \Phi) \sim \int d^4x\; \mathrm{D}'\bar{A}\,\frac{\partial g(\bar{A}, A)}{\partial \bar{A}}$; thus this term is at least bilinear in the $\mathrm{D}', \bar{A}$ fields, and hence we can use the same arguments as above to conclude that the classical solution $\mathrm{D}' = A = 0$ (which must be the case if gauge invariance and supersymmetry are unbroken classically) is still a solution at the quantum level. (A linear $A$ term would spoil this argument, but such a term cannot be written as a $d^4\theta$ integral.)

If classical gauge invariance is broken, and a $\mathrm{D}'f(A, \bar{A})$ is generated, it has been shown that for a specific class of models a supersymmetric solution ($F = \mathrm{D}' = 0$) of the quantum corrected equations exists with the $A$'s shifted from their classical values; thus, even in this case, supersymmetry is not broken by radiative corrections. However, the general situation is in need of further clarification. All of these results hold for the nonrenormalizable systems that we discuss below.

### b. Nonrenormalizable theories

We now consider more general situations. For global models coupled to supergravity (see sec. 5.5.h) the renormalizability criterion is too restrictive: The combined systems are not power-counting renormalizable, and since we can make field dependent Weyl rescalings, there is no reason to insist on polynomiality of the matter actions. However, nonderivative superfield dependent rescalings do not change the number of derivatives in the action and therefore we restrict ourselves to actions that lead to component Lagrangians with no more than two spacetime derivatives in the purely bosonic terms of the action, and no more than one spacetime derivative in the terms containing fermions; this is preserved by superfield dependent rescalings that do not involve spinor derivatives. In this subsection we discuss interacting chiral scalar superfields; we extend the discussion to gauge systems in the following subsection. The reader should review our discussion of Kähler manifolds in sec. 4.1.b.

We consider a system of $N$ chiral superfields $\Phi^i$ described by the superspace action

$$S = \int d^4x\, d^4\theta\, I\!\!K(\Phi^i, \overline{\Phi}_j) + \int d^4x\, d^2\theta\, \mathrm{P}(\Phi^i) + h.c. \quad ,$$

$$\Phi = \Phi(x,\theta,\overline{\theta}) \quad , \quad \overline{D}_{\dot\alpha}\Phi = 0 \quad , \quad \overline{\Phi} = (\Phi)^{\dagger} \quad , \quad D_\alpha \overline{\Phi} = 0 \quad . \tag{8.3.12}$$

As discussed in sec. 4.1.b, the first term of the action $S$ can be given a geometrical interpretation: $\Phi^i$, $\overline{\Phi}_j$ can be thought of as coordinates of a complex manifold with Kähler potential $I\!\!K$.

We recall that the (complex) component fields of $\Phi^i$ are defined by projection

$$A^i = \Phi^i| \quad , \quad \psi_\alpha{}^i = D_\alpha \Phi^i| \quad , \quad F^i = D^2\Phi^i| \quad . \tag{8.3.13}$$

We denote vacuum expectation values of the component fields by

$$a^i = < A^i > \quad , \quad f^i = < F^i > \quad , \quad < \psi^i > = 0 \quad . \tag{8.3.14}$$

The vacuum expectation values are obtained by solving the classical field equations for $x$-independent fields.

The action $S$ leads to the superfield equations

$$\bar{D}^2 I\!K_i + \mathrm{P}_i = 0 \tag{8.3.15}$$

and their hermitian conjugates (we use the notation of (4.1.13,25a)). Taking vacuum expectation values and evaluating at $\theta = 0$ using the definitions (8.3.13,14), we obtain in particular

$$I\!K_i{}^j(a)\bar{f}_j + \mathrm{P}_i(a) = 0 \quad . \tag{8.3.16}$$

As in the renormalizable case (sec. 8.3.a), spontaneous supersymmetry breaking occurs if $f^i = 0$ is *not* a solution to these equations. Further component equations are obtained by differentiating (8.3.15) with $D^2$ and evaluating at $\theta = 0$. We find

$$[I\!K_{ij}{}^k(a)\bar{f}_k + \mathrm{P}_{ij}(a)]f^j = 0 \quad . \tag{8.3.17}$$

After finding the vacuum solution(s), we can choose to work in normal gauge (4.1.27) at the vacuum point. In that case the vacuum equations (8.3.16,17) reduce to

$$\bar{f}_i + \mathrm{P}_i = 0 \quad , \quad \mathrm{P}_{ij}f^j = 0 \quad . \tag{8.3.18}$$

If $\mathrm{P}_{ij}(a)$ is nonsingular all $f^j = 0$ and supersymmetry is not broken. Conversely, if $f^j = 0$ is not a solution of (8.3.16,17) then supersymmetry is broken and $\mathrm{P}_{ij}(a)$ must be singular.

Returning to the action (8.3.12), we shift the fields $\Phi^i \to \Phi^i + <\Phi^i>$ and investigate fluctuations about the vacuum state. In particular we can read off the masses of the various particles from the resulting action; alternatively, we can find the mass matrices of the component fields by expanding the superfield equations (8.3.15) to linearized order in the fluctuations:

$$\bar{D}^2[< I\!K_i{}^j > \bar{\Phi}_j + < I\!K_{ij} > \Phi^j] + < \mathrm{P}_{ij} > \Phi^j = 0 \quad . \tag{8.3.19}$$

Applying $D_\alpha$ and $D^2$ and evaluating at $\theta = 0$ as in (4.1.21), we find

$$\bar{F}_i + \mathrm{P}_{ij}A^j = 0$$

$$i\partial^\alpha{}_{\dot{\alpha}}\bar{\psi}^{i\dot{\alpha}} + \mathrm{P}_{ij}\psi^{j\alpha} = 0$$

$$\Box\bar{A}_i + I\!K_{ik}{}^{jl}\bar{f}_l f^k\bar{A}_j + \mathrm{P}_{ijk}f^kA^j + \mathrm{P}_{ij}F^j = 0 \tag{8.3.20}$$

where we have dropped the <> on $\mathrm{P}_{\ldots}$ and $I\!K_{\ldots}{}^{\cdots}$ and assumed that we are in normal gauge. Eliminating the auxiliary fields we identify the fermion and boson mass matrices:

$$M_F = \mathrm{P}_{ij}(a)$$

$$M_B{}^2 = -\begin{pmatrix} I\!K_{ik}{}^{jl}\bar{f}_l f^k - \mathrm{P}_{ik}\bar{\mathrm{P}}^{kj} & \mathrm{P}_{ijk}f^k \\ \bar{\mathrm{P}}^{ijk}\bar{f}_k & I\!K_{jl}{}^{ik}\bar{f}_k f^l - \bar{\mathrm{P}}^{ik}\mathrm{P}_{kj} \end{pmatrix} \quad . \tag{8.3.21}$$

Again, as above, if supersymmetry is broken $\mathrm{P}_{ij}$ has at least one zero eigenvalue and one of the corresponding massless fermions is the Goldstino.

We evaluate the graded trace of the mass matrix squared. This supertrace gives the mass relation

$$str\, M^2 = \sum_J (-1)^{2J}(2J+1)M_J{}^2 = tr\, M_B{}^2 - 2tr\, M_F M_F{}^*$$

$$= -\,2I\!K_{ik}{}^{il}\bar{f}_l f^k \tag{8.3.22}$$

Since we are in normal gauge we can rewrite this as

$$str\, M^2 = -\,2R_k{}^l \bar{f}_l f^k \tag{8.3.23}$$

where $R_k{}^l = R_i{}^l{}_k{}^i$ is the Ricci tensor of the manifold evaluated at $\Phi^i = <\Phi^i>$ (see (4.1.28)). The result is manifestly covariant, and thus (8.3.23) holds in an arbitrary gauge. In particular, for models with conventional actions $\bar{\Phi}_i\Phi^i$ the Kähler manifold is *flat* and we obtain the simple mass formula

$$\sum_J (-1)^{2J}(2J+1)M_J{}^2 = 0 \quad . \tag{8.3.24}$$

We also observe that in contrast to renormalizable models, spontaneous supersymmetry breaking can occur in a model with a single chiral multiplet, for example with $I\!K = cos(\Phi + \bar{\Phi})$, $\mathrm{P}(\Phi) = \Phi$, where $-\pi < A < \pi$.

### c. Global gauge systems

In this section, we repeat the previous analysis but include gauge superfields $V = V^{\mathrm{A}}T_{\mathrm{A}}$. Because gauge symmetries are usually described by explicit matrix representations $(T_{\rm A})^i{}_j$ of the generators, we begin with this formulation; we then change over to a more general formulation where we describe the action of the generators by Killing vectors. As discussed in sec. 4.1.b, this allows us to choose normal *coordinates* in which the computation of the mass matrices simplifies (however, we cannot use normal *gauge*). We restrict ourselves to models where the gauged group is unbroken or *isotropic* at one or more points of the manifold of scalar fields. (The usual matrix representation assumes isotropy at the origin, i.e., the origin is kept fixed by gauge transformations.) A formulation in terms of Killing vectors should allow one to gauge groups that are realized nonlinearly at every point on the manifold of scalar fields, i.e., $\delta\Phi$ has a constant term everywhere (the constant term cannot be eliminated by shifting the scalar fields); however, the superfield description of the more general case has not been worked out.

We consider the action

$$S = \int d^4x\, d^4\theta\, [I\!K(\Phi^i, \tilde{\bar\Phi}_j) + \nu tr V]$$

$$+ \int d^4x\, d^2\theta\, [\mathrm{P}(\Phi^i) + \frac{1}{4} Q_{\rm AB}(\Phi^i) W^{\alpha\rm A} W_\alpha{}^{\rm B}] + h.\,c. \qquad (8.3.25)$$

with covariantly chiral $\tilde{\bar\Phi}$:

$$\tilde{\bar\Phi}_j = \bar\Phi_k (e^V)^k{}_j \quad , \qquad W_\alpha{}^{\rm A} = i\bar D^2 (e^{-V} D_\alpha e^V)^{\rm A} \quad . \qquad (8.3.26)$$

The chiral quantities $Q_{\rm AB} = \delta_{\rm AB} + O(\Phi)$ can generate masses for the gauge fermions contained in $V$. We have included the global Fayet-Iliopoulos term $\nu tr V$ (4.3.3).

We chose a Kähler gauge (see (4.1.26)) where $I\!K$ itself is invariant; we can always do this if the gauge group is unbroken somewhere on the manifold as discussed above. Then gauge invariance of $S$ requires

$$I\!K_j (T_{\rm A})^j{}_i \Phi^i - \tilde{\bar\Phi}_j (T_{\rm A})^j{}_i I\!K^i = 0 \quad ,$$

$$\mathrm{P}_j (T_{\rm A})^j{}_i \Phi^i = 0 \quad ,$$

$$Q_{{\rm DE},j} (T_{\rm C})^j{}_i \Phi^i + (T_{\rm C})_{\rm D}{}^{\rm A} Q_{\rm AE} + (T_{\rm C})_{\rm E}{}^{\rm A} Q_{\rm AD} = 0 \quad . \qquad (8.3.27)$$

The matrices $(T_{\rm C})_{\rm E}{}^{\rm A}$ form the adjoint representation of the generators, and are, up to an overall factor, the structure constants; thus they are independent of any special choice of coordinates on the scalar manifold.

We begin by deriving the equations for the vacuum expectation values. We define (Yang-Mills) covariant component fields by covariant projection (see (4.3.4,5))

$$A^i = \Phi^i| \quad , \quad \psi_\alpha{}^i = \nabla_\alpha \Phi^i| \quad , \quad F^i = \nabla^2 \Phi^i| \quad , \tag{8.3.28a}$$

$$\lambda_\alpha = W_\alpha| \quad , \quad f_{\alpha\beta} = \frac{1}{2} \nabla_{(\alpha} W_{\beta)}| \quad , \tag{8.3.28b}$$

$$i\nabla_\alpha{}^{\dot\alpha} \bar\lambda_{\dot\alpha} = \frac{1}{2} [\nabla^\beta , \{\nabla_\beta , W_\alpha\}]| \quad , \quad \mathrm{D}' = -\frac{i}{2} \{\nabla^\alpha , W_\alpha\}| \quad , \tag{8.3.28c}$$

where $f^{\mathrm{B}}{}_{\alpha\beta}$ is the component gauge field strength (see (4.2.85)). We also need the identity ($d^{\mathrm{A}} = < \mathrm{D}'^{\mathrm{A}} >$)

$$< \nabla^2 \bar\nabla^2 \tilde\Phi_i > | = d^{\mathrm{A}}\, \bar a_j\, (T_{\mathrm{A}})^j{}_i \quad . \tag{8.3.29}$$

The superfield equations that follow from (8.3.25) are:

$$\bar\nabla^2 I\!K_i + \mathrm{P}_i + \frac{1}{4} Q_{\mathrm{AB},i} W^{\alpha\mathrm{A}} W_\alpha{}^{\mathrm{B}} = 0$$

$$\tilde\Phi_j (T_{\mathrm{A}})^j{}_i I\!K^i - \frac{i}{2} \nabla^\alpha (Q_{\mathrm{AB}} W_\alpha{}^{\mathrm{B}}) + \frac{1}{2} i \bar\nabla^{\dot\alpha} (\bar Q_{\mathrm{AB}} \bar W_{\dot\alpha}{}^{\mathrm{B}}) + \nu tr\, T_{\mathrm{A}} = 0 \quad . \tag{8.3.30}$$

The equations for the vacuum expectation values are obtained by evaluating at $\theta = 0$ the above equations, and the equation obtained by differentiating the first one with $\nabla^2$. We find

$$I\!K_i{}^j \bar f_j + \mathrm{P}_i = 0$$

$$I\!K_{ij}{}^k \bar f_k f^j + \bar a_k (T_{\mathrm{A}})^k{}_j d^{\mathrm{A}} I\!K_i{}^j + \mathrm{P}_{ij} f^j + \frac{1}{2} Q_{\mathrm{AB},i} d^{\mathrm{A}} d^{\mathrm{B}} = 0$$

$$\bar a_j (T_{\mathrm{A}})^j{}_i I\!K^i + (Q_{\mathrm{AB}} + \bar Q_{\mathrm{AB}}) d^{\mathrm{B}} + \nu tr\, T_{\mathrm{A}} = 0 \tag{8.3.31}$$

where $I\!K$, $Q_{\mathrm{AB}}$, P are evaluated with $\Phi^i \to a^i$.

We now generalize to arbitrary coordinates by rewriting the above in terms of holomorphic Killing vectors $k_{\mathrm{A}}{}^i$. We replace the specific form of the Yang-Mills gauge transformation (4.1.35)

$$\delta\Phi^i = i\Lambda^{\mathrm{A}} (T_{\mathrm{A}})^i{}_j \Phi^j \quad , \quad \delta\tilde\Phi_i = -i\tilde\Phi_j \Lambda^{\mathrm{A}} (T_{\mathrm{A}})^j{}_i \tag{8.3.32}$$

with the more general form (4.1.31):

$$\delta\Phi^i = \Lambda^{\mathrm{A}} k_{\mathrm{A}}{}^i \quad , \quad \delta\tilde{\Phi}_i = \Lambda^{\mathrm{A}} k_{\mathrm{A}i} \tag{8.3.33}$$

where $\tilde{\Phi}_i$ is defined by analogy with (4.1.34b):

$$\tilde{\Phi}_i \equiv exp(iV^{\mathrm{A}} k_{\mathrm{A}j} \frac{\partial}{\partial \bar{\Phi}_j}) \bar{\Phi}_i \quad . \tag{8.3.34}$$

The conditions (8.3.27) that ensure gauge invariance of the action become

$$I\!K_i k_{\mathrm{A}}{}^i + I\!K^i k_{\mathrm{A}i} = 0$$

$$\mathrm{P}_i k_{\mathrm{A}}{}^i = 0$$

$$Q_{\mathrm{DE},i} k_{\mathrm{C}}{}^i + i(T_{\mathrm{C}})_{\mathrm{D}}{}^{\mathrm{A}} Q_{\mathrm{AE}} + i(T_{\mathrm{C}})_{\mathrm{E}}{}^{\mathrm{A}} Q_{\mathrm{AD}} = 0 \quad . \tag{8.3.35}$$

As discussed above, a formulation in terms of Killing vectors enables us to use normal coordinates and thus to simplify our computations. Thus, for example, we can compute the mass matrices of the various component fields and find a supertrace relation that generalizes (8.3.22). We find the linearized field equations for the component fields by expanding the covariantized form of the superfield equations (8.3.30) around the vacuum and applying the operators $1, \nabla, \nabla^2$ to the first and $1, \nabla, [\bar{\nabla}, \nabla]$ to the second of the equations and evaluating at $\theta = 0$. The result, in normal coordinates, is

$$\tilde{F}_i + \mathrm{P}_{ij} A^j = 0$$

$$i\partial^{\alpha}{}_{\dot{\alpha}} \tilde{\psi}^{\dot{\alpha}}{}_i + \lambda^{\mathrm{A}\alpha} k_{\mathrm{A}i} + \mathrm{P}_{ij} \psi^{\alpha j} - \frac{i}{2} Q_{\mathrm{AB},i} d^{\mathrm{A}} \lambda^{\mathrm{A}\alpha} = 0 \quad ,$$

$$\Box \tilde{A}_i + [id^{\mathrm{A}} k_{\mathrm{A}i}{}^{,j} + I\!K_{ik}{}^{lj} f^k \bar{f}_l - \mathrm{P}_{ik} \bar{\mathrm{P}}^{kj}] \tilde{A}_j$$

$$+ [\mathrm{P}_{ijk} f^k + \frac{1}{2} Q_{\mathrm{AB},ij} d^{\mathrm{A}} d^{\mathrm{B}}] A^j + [Q_{\mathrm{AB},i} d^{\mathrm{B}} + ik_{\mathrm{A}i}] \mathrm{D}'^{\mathrm{A}} = 0 \quad ,$$

$$(Q_{\mathrm{AB}} + \bar{Q}_{\mathrm{AB}}) \mathrm{D}'^{\mathrm{B}} + (Q_{\mathrm{AB},i} d^{\mathrm{B}} + ik_{\mathrm{A}i}) A^i + (\bar{Q}_{\mathrm{AB}}{}^{,i} d^{\mathrm{B}} - ik_{\mathrm{A}}{}^i) \tilde{A}_i = 0 \quad ,$$

$$\frac{i}{2} (Q_{\mathrm{AB}} + \bar{Q}_{\mathrm{AB}}) \partial^{\alpha}{}_{\dot{\alpha}} \tilde{\lambda}^{\mathrm{B}\dot{\alpha}} + (k_{\mathrm{A}i} - \frac{i}{2} Q_{\mathrm{AB},i} d^{\mathrm{B}}) \psi^{\alpha i} + \frac{1}{2} Q_{\mathrm{AB},i} f^i \lambda^{\mathrm{B}\alpha} = 0 \quad ,$$

$$(Q_{\mathrm{AB}} + \overline{Q}_{\mathrm{AB}})\nabla^{\alpha\dot{\alpha}} f^{\mathrm{B}}{}_{\alpha\beta} - (k_{\mathrm{A}i} k_{\mathrm{B}}{}^i + k_{\mathrm{B}i} k_{\mathrm{A}}{}^i) A^{\mathrm{B}}{}_{\beta}{}^{\dot{\alpha}} = 0 \quad . \tag{8.3.36}$$

where we have dropped $< >$ on $\mathrm{P}_{\dots}$, $I\!\!K_{\dots}{}^{\dots}$, $Q_{\mathrm{AB}\dots}$. In our normalization the vector wave equation is

$$\nabla^{\alpha\dot{\alpha}} f_{\alpha\beta} - m^2{}_V A_{\beta}{}^{\dot{\alpha}} = 0 \quad . \tag{8.3.37}$$

Eliminating the auxiliary fields we find the mass matrices from which we obtain the supertrace (in normal coordinates)

$$str\, M^2 = \, -2[id^{\mathrm{A}} k_{\mathrm{A}i}{}^{,i} + I\!\!K_{ik}{}^{li} f^k \overline{f}_l + tr(Q_i \frac{1}{Q+\overline{Q}} \overline{Q}^j \frac{1}{Q+\overline{Q}}) f^i \overline{f}_j$$

$$+ \, i(Q+\overline{Q})^{-1\mathrm{AB}} (k_{\mathrm{A}i} \overline{Q}_{\mathrm{BC}}{}^{,i} d^{\mathrm{C}} - k_{\mathrm{A}}{}^i Q_{\mathrm{BC},i} d^{\mathrm{C}})] \quad . \tag{8.3.38}$$

A covariant formula, valid in any coordinate system, is obtained by replacing $k_{\mathrm{A}i}{}^{,i}$ with $k_{\mathrm{A}i}{}^{;i}$ and $I\!\!K_{ik}{}^{li}$ with $R_k{}^l$

$$str\, M^2 = \, -2[id^{\mathrm{A}} k_{\mathrm{A}i}{}^{;i} + R_k{}^l f^k \overline{f}_l + tr(Q_i \frac{1}{Q+\overline{Q}} \overline{Q}^j \frac{1}{Q+\overline{Q}}) f^i \overline{f}_j$$

$$- \, i\, tr(\overline{Q}^i \frac{1}{Q+\overline{Q}}) k_{\mathrm{A}i} d^{\mathrm{A}}] \tag{8.3.39}$$

where we have rewritten the last term using the gauge invariance relations (8.3.35). In the coordinate system where the Yang-Mills gauge transformations are given by (8.3.32) we have $k_{\mathrm{A}i}{}^{;i} = \, -i(T_{\mathrm{A}})^i{}_i - i(T_{\mathrm{A}})^j{}_i \overline{a}_j \Gamma^i$ (cf. (4.1.29b,31,32d)).

## 8.4 Trace formulae from superspace

In the last two sections, we found the supertrace using essentially a component approach, and not taking advantage of the superfield formalism. There is a much easier way to evaluate the supertrace expression without ever computing component mass matrices: If the action is expanded in components and the one-loop effective potential is evaluated, its quadratically divergent part is proportional to the supertrace $str\, M^2$; moreover, we can easily read off this quadratically divergent term from the classical superfield action if we imagine performing a superfield one-loop calculation.

### a. Explicit breaking

We can develop the method (and derive some new mass formulae) by first considering the case of explicit soft breaking of supersymmetry. For example, we consider a massless scalar multiplet and add to it the explicit soft breaking term (8.2.3) $S_{break} = \int d^4x\, d^4\theta\, U\overline{\Phi}\Phi$. We now calculate the quadratically divergent part of the one-loop effective potential; the coefficient is the contribution of the term (8.2.3) to the supertrace. Recall that soft breaking terms are defined by the property that they give at most logarithmically divergent contributions to the effective action, and yet here we are calculating quadratic divergences; however, in sec. 8.2 we ignored vacuum diagrams (which have only spurion fields externally), whereas here that is all we are interested in.

In the calculation, we have to consider the sum of one-loop diagrams with $n$ massless chiral propagators and $n$ $U$-spurion vertices ($U = \mu^2\theta^2\overline{\theta}^2$; although the calculation simplifies if we use the explicit form of $U$, we will keep $U$ general, since then the results can be applied to other cases). At each vertex we have factors $D^2$, $\overline{D}^2$ acting on the propagators; however, each propagator is proportional to $p^{-2}$, and thus, to get a quadratic divergence, we must cancel all but one propagator with a numerator factor. This requires $n-1$ factors of $D^2\overline{D}^2 \sim -p^2$; the remaining factor is needed for the $\theta$ loop (see sec. 6.3, e.g., (6.3.28)). Hence we find

$$\Gamma_\infty = \sum_n \int \frac{d^4\theta\, d^4p}{(2\pi)^4 p^2} \frac{-1}{n} (-U)^n = \int d^4\theta\, ln(1+U) \int \frac{d^4p}{(2\pi)^4 p^2} \quad . \tag{8.4.1}$$

Therefore the supertrace is

$$
\begin{aligned}
str\, M^2 &= -\,2\int d^4\theta\, ln(1+U) = -\,2\,[D^2\bar{D}^2\, ln(1+U)]| \\
&= -\,2\,[D^2\bar{D}^2\, U]| \\
&= -\,2\mu^2 \quad .
\end{aligned} \tag{8.4.2}
$$

Comparing to the component expression in (8.2.3), we see that this is indeed the correct result: the mass of the scalar $A$ has been lowered by $\mu^2$ (the factor of 2 arises because $A$ is complex). It is clear that no diagram containing chiral self-interactions can change the result: We needed a factor of $D^2\bar{D}^2$ at each vertex, and a chiral vertex comes with only a factor $\bar{D}^2$. Such a diagram can be only logarithmically divergent. Also, since supersymmetric mass terms can be treated as interactions, our results hold in the massive case. This same argument also implies that explicit breaking terms of the types considered in (8.2.5,6,9) cannot contribute to the supertrace.

Next we consider explicit breaking terms (8.2.7) for an abelian vector field: $S_{break} = \frac{1}{2}\int d^4x\, d^4\theta\, \eta W^\alpha W_\alpha + h.c. = \frac{1}{2}\int d^4x\, d^4\theta\, (\eta+\bar{\eta})VD^\alpha\bar{D}^2D_\alpha V$. The calculation is almost identical to the above: Each propagator is still $\sim p^{-2}$, except that a vector propagator has an extra $-1$ relative to a chiral propagator, and each $\eta+\bar{\eta}$ vertex ($\eta = \mu\theta^2$) comes with a factor $D^\alpha\bar{D}^2D_\alpha$, which acts precisely in the same way as a factor $D^2\bar{D}^2$, except for a $-1$ that cancels the extra $-1$ from the propagator. Thus we find

$$
\begin{aligned}
str\, M^2 &= -\,2\int d^4\theta\, [-ln(1+\eta+\bar{\eta})] = 2\,[D^2\bar{D}^2\, ln(1+\eta+\bar{\eta})]| \\
&= -\,2\,[D^2\eta\, \bar{D}^2\bar{\eta}]| \\
&= -\,2\mu^2 \quad .
\end{aligned} \tag{8.4.3}
$$

As before, this agrees with the component expression (8.2.7) (the factor 2 comes from the two helicity components of the fermion). We can combine the explicit breaking terms (8.2.3-7) with the previous ones and find simply the sum of (8.4.2,3).

Finally, we consider (8.2.10); since this has only a factor $D^\alpha\bar{D}^2$ inside the loop at each vertex, it cannot contribute to the quadratic divergence or the supertrace.

**b. Spontaneous breaking**

In the examples of the preceding subsection, we used rather elaborate methods to derive results that can be found more easily by explicit computation of the masses; here we will apply these methods to derive results that required the somewhat lengthy calculations of sec. 8.3. We first consider the action (8.3.12). We expand $S$ to second order in quantum fields $\Phi^i$, with the coefficients evaluated at the background classical values $< \Phi^i >$:

$$S^{(2)} = \int d^4x\, d^4\theta\, \bar{\Phi}_j\, I\!K_i{}^j\, \Phi^i + \int d^4x\, d^2\theta\, X_{ij} \Phi^i \Phi^j + h.\,c. \tag{8.4.4}$$

where

$$X_{ij} = P_{ij} + \bar{D}^2 I\!K_{ij} \tag{8.4.5}$$

In complete analogy with (8.4.1), the quadratically divergent term in the one-loop effective action is

$$\Gamma_\infty = \int \frac{d^4p}{(2\pi)^4 p^2}\, d^4\theta\, tr\, [ln(I\!K_i{}^j)] \tag{8.4.6}$$

where $I\!K_i{}^j - \delta_i{}^j$ plays the role of $U$ and, as above, the chiral vertex (here $X_{ij}$) does not contribute. The supertrace is therefore

$$str\, M^2 = -\, 2 \int d^4\theta\, tr[ln(I\!K_i{}^j)] = -\, 2\, \{ D^2 \bar{D}^2 [tr\, ln(I\!K_i{}^j)] \}|$$

$$= -\, 2 [tr\, ln(I\!K_i{}^j)]_k{}^l\, \bar{f}_l f^k \tag{8.4.7a}$$

and hence, using (4.1.30b),

$$str\, M^2 = -\, 2 R_k{}^l \bar{f}_l f^k \quad , \tag{8.4.7b}$$

in agreement with (8.3.23).

For the case with gauge interactions (sec. 8.3.c), we again obtain the supertrace by examining the quadratic divergence in the one-loop effective action. To second order in quantum fields we have

$$S^{(2)} = \int d^4x\, d^4\theta\, [I\!K_i{}^j \bar{\Phi}_k (e^V)^k{}_j \Phi^i + \frac{1}{4} (Q_{AB} + \bar{Q}_{AB})\, V^A D^\alpha \bar{D}^2 D_\alpha V^B\, ] \quad , \tag{8.4.8}$$

where we have dropped terms that do not contribute to the quadratic divergence (that is, chiral interactions or terms corresponding to (8.2.10)). Now $(e^V)^k{}_j I\!\!K_i{}^j - \delta_i{}^k$ plays the role of $U$ above, and $\frac{1}{2}(Q_{AB} - \delta_{AB})$ plays the role of $\eta$. The final result is

$$str\, M^2 = -2\int d^4\theta\{tr[(V^A T_A)_i{}^j + ln(I\!\!K_i{}^j)] - tr\, ln(\frac{1}{2}[Q_{AB} + \bar{Q}_{AB}])\}$$

$$= -2[d^A(T_A)_i{}^i + d^A(T_A)_i{}^j \bar{a}_j \Gamma^i + R_k{}^l \bar{f}_l f^k$$

$$+ tr(Q_k \frac{1}{Q+\bar{Q}} \bar{Q}^l \frac{1}{Q+\bar{Q}})\bar{f}_l f^k - tr(\bar{Q}^l \frac{1}{Q+\bar{Q}})(T_A)_l{}^i \bar{a}_i d^A\,] \tag{8.4.9}$$

where we have replaced $\int d^4\theta \to \nabla^2\bar{\nabla}^2$ and used $[\bar{\nabla}^{\dot{\alpha}}, \{\bar{\nabla}_{\dot{\alpha}}, \nabla_\alpha\}] = -2iW^\alpha$, and (8.3.29). We thus recover the result (8.3.39). (We have chosen to work in the coordinate system defined by (8.3.32) simply because it is more familiar; the computation is equally straightforward in terms of Killing vectors.)

## 8.5. Nonlinear realizations

Experience with spontaneously broken internal symmetries has shown that much useful insight can be gained by studying the general theory of nonlinear realizations. The methods that have been developed can be applied quite successfully to supersymmetry.

One way to formulate a nonlinear realization of supersymmetry is to consider (nonlinearly) constrained superfields; however, it is far from obvious how to choose such constraints, and so we will return to this approach after we have studied nonlinear realizations directly.

The simplest nonlinear realization is the Volkov-Akulov model. It is found by considering a covariantly transforming set of hypersurfaces in superspace. Let

$$\omega^\alpha(x) = \theta^\alpha \tag{8.5.1}$$

define a hypersurface; it transforms as

$$\omega'(x') = \theta' \tag{8.5.2}$$

where we recall that $x' = x - \frac{i}{2}(\bar\epsilon\theta + \epsilon\bar\theta)$, $\theta' = \theta + \epsilon$. This implies

$$\omega'(x') = \theta + \epsilon = \omega(x) + \epsilon = 0 \tag{8.5.3}$$

and hence

$$\omega'(x - \frac{i}{2}[\bar\epsilon\omega(x) + \epsilon\bar\omega(x)]) = \omega(x) + \epsilon \tag{8.5.4}$$

or

$$\delta\omega^\alpha(x) = \epsilon^\alpha + \frac{i}{2}(\bar\epsilon^{\dot\beta}\omega^\beta + \epsilon^\beta\bar\omega^{\dot\beta})\partial_{\beta\dot\beta}\omega^\alpha \quad . \tag{8.5.5}$$

This gives a nonlinear realization of the algebra carried by the spinor field $\omega^\alpha(x)$; it is by no means unique, but other nonlinear realizations are related to it by field redefinitions. Note that $\delta\omega^\alpha$ (or any other equivalent nonlinear realization) contains a constant term in its transformation law, and is therefore a suitable field for describing the Goldstino.

To find an invariant action, we recall that the one-form (3.3.31)

$$s^{\alpha\dot\alpha}(x,\theta,\bar\theta) = dx^{\alpha\dot\alpha} + \frac{i}{2}(\theta^\alpha d\bar\theta^{\dot\alpha} + \bar\theta^{\dot\alpha}d\theta^\alpha) \tag{8.5.6}$$

is invariant under supersymmetry transformations. If we constrain this one-form to lie on the hypersurface $\omega(x) = \theta$, we find

$$s^{\alpha\dot{\alpha}}(x) = dx^{\alpha\dot{\alpha}} + \frac{1}{2}(\omega^{\alpha} i\overleftrightarrow{\partial}_{\underline{m}}\bar{\omega}^{\dot{\alpha}})dx^{\underline{m}} \equiv dx^{\underline{m}} v_{\underline{m}}{}^{\underline{a}} \quad , \tag{8.5.7}$$

where we have defined an "inverse vierbein" $v_{\underline{m}}{}^{\underline{a}}$. Since the one-form is invariant, the vierbein must transform covariantly, i.e., supersymmetry transformations of $\omega$ must induce coordinate transformations of $v_{\underline{a}}{}^{\underline{m}}$. Now it is easy to write down an invariant action in terms of the determinant $v = det(v_{\underline{a}}{}^{\underline{m}})$:

$$S_{\omega} = \int d^4x\, d^4\theta \;\; v^{-1}\, \delta^4(\theta - \omega(x)) = \int d^4x \;\; v^{-1} \quad . \tag{8.5.8}$$

We note that $\Psi \equiv c\, v^{-1}\delta^4(\theta - \omega(x))$ is a scalar superfield whose components are functions of $\omega$ ($c$ is an arbitrary dimensional constant that sets the scale of supersymmetry breaking; see below). We can also construct a chiral superfield $\Phi = \bar{D}^2\Psi$ out of $\omega$. Since $[\delta^4(\theta)]^2 = 0$ these superfields satisfy the nonlinear constraints

$$\Phi^2 = \Psi^2 = 0 \tag{8.5.9a}$$

$$\Phi = c^{-1}\Phi\bar{D}^2\bar{\Phi} = c^{-m}\Phi(\bar{D}^2\bar{\Phi})^m \tag{8.5.9b}$$

$$\Psi = c^{-1}D^2\Psi\bar{D}^2\Psi = \frac{1}{2}c^{-1}\Psi D^{\alpha}\bar{D}^2 D_{\alpha}\Psi \tag{8.5.9c}$$

etc. The solution to the constraints (8.5.9a,b) or (8.5.9a,c) is precisely $\Phi$ or $\Psi$ respectively.

The expectation values $<\Phi>, <\Psi>$ of $\Phi, \Psi$, that follow from $<\omega^{\alpha}> = 0$ are typical of the expectation value of a multiplet with spontaneously broken supersymmetry (as in sec. 8.3): the auxiliary components ($\theta^2$ or $\theta^2\bar{\theta}^2$ for $\Phi$ and $\Psi$ respectively) get nonvanishing expectation values $c$, and all other components can be taken to have vanishing expectation values. The fermion components at one $\theta$ −level lower than the auxiliary fields, e.g., $\eta_{\alpha} = D_{\alpha}\Phi|$ or $\eta_{\alpha} = \bar{D}^2 D_{\alpha}\Psi|$, have this constant term in their transformations as $\delta\eta_{\alpha} = c\epsilon_{\alpha} + \cdots$, confirming our identification of $c$ as the supersymmetry breaking scale.

The vierbein $v$ can be used to write down other invariant actions; *any* expression covariantized with $v$ is supersymmetric. For example, we can couple $\omega$ to a scalar field

$A$ as follows:

$$S_A = -\frac{1}{2}\int d^4x\ v^{-1}(v_{\underline{a}}{}^{\underline{m}}\partial_{\underline{m}}A)^2 \tag{8.5.10a}$$

where $A$ transforms as

$$A(x) = A'(x') = A'(x - \frac{i}{2}(\bar{\epsilon}\omega + \epsilon\bar{\omega})) \quad . \tag{8.5.10b}$$

We now discuss the prescription for describing spontaneously broken theories in terms of $\omega$. In a spontaneously broken theory, the Goldstino is one, or in general a unique linear combination, of the fermionic components of the ordinary superfields. We introduce "standard" variables by replacing the Goldstino with the Volkov-Akulov field $\omega$, and the superfields by new superfields whose components transform homogeneously as in (8.5.10b), and in particular, with no mixing of different $\theta$-components. We begin by constructing a homogeneously transforming superfield $\check{\Psi}$ out of $\omega$ and an ordinary superfield $\Psi$. Consider

$$\check{\Psi}(x,\theta,\bar{\theta}) \equiv \Psi(x + \frac{i}{2}(\bar{\omega}\theta + \omega\bar{\theta}), \theta - \omega, \bar{\theta} - \bar{\omega}) \tag{8.5.11}$$

where $\Psi(x,\theta,\bar{\theta})$ is *any* superfield. Under supersymmetry transformations $\Psi'(x',\theta',\bar{\theta}') \equiv \Psi'(x - \frac{i}{2}(\bar{\epsilon}\theta + \epsilon\bar{\theta}), \theta + \epsilon, \bar{\theta} + \bar{\epsilon}) = \Psi(x,\theta,\bar{\theta})$, we find for the transformation of $\check{\Psi}$:

$$\begin{aligned}
&\check{\Psi}\big(x - \frac{i}{2}(\bar{\epsilon}\omega + \epsilon\bar{\omega}), \theta, \bar{\theta}\big) \\
&\quad = \Psi\big(x - \frac{i}{2}(\bar{\epsilon}\omega + \epsilon\bar{\omega} - \bar{\omega}\theta - \omega\bar{\theta}), \theta - \omega, \bar{\theta} - \bar{\omega}\big) \\
&\quad = \Psi'\big(x - \frac{i}{2}(\bar{\epsilon}\omega' + \epsilon\bar{\omega}' - \bar{\omega}'\theta - \omega'\bar{\theta}) - \frac{i}{2}[\bar{\epsilon}(\theta - \omega') + \epsilon(\bar{\theta} - \bar{\omega}')], \theta - \omega' + \epsilon, \bar{\theta} - \bar{\omega}' + \bar{\epsilon}\big) \ ;
\end{aligned} \tag{8.5.12a}$$

using (8.5.4), we have

$$\begin{aligned}
&\check{\Psi}\big(x - \frac{i}{2}(\bar{\epsilon}\omega + \epsilon\bar{\omega}), \theta, \bar{\theta}\big) \\
&\qquad = \Psi'\big(x + \frac{i}{2}(\bar{\omega}\theta + \omega\bar{\theta}), \theta - \omega, \bar{\theta} - \bar{\omega}\big) \\
&\qquad = \check{\Psi}'(x,\theta,\bar{\theta}) \quad .
\end{aligned} \tag{8.5.12b}$$

Thus we see that under supersymmetry transformations $\check{\Psi}$ transforms homogeneously (different $\theta$-components do not mix) but nonlinearly with respect to $\omega$; the $x$-coordinate undergoes a translation. Therefore, the whole supersymmetry group is realized on $\check{\Psi}$ by elements of the Poincaré group. This is a general feature of nonlinear realizations: Given a group $G$ (here the supersymmetry group) and a linearly realized subgroup $H$ (here the Poincaré group), the nonlinear realizations on suitably defined fields is performed by elements of $H$. Consequently, we can impose *any translationally invariant constraint* on $\check{\Psi}$ without breaking supersymmetry. For example, if we constrain it entirely by setting it equal to $c\theta^2\bar{\theta}^2$ (or $c\theta^2$ in the chiral case), we express all the components of $\Psi$ in terms of $\omega$ and recover the previous result (8.5.9).

We now consider a model with spontaneous breakdown of supersymmetry. We can describe the model in terms of standard components, i.e., components transforming as in (8.5.4,10b), as follows:

(1) For each superfields $\Psi$ we construct the associated $\check{\Psi}$.

(2) We identify the fermionic component $\psi$ of the appropriate linear combination of $\Psi$'s that is the Goldstino, and is therefore a suitable candidate to be replaced by $\omega$, and constrain the corresponding component $\check{\psi}$ in the corresponding linear combination of $\check{\Psi}$'s to zero. This gives us the combination of the components of the $\Psi$'s that transforms as (8.5.4).

(3) We express the remaining components of the $\Psi$'s in terms of the remaining components of the $\check{\Psi}$'s.

This procedure is the analog of going to radial and angle variables for nonlinear sigma models: the remaining components of the $\check{\Psi}$'s correspond to the radial variables, whereas $\omega$ corresponds to the angle variable.

As an example, we consider the O'Raiferteaigh model of sec. 8.3.a, with three chiral superfields $\Phi_i$, $i = 0, 1, 2$. Using (8.3.18), we find that the auxiliary field $F_0$ of the multiplet $\Phi_0$ gets a nonvanishing expectation value $F_0 = c$. To describe the system in terms of standard components, we first define homogeneously transforming superfields $\check{\Phi}_i$ by introducing the Volkov-Akulov field $\omega$ as an extra variable; then we restore the number of degrees of freedom by constraining $\check{\psi}_0 = 0$ (the fermionic component of $\check{\Phi}_0$) and eliminating $\psi_0$ (the fermionic component of $\Phi_0$) in favor of $\omega$. We can do this because, examining $\check{\Phi}_0$, we find $\check{\psi}_0 = \psi_0 - c\omega + \cdots$, where $c = < F_0 >$. If spontaneous symmetry breaking did not occur (i.e., if $c = 0$), we could still define homogeneous components, but we could not remove the extra degree of freedom (the change of variables from $\psi_0$ to $\omega$ would be singular). Having eliminated $\psi_0$ in favor of $\omega$ (the standard angle variable, which transforms as (8.5.4)), we can proceed to express the remaining components of $\Phi_i$ in terms of $\omega$ and the components of the homogeneous superfields $\check{\Phi}_i$ (the standard radial variables, which transform as (8.5.10b)).

## 8.6. SuperHiggs mechanism

When supersymmetry breaks in a system coupled to supergravity, a *superHiggs* mechanism eliminates the Goldstino and gives mass to the gravitino (the Goldstino becomes its longitudinal component). To examine the superHiggs mechanism in detail, we study the locally supersymmetric analog of the constrained superfields of the previous section (8.5.9).

The basic ingredient of the superHiggs mechanism is the transformation law of the Goldstino, $\delta\eta_\alpha = c\epsilon_\alpha + \cdots$ (see previous section); when the Goldstino is coupled to supergravity, the supersymmetry parameter $\epsilon$ becomes local:

$$\delta\eta_\alpha = c\epsilon_\alpha(x) + \cdots \tag{8.6.1}$$

Consequently, the Goldstino can be completely gauged away; since the number of dynamical modes of the theory should not change, we expect the Goldstino to re-emerge somehow, and it does so by giving the gravitino a mass and becoming its longitudinal mode. To see this directly, we describe the Goldstino by a local constrained chiral field obeying the local superspace version of (8.5.9a-b). Thus we take $\Phi^2 = 0$, $\Phi = c^{-1}\Phi(\overline{\nabla}^2 + R)\overline{\Phi}$ (here we consider only minimal ($n = -\frac{1}{3}$) supergravity). These constraints have a consistent solution in terms of a single fermi component field (the Goldstino). Because of the constraints on $\Phi$, any locally supersymmetric action (without explicit derivatives, see (5.5.15)) reduces to

$$S_{higgs} = \kappa^{-2}\int d^4x\, d^2\theta\ \phi^3(\lambda + \mu\Phi) + h.\,c. \tag{8.6.2}$$

The constrained superfield $\Phi$ is a nonlinear function of the Goldstino; however, when we gauge the Goldstino away (go to *U-gauge)* it simplifies to become $\Phi = -\,\theta^2 c$ (alternatively and equivalently, we can eliminate the Goldstino by a redefinition of the gravitino. In superspace, this corresponds to rescaling $\phi$ by $(1 + \frac{\mu}{\lambda}\Phi)^{-\frac{1}{3}}$; however, it is simpler to choose U-gauge). The action (8.6.2) becomes, using (5.6.60,64)

$$S_{higgs} = \kappa^{-2}\int d^4x\ \mathrm{e}^{-1}[\,\lambda(3\overline{S} + \frac{1}{2}\,\overline{\psi}_{\alpha(\dot{\alpha}|}{}^{\dot{\alpha}}\overline{\psi}^{\alpha}{}_{|\dot{\beta})}{}^{\dot{\beta}}) + \mu c + h.\,c.\,] \tag{8.6.3}$$

where $S$ is the complex scalar auxiliary field of the supergravity multiplet; we also have the $-3\kappa^{-2}|S|^2$ term from the supergravity action (5.6.63). Eliminating $S$ by its equation of motion, we find a cosmological constant and gravitino "mass" terms:

$$S_{higgs} = \kappa^{-2} \int d^4x \, \mathrm{e}^{-1} [\, \lambda \frac{1}{2} \overline{\psi}_{\alpha(\dot{\alpha}|}{}^{\dot{\alpha}} \overline{\psi}^{\alpha}{}_{|\dot{\beta})}{}^{\dot{\beta}} + h.\,c. + 3|\lambda|^2 + \mu(c + \overline{c}) \,] \tag{8.6.4}$$

As discussed in sec. 5.7, gravitino "mass" terms when accompanied by a cosmological constant do not in general mean that the gravitino is massive. However, if

$$\mu(c + \overline{c}) = \, - \, 3|\lambda|^2 \tag{8.6.5}$$

then the cosmological term cancels, and we can unambiguously identify the $\psi\psi$ terms as mass terms.

Since any spontaneously broken theory can be described in terms of standard variables, and in particular, the Goldstino can be described in terms of $\Phi$, in any spontaneously broken theory in which the cosmological constant vanishes the gravitino mass is $Re\lambda$ when the (superspace) kinetic term has the usual normalization $-3\kappa^{-2}$. It is found by setting all matter fields to their vacuum expectation values. More generally, when (8.6.5) is not satisfied, we can still find the apparent $\psi$ mass $Re\lambda$ and the cosmological constant from the transformation of the Goldstino and by comparing to the superspace action (8.6.2).

## 8.7. Supergravity and symmetry breaking

Supersymmetry breaking in a local context can be studied directly, using the component tools of section 5.6. We can determine conditions for supersymmetry breaking and derive a mass formula analogous to (8.3.35). However, it is much more efficient to recast the problem as a *global* supersymmetry problem that can be studied using the techniques of secs. 8.3b-c and 8.4. We consider a general system of interacting scalar and vector multiplets coupled to $N = 1$ supergravity. The matter multiplets are described by chiral and (real) gauge scalar superfields $\Phi^i$, $V^{\mathrm{A}}$, respectively; the supergravity multiplet is described by the real axial-vector superfield $H^{\underline{m}}$ and (for $n = -\frac{1}{3}$) the chiral compensator $\phi$. However, $H^{\underline{m}}$ plays no direct role in the supersymmetry breaking mechanism or in the derivation of mass formulae. Therefore all the relevant information can be extracted from a global *nonrenormalizable* system described by $\phi$, $\Phi^i$, and $V^{\mathrm{A}}$.

We begin by reducing the coupled matter-supergravity system. The axial-vector real gauge superfield of supergravity $H^{\underline{m}}$ contains the graviton and gravitino physical degrees of freedom, as well as the axial vector auxiliary field $A^{\underline{m}}$ (5.2.8). In the presence of the compensator $\phi$ the supergravity gauge group consists of the full superconformal group, and we have at our disposal all of the component gauge transformations of (5.2.10). Consequently, we can go to the Wess-Zumino gauge discussed after (5.2.10), and further, use the remaining superconformal transformations to remove the graviton trace, the gravitino $\gamma$-trace, and the longitudinal part of the axial vector auxiliary field. In this gauge $H^{\underline{m}}$ contains only the traceless components of the graviton and the gravitino, and the transverse part of $A^{\underline{m}}$; the spin zero complex auxiliary field $S$, the $\gamma$-trace of the (left-handed) gravitino $(\gamma \cdot \psi)_L = \bar{\psi}^{\dot{\alpha}}{}_{,\alpha\dot{\alpha}}$, the trace of the vierbein, or equivalently, its determinant $e = det\, e_{\underline{a}}{}^{\underline{m}}$ and $\square^{-1}\partial_{\underline{m}} A^{\underline{m}}$ are contained in $\phi$ (these last two are the real and imaginary parts of the $\theta$-independent component of $\phi$). Since only these quantities are relevant for studying spontaneous supersymmetry breaking (e.g., the spin zero bosons can get vacuum expectation values and the $\gamma$-trace can mix with the matter fermions), we can ignore the $H^{\underline{m}}$ dependent terms in the Lagrangian and work entirely with $\phi$ and the matter superfields in a global setting. This simplifies the discussion enormously; however, because $Im\phi|$ replaces the divergence of $A^{\underline{m}}$, there are some subtleties associated with its contribution to the masses of the matter fields (see subsec.

8.7.a.4). We treat only $n = -\frac{1}{3}$ supergravity; analogous methods can be used for other $n$, but since $n = -\frac{1}{3}$ allows the most general coupling, it is the most interesting case.

The supergravity multiplet itself (through $\phi$) affects the pattern of symmetry breaking. At first sight, this seems strange: In the usual Higgs mechanism, we do not expect the pattern of symmetry breaking to depend on the couplings to gauge fields (at the tree level!). However, an analogous situation arises in a nonsupersymmetric context, when scalar fields are coupled to gravity. We consider the action

$$S = \int d^4x \ \sqrt{g}[-3\kappa^{-2}r(g) - \frac{1}{4} g^{\underline{mn}} G_{ij}(A)\partial_{\underline{m}} A^i \partial_{\underline{n}} A^j + rV_1(A) + V_2(A)] \quad . \tag{8.7.1}$$

To find the vacuum expectation values of the scalar fields we cannot ignore the gravitational field. In general the Ricci scalar $r$ will have a nonzero expectation value that affects the masses and scalar potential. However, we need not consider the full Einstein system; it is sufficient to look for solutions of the form $g_{\underline{mn}} = \sigma^2 \eta_{\underline{mn}}$ and to treat the system of scalar fields $\sigma$, $A_i$ (subject to the condition $\sigma \neq 0$ at all points). (This is analogous to keeping $\phi$ and ignoring $H^{\underline{m}}$.) Two possible situations can arise: If $< V_2 > \neq 0$ we have a nonzero cosmological constant, $\sigma^{-1} - 1 \sim < V_2 >^{\frac{1}{2}} x^2$, and the vacuum values and the masses of $A_i$ are shifted from their flat space values. If $< V_2 > = 0$, the cosmological constant vanishes and a consistent solution is $\sigma =$ constant. In this case the gravitational field does not modify flat space results. (This is not the case in supergravity: Even if the cosmological constant vanishes, the supergravity auxiliary fields modify global results.)

Returning to the matter-supergravity system, we consider the action (5.5.32)

$$S = \int d^4x \, d^4\theta \ E^{-1}(\phi, H) \, \{ -\frac{3}{\kappa^2} \, e^{-\frac{1}{3}\kappa^2(\nu trV \, + \, G)}$$

$$+ [\frac{1}{R} (g + \frac{1}{4} Q_{\mathbf{AB}} W^{\alpha\mathbf{A}}_{cov} W^{\mathbf{B}}_{\alpha cov}) + h.\,c.]\} \quad , \tag{8.7.2}$$

where $E^{-1}$ is the superdeterminant of the vielbein and $R$ is the scalar curvature superfield (see e.g., (5.2.74-6)). The supergravity action is given by the first term in the expansion of the exponential. Here $G(\Phi^i, \tilde{\Phi}_j)$ is an arbitrary gauge invariant function of chiral superfields, with $\tilde{\Phi}$ defined by (8.3.26), $g(\Phi^i)$ is a chiral function, and $W^{\alpha\mathbf{A}}_{cov}$ is the (supergravity covariant) Yang-Mills field strength. The function $G$ has a natural interpretation as a Kähler potential with gauge transformations $G \to G + \Lambda(\Phi^i) + \overline{\Lambda}(\overline{\Phi}_j)$ compensated by scalings of $\phi$: $\phi \to exp[\frac{1}{3}\kappa^2\Lambda(\Phi)]\phi$.

The $exp[-\frac{1}{3}\kappa^2\nu trV]$ factor is the local form of the Fayet-Iliopoulos term (4.3.3). It is gauge invariant by virtue of a combined gauge transformation of $V$ and superscale transformations of $E^{-1}$ (5.3.8-10). Its presence severely restricts the form of the $g$ terms; they must be $R$-invariant (see (3.6.14) and (4.1.15)) so that the whole action is invariant under the superscale transformations of $E^{-1}$ (see below). In the $\kappa \to 0$ limit the action (8.7.2) becomes (8.3.25), with the identification $G{\sim}I\!\!K$, $g{\sim}P$.

As discussed above, we can split off the terms independent of $H^{\underline{m}}$. Furthermore, according to the discussion following (5.5.28) the $\phi$ dependence of $W_{cov}$ can be factored out ($W_{cov} \to \phi^{-\frac{3}{2}}W$ ) so that the relevant part of (8.7.2) becomes

$$S = \int d^4x\, d^4\theta\, [-\overline{\phi}\phi e^{-(\nu trV + G)}]$$

$$+ \int d^4x\, d^2\theta\, [\phi^3 g + \frac{1}{4}Q_{\mathrm{AB}}W^{\alpha\mathrm{A}}W_\alpha{}^{\mathrm{B}}] + h.\,c. \qquad (8.7.3)$$

We have set the gravitational constant $\frac{1}{3}\kappa^2 = 1$. We will restore it when necessary.

Under the gauge transformation $trV \to tr[V + i(\overline{\Lambda} - \Lambda)]$, $G \to G$, $(\overline{D}_{\dot\alpha}\Lambda = 0)$, the action is invariant if we rescale $\phi \to \phi e^{-i\nu tr\Lambda}$. Thus the local Fayet-Iliopoulos term acts as a conventional gauge term for $\overline{\phi}\phi$. If $\nu \neq 0$, as noted above the form of $g(\Phi^i)$ is extremely restricted: $\phi^3 g$ must be gauge invariant.

We now analyze the global system (8.7.3) subject to the condition that the cosmological constant vanishes. We can then choose $Re < \phi > | = \mu =$ constant. With the identification

$$-\overline{\phi}e^{-\nu trV}\phi\, e^{-G(\Phi^i,\tilde{\Phi}_j)} = I\!\!K$$

$$\phi^3 g(\Phi^i) = \mathrm{P} \qquad (8.7.4)$$

we have the action of (8.3.25) without a *global* Fayet-Iliopoulos term. We label components of $\phi$ as

$$\phi| = \mu A \quad , \quad D_\alpha \phi| = \mu \psi_\alpha \quad , \quad D^2 \phi| = \mu S \tag{8.7.5}$$

with expectation values

$$< A > = 1 \quad , \quad < S > = s \quad . \tag{8.7.6}$$

## a. Mass matrices

We begin by explicitly computing the mass matrices for the various fields in the system. This calculation is a little lengthy, so we will simplify it as much as possible without loss of generality. Thus, we rescale $\phi$ to remove $g$ from the chiral part of the action: $\phi^3 g \to \phi^3$. We also redefine $G$: $G \to G + \frac{1}{3} ln(g\bar{g}e^{3\nu trV})$. This makes $\phi$ inert under gauge transformations, and absorbs the Fayet-Iliopoulos term into $G$. In the case when $< g > = 0$ we cannot perform this rescaling. However, this case is not interesting, since then supersymmetry is not broken even in the presence of a Fayet-Iliopoulos term (if the cosmological constant vanishes).

### a.1. Vacuum conditions

The superfield equations (8.3.30) for the action (8.7.3) are:

$$\bar{D}^2(\bar{\phi}e^{-G}) - 3\phi^2 = 0 \tag{8.7.7a}$$

$$\phi\bar{\phi}e^{-G}\overline{\nabla}^2 G_i + 3\phi^3 G_i + \frac{1}{4} Q_{\mathrm{AB},i} W^{\alpha\mathrm{A}} W_\alpha{}^{\mathrm{B}} = 0 \tag{8.7.7b}$$

$$-\phi\bar{\phi}e^{-G} k_{\mathrm{A}i} G^i + \frac{1}{2}\nabla^\alpha(Q_{\mathrm{AB}} W_\alpha{}^{\mathrm{B}}) - \frac{1}{2}\overline{\nabla}^{\dot\alpha}(\bar{Q}_{\mathrm{AB}} W_{\dot\alpha}{}^{\mathrm{B}}) = 0 \tag{8.7.7c}$$

where we have used (8.7.7a) to simplify (8.7.7b), and thrown away some terms that lead to higher order spinor and/or derivative interactions that do not enter below. We have written (8.7.7) in terms of Killing vectors by making the substitutions $(T_{\mathrm{A}})^i{}_j \Phi^j \to -ik_{\mathrm{A}}{}^i$, $\tilde{\Phi}_j (T_{\mathrm{A}})^j{}_i \to ik_{\mathrm{A}i}$. The vacuum conditions (8.3.31) found by applying spinor derivatives to (8.7.7) become

$$\bar{s} - 3\mu e^G - G^i \bar{f}_i = 0 \quad ,$$

$$G_i{}^j \bar{f}_j + 3\mu e^G G_i = 0 \quad ,$$

$$6s\mu^3 = (Q_{AB} + \overline{Q}_{AB})d^A d^B \quad ,$$

$$\mu^2 e^{-G} G^i (i\, k_{Ai}) + (Q_{AB} + \overline{Q}_{AB})d^B = 0 \quad ,$$

$$3\mu e^G G_{ij} f^j + G_{ij}{}^k \overline{f}_k f^j + G_i{}^j (i\, k_{Aj}) d^A$$

$$+ 9\mu G_i e^G (s - \mu e^G) + \frac{1}{2}\mu^{-2} e^G Q_{AB,i} d^A d^B = 0 \quad . \tag{8.7.8}$$

The assumption that the cosmological constant vanishes is equivalent to the condition that these equations have a solution for constant $\mu$. We also have the gauge invariance conditions (8.3.c13) (these hold for general values of the fields, not just at the vacuum point):

$$G^i k_{Ai} + G_i k_A{}^i = 0 \quad ,$$

$$-i\, k_C{}^i Q_{AB,i} + (T_C)_B{}^D Q_{DA} + (T_C)_A{}^D Q_{DB} = 0 \quad . \tag{8.7.9}$$

To give the gravitational action the correct normalization (5.2.72), we identify

$$\kappa^2 = 3\mu^{-2} e^G \quad . \tag{8.7.10}$$

## a.2. Gravitino mass

As discussed in sec. 8.6, we can find the spin $\frac{3}{2}$ mass by setting all *matter* fields to their vacuum expectation values, and comparing the coefficients of the $\int d^4x\, d^4\theta\, \phi\overline{\phi}$ and $\int d^4x\, d^2\theta\, \phi^3$ terms. From (8.7.3), the kinetic term has a coefficient $-\mu^2 e^{-G} = -\frac{1}{3}\kappa^2$, and the chiral term has a coefficient $\mu^3$ (the factors of $\mu$ come from the definitions of the dynamical fields (8.7.5)); hence, using (8.6.2), we find that the spin $\frac{3}{2}$ mass is

$$m = 3\mu e^G \tag{8.7.11}$$

We simplify our computation further by choosing normal coordinates $G_i{}^j = \delta_i{}^j$, $G_i{}^{j_1\cdots} = G^i{}_{j_1\cdots} = 0$. Using (8.7.10,11), and normal coordinates, we rewrite the vacuum conditions (8.7.8) as

$$\overline{s} - m - G^i \overline{f}_i = 0$$

$$\bar{f}_i + mG_i = 0$$

$$6m\kappa^{-2}s - (Q + \bar{Q})_{AB}\, d^A d^B = 0$$

$$3i\kappa^{-2}k_{Ai}G^i + (Q + \bar{Q})_{AB}\, d^B = 0$$

$$mG_{ij}f^j + i\, k_{Ai}d^A + mG_i(3s - m) + \frac{1}{6}\kappa^2 Q_{AB,i}d^A d^B = 0 \ . \tag{8.7.12}$$

### a.3. Wave equations

We now find the linearized wave equations (8.3.36) that follow from (8.7.7). As in sec. 8.3, we expand the fields in small fluctuations about their vacuum values. For the remainder of this subsection, all quantities $G_{\dots}{}^{\dots}$ and $Q_{AB\dots}$ are evaluated at $\Phi^i = a^i$ and $\tilde{\bar{\Phi}}_i = \bar{a}_i$. We find it useful to introduce shifted variables

$$A' \equiv A - G_i A^i \ \ , \ \ \psi' \equiv \psi - G_i\psi^i \ \ , \ \ S' \equiv S - G_i F^i \ \ . \tag{8.7.13}$$

From (8.7.7a) we have

$$S' - A'(s - m) - 2m\bar{A}' - 2mG^i\tilde{A}_i - (G_{ij}f^j + sG_i)A^i = 0 \tag{8.7.14a}$$

$$i\partial_\alpha{}^{\dot{\alpha}}\bar{\psi}'_{\dot{\alpha}} - k_{Ai}G^i\lambda^A{}_\alpha - 2m\psi'_\alpha - 2mG_i\psi_\alpha{}^i = 0 \ . \tag{8.7.14b}$$

From (8.7.7b), we find

$$F^i + m[(2\bar{A}' - A')G^i + (3G^iG^j + G^{ij})\tilde{A}_j + A^i] = 0 \tag{8.7.14c}$$

$$i\partial_\alpha{}^{\dot{\alpha}}\tilde{\psi}_{\dot{\alpha}i} + 2mG_i\psi'_\alpha + m(3G_iG_j + G_{ij})\psi_\alpha{}^j$$
$$+ k_{Ai}\lambda^A{}_\alpha - \frac{i}{6}\kappa^2 Q_{AB,i}d^A\lambda^B{}_\alpha = 0 \ . \tag{8.7.14d}$$

From (8.7.7c), we find

$$(Q + \bar{Q})_{AB}\,\mathrm{D}'^B + Q_{AB,i}d^B A^i + \bar{Q}_{AB}{}^{,i}d^B\tilde{A}_i$$
$$+ 3i\kappa^{-2}[k_{Ai}A^i - k_A{}^i\tilde{A}_i + k_{Ai}G^i(A' + \bar{A}')] = 0 \tag{8.7.14e}$$

$$\frac{1}{2}(Q+\bar{Q})_{AB}\, i\partial_\alpha{}^{\dot\alpha}\bar{\lambda}^B{}_{\dot\alpha} + \frac{1}{2}Q_{AB,i}(f^i\lambda^B{}_\alpha - i\,d^B\psi_\alpha{}^i)$$

$$+\,3\kappa^{-2}k_{Ai}(\psi_\alpha{}^i + G^i\psi'_\alpha) = 0 \quad . \tag{8.7.14f}$$

We are left with the equations of the physical boson fields. These simplify greatly if we use (8.7.14a,c,e); we find

$$\Box A' = 0 \tag{8.7.14g}$$

$$\Box\tilde{A}_i + \big\{3\kappa^{-2}(Q+\bar{Q})^{-1AB}(k_{Ai} - \frac{1}{3}i\kappa^2 Q_{AC,i}d^C)(k_B{}^j + \frac{1}{3}i\kappa^2\bar{Q}_{BE}{}^{,j}d^E)$$

$$+\,ik_{Ai}{}^{,j}d^A - m^2(G_{ik}G^{kj} + 3G_iG_kG^{kj} + 3G_{ik}G^kG^j - G_{ik}{}^{jl}G^kG_l)$$

$$+\,m[(3s-m)\delta_i{}^j + 3(3s-2m)G_iG^j]\big\}\tilde{A}_j$$

$$+\,\big\{3\kappa^{-2}(Q+\bar{Q})^{-1AB}(k_{Ai} - \frac{1}{3}i\kappa^2 Q_{AC,i}d^C)(k_{Bj} + \frac{1}{3}i\kappa^2 Q_{BE,j}d^E)$$

$$+\,\frac{1}{6}\kappa^2 Q_{AB,ij}d^Ad^B - m^2(3G_{ik}G^kG_j + 3G_iG_{jk}G^k + G_{ijk}G^k)$$

$$+\,m(3s-2m)(G_{ij} + 3G_iG_j)\big\}A^j = 0 \tag{8.7.14h}$$

$$(Q+\bar{Q})_{AB}\nabla^{\alpha\dot\alpha}f^B{}_{\alpha\beta} - 3\kappa^{-2}(k_{Ai}k_B{}^i + k_{Bi}k_A{}^i)A^B{}_\beta{}^{\dot\alpha}$$

$$+\,3\kappa^{-2}k_{Ai}G^iX_\beta{}^{\dot\alpha} = 0 \tag{8.7.14i}$$

where $X_{\alpha\dot\alpha} = \partial_{\alpha\dot\alpha}ImA - ImG_j\nabla_{\alpha\dot\alpha}A^j \;=\; \partial_{\alpha\dot\alpha}ImA' + (G_jk_B{}^j - G^jk_{Bj})A^B{}_{\alpha\dot\alpha}$. Here $\nabla_{\alpha\dot\alpha}$ is the Yang-Mills covariant derivative.

### a.4. Bose masses

We now discuss these results. From (8.7.14g) we see that the complex scalar $A'$ is massless. For the real part, this is no surprise: $ReA'$ is the trace of the graviton, which is massless because the cosmological term was assumed to vanish. However, the imaginary part requires some care. The pseudoscalar $Im\,A \equiv \rho$ is not recognizable as one of the fields of the supergravity multiplet; it stands for $\Box^{-1}\partial\cdot A$ where $A_{\alpha\dot\alpha}$ is the divergence of the axial vector auxiliary field. Thus the equation $\Box Im\, A' = \Box Im\, A - \Box Im\, G_i A^i = 0$ should be replaced by

$$A_{\alpha\dot\alpha} - Im\, G_i \nabla_{\alpha\dot\alpha} A^i = 0 \qquad (8.7.15)$$

For many purposes, it makes little difference whether we use $ImA$ or replace it with $\Box^{-1}\partial \cdot A$. By dimensional analysis and Lorentz invariance, $A_{\alpha\dot\alpha}$ can enter the wave equation of the scalar fields $\tilde{A}_i$ only through its divergence:

$$\Box \tilde{A}_i + cG_i \partial^{\alpha\dot\alpha} A_{\alpha\dot\alpha} + \cdots = 0 \quad . \qquad (8.7.16)$$

Substituting in (8.7.15), we find

$$\Box(\tilde{A}_i + cG_i ImG_j A^j) + \cdots = 0 \quad . \qquad (8.7.17)$$

When we have $A$ instead of $A_{\alpha\dot\alpha}$, we get the same result, since instead of (8.7.16) we have

$$\Box \tilde{A}_i + cG_i \Box ImA + \cdots = 0 \quad , \qquad (8.7.18)$$

and using the $A$ wave equation, we reobtain (8.7.17).

However, if gauge invariance is broken the gauge field wave equation can get a spurious contribution from $ImA$ that is not present when $\Box^{-1}\partial \cdot A$ is used instead. Indeed, substituting (8.7.15) into the first form of $X_{\alpha\dot\alpha}$ (with $\partial_{\alpha\dot\alpha} ImA$ replaced by $A_{\alpha\dot\alpha}$) gives a zero contribution to the spin 1 mass. When gauge invariance is unbroken, we get no contribution from the form with $A$ as well: $\partial_{\alpha\dot\alpha} ImA'$ does not affect the spin 1 mass, and the vacuum expectation value of $k_{\mathrm{B}}{}^j$ is zero. However, if gauge invariance is broken, the expectation value of $k_{\mathrm{B}}{}^j$ is not zero (equivalently, $\partial_{\alpha\dot\alpha}(G_i A^i) \neq G_i \nabla_{\alpha\dot\alpha} A^i$) and $X$ gives a spurious contribution that must be removed by hand.

### a.5. Fermi masses

The component $\psi'$ corresponds to the $\gamma$-trace of the gravitino; we define the Goldstino as that combination of matter fields that couples to $\psi'$ (it makes no essential difference whether we use $\psi$ or $\psi'$, since we are always free to add terms to the gravitino). Thus we define

$$\eta^\alpha \equiv G_i \psi^{\alpha i} + \frac{1}{2m} G^i k_{\mathrm{A}i} \lambda^{\mathrm{A}\alpha} \quad . \qquad (8.7.19)$$

We also define "transverse" fields that are orthogonal to $\eta$:

$$\psi^{iT} \equiv \psi^i - G^i \eta$$

$$\lambda^{\mathrm{A}T} \equiv \lambda^{\mathrm{A}} + \frac{i}{m} d^{\mathrm{A}} \eta \quad . \tag{8.7.20}$$

(These satisfy $G_i \psi^{iT} + \frac{1}{2m} G^i k_{\mathrm{A}i} \lambda^{\mathrm{A}T} = 0$.) In terms of these, the spinor wave equations become:

$$i\partial_\alpha{}^{\dot\alpha} \bar\psi'_{\dot\alpha} - 2m(\psi'_\alpha + \eta_\alpha) = 0 \tag{8.7.21a}$$

$$i\partial_\alpha{}^{\dot\alpha} \bar\eta_{\dot\alpha} + 2m(\psi'_\alpha + \eta_\alpha) = 0 \tag{8.7.21b}$$

$$i\nabla_\alpha{}^{\dot\alpha} \tilde{\bar\psi}_{\dot\alpha i}{}^T + m(G_{ij} + G_i G_j) \psi_\alpha{}^{jT}$$

$$+ (k_{\mathrm{A}i} - k_{\mathrm{A}j} G^j G_i - \frac{i}{6} \kappa^2 Q_{\mathrm{AB},i} d^{\mathrm{B}}) \lambda^{\mathrm{A}}{}_\alpha{}^T = 0 \tag{8.7.21c}$$

$$i\nabla_\alpha{}^{\dot\alpha} \bar\lambda^{\mathrm{B}}{}_{\dot\alpha}{}^T + [6\kappa^{-2}(Q + \bar Q)^{-1\mathrm{BA}} (k_{\mathrm{A}j} - \frac{i}{6}\kappa^2 Q_{\mathrm{AC},j} d^{\mathrm{C}}) - 2i\, d^{\mathrm{B}}] \psi_\alpha{}^{jT}$$

$$- [m(Q + \bar Q)^{-1\mathrm{BC}} Q_{\mathrm{CA},i} G^i + \frac{i}{m} d^{\mathrm{B}} k_{\mathrm{A}l} G^l] \lambda^{\mathrm{A}}{}_\alpha{}^T = 0 \quad . \tag{8.7.21d}$$

Care must be taken to ensure that the mass operator on $\psi^T, \lambda^T$ is restricted to the "transverse" subspace, i.e., preserves the orthogonality to $\eta$.

Observe that since the trace of the gravitino is a *negative norm* state, i.e., a ghost, its kinetic term has a minus sign relative to physical spinors (the same is true for the trace of the graviton; the whole $\phi$ multiplet has negative norm, as can be seen from the action (8.7.3)). Consequently, though the mass matrix in the $\psi$-$\eta$ system (which is decoupled from the other spinors) does not vanish, both eigenvalues are zero (the mass matrix is not hermitian). Actually, we did not have to explicitly find the wave equation to arrive at this result: The condition that the Goldstino can be gauged away (that we can go to a U-gauge) implies that both the Goldstino $\eta$ and the $\gamma$-trace of the gravitino $\psi'$ must have zero mass in the gauge that we use.

### a.6. Supertrace

Having found the wave equations (8.7.14g-i,a8), and understood their significance, we can evaluate the supertrace. The spin 0 contribution is (recall that we are still in normal coordinates):

$$- 2\big[ik_{\text{A}i}{}^{;i}d^{\text{A}} - 3\kappa^{-2}(Q+\bar{Q})^{-1\text{AB}}(k_{\text{A}i} - \frac{1}{3}i\kappa^2 Q_{\text{AC},i}d^{\text{C}})(k_{\text{B}}{}^i + \frac{1}{3}i\kappa^2\bar{Q}_{\text{BE}}{}^{,i}d^{\text{E}})$$

$$- m^2(G_{ij}G^{ij} + 3G_iG_jG^{ij} + 3G_{ij}G^iG^j - G_{ik}{}^{ij}G^kG_j)$$

$$+ m(3s-m)\text{N} + 3(3s-2m)(m-s)\big] \qquad (8.7.22\text{a})$$

where $\text{N} \equiv \delta_i{}^i$ is the number of chiral multiplets. The combined contribution of the spin 0 and spin $\frac{1}{2}$ fields is:

$$- 2\big[9\kappa^{-2}(Q+\bar{Q})^{-1\text{AB}}k_{\text{A}i}k_{\text{B}}{}^i + i(Q+\bar{Q})^{-1\text{AB}}d^{\text{C}}(k_{\text{A}i}\bar{Q}_{\text{BC}}{}^{,i} - k_{\text{A}}{}^iQ_{\text{BC},i})$$

$$+ ik_{\text{A}i}{}^{;i}d^{\text{A}} + G_{ij}{}^{ik}\bar{f}_kf^j - (N+1)m^2 + (N-1)3ms$$

$$+ tr(\frac{1}{Q+\bar{Q}}Q_i\frac{1}{Q+\bar{Q}}\bar{Q}^j)\bar{f}_jf^i\big] \quad . \qquad (8.7.22\text{b})$$

The spin 1 contribution, omitting the $X_{\alpha\dot{\alpha}}$ term is given by the expression $3\cdot 3\kappa^{-2}(Q+\bar{Q})^{-1\text{AB}}(k_{\text{A}i}k_{\text{B}}{}^i + k_{\text{B}i}k_{\text{A}}{}^i)$ and cancels the first term of (8.7.22b). (The normalization comes from the 3 states of a spin 1 particle and from the form (8.3.37) of the spin 1 wave equation.) Finally, the spin $\frac{3}{2}$ contribution is just $-4m^2$. Thus we get (using the gauge-invariance relations (8.7.9) to simplify some expressions)

$$str\, M^2 = -2\big[ik_{\text{A}i}{}^{;i}d^{\text{A}} + G_{ij}{}^{ik}\bar{f}_kf^j - (N-1)(m^2 - \frac{1}{2}\kappa^2(Q+\bar{Q})_{\text{AB}}d^{\text{A}}d^{\text{B}})$$

$$- i\,tr(\frac{1}{Q+\bar{Q}}\bar{Q}^i)k_{\text{A}i}d^{\text{A}} + tr(\frac{1}{Q+\bar{Q}}Q_i\frac{1}{Q+\bar{Q}}\bar{Q}^j)\bar{f}_jf^i\big] \ , \qquad (8.7.23)$$

in normal coordinates, or, in general, using coordinate invariance, we have

$$str\, M^2 = -2\big[ik_{\text{A}i}{}^{;i}d^{\text{A}} + R_i{}^j\bar{f}_jf^i - (N-1)(m^2 - \frac{1}{2}\kappa^2(Q+\bar{Q})_{\text{AB}}d^{\text{A}}d^{\text{B}})$$

$$- i\,tr(\frac{1}{Q+\bar{Q}}\bar{Q}^i)k_{\text{A}i}d^{\text{A}} + tr(\frac{1}{Q+\bar{Q}}Q_i\frac{1}{Q+\bar{Q}}\bar{Q}^j)\bar{f}_jf^i\big] \quad . \qquad (8.7.24)$$

We remind the reader that here

$$str\, M^2 \equiv \sum_{J=0}^{3/2} (-1)^{2J}(2J+1){M_J}^2 \quad . \tag{8.7.25}$$

**b. Superfield computation of the supertrace**

If our only interest is the supertrace formula (8.7.71), we can obtain it with far less work using the technique developed in sec. 8.4.b. (Of course, in general we are interested in the mass matrices themselves, and not just the supertrace). We start with

$$ln\, det(I\!K_i{}^j) = -\,(N+1)\,G + ln\, det(G_i{}^j) + N\, ln(-\phi e^{-\nu trV}\bar{\phi}) \tag{8.7.26}$$

where differentiation is with respect to $\tilde{\bar{\phi}} = \bar{\phi} e^{-\nu trV}$ and not $\bar{\phi}$.

Before adding contributions from the gravitino mass and correcting for the axial vector auxiliary field (see below), the supertrace read from (8.4.9) is

$$str\, M^2 = -\,2\big[ik_{\mathrm{A}i}{}^{;i} d^{\mathrm{A}} - (N+1)\nu\, tr\, d + R_k{}^l \bar{f}_l f^k$$

$$-\,(N+1)(G_i{}^j \bar{f}_j f^i + i\, G^i k_{\mathrm{A}i} d^{\mathrm{A}})$$

$$+\, tr(Q_k \frac{1}{Q+\bar{Q}} \bar{Q}^l \frac{1}{Q+\bar{Q}}) \bar{f}_l f^k - i\, tr(\bar{Q}^i \frac{1}{Q+\bar{Q}}) k_{\mathrm{A}i}\big] \quad . \tag{8.7.27}$$

where we use (4.1.29,30):

$$R_k{}^l = [ln\, det(G_i{}^j)]_k{}^l \quad , \quad \Gamma^l = [ln\, det(G_i{}^j)]^l \quad . \tag{8.7.28}$$

The expression (8.7.27) has not made use of the vacuum conditions (8.7.8) or (8.7.12), and does not include either the spin $\frac{3}{2}$ contribution or the axial vector auxiliary field correction to the spin 1 mass matrix discussed in subsec. 8.7.a.4. As we saw in the previous section, the spin $\frac{3}{2}$ contribution must be included separately, since the $\gamma$-trace of the gravitino cannot contribute directly: the condition for the superHiggs mechanism to occur and for the gravitino to absorb the Goldstino in U-gauge requires the Goldstino-gravitino $\gamma$-trace system to be massless. The spin 1 correction, though somewhat subtle, can also be found without extensive computation. As described in sec. 8.7.a.4, we simply subtract $\kappa^{-2}(Q+\bar{Q})^{-1\mathrm{AB}} k_{\mathrm{A}i} G^i (k_{\mathrm{B}j} G^j - k_{\mathrm{B}}{}^j G_j) = -\,2\kappa^2 (Q+\bar{Q})_{\mathrm{AB}} d^{\mathrm{A}} d^{\mathrm{B}}$ (see discussion following (8.7.22b) for an explanation of the factors).

One further point deserves comment: When we rescaled $\phi$ to remove the potential $g$ (see the beginning of sec. 8.7.a), we lost sight of the contribution of the Fayet-Iliopoulos term. When we make the shift $G \rightarrow G + \frac{1}{3} ln(g\bar{g}e^{3\nu trV})$, the $\nu\, tr\, d$ term in equation (8.7.27) is absorbed into the $iG^i k_{Ai} d^A$ term as a consequence of R-invariance of $g$; it is most straightforward to work in the coordinate system where the Killing vectors take the form of usual gauge transformations:

$$3\,\nu\,\bar{g}\,tr(T_A) - \bar{g}^i (T_A)_i{}^j \bar{a}_j = 0 \tag{8.7.29}$$

and hence

$$\nu tr(T_A) - \frac{1}{3}\,[ln(g\bar{g}e^{3\nu V})]^i (T_A)_i{}^j \bar{a}_j = 0 \tag{8.7.30}$$

Using the vacuum equations, we can substitute into the supertrace (8.7.27). Including the gravitino and the spin 1 correction term, we recover (8.7.24).

### c. Examples

We can use the supertrace formulae to study many cases of interest. In particular, in extended supergravity theories we encounter "nonminimal" $G$ and $Q$ terms. For example, in $N = 4$ supergravity, which contains one physical chiral multiplet, three vector multiplets, three $(\frac{3}{2}, 1)$ multiplets and the supergravity multiplet,

$$G \sim \; - ln(1 - \Phi\bar{\Phi}) \quad , \quad Q \sim \frac{1 - \Phi}{1 + \Phi} \quad . \tag{8.7.31}$$

We cannot treat the actual $N = 4$ theory since a description of the interacting $(\frac{3}{2}, 1)$ multiplet is not available, but (8.7.31) suggests looking at a system with one scalar multiplet and $n$ vector multiplets $V^A$, coupled to $N = 1$ supergravity, with $G$ as above and

$$Q_{AB} = \frac{1 - \Phi}{1 + \Phi}\,\delta_{AB} \quad . \tag{8.7.32}$$

We find, with

$$G'' = \frac{\partial^2}{\partial\Phi\partial\bar{\Phi}}\, G = \frac{1}{(1 - a\bar{a})^2} \tag{8.7.33}$$

the supertrace

$$\sum_{J=0}^{3/2}(-1)^{2J}(2J+1){M_J}^2 = -2(n+2)G''f\bar{f} = -2(n+2)m^2 \ . \tag{8.7.34}$$

Note that the $Q$ and $R$ terms in (8.7.17) combine because $(Q+\bar{Q})^2 = -4(G'')^{-1}[(1+\Phi)(1+\bar{\Phi})]^{-2}$. Unless a scalar potential $g(\Phi)$ is introduced, no supersymmetry breaking will occur. However, it is possible to add such a term in $N=1$ supergravity, and there exist mechanisms to generate terms that act like a potential even in $N=4$ supergravity.

For $N>4$ the analogs of $G$ and $Q$ are expressed in terms of an overcomplete set of fields. We may expect however that $Q$ and $G$ are related such that $det({G_i}^j) \sim det(Q+\bar{Q})h(\Phi^i)\bar{h}(\bar{\Phi}_i)$ where $h(\Phi^i)$ is a holomorphic function. In that case we may also expect a simple result for the supertrace.

We can also construct models with a Fayet-Iliopoulos term and vanishing cosmological constant. For example, consider

$$G = \bar{\Phi}e^V\Phi + \frac{\alpha^2}{3}\, ln[\bar{\Phi}e^V\Phi] + \bar{\chi}\chi + \frac{1}{3}\, ln[(\beta+\chi)(\beta+\bar{\chi})] \tag{8.7.35}$$

where $\Phi$ and $\chi$ are chiral fields, $\Phi$ transforming under the gauge transformation while $\chi$ is inert, and $\beta$ is chosen so as to make the cosmological constant vanish (the potential and the Fayet-Iliopoulos term are included in $G$ as the $\alpha$-term). We find a solution to (8.7.8) with $d \neq 0$ for some finite range of $\alpha$ (as can be verified by a perturbation expansion about $\alpha = 0$).

# INDEX